\documentclass[preprint]{elsarticle}

\usepackage[all,cmtip]{xy}
\usepackage{framed}
\usepackage{hyperref}
\usepackage{amsmath,amssymb,bbm,amsthm}

\newcommand{\bR}{\mathbb{R}}

\newcommand{\ga}{\mathfrak{a}}  
\newcommand{\gb}{\mathfrak{b}}  
\newcommand{\gc}{\mathfrak{c}}

\newcommand{\Id}{{\mathrm{Id}}}

\newtheorem{defi}{Definition}
\newdefinition{rmk}{Remark}
\newtheorem{thm}{Theorem}

\renewcommand{\arraystretch}{1.3}

\newcommand{\ft}[2]{{\textstyle\frac{#1}{#2}}}

\newcommand{\bea}{\begin{eqnarray}}
\newcommand{\eea}{\end{eqnarray}}
\def\beq{\begin{equation}}
\def\eeq{\end{equation}}
\def\beqa{\begin{eqnarray}}
\def\eeqa{\end{eqnarray}}

\begin{document}

\title{Special Geometry, Hessian Structures and Applications}
\author[1]{Gabriel Lopes Cardoso}
\ead{gcardoso@math.tecnico.ulisboa.pt}
\author[2]{Thomas Mohaupt \corref{cor1}}
\ead{Thomas.Mohaupt@liv.ac.uk}
\address[1]{
Center for Mathematical Analysis, Geometry and Dynamical Systems,
  Department of Mathematics, 
  Instituto Superior T\'ecnico,
Universidade de Lisboa,
  Av. Rovisco Pais, 1049-001 Lisboa, Portugal
}
\address[2]{Department of Mathematical Sciences, University of Liverpool, Peach Street, 
Liverpool L69 7ZL, UK}

\cortext[cor1]{Corresponding Author}

\begin{abstract}
The target space geometry of abelian vector multiplets in ${\cal N}= 2$ theories
in four and five space-time dimensions is called special geometry. It can be elegantly
formulated in terms of Hessian geometry. In this review, we introduce
Hessian geometry, focussing on aspects that are relevant for the special geometries
of four- and five-dimensional vector multiplets. We formulate  ${\cal N}= 2$ theories 
in terms of Hessian structures and give various concrete applications
of Hessian geometry, ranging from static BPS black holes in four and five space-time dimensions
to topological string theory, emphasizing the role of the Hesse potential. We also discuss
the r-map and c-map which relate the
special geometries of vector multiplets to each other and to hypermultiplet geometries.
By including time-like dimensional reductions, we obtain theories in Euclidean signature,
where the scalar target spaces carry para-complex versions of special geometry.
\end{abstract}


\maketitle

\newpage

\tableofcontents

\newpage

\section{Introduction}

Theories with 8 supercharges hold an interesting position between semi-realistic, but 
analytically un-tractable theories with 4 supercharges, and theories with more than
8 supercharges, which are analytically tractable, but have a two-derivative Lagrangian
which is completely determined by their matter content. In contrast, the couplings 
of theories with 8 supercharges are functions of the scalar fields, and subject to 
interesting and complicated quantum corrections. We will refer to theories with
8 conserved real supercharges as  ${\cal N}=2$ theories, irrespective of the
space-time dimension. This amounts to counting supersymmetries 
in multiples of the minimal number of supercharges of a four-dimensional theory.

Vector multiplets in ${\cal N}=2$ theories 
contain gauge fields together with scalars and fermions. We will
restrict ourselves to abelian vector fields, in which case one can take linear combinations
of vector fields. By supersymmetry this imprints itself onto the scalars, leading to 
an affine structure and a scalar geometry which is `special.' In four dimensions, where
vector fields can couple to both electric and magnetic charges, linear transformations
of vector fields and {\em electric-magnetic duality transformations} combine to a {\em symplectic
group action} on the field strengths and their duals. By supersymmetry this imprints
itself on the scalars, which in four dimensional are complex-valued, and leads to a K\"ahler
geometry with `special features.' While in rigid supersymmetry the number of 
scalar fields and vector fields is balanced, the coupling to Poincar\'e supergravity 
creates a mismatch, because the Poincar\'e supergravity multiplet contributes an additional
vector field, the graviphoton. An elegant way to handle this is to employ the 
{\em gauge equivalence} between a theory of $n$ vector multiplets coupled to Poincar\'e supergravity 
and a theory of $n+1$ superconformal vector multiplets coupled to conformal 
supergravity (the Weyl multiplet) and one additional auxiliary supermultiplet (which we will take to be a hypermultiplet). In the superconformal theory there now is a balance between $n+1$ 
scalar fields and $n+1$ vector fields. The superconformal symmetry gives the scalar geometry an additional 
{\em conical structure}. When recovering the Poincar\'e supergravity theory by imposing 
gauge fixing conditions, one scalar is eliminated, which corresponds to taking the 
{\em superconformal quotient} of the superconformal scalar manifold by a group action. In this way the
scalar geometry of vector multiplets coupled to Poincar\'e supergravity can be understood
as the projectivisation of the scalar geometry of the associated superconformal theory. 

We will refer to the scalar geometries of five- and four-dimensional vector multiplets
as {\em special geometries}.
One characteristic feature of five- and four-dimensional vector multiplets is that all couplings
of the two-derivative Lagrangians are encoded in a single function, the {\em Hesse potential}. 
In particular, the metric of the scalar manifolds of rigid vector multiplets are Hessian metrics,
that is, the metric coefficients are the second derivatives of a real function, when written
in affine coordinates with respect to a flat torsion-free connection. While the scalar metrics of local vector 
multiplets are not Hessian themselves, they can still be expressed in terms of the Hesse 
potential of the associated superconformal theory. In four dimensions, one can alternatively 
express the couplings in terms of a holomorphic function, the {\em prepotential}. This is in fact
the pre-dominant point of view in the literature. In this review we will emphasize the role of the Hesse 
potential because (i) this makes manifest the similarities between five- and four-dimensional vector multiplets, (ii) the Hesse potential of four-dimensional vector multiplets transforms 
covariantly under symplectic transformations, while the prepotential does not. As a consequence,
using the Hesse potential has advantages in many applications. 
We will review {\em Hessian geometry} and {\em special real geometry} in section \ref{sec:Hessian_geometry},
{\em electric-magnetic duality} in section \ref{sec:emduality}, and {\em special K\"ahler geometry} in section \ref{sec:secSKG}. Based on this we 
discuss {\em five-dimensional vector multiplets} in section \ref{sec:5d_VM} and
{\em four-dimensional vector multiplets} in section \ref{4dvecmult}.

In Table \ref{Table:special_geometries} we list the acronyms and defining data
of the types of special geometries relevant for five- and four-dimensional vector multiplets.
One recurrent theme is that for each type of special geometry there is an affine,
a conical and a projective version, which schematically are related like this:
\begin{equation}
\xymatrix{
\mbox{Affine} \ar[rr]^{+\mathrm{Homothety}}&& \mbox{Conical} 
\ar@<0.5ex>[rr]^{\mathrm{Quotient}}
&& \mbox{Projective}  \ar@<0.5ex>[ll]^{\mathrm{Cone}}\\
}
\end{equation}
This is meant to indicate that the conical type is a special form of the affine type, 
which is characterized by the presence of a homothetic Killing vector
field satisfying certain compatibility conditions. The projective version is obtained 
by taking the quotient of the conical version by a group action, which contains
the action generated by the homothetic Killing vector field. Conversely, the 
conical type of the geometry is realized as a cone which has the projective geometry
as its base. While the affine version corresponds to rigid vector multiplets, the conical 
version corresponds to superconformal vector multiplets, and the projective
version corresponds to vector multiplets coupled to Poincar\'e supergravity. 
The relation between conical and projective geometry reflects the gauge equivalence
between conformal supergravity and Poincar\'e supergravity.
Five- and four-dimensional vector multiplets realize a real and a complex
version of this scheme with group actions of $\mathbb{R}^{>0}$ and 
of $\mathbb{C}^*$ 
by real and by complex scale transformations, respectively. 
If we include hypermultiplets, there is as well a quaternionic version of this scheme. 
Hypermultiplets can be obtained by reduction of four-dimensional vector multiplets to
three dimensions, followed by the dualization of the three-dimensional vector fields
into scalars. Since hypermultiplets only contain scalars and fermions, their 
scalar geometry does not change under dimensional reduction, and is of the same
type in any dimension where hypermultiplets can be defined. The upper limit is
$d=6$, which is the largest dimension where a supersymmetry algebra with 8
real supercharges can be constructed. The scalar geometries of hypermultiplets are
quaternionic geometries, more precisely they are hyper-K\"ahler for rigid
hypermultiplets, hyper-K\"ahler cones (or, conical hyper-K\"ahler) for superconformal
hypermultiplets, and quaternionic K\"ahler (or, quaternion-K\"ahler) for hypermultiplets
coupled to Poincar\'e supergravity. While we will focus on vector multiplets in this review,
we will talk about hypermultiplets in the context of dimensional reduction, and regard
their scalar geometries as the quaternionic versions of special geometry. 
The real, complex and quaternionic versions of special geometry 
are related by dimensional reduction, which induce maps called the r-map
and the c-map between the scalar geometries. This is summarized in Table \ref{Table:r+c-maps}.

\begin{table}[t]
\begin{tabular}{|l|l|}\hline
ASR = affine special real & $(M,g,\nabla)$ \\ \hline
CASR = conical affine special real & $(M,g,\nabla,\xi)$ \\ \hline
PSR = projective special real & $(\bar{M}, \bar{g})\;, \bar{M} =  M / \mathbb{R}^{>0} \cong {\cal H}
\xrightarrow{\iota} M$  \\
 & 
$\bar{g} = 
\iota^*{g} $\\ \hline
ASK = affine special K\"ahler & $(M,J,g,\nabla)$ \\ \hline
CASR = conical affine special K\"ahler & $(M,J,g,\nabla,\xi)$ \\ \hline
PSK = projective special K\"ahler & $(\bar{M}, \bar{J}, \bar{g})\;, \bar{M} = M / \mathbb{C}^{*} = M//U(1) $ \\
&  
$\pi^* \bar{g} = \iota^*{g} \;,\;{\cal H} \xrightarrow{\pi} {\bar M} = {\cal H} / U(1) $ \\ 
 & ${\cal H} \xrightarrow{\iota} M$
\\ \hline
\end{tabular}
\caption{This table summarizes the acronyms we use for the various special geometries. The 
second column contains the essential geometrical data for each type of geometry. $\nabla$ indicates
a `special' connection, which in particular is flat and torsion-free. $\xi$ indicates a homothetic Killing vector field which gives the manifold locally the structure of a cone. A `bar' indicates a `projectivized' 
manifold which has been obtained by taking the orbits of a group action, which always includes the
homothetic Killing vector field $\xi$. As usual $\pi$ and $\iota$ indicate projections and immersions, respectively, and $*$ a pull-back. 
We refer
to the corresponding sections of this review for precise definitions.  
\label{Table:special_geometries}
}
\end{table}

\begin{table}[t]
\xymatrix{
\mbox{Five-dimensional vector multiplets:} &
\mbox{ASR} \ar[d]^r \ar[r]^{\mathrm{conical}} 
& \mbox{CASR}  \ar@<0.5ex>[r]^{\mathrm{Quotient}}
& \mbox{PSR} \ar[d]^{\bar{r}}  \ar@<0.5ex>[l]^{\mathrm{Cone}}\\
\mbox{Four-dimensional vector multiplets:} &
\mbox{ASK}  \ar[r]^{\mathrm{conical}} \ar[d]^{c}
& \mbox{CASK}  \ar@<0.5ex>[r]^{\mathrm{Quotient}}
& \mbox{PSK}  \ar@<0.5ex>[l]^{\mathrm{Cone}} \ar[d]^{\bar{c}}\\
\mbox{Hypermultiplets:}&
\mbox{HK}  \ar[r]^{\mathrm{conical}}
& \mbox{HKC}  \ar@<0.5ex>[r]^{\mathrm{Quotient}}
& \mbox{QK}  \ar@<0.5ex>[l]^{\mathrm{Cone}}\\
}
\caption{The real, complex and quaternionic versions of special geometry are related
by the r-map and c-map, which are induced by dimensional reduction. A `bar' indicates
the supergravity version of either map. In the quaternionic case HKC stands 
for `hyper-K\"ahler cone,' which is commonly used
instead of `conical hyper-K\"ahler', or CHK, which would be in line
with the terminology we use for vector multiplets. QK stands for quaternionic K\"ahler. 
Precise definitions are given in the respective sections of this review. 
\label{Table:r+c-maps}}
\end{table}

When discussing dimensional reduction in section \ref{sec:dim_red}, we also include dimensional reduction over time. 
This allows to construct theories with {\em Euclidean supersymmetry}. For four-dimensional
vector multiples and for hypermultiplets the special geometry of the scalar manifold
is modified, and now is of {\em para-complex} and of {\em para-quaternionic} type,
respectively.

In addition to reviewing the construction of bosonic Lagrangians and discussing the
resulting scalar geometries, we present a number of important applications:
static BPS black holes in four and in five space-time dimensions
in the presence of Weyl square interactions (sections \ref{secdic} and 
\ref{bpsentro4dbh});
deformed special K\"ahler geometry and topological string theory (section \ref{sec:hess});  
$F$-functions for point-particle Lagrangians (section \ref{sec:emduality}), for the Born-Infeld-dilaton-axion system 
(section \ref{bidaf}) and for a particular STU-model
(section \ref{FSTUmod}).
In all these applications,
the Hesse potential plays an important role: the semi-classical entropy of BPS black holes
is obtained from the Hesse potential by Legendre transformation; the holomorphic anomaly equation
of topological string theory is encoded in a Hessian structure; point-particle Lagrangians admit a reformulation
in terms of a Hesse potential; the Hesse potential approach to the STU-model yields important information about the 
function $F$ that encodes the Wilsonian Lagrangian of the model.

The topics and applications we chose to cover in this report are based on research papers and review articles which we will be referring to 
in the various sections comprising this report. The papers we chose to cite represent a small subset of the many papers that have been
published over the past decades on the subject of special geometry at large.
It would be impossible to refer to all these papers, and hence we have opted
to cite only those which we used to write this report.

Finally, we have assembled extensive appendices on the mathematics and physics background of the report, for the benefit of the reader.

\section{Hessian geometry and special real geometry\label{sec:Hessian_geometry}}

In this section we introduce Hessian geometry, focussing on the
aspects that are relevant for the special geometries of five- and
four-dimensional vector multiplets. A comprehensive treatment 
of Hessian geometry can be found in \cite{Book:Hessian}. Special
emphasis is put on {\em conical Hessian manifolds}, that is Hessian 
manifolds admitting a homothetic Killing vector field. Such manifolds
can be `projectivized', that is the space of orbits of the homothetic
Killing vector field carries a Riemannian metric, which, while not
being Hessian, is determined by the Hesse potential of the 
conical Hessian manifold. Conical Hessian manifolds admit a 
Hesse potential which is a homogeneous function. The special 
real geometry of five-dimensional vector multiplets is obtained
by restricting to Hesse potentials which are homogeneous cubic 
polynomials. The material on conical Hessian and special real
geometry is partly based on \cite{Alekseevsky:2001if,2008arXiv0811.1658A,Cortes:2011aj,Mohaupt:2011ab,Mohaupt:2012tu,Vaughan:2012}.

\subsection{Hessian manifolds}

In this subsection we provide the definition of a Hessian manifold, both in terms
of local coordinates, and coordinate-free.

\begin{defi} {\bf Hessian manifolds and Hessian metrics in terms of coordinates.}
A pseudo-Riemannian\footnote{See \ref{app:pseudo-Riemannian} for
  a review and for our conventions.} manifold $(M,g)$ is called a {\rm
  Hessian manifold} if
it admits local coordinates $q^a$, such that the metric
coefficients are the Hessian of a real function $H$, called the {\rm Hesse potential}:
\begin{equation}
\label{Hessian-local}
g_{ab} = \partial_a \partial_b H := 
\partial^2_{a,b} H := H_{ab} \;.
\end{equation}
Such metrics are called {\rm Hessian metrics}. 
\end{defi}

The relation  (\ref{Hessian-local}) is not invariant under general coordinate
transformations, but only under affine transformations. The
definition implies that the manifold can be covered by special
coordinate systems, related to each other by affine transformations, 
such that (\ref{Hessian-local}) holds in every coordinate patch. This is equivalent to the 
existence of a flat, torsion-free connection $\nabla$,
for which the special coordinates $q^a$ are affine coordinates. Equivalently, 
the differentials $dq^a$ define a parallel coordinate frame, 
$\nabla dq^a =0$.\footnote{Frames and connections are reviewed in  
\ref{app:vector-fields} and
\ref{app:connections}.
} 
The flat torsion-free connection $\nabla$ gives $M$ the structure of
an  {\em affine manifold}. By the Poincar\'e lemma, the integrability condition 
\begin{equation}
\partial_a g_{bc} = \partial_b g_{ac} = \partial_c g_{ba} 
\end{equation}
is necessary and locally sufficient for the  existence of a Hesse
potential.
Passing to general coordinates, we see 
that the rank three tensor $S=\nabla g$ must be totally symmetric. 
We thus arrive at the following coordinate-free definition:

\begin{defi}
{\bf Hessian manifolds, Hessian metrics and Hessian structures.}
A {\rm Hessian manifold} $(M,g,\nabla)$ is a pseudo-Riemannian manifold $(M,g)$ equipped with a
flat, torsion-free connection $\nabla$, such that the covariant rank three
tensor $S=\nabla g$ is totally symmetric. The pair $(g,\nabla)$ defines
a {\rm Hessian structure} on $M$, and the metric $g$ is called a 
{\em Hessian metric.}
\end{defi}

Locally, a Hessian metric takes the form
$g=\nabla d H$, where the Hesse potential $H$ is unique up to affine
transformations. When using affine coordinates we can write $g=\partial^2 H$,
since the connection $\nabla$ acts by partial derivatives. 

We note in passing that 
one example of a symmetric Hessian manifold which is prominent in 
physics is anti-de Sitter space \cite{Donets:2000sx}. Applications of 
Hessian manifolds to superconformal quantum mechanics have been 
discussed in \cite{Donets:1999jx} and 
\cite{Kozyrev:2017kiy,Kozyrev:2017rmk,Krivonos:2019pfw}. Superconformal
quantum mechanics on special K\"ahler manifolds, which as we will see later
are in particular Hessian manifolds, 
has been discussed in \cite{Bellucci:2004ky,Kozyrev:2018sgq}.

\subsection{The dual Hessian structure \label{sec:dual_hessian}}

Hessian structures always come in pairs. This will play an important role 
later when we discuss electric-magnetic duality, special K\"ahler geometry,
and black hole entropy functions. 
\begin{defi} {\bf Dual affine coordinates.}
If $q^a$ are $\nabla$-affine coordinates for a Hessian metric with Hesse potential $H$, then
\begin{equation}
q_a := H_a := \partial_a H
\end{equation}
are the associated {\em dual affine coordinates}. 
\end{defi}

Note that in general $q_a\not= H_{ab} q^b$. This reflects that 
$q^a$ and $q_a$ are functions on $M$, and not the components of a vector field
or differential form. The matrix $H^{ab}$ of metric coefficients
with respect to the dual coordinates is determined by 
\begin{equation}
g = H_{ab} dq^a dq^b = H^{ab} dq_a dq_b\;,
\end{equation}
which implies that the matrix $H^{ab}$ is 
the inverse of the matrix $H_{ab}$.

\begin{defi} {\bf Dual connection on a Hessian manifold.}
Let $(M,g,\nabla)$ be a Hessian manifold with Levi-Civita connection $D$. Then
\begin{equation}
\label{Dual_connection_Hessian}
\nabla_{\rm dual} = 2 D - \nabla \;,
\end{equation}
is called the {\rm dual connection} to $\nabla$.
\end{defi}

\begin{rmk}
{\bf Dual Hessian structures and the dual Hesse potential.}
The dual connection is flat and torsion-free, and defines a second Hessian
structure on $(M,g)$, called the {\em dual Hessian structure}. The 
$\nabla_{\rm dual}$-affine coordinates are the dual coordinates $q_a$
introduced above, and the {\em dual Hesse potential} $H_{\rm dual}$ is related
to $H$ by a Legendre transformation,
\begin{equation}
\label{Hqq}
H_{\rm dual} =q^a H_a -  H \;.
\end{equation}
The matrix of metric coefficients with respect to the dual Hessian structure is
the inverse matrix $H^{ab}$ of $H_{ab}$:
\begin{equation}
H^{ab} = \frac{\partial^2 H_{\rm dual}}{\partial q_a \partial q_b} \;.
\end{equation}
\end{rmk}
We refer to section 2.3  of \cite{Book:Hessian} for more details on the
dual Hessian structure.

\subsection{Conical Hessian manifolds  \label{sect:ConHess}}

We now consider the case where the Hesse potential is a homogeneous function. 
This is relevant for both five- and four-dimensional vector multiplet theories. 
\begin{defi} {\bf Homogeneous functions.}
\label{Def:HomFct}
A real function $H$ is {\em homogeneous
of degree} $n$ in the variables $q^a$ if  
\begin{equation}
H(\lambda q^a) = \lambda^n H(q^a) \;,\;\;\;\lambda \in \mathbb{R}^* \;.
\end{equation}
\end{defi}
This is equivalent to the {\em Euler relation}
\begin{equation}
L_\xi H = 
q^a \partial_a H = n H \;,
\end{equation}
where $\xi = q^a \partial_a$ is the so-called {\em Euler vector field} with
respect to the coordinates $q^a$, and
where $L_\xi$ is the Lie derivative.\footnote{See
  \ref{app:vector-fields} for a review and our conventions.}
The $k$-th derivative of a homogeneous function of degree $n$ 
is a homogeneous function of degree $n-k$. In local coordinates, we have the following
hierarchy of relations:
\begin{equation}
\label{Homogeneity}
q^a H_a = n H \;,\;\;\;
q^a H_{ab} = (n-1) H_b \;,\;\;\;
q^a H_{abc} = (n-2) H_{bc} \;,\ldots
\end{equation}

\begin{rmk}
{\bf Dual coordinates for homogeneous Hesse potentials.}
For a Hesse potential which is homogeneous of degree $n$, 
the dual coordinates
$q_a=H_a$ have weight $n-1$, while the metric coefficients $H_{ab}$ have
weight $n-2$, and the dual metric coefficients $H^{ab}$ have
weight $2-n$. The Legendre transform defining the dual Hesse potential 
simplifies:
\begin{equation}
\label{H_dual_homogeneous}
H_{\rm dual} = q^a H_a - H = (n-1) H \;.
\end{equation}
In particular $H_{\rm dual} = - H$ for $n=0$ and 
$H_{\rm dual}=H$ for $n=2$.
\end{rmk}

\begin{defi}{\bf Homogeneous tensor fields.}
A tensor field $T$ is called {\rm homogeneous of degree} $n$ with respect to the
action generated by a vector field $\xi$ if
\begin{equation}
L_\xi T = n T \;.
\end{equation}
We will then also say that $T$ has {\em weight} $n$. 
The case $n=0$ corresponds to the special case of an invariant tensor.
\end{defi}

The Lie derivatives 
\begin{equation}
L_\xi (\partial_a) = - \partial_a \;,\;\;\;
L_\xi (dq^a) = dq^a \;,
\end{equation}
show that derivatives $\partial_a$ have weight $-1$, while differentials $dq^a$ have weight
1. Thus the components 
of a tensor $T$ of type $(p,q)$ and weight $n$ have weight $n+p-q$.

{\bf Example:} Consider the case where the metric $g$ has weight $n$ with respect to the
Euler field $\xi$. Then
\begin{equation}
L_\xi g = n g \Leftrightarrow (L_\xi g)_{ab} = n g_{ab} 
\Leftrightarrow L_\xi(g_{ab}) = (n-2) g_{ab} \;.
\end{equation}
Here $L_\xi(g_{ab}) = \xi^c \partial_c g_{ab}$ denotes the Lie derivative of the 
components of the metric {\em considered as functions}. This is to be distinguished 
from $(L_\xi g)_{ab} = \xi^c \nabla_c g_{ab}$, which denotes the components 
of the tensor $L_\xi g$. The weight $n-2$ of the tensor components $g_{ab}$ can
be inferred from the following computation:
\begin{eqnarray}
 L_\xi g &=& 
 L_{\xi} (g_{ab} dq^a dq^b) = L_\xi (g_{ab}) dq^a dq^b 
+ g_{ab} L_\xi(dq^a) dq^b + g_{ab} dq^a L_{\xi} (dq^b) \nonumber \\
&=& (L_\xi (g_{ab})  + 2 g_{ab}) dq^a dq^b = (L_\xi g)_{ab} dq^a dq^b = n
g_{ab} dq^a dq^b \;.
\end{eqnarray}

\begin{defi}{\bf Killing vector fields and homothetic Killing vector fields.}
If the metric is a homogeneous tensor of weight $n\not=0$ with respect 
to the action generated by a vector field $\xi$,  then 
$\xi$ is called a {\em homothetic Killing vector field} of weight
$n$.  If $n=0$,  then $\xi$ is called a {\em Killing vector field}. 
\end{defi}

{\bf Example:} Let $g=\partial^2 H$ be a Hessian metric with a Hesse potential 
that is homogeneous of degree $n$. Then the Euler field $\xi$ is a homothetic
Killing vector field, and $g$ has weight $n$. This follows immediately from 
$g=H_{ab} dq^a dq^b$.

\begin{rmk}
{\bf Hypersurface orthogonality of the Euler field.} 
If $\xi$ is a homothetic Killing vector field for a Hessian metric $g$,
then $\xi$ is $g$-orthogonal to the level surfaces $H = \mathfrak{c}$ 
of the Hesse potential.  
\end{rmk}

In $\nabla$-affine coordinates this is manifest, since 
the dual coordinates are the components of a gradient:
\begin{equation}
\label{daH}
(n-1) \partial_a H = (n-1) q_a =  g_{ab} q^b  \;.
\end{equation}
Therefore the one-form $\xi^\flat = g_{ab} q^a dq^b = g(\xi, \cdot)$ dual to the
Euler field $\xi$ is exact, $\xi^\flat = (n-1) dH$. A vector $T$
is tangent to the hypersurface $H=\gc$ if and only if it is annihilated 
by the one-form $dH$ (equivalently, if it is orthogonal to the gradient of
$H$). Therefore the vector field $\xi$ is normal to the level
surfaces of $H$:
\begin{equation}
0 = (n-1) dH(T) = \xi^\flat(T) = g(\xi, T) \;.
\end{equation}
Note that the integrability condition $d\xi^\flat=0$ is a special case
of the Frobenius integrability condition for hypersurfaces, $\xi^\flat
\wedge d \xi^\flat =0$.\footnote{Hypersurface orthogonality and the
  Frobenius theorem are reviewed in \ref{app:Frobenius}}

\begin{rmk}
{\bf The case $n=1$ is to be discarded.}
Formula (\ref{daH}) shows that the case $n=1$ is special. It
corresponds to a linear  Hesse potential 
for which the metric is totally degenerate, $H_{ab}=0$. 
This case will be discarded in the following,
since we are only interested in non-degenerate metrics.
\end{rmk}

\begin{rmk}
{\bf The case $n=0$ needs to be treated separately.} The case $n=0$, where $\xi$ is 
a genuine Killing vector field, is interesting, but needs to be treated separately. In the following we will
first consider the generic case $n\not=0$ (with $n\not=1$ understood), and
then return to the case $n=0$.
\end{rmk}

We would like to have a coordinate-free characterization of Hessian manifolds which
admit homogeneous Hesse potentials. As a
first step, we consider pseudo-Riemannian manifolds equipped with a homothetic
Killing vector field which is the Euler field with respect to an affine structure. 
At this point it is not relevant whether the pseudo-Riemannian metric is Hessian 
or not. Since we admit indefinite metrics, the Euler field might become 
null, $g(\xi, \xi) =0$. We will need to divide by the function $g(\xi,\xi)$ and therefore we
require that $\xi$ is {\em nowhere isotropic}, that is $g(\xi, \xi)\not=0$ on 
the whole manifold $M$. Thus $\xi$ is either globally time-like or globally 
space-like.

\begin{defi} {\bf $n$-conical pseudo-Riemannian manifolds.}
{\em An $n$-conical pseudo-Riemannian 
manifold}  $(M,g,\nabla,\xi)$ is a pseudo-Riemannian manifold $(M,g)$ 
equipped with a flat, torsion-free connection $\nabla$ and a nowhere
isotropic vector field $\xi$, such that
\begin{equation}
D\xi = \frac{n}{2} \mbox{Id}_{TM} \;,\;\;\;
\nabla \xi = \mbox{Id}_{TM} \;.
\end{equation}
Here $D$ is the Levi-Civita connection of the metric $g$, and
$D\xi$, $\nabla \xi$ are endomorphism of the tangent bundle $TM$ of
$M$, that is, tensor fields of type $(1,1)$. Equivalently one can write
\begin{equation}
D_X \xi = \frac{n}{2} X \;,\;\;\;
\nabla_X \xi = X \;,\;\;\;\forall X\in \mathfrak{X}(M) \;,
\end{equation}
where $\mathfrak{X}(M)$ are the smooth vector fields on $M$. 
\end{defi}

The condition $\nabla_X \xi = X$ implies that $\xi$ is the Euler field with respect
to $\nabla$-affine coordinates $q^a$. Note that if this condition 
is dropped, we can change the value of $n$ by rescaling
$\xi$. One could in particular choose $n=2$, which leads to the 
standard definition of a metric cone (or Riemannian cone). 
But since we are ultimately interested in Hessian manifolds, 
we insist on  the existence of an affine structure, which prevents
us from changing the value of $n$.

By decomposition of $D_X \xi = \frac{n}{2} X$ into its symmetric and anti-symmetric
part we see that this condition is equivalent to $\xi$ being a closed, hence hypersurface
orthogonal, and homothetic Killing vector field:
\begin{equation}
D\xi = \frac{n}{2} \Id \Leftrightarrow \left\{ \begin{array}{l}
L_\xi g = n g \;,\\
d\xi^\flat = 0 \;.\\
\end{array} \right. 
\end{equation}
In general local coordinates $x^m$ this reads
\begin{equation}
D_m \xi_n = \frac{n}{2} g_{mn}
\Leftrightarrow \left\{ \begin{array}{l}
(L_\xi g)_{mn} = D_m \xi_n + D_n \xi_m
= n g_{mn} \;, \\
\partial_m \xi_n - \partial_n \xi_m = 0 \;.\\
\end{array} \right.
\end{equation}
\begin{rmk}
{\bf Standard form of an $n$-conical metric.} 
If $(M,g)$ is an $n$-conical pseudo-Riemannian manifold, then 
$g$ can 
locally be written in the from 
\begin{equation}
\label{n-conical}
g = \pm r^{n-2} dr^2 + r^n h \;,
\end{equation}
where $h$ is a pseudo-Riemannian metric on 
an immersed hypersurface $\iota\;: {\cal H} \rightarrow M$. For $n=2$ this
is the local form of a metric cone $(\mathbb{R}^{>0} \times {\cal H}, dr^2 + r^2 h)$ 
over a pseudo-Riemannian manifold $({\cal H}, h)$. 
\end{rmk}
We now give a proof following \cite{Vaughan:2012},  which generalizes the treatment of
Riemannian cones in \cite{Gibbons:1998xa}.
The vector field $\xi$ is hypersurface orthogonal and therefore 
locally the gradient of a function $H$,
$\xi_m = \partial_m H$. The level surfaces of $H$ are orthogonal to the integral lines of $\xi$.
Combining this with the homothetic Killing equation shows that $H$ is a potential for
the metric:
\begin{equation}
(L_\xi g)_{mn}  = D_m \xi_n + D_n \xi_m = n g_{mn} \Rightarrow 
D_m \partial_n H = \frac{n}{2} g_{mn} \;.
\end{equation}
Differentiating the norm\footnote{Since we work with indefinite metrics, 
we use the term `norm' for square-norm $g(\xi,\xi)$.} 
$g(\xi,\xi)$ of $\xi$ gives
\begin{equation}
\partial_p (g^{mn} \partial_m H \partial_n H) =
2 D_p(g^{mn} \partial_m H ) \partial_n H =
2 g^{mn} D_p \partial_m H \partial_n H =
n \partial_p H \;,
\end{equation}
so that upon choosing a suitable integration constant, 
\begin{equation}
g(\xi,\xi) = n H  \;.
\end{equation}We use $H$ as a coordinate along the integral lines of $\xi$,
and extend this to a local coordinate system $\{H, x^i \}$ on $M$.
For $x^i$ we choose  coordinates on the level surfaces of $H$, 
by picking any local coordinates on one level surfaces and extending
them to $M$ by the requirement 
that points on different level surfaces have the same coordinates
$x^i$ if they lie on the same integral line of $\xi$. Since the level
surfaces are orthogonal to the integral lines of $\xi$, the metric has
a block structure:
\begin{equation}
g = g_{HH} dH^2 + g_{ij} dx^i dx^j \;.
\end{equation}
Using that $dH(\xi) = g(\xi,\xi)$ and $dx^i(\xi)= 0$ we find $g_{HH} = (nH)^{-1}$, and thus
\begin{equation}
g= \frac{dH^2}{nH} + g_{ij} dx^i dx^j  \;,\;\;\;\xi = n H \partial_H \;.
\end{equation}
Introducing a new transverse coordinate $r>0$ by $r^n = \pm nH$, this becomes
\begin{equation}
g = \pm r^{n-2} dr ^2 + g_{ij} dx^i dx^j  \;,\;\;\;\xi = r \partial_r \;.
\end{equation}
Note that we have to allow a relative sign between $r$ and $H$, 
because  $H$ can be positive or negative, 
while $r$ is positive. 
Using that $L_\xi dx^i=0$, the homothetic Killing equation $L_\xi g=ng$ implies
\begin{equation}
(L_\xi g)_{ij} = r \partial_r  g_{ij} = n g_{ij} \;.
\end{equation}
Thus the functions $g_{ij}(r,x)=g_{ij}(r, x^1, \ldots, x^n)$ are homogeneous of degree $n$
in $\rho$, and therefore 
\begin{equation}
g_{ij}(r, x) = r^n  h_{ij}(x)\;,
\end{equation}
where $h_{ij}=h_{ij}(x)$ only depend on $x^i$, but not on $r$. Thus locally $g$ takes the form 
\begin{equation}
g= \pm r^{n-2} dr^2 + r^n h_{ij}(x) dx^i dx^j \;.
\end{equation}
This is the local standard form of a $n$-conical metric.  
For $n=2$ this is the metric on a pseudo-Riemannian cone, see 
\ref{app:metric_cones}.

We observe that while our derivation is not valid for $n=0$, the formula
we have obtained still makes sense, since
\begin{equation}
g = \pm \frac{dr^2}{r^2} + h_{ij} dx^i dx^j
\end{equation}
is a product metric on $\mathbb{R}^{>0} \times {\cal H}$, with 
isometric action of $\xi=r \partial_r$ by dilatation. 
Introducing a new radial coordinate $\rho$ by $d\rho = \frac{dr}{r}$,
this becomes the standard product metric
\begin{equation}
g= \pm d\rho^2 + h_{ij} dx^i dx^j 
\end{equation}
on $\mathbb{R} \times {\cal H}$, where the isometric action of $\xi
= \partial_\rho$ is now by translation.
The product form of the metric does not follow automatically from 
the $n$-conical conditions with $n=0$,  which imply $L_\xi g=0$ and
$d\xi^\flat=0$. But if we impose in addition  
that $\xi$ has constant norm, $g(\xi, \xi) =\gc\not=0$, where we used that $\xi$
is nowhere isotropic, we can show that $g$ is a product metric, as follows.
We choose a coordinate $\rho$ 
by setting $\xi = \sqrt{|\gc|} \partial_\rho$ and extend this to a local coordinate system
on $M$ by choosing coordinates $x^i$ on the hypersurfaces $\rho=\mbox{const.}$
orthogonal to the integral lines of $\xi$. In this coordinate system
\begin{equation}
g = g_{\rho \rho} d\rho^2 + g_{ij}(\rho,x) dx^i dx^j  \;.
\end{equation}
Since $g(\xi,\xi) = g_{\rho \rho} |\gc| = \gc$ it follows that $g_{\rho\rho}=\pm 1$,
depending on whether $\xi$ is time-like or space-like. Since by construction $L_\xi dx^i=0$,
the Killing equation $L_\xi g=0$ implies that $g_{ij}$ is independent of $\rho$:
\begin{equation}
(L_\xi g)_{ij} = \partial_\rho g_{ij} = 0 \;.
\end{equation}
We can therefore interpret $g_{ij}$ as a metric $h_{ij}$ on any of the hypersurfaces
$\rho = \mbox{const.}$ Thus we have shown that $g$ locally takes the form (\ref{product_local}) 
\begin{equation}
 g = \pm d\rho^2 + h_{ij} dx^i dx^j = \pm \frac{dr^2}{r^2} + h_{ij} dx^i dx^j  \;,
\end{equation}
of a product metric.

{\bf Relation to affine coordinates.}
The standard coordinates $(r,x^i)$ on an $n$-conical Riemannian manifold can be  related to a  $\nabla$-affine
coordinate $q^a$ by setting
\begin{equation}
(q^a) = (q^0,q^i) = (r, rx^i) \;.
\end{equation}
The Jacobian of this transformation is 
\begin{equation}
\frac{D(q^0, q^i)}{D(r,x^j)} = \left( \begin{array}{ll}
\frac{\partial r}{\partial r}_{|x^j} & \frac{\partial r}{\partial x^j}_{|r} \\
\frac{\partial rx^i}{\partial r}_{|x^j} & \frac{\partial rx^i}{\partial x^j}_{|r} \\
\end{array} \right) 
= \left( \begin{array}{cc}
1 & 0 \\
x^i & r \delta^i_j \\
\end{array} \right) \;,
\end{equation}
and therefore
\begin{equation}
\xi = r \frac{\partial }{\partial r} = r \left( \frac{\partial q^a}{\partial r} \frac{\partial}{\partial q^a}\right)
= q^a \frac{\partial}{\partial q^a} \;.
\end{equation}
The coordinates $q^a$ have weight
1, the derivatives $\partial_a$ have weight $-1$, and the metric coefficients $g_{ab}$ are
homogeneous functions of degree $(n-2)$ in $q^a$. We remark that the coordinates
$q^a$ can be viewed as {\em homogeneous coordinates} (also called {\em projective coordinates})
for the hypersurfaces $r=const.$, for which $x^i$ are inhomogeneous coordinates.

So far we have not required that the pseudo-Riemannian metric $g$ is Hessian. 
By adding this requirement we arrive at the following definition:
\begin{defi} {\bf $n$-conical Hessian manifolds.}
\label{n-conical_Hessian}
An {\rm $n$-conical Hessian manifold} $(M,g,\nabla, \xi)$ 
is an $n$-conical pseudo-Riemannian
$(M,g,\nabla,\xi)$ manifold which is Hessian, that is, 
$\nabla g$ is totally symmetric.
\end{defi}

\begin{rmk}
{\bf $n$-conical Hessian manifolds admit a Hesse potential which is homogeneous
of degree $n$.}
If $(M,g,\nabla,\xi)$ is an $n$-conical Hessian manifold with local affine coordinates $q^a$, then 
the function 
\begin{equation}
\label{Hesse-by-integration}
H = \frac{1}{n(n-1)} q^a q^b g_{ab}   \;,
\end{equation}
which is homogeneous of degree $n$, 
is a Hesse potential for $g$. 
\end{rmk}

The function $H$ is manifestly homogeneous of degree $n$. 
By differentiating (\ref{Hesse-by-integration}) twice 
and using the homogeneity relations (\ref{Homogeneity}) for $g_{ab}$, 
we obtain 
$H_{ab} =\partial_a \partial_b H = g_{ab}$, which shows that  $H$ is indeed 
a Hesse potential for $g_{ab}$. We remark that (\ref{Hesse-by-integration}) 
does not apply to the degenerate case $n=1$, which we discard,  
and to the interesting case $n=0$, which we will consider separately
below.

\begin{defi}{\bf Conical affine coordinates.}
Let $(M,g,\nabla,\xi)$ be an $n$-conical Hessian manifold. Then $\nabla$-affine
coordinates $q^a$ are called {\em conical $\nabla$-affine coordinates} if the
Hesse potential is homogeneous of degree $n$ in $q^a$.
\end{defi}
The homogeneity of $H$ is only preserved under linear transformations, but 
not under translations. Therefore  conical $\nabla$-affine coordinates are unique up to linear
coordinate changes. In the following it is understood that $\nabla$-affine coordinates
on a conical Hessian manifold are always chosen to be conical.

\subsection*{0-conical Hessian manifolds}

We now turn to the special case $n=0$, where the Euler field $\xi$ acts isometrically 
on the Hessian metric $g$. 
Metrics of this type can be constructed by taking
Hesse potentials of the form
\begin{equation}
\tilde{H} = \mathfrak{a} \log (\mathfrak{b}  H) \;,
\end{equation}
where $\ga,\gb$ are real constants, and where $H$ is a homogeneous
function of degree $n>1$.\footnote{Except where the determination of signatures is concerned,
we only use $n\not=0$, $n\not=1$ in the following. 
For physics applications we will need the cases $n=2$ and
$n=3$.} We will see later that certain constructions
involving vector multiplets (superconformal quotients and dimensional reduction)
naturally involve replacing a homogeneous Hesse potential by its logarithm.
The constants $\ga,\gb$ have been introduced so that we can match our
results with various conventions used in the physics literature. 

Note that the Hesse potential $\tilde{H}$ is not a homogeneous function,
since it transforms with a shift under $q^a \mapsto \lambda q^a$. However,
its $k$-th derivatives are homogeneous functions of degree $-k$ for any 
$k\geq 1$. The tensor $\tilde{g}=\tilde{H}_{ab} dq^a dq^b$ is homogeneous of 
degree zero, and defines a  0-conical Hessian metric.
The first derivatives
\begin{equation}
\tilde{q}_a := \tilde{H}_a = \partial_a \tilde{H} = \ga \frac{H_a}{H}
\end{equation}
of $\tilde{H}$ are homogeneous of degree $-1$.
 They define the coordinate
system dual to the affine coordinates $q^a$ with respect to the Hesse potential $\tilde{H}$.
The overall sign of $H$ does not have any effect on
expressions which involve derivatives of $\tilde{H}$  only, since these expressions
are invariant under $H\rightarrow -H$. In particular, the Hessian metrics
$H_{ab}=\partial^2_{a,b} H$ and $-H_{ab} = \partial^2_{a,b} (-H)$ give rise to the 
same Hessian metric $\tilde{H}_{ab}$ 
if we `take the log of the Hesse potential.'

Explicitly, the metric coefficients associated with the Hesse potential $\tilde{H}$ are:
\begin{equation}
\tilde{H}_{ab} = \partial^2_{a,b}\tilde{H} =
\ga \,\frac{ H H_{ab} - H_a H_b}{H^2} \;.
\end{equation}
The following relations implied by the homogeneity of $H$ are useful:
\begin{equation}
\label{hom_rel}
q^a q_a = q^a H_a = n H \;,\;\;\;H_{ab} q^b = (n-1) q_a \;,\;\;\;
q^a q^b H_{ab} = n(n-1) H \;,
\end{equation}
The dual affine coordinates $\tilde{q}_a$ with respect to $\tilde{H}$ satisfy
\begin{equation}
\tilde{H}_{ab} q^b = - \ga \frac{q_a}{H} = - \tilde{q}_a \;.
\end{equation}
To compare the $n$-conical metric $\partial^2 H$ and the $0$-conical
metric $\partial^2 \tilde{H}$, we evaluate them on 
the Euler field $\xi$, which is orthogonal to the level surfaces 
of $H$ and $\tilde{H}$, and on a vector field $T$, which is tangent
to the level surfaces. 
\begin{itemize}
\item
Components transversal to the foliation ${\cal H}_\gc$.
\begin{eqnarray}
g(\xi,\xi) &=& H_{ab} q^a q^b = n(n-1) H \;, \\
\tilde{g}(\xi,\xi)&=& \tilde{H}_{ab} q^a q^b = - \ga n \;.
\end{eqnarray}
The $\tilde{g}$-norm $\tilde{g}(\xi, \xi)$ 
of $\xi$ is constant on $M$, while the $g$-norm $g(\xi,\xi)$  depends on the leaf ${\cal H}_\gc$.
\item
Mixed components. If $T$ is tangent to ${\cal H}_{\gc}= \{ H = \gc \} $, then 
\begin{equation}
dH(T) = T^a H_a = T^a q_a = 0 \;.
\end{equation}
Therefore
\begin{equation}
g(T,\xi) = H_{ab} T^a q^b  =T^a q_a = 0 \;,\;\;\;\ \tilde{g}(T,\xi )= \tilde{H}_{ab} T^a q^b = 0 \;.
\end{equation}
\item
Tangential components:
\begin{equation}
\tilde{g}(T,T) =  \ga \frac{HH_{ab} T^a T^b - T^a q_a q_b
  T^b}{H^2}  = \frac{\ga}{H} H_{ab} T^a T^b  = \frac{\ga}{H} g(T,T)\;.
  \end{equation}
  These component are proportional for constant $H$. 
  \end{itemize}
Since the tangential components of both metrics are proportional for any
fixed leaf ${\cal H}_\gc$, their pullbacks
to the embedded hypersurfaces\footnote{Immersions and embeddings are
review in \ref{app:manifolds}. Since we are interested in comparing local expressions
for various tensor fields, there is no loss of generality in assuming that the hypersurfaces ${\cal H}_{gc}$ 
are embedded.} 
\begin{equation}
\iota_\gc \;:  {\cal H}_\gc = \{ q \in M | H(q) = \gc \}
\rightarrow M
\end{equation}
are proportional:
\begin{equation}
g_\gc = \iota_\gc^* \partial^2 H =  \frac{\gc}{\ga}\iota_\gc^* \partial^2 \tilde{H} \;.
\end{equation}
On the hypersurface ${\cal H} = {\cal H}_{\gc=1}$:
\begin{equation}
g_{\cal H} = \iota^* \partial^2 H = \frac{1}{\ga}  \iota^* \partial^2 \log H  \;.
\end{equation} 
By choosing $\ga=1$ we can make the pullbacks equal. Note that the
transversal components of both metrics are different. In particular both metrics
have different signatures. On a leaf ${\cal H}_\gc$ we 
have
\begin{equation}
\ga g(T,T) = \gc \tilde{g}(T,T) \;,
\end{equation}
\begin{equation}
g(\xi, \xi)= n(n-1) \gc \;,\;\;\; \tilde{g}(\xi,\xi) = \tilde{H}_{ab} dq^a dq^b = - n\ga \;.
\end{equation}
Thus if $g$ and $\tilde{g}$ have the same signature on tangent vectors, $\ga \gc >0$, then 
they have different signature in the transverse direction.\footnote{Here we use the assumption
$n>1$, which applies for the application to vector multiplets, where $n=2$ or $n=3$. 
Otherwise all expressions in this section are valid for $n\not=0$, $n\not= 1$.}

\begin{rmk}
{\bf The dual Hessian structure and dual Hesse potential for a Hessian manifold 
with logarithmic Hesse potential.}
The Hesse potential $\tilde{H}_{\rm dual}$  dual to $\tilde{H}$ is defined by
\begin{equation}
\label{dualHtilde}
\tilde{H}^{ab} = 
\frac{\partial \tilde{H}_{\rm dual}}{\partial \tilde{q}_a \partial \tilde{q}_b} \;,
\end{equation}
where $\tilde{H}^{ab}$ is the inverse of $\tilde{H}_{ab}$. 
By a straightforward computation one finds $\tilde{H}_{\rm dual} = - \tilde{H}$.\footnote{We remark 
that in the physics
literature, i.p. in \cite{Errington:2014bta}, the dual Hesse potential
was defined without minus sign. Here we use the definition given
in \cite{Book:Hessian}.}
This is consistent with (\ref{H_dual_homogeneous}), which, however, cannot
be applied directly, because $\tilde{H}$ is not a homogeneous
function of degree zero. 
\end{rmk}

\subsection{Projectivization of conical Hessian manifolds \label{Sect:rSCQ}}

The relation between the manifolds $(M,g_M)$ and $({\cal H}, g_{\cal
  H})$ can be interpreted as a quotient, and $({\cal H}, g_{\cal H})$ can 
  be viewed as the projectivization of the conical manifold $(M, g_M)$, 
  with respect to the homothetic action of $\xi$. 
     This construction is related
  to the so-called superconformal quotients in the physics literature. In particular
  the {\em real superconformal quotient} relating the scalar geometry of
  five-dimensional superconformal vector multiplets to the geometry 
  of vector multiplets coupled to Poincar\'e supergravity is a special 
  case of the quotient relating $(M,g_M)$ and $({\cal H}, g_{\cal H})$. 
  
  If $(M,g_M,\nabla, \xi)$ is a conical Hessian manifold we can consider the
  space of orbits $\bar{M} = M/ \langle \xi \rangle\cong M/\mathbb{R}^{>0}$ of the action of $\xi$ 
  on $M$. 
  We will assume that this quotient is well-behaved, so that $\bar{M}$ is a smooth
  manifold. To induce a metric $g_{\bar{M}}$  on the quotient, we need 
  a symmetric, second rank co-tensor $g^*_M$ on $M$ which is {\em projectable}, that
  is, invariant under the action of $\xi$, $L_\xi g^*_M=0$ and transversal to the action
  of $\xi$, $g^*_M(\xi,\cdot)=0$. The second condition implies that $g^*$ is not a metric 
  on $M$, because it has a kernel which contains $\xi$. In order that it projects
  to a metric $g_{\bar{M}}$ on $\bar{M}$, the kernel of $g^*_M$ must be one-dimensional,
  that is, it is spanned by $\xi$. Since the hypersurfaces ${\cal H}_\gc$ are transversal to $\xi$, any of them
  can be used as a a set of representatives for the orbit space $M/\langle \xi \rangle$,
  that is $\bar{M} \cong {\cal H}_\gc$. 
  We can view  $M$ as a real line bundle, 
  $\pi: M \rightarrow \bar{M}$ over $\bar{M}\cong {\cal H}$, and the invariant
  tensor $g^*_M$ is equal to the pull-back of $g_{\bar M} = g_{\cal H}$ 
  to $M$: 
  \begin{equation}
  g^*_M = \pi^* g_{\bar M} = \pi^* g_{\cal H} \;.
  \end{equation}

  The conical metric $g_M$ is neither invariant nor transversal with respect to the
  action of $\xi$, but there is
  a natural way to construct a projectable tensor $g^*_M$ out of $g_M$ using
  the conical Hessian structure. Moreover, the induced metric $g_{\bar{M}}$
  agrees, up to conventional normalization, with the pull-back $g_{\cal H}$ of
  the conical metric $g_M$ to ${\cal H}$. Since $g_M$ transforms with a different
  weight $n$ under $\xi$, we can obtain an invariant tensor by multiplication with 
  the appropriate power of $H$. In fact, we have seen that taking the logarithm
  of a homogeneous Hesse potential automatically associates a 0-conical 
  Hessian metric to an $n$-conical one.
  To obtain a projectable tensor, it remains to add an $\xi$-invariant symmetric rank two co-tensor
  such that the resulting tensor becomes transversal to $\xi$. For this it is helpful 
  to consider the one-form 
  \begin{equation}
  d\log H = H^{-1} H_a dq^a
  \end{equation}
  which vanishes on tangent vectors $T$ to the surfaces $H=\gc$, while
  being constant along integral lines of $\xi$:
  \begin{equation}
  d\log H(T) = 0 \;,\;\;\;
  d\log H(\xi) = H^{-1} H_a q^a = n \;.
  \end{equation}
  By taking linear combinations between the 0-conical Hessian metric and the 
  square of this one form, we obtain a family of $\xi$-invariant symmetric rank two
  co-tensors:
\begin{equation}
g_M^{(\alpha)} = \ga \frac{ H H_{ab} - \alpha H_a H_b}{H^2} \; dq^a dq^b
= \ga \left( H^{-1} H_{ab} dq^a dq^b - \alpha (d\log H)^2  \right)\;.
\end{equation}
Note that only $\alpha=1$ corresponds to a Hessian metric.  
Now we look for a critical value $\alpha^*$ of $\alpha$  where  $g_M^{(\alpha)}$ becomes 
transversal to $\xi$: 
\begin{eqnarray}
0&=& g^{(\alpha)}_M(\xi, \cdot) = \ga H^{-2} \left( HH_{ab} - \alpha H_a H_b \right)q^a dq^b=
\ga H^{-2} HH_b \left( (1-\alpha) n -1\right) dq^b \nonumber \\ 
&\Rightarrow &
\alpha = \frac{n-1}{n}  =: \alpha_*\;.
\end{eqnarray}
Note that as a function of $\alpha$ the norm $g(\xi,\xi)$ of $\xi$ changes sign
at $\alpha=\alpha^*$. Therefore $g_M^{(\alpha)}$ changes signature when 
crossing the critical value where it degenerates.

Thus we have identified the projectable tensor 
\begin{equation}
g^*_M = g_M^{(\alpha_*)} = 
\ga \left( H^{-1} H_{ab} dq^a dq^b - \frac{n-1}{n}  (d\log H)^2 \right) \;,
\end{equation}
which defines a non-degenerate metric $g_{\bar{M}}$ on the quotient space
$\bar{M}=M/\mathbb{R}^{>0}$. Since the hypersurfaces ${\cal H}_\gc$ are
transversal to the integral lines of $\xi$, we can pick any such hypersurface
to represent the quotient space.  On tangent vectors $T,S$ to ${\cal H}$, 
$g^*_M$ agrees, up to a constant factor, with $g_M$, and  therefore with the
pull-back of $g_M$ to ${\cal H}$:
\begin{equation}
g^*_M(T,S)_{\gc=1}  = \ga  H_{ab} T^a S^b = \ga g_M(T,X)_{\gc=1} = 
\ga g_{\cal H}(T,X) \;.
\end{equation}
We remark that this construction can be viewed
as a real  analogue of the construction of the Fubini-Study metric on 
complex projective spaces, which itself is a special case of the
complex version of the superconformal quotient (see for example
\cite{Cortes:2015wca}).

Finally we remark that the family $g_M^{(\alpha)}$ of $\xi$-invariant tensors can
be generalized to families of symmetric tensors with given weight $k$ under $\xi$. 
If $H$ has weight $n$ then metrics of the form
\begin{equation}
g^{(k,\alpha_1,\alpha_2)} = H^{k/n}  \left( \alpha_1 g^{*}_M 
+ \alpha_2 (d \log H)^2 \right) \;.
\end{equation}
have weight $k$. This parametrization uses three building blocks: the projectable 
invariant tensor $g^*_M$, the quadratic differential $(d\log H)^2$ which vanishes
on tangent vectors of the foliation ${\cal H}_c$, and the Hesse potential which 
determines the weight. By varying $\alpha_1$ and $\alpha_2$, the signature 
can be changed. All symmetric second rank co-tensors we need are included in this family.

\subsection{Special real geometry \label{Sect:SR}}

\subsubsection{Affine special real manifolds as Hessian manifolds}

We are now in position to define the scalar geometries
five-dimensional vector multiplets.  As we will see in section 
\ref{Sect:Lagr5drVM} the geometry of rigid five-dimensional vector multiplets
is Hessian, and the scalar fields, which
are  the lowest components of vector multiplets, are `special coordinates'
on the scalar manifold.\footnote{More
  precisely the scalar fields 
  are pullbacks from the scalar manifold
  to space-time of coordinate maps for the scalar manifold. See \ref{App:sigma_models}. }
 Here special coordinates means affine
coordinates with respect to the flat (or `special') connection defining the
Hessian structure.
Supersymmetry imposes an additional condition because it implies
the presence of a Chern-Simons term in the 
Lagrangian, whose gauge
invariance (up to surface terms) restricts the Hesse potential to be 
a cubic polynomial. This leads to the following definition:

\begin{defi}
\label{def:ASK}
{\bf Affine special real manifolds (ASR manifolds).} An {\rm affine special real manifold}
$(M,g_M, \nabla)$ 
is a Hessian manifold with a Hesse potential that is a cubic
polynomial in $\nabla$-affine coordinates.
\end{defi}

We note that this definition is independent of the choice of special
coordinates, since affine transformations preserve the degree of a 
polynomial. The $\nabla$-affine coordinates of an ASR manifold are
called {\em special real coordinates}, or special coordinates for short.

We can also define a conical version of affine special real geometry, which
turns out to be the geometry of five-dimensional rigid {\em superconformal} 
vector multiplets, to be introduced in section \ref{Sect:rSCVM}.
 
\begin{defi} 
\label{def:CASR}
{\bf Conical affine special real manifolds (CASR manifolds).} A {\rm conical affine special
real manifold} $(M,g_{M}, \nabla, \xi)$ is a 3-conical Hessian 
manifold whose Hesse potential is a homogeneous cubic polynomial in
special coordinates.
\end{defi}

Finally, we can apply the quotient construction of section \ref{Sect:rSCQ}
to a CASR manifold. In this case we will refer to the quotient as the
{\em real superconformal quotient,} because
the resulting quotient manifolds occur as scalar target spaces
for five-dimensional vector multiplet coupled to Poincar\'e supergravity, as we
will see in section \ref{Sect:SCVM+SG}.
This motivates the following definition:

\begin{defi}
{\bf Projective special real manifold (PSR manifold).} A {\rm projective special real
manifold} 
$({\bar{M}, g_{\bar{M}}})$ is a pseudo-Riemannian manifold which 
can be obtained as the real superconformal quotient of a conical 
affine special real manifold $(M,g_{M},\nabla, \xi)$. 
\end{defi}

For later use we collect some formulae, which follow from those
derived in the previous sections by specializing to the case $n=3$. 
On a CASR manifold $M$ we have the family 
\begin{equation}
\label{Family}
g^{(\alpha)}_M = \ga \frac{HH_{ab} - \alpha H_a H_b}{H^2} dq^a dq^b =
\ga \left( H^{-1} H_{ab} dq^a dq^b - \alpha (d \log \gb H)^2 \right)
\end{equation}
of $\xi$-invariant symmetric rank 2 co-tensor fields.
The following tensor fields are relevant for 
five-dimensional vector multiplet theories:
\begin{itemize}
\item
The CASR metric
\begin{equation}
g_M = H_{ab} dq^a dq^b  \;.
\end{equation}
\item
The $\xi$-invariant metric 
\begin{equation}
g^{(0)}_M = \ga H^{-1} H_{ab} dq^a dq^b \;,
\end{equation} 
which is a conformally rescaled version of the CASR metric $g_M = H_{ab} dq^a dq^b$.
\item
The 0-conical Hessian metric 
\begin{equation}
g^{(1)}_M = \ga  \partial^2 \log \gb H = \ga \frac{HH_{ab} - H_a H_b}{H^2} dq^a dq^b \;.
\end{equation}
\item
The projectable tensor field
\begin{eqnarray} 
g^{*}_M &=& \ga \left( H^{-1} H_{ab} dq^a dq^b - \frac{2}{3} (d \log H)^2\right) \\  
&=& \ga (H^{-1} H_{ab} - \frac{2}{3} H^{-2} H_a H_b) dq^a dq^b \;, \nonumber
\end{eqnarray}
where we used that $\alpha_* =\frac{2}{3}$ for $n=3$. This tensor field 
projects to the PSR metric $g_{\bar{M}} = g_{\cal H}$.
\end{itemize} 
We also note the norms of $\xi$ with respect to these metrics:
\begin{equation}
\label{Norms}
g_M(\xi, \xi) = 6 H \;,\;\;\;
g^{(0)}_M(\xi, \xi) = 6 \ga  \;,\;\;\;
g_M^{*}(\xi, \xi) = 0 \;,\;\;
g^{(1)}_M (\xi,\xi) = - 3 \ga \;.
\end{equation}
As observed before, the signature of $g_M^{(\alpha)}$ changes 
at $\alpha = \alpha_* = \frac{2}{3}$.

\subsubsection{Projective special real manifolds as centroaffine hypersurfaces \label{sect:PAHS}}

The original construction of five-dimensional vector multiplets
coupled to supergravity \cite{Gunaydin:1983bi} did not make use
of the superconformal formalism. Instead the Poincar\'e supergravity
Lagrangian and on-shell supertransformations were constructed 
directly. The resulting scalar manifold $\bar{M}$ was interpreted
as a cubic hypersurface in $\mathbb{R}^{n+1}$, with a metric
determined by the homogeneous cubic polynomial defining 
the embedding. We will not follow \cite{Gunaydin:1983bi} in detail,
but instead review the construction of \cite{Alekseevsky:2001if}, which
realizes $\bar{M}$ as a so-called {\em centroaffine hypersurface} and 
allows to recover the local formulae of \cite{Gunaydin:1983bi}.

We start with $\mathbb{R}^{n+1}$ equipped with its standard flat connection
$\partial$. Note that we do not introduce a metric on $\mathbb{R}^{n+1}$ so that
the construction is done within the framework of affine differential geometry. The {\em position vector 
field} $\xi$ is defined by $\xi(p) = p$ for all $p\in \mathbb{R}^{n+1}$. For
linear coordinates $h^I$ on $\mathbb{R}^{n+1}$ and $\xi$ is the corresponding Euler field,
$\xi = h^I \partial_I$.

\begin{defi} {\bf PSR manifolds as centroaffine hypersurfaces.}
A PSR manifold $\bar{M}$ is a connected immersed hypersurface
\begin{equation}
\iota \;:\bar{M} \rightarrow {\cal H} := \{ {\cal V} = 1 \} \subset \mathbb{R}^{n+1}
\end{equation}
where the homogeneous cubic polynomial
\begin{equation}
{\cal V} := C_{IJK} h^I h^J h^K 
\end{equation}
is assumed to be non-singular in a neighbourhood 
\begin{equation}
U = U_\epsilon = \{ {\cal V} = c | 1-\epsilon  < c < 1 + \epsilon \} \subset \mathbb{R}^{n+1} 
\end{equation}
of the hypersurface ${\cal H}$ 
for some $\epsilon >0$. 
\end{defi}
We will assume that $\bar{M}$ is an embedded submanifold, so that we can
identify $\bar{M}$ and ${\cal H}$. Let us verify that we can recover the
alternative Definition \ref{def:ASK}.
For a homogeneous cubic polynomial, the position vector field $\xi$ is everywhere 
transversal to ${\cal H}$. This allows to define a metric $g_{\cal H}$ and a 
torsion-free connection $\nabla$ on ${\cal H}$ by decomposing the connection $\partial$,
acting on tangent vectors $X,Y \in T_p {\cal H}$, $p\in {\cal H}$, into a tangent
and a transversal component:
\begin{equation}
\label{PAHS1}
\partial_X Y = \nabla_X Y + \frac{2}{3} g_{\cal H} (X,Y)\xi \;.
\end{equation}
The factor $\frac{2}{3}$ is conventional. This construction is a special case 
of the construction of a {\em centroaffine hypersurface}, see \ref{App:affine_hyperspheres}
for more details.

It is useful to introduce the totally symmetric trilinear from 
\begin{equation}
C = C_{IJK} dh^I dh^J dh^K \;.
\end{equation}
By contracting with the position vector $\xi$ we obtain the following tensors:
\begin{enumerate}
\item
The function 
\begin{equation}
C(p,p,p) = C_{IJK} h^I h^J h^K = {\cal V} \;,
\end{equation}
which defines the embedding.
\item
The one-form
\begin{equation}
C(p,p,\cdot) = C_{IJK} h^I h^J dh^K = \frac{1}{3} d{\cal V} \;,
\end{equation}
which is proportional to the differential of ${\cal V}$, and which therefore vanishes
precisely on tangent vectors of ${\cal H}$. 
\item
The symmetric two-form 
\begin{equation}
C(p,\cdot, \cdot) = C_{IJK} h^I dh^J dh^K = \frac{1}{6} \partial d {\cal V}\;,
\end{equation}
which is proportional to the Hessian of the function ${\cal V}$. If this two-form 
is non-degenerate, it defines a Hessian metric on $U\subset \bR^{n+1}$. 
\end{enumerate}
Since $U$ is equipped with a 3-conical Hessian metric, we can identify it with
the CASR manifold  $M$ of the previous section. 

One defines the conjugate or dual coordinates
\begin{equation}
h_I := C_{IJK} h^J h^K \;,
\end{equation}
so that ${\cal V}=h_I h^I$, $d{\cal V} = 3 h_I dh^I$. The dual coordinates $h_I$ are,
up to a numerical factor, the dual affine coordinates of the Hessian structure defined
by $C(p,\cdot, \cdot)$. 

We claim that $g_{\cal H}$ is 
proportional to the pullback of the Hessian metric 
$\partial d {\cal V}$ to ${\cal H}$: 
\begin{equation}
g_{\cal H}(X,Y)_p  = - 3 C(p,X,Y) = - \frac{1}{2} (\partial^2_{X,Y} {\cal V})_{|p}  \;,
\end{equation}
for all tangent vectors $X,Y=T_p {\cal H}$. 
To show this we extend the tangent vector fields $X,Y$ to a neighbourhood
$U=U_\epsilon$ of ${\cal H} \subset \mathbb{R}^{n+1}$, such that 
$X({\cal V})=Y({\cal V})=0$. In other words the extended vector fields $X,Y$ are 
tangent to the local foliation of $\mathbb{R}^{n+1}$ by hypersurfaces
${\cal H}_\gc  = \{ {\cal V} =\gc \} $. 
The Hessian of the function ${\cal V}$ is\footnote{We refer to \ref{App:ConnVB} for the
definition of higher covariant derivatives with respect to vector fields, and the definition 
of the Hessian of a function with respect to a general linear connection.}
\begin{equation}
\partial^2_{X,Y} {\cal V} = X(Y ({\cal V})) - (\partial_X Y) ({\cal V} ) = X^I Y^J {\cal V}_{IJ} 
\end{equation}
so that on tangent vector fields of ${\cal H}$:
\begin{equation}
(\partial^2_{X,Y}{\cal V})_{p} =( - \partial_X Y) ({\cal V})_p = - 3 C(p,p,\partial_X Y) \;,\;\;p\in {\cal H}\;.
\end{equation}
In the second step we used the formula 
\begin{equation}
Z ({\cal V})_p = Z^L \partial_{L} (C_{IJK} h^I h^J h^K)  = 3 Z^L C_{LJK} h^J h^K = 3 C(p,p,Z) \;.
\end{equation}
Using (\ref{PAHS1}) we obtain
\begin{equation}
(\partial^2_{X,Y}{\cal V})_{p} =-  3 C(p,p,\nabla_X Y) - 2 g_{\cal H}(X,Y) C(p,p,p) = - 2 g_{\cal H}(X,Y) \;,
\end{equation}
where we used that $C(p,p,\cdot)$ vanishes on tangent vector of ${\cal H}$, and that
$C(p,p,p)=1$ for $p\in {\cal H}$. 
Thus $g_{\cal H}$ agrees with $-\frac{1}{2} \partial d {\cal V}= - 3 C(p,\cdot, \cdot)$ on tangent vectors,
and we can therefore extend $g_{\cal H}$ to a Hessian metric $g = -\frac{1}{2} \partial d {\cal V}$ 
with Hesse potential $-\frac{1}{2}{\cal V}$ 
in a neighbourhood $U_\epsilon$ of  ${\cal H}$. The metric $g=h_{IJ} dh^I dh^J$
is the 3-conical ASR metric denoted $g_M$, which occurred previously  in 
the superconformal quotient construction. In local coordinates 
\begin{equation}
 h_{IJ} = -\frac{1}{2} \partial^2_{I,J} {\cal V}
 = - 3 C_{IJK} h^K \;.
 \end{equation}
The torsion-free connection $\nabla$ is not the Levi-Civita connection $D$ of the 
metric $g_{\cal H}=\iota^* g$. The connections $\nabla$ and $D$  can be related 
using a tensor $S$, which is defined 
in terms of the  trilinear form $C$:
\begin{equation}
\label{gC}
g(S_XY, Z) = \frac{3}{2} C(X,Y,Z) \;,
\end{equation}
where $X,Y,Z$ are vector fields tangent to ${\cal H}$. 
Now we define a new connection $D$ by\footnote{Note that 
compared to \cite{Alekseevsky:2001if} the symbols $D$ and $\nabla$ have been exchanged.}
\begin{equation}
D = \nabla - S \;.
\end{equation}
To show that $D$ is the Levi-Civita connection of $g_{\cal H}$ 
we must prove that $D$ is metric and torsion-free. The total symmetry of the
trilinear form implies $S_XY=S_YX$, and since $\nabla$ is torsion-free, it follows
that $D$ is torsion-free.  It remains to show $D$ is metric, that is 
\begin{equation}
(D_X g)(Y,Z) = X g(Y,Z) - g(D_XY,Z) - g(Y, D_X Z) = 0 \;,
\end{equation}
where $X,Y,Z$ are tangent to ${\cal H}$.  We extend $X,Y,Z$ to $U =U_\epsilon$
such that $X({\cal V})=Y({\cal V})=Z({\cal V})=0$. 
Substituting in $D=\nabla - S$ and using (\ref{PAHS1}), together with
the fact that $\xi$ is $g$-orthogonal to tangent vectors, we find
\begin{eqnarray}
(D_X g)(Y,Z) &=& X g(Y,Z) - g(\partial_X Y, Z) - g(Y,\partial_X Z)  + g(S_XY,Z)
+ g(Y,S_X Z)  \nonumber  \\
&=& (\partial_X g)(Y,Z) + 3 C(X,Y,Z) \;, 
\end{eqnarray}
where we used the relation between the difference tensor $S$ and the trilinear form $C$
in the second step. Now we use that $g$ is Hessian:
\begin{equation}
(\partial_X g)(Y,Z) 
= -\frac{1}{2} \partial^3_{X,Y,Z} {\cal V} = - 3 C(X,Y,Z) \;.
\end{equation}
and therefore $(D_X g)(Y,Z)=0$, as required to show that $D$ is the Levi-Civita
connection of $g_{\cal H}$. 
We remark that the metric $g_{\bar{M}}=g_{\cal H}$ defined on the hypersurface $\bar{M}={\cal H}$
is not a Hessian metric. Moreover, the connections $\nabla$ and $D$ do not 
define flat connections on ${\cal H}$.

\subsection{Conical and projective special real geometry in local coordinates \label{sect:special-real-local-coordinates}}

In this section we derive explicit expressions for various quantities
in terms of local coordinates on the CASR manifold $M$ and on the
PSR manifold $\bar{M} \cong {\cal H}$. Since we are interested in 
local expressions we assume that ${\cal H}$ is embedded into $M$,
rather than only immersed, and take $M$ to be foliated by hypersurfaces
${\cal H}_{\gc}$. 
We will relate the 
notation and convention used in the previous sections to those of
\cite{Gunaydin:1983bi}, where the geometry of five-dimensional
vector multiplets coupled to Poincar\'e supergravity was derived
originally. 

As in section \ref{sect:PAHS} and in \cite{Gunaydin:1983bi} affine coordinates
on $M\cong U \subset \mathbb{R}^{n+1}$ are denoted $h^I$, $I=0, \ldots, n$,
local coordinates on ${\cal H}$ are denoted $\phi^x$, $x=1, \ldots, n$ and
the Hesse potential is denoted ${\cal V}$. In section  \ref{sect:ConHess}
these quantities were denoted $q^a$, $x^i$ and $H$, respectively. 
On $M$ we are using a second
coordinate system, which consists of a coordinate along the integral lines
of the Euler field $\xi$, 
together with coordinates on the level surfaces of
the Hesse potential. 
Since the Euler field is transversal to ${\cal H}$, the CASR manifold $M$ is foliated
by hypersurfaces ${\cal H}_\gc = \{ {\cal V} = \gc \}$. We can extend the coordinates
$\phi^x$ to $M$ by imposing that two points $p\in {\cal H}$ and $p' \in {\cal H}_\gc$ 
have the same coordinates $\phi^x$ is they lie on the same integral line of $\xi$. 
With regard to the transversal coordinate, the two natural choices are $\rho$ and 
$r= e^\rho$, defined by
\begin{equation}
\xi = h^I \partial_I = \partial_\rho = r  \partial_r \;.
\end{equation}
The differential $\frac{\partial h^I}{\partial \phi^x}$ of the embedding
\begin{equation}
\iota_\gc \;:  {\cal H}_\gc \ni \phi^x \mapsto h^I \in U 
\end{equation}
allows to pull-back tensor components to ${\cal H}_\gc$. 
Following \cite{Gunaydin:1983bi} we define the rescaled quantities
\begin{equation}
\label{hIx}
h^I_x = - \sqrt{\frac{3}{2}} \partial_x h^I \;, \;\;\;
h_{Ix} = \sqrt{\frac{3}{2}} \partial_x h_I 
\end{equation}
for later convenience. Given the definitions
\begin{equation}
\label{Vcasr}
{\cal V} = C_{IJK} h^I h^J h^K \;,\;\;\; h_I = C_{IJK} h^J h^K 
\end{equation}
for the Hesse potential and for the dual coordinates,\footnote{Remember that the $h_I$ then 
differ from the standard dual coordinates of Hessian geometry by a factor. }
we note the following relations:
\begin{equation}
h^I h_I = {\cal V} \Rightarrow h^I_x h_I = 0  =  h^I h_{Ix} \;.
\end{equation}
The second relation follows because derivatives $\partial_x$ are taken
along hypersurfaces ${\cal H}_\gc$. Note that here and in the following
some of our relations will differ from those found in  \cite{Gunaydin:1983bi}
by factors of ${\cal V}$. The reason is that the relations given in 
 \cite{Gunaydin:1983bi} are valid on ${\cal H}$, that is for ${\cal V}=1$, whereas
 we extend these relations to all of $M$. We now specify the relevant rank two symmetric
 tensor fields on $M$.
 \begin{itemize}
 \item
 The CASR metric on $M$ is
 \begin{equation}
 g_M = - \frac{1}{2} \partial^2 {\cal V} = h_{IJ} dh^I dh^J \;,\;\;\;
 h_{IJ} = - \frac{1}{2} \partial^2_{I,J} {\cal V} = - 3 C_{IJK} h^K \;.
 \end{equation}
 Compared to section \ref{sect:ConHess} this corresponds to the choice
 $H=-\frac{1}{2} {\cal V}$ while identifying the coordinates $h^I$ with the
 coordinates $q^a$. 
 \item
The 0-conical metric on $M$ is 
 \begin{eqnarray}
g^{(1)}_M &=& - \frac{1}{3} \partial^2 \log {\cal V} = a_{IJ} dh^I dh^J \;, \\
a_{IJ} &=& -\frac{1}{3} \partial^2_{I,J} \log {\cal V} = 
 \frac{-2 C_{IJK} h^K {\cal V} + 3 h_I h_J}{ {\cal V}^2} \;.
\end{eqnarray}
 Compared to section \ref{sect:ConHess} this corresponds to the choices
 $\ga = - \frac{1}{3}$ and $\gb = -\frac{1}{2}$. We note that with this convention 
 $\xi$ has unit norm, $g_M^{(1)}(\xi, \xi)=1$, while on tangent vectors $T,S$ we find
 $g_M^{(1)}(T,S) = \frac{3}{2{\cal V}} g_M(T,S)$. 
 \item
 The projectable tensor on $M$ is
 \begin{equation}
 g^*_M = \frac{-2 C_{IJK} h^K {\cal V} + 2 h_I h_J}{{\cal V}^2} dh^I dh^J \;,
 \end{equation}
 since
 \begin{equation}
 g^*_M(\xi, \cdot) = \frac{-2C_{IJK} h^I h^K {\cal V} + 2 h^I h_I h_J}{{\cal V}^2} = 0 \;.
 \end{equation}
 Note that
 \begin{equation}
 g_M^{(1)} = g^*_M  + \frac{h_I h_J}{{\cal V}^2} dh^I dh^J 
 \end{equation}
 is the product decomposition of the 0-conical metric into the projectable
 tensor and the square of a one-form dual to the Euler field $\xi$.  
   \item
 The PSR metric $g_{\cal H}$ is the pullback of the CASR metric $g_M$ to 
${\cal H}$, but differs by a factor $\frac{3}{2}$ from the pullback of the 0-conical
 metric $g^{(1)}_M$, which makes the definition (\ref{hIx}) convenient:
 \begin{equation}
g_{xy} = h_{IJ} \partial_x h^I \partial_y h^J = a_{IJ} h^I_x h^J_y \;.
\end{equation}
\end{itemize}
We would also like to give expressions for the horizontal lifts of tensors from 
${\cal H}$, or more generally from ${\cal H}_\gc$, to $M$. For this it is
useful to note that 
\begin{equation}
h_I = {\cal V}  a_{IJ} h^J \;,\;\;\; h_{Ix} = {\cal V} a_{IJ} h^J_x \;,\;\;\;
h_{Ix} h^I_y = {\cal V} a_{IJ} h^I_x h^J_y  = {\cal V} g_{xy}\;.
\end{equation}
We also define
\begin{equation}
h^x_I = g^{xy} h_{Iy} \;,\;\;\;h^{Ix} = g^{xy} h^I_y \;.
\end{equation}
Then the quantities $h^x_I$ can be used to lift tensors from ${\cal H}$ to $M$, and to 
convert tensors from coordinates $(\rho, \phi^x)$ to coordinates $h^I$. For example, the
components of the horizontal lift of $g_{xy}$ to $M$ are
\begin{equation}
\frac{3}{2} \frac{1}{ {\cal V}^2 } g_{xy} h^x_I h^y_J =
\frac{3}{2} g^*_{IJ} = \frac{3}{2} \left( \frac{ - 2 C_{IJK} h^K {\cal V} + 2 h_I h_J}{{\cal V}^2}   \right)\;.
\end{equation}
To verify this we evaluate the tensor on the left hand side on the coordinate frame
\begin{equation}
\xi = h^I \partial_I = \partial_\rho  \;,\;\;\; \partial_u = \partial_u h^I \partial_I \;.
\end{equation}
Firstly, $g^*_{IJ} h^J = 0$, so that $\xi$ is in the kernel. On tangent vectors we find
\begin{equation}
\left( \frac{3}{2 {\cal V}^2 } g_{xy} h^x_I h^y_J  \right) \partial_u h^I \partial_v h^J = 
\frac{1}{{\cal V}^2} g_{xy} h^x_I h^J_y h^I_u h^J_v = g_{uv} \;,
\end{equation}
as required. 

Similarly, the 0-conical metric $a_{IJ}$ can be decomposed into a term
proportional to the horizontal lift of $g_{xy}$ and an orthogonal complement:
\begin{equation}
\label{agh}
a_{IJ} = \frac{g_{xy} h^x_I h^y_J + h_I h_J}{{\cal V}^2} \;.
\end{equation}
This can again be verified by evaluation on the coordinate frame $\partial_\rho, \partial_x$.
We find
\begin{equation}
a_{IJ} h^I h^J  = \frac{h_I h^I h_J h^J}{ {\cal V}^2} = 1 = g^{(1)}_M(\xi, \xi) 
\end{equation}
and  
\begin{equation}
a_{IJ} \partial_u h^I \partial_v h^J = 
\frac{2}{3 {\cal V}^2} g_{xy} h^x_I h^y_J  h^I_u h^J_v  = \frac{2}{3} g_{uv} = g^{(1)}_M(\partial_u, \partial_v)\;,
\end{equation}
while $a_{IJ} h^I h^J_x=0 \Rightarrow g^{(1)}_M(\xi, \partial_x)=0 $, thus verifying that (\ref{agh}) 
are the coefficients of the 0-conical metric $g^{(1)}_M$. To convert these coefficients
from the linear coordinates $h^I$ to the coordinates $(\rho, \phi^x)$, 
we compute
\begin{eqnarray}
a_{IJ} dh^I dh^J &=& {\cal V}^{-2} g_{xy} h^x_I h^y_J \frac{2}{3} h^I_u h^J_v d\phi^u d\phi^v 
+ {\cal V}^{-2}  h_I h_J h^I h^J d\rho^2 \nonumber \\
&=& \frac{2}{3} g_{xy} d\phi^x d\phi^y + d\rho^2 \;.
\end{eqnarray}
Here we have substituted in $a_{IJ}$ and used that $h^x_I h^I =0$ to simplify the first term. 
In the second step we used 
\begin{equation}
\xi = h^I \partial_I = \partial_\rho \Rightarrow \xi^\flat =d\rho = {\cal V}^{-1} h_I dh^I \;.
\end{equation}

Next, we express the connections $D$ and $\nabla$ in local coordinates $\phi^x$
on ${\cal H}$, following \cite{Alekseevsky:2001if}. 
Let $X$ be a vector field tangent to ${\cal H}$. Then
\begin{equation}
X = X^I \partial_I = X^x \partial_x \Rightarrow X^I = X^x \partial_x h^I \;.
\end{equation}
Equation (\ref{PAHS1}) becomes
\begin{equation}
\label{Decomp_local}
\partial_x ( Y^y \partial_y h^I) = (\nabla_x Y^y) (\partial_y h^I) + \frac{2}{3}g_{xy} h^I  Y^y \;.
\end{equation}
Rewriting (\ref{gC}) in local coordinates we obtain
the relation 
\begin{equation}
\label{gC_local}
\nabla_x Y^y = D_x Y^y  + \frac{3}{2} C_{xz}^y Y^z
\end{equation} 
between the connections $D$ and $\nabla$ evaluated on tangent vectors $X,Y$,
where 
\begin{equation}
C_{xyz} := C_{IJK} \partial_x h^I \partial_y h^J \partial_z h^K \;.
\end{equation}
is the pullback of the trilinear form to ${\cal H}$.
Combining (\ref{Decomp_local}) and (\ref{gC_local}) we obtain
\begin{eqnarray}
&& \partial_x (Y^y \partial_y h^I) = 
(D_x Y^y) (\partial_y h^I) + Y^y D_x \partial_y h^I =
(\nabla_x Y^y) (\partial_y h^I) + \frac{2}{3} g_{xy} h^I Y^y  \nonumber \\
& \Rightarrow & 
Y^y D_x \partial_y h^I = (\nabla_x Y^y - D_x Y^y) \partial_y h^I + h^I g_{xy} Y^y =
\frac{3}{2} C^z_{xy} Y^y \partial_z h^I + \frac{2}{3}  h^I g_{xy} Y^y  \nonumber \\
& \Rightarrow & 
D_x \partial_y h^I = \frac{3}{2} C^z_{xy} \partial_z h^I +  \frac{2}{3} h^I g_{xy} \;.   \label{Ddh}
\end{eqnarray}

The corresponding formula (2.16)  in \cite{Gunaydin:1983bi} is 
\begin{equation}
D_x h^I_y = - \sqrt{\frac{2}{3}} \left( g_{xy} h^I + T_{xy}^z h^I_z \right) 
\Leftrightarrow
D_x \partial_y h^I = \frac{2}{3} g_{xy} h^I - \sqrt{\frac{2}{3}} T_{xy}^z \partial_z h^I \;.
\end{equation}
Matching with our formula requires
\begin{equation}
\frac{3}{2} C_{xyz} = - \sqrt{\frac{2}{3}} T_{xyz} 
\Rightarrow
T_{xyz} = - \left( \frac{3}{2} \right)^{3/2} C_{IJK} \partial_x h^I \partial_y h^J \partial_z h^K \;.
\end{equation}
The constant tensor $C_{IJK}$ on $M$ can be decomposed as
\begin{equation}
C_{IJK} = \frac{5}{2 {\cal V}^2} h_I h_J h_K + \frac{3}{2} a_{(IJ} h_{K)} + \frac{1}{{\cal V}^2} 
T_{xyz} h^x_I h^y_J h^z_K  \;.
\end{equation}
To verify this decomposition we contract $C_{IJK}$ with the vectors of the frame $\xi=h^I\partial_I
=\partial_\rho$ and $\partial_x = \partial_x h^I \partial_I$. 
\begin{itemize}
\item
Contraction with three tangent vectors gives precisely the pullback of $C_{IJK}$ to ${\cal H}_{\gc}$
\begin{equation}
C_{IJK} \partial_x h_I \partial_y h_J \partial_z h_K = 
C_{xyz} = C(\partial_x, \partial_y, \partial_z) \;.
\end{equation}
\item
Contracting once with the Euler field $\xi$ we obtain the two-form $C(\xi, \cdot, \cdot)$
with components $C_{IJK} h^K$ on the left hand side. When applying the same contraction on 
the right hand side the third term does not contribute, and the contributions from the first and
second term combine in $C_{IJK} h^K$.  
\end{itemize}
We remark that the corresponding  formula (2.12) of \cite{Gunaydin:1983bi} is recovered
for ${\cal V}=1$. In \cite{Gunaydin:1983bi} one can also find expressions for the 
curvature tensors of the CASR metric  $g_M$ and of the PSR metric $g_{{\cal H}}$, but we 
will not need these for our applications.

\section{Five-dimensional vector multiplets \label{sec:5d_VM}}

\subsection{Rigid vector multiplets \label{Sect:Lagr5drVM}}

In this section we present rigid five-dimensional vector multiplets,
focussing on the bosonic part of the Lagrangian. We follow
\cite{Cortes:2003zd}, where an off-shell realization has been worked out,
based on the work of \cite{Bergshoeff:2001hc} on the superconformal case. 
The components of a five-dimensional rigid off-shell vector multiplet are
\begin{equation}
(A_\mu, \lambda^i, \sigma, Y^{ij}) \;,
\end{equation}
where $\mu=0,1,2,3,4$ is the Lorentz index, and $i,j=1,2$ is 
an internal index, transforming in the fundamental representation 
of the R-symmetry group $SU(2)_R$. R-symmetry indices $i,j$
are raised and lowered using 
\begin{equation}
(\varepsilon_{ij})  = \left( \begin{array}{cc}
0 & 1 \\
-1 & 0 \\
\end{array} \right)
\end{equation}
and $\varepsilon^{ij} := \varepsilon_{ij}$.\footnote{Note that $(\varepsilon^{ij})$ is
minus the inverse of $(\varepsilon_{ij})$. This choice is consistent with the 
NW-SE convention for the $SU(2)_R$ indices.} 
$A_\mu$ is a vector field,
$\lambda^i$, $i=1,2$ is an $SU(2)_R$  doublet of symplectic Majorana spinors,
$\sigma$ is a real scalar, and $Y^{ij}=Y^{ji}$ are auxiliary
fields, subject to the reality condition
\begin{equation}
(Y^{ij})^* = Y^{kl} \varepsilon_{ki} \varepsilon_{lj} = Y_{ij} \;.
\end{equation}
Thus $Y^{ij}$ has three independent real components. 
Taking into account the reality conditions, a vector multiplet has 
8 bosonic and 8 fermionic off-shell degrees of freedom. These
reduce  to $4+4$ on-shell
degrees of freedom upon imposing the equations of motion.

We consider an arbitrary number of vector multiplets, labelled by 
$I=1, \ldots, n$. The bosonic part of the Lagrangian 
worked out in \cite{Cortes:2003zd} is
\begin{eqnarray}
{L} &=& h_{IJ} \left( - \frac{1}{2} \partial_\mu
             \sigma^I \partial^\mu \sigma^J - \frac{1}{4} F^I_{\mu
             \nu} F^{J \; \mu \nu}
                          + Y^I_{ij} Y^{J\;ij} \right) \nonumber  \\
&& - h_{IJK}  \frac{1}{24} \epsilon^{\mu \nu \lambda \rho
   \sigma}
A^I_{\mu} F^{J}_{\nu \lambda} F^K_{\rho \sigma} \;.
\label{5d_rigid_VM_Lagrangian}
\end{eqnarray}
Here $h_I, h_{IJ}, h_{IJK}$ denote derivatives of a function $h$ of
the  scalar fields $\sigma^I$, 
\begin{equation}
h_I =\partial_I h \;,\;\;\;
h_{IJ} =\partial^2_{I,J} h \;,\;\;\;
h_{IJK} =\partial^2_{I,J,K} h \;.
\end{equation}

Since the Chern-Simons term must be gauge invariant up to boundary
terms, $h_{IJK}$ must be constant, which implies that $h$ must be
a cubic polynomial. The special case where $h$ is a quadratic polynomial
corresponds to a free theory, while lower degrees of $h$ lead to
degenerate kinetic terms and can be discarded. Thus the scalar manifold 
of a theory of five-dimensional rigid vector multiplets is an affine
special real manifold, as defined 
in section \ref{Sect:SR}, see Definition \ref{def:ASK}. 

We remark that 
compared to \cite{Cortes:2003zd} we have changed the definition 
of the $\epsilon$-tensor by a sign, but we have kept the relation 
$\gamma_{\mu \nu \rho \sigma \tau}=
+ i \epsilon_{\mu \nu \rho \sigma \tau}\mathbbm{1}$,
which determines the sign of the Chern-Simons term,
by simultaneously changing the representation of the Clifford algebra. 
We refer to \cite{Gall:2018ogw} for a systematic discussion of the
relative factors and signs between the terms in the supersymmetry variations
and in the Lagrangians of five-dimensional vector multiplets. Note that in 
\cite{Gall:2018ogw}, the same convention  $\epsilon_{01235}=1$ for the
$\epsilon$-tensor was used as in this review, but in combination with 
a different sign in the relation between $\gamma_{\mu \nu \rho \sigma\tau}$ 
and the $\epsilon$-tensor (that is,  
$\gamma_{\mu \nu \rho \sigma \tau}=
- i \epsilon_{\mu \nu \rho \sigma \tau}\mathbbm{1}$) this resulted in a Chern-Simon
term with opposite sign compared to \eqref{5d_rigid_VM_Lagrangian}.
The choices made in this review are more convenient for matching with the supergravity
literature.

\subsection{Rigid superconformal vector multiplets \label{Sect:rSCVM}}

We next specialize to the case where the vector multiplet theory 
is superconformal, following \cite{Bergshoeff:2001hc}.
Superconformal invariance
implies the Hesse potential must be a {\em homogeneous} cubic polynomial,
which makes the scalar manifold a conical affine special real manifold
in the sense of Definition \ref{def:CASR}.
For later convenience we choose the Hesse potential
\begin{equation}
h=- \frac{1}{2} C_{IJK} \sigma^I \sigma^J \sigma^K \;,
\end{equation}
where $C_{IJK}$ are constants. Then
\begin{equation}
h_{IJ} =  - 3 C_{IJK} \sigma^K  \;,\;\;\;h_{IJK} = - 3 C_{IJK} \;,
\end{equation}
and the rigid superconformal vector multiplet Lagrangian is:
\begin{eqnarray}
\label{5dRigidSuperconformal}
L &=& 3 C_{IJK} \sigma^K 
\left( \frac{1}{4} F^I_{\mu \nu} F^{J\;\mu\nu}  +
\frac{1}{2} \partial_\mu \sigma^I \partial^\mu \sigma^J 
- Y^I_{ij} Y^{J ij} 
\right) \nonumber \\
&& + \frac{1}{8} \epsilon^{\mu \nu \rho \sigma \lambda} C_{IJK}  A^I_\mu F^J_{\nu \rho}
F^K_{\sigma \lambda} \;,
\end{eqnarray}
where we omitted all fermionic terms.

\subsection{Superconformal matter multiplets coupled to 
superconformal gravity \label{Sect:SCVM+SG}}

We will follow the superconformal approach to construct a theory 
of $n$ vector multiplets coupled to Poincar\'e supergravity. A comprehensive
review of the superconformal approach can be found in the textbook 
\cite{Freedman:2012zz}, and the elements relevant for this review have
been collected in \ref{sec:5dscg}.
The  superconformal approach is based on the observation that a theory of $n$ vector multiplets and
$n_H$ hypermultiplets coupled to
Poincar\'e supergravity is {\em gauge equivalent} to a theory of $n+1$ 
superconformal vector multiplets and $n_H+1$ superconformal hypermultiplets
coupled to conformal supergravity. Gauge equivalence means that the
Poincar\'e supergravity theory is obtained from the superconformal theory
by gauge fixing those superconformal symmetries that do not belong
to the Poincar\'e supersymmetry algebra. Conversely, a
Poincar\'e supergravity theory can be extended to a 
superconformal theory by adding one 
vector and one hypermultiplet which act as superconformal compensators.
That is, the additional symmetries are introduced by adding new degrees of freedom.

\subsubsection{Coupling of vector multiplets}

The bosonic Lagrangian for a rigid superconformal vector multiplet theory
was given in (\ref{5dRigidSuperconformal}). Since we need to start with $n+1$
superconformal vector multiplets we change the range of the indices $I,J, \ldots$ 
to $I,J=0,1, \ldots, n$. The next step is to promote the superconformal symmetry to
a local symmetry, and to add at least one hypermultiplet. Gauging the superconformal
symmetry involves replacing partial derivatives by superconformal covariant derivatives,
which contain the superconformal connections, or, in physics terminology, the superconformal
gauge fields. The superconformal gauge fields belong to the so-called
Weyl multiplet, together with certain auxiliary fields. We refer to 
\ref{sec:5dscg} for an overview. Our presentation will follow \cite{deWit:2009de},
but we will only retain the
connections and auxiliary fields which are relevant for the bosonic vector 
multiplet Lagrangian. The bosonic part of the locally
superconformally invariant vector multiplet Lagrangian can be brought to the form
\begin{eqnarray}
L_V &=& 3 C_{IJK} \sigma^K \left[ \frac{1}{2} {\cal D}_\mu \sigma^I 
{\cal D}^\mu \sigma^J + \frac{1}{4} F^I_{\mu \nu} F^{\mu \nu J} - Y^I_{ij} Y^{ijJ} 
-  3 \sigma^I F^J_{\mu \nu} T^{\mu \nu} \right] \nonumber \\
 && +  \frac{1}{8} C_{IJK} e^{-1} \epsilon^{\mu \nu \rho \sigma \tau} A^I_\mu F^J_{\nu\rho}
F^K_{\sigma \tau}\nonumber \\
&& + C_{IJK} \sigma^I \sigma^J \sigma^K \left( \frac{1}{8} R + 4 D + \frac{39}{2} T_{\mu \nu} T^{\mu 
\nu} \right) \;. \label{SCL}
\end{eqnarray}
Here 
\begin{equation}
{\cal D}_\mu \sigma^I = (\partial_\mu - b_\mu)  \sigma^I  \;,
\end{equation}
where $b_\mu$ is the gauge field for dilatations. $T_{\mu \nu}$ and $D$ are auxiliary fields belonging
to the Weyl multiplet. In the so-called K-gauge, to be introduced below, 
$R$ becomes the Ricci scalar associated to the 
space-time metric $g_{\mu \nu}$ with vielbein $e^a_{\mu}$ and vielbein determinant $e$. 
We refer to \ref{sec:5dscg} for details regarding the vielbein ad Ricci scalar.
In (\ref{SCL}) we have adapted the Lagrangian of \cite{deWit:2009de} to our conventions.
This changes the sign in front of the Ricci tensor and removes a factor $-i$ from 
the Chern-Simons term.\footnote{Note that \cite{deWit:2009de} use an imaginary totally antisymmetric
tensor defined by $\varepsilon_{01235} = i = i \epsilon_{01235}$. Taking this into account
the relation which determines the sign of the Chern-Simons term is the same:
$\gamma_{\mu \nu \rho \sigma \tau} = \varepsilon_{\mu \nu \rho \sigma \tau} \mathbbm{1} =
i \epsilon_{\mu \nu \rho \sigma \tau} \mathbbm{1}$.}

\subsubsection{Coupling of hypermultiplets}

The bosonic part of the locally superconformal hypermultiplet Lagrangian is
\begin{eqnarray}
L_H &=& - \frac{1}{2} \varepsilon^{ij} \Omega_{\alpha \beta} {\cal D}_\mu A^\alpha_i 
A^\beta_{j} 
+ \chi \left( - \frac{3}{16} R + 2 D + \frac{3}{4} T_{\mu \nu} T^{\mu \nu} \right) \;.
\end{eqnarray}
Here $A^\alpha_i$, where $\alpha=1, \ldots, 2n_H +2$ and $i,j=1,2$ encode the
$4n_H+4$ scalar degrees of freedom of the hypermultiplets. 
The quantity $\chi$ is the so-called {\em hyper-K\"ahler potential} and satisfies 
\begin{equation}
\label{HKpotential}
\varepsilon_{ij} \chi = \Omega_{\alpha \beta} A_i^\alpha A_j^\beta\;.
\end{equation}
We refer to  \ref{sec:5dscg} for explicit expressions for the covariant 
derivative ${\cal D}_\mu A^\alpha_i$ and the quantity $\Omega_{\alpha \beta}$.
The scalar geometry of rigid hypermultiplets is hyper-K\"ahler. If superconformal
symmetry is imposed the scalar multiplet is a hyper-K\"ahler cone, that is,
it admits a holomorphic and homothetic action of the group $\mathbb{H}^*$  of invertible quaternions. 
The relevant concepts of hyper-K\"ahler geometry are briefly reviewed in \ref{app:epsilon_quat}.

\subsubsection{Poincar\'e supergravity}

Combining the bosonic vector multiplet and hypermultiplet Lagrangians, we obtain:
\begin{eqnarray}
L &=& 3 C_{IJK} \sigma^K \left[ \frac{1}{2} {\cal D}_\mu \sigma^I 
{\cal D}^\mu \sigma^J + \frac{1}{4} F^I_{\mu \nu} F^{\mu \nu J} - Y^I_{ij} Y^{ijJ} 
- 3 \sigma^I F^J_{\mu \nu} T^{\mu \nu} \right] \nonumber \\
 && +  \frac{1}{8} C_{IJK} e^{-1} \epsilon^{\mu \nu \rho \sigma \tau} A^I_\mu F^J_{\nu\rho}
F^K_{\sigma \tau}\nonumber  \\
&& + \frac{1}{8} R \left( C_{IJK} \sigma^I \sigma^J \sigma^K - \frac{3}{2} \chi \right)
+ D \left(2\chi + 4 C_{IJK} \sigma^I \sigma^J \sigma^K \right) \nonumber \\
&& + T^{ab} T_{ab}  \left( \frac{3}{4} \chi + \frac{39}{2} C_{IJK} \sigma^I \sigma^J \sigma^K\right)\nonumber \\
&& - \frac{1}{2} \varepsilon^{ij} \Omega_{\alpha \beta} {\cal D}_\mu {\cal A}_i^a {\cal D}^\mu {\cal A}_j^\beta\;.
\end{eqnarray}

The auxiliary field $Y^I_{ij}$ has the field equation $Y^I_{ij}=0$ and can be 
eliminated trivially. The algebraic field equation for the auxiliary field $D$
can be used to eliminate $\chi$:
\begin{equation}
\chi = - 2 C_{IJK} \sigma^I \sigma^J \sigma^K \;.
\end{equation}
Substituting this back into the Lagrangian, we obtain
\begin{eqnarray}
L &=& 3 C_{IJK} \sigma^K \left[ \frac{1}{2} {\cal D}_\mu \sigma^I 
{\cal D}^\mu \sigma^J +\frac{1}{4} F^I_{\mu \nu} F^{\mu \nu J} 
- 3 \sigma^I F^J_{\mu \nu} T^{\mu \nu} \right]\nonumber  \\
 &&+ \frac{1}{8} C_{IJK} e^{-1} \epsilon^{\mu \nu \rho \sigma \tau} A^I_\mu F^J_{\nu\rho}
F^K_{\sigma \tau}\nonumber  \\
&& + \frac{1}{2} R C_{IJK} \sigma^I \sigma^J \sigma^K\nonumber  \\
&& +18 T^{ab} T_{ab}  C_{IJK} \sigma^I \sigma^J \sigma^K \nonumber \\
&& - \frac{1}{2} \varepsilon^{ij} \Omega_{\alpha \beta} {\cal D}_\mu {\cal A}_i^a {\cal D}^\mu {\cal A}_j^\beta\;.
\label{LVH}
\end{eqnarray}
In the next step we gauge-fix those superconformal transformations which are not
super-Poincar\'e transformations. Local dilatations are gauge-fixed by the so-called
D-gauge which imposes 
that the Einstein-Hilbert term acquires its canonical form:
\begin{equation}
C_{IJK} \sigma^I \sigma^J \sigma^K = \kappa^{-2} \;,
\end{equation}
where $\kappa = \sqrt{8\pi G_N}$ is the gravitational coupling constant and
$G_N$ is Newton's gravitational constant. 
This implies that $\chi = -2 \kappa^{-2}$, which because of  (\ref{HKpotential}) 
removes one real scalar degree of freedom from the
hypermultiplet sector. The superconformal symmetries include an $SU(2)$
symmetry which acts in the adjoint representation on the hypermultiplet 
scalars. Gauge fixing this symmetry removes another three real scalar degrees
of freedom. If we consider only one hypermultiplet 
at the superconformal level, i.e. $n_H=0$, then
all bosonic hypermultiplet degrees of freedom are removed and we can drop the last line
in (\ref{LVH}).\footnote{As 
required for consistency, gauge fixing fermionic superconformal symmetries
removes the fermionic partners of the four hypermultiplet scalars.}
Since we are interested in the vector multiplet Lagrangian, we will assume this here.
Note that since $\chi\not=0$, consistency of the procedure requires that at least
one superconformal hypermultiplet is present. This hypermultiplet is needed as a superconformal
compensator. 

For completeness we briefly mention what happens for $n_H>0$. The gauge
fixing removes one hypermultiplet, leaving a theory with $n_H$ hypermultiplets. 
The resulting scalar manifold of dimension $4n_H$ is a quaternion-K\"ahler manifold.
It was shown in \cite{Bagger:1983tt} that the scalar geometry of hypermultiplets
coupled to supergravity is quaternion-K\"ahler. In the superconformal approach 
the quaternion-K\"ahler manifold arises as the superconformal quotient of 
a hyper-K\"ahler cone \cite{deWit:1984rvr,Galicki:1992tm,deWit:1998zg,deWit:1999fp}.
We remark that the hypermultiplet Lagrangian
only couples gravitationally to the vector multiplet Lagrangian, and thus can
always be truncated out consistently. We now return to the case $n_H=0$.

The special superconformal transformations are gauge-fixed by the so-called
K-gauge, which eliminates the dilatation gauge field: $b_\mu=0$. This
replaces the covariant derivatives ${\cal D}_\mu\sigma^I$ by partial derivatives
$\partial_\mu \sigma^I$. Then
the bosonic Lagrangian is
\begin{eqnarray}
L &=& 3 C_{IJK} \sigma^K \left(
\frac{1}{2} \partial_\mu \sigma^I \partial^\mu \sigma^J
+ \frac{1}{4} F^I_{\mu \nu} F^{J \mu \nu} - 3 \sigma^I F^J_{\mu \nu} T^{\mu \nu} \right) \nonumber \\
&& + \frac{1}{8} C_{IJK} e^{-1} \epsilon^{\mu \nu \rho \sigma \tau} A^I_\mu F^J_{\nu\rho} 
F^K_{\sigma \tau} \nonumber \\
&& + \frac{1}{2\kappa^2} R + \frac{18}{\kappa^2} T_{\mu \nu} T^{\mu \nu}  \;.
\end{eqnarray}
Now we eliminate the auxiliary field $T_{\mu \nu}$ using its algebraic equation of motion
\begin{equation}
T_{\mu \nu} = \frac{\kappa^2}{4} C_{IJK} \sigma^I \sigma^J F^K_{\mu \nu} \;,
\end{equation}
resulting in 
\begin{eqnarray}
L &=& \frac{3}{2} C_{IJK}\sigma^K \partial_\mu \sigma^I \partial^\mu \sigma^J
+ \frac{1}{2\kappa^2} R + \frac{1}{8} C_{IJK} e^{-1} \epsilon^{\mu \nu \rho \sigma \tau} 
A^I_\mu F^J_{\nu\rho} F^K_{\sigma\tau} \nonumber \\
&& - \frac{3}{8} (-2 C_{IJK} \sigma^K + 3 \kappa^2 C_{IAB} \sigma^A \sigma^B
C_{JCD} \sigma^C \sigma^D) F^I_{\mu \nu} F^{J\mu \nu} \;.
\end{eqnarray}
The scalar fields $\sigma^I$ and couplings $C_{IJK}$ 
are dimensionful. We define dimensionless scalars $h^I$  and couplings ${\cal C}_{IJK}$ 
by 
\begin{equation}
h^I := \kappa \sigma^I \;,\;\;\; {\cal C}_{IJK} :=\frac{1}{\kappa} C_{IJK} \;.
\end{equation}
The scalars $h^I$ satisfy
\begin{equation}
\label{Constraint_h}
{\cal C}_{IJK} h^I h^J h^K = 1 \;.
\end{equation}
It is convenient to define
\begin{equation}
h_I = {\cal C}_{IJK} h^J h^K \;.
\end{equation}
In these new variables the Lagrangian becomes
\begin{eqnarray}
L &=&  \frac{1}{2\kappa^2} R + \frac{3}{2\kappa^2} {\cal C}_{IJK} h^K \partial_\mu h^I 
\partial^\mu h^J \nonumber \\
&& - \frac{3}{8} \left( -2 {\cal C}_{IJK} h^K + 3 h_I h_J\right) F^I_{\mu \nu} F^{J\mu \nu}  \nonumber\\
&& + \frac{\kappa}{8} {\cal C}_{IJK} e^{-1} \epsilon^{\mu \nu \rho \sigma \tau} 
A^I_\mu F^J_{\nu\rho} F^K_{\sigma\tau}\;.
\label{lagpoinc5d}
\end{eqnarray}
We would like to verify that this Lagrangian, which has been obtained using 
the superconformal approach, agrees with the bosonic part of the 
on-shell Poincar\'e supergravity Lagrangian constructed in \cite{Gunaydin:1983bi}.
The scalars $h^I$ already have the same normalization. Following \cite{Gunaydin:1983bi}
we define
\begin{equation}
\label{aijCh}
\stackrel{\circ}{a}_{IJ} := -2 {\cal C}_{IJK} h^K + 3 h_I h_J \;,
\end{equation}
and note that
\begin{equation}
h_I = \stackrel{\circ}{a}_{IJ} h^J \;.
\end{equation}
To obtain the same normalization of the vector fields as in  \cite{Gunaydin:1983bi}
we define
\begin{equation}
\tilde{A}^I_\mu := \sqrt{\frac{3}{2}} A^I_\mu  \;.
\end{equation}
We also note that the scalar fields $h^I$ are not independent, because 
they satisfy the constraint (\ref{Constraint_h}). This implies that 
$h_I \partial_\mu h^I=0$. Using this, the bosonic Lagrangian takes the form
\begin{eqnarray}
L &=& \frac{1}{2\kappa^2} R - \frac{3}{4\kappa^2} \stackrel{\circ}{a}_{IJ} \partial_\mu h^I \partial^\mu h^J
- \frac{1}{4} \stackrel{\circ}{a}_{IJ} \tilde{F}^I_{\mu \nu} \tilde{F}^{J\mu \nu} \nonumber \\
&& + \frac{\kappa}{6 \sqrt{6}} 
e^{-1} {\cal C}_{IJK} \epsilon^{\mu \nu \rho \sigma \lambda} \tilde{A}^I_\mu 
\tilde{F}^J_{\nu \rho} \tilde{F}^K_{\sigma\lambda}\;. \label{Lagr5dsugra1}
\end{eqnarray}
Finally, we introduce independent scalars $\phi^x$, $x=1, \ldots, n$
by solving the constraint (\ref{Constraint_h}). 
The metric $g_{xy}$ for the target space of the scalars $\phi^x$ 
is obtained by re-writing
the scalar term in the Lagrangian. The normalization chosen in 
\cite{Gunaydin:1983bi} is such that
\begin{equation}
\label{g2a}
- \frac{1}{2} g_{xy} \partial_\mu \phi^x \partial^\mu \phi^y =
- \frac{3}{4} \stackrel{\circ}{a}_{IJ} \partial_\mu h^I \partial^\mu h^J =
- \frac{3}{4} \stackrel{\circ}{a}_{IJ} \frac{\partial h^I}{\partial \phi^x} 
\frac{\partial h^J}{\partial \phi^y} \partial_\mu \phi^x \partial^\mu \phi^y \;.
\end{equation}
The resulting Lagrangian, 
\begin{eqnarray}
L &=& \frac{1}{2\kappa^2} R - \frac{1}{2\kappa^2} g_{xy} \partial_\mu 
\phi^x \partial^\mu \phi^y
- \frac{1}{4} \stackrel{\circ}{a}_{IJ} \tilde{F}^I_{\mu \nu} \tilde{F}^{J\mu \nu}  \nonumber \\
&& + \frac{\kappa}{6 \sqrt{6}} 
e^{-1} {\cal C}_{IJK} \epsilon^{\mu \nu \rho \sigma \lambda} \tilde{A}^I_\mu 
\tilde{F}^J_{\nu \rho} \tilde{F}^K_{\sigma\lambda}\;, \label{Lagr5dsugra2}
\end{eqnarray}
agrees with the corresponding terms in (2.7) of \cite{Gunaydin:1983bi} upon setting $\kappa=1$, and
taking into account a relative sign in the definition of the Riemann tensor.
When setting $\kappa=1$ we see that $g_{xy}$ is the PSR metric $g_{ {\cal H}}$
associated with the Hesse potential ${\cal V} = {\cal C}_{IJK} h^I h^J h^K$  and 
$\stackrel{\circ}{a}_{IJ}$ the restriction of the corresponding 0-conical metric 
$g^{(1)}_M = a_{IJ} dh^I dh^J$ 
to ${\cal H} = \{ {\cal V}=1 \}$, with 
the same normalization as in section \ref{sect:special-real-local-coordinates}.
Note that for $\kappa=1$ we have $h^I = \sigma^I$ and $C_{IJK}={\cal C}_{IJK}$.

The decomposition (\ref{agh}) of $a_{IJ}$ can be used to rewrite the Maxwell term\footnote{We set ${\cal V}=1$.}
\begin{eqnarray}
&& - \frac{1}{4} \stackrel{\circ}{a}_{IJ} F^I_{\mu \nu} F^{J\mu \nu} = 
- \frac{1}{4} g_{xy} h^x_I h^y_J F^I_{\mu \nu} F^{J\mu \nu} 
-\frac{1}{4} h_I h_J F^I_{\mu \nu} F^{J\mu \nu}  \\
&=&
-\frac{1}{4} g_{xy} {\cal F}^x_{\mu \nu} {\cal F}^{y\mu \nu} 
- \frac{1}{4} {\cal F}_{\mu \nu} {\cal F}^{\mu \nu} \;, \nonumber
\end{eqnarray}
where we have defined
\begin{equation}
{\cal F}_{\mu \nu} = h_I F^I_{\mu \nu} \;,\;\;\;
{\cal F}^x_{\mu \nu} = h^x_I F^I_{\mu \nu} \;.
\end{equation}
The $n$ field strengths ${\cal F}^x_{\mu \nu}$ belong to vector fields
${\cal A}^x_\mu$ which are the superpartners of the scalars $\phi^x$
under Poincar\'e supersymmetry.
The additional field strength ${\cal F}_{\mu \nu}$ belongs to a vector
field which is part of the Poincar\'e supergravity multiplet. In contrast, 
the $F^I_{\mu \nu}$ correspond to vector fields in the $n+1$ superconformal
vector multiplets. Thus the decomposition into components tangential and
orthogonal to ${\cal H}$ corresponds to mapping components of superconformal
multiplets to the corresponding Poincar\'e vector multiplets. In the superconformal
description there is a manifest linear action of the group $GL(n+1,\mathbb{R})$
on the field strength $F^I_{\mu \nu}$, and an associated action of the affine
group $GL(n+1,\mathbb{R}) \ltimes \mathbb{R}^{n+1}$ on the scalars 
$h^I$. In the gauge-fixed description this  is no longer manifest, because
there are only $n$ independent scalars, but $n+1$ vector fields. For this reason it 
is often advantageous to work in the superconformal formulation of the theory.

We can now decide which signature we should choose for the CASR metric
defining the superconformal theory. In the Poincar\'e theory, $\stackrel{\circ}{a}_{IJ}$ and $g_{xy}$ 
must be positive definite, in order that the vector and scalar fields have positive
kinetic energy. From the above decomposition of the Maxwell term it is clear
that $\stackrel{\circ}{a}_{IJ}$ is positive definite if and only if $g_{xy}$ is positive definite. 
Using the relations \eqref{Family} -- \eqref{Norms} between the metrics,
we see that $h_{IJ}$ must have Lorentz signature with the time-like direction 
along the integral lines of the Euler field $\xi$.\footnote{As we have seen, the
overall sign of the CASR metric is not relevant.} The direction normal to ${\cal H}$
corresponds to the extra `compensating' vector multiplet, which shows that
the kinetic term of the compensator has a flipped sign.

\subsection{$R^2$-terms in five dimensions}

We briefly describe the coupling of vector multiplets to 
$R^2$-interactions encoded in the square of the Weyl multiplet using the superconformal approach \cite{Hanaki:2006pj,deWit:2009de}.

The bosonic part of the Lagrangian containing the higher-derivative couplings reads, in the notation used in  \cite{deWit:2009de},\footnote{Note that our definition of the Riemann tensor differs from the one in 
 \cite{deWit:2009de} by an overall minus sign.}
\begin{eqnarray}
L_{R^2} &=& \tfrac{1}{64} \, c_I \, \sigma^I \, {R}_{ab}{}^{cd} (M) {R}_{cd}{}^{ab} (M) 
\nonumber\\
&&- \tfrac{3}{16} c_I \left(10 \sigma^I T_{ab} - F_{ab}^I \right) \, {R}_{cd}{}^{ab} \, T^{cd} + \tfrac32 c_I \sigma^I \, T^{ab} [ {\cal D}^c, {\cal D}_a]  T_{bc} \nonumber\\
&& -  c_I \, \sigma^I \, R_{ab}  \left( T^{ac} T^b{}_c - \tfrac12 \eta^{ab} T^{cd}ÊT_{cd} \right) + \dots \;,
\end{eqnarray}
where ${R}_{ab}$ denotes the Ricci tensor \eqref{Ricci}, and 
where we have only displayed the terms that are relevant for computing 
Wald's entropy of static BPS black holes, see section \ref{secdic}.
We refer to  \cite{deWit:2009de}
for the complete set of bosonic terms.
The $c_I$ denote arbitrary real constants.

Using \eqref{confconst5d} and \eqref{f-FP}, we obtain 
\begin{equation}
{R}_{ab}{}^{cd} (M) = 
R_{ab}{}^{cd} -  \tfrac43 \left( R_{[a}{}^{[c} - \tfrac18 R \, \delta_{[a}{}^{[c} \right) \delta_{b]}{}^{d]} \;,
\end{equation}
which,  in the K-gauge $b_{\mu} =0$, denotes the Weyl tensor in five dimensions.

For future reference, we collect the bosonic terms in the $R^2$-corrected Lagrangian that are relevant for computing
the entropy of static BPS black holes 
using Wald's definition of black hole
entropy \eqref{entronoether},
\begin{eqnarray}
L&=&  3 \, C_{IJK} \,  \sigma^K \left[ \tfrac12 \, {\cal D}_{\mu} \sigma^I {\cal D}^{\mu} \sigma^J+ \tfrac14
F_{\mu \nu}^I F^{\mu \nu J} - 3\,  \sigma^I F_{\mu \nu}^J \, T^{\mu \nu}
\right] \nonumber\\
&& + \tfrac18 \, R \left( - \tfrac32 \, \chi + C_{IJK} \, \sigma^I \sigma^J  \sigma^K \right) 
 \nonumber\\
&& + T_{ab} T^{ab} \left( \tfrac34 \, \chi + \tfrac{39}{2} \, C_{IJK} \, \sigma^I  \sigma^J  \sigma^K \right)  \nonumber\\
&& +  \tfrac{1}{64} \, c_I \, \sigma^I \, {R}_{ab}{}^{cd} (M) {R}_{cd}{}^{ab} (M) 
\nonumber\\
&&- \tfrac{3}{16} c_I \left(10 \sigma^I T_{ab} - F_{ab}^I \right) \, {R}_{cd}{}^{ab} \, T^{cd} + \tfrac32 c_I \sigma^I \, T^{ab} [ {\cal D}^c, {\cal D}_a]  T_{bc} \nonumber\\
&& -  c_I \, \sigma^I \, R_{ab}  \left( T^{ac} T^b{}_c - \tfrac12 \eta^{ab} T^{cd}ÊT_{cd}  \right) \;.
\label{lagR2}
\end{eqnarray}

\section{Electric-magnetic duality \label{sec:emduality}} 

Electric-magnetic duality in four dimensions is a characteristic feature of Maxwell's equations in vacuum. It describes
the invariance of the combined system of equations of motion and Bianchi identities for the Maxwell gauge
field $A_{\mu}$ under rotations of the electric field into the magnetic field and vice-versa.  
Electric-magnetic duality is also present in ${\cal N}=2$ supergravity theories coupled to abelian ${\cal N}=2$ vector multiplets
in four dimensions \cite{Gaillard:1981rj,deWit:1984wbb}, and continues to hold when allowing for the coupling to a chiral background $\eta$ \cite{deWit:1996gjy}.
Theories of this type are based on holomorphic functions $F(X, \eta)$, and electric-magnetic duality is 
defined in terms of a symplectic vector constructed from $F(X, \eta)$. This will be reviewed in the following subsections.

Non-holomorphic functions $F$ are also of relevance and occur in various types of models 
\cite{Cardoso:2012nh,Cardoso:2014kwa}.
We will discuss three applications thereof, namely to 
point-particle Lagrangians 
that depend on coordinates and velocities, as well as on parameters $\eta$, in section \ref{sec:ubiquity} below, 
to topological string theory in section \ref{sec:hess}, and to
the Born-Infeld-dilaton-axion system in section \ref{bidaf}.

We begin by reviewing the formulation of point-particle Lagrangians in terms of a function $F$
given in \eqref{FOm} below, following \cite{Cardoso:2012nh}.  When
passing over to the Hamiltonian description, one obtains a description based on a real Hesse potential 
associated to $F$.  In this context, canonical transformations on phase space play a similar role to electric-magnetic
duality transformations in Maxwell-type theories.
Then
we turn to electric-magnetic duality in Maxwell-type theories at the two-derivative level which arise in the ${\cal N}=2$ 
supergravity context, and subsequently we allow for the presence of a chiral background.

\subsection{Point-particle models and $F$-functions}
\label{sec:ubiquity}

In the following, we will review \cite{Cardoso:2012nh} how general point-particle Lagrangians (that depend on coordinates
and velocities, as well as on real parameters $\eta$) can be recast in terms of a function 
$F$ of the form
\beq
F(x,{\bar x}, \eta) = F^{(0)}(x) + 2 i \, \Omega (x, {\bar x}, \eta ) \;,
\label{FOm}
\eeq
where $\Omega$ is real.
This is achieved with the help of a theorem that states that
the dynamics of these models can be reformulated in terms of a 
symplectic vector $(X, \partial F/\partial X)$ constructed out of a 
complex function $F$ of the form \eqref{FOm}, whose real part
comprises the 
canonical variables of the associated Hamiltonian.

Let us consider a point-particle model 
described by a Lagrangian ${L}$
with  $n$
coordinates $\phi^i$ and $n$ velocities $\dot \phi^i$.  The associated canonical momenta $\partial L /
\partial \dot \phi^i$
will be denoted by $\pi_i$.
The Hamiltonian ${H}$ of the system, which follows from $L$ by Legendre transformation,
\begin{equation}
{H}(\phi,\pi)= \dot \phi^i\, \pi_i - {L}(\phi,\dot \phi) \;,
\label{eq:leg-l-h}
\end{equation}
depends on $(\phi^i, \pi_i)$, which are called canonical variables, since they
satisfy the canonical Poisson bracket relations. The variables $(\phi^i, \pi_i)$
can be interpreted as 
local coordinates on 
a symplectic manifold called the classical phase 
space of the system.  
In these coordinates, the symplectic 2-form is $d \pi_i \wedge d \phi^i$.
This 2-form is preserved under canonical transformations of $(\phi^i,\pi_i)$ given by
\begin{equation}
  \label{eq:symplectic-pq}
  \begin{pmatrix} 
    \phi^i\\[2mm] \pi_i
  \end{pmatrix}
  \mapsto
  \begin{pmatrix} 
    \tilde \phi^i\\[2mm] \tilde \pi_i
  \end{pmatrix}
  =
  \begin{pmatrix} 
    U^i{}_j& Z^{ij}\\[2mm]  W_{ij} & V_i{}^j 
  \end{pmatrix}
  \begin{pmatrix} 
    \phi^j\\[2mm] \pi_j
  \end{pmatrix}\;,
\end{equation}
where $U, V, Z$ and $W$ denote $n \times n$ matrices that satisfy the relations 
\begin{eqnarray}
U^T \, V - W^T \, Z = V^T \, U - Z^T \, W = \mathbbm{1} \;,\nonumber\\
U^T \, W = W^T \, U \;\;\;,\;\;\; Z^T \, V = V^T \, Z \;.
\label{eq:matrix-sympl}
\end{eqnarray}
Thus, the transformation \eqref{eq:symplectic-pq}
constitutes an element of 
${\rm Sp} (2n, \mathbb{R})$.  This transformation leaves the Poisson brackets invariant. 
The Hamiltonian transforms
as a function under symplectic transformations, i.e. $\tilde{{H}} (\tilde \phi, \tilde \pi)
= {H} (\phi, \pi)$.  When the Hamiltonian is invariant under a subset of 
${\rm Sp} (2n, \mathbb{R})$ transformations, this subset describes a symmetry of the system.  This 
invariance is often called
duality invariance.

Now we give the theorem of \cite{Cardoso:2012nh} that states  that the Lagrangian $L(\phi, \dot \phi)$
can be reformulated in terms of a complex function $F(x, \bar x)$
based on complex variables $x^i$, such that the 
canonical coordinates $(\phi^i, \pi_i)$ coincide with (twice) the real part of $(x^i, F_i)$, where
$F_i = \partial F(x, \bar x)/\partial x^i$.  

{\thm {\bf Point-particle Lagrangians and $F$-functions.}
Given a Lagrangian ${L}(\phi,\dot \phi)$ depending on $n$
coordinates $\phi^i$ and $n$ velocities $\dot \phi^i$, with corresponding
Hamiltonian ${H}(\phi,\pi)= \dot \phi^i\, \pi_i -{L}(\phi,\dot \phi)$,
there exists a description in terms of complex coordinates
$x^i=\tfrac12(\phi^i+i \dot \phi^i)$ and a complex function
$F(x,\bar x)$, such that,
\begin{align}
  \label{eq:theorem-prop}
  2\, \mathrm{Re}\,x^i =&\, \phi^i\,,\nonumber\\
    2\, \mathrm{Re}\,F_i(x,\bar x) =&\, \pi_i\,, \quad\mbox{where}\;\; F_i =
    \frac{\partial F(x,\bar x)}{\partial x^i}\,.
\end{align}
The function $F(x,\bar x)$ 
can be decomposed as
\begin{equation}
  \label{eq:F(x)}
  F(x,\bar x) = F^{(0)}(x) + 2i \,  \Omega(x,\bar x)\,,
\end{equation}
where $\Omega$ is real.
The decomposition \eqref{eq:F(x)} may be subjected to the following equivalence
transformation,
\begin{equation}
  \label{eq:ambiguity}
  F^{(0)}\mapsto F^{(0)} + g(x)\,, \qquad \Omega\mapsto \Omega- \mathrm{Im}\,
  g(x)\,, 
\end{equation}
which results in $F(x,\bar x)\mapsto F(x,\bar x)+\bar g(\bar x)$, and which leaves $(x^i, F_i)$ invariant.
The Lagrangian and Hamiltonian can then be expressed in terms of
$F^{(0)}$ and $\Omega$ as,
\begin{align}
  \label{eq:H-sympl}
  {L}=&\, 4 [\mathrm{Im}\, F -\Omega] \,, \\
   {H} =&\, -\mathrm{i}(x^i \,\bar F_{\bar \imath} -\bar
  x^{\bar\imath} \,F_i)
  -4\,\mathrm{Im} [F-\tfrac12 x^i\,F_i] +4\, \Omega \nonumber\\
  =&\, -\mathrm{i}(x^i \,\bar F_{\bar \imath} -\bar
  x^{\bar\imath} \,F_i) -4\,\mathrm{Im}
  [F^{(0)}-\tfrac12 x^i\,F^{(0)}_i] -2(2\,\Omega
  -x^i\Omega_i -\bar x^{\bar\imath} \Omega_{\bar\imath}) \,, \nonumber
\end{align}
with $F_i = \partial F/\partial x^i , F^{(0)}_i = \partial F^{(0)}/\partial x^i , 
\Omega_i = \partial \Omega / \partial x^i$, 
and similarly for ${\bar F}_{\bar \imath}, {\bar F}^{(0)}_{\bar \imath}$ and $\Omega_{\bar \imath}$.

Furthermore, the $2n$-vector $(x^i,F_i)$ 
denotes a complexification of the phase space coordinates
 $(\phi^i, \pi_i)$ and
transforms precisely as 
$(\phi^i, \pi_i)$ under symplectic transformations, i.e.
\begin{equation}
  \label{eq:symplectic}
  \begin{pmatrix} 
    x^i\\[2mm] F_i(x,\bar x)
  \end{pmatrix}
  \mapsto
  \begin{pmatrix} 
    \tilde x^i\\[2mm] \tilde F_i(\tilde x,\bar{\tilde x})
  \end{pmatrix}
  =
  \begin{pmatrix} 
    U^i{}_j& Z^{ij}\\[2mm]  W_{ij} & V_i{}^j 
  \end{pmatrix}
  \begin{pmatrix} 
    x^j\\[2mm] F_j(x,\bar x)
  \end{pmatrix}\;.
\end{equation}
The equations (\ref{eq:symplectic}) are integrable: the
symplectic transformation yields a new function $\tilde F (\tilde x, \bar{\tilde{x}}) = 
\tilde F^{(0)}(\tilde x) + 2i \, \tilde\Omega  (\tilde x, \bar{\tilde{x}})$, with 
 $\tilde\Omega$ real.
 }

 \begin{proof}

 We refer to \cite{Cardoso:2012nh} for the proof of the theorem. We note the following relations,
 \begin{eqnarray}
  \label{eq:def-x-y}
  x^i &=&
    \tfrac12\Big(\phi^i +
  i \frac{\partial {H}}{\partial \pi_i} \Big)\,,\nonumber\\
  y_i &=& 
  \tfrac12\Big(\pi_i -
  i \frac{\partial {H}}{\partial \phi^i} \Big) = \frac{\partial F(x, \bar x)}{\partial x^i} \;.
\end{eqnarray}

\end{proof}

We close this subsection with the following comments.  Firstly, we note that since both ${H}$ and 
$F^{(0)} - \tfrac12 \, x^i \, F^{(0)}_i$ transform as functions under symplectic transformations,
so does the following combination that appears in \eqref{eq:H-sympl},
\begin{equation}
2\,\Omega
  -x^i\Omega_i -\bar x^{\bar\imath} \Omega_{\bar\imath} \;.
  \label{eq:ex-sympl-func}
\end{equation}
Secondly, the transformation law of $2 \mathrm{i} \Omega_i = F_i - F_i^{(0)}$ under symplectic transformations
is determined by the transformation behaviour
of $F_i$ and $F_i^{(0)}$, as described above. 
 The transformation law of $2 \mathrm{i} \Omega_{\bar \imath} = F_{\bar \imath}$, on the other hand, follows from the reality of $\tilde \Omega$, 
\begin{eqnarray}
\tilde{\Omega}_{\bar \imath} = (\overline{\tilde{\Omega}_i}  )       \;.
\label{eq:rel-omega-bari_cond}
\end{eqnarray}
Thirdly, as indicated in \eqref{FOm},
the function $F(x, \bar x)$ may, in general, depend on a number of real parameters $\eta$
that are inert under symplectic transformations. 
Without loss of generality, we may take 
$\eta$ to be solely encoded in $\Omega$, and, upon transformation, in  $\tilde \Omega$
(we can use the equivalence relation \eqref{eq:ambiguity} to achieve this).
As discussed below in subsection \ref{sec:emchirback},
$\partial_{\eta} F = \partial F / \partial \eta$ transforms as a function under symplectic transformations
\cite{Cardoso:2008fr}.


\subsection{Homogeneous $F(x, \bar x, \eta)$ \label{sec-F-hom}}

The theorem in subsection \ref{sec:ubiquity} did not assume any homogeneity properties for $F$.
Here we will focus on the case when $F$ is homogeneous of degree two and discuss some of the consequences
of homogeneity \cite{Cardoso:2012nh}.
This is the case that is relevant when coupling vector multiplets to supergravity.  Moreover, it also
covers other interesting systems, such as the Born-Infeld dilaton-axion system in an $AdS_2 \times S^2$ background,
as we will explain in section \ref{bidaf}.

Let us consider a function $F(x, \bar x, \eta) = F^{(0)}(x) + 2 i \Omega (x, \bar x, \eta)$
that depends on a real parameter $\eta$, and let us discuss its behaviour under the scaling
\beq
x \mapsto \lambda \, x \;,\; \eta \mapsto \lambda^m \, \eta
\label{xeta}
\eeq
with $\lambda \in \mathbb{R}\backslash\{0\}$.
We take $F^{(0)}(x)$ to be quadratic in $x$, so that $F^{(0)}$ scales as 
$F^{(0)}(\lambda\, x) = \lambda^2  \, F^{(0)}(x)$.  This scaling behaviour can be extended to the full
function $F$ if we demand that the canonical pair $(\phi,\pi)$ given in \eqref{eq:theorem-prop}
scales uniformly as
$(\phi,\pi) \mapsto \lambda \, (\phi,\pi)$.  Then we have
\begin{equation}
F(\lambda\, x, \lambda \, \bar x, \lambda^m \, \eta) = \lambda^2  \, F(x, \bar x, \eta) \;,
\label{eq:resc-F}
\end{equation}
which results in the homogeneity relation
\begin{equation}
2 \,F = x^i \, F_i + {\bar x}^{\bar \imath} \, F_{\bar \imath} + m \, \eta \, F_{\eta} \;,
\label{eq:homog-F}
\end{equation}
where $F_{\eta} = \partial F / \partial \eta$.
Inspection of \eqref{eq:leg-l-h}
shows 
that the associated Hamiltonian $H$ scales with weight two as
\begin{equation}
{H} (\lambda \, \phi, \lambda \, \pi, \lambda^m \, \eta) = \lambda^2 \, {H} (\phi,\pi, \eta)\;,
\end{equation}
so that ${H}$ satisfies
the homogeneity relation,
\begin{equation}
2 \, {H} = 
\phi \, \frac{\partial { H}}{\partial \phi} + \pi \, 
\frac{\partial {H}}{\partial \pi}
+ m \, \eta \, \frac{\partial {H}}{\partial \eta} \;.
\label{eq:homog-H}
\end{equation}
Using \eqref{eq:def-x-y}, this can be written as 
\begin{equation}
\label{eq:H-hom-om}
{H} = i \left( {\bar x}^{\bar \imath} \, F_i - x^i \, {\bar F}_{\bar \imath}
\right)  
+ \frac{m}{2} \, \eta \, \frac{\partial { H}}{\partial \eta} \;.
\end{equation}
Next, using that the dependence on $\eta$ is solely contained in $\Omega$, we obtain
\begin{equation}
\frac{\partial {H}}{\partial \eta}|_{\phi, \pi} = -  \frac{\partial {L}}{\partial \eta}|_{\phi, \dot \phi}
= - 4 \Omega_{\eta} \;,
\end{equation}
where $\Omega_{\eta} = \partial \Omega / \partial \eta$.
Thus, we can express \eqref{eq:H-hom-om} as
\begin{equation}
{H} = i \left( {\bar x}^{\bar \imath} \, F_i - x^i \, {\bar F}_{\bar \imath}
\right)  
-  2 \, m \,
\eta \, \Omega_{\eta} \;.
\label{eq:hamil-sympl}
\end{equation}
This relation is in accordance with \eqref{eq:H-sympl} upon substitution of the homogeneity relations
$2 F^{(0)}(x) = x^i \, F^{(0)}_i $ and 
$2\,\Omega = x^i\Omega_i + \bar x^{\bar\imath} \Omega_{\bar\imath} + m \, \eta \, \Omega_{\eta} $
that follow from \eqref{eq:homog-F}. 

The Hamiltonian transforms as a function under symplectic transformations.  Since the first term in 
\eqref{eq:hamil-sympl} transforms as a function, it follows that $\Omega_{\eta}$ also transforms as a function.
This is in accordance with the general result quoted at the end of subsection \ref{sec:ubiquity} which states that $\partial_{\eta}
F$ transforms as a function.

In certain situations, such as in the study of 
BPS black holes in ${\cal N} = 2$ supergravity theories \cite{LopesCardoso:1998tkj}, the discussion needs to be extended
to a parameter $\eta$ that is complex, so that now we consider a function 
$F(x, \bar x, \eta, \bar \eta) = F^{(0)}(x) + 2 i\,  \Omega (x, \bar x, \eta, \bar \eta)$
 that scales as follows
(with $\lambda \in \mathbb{R}\backslash\{0\}$),
\begin{equation}
F(\lambda \, x, \lambda \, \bar x, \lambda^m \, \eta, 
\lambda^m \, \bar \eta) = \lambda^2 \, F(x, \bar x, \eta, \bar \eta) \;.
\end{equation}
The extension to a complex $\eta$
results in the presence of an additional term
on the right hand side of \eqref{eq:homog-F} and \eqref{eq:homog-H},
\begin{align}
2 \, F =&\, x^i \, F_i + {\bar x}^{\bar \imath} \, F_{\bar \imath} + m  \left( \eta \, F_{\eta} + \bar \eta F_{\bar \eta} \right)
 \;, \nonumber\\
2 \, {H} =&\,
\phi \, \frac{\partial {H}}{\partial \phi} + \pi \, 
\frac{\partial { H}}{\partial \pi}
+ m  \left( \eta \, \frac{\partial {H}}{\partial \eta} + \bar \eta \, \frac{\partial {H}}{\partial \bar
\eta}
\right)\;,
\label{eq:homog-ceta}
\end{align}
and hence
\begin{equation}
{H} = i \left( {\bar x}^{\bar \imath} \, F_i - x^i \, {\bar F}_{\bar \imath}
\right)  
+ \frac{m}{2} \left( \eta \, \frac{\partial {H}}{\partial \eta} + 
\bar \eta \, \frac{\partial {H}}{\partial \bar
\eta}
\right)\;.
\end{equation}
Then, since the dependence on $\eta$ and $\bar \eta$ is solely contained in $\Omega$, 
we obtain
\begin{equation}
{H} = i  \left( {\bar x}^{\bar \imath} \, F_i - x^i\, {\bar F}_{\bar \imath}
\right)  
-  2 \, m \left( 
\eta \, \Omega_{\eta} + \bar \eta \, \Omega_{\bar \eta} \right)\;.
\end{equation}
This is in accordance with \eqref{eq:H-sympl} upon substitution of the homogeneity relations
$2 F^{(0)}(x) = x^i \, F^{(0)}_i $ and 
$2\,\Omega = x^i\Omega_i + \bar x^{\bar\imath} \Omega_{\bar\imath} + m \, 
(\eta \, \Omega_{\eta} +  \bar \eta \, \Omega_{\bar \eta} )$
that follow from \eqref{eq:homog-ceta}.

The above extends straightforwardly to the case of multiple real or complex parameters.

\subsection{Duality covariant complex variables \label{sec:new-var}}

The Hamiltonian \eqref{eq:H-sympl} 
is given in terms of complex fields $x^i$ and $\bar x^{\bar \imath}$. It may also depend on parameters $\eta$,
in which case the transformation law of $x^i$ 
under symplectic transformations  \eqref{eq:symplectic}
will
depend on $\eta$.  
It is therefore convenient to 
introduce duality covariant complex variables $t^i$, whose symplectic transformation law is independent of $\eta$.
These variables ensure
that when expanding the Hamiltonian 
in powers of $\eta$, the resulting expansion coefficients transform covariantly under symplectic 
transformations. This expansion can also be organized by employing a suitable covariant derivative.
We review these aspects following \cite{Cardoso:2012nh}.

We take the Hamiltonian \eqref{eq:H-sympl} to depend on a single real parameter $\eta$ that is inert under
symplectic transformations. The discussion can be extended to the case of
multiple real external parameters in a straightforward manner.
We define
complex variables $t^i$ by \cite{Cardoso:2010gc},
\begin{align}
  \label{eq:canon-t-rel}
 2 \, {\rm Re} \, t^i = &\ \phi^i \;, \nonumber\\
 2 \, {\rm Re} \,   F_i^{(0)} (t)  = &\, \pi_i \;.
  \end{align}
Then, the vector $(t^i, F_i^{(0)} (t))$ describes a complexification of 
$(\phi^i, \pi_i)$ that transforms as in \eqref{eq:symplectic-pq}
under symplectic transformations.
This yields the transformation law
  \begin{equation}
    \tilde{t}^i= U^i{}_j\,t^j + Z^{ij}  F^{(0)}_j(t) \;,
    \end{equation}
which is independent of $\eta$.  The new variables $t^i$ are related to the 
$x^i$ by (c.f. \eqref{eq:theorem-prop})
\begin{align}
  \label{eq:x-t-rel}
  2 {\rm Re} \, t^i    =&\, 2 {\rm Re} \, x^i  \;, \nonumber\\
  2 {\rm Re} \,   F_i^{(0)} (t) =&\,  2 {\rm Re} \, F_i(x, \bar x, \eta)  \;.
  \end{align}
 We may now view $H$ either as a function of $t^i$ and $\bar{t}^{\bar\imath}$, or as a function 
of $x^i$ and  $\bar{x}^{\bar\imath}$. Differentiating $H(\phi, \pi (x, \bar x, \eta), \eta)$
with respect to $\eta$ yields
\begin{eqnarray}
\left. \frac{\partial H}{\partial \eta} \right|_{x, \bar x} &=&
\left. \frac{\partial H}{\partial \eta} \right|_{\phi, \pi} +
\left. \frac{\partial H}{\partial \pi_k} \right|_{{\rm Re} \, x} 
\frac{\partial \pi_k}{\partial \eta} = \left. \frac{\partial H}{\partial \eta} \right|_{t,\bar t } +
\left. \frac{\partial H}{\partial \pi_k} \right|_{{\rm Re} \, x}  \left( F_{k \eta} + \bar{F}_{ \bar{k}  \eta} \right) \;,
\nonumber\\
\end{eqnarray}
where $F_{\eta k} = \partial^2 F/\partial \eta
\partial x^k$, etc., and
where on the right hand side we used $\pi_k = 2 {\rm Re} \, F_k(x, \bar x, \eta)$. 
Next, 
 we use  the conversion formula
\begin{eqnarray}
\left. \frac{\partial H}{\partial \pi_k}  \right|_{{\rm Re} \, x} = 
\left. \frac{\partial H}{\partial {\rm Im} \, x^i} \right|_{{\rm Re} \, x}
\frac{\partial {\rm Im} \, x^i }{\partial \pi_k} 
= -
\left. \frac{\partial H}{\partial {\rm Im} \, x^i} \right|_{{\rm Re} \, x}
\, 
 \hat{N}^{ik} \;,
\end{eqnarray}
where  $\hat{N}^{ik} $ denotes the inverse of 
\begin{equation}
-
\frac{\partial \pi_k}{ \partial {\rm Im} \, x^i } =- i  \left( \frac{\partial}{\partial x^i} -  \frac{\partial}{\partial \bar{x}^{\bar \imath}}
\right) \left( F_k + \bar{F}_{\bar k} \right) = -  i \left[ F_{ik} - {\bar F}_{\bar\imath \bar k} - F_{k \bar\imath} + {\bar F}_{\bar k i} \right]  =
\hat{N}_{ik} \;.
\label{eq:hat-N-ij}
\end{equation}
Note that $\hat{N}_{ik}$ is a real symmetric matrix.,
\begin{equation}
\hat{N}_{ik} =  -  i \left[ F^{(0)}_{ik} - {\bar F}^{(0)}_{\bar\imath \bar k}  \right]  
+ 2 \left( \Omega_{ik} + \Omega_{\bar \imath \bar k} - 
\Omega_{k \bar\imath} - \Omega_{i \bar k } 
\right) \;.
\end{equation}
We obtain
\begin{eqnarray}
\left. \partial_{\eta} H  \right|_{t, \bar t} &=&
\left. {\cal{D}}_{\eta} H \right|_{x, \bar x} \;,
\end{eqnarray}
where ${\cal D}_{\eta}$ is given by 
\begin{equation}
{\cal D}_{\eta} = \partial_{\eta} + i \, {\hat N}^{ij} \left( F_{\eta j} + {\bar F}_{\eta \bar\jmath} \right) \left(\partial_i
- \partial_{\bar\imath} \right) \;.
\label{eq:cov-der-multiple}
\end{equation}

${\cal D}_{\eta}$ acts as a covariant derivative for symplectic transformations.  
Applying multiple covariant derivatives $\mathcal{D}_\eta$ on
any symplectic function depending on $x^i$ and $\bar x^{\bar \imath}$, will
again yield a symplectic function. For instance, consider
applying 
$\mathcal{D}_\eta^{^2}$ on $H(\phi, \pi (x, \bar x, \eta), \eta)$ given in \eqref{eq:H-sympl},
\begin{align}
  \label{eq:D-eta-H}
    \mathcal{D}_\eta^{^2} {H} (x, \bar x, \eta)
  =&\,- 4 \left[  \partial_\eta^{^2}\Omega -2\,\hat N^{ij} \,
   \partial_\eta \left( \Omega_i-\Omega_{\bar \imath} \right)
  \,\partial_\eta \left( \Omega_j-\Omega_{\bar \jmath} \right)
    \right]\,.
\end{align}
As discussed in 
section \ref{sec:emchirback}, while 
$\partial_\eta^{^2}\Omega$ does not transform as a function under symplectic transformations, there exists
a modification of it, given by \eqref{eq:D-eta-H}, such that the modified expression
transforms as a function.

\subsection{Maxwell-type theories}

Now we turn to Maxwell-type theories in four dimensions, namely, we consider the Maxwell sector
of ${\cal N}=2$ supergravity theories coupled to abelian ${\cal N}=2$ vector multiplets. 
Below we will use some of the ingredients that go into the construction of these theories.
We refer to section \ref{4dvecmult} for a detailed description of these theories.
In the following, we review electric-magnetic duality in these theories, first
at the two-derivative level, and then in the presence of an arbitrary chiral background field.

\subsubsection{Electric-magnetic duality at the two-derivative level  \label{sec:em-dual-2d}}

The Wilsonian effective action is a local action that describes the effective dynamics at long distances \cite{Kaplunovsky:1994fg}.
 The Wilsonian effective action
describing the coupling of $n$ abelian ${\cal N}=2$ vector supermultiplets
to four-dimensional ${\cal N}=2$ supergravity at the two-derivative level is encoded in a holomorphic
function $F(X)$, called the prepotential, which depends on $n+1$ complex scalar fields $X^I$ ($I=0, 1, \dots n$)
and which is a homogeneous function of degree two under complex rescalings \cite{deWit:1984wbb},
\begin{equation}
F(\lambda X) = \lambda^2 \, F(X) \;\;\;,\;\;\; \lambda \in 
\mathbb{C} \backslash\{0\} \;,
\end{equation}
from which one infers the relations
\begin{eqnarray}
2 F  &=& F_I  \, X^I\;, \nonumber\\
F_I  &=& F_{IJ}  \, X^J \;, \nonumber\\
0 &=& F_{IJK}  \, X^K  \;,
\label{homprep}
\end{eqnarray}
where $F_I = \partial F/\partial X^I$, $F_{IJ} = \partial^2 F/\partial X^I \partial X^J$, $F_{IJK} = \partial^3 F/\partial X^I \partial X^J \partial X^K$.

The resulting equations of motion for the abelian gauge fields $A_{\mu}^I$ only involve their field strengths $F_{\mu \nu}^I$. 
The combined system of equations of motion and Bianchi
identities for the abelian gauge fields are invariant under so-called electric-magnetic duality
transformations, which constitute symplectic $Sp(2n+2, \mathbb{R})$ transformations \cite{deWit:1984wbb}.
These transformations also induce $Sp(2n+2, \mathbb{R})$ transformations of the symplectic vector $(X^I, F_I)$, as follows 
\cite{deWit:1996gjy}.

Consider the following Lagrangian for Maxwell fields $A_{\mu}^I$, 
\begin{equation}
{L} = - \frac{i}{4} \, \Big( 
\bar F_{IJ} 
\,F_{\mu\nu}^{+I}F^{+\mu\nu J}
+ 2 
\,{\cal O}_{\mu\nu I}^{+ } \, F^{+\mu\nu I}
-
F_{IJ} 
\,F_{\mu\nu}^{-I} F^{-\mu\nu J} 
-
2 
\,{\cal O}_{\mu\nu I }^{-} \, F^{-\mu\nu I}
\Big) 
\,, 
\label{Maxwelllag} 
\end{equation}
where $F_{\mu\nu}^{\pm I}$ denote the (anti-)selfdual field strengths (c.f. \eqref{dualF}), and 
where we allow for a linear coupling of the field strengths $F_{\mu\nu}^{\pm I} $ to tensors  $\,{\cal O}_{\mu\nu I }^{\pm}$.
A Lagrangian of this form arises when considering the part of the ${\cal N}=2$ Wilsonian effective Lagrangian 
that describes the coupling of
vector multiplets to ${\cal N}=2$ supergravity at the two-derivative level, c.f. \eqref{efflagtotal}.

We define the dual field strength by 
\begin{equation}
  \label{eq:dual-F}
  G_{\mu\nu I} =  \sqrt{-g}\,
  \varepsilon_{\mu\nu\rho\sigma}\, 
  \frac{\partial {L}}{\partial F_{\rho\sigma}{}^I} \,.
\end{equation}
Decomposing it into (anti-)selfdual parts $G^{\pm}_{\mu\nu I}$,
\begin{equation}
  \label{eq:Gpm}
  G^{\pm}_{\mu\nu I} = \pm 2 i \, 
   \frac{\partial {L}}{\partial F^{\pm}_{\rho\sigma}{}^I} \,,
\end{equation}
we obtain
\begin{equation}
G^+_{\mu\nu I}= {\bar F}_{IJ}
F^{+J}_{\mu\nu} + {\cal O}_{\mu\nu I }^{+} 
\,,\quad G^-_{\mu\nu
I}= F_{IJ}F^{-J}_{\mu\nu}
+ {\cal O}_{\mu\nu I }^{-} 
\,. \label{defG}
\end{equation}
The Bianchi identities and equations of motion
for the abelian gauge fields take the form
\begin{equation}
\partial^\mu \big(F^{+I}_{\mu\nu} -F^{-I}_{\mu\nu}\big)
=0\,,\qquad
\partial^\mu \big(G_{\mu\nu I}^+ -G^-_{\mu\nu I}\big) =0\,.
\label{Maxwell}
\end{equation}
The combined system \eqref{Maxwell} is
invariant under the transformation 
\begin{eqnarray}
\begin{pmatrix}
F^{\pm I}_{\mu\nu} \\ G^\pm_{\mu\nu I}
\end{pmatrix}
\mapsto
\begin{pmatrix}
{\tilde F}^{\pm I}_{\mu\nu} \\ {\tilde G}^\pm_{\mu\nu I}
\end{pmatrix}
=
\begin{pmatrix} 
U^I_{\,J} &Z^{IJ} \\
 W_{IJ}&V_I^{\,J}
 \end{pmatrix}
 \begin{pmatrix} 
 F^{\pm J}_{\mu\nu} \\
   G^\pm_{\mu\nu J}
  \end{pmatrix}
     \,,\label{FGdual}
\end{eqnarray}
where $U^I_{\,J}$, $V_I^{\,J}$, $W_{IJ}$ and $Z^{IJ}$ are 
constant real  $(n+1)\times(n+1)$ submatrices. We demand the transformation matrix in 
\eqref{FGdual}
to be invertible. Since we may rescale the field strengths  $F^{I}_{\mu\nu}$ by a real constant, we
impose the normalization $\det (U^T V - W^T Z ) = 1$.
Thus, the transformation matrix in \eqref{FGdual}
belongs to $SL(2n+2, \mathbb{ R})$.

Next, decomposing the transformed field strenghts $\tilde{F}^{\pm I}_{\mu\nu} , \; \tilde{G}^\pm_{\mu\nu I}$
as in  \eqref{defG},
\begin{equation}
\label{dcompGtil}
\tilde{G}^+_{\mu\nu I}= \tilde{\bar F}_{IJ}
\tilde{F}^{+J}_{\mu\nu} + \tilde{\cal O}_{\mu\nu I }^{+} 
\,,\quad \tilde{G}^-_{\mu\nu
I}= \tilde{F}_{IJ} \tilde{F}^{-J}_{\mu\nu}
+\tilde{\cal O}_{\mu\nu I }^{-} \;,
\end{equation}
we infer that under \eqref{FGdual}, $F_{IJ}$ transforms as 
\begin{equation}
\tilde{F}_{IJ} = (W_{IL} + V_I{}^K F_{KL})\, \big[{\cal S}^{-1}\big]^L{}_J  \;\;\;,\;\;\;
{\cal S}^I{}_J = U^I{}_J + Z^{IK} F_{KJ}  \,.\label{Nchange}
\end{equation}
Then, demanding that $\tilde{F}_{IJ}$ is a symmetric matrix yields the condition
\begin{eqnarray}
U^T W - W^T U + (U^T V - W^T Z) \, F - F \, (U^T V - W^T Z )^T \nonumber\\
 +
F\, (Z^T V - V^T Z) \, F=0 \;,
\end{eqnarray}
where in this equation $F$ denotes the matrix $F_{IJ}$.
By comparing terms with the same power of $F_{IJ}$, we infer the conditions
$U^T W = W^T U $ and $Z^T V =V^T Z$. In addition, 
the combination $U^T V - W^T Z$ needs to be proportional to
the identity matrix, since the terms linear in $F_{IJ}$ need to cancel for general $F_{IJ}$ \cite{deWit:1992wf}.
These conditions, 
when combined with the property that the transformation matrix 
belongs to $SL(2n+2, \mathbb{ R})$, imply  that 
the transformation matrix in 
(\ref{FGdual}) must be an element of $Sp(2n+2, \mathbb{ R})$.
Indeed, defining
\begin{equation}
{\Delta} =
\left(\begin{array}{cc} U & Z \\[1mm] W & V \end{array}\right) \;,
\end{equation}
and demanding $\Delta$ to be a symplectic matrix, i.e.
\begin{equation}
{\Delta }^{-1} = \Omega\, {\Delta}^{\rm T} \,\Omega^{-1} \qquad
\mbox{where} \quad \Omega =\left(\begin{array}{cc}
0&\mathbbm{1} \\[1mm]
-  \mathbbm{1} & 0
\end{array}\right) ,
\label{symplec}
\end{equation}
gives 
\begin{eqnarray}
U^{\rm T} V- W^{\rm T} Z = V^{\rm T}U - Z^{\rm T}W =
\mathbbm{1} \;,\; 
U^{\rm T}W = W^{\rm T}U \;,\;Z^{\rm T}V= V^{\rm T}Z
\label{UVWZ}
\end{eqnarray}
as a consequence of $\Delta^{-1} \, \Delta = \mathbbm{1}$, and 
\begin{eqnarray}
U V^{\rm T} - Z W^{\rm T}  = V U^{\rm T} - W Z^{\rm T} =
\mathbbm{1} \;,\; 
U Z^{\rm T} = Z U^{\rm T} \;,\;W V^{\rm T}= V W^{\rm T}
\label{UVWZ2}
\end{eqnarray}
as a consequence of  $\Delta\, \Delta^{-1} = \mathbbm{1}$.

Furthermore, we infer from \eqref{dcompGtil} that under  \eqref{FGdual}, 
the tensors  ${\cal O}_{\mu\nu I }^{\pm} $ transform as
\begin{equation}
\tilde{\cal O}_{\mu\nu I }^{+} = \tilde{\cal O}_{\mu\nu J }^{+} \, \big[\bar{\cal S}^{-1}\big]^J{}_I \;\;\;,\;\;\;
\tilde{\cal O}_{\mu\nu I }^{-} = \tilde{\cal O}_{\mu\nu J }^{-} \, \big[{\cal S}^{-1}\big]^J{}_I \;.
\end{equation}

Next, we note that 
the transformation  \eqref{Nchange} of $F_{IJ}$ is induced by 
the following transformation
of the scalar 
fields $X^I$,
\begin{eqnarray}
\label{symXF}
\begin{pmatrix}
X^{I} \\ F_{ I}
\end{pmatrix}
\mapsto
\begin{pmatrix}
{\tilde X}^{I} \\ {\tilde F}_{ I} 
\end{pmatrix} = 
\begin{pmatrix} 
U^I_{\,J} &Z^{IJ} \\
 W_{IJ}&V_I^{\,J}
 \end{pmatrix}
 \begin{pmatrix} 
 X^{J}\\
   F_{J}
  \end{pmatrix} \;,
   \end{eqnarray}
which is the aforementioned $Sp(2n+2, \mathbb{ R})$ transformation of the 
vector $(X^I, F_I)$. Indeed, using \eqref{symXF}, one derives
\begin{equation}
\frac{\partial {\tilde F}_I}{ \partial X^J} = {\tilde F}_{IK} \left(U^K{}_J+ Z^{KL}Ê\, F_{LJ}Ê\right) = W_{IJ}Ê+ V_I{}^L \, F_{LJ} \;.
\end{equation}
For $N_{IJ} \equiv 2 {\rm Im} \, F_{IJ} $, one obtains the transformation law
\begin{eqnarray}
{\tilde N}_{IJ} &=& N_{KL} \, \big[ \bar{\cal S}^{-1} \big]^K{}_I \, \big[ {\cal S}^{-1} \big]^L{}_J \;, \nonumber\\
{\tilde N}^{IJ} &=& \bar{\cal S}{}^I{}_K \, {\cal S}{}^J{}_L  \, N^{KL} =
{\cal S}{}^I{}_K \, {\cal S}{}^J{}_L \left( N^{KL} - i {\cal Z}^{KL} \right) \;,
\label{transfN}
\end{eqnarray}
where
\begin{eqnarray}
{\cal Z}^{IJ} = [{\cal S}^{-1}]^I{}_K Z^{KJ} \;.
\label{SZ}
\end{eqnarray}
Note that ${\cal Z}$ is a symmetric matrix by virtue of \eqref{UVWZ2}.

Owing to the symplectic condition \eqref{symplec}, the quantities
$\tilde F_I$ can be written as the derivative of a new function
$\tilde F(\tilde X)$ 
with respect to the new coordinates $\tilde X^I$,
\begin{eqnarray}
\label{Ftilde}
\tilde F(\tilde X)
&=& \tfrac12 \big(U^{\rm T}W\big)_{IJ}X^I X^J
+ \tfrac12
\big(U^{\rm T}V + W^{\rm T}  Z\big)_I{}^J X^IF_J
+ \tfrac12 \big(Z^{\rm T}V\big){}^{IJ}F_I F_J \nonumber\\
&=& F(X) + \tfrac12 \big(U^{\rm T}W\big)_{IJ}X^I X^J
+ \big( W^{\rm T}  Z\big)_I{}^J X^IF_J
+ \tfrac12 \big(Z^{\rm T}V\big){}^{IJ}F_I F_J \;, \nonumber\\
\end{eqnarray}
where we made use of the homogeneity property \eqref{homprep}. Note that $F(X)$
does not transform as a function under symplectic transformations \eqref{symXF}, i.e. $ \tilde F(\tilde X)
\neq F(X)$. Its geometrical meaning will be discussed in subsection \ref{sect:SK_vector_bundle}.

Two
${\cal N}=2$ Wilsonian effective Lagrangians that are encoded in
$F(X)$ and $\tilde F(\tilde X)$, respectively, represent equivalent vector multiplet theories
coupled to ${\cal N}=2$ supergravity. On the other hand, symplectic transformations that constitute a symmetry of 
the theory are transformations \eqref{symXF} for which
\begin{equation}
\tilde F(\tilde X) = F(\tilde X)\,, 
\label{Finv}
\end{equation}
since they leave the field equations invariant. Differentiating \eqref{Finv} with respect to ${\tilde X}^I$
gives ${\tilde F}_I (\tilde{X}) = \partial F(\tilde{X}) / \partial \tilde{X}^I$,
which means that the transformation law for $F_I (X)$ given in \eqref{symXF} is induced by substituting $\tilde{X}$ for $X$ 
in $F_I(X)$. This yields a practical way for checking whether a symplectic transformation
constitutes an invariance of the theory. Note that the property \eqref{Finv} does not imply that $F(X)$ is an invariant 
function; inspection of \eqref{Ftilde} shows that $ \tilde F(\tilde X)
\neq F(X)$, and hence $F(\tilde X) \neq F(X)$.

\subsubsection{Electric-magnetic duality in a chiral background \label{sec:emchirback}}

Let us now briefly summarize various features of electric-magnetic duality in the presence
of a chiral background field \cite{deWit:1996gjy}. We refer to sections \ref{sec:couchiralback} and \ref{sec:hess} for an extensive
discussion of supergravity theories in the presence of a chiral background field, and for the relation
with Hessian geometry.

We consider the Wilsonian effective action describing the coupling of ${\cal N}=2$ supergravity to 
abelian vector multiplets in the presence of a chiral background field
$\hat A$. The action is now encoded in a holomorphic function $F(X, \hat A)$ which is 
homogeneous of degree two under complex rescalings, i.e.
\begin{equation}
\label{homFA}
F(\lambda X, \lambda^w \hat A) = \lambda^2 \, F(X, \hat A) \;\;\;,\;\;\; \lambda \in 
\mathbb{C} \backslash\{0\} \;,
\end{equation}
where $w$ denotes the scaling weight of $\hat A$, which we take to be non-vanishing.
{From} \eqref{homFA} one infers the relation
\begin{equation}
2 F (X,  \hat A) = X^I \, F_I (X,  \hat A) + w {\hat A}\, F_{A} (X,  \hat A) \;,
\label{scalingrelA}
\end{equation}
where we introduced the notation $F_I (X, \hat A) = \partial F (X, \hat A)/\partial X^I
, \, F_A  (X, \hat A) = \partial F (X, \hat A)/\partial \hat{A}$.
Symplectic transformations act on $(X^I, F_I(X, \hat A))$
as in \eqref{symXF},
\begin{eqnarray}
\tilde X{}^I&=&U^I_{\ J}\,X^J + Z^{IJ}\,
F_J (X, \hat A),\nonumber\\
\tilde F{}_I (\tilde{X}, \hat A) &=&V_I{}^J\,F_J (X, \hat A) + W_{IJ}\,X^J \;,
\label{XFXtransfA}
\end{eqnarray}
and they leave $\hat A$ inert. We will now show that $F_A  (X, \hat A)$ transforms as a function
under symplectic transformations. It follows that the combination 
$F(X,\hat A)- \tfrac12 X^I 
F_I(X,\hat A)$ also transforms as a function due to the relation \eqref{scalingrelA}, 
\begin{equation}
\tilde F(\tilde X,\hat A)- \tfrac12 \tilde X^I \tilde 
F_I(\tilde X,\hat A) =
 F(X,\hat A)- \tfrac12 X^I 
F_I(X,\hat A) \,.
\end{equation}

We start from the second relation in \eqref{XFXtransfA} and 
differentiate with
respect to $X^J$ keeping $\hat A$ fixed. This gives
\begin{equation}
{\tilde F}_{IK} = \left( W_{IP}Ê+ V_I{}^L \, F_{LP} \right) \big[ {\cal S}^{-1} \big]^P{}_K 
\;,
\label{Ftilde12}
\end{equation}
where
\begin{equation}
\frac{ \partial \tilde{X}^I}{\partial X^J} \equiv 
 {\cal S}^I{}_J = U^I{}_J + Z^{IK} \, F_{KJ} (X, \hat A) \;.
  \end{equation}
Taking the transposed of this equation, one verifies that $\tilde{F}_{IK}$ is symmetric in $I$ and $K$, i.e. $\tilde{F}_{IK} = \tilde{F}_{KI}$.

Next, we differentiate the second relation in \eqref{XFXtransfA} with
respect to 
$\hat A$, keeping $X^I$ fixed. This yields
\begin{equation}
\tilde{F}_{I A} (\tilde X,\hat A) = \left(V_I{}^K - \tilde{F}_{IL} \, Z^{LK} \right) {F}_{K A} (X,\hat A) \;.
\label{FtilIA}
\end{equation}
Using \eqref{Ftilde12}, we obtain for the transposed of the matrix on the right hand side of \eqref{FtilIA},
\begin{equation}
V^T  - Z^T \tilde{F} =  {\cal S}^{-1} \;,
\end{equation}
where here $\tilde F$ denotes the symmetric matrix $\tilde{F}_{IJ}$. Hence, 
\begin{equation}
\tilde{F}_{I A} (\tilde X,\hat A) = {F}_{K A} (X,\hat A) \big[ {\cal S}^{-1} \big]^K{}_I \;,
\end{equation}
With this result, and using $ {F}_{K A} (X,\hat A) \big[ {\cal S}^{-1} \big]^K{}_I = \partial ( F_{A} (X,\hat A) ) / \partial \tilde{X}^I$, 
we obtain
\begin{equation}
\tilde{F}_{A} (\tilde X,\hat A) = {F}_{A} (X,\hat A) \;,
\label{tilFAFA}
\end{equation}
up to terms that are independent of $X^I$, and which we drop, since they are not relevant for the vector multiplet
Lagrangian. Thus, ${F}_{A} (X,\hat A)$ transforms as a function under symplectic transformations.

Defining $N_{IJ} \equiv 2 {\rm Im} \, F_{IJ} $ and $N^{IJ}\equiv \big[N^{-1}\big]^{IJ}$, 
and using \eqref{Ftilde12}, one obtains the transformation laws
\begin{eqnarray}
\label{trafN}
\tilde N_{IJ} &=& N_{KL}\, \big[\bar{\cal 
S}^{-1}\big]^K{}_{\!I}\,  \big[{\cal S}^{-1}\big]^L{}_{\!J}\,, 
\nonumber\\ 
\tilde N^{IJ} &=& N^{KL}\, \bar{\cal 
S}^I{}_{\!K}\,{\cal S}^J{}_{\!L}\,, \\  
\tilde F_{IJK} &=& F_{MNP}\, \big[{\cal S}^{-1}\big]^M{}_{\!I} \,
    \big[{\cal S}^{-1}\big]^N{}_{\!J}\, \big[{\cal 
S}^{-1}\big]^P{}_{\!K} \,. \nonumber
\end{eqnarray}
Using \eqref{tilFAFA}, one finds
\begin{equation}
\tilde F_{A A}(\tilde X,\hat A)= F_{ A A } (X,\hat A)
 -F_{A I}(X,\hat 
A)\,F_{A J}(X,\hat A)\, {\cal Z}^{IJ}\,,
\end{equation}
where
\begin{equation}
{\cal Z}^{IJ}\,\equiv \,[{\cal S}^{-1}]^I{}_K\, Z^{KJ}\,  ,\\ 
\end{equation}
which is symmetric in $I$ and $J$, see below \eqref{UVWZ2}.
This shows that, while 
$F_A$ transforms as a function under symplectic 
transformations, higher derivatives of $F$ with respect to $\hat A$, such as $F_{AA}$, 
do not transform as functions under symplectic transformations. 
Combinations that do transform as symplectic functions can be 
generated 
systematically, as follows \cite{deWit:1996gjy}. Assume that $G(X,\hat A)$ transforms as a 
function under symplectic transformations. Then, also ${\cal D}G(X,\hat A)$ transforms as a symplectic 
function (c.f. \eqref{eq:cov-der-multiple}), where
\begin{equation}
{\cal D}\,\equiv\, {\partial \over \partial \hat A} +
iF_{AI}N^{IJ}{\partial \over \partial X^J} \,,
\label{cov-der-A}
\end{equation}
as one readily verifies using \eqref{trafN}.
Consequently one can introduce a hierarchy of symplectic functions $F^{(n)} (X,\hat A) $, which are modifications of 
$F_{A\cdots A}$,
\begin{equation}
F^{(n)} (X,\hat A) \,\equiv \, {1\over n!}{\cal D}^{n-1} F_A(X,\hat A) \;\;\;,\;\;\; n \geq 1 \;.
\label{Fn}
\end{equation}
While $F^{(1)}$ is holomorphic, all the higher 
$F^{(n)}$ (with $n \geq 2$ )
are non-holomorphic. This lack of holomorphy is governed 
by the following equation (with $n \geq 2 $),
\begin{equation}
{\partial F^{(n)} \over \partial \bar X^I}= \tfrac12 \bar F_I{}^{JK} 
\sum_{r=1}^{n-1}\; {\partial F^{(r)} \over \partial  X^J}\,{\partial F^{(n-r)} 
\over \partial  X^K}\,,
 \label{buckow_anom}
\end{equation}
where $\bar F_I{}^{JK}= \bar F_{ILM}\,N^{LJ}N^{MK}$. 

In section \ref{sec:hess} we will relate the covariant derivative \eqref{cov-der-A} and the holomorphic
anomaly equation \eqref{buckow_anom} to properties of Hessian structures in the presence of a chiral background
field, (c.f. \eqref{D_Upsilon_holo} and \eqref{A0}).


\section{Special K\"ahler geometry \label{sec:secSKG}}

In this section we discuss special K\"ahler geometry from the
mathematical point of view. The definition is ultimately motivated
by physics: special K\"ahler geometry is the geometry of ${\cal N}=2$
vector multiplets. As we have seen in the previous section, the field
equations of theories of abelian vector fields are invariant under
symplectic transformations, which generalize the electric-magnetic
rotations of Maxwell theory. In ${\cal N}=2$ vector multiplets, 
which contain scalars and fermions together with vector fields,
this extends to an action of the symplectic group on all fields,
which imposes strong constraints on the scalar geometry. 
In short, special K\"ahler manifolds
are K\"ahler manifolds equipped with a flat connection $\nabla$ which 
is compatible with the symplectic structure, in the sense that
symplectic transformations act linearly on $\nabla$-affine coordinates.
Moreover, the K\"ahler metric is Hessian with $\nabla$ as the
associated flat connection.

Special K\"ahler geometry has undergone various re-formulations 
over the past 30 years. Our approach blends the original 
definition \cite{deWit:1984pk} in terms of special coordinates
and using the superconformal calculus with the intrinsic 
construction of \cite{Freed:1997dp} and the universal construction 
of \cite{Alekseevsky:1999ts}, which allows to relate the former two
approaches. Other formulations of special K\"ahler geometry will be
discussed in section \ref{sec:other_SK}.

\subsection{Affine special K\"ahler geometry}

We will first present an intrinsic definition, and introduce special
real and special holomorphic coordinates, the Hesse potential
and the holomorphic prepotential. Then we give two extrinsic constructions,
firstly as a K\"ahlerian Lagrangian immersion into a complex symplectic 
vector space, secondly as  a parabolic affine hypersphere immersed into a real space.
The holomorphic prepotential and the Hesse potential are the generating 
functions for these two immersions.

\subsubsection{The intrinsic definition \label{sect:intrinsic_ASK}}

We start with the relatively recent definition given in \cite{Freed:1997dp}, which is
intrinsic in the sense of only using data
involving the tangent bundle and associated bundles. 
Our presentation is based on \cite{Freed:1997dp} and 
\cite{Alekseevsky:1999ts}.

\begin{defi}
{\bf Affine special K\"ahler manifolds (ASK manifolds).}
An affine special K\"ahler manifold $(M,J,g,\nabla)$ is 
a K\"ahler manifold $(M,J,g)$ endowed with a flat, torsion-free
connection $\nabla$, such that 
\begin{enumerate}
\item
$\nabla$ is symplectic, that is, the K\"ahler form $\omega=g(\cdot,J\cdot)$ is parallel: $\nabla \omega = 0$. 
\item
$\nabla J$ is covariantly closed, $d_\nabla J =0$. 
\end{enumerate}
\end{defi}

In the second condition, $J \in \Gamma(\mbox{End}(TM)) \cong
\Gamma(TM \otimes T^*M)$ is regarded as a vector valued one-form,
$J \in \Omega^1(M,TM)$. This condition can be rephrased as 
$\nabla J \in {\cal T}^1_{\;2}(M) =\Gamma(TM \otimes T^*M \otimes T^*M)$ being
symmetric:
\begin{equation}
(\nabla_X J)(Y) = (\nabla_Y J)(X) \;,\;\;\;\forall X,Y \in \mathfrak{X}(M) \;.
\end{equation}
The definition implies that $\nabla g \in {\cal T}^0_{\;3}(M)$ is totally symmetric, and therefore
ASK manifolds are Hessian. On a Hermitian manifold any two of the
three tensor fields $g$, $J$ and $\omega$ determine the third,\footnote{See \ref{App:Herm_Mfd}} 
and this allows to 
replace condition 2 by the alternative condition
\begin{itemize}
\item[2'.] $\nabla g\in {\cal T}^0_{\;3}(M)$ is completely symmetric.
\end{itemize}
Thus we may say that an ASK manifold is a K\"ahler manifold with a 
compatible Hessian structure. The associated flat connection $\nabla$
 is called the {\em special connection}.  If we impose that $\nabla$-affine
coordinates on $M$ are $\omega$-Darboux
coordinates, this restricts our freedom of making affine transformations
to those where the linear part is symplectic. We will call the corresponding
group 
$\mbox{Aff}_{Sp(\mathbb{R}^{2n})}(\mathbb{C}^{2n}) =
Sp(\mathbb{R}^{2n}) \ltimes \mathbb{C}^{2n} \subset \mbox{Aff}(\mathbb{C}^{2n})$
the {\em affine symplectic group}.

We will now verify the statements made in the preceding paragraphs using
special real coordinates. Since the connection $\nabla$ is flat and 
torsion-free, we can choose local $\nabla$-affine coordinates $q^a$
which define a parallel coframe $e^a = dq^a$, $\nabla e^a=0$ and
a parallel frame $e_a =\partial_a=\frac{\partial}{\partial q^a}$, $\nabla e_a=0$, see 
\ref{app:flat-connections}. Such coordinates are unique up to affine
transformations. 
The connection $\nabla$ is
symplectic, and therefore 
\begin{equation}
\nabla \omega = \nabla \left( \frac{1}{2} \omega_{ab} e^a \wedge e^b\right)=
\frac{1}{2} \partial_c \omega_{ab} e^c \otimes e^a
\wedge e^b + \frac{1}{2} \omega_{ab} \nabla (e^a \wedge e^b) =0\;.
\end{equation}
In $\nabla$-affine coordinate the second term vanishes, and the symplectic
form $\omega$ has constant coefficients:
\begin{equation}
\nabla \omega =0 \Rightarrow \partial_c \omega_{ab} = 0  \;.
\end{equation}
We can fix a standard form for the constant antisymmetric
matrix $\omega_{ab}$. The conventional choice we make is
\begin{equation}
\omega = \frac{1}{2} \omega_{ab} dq^a \wedge dq^b =
\Omega_{ab} dq^a \wedge dq^b = 2 dx^I \wedge dy_I \;,
\end{equation}
where 
\begin{equation}
\Omega_{ab} = \left( \begin{array}{cc}
0 & \mathbbm{1} \\
- \mathbbm{1} & 0 \\
\end{array} \right) \;.
\end{equation}
The coordinates 
$q^a = (x^I, y_I)$ are called {\em special real coordinates}, and the 
splitting of the $q^a$ into $x^I$ and $y_I$ corresponds to the choice of
a {\em polarization}, that is, a splitting of the symplectic vector space
$T_p M$, $p \in M$ into two maximally isotropic subspaces. The special real 
coordinates $q^a$ are $\omega$-Darboux coordinates, but differ from standard 
Darboux coordinates by a factor $\sqrt{2}$.\footnote{Darboux
  coordinates are usually normalized such that $\omega = \frac{1}{2}
  \Omega_{ab} 
d\tilde{q}^a \wedge d \tilde{q}^b = d\tilde{x}^I \wedge
d\tilde{y}_I$. }  
Choosing special real coordinates  restricts our freedom to perform coordinate
transformations to affine symplectic transformations,
\begin{equation}
\left( \begin{array}{c}
x \\  y \\
\end{array} \right) \mapsto
M \left( \begin{array}{c}
x \\  y \\
\end{array} \right) + 
\left( \begin{array}{c}
a \\ b \\
\end{array} \right) \;,\;\;\;M \in Sp(2n,\mathbb{R}) \;,\;\;a,b\in \mathbb{R}^n \;.
\end{equation}

Next, we evaluate the condition $d_\nabla J=0$ in special real
coordinates, using the rules for the covariant exterior derivative 
from \ref{App:ConnVB}:
\begin{equation}
0 = d_\nabla J = d( J^a_b e^b) \otimes e_a + J^a_b e^b \otimes \nabla e_a \;.
\end{equation}
In the co-frame $e^a =dq^a$ 
this condition reduces to
\begin{equation}
\label{d-nabla-J-local}
d_\nabla J = (\partial_c J^a_{\;b}) (dq^c \wedge dq^b) \otimes \partial_a = 0 
 \Rightarrow \partial_{[c} J^a_{\;b]} = 0 \;.
\end{equation}
To relate this to $\nabla J$ being symmetric, note that 
\begin{equation}
\nabla_X J = X^a (\partial_a J^c_{\;b}) e^b \otimes e_c +  J^c_{\;b} \nabla_X(e^b\otimes e_c)
\end{equation}
reduces in special real coordinates to
\begin{equation}
\nabla_X J = (X^a \partial_a J^c_{\;b}) dq^b \otimes \partial_c
\end{equation}
so that
\begin{equation}
(\nabla_X J)(Y) = X^a Y^b (\partial_a J_{\;b}^c) \partial_c  \;.
\end{equation}
Using (\ref{d-nabla-J-local}) we see that
\begin{equation}
d_\nabla J=0 \Leftrightarrow (\nabla_X J)(Y) = (\nabla_Y J)(X) \;,\;\;\ \forall X,Y\in \mathfrak{X}(M)\;,
\end{equation}
that is, $\nabla J$ is symmetric, $\nabla J \in \Gamma( \mbox{Sym}^2(T^*M) \otimes TM)$. Metric and K\"ahler
form are related by 
\begin{equation}
\omega(X,Y) = g(X,JY) \Leftrightarrow g(X,Y) = - \omega(X,JY) \;.
\end{equation}
In local coordinates this implies
\begin{equation}
\omega_{ab} = g_{ac} J^c_{\;b} \Leftrightarrow g_{ab} = -\omega_{ac}J^c_{\;b}
\Leftrightarrow J^a_{\;b} = g^{ac} \omega_{cb} 
\;.
\end{equation}
In special real coordinates, 
\begin{equation}
(\nabla_X g) (Y,Z) = \partial_c g_{ab} X^c Y^a Z^b  \;.
\end{equation}
Expressing $g_{ab}$ in terms of $\omega_{ab}$ and $J^a_{\;b}$, and using
that $\omega_{ab}$ is constant in special real coordinates, we find
\begin{equation}
\partial_c g_{ab} = - \omega_{ad} \partial_c J^d_{\;b} \;.
\end{equation}
Using (\ref{d-nabla-J-local}) we obtain that $\partial_c g_{ab}$ is
totally symmetric, which shows that $g$ is Hessian. It is also clear
that for a flat, torsion-free and symplectic connection 
$\nabla$, $g$ being Hessian implies that $\nabla J$ is symmetric, so
that condition 2 in the definition of an ASK manifold can be replaced
by condition 2'.

For later use we collect further local formulae in special real coordinates.
The metric is Hessian,
\begin{equation}
g= H_{ab} dq^a dq^b \;,\;\;\; H_{ab} = \frac{\partial^2 H}{\partial q^a \partial q^b}\;.
\end{equation}
We denote the inverse metric coefficients by $H^{ab}$. 
The inverse of $\Omega_{ab} = \frac{1}{2} \omega_{ab}$ is
\begin{equation}
\Omega^{ab} = 2 \omega^{ab} = \left( \begin{array}{cc}
0 & - \mathbbm{1} \\
\mathbbm{1} & 0 \\
\end{array} \right) \;.
\end{equation}
Using that $J^a_{\;\;b} = H^{ac} \omega_{cb}$ and $J^a_{\;\;c} J^c_{\;\;b}
= - \delta^a_b$ we obtain,
\begin{equation}
\frac{1}{2} \Omega^{ab} = - 2 H^{ac}H^{bd} \Omega_{cd}
\Leftrightarrow H_{ab} \Omega^{bc} H_{cd} = - 4\Omega_{ad} \;,
\end{equation}
where the numerical factors are due to the normalization of $\Omega_{ab}$.  
The components of the complex structure in terms of $H_{ab}$ and $\Omega_{ab}$ are:
\begin{equation}
J^{a}_{\;\;b} = 2 H^{ac} \Omega_{cb} = - \frac{1}{2} \Omega^{ac} H_{cb} \;.
\end{equation}
As on any Hessian manifold, there is a dual special connection 
$\nabla_{\rm dual} = 2 D - \nabla$, whose affine coordinates are
the dual special real coordinates,
\begin{equation}
q_a := H_a := \frac{\partial H}{\partial q^a} \;.
\end{equation}
As discussed in section \ref{sec:dual_hessian},
the metric coefficients with respect to $q_a$ are given
by the inverse matrix $H^{ab}$:
\begin{equation}
g = H^{ab} dq_a dq_b \;,\;\;\;H^{ac}H_{cb} = \delta^a_c \;,
\end{equation}
and the dual Hesse potential is obtained by a Legendre transformation:
\begin{equation}
H^{ab} = \frac{\partial^2 H_{\rm dual}}{\partial q_a \partial q_b} \;,\;\;\;
H_{\rm dual} = q^a q_a - H \;.
\end{equation}
The dual special coordinates $q_a$ are $\omega$-Darboux coordinates:
\begin{equation}
\omega = \Omega_{ab} dq^a \wedge dq^b = 2 dx^I \wedge dy_I =
- \frac{1}{4} \Omega^{ab} dq_a \wedge dq_b = 2 du^I \wedge dv_I \;,
\end{equation}
corresponding to the dual polarization
\begin{equation}
\label{dual_coordinates}
q_a =: (2v_I, - 2 u^I)  \;.
\end{equation}
Special real coordinates are adapted to the symplectic and Hessian
structure of an ASK manifold. We now turn to the complex aspects
of ASK geometry, following \cite{Freed:1997dp}. 
The complexified tangent bundle $T_{\mathbb{C}}M$ of
$M$ decomposes into the holomorphic tangent bundle $T^{(1,0)}M$ and
the anti-holomorphic tangent bundle $T^{(0,1)}M$,\footnote{See \ref{App:cplxmfds} for some
background on complex manifolds.} 
\begin{equation}
T_\mathbb{C}M = T^{(1,0)}M  \oplus T^{(0,1)}M \;,
\end{equation}
which can be characterized as the eigendistributions of the complex
structure $J$,
\begin{equation}
TM^{(1,0)} = \mbox{ker}(J - i \mathbbm{1}) \;,\;\;\;
TM^{(0,1)} = \mbox{ker}(J + i \mathbbm{1}) \;.
\end{equation}
Similarly, the complexified cotangent bundle decomposes as
$T^*_{\mathbb{C}}M = T^{*(1,0)}M \oplus T^{*(0,1)}M$. 
Since $d_\nabla J=0$, the projection operator
\begin{equation}
\Pi^{(1,0)} = \frac{1}{2} \left( \mathbbm{1} + i J \right) \in
\Gamma( T^* _\mathbb{C} M \otimes T^{(1,0)} M ) \;: \;\;\;
T_{\mathbb{C}}M \rightarrow
T^{(1,0)}M 
\end{equation} 
satisfies $d_\nabla \Pi^{(1,0)}=0$. Hence locally 
$\Pi^{(1,0)} = d_\nabla \zeta = \nabla \zeta$,
where $\zeta$ is a complex, not necessarily holomorphic vector field,
which is unique up to a flat complex vector field.\footnote{The relevant properties of
the exterior covariant derivative $d_\nabla$ are reviewed in \ref{App:ConnVB}.} 
In special real coordinates $\zeta$ has an expansion\footnote{Compared to \cite{Freed:1997dp} we have changed the relative sign
between the two terms of $\zeta$ to be consistent with our conventions.}
\begin{equation}
\zeta = X^I \frac{\partial}{\partial x^I} + W_I
\frac{\partial}{\partial y_I}\;,
\end{equation}
where $X^I, W_I$ are complex functions on $M$. Then
\begin{equation}
\Pi^{(1,0)} = dX^I \otimes \frac{\partial}{\partial x^I} + dW_I \otimes
\frac{\partial}{\partial y_I} \;,
\end{equation}
where $dX^I, dW_I \in T^{* (1,0)}M$, which implies that the functions
$X^I, W_I$ are holomorphic. Since $\mbox{Re}(\Pi^{(1,0)})=\mbox{Id}_{TM}$ it follows
that
\begin{equation}
\mbox{Re}(dX^I) = dx^I \;,\;\;\;\mbox{Re}(dW_I) = dy_I \;.
\end{equation}
Using that the real differentials $dx^I$ are linearly independent, it can 
be shown that the differentials $dX^I$ are linearly 
independent over $\mathbb{C}$, and therefore the holomorphic 
functions $X^I$ define a local holomorphic coordinate system on 
$M$ \cite{Freed:1997dp,Alekseevsky:1999ts}. 
These are the so-called {\em special holomorphic coordinates}, often
simply called special coordinates. Similarly, the functions $W_I$
define another holomorphic coordinate system on $M$, which is called
the dual (holomorphic) special coordinate system.

Since $\frac{\partial}{\partial X^I}$ is of type $(1,0)$, that is 
$\Pi^{(1,0)} \frac{\partial}{\partial X^I}= \frac{\partial}{\partial
  X^I}$, it follows that 
\begin{equation}
\frac{\partial}{\partial X^I} = 
  \frac{\partial}{\partial x^I} + \frac{\partial W_K}{\partial X^I}
  \frac{\partial}{\partial y_K} \;.
\end{equation}
The K\"ahler form $\omega = 2 dx^I \wedge dy_I = \frac{1}{2} (dX^I + d\bar{X}^I)
\wedge (dW_I + d\bar{W}_I)$ must be a $(1,1)$-form, therefore
\begin{equation}
0= dX^I \wedge dW_I = dX^I \wedge \frac{\partial W_I}{\partial X^J}
dX^J  \Rightarrow \frac{\partial W_I}{\partial X^J} = \frac{\partial
  W_J}{\partial X^I} \;.
\end{equation}
This implies that locally $W_I$ is the holomorphic gradient of a
function $F(X^I)$, called the {\em prepotential}, which is determined up to
a constant:
\begin{equation}
W_I  = \frac{\partial F}{\partial X^I} =: F_I \;,\;\;\;
\frac{\partial W_I}{\partial X^J}  = \frac{\partial^2 F}{\partial
    X^I \partial X^J } =: F_{IJ} \;.
\end{equation}
The K\"ahler form can be expressed in terms of the prepotential as
\begin{equation}
\omega = - \frac{i}{2} N_{IJ} dX^I \wedge d\bar{X}^J \;,
\end{equation}
where
\begin{equation}
\label{NIJdef}
N_{IJ} = 2 \mbox{Im} F_{IJ} = - i (F_{IJ} - \bar{F}_{IJ}) \;,
\end{equation}
and where $\bar{F}_{IJ}$ is the complex conjugate of $F_{IJ}$. The
corresponding K\"ahler metric and Hermitian form are
\begin{equation}
g = N_{IJ} dX^I d\bar{X}^J \;,\;\;\;
\gamma = g + i \omega = N_{IJ} dX^I \otimes d\bar{X}^J \;.
\end{equation}
Since
\begin{equation}
\label{kaehmetg}
N_{IJ} = \frac{\partial^2 K}{\partial X^I \partial \bar{X}^J}
\;,\;\;\;
K = i (X^I \bar{F}_I - F_I \bar{X}^I)  \;,
\end{equation}
the function $K$ is a K\"ahler potential. The choice of the sign 
in the definition of $K$ is conventional. Sometimes $N_{IJ}$ and $K$
are defined with an additional minus sign. Note that to obtain 
a model where $N_{IJ}$ is positive definite, or more generally 
is non-degenerate and carries a specific signature, one may need
to restrict the coordinates $X^I$ to a suitable domain. This has to
be analysed model by model.

We have now recovered the original 
definition of ASK manifolds in terms of local formulae in 
special coordinates \cite{deWit:1984pk}: {\em an ASK manifold is a 
K\"ahler manifold where the K\"ahler potential admits a holomorphic
prepotential.}\footnote{Note that K\"ahler potentials are only
  determined up to K\"ahler transformations, and the formula expressing
  $K$ in terms of $F$ 
  provides only a subclass of the K\"ahler potentials for a given ASK
  metric.}

\subsubsection{Extrinsic construction as a K\"ahlerian Lagrangian immersion \label{sec:kaehimm}}

The intrinsic definition of \cite{Freed:1997dp} has an extrinsic counterpart:
every simply connected ASK manifold can be realized as a 
K\"ahlerian Lagrangian immersion 
into the standard complex symplectic vector space $V=T^*\mathbb{C}^n \cong
\mathbb{C}^{2n}$ \cite{Alekseevsky:1999ts}.
Lagrangian immersions have a potential, which for ASK manifolds is 
the holomorphic prepotential.

We start with
the standard complex symplectic vector space
$V=T^*\mathbb{C}^n$ equipped with complex Darboux coordinates 
$(X^I, W_I)$, the standard complex symplectic form $\Omega = dX^I \wedge dW_I$, 
and the standard real structure defined by complex conjugation
$\tau: V \rightarrow V,\;v \mapsto \tau v = \bar{v}$.\footnote{See \ref{app:complex_symplectic} for
a few additional remarks regarding complex symplectic manifolds.} 
The set of fixed points of the real structure $\tau$ are the real points
$V^\tau = T^*\mathbb{R}^n\cong \mathbb{R}^{2n} \subset \mathbb{C}^{2n}$. Given these data we can define the Hermitian
form 
\begin{equation}
\label{hermV}
\gamma_V =i \Omega(\cdot, \tau \cdot) = i \left( dX^I \otimes d \overline{W}_I 
- dW_I \otimes d\overline{X}^I \right)  = g_V + i \omega_V \;,
\end{equation}
which has complex signature $(n,n)$. Its real part defines a flat K\"ahler metric
of real signature $(2n,2n)$, with associate K\"ahler form $\omega_V$, and
complex structure $I_V$. 

Let $M$ be a connected complex manifold of complex dimension $n$. A 
holomorphic immersion $\phi: M \rightarrow V$ is called
{\em  non-degenerate} if $g_M := \phi^* g_V$ is non-degenerate, where
$\phi^* g_V$ denotes the pull-back of the metric $g_V$ by $\phi$ to $M$, 
see \ref{App:Pull_back}.
In this 
case $g_M$ is a K\"ahler metric on $M$, which in general has indefinite signature. 
Therefore non-degenerate holomorphic immersions are also called
{\em K\"ahlerian} immersions. One can show that $\phi^* g_V$ being non-degenerate
is equivalent to $\omega_M:= \phi^* \omega_V$ being non-degenerate, 
and also to $\gamma_M := \phi^* \gamma_V$ being non-degenerate. 

A holomorphic immersion $\phi: M \rightarrow V$ 
is called {\em Lagrangian} if $\phi^* \Omega =0$. 
It has been shown in \cite{Alekseevsky:1999ts} that a  K\"ahlerian Lagrangian
immersion $M \rightarrow V$ induces on $M$ the structure of an affine special
K\"ahler manifold. Conversely every simply connected affine special K\"ahler manifold admits 
a K\"ahlerian Lagrangian immersion which induces its ASK structure. The immersion
is unique up to transformations of $V$ which leave the data $(I_V, \Omega, \tau)$
invariant. These transformations act on complex Darboux coordinates as
\begin{equation}
\left( \begin{array}{c}
X^I \\ W_I \\
\end{array} \right) \mapsto 
M \left( \begin{array}{c}
X^I \\ W_I \\
\end{array} \right) + 
 \left( \begin{array}{c}
 A^I \\ B_I \\
 \end{array} \right) \;,\;\;\;
 M \in \mbox{Sp}(2n,\mathbb{R}) \;,\;\;\;A^I, B_I \in \mathbb{C} \;,
 \end{equation}
and belong to the subgroup $\mbox{Aff}_{Sp(\mathbb{R}^{2n})} (\mathbb{C}^{2n}) 
= Sp(2n,\mathbb{R}) \ltimes \mathbb{C}^{2n}$ of the complex affine group
$GL(2n,\mathbb{C}) \ltimes \mathbb{C}^{2n}$. 

The {\em Liouville form} $\lambda = W_I dX^I$ of $V$ is a potential for
the symplectic form: $d \lambda = - \Omega$. Therefore its pullback 
$\phi^* \lambda$ under the Lagrangian immersion $\phi$ is locally 
exact and admits a holomorphic potential $F$, defined on some 
domain $U\subset M$:
\begin{equation}
dF = \phi^* \lambda \;.
\end{equation}
The pullbacks $\tilde{X}^I =\phi^* X^I$, $\tilde{W}_I = \phi^* W_I$ 
are holomorphic functions on $M$. Since $\phi$ is non-degenerate
one can pick $n$ independent functions and use them as local holomorphic
coordinates on $M$. By applying a symplectic transformation if necessary one can
always arrange that $\tilde{X}^I$ are local holomorphic coordinates on $M$.
In this case the functions $\tilde{W}_I$ form a second `dual' holomorphic coordinate
system, which we will discuss in more detail in section \ref{sect:dual_coord}.
We can always choose $U\subset M$ small enough so that $\phi$ becomes an embedding.
In this case we do not need to distinguish by notation between $(X^I, W_I)$ and 
$(\tilde{X}^I, \tilde{W}_I)$. If we use special coordinates $X^I$ on $M$
then $dF = W_I dX^I$, implying  $W_I = F_I =\frac{\partial F}{\partial X^I}$. 
Note that the integrability condition $F_{IJ} = \partial_I W_J = \partial_J W_I = F_{JI}$
is satisfied since $\phi$ is Lagrangian. The immersion 
$\phi$ locally takes the form
\begin{equation}
\mathbb{C}^n \supset U \ni (X^I) \mapsto (X^I, W_I) \in T^*U \subset \mathbb{C}^{2n}\;,
\end{equation}
where we identify $U$ with a domain in $\mathbb{C}^n$ using the coordinates $X^I$. 
We can also identify $\phi$ with $dF$ and $\phi(U)$ with the graph
\begin{equation}
\left\{ (X^I, W_I) \in \mathbb{C}^{2n} | (X^I) \in U , W_I = \frac{\partial F}{\partial X^I} \right\}
\end{equation}
of $dF$ over $U$. With these properties and identifications $U\subset \mathbb{C}^n$
is called an {\em affine special K\"ahler domain}. 

We proceed by deriving local expressions for the metric $g_M$, K\"ahler form
$\omega_M$ and special connection $\nabla$ on $M$. 
We decompose the complex Darboux coordinates on $V$
into their real and imaginary parts:
\begin{equation}
X^I = x^I + i u^I \;,\;\;\;
W_I = y_I  + i v_I  \;.
\end{equation}
Then 
\begin{equation}
\gamma_V = g_V + i \omega_V = 
\frac{1}{2} \left( dx^I dv_I - du^I dy_I \right) 
+ i \left( dx^I \wedge dy_I + du^I \wedge dv_I \right)
\end{equation}
and 
\begin{equation}
\Omega = dx^I \wedge dy_I - du^I \wedge dv_I + i \left(
du^I \wedge dy_I + dx^I \wedge dv_I \right) \;.
\end{equation}
By pullback we define the functions 
$\tilde{x}^I = \mbox{Re} (\phi^* x^I)$, $\tilde{y}_I = \mbox{Re} (\phi^* y_I)$
on $M$. 
Since the immersion is Lagrangian,
\begin{equation}
\mbox{Re}( \phi^*  \Omega) = 0 \Rightarrow
d\tilde{x}^I \wedge d\tilde{y}_I = d\tilde{u}^I \wedge d\tilde{v}_I  \;,
\end{equation}
and therefore
\begin{equation}
\omega_M = d\tilde{x}^I \wedge d\tilde{y}_I + d\tilde{u}^I \wedge d\tilde{v}_I 
= 2 (d\tilde{x}^I \wedge d\tilde{y}_I ) \;.
\end{equation}
For a simply connected ASK manifold $M$, 
$(\tilde{x}^I, \tilde{y}_I)$ are globally defined functions, but they are  
only global coordinates if the immersion $\phi$ is an embedding. 
By restricting to a domain $U\subset M$ where $\phi$ becomes an
embedding, we can use $(\tilde{x}^I, \tilde{y}_I)$ as coordinates and do 
not need to distinguish them from $(x^I, y_I)$ by notation. They are 
Darboux coordinates for the K\"ahler form $\omega_M$, and define a
flat, torsion-free, symplectic connection $\nabla$ by $\nabla d{x}^I=0$, 
$\nabla d{y}_I=0$. One can show that $\nabla$ is the special connection 
occurring in the intrinsic definition, 
and
that $({x}^I, {y}_I)$ are the corresponding 
special real coordinates. 

Next, we work out some expressions in terms of special holomorphic
coordinates.
The pull-back of the Hermitian form $\gamma_V$ is
\begin{equation}
\gamma_M = \phi^* \gamma_V = i \left( dX^I \otimes d\bar{F}_I - dF_I \otimes
d\bar{X}^I\right) = N_{IJ} dX^I \otimes d\bar{X}^J \;,
\end{equation}
where
\begin{equation}
N_{IJ}  = 2 \mbox{Im}F_{IJ} = \frac{\partial^2 K}{\partial X^I \partial \bar{X}^J} \;,\;\;
K = i (X^I \bar{F}_I - F_I \bar{X}^I )\;.
\end{equation} 
By decomposing $\gamma_M = g_M + i \omega_M$ 
we obtain a non-degenerate, in general indefinite K\"ahler metric 
\begin{equation}
g_M = N_{IJ} dX^I d\bar{X}^J \;,
\end{equation}
with associated K\"ahler form
\begin{equation}
\omega_M = - \frac{i}{2} N_{IJ} dX^I \wedge d\bar{X}^J \;.
\end{equation}
Thus we have recovered all the local expressions of section \ref{sect:intrinsic_ASK}. 

We note that the characteristic property of a K\"ahlerian immersion, 
the non-degeneracy of $g_M = \phi^* g_V$ corresponds in special
coordinates to $F_{IJ} = \partial^2_{IJ} F$ having an invertible
imaginary part. A holomorphic one-form $\phi=dF$ is called 
regular if $\det (\mbox{Im} F_{IJ}) \not=0$. It follows that locally 
every regular closed holomorphic one-form defines 
a Lagrangian K\"ahlerian immersion.

We conclude this section by expanding on some details.  
Firstly, the image $\phi(U)$ of $U\subset M$ is not automatically 
a graph, although this is the generic situation. For special choices 
of $\phi$ the functions $X^I$ on $U$ are not independent and do 
not define a holomorphic coordinate system on $U$. This can 
be detected by $W_I$ not satisfying the integrability condition
for the existence of of a prepotential $F$ with gradient $F_I=W_I$. 
In this situation one can choose local holomorphic
coordinates $z^I$ on $U$ and work with the functions
$(X^I(z), F_I(z))$. As we will discuss in section \ref{sec:formlinebund}, the map
\begin{equation}
z \mapsto (X^I(z), F_I(z)) 
\end{equation}
can be interpreted as a holomorphic section of a line bundle 
over $M$. We will discuss definitions of ASK geometry based on 
line bundles in section \ref{sect:comparison}.
Finally, only simply connected ASK manifolds admit a global immersion into
$V\cong T^*\mathbb{C}^n$. As far as the local description is concerned 
this is not an issue, as we can restrict to simply connected submanifolds
$U\subset M$. In order to obtain a global construction of general, not necessarily
simply connected, ASK manifolds, the vector space $V$ must be replaced by 
an affine bundle with fibre $V$. This will be discussed in section 
\ref{sect:SK_vector_bundle}.

\subsubsection{Extrinsic construction as a parabolic affine hypersphere  \label{Sect:Aff_HS}}

Affine special K\"ahler manifolds admit a second extrinsic construction, which
is real rather than complex, with the Hesse potential as generating function. Our presentation follows \cite{1999math.....11079B,Cortes:2001bta}.

In this construction the ASK manifold $M$ is immersed into $\mathbb{R}^{2n+1}$ 
as a hypersurface
\begin{equation}
\varphi: M \rightarrow \bR^{2n+1}  \;.
\end{equation}
Using the standard connection $\partial$ (defined by the partial derivative
with respect to linear coordinates) on $\mathbb{R}^{2n+1}$
and a vector field $\xi$ which is transversal to $M$, one can give $M$
the structure of an {\em affine hypersphere}, see \ref{App:affine_hyperspheres}.
The decomposition 
\begin{equation}
\partial_X Y = \nabla_X Y + g(X,Y) \xi
\end{equation}
of derivatives of vector fields $X,Y$ tangent to $M$ defines a torsion-free connection
$\nabla$ and a so-called Blaschke metric $g$ on $M$. The connection
$\nabla$ is flat if the vector field $\xi$ is chosen such that
its integral lines are parallel on $\mathbb{R}^{2n+1}$,
$\partial \xi =0 $, and thus do not intersect at finite points. This
makes $M$ a {\em parabolic} (or improper) {\em affine hypersphere}.
A parabolic affine hypersphere is called {\em special} if there exists an almost complex
structure $J$ on $M$ such that 
$J$ is skew with respect to the Blaschke metric $g$,
and such that the fundamental form $\omega=g(\cdot, J \cdot)$ is $\nabla$-parallel. 
It has been shown in \cite{1999math.....11079B} that 
if $\varphi: M \rightarrow \mathbb{R}^{2n+1}$ is a special parabolic affine 
hypersphere with data $(J,\omega,\nabla)$, then $(M,J,g,\nabla)$ is an affine
special K\"ahler manifold. Conversely, any simply connected ASK manifold
admits an immersion as a  special parabolic affine hypersphere. 
The immersion is unique up to unimodular affine transformations
of $\mathbb{R}^{2n+1}$. In terms of $\nabla$-affine coordinates
$(x^I,y_I)$ on $M$, the immersion takes the form
\begin{equation}
\varphi \;: M \rightarrow \mathbb{R}^{2n+1} \;,
(x^I, y_I) \mapsto 
\varphi_F = (x^I, y_I, H(x,y) ) \;,
\end{equation}
where $H$ is the Hesse potential of the ASK manifold. 

Since any ASK manifold can also be characterized locally by a holomorphic 
prepotential $F$, the Hesse potential $H$ and the prepotential $F$ 
determine each other. It has been shown in \cite{1999math.....11079B} 
that their relation is
\begin{equation}
\label{HFleg}
H(x,y) = 2 \mbox{Im}(F(X(x,y))) - 2 \mbox{Re}(F_I(x,y)) \mbox{Im} X^I(x,y) \;,\;\;\;
\end{equation}
where
$x^I=\mbox{Re}(X^I)$ and $y^I = \mbox{Re}(F_I)$.
That is, the Hesse potential is
twice the Legendre transform of the imaginary 
part of the prepotential. Note that compared to the `full'
Legendre transformation which replaces all affine coordinates
by their duals, $q^a \mapsto q_a$, this is a `partial' 
Legendre transformation, where $u^I=\mbox{Im}(X^I)$ is replaced by 
$y_I=\mbox{Re}(F_I)$ as an independent variable,
$(x^I,u^I) \mapsto (x^I,y_I)$.

\subsubsection{Dual coordinate systems \label{sect:dual_coord}}

In section \ref{sec:dual_hessian} we have seen that Hessian 
structures always come in pairs, with associated dual affine
coordinate systems $q^a$ and $q_a$. This extends to ASK
manifolds through the existence of a dual (or conjugate) special connection,
$\nabla^{(J)}$, 
which coincides with the dual connection in the Hessian sense.
Consequently, apart from the special holomorphic coordinates
$X^I= x^I+iu^I$ and the special real coordinates $(x^I,y_I)$ an
ASK manifold has dual special holomorphic coordinates $W_I = F_I 
= y_I + i v_I$ and dual special real coordinates $(2 v_I, -2 u^I)$, c.f. 
\eqref{dual_coordinates}.
For the discussion of dual special connections 
we follow \cite{Alekseevsky:1999ts}.

Given a connection $\nabla$ and an invertible endomorphism field 
$A\in \Gamma(\mbox{End}(TM))$ one can define a new connection
by
\begin{equation}
\nabla^{(A)} X = A \nabla (A^{-1} X) \;.
\end{equation}
For a flat connection on a complex manifold $(M,J)$ one can in
particular define the one-parameter family of flat connections
$\nabla^\theta := \nabla^{\exp(\theta J)}$. By Taylor expanding 
$\exp (\theta J)$ and using that $J^2=-\Id$, we find that the 
connections $\nabla$
and $\nabla^\theta$ are related by
\begin{equation}
\nabla^\theta =\nabla + A^\theta \;,\;\;\;\mbox{where}\;\;\;
A^\theta = e^{\theta J} \nabla (e^{-\theta J}) = -\sin \theta e^{\theta
  J}  \nabla J \;.
\end{equation}
Note that this family of connections is periodic in $\theta$ and thus is
parametrized by $S^1$. If $(M,J,\omega,\nabla)$ is
an ASK manifold with special connection $\nabla$, then
$(M,J,\omega,\nabla^\theta)$ is an ASK manifold with special connection $\nabla^\theta$,
for any value of $\theta$. As
K\"ahler manifolds such manifolds are identical. In the physics
literature ASK manifolds are usually identified if their special
connections differ by $A=e^{\theta J}$, see section \ref{sect:comparison}.

The connection
\begin{equation}
\nabla^{\pi/2} = \nabla^{(J)} = \nabla - J \nabla J
\end{equation}
is called the connection conjugate to $\nabla$.
 The convex combination 
\begin{equation}
\label{Dnablanabladual}
D:= \frac{1}{2} (\nabla + \nabla^{(J)}) 
\end{equation}
of special connections satisfies $DJ=0$. For ASK manifolds the connection 
$D$ is metric compatible, and since it is by construction also torsion-free, 
$D$ is  the Levi-Civita connection. 

This implies that for ASK 
manifolds where the complex structure is $\nabla$-parallel,  $\nabla J=0$ 
(rather than just $d_\nabla$-closed), the K\"ahler metric is flat:
 $\nabla J =0$ implies 
$\nabla = \nabla^{(J)} = D$, so that the Levi-Civita connection $D$ is flat. In local special coordinates,
this corresponds to the case where the Hesse potential and the 
prepotential are quadratic polynomials. Physics-wise, these are free theories. 

Comparing (\ref{Dnablanabladual}) to (\ref{Dual_connection_Hessian})
shows that the conjugate 
connection $\nabla^{(J)}$ coincides with the dual connection
$\nabla_{\rm dual}$ in the Hessian hence. This implies 
that the special real 
coordinates with respect to $\nabla^{(J)}$ are the dual special 
real coordinates $q_a= H_a=(2v_I, -2u^I)$. The corresponding 
dual holomorphic coordinates are $W_I= y_I + i v_I$.

\subsubsection{Symplectic transformations for special complex and special real coordinates}

In this section we derive explicit formulae which relate the 
local expressions for the metric and other tensors in complex 
and real special coordinates. We also study how 
various quantities transform under symplectic transformations.

We start with comparing the coefficients 
of the metric in special holomorphic coordinates
$X^I$ and in special real coordinates $q^a=(x^I,y_I)$:
\begin{equation}
g_M = N_{IJ} dX^I d\bar{X}^J = H_{ab} dq^a dq^b \;.
\end{equation}
We need to express the Hessian  $(H_{ab})$ of $H$
in terms of the matrices
$R = (R_{IJ}) = (2\mbox{Re}(F_{IJ}))$
and $N=(N_{IJ}) = (2\mbox{Im}(F_{IJ}))$,
which are twice the real and imaginary part, respectively, 
of the holomorphic Hessian $(F_{IJ})$  of the prepotential $F$. 
This amounts to performing a coordinate transformation from the real coordinates
$(x^I,u^I)$ underlying the complex coordinates $X^I = x^I + i u^I$ to the
special real coordinates $q^a=(x^I,y_I)$. By taking derivatives of the relations
\begin{eqnarray}
X^I &=& x^I + i u^I(x,y)  \;,\nonumber \\
F_I &=& y_I + i v_I (x,y)  \;,
\label{XFxy}
\end{eqnarray}
we obtain the components of the Jacobians 
of the coordinate transformations
\begin{equation}
(x,u) \mapsto (x,y) \;,\;\;\;(x,y) \mapsto (x,u) \;.
\end{equation}
When taking derivatives of a function of the form $\tilde{f}(x,u)= f(x,y(x,u))$ 
we need to employ the chain rule
\begin{equation}
\tilde{f}_{x^I}   = f_{x^I} + f_{y_K} \frac{\partial y_K}{\partial x^I} \;,\;\;
\tilde{f}_{u^I} = f_{y_K} \frac{\partial y_K}{\partial u^I}  \;,
\end{equation}
where we use the short-hand notation $f_{x^I} = \frac{\partial f}{\partial x^I}$, etc.

Using this we obtain the Jacobians
\begin{equation}
\frac{D(x,u)}{D(x,y)} = \left( \begin{array}{cc}
\mathbbm{1}&0 \\
\left. \frac{\partial u}{\partial x}  \right|_{y} &
\left. \frac{\partial u}{\partial y}  \right|_{x} \\
\end{array} \right) =
\left( \begin{array}{cc}
\mathbbm{1} & 0 \\
N^{-1} R & -2 N^{-1} \\
\end{array} \right) 
\end{equation}
and 
\begin{equation}
\frac{D(x,y)}{D(x,u)} = \left( \begin{array}{cc}
\mathbbm{1}&0 \\
\left. \frac{\partial y}{\partial x}  \right|_{u} &
\left. \frac{\partial y}{\partial u}  \right|_{x} \\
\end{array} \right) =
\left( \begin{array}{cc}
\mathbbm{1} & 0 \\
\frac{1}{2}  R & -\frac{1}{2} N \\
\end{array} \right)  \;.
\end{equation}
Together with further relations given in \ref{jacob_conv}
one obtains 
\begin{equation}
\label{Hessian_RN}
(H_{ab}) = \left( \begin{array}{cc}
N + R N^{-1} R & -2 RN^{-1} \\
-2 N^{-1} R & 4 N^{-1} \\
\end{array} \right)  \;,
\end{equation}
where $N^{-1}=(N^{IJ})$  is the inverse of $N=(N_{IJ})$. 
As discussed in section \ref{Sect:Aff_HS}
the Hesse potential $H$ is related to the imaginary part of the prepotential by 
a Legendre transformation: 
\begin{equation}
H(q) = H(x,y) = 2 \mbox{Im} F(x+iu(x,y)) - 2 y_I u^I(x,y)  \;.
\end{equation} 
We can also express the metric in dual special real coordinates $q_a$:
\begin{equation}
g = H_{ab} dq^a dq^b = H^{ab} dq_a dq_b \;,
\end{equation}
where, as for any Hessian metric,  
the metric coefficients $H^{ab}$  with respect to the dual coordinates
are the inverse of $H_{ab}$, hence
\begin{equation}
(H^{ab}) = \left( \begin{array}{cc}
N^{-1} & \frac{1}{2} N^{-1} R \\
\frac{1}{2} R N^{-1} & \frac{1}{4} ( N + R N^{-1} R) \\
\end{array} 
\right) \;.
\end{equation}
The dual Hesse potential $H_{\rm dual}$, 
\begin{equation}
H^{ab} = \frac{\partial^2  H_{\rm dual}}{ \partial q_a \partial q_b}
\end{equation}
is related to the Hesse potential by a full Legendre transformation\footnote{We call
this a `full' Legendre transformation because it involves all of the variables. In 
contrast the Legendre transformation relating the Hesse potential and the prepotential only 
involves half of the coordinates, $(x^I,y_I) \mapsto (x^I,u^I)$, and therefore we will call it a
`partial' Legendre transformation.}
\begin{equation}
H_{\rm dual} =q^a q_a - H \;,
\end{equation} 
as discussed in section \ref{sec:dual_hessian}.

Special real coordinates are unique up to affine transformations
with linear part in $\mbox{Sp}(\mathbb{R}^{2n}) = \mbox{Sp}(2n, \mathbb{R})$. 
In the following we discard translations and focus on linear symplectic transformations,
under which
the  coordinates $q^a$ transform as 
\begin{equation}
q^a \mapsto {\cal O}^a_{\;\;b} q^b \;,\;\;\;
\end{equation}
where ${\cal O}=({\cal O}^a_{\;\;b})$ is a symplectic matrix:
\begin{equation}
{\cal O}^b_{\;\;a} \Omega_{bc} {\cal O}^c_{\;\;d} = \Omega_{ad}
\Leftrightarrow 
{\cal O}^T \Omega {\cal O} = \Omega \;.
\end{equation}
We will call any object transforming in the fundamental representation of 
$\mbox{Sp}(2n,\mathbb{R})$ a {\em symplectic vector}. 
Objects $p_a$ which transform in the contragradient representation,
\begin{equation}
p_a \mapsto {\cal O}_a^{\;\;b} p_b \;,
\end{equation}
where
\begin{equation}
{\cal O}_a^{\;\;b} {\cal O}^c_{\;\;b} =\delta^a_c 
\Leftrightarrow
({\cal O}_a^{\;\;b}) = {\cal O}^{T,-1}\;,
\end{equation}
will be called {\em symplectic co-vectors}. The matrix $\Omega$ 
intertwines the two representations: if $q^a$ is a symplectic vector, then
$\Omega_{ab} q^b$ is a symplectic co-vector. Similarly, we 
define {\em symplectic tensors} as objects which have components with 
several upper and lower indices,  such that each upper index transforms in the 
fundamental and each lower index
transforms in the contragradient representation.

As an example, the metric $g=H_{ab} dq^a dq^b$ is an invariant symmetric 
rank two co-tensor, and since 
$dq^a$ transform in the fundamental representation, the components $H_{ab}$ of $g$ 
transform as follows:
\begin{equation}
H_{ab} \mapsto {\cal O}_{a}^{\;\;b} {\cal O}_c^{\;\;d} H_{cd} \;.
\end{equation}
Therefore 
\begin{equation}
H^{ab} \mapsto {\cal O}^{a}_{\;\;b} {\cal O}^c_{\;\;d} H^{cd} \;,
\end{equation}
which implies that the dual coordinates $q_a = H_a$ transform 
contragradiently, 
\begin{equation}
q_a \mapsto {\cal O}_a^{\;\;b} q_b  \;.
\end{equation}
Consistency requires that the Hesse potential $H$ must be a symplectic function since 
$H_{ab} =\partial^2_{a,b} H$. The tensor $\Omega_{ab}$ is by definition
an invariant tensor, and the complex structure $J^a_{\;b}$ is a symplectic 
tensor of type $(1,1)$. Therefore all quantities we have defined using special
real coordinates and dual special real coordinates are tensor components which
transform as indicated by their indices.

In contrast, quantities expressed in terms of special holomorphic 
coordinates do not transform as tensor components in general.  
Since $X^I=x^I + i u^I$, $W_I = y_I + i v_I$, where $q^a=(x^I, y_I)$ and 
$q_a = (2v_I, -2 u^I)$, it is clear that $(X^I, F_I)^T$ is a complex linear combination
of symplectic vectors and therefore a complex symplectic  
vector. As in section \ref{sec:em-dual-2d}  we set
\begin{equation}
{\cal O} = \left( \begin{array}{cc}
U & Z \\
W & V\\
\end{array} \right) \;,
\end{equation}
so that
\begin{equation}
U^T V - W^T Z = V^T U - Z^T W = \mathbbm{1} \;,\;\;\;
U^T W = W^T U \;,\;\;\;Z^T V = V^T Z 
\end{equation}
and
\begin{eqnarray}
X^I & \mapsto & U^I_{\;\;J} X^J + Z^{IJ} F_J \;, \label{SympTransfXF}\\
F_I & \mapsto & V_I^{\;\;J} F_J + W_{IJ} X^J \;.\nonumber
\end{eqnarray}
The special holomorphic coordinates $X^I$ comprise half of the components
of a symplectic vector and therefore do not define a symplectic tensor by themselves. 
We have already seen in section \ref{sec:em-dual-2d} that the holomorphic prepotential 
$F(X^I)$ is not a symplectic function. There we worked out the explicit transformation 
formula for the special case of prepotentials which are 
homogeneous of degree two.
We will provide a general formula for the transformation of the prepotential
together with a geometrical interpretation in section \ref{sect:SK_vector_bundle}

Similarly, $N_{IJ}$, $N^{IJ}$ and 
other expressions involving holomorphic indices do not transform
as symplectic tensors, as we have already seen in section \ref{sec:em-dual-2d}.
By contracting
the symplectic vector $(X^I, F_I)^T$ with its complex conjugate,
we obtain a symplectic function, namely the K\"ahler potential:
\begin{equation}
K = i(X^I \bar{F}_I - F_I \bar{X}^I ) \;.
\end{equation}

Comparing to section \ref{sec:emduality} we see that 
the holomorphic and real formalism of special geometry are
related in a way similar to the relation between the Lagrangian
and Hamiltonian formalism of mechanics. In particular, the real (Hamiltonian)
formalism is covariant with respect to symplectic transformations, whereas
the holomorphic (Lagrangian) formalism is not. 
We remark that the functions $(q^a)=(x^I, y_I)$, 
always define local coordinates on $M$,
irrespective of whether the
`symplectic frame' $(X^I,F_I)$ allows a prepotential or not.
For simply connected ASK manifolds $q^a$ are in fact globally defined functions,
since the immersion $\phi$ is global. Note, however that they only define a global 
coordinate system on $M$ if $\phi$ is a global embedding, which need not be
the case even if $\phi$ is a global immersion. In contrast, $X^I$, which are
half of a set of complex coordinates $(X^I,F_I)$ on $V$, only define local complex 
coordinates on $U\subset M$ if $\phi(U)\subset V$ is the graph of a 
map $V \rightarrow V\;: X^I \mapsto W_I = F_I(X)$.

\subsection{Conical affine special K\"ahler geometry \label{caskg}}

When extending ${\cal N}=2$ supersymmetry to ${\cal N}=2$ 
superconformal symmetry, two additional bosonic symmetries
become relevant for vector multiplets: dilatations $\mathbb{R}^{>0}$
and phase transformations $U(1)$. On the scalar fields these are 
realized as a holomorphic homothetic action of $\mathbb{C}^* \cong
\mathbb{R}^{>0} \times U(1)$. To obtain a superconformal Lagrangian,
the prepotential must be homogeneous
of degree two under complex scale transformations $X^I \mapsto
\lambda X^I$, $\lambda \in \mathbb{C}^*$, while the Hesse potential
must be homogeneous of degree two under real scale transformations
$q^a \rightarrow \lambda q^a$, $\lambda \in \mathbb{R}^{>0}$, and
invariant under $U(1)$ transformations.
We will follow \cite{Alekseevsky:1999ts,Cortes:2009cs}.

\begin{defi}
{\bf Conical affine special K\"ahler manifolds (CASK manifolds).}  A conical affine special
K\"ahler manifold $(M,g,\omega,\nabla,\xi)$ is an affine
special K\"ahler manifold $(M,g,\omega,\nabla)$ equipped with a nowhere 
null vector field $\xi$, such that
\begin{equation}
\label{xi_CASK}
D \xi = \nabla \xi = \mathrm{Id}_{TM} \;,
\end{equation}
where $D$ is the Levi-Civita connection of $g$.
\end{defi}

From section \ref{sect:ConHess} we know that (\ref{xi_CASK})
implies that $(M,g,\nabla,\xi)$ is a 2-conical Riemannian manifold,\footnote{As usual
we admit indefinite signature.}
hence a Riemannian cone in the standard sense. Since $(M,g,\nabla)$ is in addition Hessian,
it is a 2-conical Hessian manifold in the sense of Definition \ref{n-conical_Hessian}
given in section \ref{sect:ConHess},
and admits a Hesse potential which is 
homogeneous of degree 2 
under the $\mathbb{R}^{>0}$-transformations generated by $\xi$:
\begin{equation}
L_\xi g = 2 g \;,\;\;\;L_\xi H = 2 H \;.
\end{equation}
In addition $M$ is K\"ahler, and  the ASK conditions imply that the
vector field $J\xi$ is isometric, $L_{J\xi}g=0$, and preserves the homogeneous Hesse
potential, $L_{J\xi} H = 0$. The two vector fields $\{\xi, J\xi\}$ commute and generate a holomorphic,
homothetic $\mathbb{C}^*$-action on $M$. On a CASK manifold one may choose, at least 
locally, {\em conical special real coordinates} $q^a = (x^I, y_I)$ such that the homothetic 
Killing vector field takes the form
\begin{equation}
\xi = q^a \frac{\partial}{\partial q^a} = x^I \frac{\partial}{\partial x^I} + y_I \frac{\partial}{\partial y_I} \;.
\end{equation}
Such coordinates are unique up to symplectic transformations, 
since the compatibility with the conical structure prevents us from admitting 
translations. 
A holomorphic immersion 
$\phi: M \rightarrow V\cong T^*\mathbb{C}^{n}$ is called {\em conical} if the position vector field
$\xi^V$ on $V$ is 
tangent along $\phi$. If $\phi: M \rightarrow V$ is a conical K\"ahlerian Lagrangian
immersion of a complex connected manifold $(M,J)$ with induced data $(g,\nabla, \xi)$, 
then $(M,J,g,\nabla,\xi)$ is a conical affine special K\"ahler manifold. Conversely, 
any simply connected CASK manifold can be realized as a conical K\"ahlerian Lagrangian
immersion \cite{Cortes:2009cs}. 

By considering an open subset $U\subset M$ if necessary, we can assume that $\phi$ is
an embedding. Using this we can easily verify those local formulae that do not follow
from previous results on Hessian manifolds using conical special real coordinates. 
For reference we first collect some useful relations following from homogeneity:
\begin{equation}
q^a H_a = 2 H \;,\;\;\;
q^a H_{ab} = H_b = q_b \;,\;\;\;q^a q^b H_{ab} = 2 H \;,\;\;\;
q^a H_{abc} = 0 \;.
\end{equation}
For CASK manifolds the
special real coordinates $q^a$ and dual special real coordinates
$q_a=H_a$ are related by $q_a  = H_{ab} q^b$, $q^a = H^{ab} q_b$. 
This is a special feature of Hesse potentials which are homogeneous of
degree two, compare \eqref{hom_rel}.

Using that $J^a_{\;b}= -\frac{1}{2} \Omega^{ac} H_{cb}$, 
and the above homogeneity properties, the components of
$J\xi$ are
\begin{equation}
J\xi = J^a_{\;b} q^b \partial_a = \frac{1}{2} H_b \Omega^{ba} \partial_a \;.
\end{equation}
From this we see immediately that $\xi$ and $J\xi$
commute, and therefore generate an abelian transformation group
\begin{equation}
[\xi, J\xi ] =  L_\xi J\xi = - L_{J\xi} \xi = 0 \;.
\end{equation}
The Lie derivatives of the differentials are
\begin{equation}
L_\xi dq^a = \frac{\partial q^a}{\partial q^b} dq^b=dq^a \;,\;\;\;
L_{J\xi} dq^a = \frac{\partial (J^a_c q^c)}{\partial q^b} dq^b=-\frac{1}{2}
\Omega^{ab}H_{bc} dq^c \;.\;\;\;
\end{equation}
The Lie derivatives of the Hesse potential 
\begin{equation}
L_\xi H = 2 H \;,\;\;\;
L_{J\xi} H = \frac{1}{2} H_a \Omega^{ab} H_b = 0 
\end{equation}
show that $H$ is $J\xi$-invariant. We also list the Lie derivatives
of $q_a=H_a$
\begin{equation}
L_\xi H_a = H_a \;,\;\;\;
L_{J\xi} H_a = \frac{1}{2} H_c \Omega^{cb} H_{ba} = 2 \Omega_{ab} q^b \;,
\end{equation}
and the Lie derivatives of the second derivatives of $H$
\begin{equation}
L_\xi H_{ab} = 0 \;,\;\;\;
L_{J\xi} H_{ab} = \frac{1}{2} H_c \Omega^{cd} H_{dab} = 0 \;.
\end{equation}
The last equality follows from differentiating 
$H_{ab} \Omega^{bc} H_{cd} = - 4 \Omega_{ad}$ upon 
contraction with $q^a$ and using homogeneity. Combining 
results, we find that $J\xi$ is a Killing vector field, 
$L_{J\xi} g=L_{J\xi} (H_{ab} dq^a dq^b )= 0$.

In summary we have the following infinitesimal $\mathbb{C}^*$-action:
\begin{equation}
[\xi, J\xi]=0 \;,\;\;\; L_\xi g = 2 g \;,\;\;\; L_{J\xi} g = 0 \;.
\end{equation}
Moreover, the action of $J\xi$ is $\omega$-Hamiltonian:
\begin{eqnarray}
\omega(J\xi , X) &=& g(J\xi, JX) = g(\xi, X) = q^a H_{ab} X^b = H_a X^a  \nonumber \\
&=& X^a \partial_a H = X(H) = dH(X)  \;,\;\;\;\forall X\in \mathfrak{X}(M)\;, \label{Moment_map_Jxi}
\end{eqnarray}
hence
\begin{equation}
\omega(J\xi, \cdot) = dH(\cdot)  \;,
\end{equation} 
with moment map\footnote{See \ref{app:symp_man} for a brief review of Hamiltonian
vector fields and moment maps.} 
\begin{equation}
H  = \frac{1}{2} H_{ab} q^a q^b =\frac{1}{2} g(\xi, \xi)\;.
\end{equation}

At each point, the vector fields $\xi$ and $J\xi$ define two distinguished 
directions, which correspond to the radial and angular direction of a complex
cone whose base is spanned by the vector fields in $\langle \xi, J\xi\rangle^\perp$.
It is useful for the following discussion 
of project special K\"ahler manifolds to decompose the metric and other 
tensors into tangential and transversal parts with respect to the $\mathbb{C}^*$-action. 
For this purpose we introduce the one-forms
\begin{equation}
\alpha = dH= H_a dq^a \;,\;\;\;\beta = q^a \Omega_{ab} dq^b\;,
\end{equation}
which, up to normalization, are dual to the vector fields $\xi$ and $J\xi$:
\begin{eqnarray}
\alpha(\xi) = 2H\;, && \alpha(J\xi) = 0 \;, \;\;\;\alpha(X)=0 \;, \nonumber \\
\beta(\xi) = 0\;, &&  \beta(J\xi) = H  \;,\;\;\;\beta(X)=0\;,
\end{eqnarray}
for all $X \in \langle \xi, J\xi\rangle^{\perp}$.\footnote{Here $\perp$ denotes orthogonality
with respect to $g_M=H_{ab} dq^a dq^b$.} 
The forms $\alpha,\beta$ carry weight 2 under the dilatations generated by $\xi$ and are 
invariant under the $U(1)$ transformations generated by $J\xi$:
\begin{eqnarray}
L_\xi \alpha = 2\alpha \;, && L_{J\xi} \alpha = 0 \;, \nonumber \\
L_\xi \beta = 2\beta\;, && L_{J\xi} \beta = 0 \;.
\end{eqnarray}
Note that the scaling weight of any tensor which transforms homothetically
under $\mathbb{C}^*$ can be changed by multiplying it with the appropriate
power of the Hesse potential. In particular any tensor transforming with 
a definite scaling weight can be made invariant, and
\begin{eqnarray}
\tilde{g}_M^{(A,B,C)} &=& A H^{-1} g_M + B H^{-2} \alpha^2 + C H^{-2} \beta^2 \nonumber \\
&=& \left( A \frac{H_{ab}}{H} + B \frac{H_a H_b}{H^2} + C \frac{\Omega_{ac} q^c 
\Omega_{bd} q^d}{H^2} \right) dq^a dq^b \;
\end{eqnarray}
is a family of $\mathbb{C}^*$-invariant symmetric rank two co-tensor fields
which includes the conformally rescaled metric $H^{-1}g_M$ as  the special case
$A=1, B=C=0$.  We can obtain a tensor field which is transversal to the 
$\mathbb{C}^*$-action by imposing
\begin{equation}
\tilde{g}_M^{(A,B,C)} (\xi, \cdot) = 
\tilde{g}_M^{(A,B,C)} (J\xi, \cdot) = 0 
\Rightarrow B = -\frac{A}{2} \;,\;\;\;C=-2 A\;.
\end{equation}
Thus the transversal part, which has a two-dimension kernel spanned by $\{\xi, J\xi \}$
 is
\begin{equation}
\label{H_0}
\tilde{g}^{(0),A'}_M = A' H_{ab}^{(0)} dq^a dq^b \;,
\end{equation}
where 
\begin{equation}
\label{H0-decomposition}
{H}^{(0)}_{ab} = -  \frac{1}{2H} H_{ab} + \frac{1}{4H^2} H_a H_b 
+ \frac{1}{H^2} \Omega_{ac} q^c \Omega_{bd} q^d \;, \;\;\;\;A'=-2A \;.
\end{equation}
Solving (\ref{H0-decomposition}) for the CASK metric $g_M$ we obtain
\begin{equation}
H_{ab} = (-2H) H_{ab}^{(0)} + \frac{1}{2H} H_a H_b + \frac{2}{H} 
\Omega_{ac} q^c \Omega_{bd} q^d \;.
\end{equation}
This is an orthogonal decomposition of $H_{ab}$ into projections
onto the distributions $\langle \xi, J\xi \rangle^\perp$, $\langle\xi \rangle$ and $\langle J\xi \rangle$.
The signatures of $H_{ab}$ and
$H^{(0)}_{ab}$ are related: if $H^{(0)}_{ab}$ is positive
definite on $\langle \xi, J\xi\rangle^\perp$,
then the CASK metric $g_M$ has complex 
Lorentz signature $(\mp \mp \pm \cdots \pm)$.
The overall sign of 
$H_{ab}$ depends on the sign of the Hesse potential $H$. 

For later use we define the tensor field 
\begin{equation}
\label{hatg}
\hat{g}_M = \hat{H}_{ab} dq^a dq^b \;,\;\;\;
\hat{H}_{ab} = -\frac{1}{2} H_{ab} + \frac{2}{H} 
\left(\frac{1}{4} H_a H_b + \Omega_{ac} q^c \Omega_{bd} q^d\right) \;,
\end{equation}
which differs from the CASK metric by an overall factor $-\frac{1}{2}$ and
a sign flip along the distribution spanned by $\xi$ and $J\xi$. 
Thus $\hat{g}_M$ is positive or negative definite if the CASK metric
$g_M$ has complex Lorentz signature. The tensor $\hat{H}_{ab}$ and 
its inverse $\hat{H}^{ab}$ are related to the complex symmetric matrix
\begin{equation}
{\cal N}_{IJ} = \bar{F}_{IJ} + i \frac{N_{IK} X^K N_{JL} X^L}{X^M N_{MN} X^N}
\end{equation}
by
\begin{equation}
\label{H_hat_cal_N}
\hat{H}_{ab} = \left( \begin{array}{cc}
{\cal I} + {\cal R}{\cal I}^{-1} {\cal R} & - {\cal R}{\cal I}^{-1} \\
- {\cal I}^{-1} {\cal R} & {\cal I}^{-1} \end{array} \right) \;,\;\;\;
\hat{H}^{ab} = \left( \begin{array}{cc}
{\cal I}^{-1} & {\cal I}^{-1} {\cal R} \\
{\cal R} {\cal I}^{-1}  & {\cal I} + {\cal R}{\cal I}^{-1} {\cal R} \\
\end{array} \right) \;,
\end{equation}
where ${\cal N}_{IJ} = {\cal R}_{IJ} + i {\cal I}_{IJ}$. We will see in section  
\ref{4dvecmult} that ${\cal N}_{IJ}$ is the coefficient matrix of
the terms quadratic in the abelian field strengths in the Lagrangian for four-dimensional
vector multiplets coupled to Poincar\'e supergravity. While its real version $\hat{H}_{ab}$ is
a symplectic tensor, the complex matrix ${\cal N}_{IJ}$ transforms fractionally linearly
under symplectic transformations,
\begin{equation}
\label{transf_cal_N}
{\cal N} \mapsto (W+ V {\cal N})(U + Z {\cal N})^{-1} \;.
\end{equation}
Observe that the relation between $\hat{H}_{ab}$ and ${\cal N}_{IJ}$ is
analogous to the one between $H_{ab}$ and $F_{IJ}$, in particular both 
$F_{IJ}$ and ${\cal N}_{IJ}$ transform fractionally linearly.

Another natural symmetric tensor field on $M$ is the 0-conical Hessian metric that we obtain
by taking the logarithm of the Hesse potential $H$ as a  Hesse potential. 
Choosing the normalization
\begin{equation}
\tilde{H} = -\frac{1}{2} \log |H| \;,
\end{equation}
we obtain
\begin{eqnarray}
\tilde{g} &=& \tilde{H}_{ab} dq^a dq^b = \tilde{g}_M^{(-1/2, 1/2,0)}\;, \\
\tilde{H}_{ab}&=& \partial^2_{a,b} \tilde{H} = 
- \frac{1}{2H} H_{ab} + \frac{1}{2H^2} H_a H_b   \label{Htilde_H}\\
&=& H_{ab}^{(0)} + \frac{1}{4H^2} H_a H_b - \frac{1}{H^2} \Omega_{ac}q^c 
\Omega_{bd}q^d \nonumber  \\
&=& \frac{1}{H} \hat{H}_{ab} - \frac{2}{H^2} \Omega_{ac}q^c \Omega_{bd} q^d \;.
\nonumber
\end{eqnarray}
This tensor differs from the CASK metric by an overall factor $-2H$, which makes
it $\mathbb{C}^*$-invariant. Its signature differs from the one of $H_{ab}$ by 
a sign flip along $\xi$. Thus if the CASK metric has complex Lorentz signature
$(\mp, \mp, \pm, \ldots, \pm)$,
then $\tilde{H}_{ab}$ has real Lorentz signature
$(\pm, \mp, \pm, \ldots, \pm)$, with the time-like direction
generated by $J\xi$.

\subsection{Projective special K\"ahler geometry\label{sec:PSR}}

In sections \ref{Sect:rSCQ} and \ref{Sect:SR} we have discussed 
the real superconformal quotient, which relates affine and projective
special real geometry. Similarly, given a CASK manifold $M$ we can 
obtain a {\em projective special K\"ahler manifold} $\bar{M}$ 
by a complex quotient construction.
To do this we construct a K\"ahler metric
on the orbit space $\bar{M}=M/\mathbb{C}^*$, of the 
$\mathbb{C}^*$-action on a CASK manifold $M$.
Since the CASK metric $g_M =H_{ab} dq^a dq^b$ transforms
homothetically, we can make it $\mathbb{C}^*$-invariant 
through multiplication by a multiple $H^{-1}$. To obtain 
a projectable tensor $\tilde{g}^{(0)}_M$, we then take the transversal part:
\begin{equation}
\tilde{g}^{(0)}_M =  H_{ab}^{(0)} dq^a dq^b \;,\;\;\;
 {H}^{(0)}_{ab} = -  \frac{1}{2H} H_{ab} + \frac{1}{4H^2} H_a H_b 
+ \frac{1}{H^2} \Omega_{ac} q^c \Omega_{bd} q^d  \;.
\end{equation}
As in \eqref{H_0} we have chosen  $A'=1\Leftrightarrow A=-\frac{1}{2}$, to be consistent
with supergravity conventions. 
By projection onto orbits 
$\tilde{g}^{(0)}_M$ defines a non-degenerate metric $\bar{g}_{\bar{M}}$ on $\bar{M}$, 
which conversely lifts to $\tilde{g}^{(0)}_M$ under the 
pullback of the projection $\pi: M \rightarrow \bar{M}$, that is $\tilde{g}^{(0)}_M = \pi^* \bar{g}_{\bar{M}}$.

The quotient by the holomorphic homothetic $\mathbb{C}^*$-action will be referred to 
as the (complex) superconformal quotient. In order for $\bar{g}_{\bar{M}}$ to be well
defined, we need that $g(\xi,\xi)=-2H\not=0$. Moreover, we need to assume that
the quotient by the $\mathbb{C}^*$-action is well behaved.  This gives rise to the following 
definitions \cite{Alekseevsky:1999ts,Cortes:2009cs}:

\begin{defi}
{\bf Regular conical affine special K\"ahler manifold.} A conical affine special K\"ahler manifolds
$(M,J,g,\nabla,\xi)$ is called {\em regular} if the function $g(\xi,\xi) = - 2 H$ is nowhere vanishing
on $M$, and if the canonical quotient $\pi: M \rightarrow \bar{M}$ onto the space of orbits
of $\mathbb{C}^*$ on $M$ is a holomorphic submersion onto a Hausdorff manifold.
\end{defi}

\begin{defi}
{\bf Projective special K\"ahler manifold (PSK manifold).} A projective special K\"ahler manifold
$(\bar{M}, g_{\bar{M}})$ is a (possibly indefinite) K\"ahler manifold which can be obtained
as the superconformal quotient of a regular CASK manifold $(M,g_M,\nabla, \xi)$. 
\end{defi}

In supergravity applications, $\bar{g}_{\bar{M}}$ is the metric on the manifold parametrized 
by the physical scalar fields, and therefore must be positive definite. The results of the
preceding section imply that the underlying CASK metric must then have 
complex Lorentz signature $(\mp, \mp, \pm, \ldots, \pm)$, where the time-like
directions are along the orbits of the $\mathbb{C}^*$-action generated by $\langle \xi, J\xi\rangle$.
In physics these directions correspond to an additional vector multiplet acting as a 
conformal compensator. 
Note that an overall sign flip of $g_M$ does not change $\bar{g}_{\bar{M}}$. 
The  tensor field $\hat{g}_M$ defined in \eqref{hatg} also plays a role in physics. 
It is proportional to the vector field metric and therefore must have definite signature.
This is automatic if  $\bar{g}_{\bar{M}}$ is positive definite.

The superconformal quotient can be interpreted as a K\"ahler quotient, that is
as a symplectic quotient consistent with a K\"ahler structure, see also 
\ref{app:Kaehler}.
To see how this work we use the holomorphic parametrization of the
CASK manifold and follow the original construction of \cite{deWit:1984pk}.
When using special coordinates $X^I$, the 
homothetic Killing vector fields take the form
\begin{equation}
\xi = X^I \frac{\partial}{\partial X^I} + cc \;,\;\;\;
J\xi = i X^I \frac{\partial}{\partial X^I} + cc\;.
\end{equation}
The superconformal quotient proceeds in two steps. 
First the coordinates $X^I$ are restricted to the 
hypersurface 
\begin{equation}
S = \{ X^I \in M | i(X^I \bar{F}_I - F_I \bar{X}^I) = -1\}\;.
\end{equation}
In physics the condition
$i(X^I \bar{F}_I - F_I \bar{X}^I) = -1$ is called the D-gauge,
because it fixes the local dilatation symmetry which is
part of the superconformal group. We will discuss the physics
aspects in section \ref{4dvecmult}, while a review of the superconformal
formalism can be found in \ref{sec:4dscg}.

Since $J\xi$ acts isometrically on $S$, one can take the quotient
with respect to the $U(1)$ group action and obtains $\bar{M} = S/U(1)$.
To recognize this construction as a K\"ahler quotient, we note that
\begin{equation}
\label{Relations_CASK_Kaehlerpot}
i(X^I \bar{F}_I - F_I \bar{X}^I) = N_{IJ} X^I \bar{X}^J = K_{CASK}=
H_{ab} q^a q^b = 2 H = 
g(\xi,\xi) 
\end{equation}
is the norm of the homothetic Killing vector field $\xi$, which 
is proportional to the Hesse potential, which is the moment map 
for the Hamiltonian isometric $U(1)$ action on the CASK manifold $M$,
see \eqref{Moment_map_Jxi}. This shows that 
$S/U(1) = M//U(1)$ is a symplectic quotient with respect to the Hamiltonian
isometric action of $J\xi$ on $M$. Moreover $(M,g_M)$ is a Riemannian
cone over $(S,g_S)$ with $g_M$ and $g_S$ related by $g_M = dr^2 + r^2 g_S$. 
Since $(M,g_M)$ is K\"ahler, it follows that $(S,g_S)$ is Sasakian.\footnote{Sasakian
geometry is reviewed in \ref{app:sasakian}.} The metric
induced by $g_S$ on the quotient $\bar{M} = S/U(1) = M// U(1)$ is K\"ahler,
as we will show below, and therefore $M//U(1)$ is a K\"ahler quotient. 

To show that the metric is K\"ahler, we express the projectable tensor 
$H^{(0)}_{ab} dq^a dq^b$ in holomorphic coordinates. Rather than performing
the coordinate transformation from special real to special holomorphic coordinates, 
we start with $g_M = N_{IJ} dX^I d\bar{X}^J$ 
and construct a tensor which is projectable onto the orbits of the 
$\mathbb{C}^*$-action. The resulting tensor $\tilde{g}_M^{(0)}= N^{(0)}_{IJ} dX^I d\bar{X}^J$ 
has the components 
\begin{equation}
\label{N0}
N^{(0)}_{IJ} = - \frac{ N_{IJ}}{N_{MN}X^M \bar{X}^N} 
+ \frac{N_{IK} \bar{X}^K N_{JL} X^L}{(N_{MN} X^M \bar{X}^N)^2} \;.
\end{equation}
To see that this is correct we note that the components $N^{(0)}_{IJ}$ are homogeneous
of degree $-2$, so that $\tilde{g}_M^{(0)}$ is invariant under $\xi$. Moreover 
\begin{equation}
N^{(0)}_{IJ} X^I = 0 = N_{IJ}^{(0)} \bar{X}^J \;,
\end{equation}
which shows that $\tilde{g}_M^{(0)}$ is transversal to the actions generated by $\xi$ and $J\xi$ .
Therefore this tensor field is projectable. The CASK K\"ahler potential $K=g(\xi,\xi)$ is a global function,
and we can use it provide a global expression for $\tilde{g}^{(0)}_M$:
\begin{equation}
\label{Projectable_global}
\tilde{g}^{(0)}_M = - \frac{\partial \bar{\partial} K}{K} + \frac{\partial K \bar{\partial} K}{K^2} \;,
\end{equation}
where $\partial \bar{\partial} K=g_M$ is the CASK metric. By inspection
\begin{equation}
\tilde{g}^{(0)}_M = \partial \bar{\partial} \left( - \log (-K) \right) \;,
\end{equation}
so that the degenerate symmetric rank two co-tensor field $\tilde{g}^{(0)}_M$ has
a `K\"ahler like' potential
\begin{equation}
\label{KL_potential}
{\cal K}(X,\bar{X}) = -\log (-K) = 
- \log (-N_{IJ} X^I \bar{X}^J) = - \log( -i(X^I \bar{F}_I - F_I \bar{X}^I)) \;.
\end{equation}
Upon projection onto $\bar{M}$ the `K\"ahler like' potential $K(X,\bar{X})$ 
becomes a genuine K\"ahler potential for $\bar{g}_{\bar{M}}$. We will obtain 
an expression in terms of local coordinates on $\bar{M}$ in 
section \ref{sec:formlinebund}, see
\eqref{KpotPSK}.

Rather than viewing $\bar{M}$ as an abstract quotient, one usually prefers
to realize it concretely as a submanifold $\bar{M}\subset S\subset M$.
In physics describing $\bar{M}$ as a submanifold corresponds to 
imposing a $U(1)$ gauge on top of the D-gauge. There is no canonical
choice for a $U(1)$ gauge. The only canonical choice would be to take a
 hypersurface in $S$ which is orthogonal at every point to orbits of the
 $U(1)$-action. However, $S$ is a contact manifold and the distribution 
 defined by this condition is a contact distribution, and therefore not
 integrable.\footnote{Contact structures and their relation to integrability are reviewed in 
 \ref{App:ContactGeometry}.} This situation is different from the first step, where we defined $S$
 as the level set of the symplectic function $g(\xi,\xi)$, which is a moment map for 
 the Hamiltonian $U(1)$ action generated by $J\xi$, see \eqref{Moment_map_Jxi}, 
 \eqref{Relations_CASK_Kaehlerpot}. 
 There are two ways to proceed. One can choose
  a gauge, for example by imposing that one of the special holomorphic
 coordinates is real, such as $\bar{X}^0=X^0$. This will
 always break the full symplectic covariance that we have preserved
 so far, because the orthogonality to the $U(1)$-orbits was the only
 remaining symplectically invariant equation involving $\xi$ and $J\xi$. 
 Alternatively, one can work `upstairs,' on $S$ or $M$
 using $U(1)$-invariant quantities or $\mathbb{C}^*$-invariant quantities,
 respectively. This has the advantage of preserving symplectic 
 covariance, and we will see how it is done in the following.

\subsection{Other formulations of special K\"ahler geometry \label{sec:other_SK}}

\subsubsection{Formulation in terms of line bundles \label{sec:formlinebund}}

In the physics literature, special K\"ahler geometry is often presented in a slightly
different language where the quantities
 $(X^I,F_I)$ are interpreted as sections of a line bundle ${\cal U}^{\bar{M}} \rightarrow \bar{M}$. 
In this section we explain how this formulation can be recovered from the
immersion $M \rightarrow T^*{\mathbb C}^{n+1}$ discussed in section \ref{sec:kaehimm}, 
following \cite{Cortes:2009cs}.

\subsubsection*{The universal line bundle}

We start by recalling that on a holomorphic Hermitian vector bundle there is a unique
connection, called the Chern connection, which is simultaneously holomorphic and Hermitian,
see \ref{App:ComplexVBdl}. Consider the open set of non-isotropic vectors
$V' = \{ v \in V | \gamma(v,v) \not=0 \}\subset V = T^* {\mathbb C}^{n+1}$ of the vector
space $V$. The space of complex lines $P(V') = \{ [v]= \mathbb{C} v | v \in V'\}$ 
is the projectivization of  $V$. Then the trivial vector bundle ${\cal V}:= P(V') \times V \rightarrow P(V')$
equipped with the standard Hermitian metric $\gamma = \Omega(\cdot, \bar{\cdot})$ is 
a holomorphic Hermitian vector bundle, with Chern connection 
 $d = \partial + \bar{\partial}$. The {\em universal bundle}
${\cal U} \rightarrow P(V')$ is defined as the holomorphic line sub-bundle of ${\cal V}$ whose
fibre ${\cal U}_p$ over $p =[v] \in P(V')$ is the corresponding line $\mathbb{C}v$. The Chern connection
on ${\cal U}$ is given by the $\gamma$-orthogonal projection of the flat Chern connection $d$ 
of ${\cal V}$:
\begin{equation}
{\cal D}_X v := \pi_{\cal U} d_X v = \frac{\gamma(d_X v, v)}{\gamma(v,v)} v \;,
\end{equation}
where $X$ is a complex vector field on $P(V')$ and $v$ a section of ${\cal U} \subset {\cal V}$.

\subsubsection*{Pull-back of universal line bundle to the CASK manifold $M$}

If $(M,J,g,\nabla,\xi)$ is a regular CASK manifold, then we have the following commutative
diagram:
\begin{equation}
\xymatrix{
M \ar[r]^{\phi}\ar[d]^{\pi} & V'  \ar[d]^{\pi_V}\\
\bar{M} \ar[r]^{\bar{\phi}}& P(V')
}
\end{equation}

\begin{rmk}
The projectivization 
$P(V')$ of the symplectic manifold $V'$ is a {\em contact manifold}, see \ref{App:ContactGeometry}. 
The holomorphic
map $\bar{\phi}$ is a {\em Legendrian immersion} induced 
by the holomorphic Lagrangian immersion $\phi$. 
\end{rmk}

The map $f = \bar{\phi} \circ \pi= \pi_V  \circ {\phi}: M \rightarrow P(V')$ defines
the pull-back $({\cal U}^M, {\cal D})$ of the universal bundle $({\cal U}, {\cal D})$,
where we use the same symbol ${\cal D}$ for  the Chern connection 
on ${\cal U}$ and its pull-back to ${\cal U}^M$:
\begin{equation}
\xymatrix{
{\cal U}^M \ar[rr] \ar@<0.5ex>[d]& & {\cal U} \ar[d]^{\pi_{\cal U}}\\
M \ar@<0.5ex>[u]^{\phi}\ar[rr]^{\bar{\phi} \circ \pi= \pi_V \circ  {\phi}  } &  & P(V') \\
}
\end{equation}
The holomorphic Lagrangian immersion $\phi:M \rightarrow V$ can be regarded as a 
holomorphic section of ${\cal U}^M$, as follows:
according to  \ref{App:Pull_back}
the pull-back bundle ${\cal U}^M$ is defined as
\begin{equation}
{\cal U}^M = \{ (m,u) \in M \times {\cal U} | (\pi_V \circ \phi)(m) = \pi_{\cal U}(u) \} \;.
\end{equation} 
A section of $s: P(V') \rightarrow {\cal U}$ of 
${\cal U} \rightarrow P(V')$ has the form $s(p) = v_p$, where $v_p\in V$ is
a vector such that $[v_p] = p \in P(V')$. For $v_p$ we can choose any vector on the line
$p$.
The pull-back section ${\cal U}^M \rightarrow M$
has the form $((\pi_V \circ \phi)^* s)(m) = v_{(\pi_V \circ \phi)(m)}$, where 
$ v_{(\pi_V \circ \phi)(m)}$ is a vector
on the same line in $V$ as $\phi(m)$. If we choose a section  $s$ of ${\cal U}$
such that  $s(p) = \phi(m)$ for $p=\pi_V \phi(m)$, then the corresponding
pull-back section $m \mapsto \phi(m)$ can be identified with $\phi$. 
The local form of $\phi$, regarded as a section of ${\cal U}^M \rightarrow M$  is 
\begin{equation}
\phi : M \rightarrow {\cal U}^M  : (X^I) \mapsto (X^I, F_I(X))  \;,
\end{equation}
 where $X^I$ are special holomorphic coordinates on $M$, $F_I = \partial{F}/\partial X^I$,
 and where $F$ is the prepotential of $\phi$ regarded as a 
 local holomorphic Lagrangian immersion  $\phi: M \rightarrow V$. 
 
Since the Chern connection ${\cal D}$ on the universal line bundle 
${\cal U}$ is defined by orthogonal projection, 
the pull-back connection satisfies 
\begin{equation}
{\cal D}_I  \phi = i A^h_I \phi \;,\;\;\;{\cal D}_{\bar{I}} \phi = 0  \;,
\end{equation}
where ${\cal D}_I := {\cal D}_{\partial_I}$ and ${\cal D}_{\bar{I}} = {\cal D}_{\partial_{\bar{I}}}$,
and where the components $A^h_I$ of the connection one-form $i A^h_I dX^I$ 
of the pull-back connection are:
\begin{eqnarray}
i A_I^h  &=& \frac{\gamma(\partial_I \phi, \phi)}{\gamma(\phi, \phi)}  =
\frac{i(\partial_I X^J \bar{F}_J - \partial_I F_J \bar{X}^J)}{i(X^K \bar{F}_K - F_K \bar{X}^K)} \;,\\
i A_{\bar{I}}^h&=& \frac{\gamma(\partial_{\bar{I}} \phi, \phi)}{\gamma(\phi, \phi)}  =
\frac{i(\partial_{\bar{I}} X^J \bar{F}_J - \partial_{\bar{I}} F_J \bar{X}^J)}{i(X^K \bar{F}_K - F_K \bar{X}^K)}
=
 0  \;.
\end{eqnarray}
We can also express the pull-back connection with respect to a unitary (unit norm)
section $\phi_1 = \phi/||\phi||$, $|| \phi || := \sqrt{ |\gamma(\phi,\phi)|}$, where
$\gamma(\phi, \phi) = i (X^I \bar{F}_I - F_I \bar{X}^I)$. The unitary section $\phi_1$
can be interpreted as a section of a principal $U(1)$ bundle ${\cal P}^M \rightarrow 
M$, to which the holomorphic line bundle ${\cal U}^M \rightarrow M$ is associated.
Let ${\cal D}$ be a principal connection on ${\cal P}^M$ with connection
one-form $i A_I dX^I + i A_{\bar{I}} d\bar{X}^I$, so that covariant derivatives 
of sections of ${\cal P}^M$ take the form 
\begin{equation}
\label{Section_P}
{\cal D}_I \phi_1 = i A_I \phi_1 \;,\;\;\; {\cal D}_{\bar{I}} \phi_1 = i A_{\bar{I}} \phi_1 \;.
\end{equation}
We require that the pull-back connection on ${\cal U}^M$ is induced by 
this principal connection. Then we can read off the components $(A_I, A_{\bar{I}})$ of
the connection one-form of ${\cal P}^M$ by comparing \eqref{Section_P} to the covariant derivatives
of unit sections of ${\cal U}^M$. Using that
\begin{equation}
D_I \phi_1 = \frac{D_I \phi}{|| \phi ||} - \frac{\phi}{||\phi||^2} \partial_I || \phi || \;,\;\;\;
D_{\bar{I}} \phi_1 = - \frac{\phi}{||\phi||^2} \partial_{\bar{I}}  || \phi || \;,\;\;\;
\end{equation} 
we find:
\begin{equation}
A_I= \frac{1}{2} A_I^h \;,\;\;\;A_{\bar{I}} = \frac{1}{2} \overline{ A^h_I } \;.
\end{equation}

\subsubsection*{Pull back of the universal line bundle to the PSK manifold $\bar{M}$}

By choosing a section $s: \bar{M} \rightarrow M$ of the $\mathbb{C}^*$-bundle $\pi: M \rightarrow 
\bar{M}$ we can regard $\bar{M}$ as an embedded submanifold, at least locally. We can 
also use $\bar{\phi} : \bar{M} \rightarrow P(V')$ to obtain the pull-back bundle 
$({\cal U}^{\bar{M}}, {\cal D})$ of the universal bundle $({\cal U}, {\cal D})$:
\begin{equation}
\xymatrix{
{\cal U}^{\bar{M}} \ar[r] \ar@<0.5ex>[d] & 
{\cal U}^M \ar[rr] \ar@<0.5ex>[d]& & {\cal U} \ar[d]\\
\bar{M} \ar[r]^{s} \ar@<0.5ex>[u]^{s^*\phi}  \ar@/_/[rrr]_{\bar{\phi}}& 
M \ar@<0.5ex>[u]^{\phi}\ar[rr]^{\pi_V \circ {\phi} = \bar{\phi} \circ \pi} &  & P(V') \\
}
\end{equation}
The local form of the pull-back of $\phi: M \rightarrow {\cal U}^M$ by
$s: \bar{M} \rightarrow M$ to a section $s^* \phi : \bar{M} \rightarrow {\cal U}^{\bar{M}}$ 
is
\begin{equation}
s^* \phi : \bar{M} \rightarrow {\cal U}^{\bar{M}} \;\;(\zeta^a) \mapsto (X^I(\zeta), F_I(\zeta))  \;,
\end{equation}
where $\zeta^a$ are local holomorphic coordinates on $\bar{M}$, and where $X^I$ and $F_I$
depend holomorphically on $\zeta^a$. 
Evaluating the pull-back connection on a holomorphic section $s:\bar{M} \rightarrow M$
we obtain 
\begin{equation}
\label{connsecMbarM}
{\cal D}_a s = i A^h_a s = i \partial_a X^I A^h_I s = 
\frac{\gamma(\partial_a \phi, \phi)}{\gamma(\phi,\phi)} s= 
\frac{\partial_a X^I \bar{F}_I - \partial_a F_I \bar{X}^I}{X^I \bar{F}_I - F_I \bar{X}^I} s \;,\;\;\;
{\cal D}_{\bar{a}} s = 0 \;.
\end{equation}
On a unitary section $s_1 = s/||s||$ we pull back the principal connection 
$(A_I, A_{\bar{I}})$ of ${\cal P}^M$ to obtain a principal connection with 
components
\begin{equation}
A_a = \partial_a X^I A_I  + \partial_a \overline{X^I} A_{\bar{I}} \;,\;\;
A_{\bar{a}} = \partial_{\bar{a}} X^I A_ I + \partial_{\bar{a}} \overline{X^I} A_{\bar{I}} \;,\;\;
\end{equation}
The local components $(X^I(\zeta, \bar{\zeta}), F_I(\zeta, \bar{\zeta}))$ of the 
pull-back of $\phi$ by a unit section $s_1$ satisfy 
\begin{equation}
\gamma(\phi, \phi) =
\gamma(\phi(s_1), \phi(s_1)) = 
i(X^I(\zeta,\bar{\zeta}) \bar{F}_I(\zeta,\bar{\zeta}) - F_I(\zeta,\bar{\zeta})
 \bar{X}^I(\zeta, \bar{\zeta}) ) = \pm 1
 \end{equation}
 and
depend non-holomorphically on the
local holomorphic coordinates $\zeta^a$. 
Evaluating the connection on a unit section $s_1$ we find
\begin{eqnarray}
\label{connunit}
{\cal D}_a s_1 &=& i A_a s_1 = \frac{\gamma(\partial_a \phi, \phi) - \gamma(\phi, \partial_{\bar{a}} \phi)}{2 \gamma(\phi, \phi)} s_1 \;,\;\;\;\\
{\cal D}_{\bar{a}} s_1 &=& i A_{\bar{a}} s_1 = 
\frac{ \gamma(\partial_{\bar{a}} \phi, \phi) - \gamma(\phi, \partial_{{a}} \phi)}{2 \gamma(\phi, \phi)}
s_1 \;.
\end{eqnarray}
Note that for a unitary section $\gamma(\phi,\phi) =\pm 1$ and $\gamma(\partial_a \phi, \phi) = - \gamma(\phi, 
\partial_{\bar{a}} \phi)$. Also note that $A_{\bar{a}} = \overline{A_a}$. 
In terms of components $(X^I(\zeta, \bar{\zeta}), F_I(\zeta, \bar{\zeta}))$
the components of the connection one form are
\begin{equation}
\label{connAa}
i A_a = - i A_{\bar{a}} = - \frac{i}{2}
\frac{X^I \stackrel{\leftrightarrow}{\partial}_a \bar{F}_I 
- F_I \stackrel{\leftrightarrow}{\partial}_a \bar{X}^I }{
i(X^I \bar{F}_I - F_I \bar{X}^I)} \;,
\end{equation}
where $i(X^I \bar{F}_I - F_I \bar{X}^I) =\pm 1$, and 
where we use the notation $a \stackrel{\leftrightarrow}{\partial} b =\left( a \partial b 
- (\partial a)b\right)$.

\subsubsection*{Pull-back of the universal line bundle to space-time $N$}

Our last step is to consider the situation where a physical theory defined on 
space-time $N$ contains massless scalar fields with values in a PSK manifold 
$\bar{M}$. The Lagrangian description of such scalar fields is given by 
a non-linear sigma model, see \ref{App:sigma_models} for details. The scalar
fields are the components of a map ${\cal Z}: N \rightarrow \bar{M}$ from space-time $N$ into
a PSK manifold $\bar{M}$. This defines a further pull-back $({\cal U}^N, {\cal D})$
of the universal bundle to a line bundle over space-time. 
\begin{equation}
\xymatrix{
{\cal U}^N \ar[r] \ar@<0.5ex>[d]& 
{\cal U}^{\bar{M}} \ar[r] \ar@<0.5ex>[d]  & 
{\cal U}^M \ar[rr] \ar@<0.5ex>[d]& & {\cal U} \ar[d]\\
N \ar[r]^{ {\cal Z} }  \ar@<0.5ex>[u]^{ {\cal X}^* \phi} \ar@/_/[rr]_{ {\cal X} }& 
\bar{M} \ar[r]^{s} \ar@<0.5ex>[u]^{s^*\phi}  \ar@/_/[rrr]_{\bar{\phi}}& 
M \ar@<0.5ex>[u]^{\phi}\ar[rr]^{\pi_V \circ {\phi} = \bar{\phi} \circ \pi} &  & P(V') \\
}
\end{equation}
Introducing local coordinates $x^\mu$ on space-time, sections of the 
pull-back of the universal bundle by a holomorphic section take the following form
in terms of components:
\begin{eqnarray}
{\cal X}^* \phi: N \rightarrow {\cal U}^N && (x^\mu) \mapsto (X^I(\zeta(x)), F_I(\zeta(x)))
\end{eqnarray}
Given a set of local holomorphic coordinates $z^a$ on $\bar{M}$, we can 
choose a local holomorphic non-vanishing function $h$ on $\bar{M}$ and set 
$X^0 = h(z)$. Then $X^a(z) = h(z) z^a$, and we can interpret
the conical holomorphic special coordinates $X^I$ as local functions on $\bar{M}$.
Since $z^a = X^a/X^0$, the local holomorphic coordinates $z^a$ are the
`inhomogeneous' special holomorphic coordinates on $\bar{M}$ associated to the 
special holomorphic coordinates $X^I$ on $M$, which can be viewed as
projective coordinates (or homogeneous coordinates) on $\bar{M}$.\footnote{The terms
`inhomogeneous coordinate' and `homogeneous/projective coordinate' 
are used here as in projective geometry, for example for coordinates on the complex projective space 
$P_n(\mathbb{C})$.}

This construction provides us with a section $s: \bar{M} \rightarrow M: z^a \mapsto X^I(z)$. 
By making a 
holomorphic coordinate transformation $z^a \mapsto \zeta^a$ on $\bar{M}$ 
we can then go from special
holomorphic coordinates $z^a$ to general holomorphic coordinates $\zeta^a$. 
Given a holomorphic section $(X^I(\zeta), F_I(\zeta))$ of ${\cal U}^{\bar{M}}\rightarrow \bar{M}$, 
we can obtain an expression for the K\"ahler metric $g_{\bar{M}}$. Firstly, we locally 
identify
$\bar{M}$ with an embedded complex submanifold of $M$ using the section
$s:\bar{M} \rightarrow M\;:  \zeta^a \mapsto X^I(\zeta)$. The
metric $g_{\bar{M}}$ is obtained by  pulling back the projectable tensor 
$\tilde{g}^{(0)}_{{M}}$, see \eqref{Projectable_global},  that we have built out of the K\"ahler metric 
$g_M$. According to \eqref{KL_potential} the tensor $\tilde{g}^{(0)}_{{M}}$ 
has a `K\"ahler like' potential ${\cal K}(X,\bar{X})$. 
Since $(X^I(z), F_I(z))$ are local holomorphic functions on $\bar{M}$, 
it follows that $g_{{M}} = \iota^* \tilde{g}^{(0)}_M$ is a K\"ahler metric
$g_{\bar{M}} = \partial \bar{\partial} {\cal K}$ with K\"ahler potential
\begin{equation}
 \label{KpotPSK}
 {\cal K} = - \log ( -i (X^I(\zeta) \bar{F}_I (\bar{\zeta}) - F_I(\zeta) \bar{X}^I(\bar{\zeta})  ))\;.
 \end{equation}

We also note that
the pullback of the Chern connection to the pull back bundle
${\cal U}_N \rightarrow N$ over space-time by a unitary section is
\begin{eqnarray}
A_\mu(x)&=& \partial_\mu \zeta^a(x) A_a(\zeta(x), \bar{\zeta}(x)) + \partial_\mu \bar{\zeta}^{\bar{a}}(x) 
A_{\bar{a}}(\zeta(x),\bar{\zeta}(x))  \label{Chern_conn_ST} \\
&=& - \frac{1}{2} \frac{
X^I \stackrel{\leftrightarrow}{\partial}_\mu \bar{F}_I - F_I \stackrel{\leftrightarrow}{\partial}_\mu 
\bar{X}^I }{ i (X^I \bar{F}_I - F_I \bar{X}^I)}  = 
-\frac{i}{2} \frac{N_{IJ} ((\partial_\mu X^I)\bar{X}^J - X^I \partial_\mu \bar{X}^J)}{
N_{KL} X^K \bar{X}^L} \;, \nonumber
\end{eqnarray}
where $i(X^I \bar{F}_I - F_I \bar{X}^I) = N_{IJ} X^I 
\bar{X}^J = \pm 1$. 
We will see in section \ref{4dvecmult}
that this pull back connection 
is equal, {\em up to an overall minus sign}, to the $U(1)$ connection used in the 
superconformal calculus (see also \ref{sec:4dscg}).

\subsubsection{Formulation in terms of an affine bundle, and why
the prepotential transforms as it does  \label{sect:SK_vector_bundle}}

In this section we elaborate on the following two points:
\begin{enumerate}
\item
The extrinsic realization of ASK manifolds \cite{Alekseevsky:1999ts}
which we have described in section \ref{sec:kaehimm} only provides
a global construction for simply connected ASK manifolds. It is 
desirable to have a generalization which allows the global construction
of general ASK manifolds. 
\item
The transformation properties of the holomorphic prepotential under
symplectic transformations are complicated, and their geometric 
origin remains obscure. The prepotential is not a symplectic function,
and when deriving its transformation formula by integrating the
transformation formula \eqref{SympTransfXF} for the symplectic 
vector $(X^I,F_I)$, this leaves the integration constant undetermined. 
For a homogeneous prepotential this constant is absent, and 
for degree two we found the complicated looking expression \eqref{Ftilde}.
\end{enumerate}

We will now report how these issues have been resolved in 
\cite{Cortes:2017utn,Dietrich:2017}. The prepotential $F$
can be defined as the potential of the Liouville form
$\lambda = W_I dX^I$, restricted to a special Lagrangian
submanifold $L\subset V=T^*\mathbb{C}^n$, where 
$d\lambda_{|L} = - \Omega_{|L} = 0$ and $\lambda_{|L}  = F_I dX^I = dF$.
Here we assume that the complex symplectic coordinates $(X^I,W_I)$ on 
$V$ have been chosen such that $L$ is a graph. Then $W_I =F_I = \partial F/
\partial X^I$ on $L$. From these expressions it is clear that neither 
$\lambda$ nor $F$ is invariant under symplectic transformations. 
However the one-form 
\begin{equation}
\eta = X^I dW_I -W_I dX^I \;
\end{equation}
is symplectically invariant, and, like the Liouville form, a potential
for the complex symplectic form $\Omega$, hence closed 
when restricted to a Lagrangian submanifold $L\subset V$:
\begin{equation}
d\eta = 2 \Omega_{|L} = 0 \;.
\end{equation}
Consequently $\eta$ is locally exact on $L$ and admits a potential $f$, which 
is a symplectic function, and which 
is unique up to an additive constant,
\begin{equation}
\eta_{|L} = - df \;.
\end{equation}
We will call the potential $f$ a {\em Lagrange potential}, and note
that Lagrange potentials and prepotentials are related by
\begin{equation}
\label{Ff}
2F = f + X^I F_I \Leftrightarrow f = 2 F - X^I F_I \;. 
\end{equation}
From the physics literature it is well known that the
combination $F-\frac{1}{2} X^IF_I$ is a symplectic function \cite{deWit:1996gjy}.
We now see that this function is, up to normalization, 
the Lagrange potential associated to $F$.

Let now $M$ be a connected, but not necessarily simply connected
ASK manifold. Then the above applies locally, if we choose
a domain $U\subset M$ which is small enough to admit
a Lagrangian K\"ahlerian embedding  $\phi: U \rightarrow V\cong T^*\mathbb{C}^n
\cong \mathbb{C}^{2n}$. Such an embedding identifies $U$ with a 
Lagrangian submanifold $L\subset \mathbb{C}^{2n}$. On each such $L$ 
we have a symplectically invariant one-form
$\eta = X^I dW_I -W_I dX^I$ and can choose a Lagrange potential $f$. 
Then $(L,f)$ is called a {\em Lagrangian 
pair}, and a Lagrangian pair $(L,f)$ is called {\em K\"ahlerian} if the restriction 
of the Hermitian form $\gamma= i \Omega(\cdot, \overline{\cdot})$ is
non-degenerate. 
Lagrangian pairs are related to each other by a group action. The relevant
group is $G_\mathbb{C} := Sp(\mathbb{C}^{2n} )\ltimes \mbox{Heis}_{2n+1}(\mathbb{C})$,
where $\mbox{Heis}_{2n+1}(\mathbb{C})$ is the $(2n+1)$-dimensional complex
Heisenberg group. The group $G_\mathbb{C}$ is a central extension of the
complex symplectic affine $Sp(\mathbb{C}^{2n}) \ltimes \mathbb{C}^{2n} \subset
\mbox{Aff}(\mathbb{C}^{2n})$. We will see that the central extension is needed 
to include the freedom of shifting Lagrange potentials and  prepotentials by a constant, and 
refer to \ref{app:groups} for further details about the group $G_\mathbb{C}$,
its subgroups and its representations. The group $G_\mathbb{C}$
maps a given Lagrangian pair $(L,f)$ to the new pair
\begin{equation}
g \cdot (L,f) = (\bar{\rho}(g) L , g\cdot f) \;,
\end{equation}
where
$g=(M,s,v)\in G_\mathbb{C}$, with $M\in Sp(\mathbb{C}^{2n})$, $s\in \mathbb{C}$ central,
$v\in \mathbb{C}^{2n}$ a translation, where $\bar{\rho}$ is the affine representation 
of $G_\mathbb{C}$ obtained by `forgetting the centre', that is by the 
natural action of $(M,v) \in Sp(\mathbb{C}^{2n}) \ltimes \mathbb{C}^{2n}$, and where
\begin{equation}
\label{TransfLagrange}
g\cdot f = f \circ g^{-1} + \Omega(\cdot, v) -2 s
\end{equation}
is the new Lagrange potential. While the first term is the natural action of 
the affine group on functions, the second and third term 
correspond  to translations and to central transformations, 
respectively. In particular, the third term, which represents the action of the
centre of the group $G_\mathbb{C}$,
corresponds to shifting the Lagrange
potential, and the associated prepotential, by a constant. 

To describe the local embedding of an ASK manifold, we can only admit 
Lagrangian pairs which are K\"ahlerian. The subgroup of $G_\mathbb{C}$ 
acting on K\"ahlerian Lagrangian pairs is 
$G_{SK} = Sp(\mathbb{R}^{2n}) \ltimes
\mbox{Heis}_{2n+1}(\mathbb{C})  \subset G_\mathbb{C}$, which is a central 
extension of the 
affine symplectic group $\mbox{Aff}_{Sp(\mathbb{R}^{2n})}(\mathbb{C}^{2n})=
Sp(\mathbb{R}^{2n}) \ltimes \mathbb{C}^{2n}$ 
which we have encountered before. 

We need a further definition. 
A {\em special 
K\"ahler pair} $(\phi,F)$ is a
K\"ahlerian Lagrangian embedding $\phi: U \rightarrow \phi(U) \subset V$,
which induces on $U$ the restriction of the ASK structure of $M$,  
together with the choice of a prepotential $F$. For each 
$U$, one denotes by ${\cal F}(U)$ the set of all special K\"ahler pairs, where 
only domains $U$ are admitted where ${\cal F}(U) \not= \emptyset$. 
A K\"ahlerian Lagrangian pair $(\phi, F)$ determines a Lagrangian pair
$(L,f)$ with Lagrangian submanifold  $L:= \phi(U)$ and Lagrange potential
$f$ given by
\begin{equation}
\label{Lagrange2Prepo}
\phi^* f = 2 F - X^I W_I \;,
\end{equation}
where $\phi$ has components $\phi=(X^I, W_I)$. Formula \eqref{Lagrange2Prepo}
relates Lagrange potentials and prepotentials. By assumption the 
functions $X^I$ define special
coordinates on $U$ and if we identify 
$U$ with $\phi(U)$ we can omit $\phi^*$ in \eqref{Lagrange2Prepo}
and relate $F$ and $f$ as functions of holomorphic special coordinates.
Then we are back to \eqref{Ff}.

The group $G_{SK}$ acts on the special K\"ahler pairs ${\cal F}(U)$ by
\begin{equation}
g \cdot (\phi, F) := (g\phi \ , g\cdot F) \;,
\end{equation}
where
\begin{equation}
g\phi = \bar{\rho}(g) \circ \phi \;,
\end{equation}
with $\bar{\rho}$ the same representation of $G_{SK}$ as above and 
\begin{equation}
\label{TransfPrepo}
g\cdot F := F - \frac{1}{2} X^I W_I + \frac{1}{2} X'^I W'_I + \frac{1}{2}
(g\phi)^* \Omega(\cdot, v) - s \;,
\end{equation}
where $\phi=(X^I, W_I)$ and $g\phi = (X'^I, W'_I)$ are the local expressions
for $\phi$ and $g\phi$. The somewhat complicated 
transformation formula \eqref{TransfPrepo} for prepotentials follows
from the formula \eqref{TransfLagrange} for Lagrange potentials 
together with \eqref{Lagrange2Prepo}.

By specialization to the subgroup $Sp(\mathbb{R}^{2n})\subset G_{SK}$
we see that under symplectic transformations $g=(M,0,0)$:
\begin{equation}
F \rightarrow F' - \frac{1}{2} X^I W_I + \frac{1}{2} X'^I W'_I \Leftrightarrow 
F' - \frac{1}{2} X'^I W'_I = F - \frac{1}{2} X^I W_I \;.
\end{equation}
This is the standard formula for the transformation of the prepotential,
now derived without the ambiguity of adding a constant. The
observation that $F-\frac{1}{2} X^I F_I$ is a symplectic function
is now explained by this function being proportional to the associated
Lagrange potential. For CASK manifolds, symplectic transformations act on 
the set of homogeneous prepotentials of degree two. Note that two is
the only degree of homogeneity for the prepotential, 
where $F_I$ has the same degree
of homogeneity as $X^I$, so that a linear combination of $X^I$ and $F_I$
transforms homogeneously.

Let us now turn our attention to how a global construction of ASK manifolds
can be achieved by glueing together special K\"ahler pairs. We will only
give a summary and refer the interested reader to \cite{Cortes:2017utn,Dietrich:2017} for details. 
The group $G_{SK}$ acts simply transitively on the set ${\cal F}(U)$ of
special K\"ahler pairs for fixed $U\subset M$. One can show that by letting
$U$ vary over $M$ one obtains a $G_{SK}$ principal bundle $P\rightarrow M$,
called the {\em bundle of special K\"ahler pairs}, which comes equipped with 
a flat connection. The group 
$G_{SK}$ admits a linear representation $\rho: G_{SK} \rightarrow Sp(\mathbb{R}^{2n})$
which defines a flat real symplectic vector bundle $(V_\mathbb{R},\Omega, \nabla)$ of rank $2n$,
such that $\nabla \Omega=0$. By complex linear extension $V_\mathbb{C} = V_\mathbb{R} \otimes 
\mathbb{C}$  we obtain a flat symplectic holomorphic
vector bundle $(V_\mathbb{C}, \Omega, \nabla)$, with $\nabla \Omega=0$, where we use
the same symbol for $\nabla, \Omega$ and their extensions. 
The complex symplectic form $\Omega$ on $V_\mathbb{C}$ defines a 
Hermitian metric $\gamma = i \Omega(\cdot, \tau \cdot)$, where $\tau$ is 
complex conjugation. Since the group $Sp(\mathbb{R}^{2n})$ acts on $\mathbb{C}^{2n}$, 
the complex vector bundle $V_\mathbb{C}$ is associated to the $G_{SK}$ principal
bundle of special K\"ahler pairs through the (extension of the) linear representation
$\rho$. 

One can further show that $M$ being an ASK manifold implies 
that $V_\mathbb{C}$ admits a global holomorphic section $\Phi: M \rightarrow
V_\mathbb{C}$, such that
\begin{eqnarray}
(\nabla \Phi)^* \Omega &=& 0  \;, \label{Cond1} \\
(\nabla \Phi)^* \gamma &\mbox{is}& \mbox{non-degenerate}.\label{Cond2}
\end{eqnarray} 
The map $\nabla \Phi: TM \rightarrow V_\mathbb{C}$ is a morphism of
holomorphic vector bundles. The global section $\Phi$ generalizes the 
global immersion $M\rightarrow V$ of simply connected 
ASK manifolds, with conditions \eqref{Cond1} and \eqref{Cond2}
corresponding to the requirements that $\phi$ must be symplectic
($\phi^* \Omega=0$) and K\"ahlerian ($\phi^* \gamma$ non-degenerate). 
This construction does not yet encode the freedom of making translations.
To include these we need to introduce a flat complex affine bundle $A\rightarrow M$
modelled on $V_\mathbb{C}$,\footnote{See \ref{app:manifolds} for the definition of an affine bundle.}  which can also be defined as the
affine bundle associated to the principal bundle $P\rightarrow M$ 
of special K\"ahler pairs by the affine representation  $\bar{\rho}: G_{SK} \rightarrow \mbox{Aff}_{\mathrm{Sp}(\mathbb{R}^{2n})}(\mathbb{C}^{2n})$ on $\mathbb{C}^{2n}$.

One then obtains the following theorem, which generalizes the 
construction of \cite{Alekseevsky:1999ts}:
\begin{thm}
{\bf Extrinsic construction of general affine special K\"ahler manifolds
(Theorem 3.5.4 of \cite{Dietrich:2017}).}
Let $M$ be a complex manifold, and $A\rightarrow M$ be a flat complex affine
bundle modelled on the complex vector bundle $V_\mathbb{C} = V_\mathbb{R} \otimes 
\mathbb{C}$, where $(V,\Omega, \nabla)$ is a flat real symplectic vector bundle such that 
$\nabla \Omega=0$. If there is a global holomorphic section $\Phi: M \rightarrow A$
such that the conditions \eqref{Cond1} and \eqref{Cond2} are satisfied, then $M$
carries the structure of an affine special K\"ahler manifold, and $A$ is associated to
the principal $G_{SK}$ bundle of special K\"ahler pairs by the affine 
representation $\bar{\rho}: G_{SK} \rightarrow \mbox{Aff}_{\mathrm{Sp}(\mathbb{R}^{2n})}(\mathbb{C}^{2n})$ acting on $\mathbb{C}^{2n}$. 

Conversely, if $M$ is an affine special K\"ahler manifold, then the associated 
complex affine bundle $A\rightarrow M$ corresponding to the affine 
representation $\bar{\rho}: G_{SK} \rightarrow \mbox{Aff}_{\mathrm{Sp}(\mathbb{R}^{2n})}(\mathbb{C}^{2n})$ acting on $\mathbb{C}^{2n}$ has a global section $\Phi: M \rightarrow A$, which satisfies the conditions \eqref{Cond1} and \eqref{Cond2}.
\end{thm}

\subsubsection{Comparison to the literature \label{sect:comparison}}

In this section we will compare the definitions we have given for affine
and projective special K\"ahler geometry with other definitions in the 
literature. So far we have covered the original definition \cite{deWit:1984pk} of PSK geometry,
which was expressed in terms of special holomorphic coordinates and 
based on the superconformal tensor calculus;  the intrinsic definition of \cite{Freed:1997dp},
and the extrinsic construction of \cite{Alekseevsky:1999ts}, which has extended 
the earlier work \cite{Cortes:1996,Cortes:1998} into the framework of {\em special complex
geometry,} which contains special K\"ahler geometry as a subset. An alternative 
`bilagrangian' extrinsic construction of ASK manifolds has been given in \cite{1999math......1069H}.

In between \cite{deWit:1984pk} and \cite{Freed:1997dp} various other formulations 
of special K\"ahler geometry have been presented in the physics literature. Common themes
in these approaches are: (i) to have manifest holomorphic coordinate invariance of the
formalism, that is, 
to use general holomorphic coordinates instead of special holomorphic coordinates,
and (ii) to avoid using the prepotential explicitly, because the prepotential is not a 
symplectic function, and because there are (non-generic)
symplectic frames where no prepotential exists. This leads one to work with a 
collection  $\Phi(z)=(X^I(z), F_I(z))$ of holomorphic functions defined on 
local coordinate charts, which are glued together by transition functions,
and which are are interpreted as defining a global section of a vector bundle. 
Equivalently, one can use a unit section $\Phi_1(z,\bar{z})=(X^I(z,\bar{z}), F_I(z,\bar{z}))$, which 
then is not holomorphic. In this setting special K\"ahler geometry is defined by imposing
suitable conditions on this section which allow to define a non-degenerate special
K\"ahler metric, 
and,  more generally, to obtain all the local expressions needed to have a well
defined vector multiplet Lagrangian. Since these approaches are covered 
by excellent reviews, articles and books 
including \cite{Andrianopoli:1996vr,Andrianopoli:1996cm,Craps:1997gp,Lledo:2006nr,Freedman:2012zz}, 
which contain comprehensive bibliographies, 
we only mention a few
selected papers in the following. The work of 
\cite{Strominger:1990pd} gave a geometric definition of PSK manifolds 
in terms of holomorphic vector bundles, which  was motivated by the insight that special K\"ahler
geometry plays an important role in the geometry of moduli spaces of
Calabi-Yau compactifications of string theory, see also section \ref{sec:CY3}. 
The so-called rheonomic approach to supergravity, see
\cite{Castellani:1991eu} for a review, was applied to 
${\cal N}=2$ vector multiplets in \cite{Castellani:1990zd,Castellani:1990tp}
to obtain a formulation based on general holomorphic coordinates. 
Issues relating to the (non-)existence of a prepotential were discussed in 
\cite{Ceresole:1994cx}. This is particularly relevant for gauged supergravity, that is 
for supergravity theories with non-abelian gauge symmetries or charged
matter multiplets, because the gauging breaks the continuous symplectic 
symmetry and distinguishes a discrete subset of frames. Gauged 
supergravity is outside the scope of this review. The formulation 
of special K\"ahler geometry in terms of 
real symplectic coordinates was discussed in \cite{Ferrara:2006js,Ferrara:2006at}.

For a more detailed comparison between the approach presented in this 
review and alternative formulations, we use \cite{Craps:1997gp},
where various definitions of special K\"ahler geometry
have been collected and compared
to each other, and \cite{Lledo:2006nr}, which has extended these
definitions to arbitrary target space signature.

For ASK manifolds, the transition functions given 
in \cite{Craps:1997gp} take the form
\begin{equation}
\label{TransDef1vP}
\left( \begin{array}{c}
X^I(z_{(i)}) \\ F_I(z_{(i)}) \\
\end{array}\right) = e^{ic_{(ij)}} M_{(ij)} 
\left( \begin{array}{c}
X^I(z_{(j)})  \\ F_I(z_{(j)}) \\
\end{array}\right) + b_{(ij)} \;.
\end{equation}
Here the indices $i,j$ refer to two overlapping patches $U_i, U_j \subset M$,
$(M_{(ij)}, b_{(ij)})\in \mbox{Sp}(\mathbb{R}^{2n}) \times \mathbb{C}^{2n}$, 
are transition functions corresponding to affine symplectic transformations,
and
$e^{ic_{(ij)}} \in U(1)$ are constant $U(1)$ phases. While 
$(M_{(ij)}, b_{(ij)})$ realize the affine representation $\bar{\rho}$
of the group $G_{SK}$, 
and therefore can be interpreted as transition functions of the 
complex affine bundle $A\rightarrow M$, the phases $e^{ic_{(ij)}}$ 
reflect an additional freedom which is not present 
in \cite{Freed:1997dp}, \cite{Alekseevsky:1999ts},
where the special connection $\nabla$ is part of the data 
defining an ASK manifold. As discussed
in section \ref{sect:dual_coord}, special connections always come in 
$S^1$-families. While the underlying K\"ahler manifold is the same,
ASK manifolds with different special connections from the same 
$S^1$-family are considered distinct according to the definitions 
in \cite{Freed:1997dp}, \cite{Alekseevsky:1999ts}. 
However,
this choice does not influence the K\"ahler metric and other data
needed to build a vector multiplet Lagrangian, and therefore 
definitions in the physics literature do not require to fix the special 
connection. 
The phases $e^{ic_{(ij)}}$ in the transition functions \eqref{TransDef1vP}
reflect the freedom of choosing different special
connections $\nabla_{(i)}$ and $\nabla_{(j)}$ from the same $S^1$-family
on $U_i$ and $U_j$. Thus compared to the complex affine bundle $A\rightarrow M$
transition functions of the form \eqref{TransDef1vP} define a bundle which 
 is modified by  a `twist.' It would be interesting to describe this twist within the
framework of \cite{Alekseevsky:1999ts,Dietrich:2017}. Moreover, since we are
not aware of explicit examples where the additional freedom of rotating
the special connection is actually used, it would be interesting 
to find explicit examples.

Let us also have a look at the definition of PSK manifolds given
in \cite{Craps:1997gp}. In this case the transition functions 
between patches $U_i,U_j \subset \bar{M}$ are of the form
\begin{equation}
\left( \begin{array}{c}
X^I(z_{(i)} ) \\ F_I(z_{(i)})   \\
\end{array} \right)
 = e^{f_{(ij)}(z)} M_{(ij)} 
\left( \begin{array}{c}
X^I(z_{(j)}) \\ F_I(z_{(j)}) \\
\end{array} \right) \;,
\end{equation}
where $f_{(ij)}(z)$ are holomorphic functions and 
$M_{(ij)} \in Sp(2n+2, \mathbb{R})$. Such transition functions
correspond to a product bundle ${\cal L}\otimes {\cal H} \rightarrow \bar{M}$,
where ${\cal L}$ is a holomorphic line bundle and ${\cal H}$ is
a flat symplectic vector bundle. If ${\cal H} \rightarrow \bar{M}$ is trivial
we can identify ${\cal L}$ with 
the pull-back line bundle ${\cal U}^{\bar{M}} \rightarrow \bar{M}$. 
If ${\cal H}$ is non-trivial, we expect that this bundle will arise when applying the construction 
of section \ref{sec:formlinebund}
to the complex affine bundle $A\rightarrow M$.
We remark that the special connection $\nabla$ on $M$ does not induce
a flat connection on $\bar{M}$, since the superconformal quotient 
includes dividing out the isometric $U(1)$-action which acts by 
rotation on the $S^1$-family of special connections. It would be 
interesting to have an intrinsic characterization of PSK manifolds,
which then could be related to the constructions in terms of 
line bundles and vector bundles.

Finally, another global condition which is included explicitly in the definition 
 \cite{Craps:1997gp} of PSK manifolds is that $\bar{M}$ should
 be a {\em K\"ahler-Hodge manifold}. In the mathematical literature
 a K\"ahler manifold $\bar{M}$ is called a K\"ahler-Hodge manifold or
 K\"ahler manifold of restricted type if its K\"ahler form $\omega$ 
 defines an integral cohomology class, $[\omega] \in H^2(\bar{M}, \mathbb{Z})$. 
 For compact $\bar{M}$ this implies that $\bar{M}$ is a projective 
 variety, that is, embeddable into complex projective space. In supergravity 
 a normalization condition for the K\"ahler form arises since the fields
 transform under the local action of the group $U(1)$, which in the superconformal 
 approach is part of the superconformal group. One must therefore impose that
 these transformations are globally well defined on the scalar manifold. 
 This also applies to ${\cal N}=1$ supergravity, which like
 ${\cal N}=2$ has a local $U(1)$ group action on its scalar manifold 
 $\bar{M}$. For compact $\bar{M}$ it was shown in \cite{Witten:1982hu}
that this implies that the K\"ahler form must define an even integer class
in $H^2(\bar{M}, \mathbb{Z})$. That the condition is even-ness rather than
integrality results from the normalization of the $U(1)$ charges, which are 
half-integer valued for fermions. In the physics literature the term 
K\"ahler-Hodge is used for K\"ahler manifolds which are target spaces of supermultiplets
that can be coupled consistently to supergravity. 
Most standard examples for PSK are open domains which 
have trivial topology, so that $[\omega]=0$, and the K\"ahler-Hodge condition is automatically
satisfied. For non-compact scalar manifolds with 
non-trivial topology the global well-definedness of $U(1)$-transformations can
impose non-trivial conditions. 
A recent comprehensive 
analysis has shown that a scalar manifold $\bar{M}$ is an admissible
target space for chiral supermultiplets coupled to ${\cal N}=1$ supergravity 
if it admits a so-called chiral triple \cite{Cortes:2018lan}. If space-time is a spin manifold, 
then every K\"ahler-Hodge manifold admits a chiral triple, irrespective
of whether it is compact or non-compact  \cite{Cortes:2018lan}. 
In \cite{Lledo:2006nr} it was shown that `projective K\"ahler manifolds', 
that is scalar manifolds constructed as K\"ahler quotients using
the superconformal calculus  are automatically K\"ahler-Hodge. 
 Since we have defined
 PSK manifolds as K\"ahler quotients of CASK manifolds, there is no
 need to require the K\"ahler-Hodge property explicity.

\subsection{Special geometry and Calabi-Yau three-folds \label{sec:CY3}}

The geometry of moduli spaces of Calabi-Yau three-folds provides natural realizations
of special real and special K\"ahler geometry. These moduli spaces appear in 
compactifications of supergravity and of string and M-theory on Calabi-Yau three-folds.
The scalar manifolds in physical applications usually combine moduli which 
correspond to deformations of the Calabi-Yau metric with moduli associated with
the deformations of antisymmetric tensor fields.
We start with the discussion of the moduli of the Calabi-Yau metric, and then turn 
to the moduli spaces of string compactifications. In this section we assume
knowledge of some mathematical concepts, including holonomy, de Rham and Dolbeault
cohomology,\footnote{Some aspects of Dolbeault cohomology are 
presented in \ref{App:cplxmfds}.} Hodge numbers,  homology, Poincar\'e  duality, the cup and intersection
product. Since this material is not needed in other parts of this review, we will not
explain these concepts in detail, but refer the readers to \cite{GSW} Vol 2 and
\cite{Candelas:1987is,Huebsch:1991}, on which this
section is partly based.

A {\em Calabi-Yau $n$-fold} $X$ is a $2n$-dimensional compact Riemannian manifold
with  holonomy group contained in $SU(n) \subset U(n) \subset SO(2n)$. 
This implies that $X$ is K\"ahler, but it is more restrictive
than that, by excluding a subgroup  $U(1) \subset U(n)$ from the holonomy, which 
implies that the metric is Ricci-flat. Therefore a Calabi-Yau manifold can
alternatively be defined as a K\"ahler manifold admitting a Ricci-flat metric.\footnote{For
string theory compactifications the metric is only Ricci-flat to leading order in $\alpha'$
for $n>2$, but
this does not affect the following discussion.}  

We now specialize to Calabi-Yau three-folds. In the following it is 
understood that  the holonomy group is not contained
in $SU(2) \subset SU(3)$, thus excluding the cases where $X = K3 \times T^2$
(which is hyper-K\"ahler with holonomy $SU(2)$), and where $X=T^6$ (which is
flat). 
The moduli space arising when dimensionally reducing the Einstein-Hilbert 
term on $X$ is the space ${\cal M}_{\mathrm{Ricci}}$ of Ricci-flat metrics on $X$. 
If the field equations of a higher-dimensional theory of gravity and matter
admit a solution where space-time takes
the form $\mathbb{R}^{1,3} \times X$ with a metric $\eta_{1,3} \times g$, which
is the product of the four-dimensional Minkowski metric $\eta_{1,3}$ with 
a Ricci-flat metric $g$ on $X$, then the four-dimensional massless fields
corresponding to zero modes of the higher-dimensional metric are: (i) 
the four-dimensional graviton, equivalently, linearized fluctuations of the
Minkowski metric $\eta_{1,3}$, (ii) four-dimensional vector fields in the adjoint representation 
of the isometry group of $g$, (iii) four-dimensional scalars in one-to-one
correspondence with linearly independent solutions of  the linearized Einstein 
equation on $X$. If $X$ is a Calabi-Yau three-fold, then there are no
continuous isometries and hence no massless vector fields descending from 
the higher-dimensional metric. A Ricci-flat metric on $X$ is consistent with
the field equations if the energy-momentum tensor
has no non-zero components along $X$. In this case  the Einstein equations reduce
to the condition that $X$ is Ricci-flat, and scalar zero modes of the metric
parametrize
the moduli space of Ricci-flat metrics on $X$.\footnote{In string theory the Einstein
equations receive $\alpha'$-corrections. This leads to corrections to the 
metric on the moduli space, which can be computed using two topologically twisted
versions of string theory \cite{Huebsch:1991}.}
The linearized form of the Ricci-flatness condition is a Laplace-type equation 
for the so-called Lichnerowicz Laplacian, whose zero modes are the moduli
scalars. They enter
into the low-energy effective four-dimensional action through a non-linear
sigma model with target space ${\cal M}_{\mathrm{Ricci}}$, equipped with 
the metric
\begin{equation}
G(\delta g_{(1)}, \delta g_{(2)}) = \frac{1}{V} \int_X 
\delta g_{(1)mn} \delta g_{(2)pq} g^{mp} g^{nq} \sqrt{g} d^6 x \;,
\end{equation}
where $x^m, m=1,\ldots, 6$ are coordinates on $X$, where
$g=(g_{mn})$ is the metric on $X$, where $\delta g_{(i)mn}$ $i=1,2$ are
infinitesimal deformations of the metric, and where $V$ is the volume of $X$.

For Calabi-Yau three-folds ${\cal M}_{\mathrm{Ricci}}$ is locally isometric
to the product of the 
moduli space ${\cal M}_{\mathrm{cplx}}$ of complex structures on $X$ and the 
moduli space ${\cal N}_{\mathrm{Kahler}}$ of K\"ahler structures on $X$,
\begin{equation}
{\cal M}_{\mathrm{Ricci}} \cong {\cal M}_{\mathrm{cplx}} \times {\cal N}_{\mathrm{Kahler}} \;.
\end{equation}
This factorization is a special feature of Calabi-Yau three-folds.  
The definition of a K\"ahler form requires the choice of  a complex structure, 
and in general the space ${\cal N}_{\mathrm{Kahler}}$ of K\"ahler forms 
 of a complex manifold is fibred over 
its space ${\cal M}_{\mathrm{cplx}}$ of complex structures. However, for Calabi-Yau
three-folds the K\"ahler structure and complex structure can locally be varied independently.
From the physics perspective this is predicted by supersymmetry, since in ${\cal N}=2$
theories both types of moduli belong to different types of multiplets.
In terms of complex coordinates $u^a$, $a=1,2,3$ on the Calabi-Yau three-fold 
$X$ with metric $g$, complex structure $J$ and K\"ahler form $\omega$,
deformations of the
complex structure $J$ correspond to deformations of the K\"ahler metric $g_{a\bar{b}}$
which have the form $\delta g_{ab}, \delta g_{\bar{a}\bar{b}}$
 and therefore are not Hermitian
with respect to the undeformed complex structure $J$. In contrast 
deformations of the K\"ahler structure correspond to deformations $\delta g_{a\bar{b}}$
of the metric, which are Hermitian with respect to the complex structure $J$ but change
the K\"ahler form $\omega$ of $X$. 

Infinitesimal deformations of a complex structure $J \in \Gamma(\mbox{End}(TX))$
are generated by holomorphic vector-valued one forms $\tau = \tau^a_{\;\;\bar{b}} \partial_a \otimes
d\bar{u}^{\bar{b}}$, $\bar{\partial} \tau =0$.
Two such forms generate equivalent deformations if they differ by an $\bar{\partial}$-exact
form, therefore complex structure deformations are classified by $H^1(X,T_\mathbb{C} X)$, the  first Dolbeault cohomology group of $X$ with values in the complexified tangent bundle. On a  Calabi-Yau three-fold there exists a holomorphic, covariantly constant $(3,0)$-form $\Omega$, called the holomorphic 
top-form, which is unique up to complex rescalings $\Omega \rightarrow \lambda \Omega$, where $\lambda \in \mathbb{C}^*$. This provides an isomorphism
between $T_\mathbb{C} X$ and $\Lambda^{2} T^*_\mathbb{C} X$ by
\begin{equation}
\phi^a \mapsto \psi_{bc} = \Omega_{abc} \phi^a \;,
\end{equation}
which implies the relation
\begin{equation}
H^1(X,T_\mathbb{C} X) \cong  H^1(X,\Lambda^2 T^*_\mathbb{C} X) 
\cong H^{2,1}_{\bar{\partial}}(X) \;,
\end{equation}
so that complex structure deformations of Calabi-Yau three-folds are
parametrized by the Dolbeault cohomology group $H^{2,1}_{\bar{\partial}}(X)$,
that is, by equivalence classes of $\bar{\partial}$-closed $(2,1)$-forms modulo
$\bar{\partial}$-exact forms, which are related to vector-valued one-forms by the
isomorphism, 
\begin{equation}
\phi_{ab \bar{c}} = \Omega_{abd} \tau^d_{\;\;\bar{c}} \;.
\end{equation}
The dimension of $H^{2,1}(X)$ (considered as a vector space) is given by 
the Hodge number $h^{2,1}\geq 0$, which is a topological invariant of $X$. 
Since there is a one-to-one correspondence between linearly independent
harmonic $(p,q)$-form on $X$ and elements of $H^{p,q}(X)$, one can
choose harmonic $(2,1)$-forms to generate the complex structure 
deformations. The expansion of a general harmonic $(2,1)$-form $\phi$ 
in a basis $\phi_A$, $A=1,\ldots, h^{2,1}$,
\begin{equation}
\phi = z^A \phi_A \;, \;\;\;z^A \in \mathbb{C}
\end{equation}
provides local coordinates $z^A$ on ${\cal M}_{\mathrm{cplx}}$. The metric 
on ${\cal M}_{\rm{cplx}}$ is induced by the standard scalar product
$(\alpha, \beta)=\int_X \alpha \wedge * \beta$
between $(2,1)$-forms. To see that this metric is K\"ahler, and more
specifically projective special K\"ahler, one uses the relation between
complex structures on $X$ and the periods of the holomorphic top-form
$\Omega$. Choosing a complex structure
on $X$ is equivalent to specifying a decomposition of the third de-Rham
cohomology group into Dolbeault cohomology groups,
\begin{equation}
H^3(X) = H^{3,0}_{\bar{\partial}} (X) \oplus
 H^{2,1}_{\bar{\partial}} (X) \oplus
  H^{1,2}_{\bar{\partial}} (X) \oplus
   H^{0,3}_{\bar{\partial}} (X) \;.
\end{equation}
Such a decomposition is obtained by picking one of the $b_3=1+h^{2,1}+h^{2,1}+1$
harmonic forms, where $b_3$ is the third Betti number, and declaring it to be 
the holomorphic top form. More precisely, the complex structure does not
depend on the explicit choice of $\Omega$, but only on the corresponding
`complex direction', since we can rescale $\Omega \mapsto \lambda \Omega$,
$\lambda \in \mathbb{C}^*$. 

We now choose a basis $(A^I, B_I)$, $I=0, \ldots, h^{2,1}$ of the 
third homology group $H_3(X,\mathbb{Z})$ of $X$, with normalization
\begin{equation}
A^I \cdot B_J = \delta^I_J = - B_J \cdot A^I\;,
\end{equation}
where $\cdot$ denotes the intersection product (which is defined
by counting intersection points between submanifolds, weighted with
orientation). The periods 
\begin{equation}
X^I(z) := \int_{A^I} \Omega \;,\;\;\;
F_I(z) := \int_{B_I} \Omega \;,
\end{equation}
of the holomorphic top-form depend holomorphically on the 
complex coordinates $z^A$ on ${\cal M}_{\mathrm{cplx}}$. 
Poincar\'e duality provides an isomorphism between the
homology groups $H_p(X,\mathbb{Z})$ and the cohomology groups
$H^{6-p}(X,\mathbb{Z})$,\footnote{The ring $\mathbb{Z}$ can be replaced by $\mathbb{R}$ or
$\mathbb{C}$.} 
\begin{equation}
C \mapsto [C] \;,\;\;\mbox{such that}\;\;\;
\int_C \beta = \int_X [C]\wedge \beta \;,
\end{equation}
for all $\beta \in \Omega^p(X)$. Poincar\'e duality maps the intersection 
product of cycles (defined by counting intersection points weighted by
orientation) to the cup product of cohomology cycles (induced by the
wedge product of forms). This allows to define a basis $(\alpha_I, \beta^I)$ 
of $H^3(X,\mathbb{Z})$ dual to the basis $(A^I,B_I)$ of $H_3(X,\mathbb{Z})$:
\begin{equation}
\int_{A^J} \alpha_I = \int \alpha_I \wedge \beta^J = \delta^I_J \;,\;\;\;
\int_{B_J} \beta^I = \int \beta^I \wedge \alpha_J = - \delta_J^I \;.
\end{equation}
In terms of this basis the top-form has the expansion
\begin{equation}
\Omega = X^I \alpha_I - F_I \beta^I  \;.
\end{equation}
Since only half of the periods are independent, $X^I$ can be chosen to
parametrize the possible choices of a top-form out of the harmonic three-forms.
It follows that the $X^I$ can be used as projective coordinates for 
${\cal M}_{\mathrm{cplx}}$. At this point the relation to CASK and PSK
manifolds becomes obvious. It turns out that the metric on ${\cal M}_{\mathrm{cplx}}$,
which is defined by the scalar product between $(2,1)$ forms,
is a K\"ahler metric with K\"ahler potential 
\begin{equation}
K = - \log\left[ i \int_{X} \Omega \wedge \bar{\Omega} \right]= - \log\left[- i \left( 
X^I(z) \bar{F}_I(z) - F_I (z)\bar{X}^I(z) \right) \right]\;.
\end{equation}
This is a PSK metric, given in terms of a holomorphic section 
$(X^I(z), F_I(z))$ of the complex line bundle ${\cal L} \rightarrow \bar{M} = {\cal M}_{\mathrm{cplx}}$. 
The associated CASK metric also has a natural interpretation. If we 
do not only choose a complex structure, but in addition a specific top-form
compatible with this structure, the resulting space, which is 
parametrized by the independent periods $X^I$,  is a complex cone
over ${\cal M}_{\mathrm{cplx}}$ which carries the structure of a CASK 
manifold.

We now turn to infinitesimal 
deformations $\delta g_{a\bar{b}}$ of the Ricci flat metric which preserve the
complex structure. In local complex coordinates $u^a$ on $X$ the K\"ahler forms is 
given by $\omega= i g_{a\bar{b}}du^a \wedge d\bar{u}^{\bar{b}}$, and therefore
such deformations change the K\"ahler form. Since the K\"ahler form is a closed
real $(1,1)$-form, it defines a class in 
\begin{equation}
H^{1,1}(X,\mathbb{R}) := H^2(X,\mathbb{R}) \cap H^{1,1}(X,\mathbb{C}) \;,
\end{equation}
which labels K\"ahler structures on $X$. Changes of the K\"ahler structure
are changes of the K\"ahler form by a real $(1,1)$-form which is closed but
not exact. As representatives one can choose $h^{1,1}$ linearly independent
harmonic $(1,1)$-forms $\omega_x$, $x=1, \ldots h^{1,1}$. Then the expansion 
of the K\"ahler form in terms of this basis,
\begin{equation}
\omega = t^x \omega_x \;,\;\;\;t^x \in \mathbb{R}\;,
\end{equation}
provides real coordinates on the space of ${\cal N}_{\mathrm{Kahler}}$ 
of K\"ahler structures. We remark that ${\cal N}_{\mathrm{Kahler}} \subsetneq
H^{1,1}(X,\mathbb{R}) \cong \mathbb{R}^{h^{1,1}}$, since when deforming 
the K\"ahler form we need to preserve the positivity of the metric $g$ on $X$.
This can be expressed by the requirement that the volumes of $X$ and
 of all its complex submanifolds must be positive. The top exterior power
of the K\"ahler form is proportional to the volume form of $(X,g)$. The volume
$V$ of $X$ is given by
\begin{equation}
V = \frac{1}{3!} \int_X \omega \wedge \omega \wedge \omega \;.
\end{equation}
Moreover the K\"ahler form is a so-called {\em calibrating form} for holomorphic curves $C$ and
holomorphic surfaces $S$ in $X$, that is 
\begin{equation}
\mbox{Vol}(C) = \int_C  \omega \;,\;\;\;
\mbox{Vol}(S) = \frac{1}{2} \int_S \omega \wedge \omega  \;.
\end{equation}
The conditions 
\begin{equation}
\int_C \omega >0 \;,\;\;\;
\int_S \omega \wedge \omega > 0\;,\;\;\;
\int_X \omega \wedge \omega \wedge \omega > 0
\end{equation}
define the {\em K\"ahler cone} of $X$, the space of positive K\"ahler classes, which 
is ${\cal N}_{\mathrm{Kahler}}$. 

Using the basis $\omega_x$, the volume takes the form 
\begin{equation}
\label{volume}
V = \frac{1}{3!} C_{xyz} t^x t^y t^z\;,
\end{equation}
where the quantities
\begin{equation}
C_{xyz} := \int_X \omega_x \wedge \omega_y \wedge \omega_z 
\end{equation}
are topological invariants, called {\em triple intersection numbers}. 
To explain this name, we use the isomorphism
$H^2(X,\mathbb{Z}) \cong H_4(X,\mathbb{Z})$ provided by Poincar\'e duality, which maps
the cup product of cohomology classes of closed differential forms to the intersection product
of homology classes of closed submanifolds. This implies that 
\begin{equation}
C_{xyz} = D_x \cdot D_y \cdot D_z\;,
\end{equation}
where $D_x$, $x=1, \ldots h^{1,1}=b_2$ is the basis of $H_4(X,\mathbb{Z})$
dual to the basis $\omega_x$ of $H^2(X,\mathbb{Z})$, and 
where $\cdot$ is the intersection product between homological four-cycles.\footnote{$b_2$ is the second Betti number of $X$. Note that for Calabi-Yau three-folds $h^{2,0}=h^{0,2}=0$, hence $b_2=h^{1,1}$.}

 Since the moduli dependence of the volume $V$ is given by a homogeneous symmetric polynomial in the K\"ahler moduli
 $t^x$, we can use it to define a 3-conical metric, in fact an ASR metric,
 on ${\cal N}_{\mathrm{Kahler}}$. Its logarithm $\log V$ defines the associated 
 0-conical Hessian metric. The metric on ${\cal N}_{\mathrm{Kahler}}$ 
 obtained by dimensional reduction of the Einstein-Hilbert action is the
 metric induced by the scalar product of $(1,1)$ forms. Its metric coefficients with
 respect to the basis $\omega_x$ can be shown to be of the form
 \begin{equation}
 G_{xy} = G(\omega_x, \omega_y) = \frac{\partial^2 \log V}{\partial t^x \partial t^y} \;.
 \end{equation}
 Thus the metric on the K\"ahler cone is 0-conical with a Hesse potential given
 by the logarithm of the volume. The associated PSR metric on hypersurfaces
 of constant volume also has a natural interpretation. It is the metric on the 
 moduli space ${\cal N}_{\mathrm{Kahler}}^V$ of K\"ahler structures at fixed volume.
 As we have seen in 
 section \ref{Sect:SR} 
 this metric is obtained by 
 pulling back either the 3-conical metric $\partial^2 V$ or the 0-conical metric 
 $\partial^2 \log V$ to the hypersurface 
${\cal N}^V_{\mathrm{Kahler}} \subset {\cal N}_{\mathrm{Kahler}}$. 

In physics applications it is ${\cal N}^V_{\mathrm{Kahler}}$ 
rather than ${\cal N}_{\mathrm{Kahler}}$
which appears as the target space of a sigma model, and therefore it must
carry a positive definite metric. From section \ref{Sect:SR}
we know that the PSR metric is positive definite if the 0-conical metric
$\partial^2 \log V$ is positive definite, and equivalently if 
the 3-conical metric $\partial^2 V$ has real Lorentz signature $(1,h^{1,1}-1)$.
These conditions are indeed satisfied in Calabi-Yau compactifications.
The distinction between time-like and space-like directions 
with respect to $\partial^2V$
in the space
of $(1,1)$-forms  corresponds to the so-called Lefschetz decomposition 
of $H^2(X,\mathbb{R})$ into `primitive forms', which are orthogonal to 
the K\"ahler form $\omega$, and the direction parallel to the K\"ahler form. 

A related but different question is to determine the maximal domain
in $H^{1,1}(X,\mathbb{R}) \cong \mathbb{R}^{h^{1,1}}$ where the
PSR metric is positive definite. 
The boundary of this region
can be characterized using the simpler 3-conical metric $\partial^2 V$
by $\det( \partial^2 V)=0$. Note that the region where the PSR metric
is positive definite is in general larger than the K\"ahler cone. 
Therefore it is important to keep track of K\"ahler cone of 
the underlying Calabi-Yau manifold when working with an effective 
supergravity theory. For example, in \cite{Kallosh:2000rn,Mayer:2003zk}
it has been shown
that naked singularities which are naively present in some solutions 
of five-dimensional supergravity are unphysical if the theory is obtained
as a Calabi-Yau compactification of eleven-dimensional supergravity,
because at the singularity the scalar fields take values which are 
inside the domain where the PSR metric is positive definite, but 
outside the K\"ahler cone of the underlying Calabi-Yau manifold. 
If the theory is considered as embedded into M-theory 
one needs to modify the effective Lagrangian when the boundary 
of the K\"ahler cone is reached, even though all data in the Lagrangian,
and solutions with space or time dependent moduli, remain smooth
at this point. The modification of the Lagrangian corresponds to 
continuing 
into the K\"ahler cone of another Calabi-Yau manifold, which differs
from the original one by a transition which changes the topology. At the boundary 
of the K\"ahler cone additional massless vector or hypermultiplets are present.
Integrating out these multiplets induces threshold
corrections to the couplings in the effective Lagrangian for the remaining
modes, which for five-dimensional vector multiplets induce finite
shifts of the coefficients of the Hesse potential 
\cite{Witten:1996qb,Kallosh:2000rn,Mayer:2003zk}. The proper treatment of this
subtlety removes naked singularities which are naively present
in domain and black hole solutions with non-constant scalars.
In this sense, the K\"ahler cone acts as a cosmic censor.

So far we have been discussing the moduli space of Ricci-flat metrics on $X$. 
In supergravity and string compactifications, there are additional moduli
resulting from the dimensional reduction of various $p$-form fields. Massless scalar
fields arise whenever the components of such a $p$-form along $X$ are harmonic
forms on $X$. The number of massless scalars is given by the corresponding Hodge number. 
Such massless scalars are moduli, unless the effective theory contains a potential
for them, which is not the case for Calabi-Yau compactifications in the absence of
fluxes.  A particular role is played by the Kalb-Ramond two-form field $B$ of string theory. 
When reducing a type-II string theory on a Calabi-Yau three-fold, the $B$-field
gives rise to $h^{1,1}$ real moduli, which naturally combine with the moduli of 
the K\"ahler structure. Defining a complexified K\"ahler form and expanding 
in the basis $\omega_A$ of $H^{1,1}(X,\mathbb{R})$, 
\begin{equation}
\omega_\mathbb{C} = B + i \omega = z^A \omega_A \;,\;\;\;
z^A \in \mathbb{C} \;,\;\;A=1,\ldots h^{1,1}\;,
\end{equation}
we obtain complex coordinates $z^A$ on the moduli space
${\cal M}_{\mathrm{Kahler}}$ of {\em complexified K\"ahler structures}. 
This space turns out to be a K\"ahler manifold with a K\"ahler potential that is
obtained from the Hesse potential $\log V$, as follows. 

Generally, given a Hessian manifold $N$ of dimension $n$ 
with local coordinates $t^A$ and Hesse
potential $H$, we can extend this to a complex manifold $M \cong \mathbb{R}^n \times N$, with 
coordinates $z^A= s^A + i t^A$. The Hessian metric on $N$ can be extended to a 
K\"ahler metric on $M$ with  K\"ahler potential 
\begin{equation}
K(z,\bar{z}) = H (\mbox{Im}(z)) \;.
\end{equation}
This defines a K\"ahler metric with metric coefficients\footnote{We do not correct
for the factor 4, which comes from the Jacobian, so that the two metrics 
differ by a constant factor. This is does not matter here, since we only want
to illustrate the principle. In applications the
normalization is fixed by the Lagrangian of the explicit model one considers. }
\begin{equation}
g_{A\bar{B}} = \frac{\partial^2 K}{\partial z^A \partial \bar{z}^{\bar{B}}} 
= 4 \frac{\partial^2 H}{\partial t^A \partial t^B} \;,
\end{equation}
which has an isometry group which contains the $n$ commuting shifts 
$z^A \mapsto z^A + r^A$, $r^A \in \mathbb{R}$. 

In the case at hand, we can use the Hesse potential $-\log V$ of a 0-conical
Hessian metric on ${\cal N}_{\mathrm{Kahler}}$ 
as a K\"ahler potential for a K\"ahler metric ${\cal M}_{\mathrm{Kahler}}$.\footnote{The
minus sign is introduced for consistency with the literature.}  
To see that this metric is actually a PSK metric, we introduce projective
coordinates $X^I$, $I=0, \ldots h^{1,1}$ on ${\cal M}_{\mathrm{Kahler}}$
by choosing local holomorphic functions $X^I(z)$ such that 
$X^A/X^0 = z^A$. 
Then we define the holomorphic function, homogeneous of degree two,
\begin{equation}
\label{PrepoCY3}
F = \frac{1}{3!} \frac{C_{ABC} X^A X^B X^C}{X^0} \;.
\end{equation}
It is straightforward to see that
\begin{eqnarray}
&& -i \left( X^I \bar{F}_I - F_I \bar{X}^I \right) \nonumber \\
&=&
-\frac{i}{3!}|X^0|^2  C_{ABC} (z^A -\bar{z}^{\bar{A}})
(z^B-\bar{z}^{\bar{B}})(z^C -\bar{z}^{\bar{C}}) = 8 |X^0|^2 V \;.
\end{eqnarray}
Therefore the K\"ahler potentials $-\log V$ and $-\log\left( - i (X^I\bar{F}_I - F_I \bar{X}^I) \right)$
differ by a K\"ahler transformation and define the same K\"ahler metric on 
${\cal M}_{\mathrm{Kahler}}$.\footnote{Note that the domains where the argument of the
logarithm is positive agree. Therefore both K\"ahler potentials are defined over the same
domain.}
Therefore the metric on ${\cal M}_{\mathrm{Kahler}}$ 
is a PSK metric with prepotential \eqref{PrepoCY3} and K\"ahler potential
\begin{equation}
K = -\log \left( -i (X^I \bar{F}_I - F_I \bar{X}^I )\right) \;.
\end{equation}
We remark that in string theory the `very special' cubic form of the 
prepotential only holds to leading order in perturbation theory and is subject 
to complicated corrections.

We conclude by indicating how the moduli of Calabi-Yau compactifications of 
eleven-dimensional supergravity and of type-II string theories are distributed among
five- and four-dimensional ${\cal N}=2$ supermultiplets. Moduli are either
allocated to vector multiplets, where the geometry of the target space is PSR or
PSK, or to hypermultiplets, where the geometry is quaternionic K\"ahler, 
denoted QK in the tables. The dimension of a quaternionic K\"ahler manifold
is divisible by four. The maximal dimension of a K\"ahler submanifold of a QK manifold
is half of the total dimension \cite{Alekseevsky:2001}. Hypermultiplets contain a mixture of 
moduli of the metric, moduli resulting from reducing $p$-form gauge fields, and, for
type II 
string theory, the dilaton and the axion obtained from dualizing the Kalb-Ramond
two-form. The PSK spaces ${\cal M}_{\mathrm{cplx}}$ and ${\cal M}_{\mathrm{Kahler}}$
are K\"ahler submanifolds of hypermultiplet target manifolds, at least to lowest
order in $\alpha'$.

\renewcommand{\arraystretch}{1.5}
\begin{table}
\begin{tabular}{l|l|l|l}
\mbox{Theory} & Number of vector multiplets & Moduli space & Geometry \\ \hline
M-theory & $h^{1,1}-1$ & ${\cal N}_{\mathrm{Kahler}}^V$ & PSR \\
II A & $h^{1,1}$ & ${\cal M}_{\mathrm{Kahler}}$ & PSK \\
II B & $h^{2,1}$ & ${\cal M}_{\mathrm{cplx}}$ & PSK \\
\end{tabular}
\caption{This table shows which moduli of a Calabi-Yau compactification 
sit in vector multiplets. PSR=projective special real, PSK = projective special
K\"ahler \label{fig:VM}}
\end{table}

\begin{table}
\begin{tabular}{l|l|l|l}
\mbox{Theory} & Number of hypermultiplets & Moduli space & Geometry \\ \hline
M-theory/IIA & $h^{2,1}+1$ & ${\cal M}_{\mathrm{cplx}} \subset {\cal M}_{HM}$ & PSK $\subset$ QK \\
II B & $h^{1,1}+1$ & ${\cal M}_{\mathrm{Kahler}} \subset {\cal M}_{HM}$ &
PSK $\subset$ QK \\
\end{tabular}
\caption{This table shows which moduli of a Calabi-Yau compactification sit in 
hypermultiplets. PSK = projective special K\"ahler, QK = quaternionic K\"ahler.
\label{fig:HM}}
\end{table}

Table \ref{fig:VM} lists vector multiplet 
moduli, Table \ref{fig:HM} lists hypermultiplet moduli. 
In compactifications from eleven to five dimensions, the moduli of the 
real K\"ahler form split: the volume modulus sits in a hypermultiplet,
the remaining K\"ahler moduli parametrizing the fixed volume hypersurface
in the K\"ahler cone
sit in vector multiplets. Note that this split is required in order 
to obtain a PSR manifold. The volume modulus
and the complex structure modulus sit in hypermultiplets together with
moduli coming from reducing $p$-form gauge fields. In compactifications
from ten to four dimensions the moduli of the K\"ahler form and those of 
the Kalb-Ramond $B$-field combine into complex moduli. Depending
on whether one considers the IIA or IIB theory, either the moduli of the complexified K\"ahler 
form or the moduli of the complex structure moduli sit in vector multiplets.
The remaining moduli of the metric 
sit in hypermultiplets together with the dilaton, 
the axion obtained by dualizing the Kalb-Ramond field, and moduli associated
to $p$-form gauge fields in the Ramond-Ramond sector.

\section{Four-dimensional vector multiplets \label{4dvecmult}}

\subsection{Rigid vector multiplets}

The field content of a four-dimensional rigid abelian vector multiplet is $(X, \Omega_i, A_{\mu}, Y_{ij})$ \cite{Fayet:1975yi}.
$X$ denotes a complex scalar field;
$A_{\mu}$ denotes an abelian gauge field with field strength $F=dA$; $\Omega_i$ denotes an $SU(2)_R$ doublet of chiral
fermions; $ Y_{ij}$ denotes an $SU(2)_R$ triplet of scalar fields, i.e. $Y_{ij}$ is a symmetric
matrix satisfying the reality condition
\begin{equation}
Y_{ij} = \varepsilon_{ik} \, \varepsilon_{jl} \, Y^{kl} \;\;\;,\;\;\; Y^{ij} = (Y_{ij} )^* \;.
\end{equation}
Thus, off-shell, an abelian vector multiplet has
eight bosonic and eight fermionic real degrees of freedom.

We are interested in the Lagrangian describing the 
dynamics of $n$ abelian vector multiplets. These vector multiplets will be labelled by an index $I = 1, \dots, n$.
The Lagrangian  is encoded \cite{deWit:1984wbb} in a holomorphic function
$F(X)$, called the prepotential.  We denote holomorphic derivatives of $F(X)$ with respect to $X^I$ by 
$F_I = \partial F / \partial X^I, \; F_{IJ}  = \partial^2 F / \partial X^I \partial X^J$, etc.
We denote the complex conjugate of $X^I$ by ${\bar X}^I$, and
anti-holomorphic 
derivatives of ${\bar F} ({\bar X})$ by ${\bar F}_I = \partial {\bar F}/\partial
{\bar X}^I$, etc.

The bosonic part of the Lagrangian reads
\begin{eqnarray}
L = - N_{IJ} \, \partial_{\mu} X^I \, \partial^{\mu} {\bar X}^J + \left(\tfrac14 i \, F_{IJ} \, 
F_{\mu \nu}^{-I} F^{\mu \nu -J} - \tfrac18 i  \, F_{IJ} \, Y^I_{ij} Y^{J \, ij} 
+ {\rm h.c.} \right) \;,
\label{actbosrig}
\end{eqnarray}
where
\begin{equation}
F_{\mu \nu}^{I} = 2 \partial_{[\mu} A^I_{ \nu]} \;,
\end{equation}
and where $N_{IJ}$ is given by \eqref{NIJdef}.
Note that the 
kinetic terms for the scalar fields and for the abelian gauge fields are 
determined in terms of $N_{IJ}$,
\begin{equation}
L_{\rm kin}=  - N_{IJ} \, \partial_{\mu} X^I \, \partial^{\mu} {\bar X}^J - \tfrac18  \, N_{IJ} \, 
F_{\mu \nu}^{I} F^{\mu \nu J} \;.
\end{equation}
The kinetic term for the scalar fields describes a sigma-model, whose target space is an affine special
K\"ahler (ASK) manifold. This is a Riemannian manifold with K\"ahler metric 
$N_{IJ} = \partial^2 K(X, \bar X)/\partial X^I \partial {\bar X}^J$ and K\"ahler potential \eqref{kaehmetg}.

As discussed in subsection \ref{sect:intrinsic_ASK}, the metric $g$ of an ASK manifold, when expressed in terms of 
 special real coordinates $q^a= (x^I, y_I) = ( {\rm Re} \, X^I ,  {\rm Re} \, F_I )$, is Hessian,
 \begin{equation}
g = N_{IJ} \, dX^I d {\bar X}^J = H_{ab} dq^a dq^b \;\;\;,\;\; a, b = 1, \dots, 2n \;,
\end{equation}
where $H_{ab} = \partial^2 H /\partial q^a \partial q^b$ is determined in terms of the real Hesse potential
$H$. The Hesse potential $H$ is related to the prepotential $F$ by Legendre transformation, c.f. \eqref{HFleg}.
As in subsection \ref{sect:dual_coord}, 
we decompose $(X^I, F_I)$ into real and imaginary parts, 
\begin{eqnarray}
X^I &=& x^I + i u^I \;, \nonumber\\
F_I &=& y_I + i v_I \;.
\end{eqnarray}
Next, 
we perform the Legendre transform of the imaginary part of $F$ with respect to $u^I$, thereby
replacing $u^I$ by $y_I$ as independent variables,
\begin{equation}
H(x,y) = 2 \, {\rm Im} \,  F(x + i u) - 2 \, y_I \, u^I  \;,
\label{hess1}
\end{equation}
where
\begin{equation}
\frac{\partial  {\rm Im} \,  F  }{\partial u^I} = y_I \;.
\end{equation}
The latter expresses $u$ as a function of $(x,y)$, locally, and inserting this expression
on the right hand side of \eqref{hess1} yields $H(x,y)$.

\subsection{Rigid superconformal vector multiplets}

Next, we specialize  to the case where the vector multiplet theory is superconformal.
This implies that $F(X)$ must be 
homogeneous of degree $2$
under complex scalings,
\begin{equation}
F(\lambda X) = \lambda^2 \, F(X) \;\;\;,\;\;\; \lambda \in \mathbb{C}^*  \;,
\end{equation}
from which one infers the relations \eqref{homprep}. The associated Hesse potential is homogeneous of degree $2$,
and the scalar manifold is a conical affine special K\"ahler manifold.

\subsection{Superconformal matter multiplets coupled to conformal supergravity}

As in the five-dimensional case, 
we will follow the superconformal approach to construct a theory 
of $n$ abelian vector multiplets coupled to Poincar\'e supergravity.
This is based on the fact that a theory of $n$ vector multiplets and
$n_H$ hypermultiplets coupled to
Poincar\'e supergravity is {\em gauge equivalent} to a theory of $n+1$ 
superconformal vector multiplets and $n_H+1$ superconformal hypermultiplets
coupled to conformal supergravity.

\subsubsection{Coupling of vector multiplets}

First, we consider the coupling of $n+ 1$ abelian vector multiplets to conformal
supergravity at the two-derivative level.  The index $I$ labelling these abelian vector multiplets
now runs over $I=0, 1, \dots, n$.  The component fields of the abelian vector multiplets
carry the Weyl and chiral weights given in Table \ref{vechyp}. Then, using \eqref{covphi}, we have
\begin{equation}
{\cal D}_{\mu}ÊX^I= \left( \partial_{\mu} -  \, b_{\mu} + i  \, A_{\mu} \right) X^I \;.
\label{covX}
\end{equation}

The bosonic part of the Lagrangian describing the coupling of abelian vector multiplets to conformal
supergravity 
reads,
\begin{eqnarray}
L &=&  \Big[
 i {\cal D}^{\mu} F_I \, {\cal D}_{\mu} \bar X^I   - i F_I\,\bar X^I 
 (- \ft16  R - D) 
-\ft18i  F_{IJ}\, Y^I_{ij} Y^{Jij} 
 \nonumber\\
&&+\ft14 i F_{IJ} (F^{-I}_{\mu \nu} -\ft 12 \bar X^I 
T_{\mu\nu}^{-})(F^{\mu\nu \, - J} -\ft12 \bar X^J 
T^{\mu\nu-})  \nonumber\\
&&-\ft14 i F_I(F^{+I}_{\mu\nu} -\ft12  X^I 
T_{\mu\nu}^+) T^{\mu\nu +} 
-\ft1{8} iF \,T_{\mu\nu}^+ T^{\mu\nu+} + {\rm h.c.} \Big] \;.
\end{eqnarray}
This equals
\begin{eqnarray}
\label{efflag}
L = - N_{IJ}
 {\cal D}^{\mu} X^I \, {\cal D}_{\mu} \bar X^J   - i \left( F_I\,\bar X^I - X^I {\bar F}_I\right)
 (- \ft16  R - D) 
+\ft18 \, N_{IJ}\, Y^I_{ij} Y^{Jij} \nonumber\\
+ \left( - \ft14 i {\bar F}_{IJ} F^{+I}_{\mu\nu} F^{\mu\nu \, +J } 
- \tfrac{1}{16} N_{IJ} X^I X^J T_{\mu\nu}^+ T^{\mu\nu +} 
+ \ft14 \, N_{IJ} X^I \, F^{+J}_{\mu\nu} 
T^{\mu\nu\, +} + {\rm h.c.} \right) \;. \nonumber\\
\end{eqnarray}

\subsubsection{Coupling of hypermultiplets \label{sec:hyper}}
We consider the coupling of $r = n_H + 1$
hypermultiplets that are neutral with
respect to the gauge symmetries of the vector multiplets. We  
follow the presentation given in \cite{deWit:1999fp},
which 
is based on sections $A_i{}^{\alpha}(\phi)$
 of an ${\rm Sp} (r) \times {\rm Sp} (1)$ bundle ($\alpha = 1, \dots, 2r; i = 1,2$) 
which depend on scalar
fields $\phi^A$, defined in the context of a so-called 
hyper-K\"ahler cone of dimension  $4r$.

The bosonic part of the Lagrangian describing the coupling of hypermultiplets to 
conformal supergravity is given by
\begin{equation}
- \ft12 \varepsilon^{ij}\, \bar \Omega_{\alpha \beta} \,
{\cal D}_{\mu} A_i{}^{\alpha} \,{\cal D}^{\mu}  A_j{}^{\beta} +\chi (- \ft16 R+  \ft12 D) \;,
\label{laghyper}
\end{equation}
where
 the hyper-K\"ahler potential  $\chi$ and the covariant
derivative ${\cal D}_{\mu} A_i{}^{\alpha}(\phi)$ are given in \eqref{Dhyp4d}.

\subsubsection{Poincar\'e supergravity}

Combining the bosonic Lagrangians \eqref{efflag} and \eqref{laghyper}, we obtain
\begin{eqnarray}
\label{efflagtotal}
L &=&  \left[ i \left( \bar X^I \, F_I  - X^I {\bar F}_I\right) - \chi \right] \tfrac16 \, R
+  \left[ i \left( \bar X^I \, F_I  - X^I {\bar F}_I\right) + \tfrac12 \chi \right] D \nonumber\\
&& - N_{IJ}
 {\cal D}^{\mu} X^I \, {\cal D}_{\mu} \bar X^J   
 +\ft18 \, N_{IJ}\, Y^I_{ij} Y^{Jij} \\
&&+ \left( - \ft14 i {\bar F}_{IJ} F^{+I}_{\mu\nu} F^{\mu\nu \, +J } 
- \tfrac{1}{16} N_{IJ} X^I X^J T_{\mu\nu}^+ T^{\mu\nu +}  \right. \nonumber\\
 && \left. \qquad + \ft14 \, N_{IJ} X^I \, F^{+J}_{\mu\nu} 
T^{\mu\nu \, +} + {\rm h.c.} \right) 
 - \ft12 \varepsilon^{ij}\, \bar \Omega_{\alpha \beta} \,
{\cal D}_{\mu} A_i{}^{\alpha} \,{\cal D}^{\mu}  A_j{}^{\beta} \;. \nonumber
\end{eqnarray}
Note that the field $D$ does not have a kinetic term: it appears as a multiplier.
Its field equation yields the condition
\begin{equation}
\chi = -2  i \left( \bar X^I \, F_I - X^I {\bar F}_I\right) \;.
\label{D}
\end{equation}
Similarly, the field equation for $Y_{ij}$ is simply
\begin{equation} 
Y_{ij}Ê= 0 \;.
\label{Yij}
\end{equation}
Inserting \eqref{D} and \eqref{Yij} into \eqref{efflagtotal} yields
\begin{eqnarray}
\label{efflagtotal_DY}
L &=&   i \left( \bar X^I \, F_I  - X^I {\bar F}_I\right)  \tfrac12 \, R
 - N_{IJ}
 {\cal D}^{\mu} X^I \, {\cal D}_{\mu} \bar X^J   
  \nonumber\\
&&+ \left( - \ft14 i {\bar F}_{IJ} F^{+I}_{\mu\nu} F^{\mu\nu \, +J } 
- \tfrac{1}{16} N_{IJ} X^I X^J T_{\mu\nu}^+ T^{\mu\nu +} \right. \\
&& \left. \qquad + \ft14 \, N_{IJ} X^I \, F^{+J}_{\mu\nu} 
T^{\mu\nu \, +} + {\rm h.c.} \right) 
 - \ft12 \varepsilon^{ij}\, \bar \Omega_{\alpha \beta} \,
{\cal D}_{\mu} A_i{}^{\alpha} \,{\cal D}^{\mu}  A_j{}^{\beta} \;. \nonumber
\end{eqnarray}
Next, we use the symmetries of conformal supergravity to impose gauge conditions.
We begin by fixing the freedom to perform dilations (whose generator is $D$, see Table \ref{conf_gen_fie_N2}),
 by picking
\begin{equation}
 i \left( \bar X^I \, F_I  - X^I {\bar F}_I\right) = \kappa^{-2} \;\;\;,\;\;\; {\rm D-gauge} \;.
 \label{D-gauge}
 \end{equation}
This is the so-called D-gauge. Here $\kappa^{2} = 8 \pi G_N$, where $G_N$  denotes the Newton's constant. 
With this choice, we obtain the Einstein-Hilbert term \eqref{EH_action}.
In the following, we set $\kappa^2 = 1$.
Note that with the choice \eqref{D-gauge}, we obtain
\begin{equation}
\chi = -2 \;,
\label{chi_hyper}
\end{equation}
which shows that at least one hypermultiplet is needed in order to obtain the Einstein-Hilbert term \eqref{EH_action}.
The condition \eqref{chi_hyper} removes one real bosonic degree of freedom in the hypermultiplet sector.
Fixing the freedom under $SU(2)_R$ transformations (c.f. \eqref{Dhyp4d})
removes three additional real degrees of freedom
in the hypermultiplet sector, so that in total, we have removed four real degrees of freedom.
This amounts to removing the bosonic degrees of freedom of one hypermultiplet. There are
then $r-1 = n_H$ physical hypermultiplets left. We will not consider them any further, and hence we drop 
them in what follows.

Now we pick the K-gauge \eqref{b0K}, which removes the dilational connection $b_{\mu}$
from the covariant derivatives \eqref{covX} and \eqref{Dhyp4d}. Next, 
varying with respect to the $U(1)$ connection $A_{\mu}$ gives (c.f. \eqref{Chern_conn_ST})
\begin{equation}
A_{\mu} = \tfrac12 i \frac{N_{IJ} \left( (\partial_{\mu} X^I )\, {\bar X}^J - X^I \, \partial_{\mu}  {\bar X}^J 
\right)}{N_{KL} X^K {\bar X}^L} \;.
\label{A2der}
\end{equation}
In the D-gauge \eqref{D-gauge}, where $N_{KL} X^K {\bar X}^L = -1$, this becomes
\begin{eqnarray}
A_{\mu} &=& \tfrac12 i \, N_{IJ} \left(  X^I \, \partial_{\mu}  {\bar X}^J - (\partial_{\mu} X^I) \, {\bar X}^J
\right)_{ \vert_{N_{KL} X^K {\bar X}^L = -1}}
\;, \nonumber\\
&=& -  \tfrac12  i \left(\partial_a K \, \partial_{\mu} z^a - \partial_{\bar a} K
\partial_{\mu} {\bar z}^a \right)  \nonumber\\
&=&    A_a \, \partial_{\mu} z^a - A_{\bar a} 
\partial_{\mu} {\bar z}^a  \;,
\label{U1_fix}
\end{eqnarray}
where $z^a = X^a/X^0$ denote complex physical scalar fields ($a = 1, \dots, n)$,
and where $K(z, \bar z)$ denotes the K\"ahler potential given in 
\eqref{kaehpot}. This is in agreement with \eqref{eq:kah-conn}. 
The connection $A_{\mu}$ is the pull-back to space-time of the connection $A_a$ given in 
\eqref{connAa}.
Finally, varying with respect to $T_{\mu\nu}^+$ gives
\begin{equation}
T_{\mu\nu}^+= 2 \frac{N_{IJ} X^I}{N_{KL} X^K X^L} \, F_{\mu\nu}^{+ J} \;.
\label{T_2d}
\end{equation}
Thus, at the two-derivative level, the fields $A_{\mu}$ and $T_{\mu\nu}^{\pm}$ are auxiliary fields.

Inserting these various expressions into \eqref{efflagtotal_DY}, using the relation \eqref{eq:relgN}
and dropping terms that involve
physical hypermultiplets, we obtain the following gauge fixed Lagrangian,
\begin{eqnarray}
\label{efflag_poincare}
L  &=&    \tfrac12 \, R 
- g_{a \bar b} \, \partial_{\mu} z^a \,
\partial^{\mu} {\bar z}^b 
+ \left( - \ft14 i {\cal N}_{IJ} F^{+I}_{\mu\nu} F^{\mu\nu \, +J } 
+ {\rm h.c.} \right)  \nonumber\\
&=& \ft12  R - g_{a \bar b} \, \partial_{\mu} z^a \,
\partial^{\mu} {\bar z}^b 
+ \tfrac14 {\rm Im} {\cal N}_{IJ}
\, F_{\mu \nu}^I \, F^{\mu \nu J} - \tfrac{i}{4} {\rm Re} {\cal N}_{IJ}
\, F_{\mu \nu}^I \, {\tilde F}^{\mu \nu J} ,
\end{eqnarray}
where 
\begin{equation}
{\cal N}_{IJ}Ê= {\bar F}_{IJ} + i \frac{ N_{IP} X^P \, N_{J Q} X^Q}{N_{KL} X^K X^L} 
\label{calN-N}
\end{equation}
with  $N_{KL} X^K {\bar X}^L = -1$.  
The resulting Lagrangian
describes the
bosonic part of the action for vector multiplets coupled to Poincar\'e supergravity.
It is obtained from the action for matter multiplets coupled to conformal supergravity by
using two compensating multiplets: one vector multiplet, and one hypermultiplet.

\subsection{Coupling to a chiral background \label{sec:couchiralback}}

The construction of the action \eqref{efflagtotal} describing the coupling of abelian vector multiplets and 
neutral hypermultiplets to conformal supergravity at the two-derivative level can be extended, within
the superconformal approach,
to allow for the presence of a chiral background field \cite{deWit:1996gjy}. This is achieved by allowing the function 
$F(X)$ that enters in the construction of \eqref{efflagtotal}, to depend on an additional holomorphic
field $\hat A$, so that now $F(X, \hat A)$. The background field $\hat A$ is introduced as the lowest component of a chiral supermultiplet. 
Compatibility with 
superconformal symmetry determines the scaling behaviour
of the chiral multiplet, while insisting on a local supersymmetric
action implies that the dependence on the chiral multiplet is holomorphic.
Therefore, the function $F$ has to be (graded) homogeneous of degree two, that is
\begin{equation}
F(\lambda X, \lambda^w \hat A) = \lambda^2 F(X,\hat A) \;,\;\;\;
\lambda \in \mathbb{C}^* \;,
\label{Fback}
\end{equation}
where $w$ is the weight of $\hat A$ under scale transformations.
It follows that $F$
satisfies the relation,
\begin{equation}
X^I F_I + w \hat A \,F_A = 2 F\,.
\end{equation}
Here $F_I$ and $F_A$ denote the derivatives of $F(X,\hat A)$ with
respect to $X^I$ and $\hat A$, respectively.

We denote the component fields of the chiral background 
superfield
with a caret.  We focus on the bosonic component fields, which we
denote by
 ${\hat A}$,
${\hat B}_{ij}$, ${\hat F}_{ab}^-$ and by ${\hat C}$.  Here ${\hat A}$
and ${\hat C}$ denote complex scalar fields, appearing at the
$\theta^0$- and $\theta^4$-level of the chiral background superfield,
respectively, while the symmetric complex SU(2) tensor ${\hat B}_{ij}$
and the anti-selfdual Lorentz tensor ${\hat F}_{ab}^-$ reside at the
$\theta^2$-level.  

In the presence of the chiral background, the action \eqref{efflagtotal} becomes encoded
in $F(X,\hat A)$, and reads as follows,
\begin{eqnarray}
\label{action_back}
L  &=&  \Big[
 i {\cal D}^{\mu} F_I \, {\cal D}_{\mu} \bar X^I   - i F_I\,\bar X^I 
 (- \ft16  R - D) 
-\ft18i  F_{IJ}\, Y^I_{ij} Y^{Jij} - \ft14 i \hat 
B_{ij}\,F_{{ A}I}  Y^{Iij}   \nonumber\\
&&+\ft14 i F_{IJ} (F^{-I}_{ab} -\ft 12 \bar X^I 
T_{ab}^{-})(F^{ab -J} -\ft12 \bar X^J 
T^{ab-})  \nonumber\\
&&-\ft18 i F_I(F^{+I}_{ab} -\ft12  X^I 
T_{ab}^+ ) T^{ab +}
+\ft12 i \hat F^{-ab}\, F_{{ A}I} (F^{-I}_{ab} - \ft12  \bar X^I 
T_{ab}^{-} )   \nonumber \\
&&+\ft12 i F_{A}
\hat C -\ft18 i F_{ A A}(\varepsilon^{ik}
\varepsilon^{jl}  \hat B_{ij} 
\hat B_{kl} -2 \hat F^-_{ab}\hat F^{-ab}) 
-\ft1{8} iF \,T_{ab}^+ T^{ab+} + {\rm h.c.} \Big] \nonumber \\ 
&& - \ft12 \varepsilon^{ij}\, \bar \Omega_{\alpha \beta} \,
{\cal D}_{\mu} A_i{}^{\alpha} \,{\cal D}^{\mu}  A_j{}^{\beta} +\chi (- \ft16 R+  \ft12 D) \;.
\end{eqnarray}
The last line pertains to the hypermultiplets, as discussed in subsection \ref{sec:hyper}.

\subsubsection{Coupling to $R^2$ terms}

When identifying the chiral background superfield with the square of the Weyl superfield,
the action \eqref{action_back} will contain
higher-derivative curvature terms proportional to
the square of the Weyl tensor. In this case the chiral weight $w$ in \eqref{Fback} equals $w=2$,
and the bosonic fields of the chiral background superfield becomes identified with
\begin{eqnarray}
\label{background-def}
\hat A   &=& 4\,  (T^{-}_{ab})^2\,, \\
\hat B_{ij}  &=& -32 \,\varepsilon_{k(i}R({\cal V})^k{}_{j)ab} \,
T^{ab -}
  \,,\nonumber\\
\hat F^{-ab}  &=& 32 \,{\cal R}(M)_{cd}{}^{\!ab} \,
T^{cd -}  
 \,,\nonumber\\
\hat C &=&  64\, {\cal R}(M)^{-cd}{}_{\!ab}\, {\cal 
R}(M)^-_{cd}{}^{\!ab}  + 32\, R({\cal V})^{-ab\,k}{}_l^{~} \, 
R({\cal V})^-_{ab}{}^{\!l}{}_k^{~} - 64\, T^{ab -} \, D_a \,D^cT_{cb}^+    \,.   \nonumber
\end{eqnarray}
In these expressions, we have suppressed all terms that involve fermionic fields.
The curvatures appearing in \eqref{background-def} are given by 
\beqa
R({\cal V})_{\mu\nu}\,^i{}_j &=& 2\partial_{[\mu} {\cal V}_{\nu ]}^i{}_j 
     + {\cal V}_{[\mu }^i{}_k{\cal V}_{\nu ]}^k{}_j \nonumber\\
{\cal R}(M)_{ab}{}^{\!cd} & = &    R_{ab}\,^{cd} + 8f_{[a}{}^{[c} \delta_{b]}{}^{d]} - \ft18\Big( 
T^{cd+}\,T_{ab}^- + T^{+}_{ab}\, T^{cd-}  \Big)\,,
\label{RT2}
\eeqa
where we recall that $R_{ab}\,^{cd} $ is computed using the spin connection \eqref{om-b-f}.
Note that the $T^2$-modification in \eqref{RT2} exactly cancels 
 the $T^2$-terms contained in 
$f_{\mu}{}^a$, as can be verified by using the relation \eqref{f-FP4},
\beqa
{\cal R}(M)_{ab}{}^{\!cd}  =     C_{ab}\,^{cd} - D \, \delta_{[a}{}^{[c} \delta_{b]}{}^{d]} - 2i \,
{\tilde R}_{[a}{}^{[c} (T) \,
\delta_{b]}{}^{d]} \;,
\eeqa
where 
\beq
C_{ab}{}^{cd} = R_{ab}{}^{cd} - 2 \left( R_{[a}{}^{[c} - \tfrac16 R \, \delta_{[a}{}^{[c} \right) \delta_{b]}{}^{d]}
\;.
\eeq
In the K-gauge \eqref{b0K}, $C_{ab}{}^{cd}$ denotes the Weyl tensor, and $\hat{C}$ includes a term
proportional to the square of the anti-selfdual part of the Weyl tensor,
\begin{equation}
{\hat C} = 64\, C^{-cd}{}_{\!ab}\, C^-_{cd}{}^{\!ab}  + \dots
\end{equation}
The term  $T^{ab -} \, D_a \,D^cT_{cb}^+ $ in \eqref{background-def}
 is written out in \eqref{TboxT}.

Observe that the $U(1)$ connection $A_{\mu}$ and the 
field $T_{ab}^-$ cannot any longer be eliminated in closed form, as in \eqref{A2der}
as in \eqref{T_2d} at the two-derivative level, but only iteratively.  In particular, 
$T_{ab}^-$
can be eliminated iteratively by means of an expansion of $F(X, \hat A)$ in powers of $\hat A$,
\begin{equation}
F(X, \hat A) = \sum_{n=0}^{\infty} F^{(n)} (X) \, {\hat A}^n \;,
\end{equation}
which generates an expansion with infinitely many higher-derivative
terms that are all proportional to ${\hat C}$. This results in an action that contains
infinitely many higher-derivative terms that are proportional 
to the square of the anti-selfdual part of the Weyl tensor.
Such an action is naturally interpreted as a Wilsonian
effective action.

\section{Hessian geometry in the presence of a chiral background \label{sec:hess}}

In this section, we discuss the geometric meaning of deformations 
of the 
prepotential function $F^{(0)} (X)$ by chiral background fields, such as in \eqref{Fback}.
We begin by considering holomorphic deformations of $F^{(0)} (X)$.
We use the description of
affine special K\"ahler manifolds as immersions, to 
introduce the notion 
of deformed affine special K\"ahler manifolds \cite{Cardoso:2015qhq}. We then discuss the existence of a Hessian structure
on these deformed manifolds, and relate the Hessian structure to the holomorphic 
anomaly equation for a hierarchy of symplectic functions. 

Subsequently, we turn to non-holomorphic deformations of $F^{(0)} (X)$.
We follow \cite{Cardoso:2015qhq}.

\subsection{Holomorphic deformation of the immersion \label{sec:holo-def-prep}}

We deform the 
prepotential $F^{(0)} (X)$ by allowing for the presence of a complex deformation parameter $\Upsilon$.
The prepotential $F^{(0)} (X)$ gets replaced by
the generalized prepotential 
$F(X,\Upsilon)$, which is holomorphic in $X^I$ and $\Upsilon$.

\subsubsection{Holomorphic family of immersions }

The geometric model for the deformation parametrized by 
$\Upsilon$ is a map \cite{Cardoso:2015qhq}
\begin{equation}
\label{phi_holo}
\phi: \hat{M} := M \times \mathbb{C} \rightarrow V \;,\;\;\;
(X^I, \Upsilon) \mapsto (X^I, F_I (X, \Upsilon)) \;,
\end{equation}
which can be interpreted as a holomorphic family of immersions
$\phi_\Upsilon :  M \rightarrow V \;,\;\; (X^I) \mapsto (X^I,F_I(X,\Upsilon))$,
that define a family of affine special K\"ahler structures on $M$.

Next, we define a metric and a two-form on $\hat{M}=M\times\mathbb{C}$
by pulling back the canonical Hermitian form $\gamma_V$ given in \eqref{hermV}, 
\begin{equation}
\gamma = \phi^* \gamma_V = g + i \omega = 
N_{IJ} dX^I \otimes d\bar{X}^J + i \bar{F}_{I\Upsilon}
dX^I \otimes d\bar{\Upsilon} - i F_{I\Upsilon} d\Upsilon \otimes
d\bar{X}^I \;,
\end{equation}
where $N_{IJ}= -i (F_{IJ} - {\bar F}_{IJ})$, and where $F_{I \Upsilon} = \partial_I \partial_{\Upsilon} F$. 
We assume that $\gamma$ is non-degenerate.
Denoting the holomorphic coordinates on $\hat{M}$ by
$(v^A) = (X^I,\Upsilon)$,
we obtain for the 
metric on $\hat{M}$,
\begin{equation}
g = g_{A \bar B} \, d v^A d\bar{v}^{{B}} =
N_{IJ} dX^I d\bar{X}^J 
+ i \bar{F}_{I\Upsilon} dX^I d\bar{\Upsilon}
- i F_{J \Upsilon} d \Upsilon d \bar{X}^J \;,
\label{kahgups}
\end{equation}
which is a K\"ahler metric  $g_{A \bar B} =  \partial_A \partial_{\bar B}ÊK$
with 
K\"ahler potential 
\begin{equation}
K = - i \left( \bar{X}^I F_I(X,\Upsilon) -  X^I \, \bar{F}_I(\bar{X}, \bar{\Upsilon})  \right) \;.
\label{K_ups}
\end{equation}
The associated K\"ahler form is
\begin{equation}
\omega = -\frac{i}{2} N_{IJ} dX^I \wedge d\bar{X}^J + \frac{1}{2}
\bar{F}_{I\Upsilon} dX^I \wedge d\bar{\Upsilon} - \frac{1}{2}
F_{I\Upsilon} d\Upsilon \wedge d \bar{X}^I \;.
\label{kahformups}
\end{equation}
The K\"ahler metric $g_{A \bar B}$
has occured in the deformed sigma model \cite{Cardoso:2012mc}, 
which provides a field theoretic realization of the set-up just described.

For latter use, we introduce the decomposition 
\begin{equation}
2 F_{IJ} = R_{IJ} + i N_{IJ} \;,
\end{equation}
where $R_{IJ} = 2 \, {\rm Re} \, F_{IJ}, \;  N_{IJ} = 2 \,  {\rm Im} \, { F}_{IJ} $.
We denote the inverse of $N_{IJ}$ by $N^{-1} = (N^{IJ})$.

\subsubsection{
The Hesse potential \label{sec:real-hess-pot}}

We now define special real
coordinates and a Hesse potential in presence of the deformation.
We then show that the K\"ahler metric $g$ on $\hat M$ given in \eqref{kahgups}
is no longer Hessian. There is, however, another metric on $\hat M$ that is Hessian.
We denote this metric by $g^H$. We show that $\hat M = M \times \mathbb{C} $ can be equipped
with a Hessian structure $(\nabla, g^H)$, where $g^H \neq g$.

We introduce real coordinates $(q^a) = (x^I, y_I)$ as in \eqref{XFxy}
\begin{equation}
X^I = x^I + i u^I(x,y,\Upsilon, \bar{\Upsilon}) \;,\;\;\;
F_I = y_I + i v_I(x,y, \Upsilon, \bar{\Upsilon}) \;.
\label{XFxyups}
\end{equation}
Then, 
the (generalized) Hesse potential is defined by a Legendre transform
of the 
generalized prepotential $F(X, \Upsilon)$,
\begin{equation}
H(x,y,\Upsilon, \bar{\Upsilon}) = 2 \, {\rm Im}Ê\, F(x + i u(x,y,\Upsilon,\bar{\Upsilon}), \Upsilon)
- 2 y_I u^I(x,y,\Upsilon,\bar{\Upsilon}) \;.
\label{hesseups}
\end{equation}
Note that $H$ is homogeneous of degree two.

We will be interested in the coordinate transformations
\begin{eqnarray}
(x,u,\Upsilon, \bar{\Upsilon}) &\mapsto&
(x,y, \Upsilon, \bar{\Upsilon}) \;, \nonumber\\
(x,y,\Upsilon,\bar{\Upsilon}) &\mapsto& 
(x,u, \Upsilon, \bar{\Upsilon}) \;.
\label{conv_coordinv}
\end{eqnarray}
To convert from one coordinate system to the other one, we use 
the following formulae when differentiating
a function $\tilde{f} (x, u,  \Upsilon, \bar{\Upsilon})= f (x, y(x, u, \Upsilon, \bar{\Upsilon}), \Upsilon, \bar{\Upsilon})$,
\begin{eqnarray}
\left. \frac{\partial \tilde{f}}{\partial x^I} \right|_{u} &=& 
\left. \frac{\partial f}{\partial x^I} \right|_{y} +
\left. \frac{\partial f}{\partial y_K} \right|_{x} 
\frac{\partial y_K}{\partial x^I}\;,  \nonumber\\
\left. \frac{\partial \tilde{f}}{\partial u^I} \right|_{x} &=&
\left. \frac{\partial f}{\partial y_K}  \right|_{x} 
\frac{\partial y_K}{\partial u^I} \;,  \nonumber\\
\left. \frac{\partial \tilde{f}}{\partial \Upsilon} \right|_{x,u} &=&
\left. \frac{\partial f}{\partial \Upsilon} \right|_{x,y} +
\left. \frac{\partial f}{\partial y_K} \right|_{x} 
\frac{\partial y_K}{\partial \Upsilon} \;.
\end{eqnarray}
We refer to \ref{jacob_conv}, where we have collected various formulae with
details on the conversion \eqref{conv_coordinv}.

The K\"ahler metric $g$ in \eqref{kahgups}, when
expressed in coordinates $(q^a, \Upsilon, \bar{\Upsilon})$, takes the 
form
\begin{equation}
g = \frac{\partial^2 H}{\partial q^a \partial q^b} dq^a dq^b
+ \frac{\partial^2 H}{\partial q^a \partial \Upsilon} dq^a d\Upsilon
+ \frac{\partial^2 H}{\partial q^a \partial \bar{\Upsilon}} dq^a 
d\bar{\Upsilon} \;,
\label{gwithups}
\end{equation}
where 
\begin{equation}
\left( \frac{\partial^2 H}{\partial q^a \partial q^b} \right)
= \left( \begin{array}{cc}
N + RN^{-1} R \; & \;  -2 RN^{-1} \\
- 2 N^{-1} R \; & \;  4 N^{-1} \\
\end{array} \right) \;,
\end{equation}
and 
\begin{eqnarray}
\frac{\partial^2 H}{\partial x^I \partial \Upsilon} &=&
2 \bar{F}_{IM} N^{MN} F_{N\Upsilon} \;,\;\;\;
\frac{\partial^2 H}{\partial x^I \partial \bar{\Upsilon}} =
2 {F}_{IM} N^{MN} \bar{F}_{N\Upsilon} \;,\;\;\; \nonumber\\
\frac{\partial^2 H}{\partial y_I \partial \Upsilon} &=&
-2 N^{IJ} F_{J\Upsilon} \;,\;\;\;
\frac{\partial^2 H}{\partial y_I \partial \bar{\Upsilon}} =
-2 N^{IJ} \bar{F}_{J\Upsilon} \;.
\label{metricceffH}
\end{eqnarray}
In the undeformed case ($\Upsilon =0$), the K\"ahler metric is also Hessian.
In the deformed case ($\Upsilon \neq 0$), this is not any longer the case. This can be seen as follows.

First, we note that $\hat{M}$
can be equipped with a Hessian structure $(\nabla, g^H)$.
This  requires the existence of a flat, torsion-free
connection $\nabla$, which can be constructed as follows.
For fixed $\Upsilon$, the map $\phi_\Upsilon : M \rightarrow V$ induces an
affine special K\"ahler structure,
with special connection $\nabla$ and $\nabla$-affine coordinates
$(q^a) = (x^I,y_I)$.
We can extend $\nabla$ to a
flat, torsion-free connection on $\hat{M}=M\times \mathbb{C}$ by
imposing 
\begin{equation}
\nabla dx^I = 0 \;,\;\;\;
\nabla dy_I= 0 \;,\;\;\;
\nabla d\Upsilon = 0 \;,\;\;\;
\nabla d\bar{\Upsilon} = 0 \;.
\label{nabdv}
\end{equation}
Since $M$ can be covered by special real coordinate systems, 
we may extend these relations to $\hat{M}$, providing it with the affine structure required
to define a flat, torsion-free connection on $\hat{M}$.

Now we define the metric $g^H$ to be the Hessian metric of the (generalized) Hesse potential
\eqref{hesseups}.
Upon computing its components 
explicitly,
we find that $g^H$ differs from the K\"ahler metric $g$ by
\begin{equation}
g^H - g =\partial^2 H_{|x,y} = 
\frac{\partial^2 H}{\partial \Upsilon \partial \Upsilon} d\Upsilon d\Upsilon
+ 2 \frac{\partial^2 H}{\partial \Upsilon \partial \bar{\Upsilon}} 
d\Upsilon d\bar{\Upsilon} 
+ \frac{\partial^2 H}{\partial \bar{\Upsilon} \partial \bar{\Upsilon}} 
d\bar{\Upsilon} d\bar{\Upsilon} \;,
\end{equation}
where
\begin{eqnarray}
\frac{\partial^2 H}{\partial \Upsilon \partial \bar{\Upsilon}} &=&
N^{IJ} F_{I\Upsilon} \bar{F}_{J\Upsilon} \;,\;\;\;
\frac{\partial^2 H}{\partial \Upsilon \partial \Upsilon} =
-i F_{\Upsilon \Upsilon} + N^{IJ} F_{I\Upsilon} F_{J\Upsilon} \;,\;\;\;
\nonumber\\
\frac{\partial^2 H}{\partial \bar{\Upsilon} \partial \bar{\Upsilon}} &=&
i \bar{F}_{\Upsilon \Upsilon} + N^{IJ} \bar{F}_{I\Upsilon} \bar{F}_{J\Upsilon} 
\;.
\label{metrigHuu}
\end{eqnarray}
We remark that these metric coefficients are symplectic 
functions \cite{deWit:1996gjy}, which is necessary 
in order that $g^H - g$ is a well defined tensor field
(which we know to be the case, because $g^H$ and $g$ are 
both metric tensors). We further remark that
\begin{equation}
2 H = K -2i  \Upsilon F_\Upsilon + 2i \bar{\Upsilon} \bar{F}_\Upsilon
\label{relKH}
\end{equation}
differs from the K\"ahler potential (\ref{K_ups}) 
by a K\"ahler transformation. Therefore $2H$, taken as a 
K\"ahler potential, defines the same K\"ahler metric 
$g=g^K$ 
as $K$. However, when taking $K$ as a Hesse potential
one does not get the Hessian metric $g^H$.
Note that the Hesse potential \eqref{relKH} is the sum of 
two symplectic functions, namely $K$ and ${\rm Im} \left( \Upsilon F_\Upsilon
\right)$, c.f. subsection \ref{sec:ubiquity}.

Thus, the K\"ahler metric $g$ on $\hat M$ is not Hessian with respect to the affine
structure that we have defined on $\hat{M}$ , i.e. $g \neq g^H$.

\subsubsection{Deformed affine special K\"ahler geometry}

Next we show
that $\hat M$ carries a deformed version of affine special K\"ahler geometry. Namely, we
show that
$(\hat{M} = M \times \mathbb{C}, J, g)$ is a K\"ahler 
manifold with K\"ahler form $\omega$, 
equipped with a flat, torsion-free connection $\nabla$ for which $\nabla \omega \neq 0$
and $d_\nabla J \neq 0$. The non-vanishing of $\nabla \omega$ and $d_\nabla J$
 is controlled by the symplectic function
$F_{\Upsilon}$.

We will call such manifolds deformed affine special K\"ahler manifolds.
Since our definition involves the map $\phi$ defined in \eqref{phi_holo}, this is not an intrinsic
definition, but the name for a specific construction.

We have already established that $g$ is a K\"ahler metric with 
K\"ahler form $\omega$, c.f. \eqref{kahgups} and \eqref{kahformups}.
To compare the latter with the two-form 
$2 dx^I \wedge dy_I$, which is the K\"ahler form on $M$,
we compute
\begin{eqnarray}
2 dx^I \wedge dy_I &=&
- \frac{i}{2} N_{IJ} dX^I \wedge d\bar{X}^J -
\frac{1}{2} F_{I\Upsilon} d \Upsilon \wedge d\bar{X}^I +
\frac{1}{2} \bar{F}_{I \Upsilon} dX^I \wedge d\bar {\Upsilon} 
\nonumber\\
&& +
\frac{1}{2} F_{I \Upsilon} dX^I \wedge d\Upsilon +
\frac{1}{2} \bar{F}_{I\Upsilon} d\bar{X}^I \wedge d \bar{\Upsilon} \;,
\label{xyold}
\end{eqnarray}
and therefore the K\"ahler form can be written as
\begin{equation}
\omega = 2 dx^I \wedge dy_I 
- \frac{1}{2} F_{I\Upsilon} dX^I \wedge d\Upsilon
- \frac{1}{2} \bar{F}_{I \Upsilon} d\bar{X}^I \wedge d\bar{\Upsilon} \;.
\end{equation}
This shows that $2 dx^I \wedge dy_I$, when considered
as a two-form on $\hat{M}$,
is not of type $(1,1)$ (since $\omega$ is, and both differ by pure forms). Using
the rewriting
\begin{equation}
F_{I \Upsilon} dY^I \wedge d \Upsilon =
d F_\Upsilon \wedge d \Upsilon = 
- d (\Upsilon d F_{\Upsilon}) \;,
\end{equation}
we find
\begin{equation}
\label{Diff_holo}
\omega = 2 dx^I \wedge dy_I  + \frac{1}{2} d(\Upsilon d F_\Upsilon) +
\frac{1}{2} d ( \bar{\Upsilon} d\bar{F}_{\Upsilon}) \;.
\end{equation}
Thus the difference between the K\"ahler forms
$\omega$ of $\hat{M}$ and $2dx^I \wedge dy_I$ of $M$ is 
exact. The deformation 
involves the function $F_\Upsilon=\partial_\Upsilon F$. The latter is 
a symplectic function, c.f. \eqref{tilFAFA}.
It contains all the information about the deformation.

Next we compute
\begin{equation}
\label{DeformOmega}
\nabla \omega = - \frac{1}{2} d (F_{I\Upsilon}) \otimes (dX^I \wedge
d \Upsilon) +  c.c. \;,
\end{equation}
and hence,  $\omega$ is not parallel. Thus, the connection $\nabla$
is not a symplectic connection on $\hat{M}$. This shows that while
$(\hat{M}, g, \omega, \nabla)$ is K\"ahler, it is not special 
K\"ahler. The deformation is controlled by an exact form, which 
is determined by the symplectic function $F_\Upsilon$.

Next, we show that the complex structure $J$ is not covariantly closed, i.e.
$d_{\nabla} J  \neq 0$.
To compute the exterior covariant derivative of the complex structure
$J$, we note that the vector fields $\partial_{x^I}, \partial_{y_I},
\partial_\Upsilon , \partial_{\bar{\Upsilon}}$ define a $\nabla$-parallel
frame which is dual to the $\nabla$-parallel co-frame 
$dx^I, dy_I, d\Upsilon, d\bar{\Upsilon}$. Using this one obtains
\begin{equation}
\nabla \frac{\partial}{\partial X^I} = 
\nabla \left( \frac{1}{2} \frac{\partial}{\partial x^I} +
\frac{1}{2} F_{IJ} \frac{\partial}{\partial y_J} \right) =
\frac{1}{2} d F_{IJ} \otimes \frac{\partial}{\partial y_J} \;.
\end{equation}
Using that $d_\nabla J = dJ^a e_a - J^a \wedge d_\nabla e_a$
where $e_a$ is any basis of sections of $T\hat{M}$, 
so that $d_\nabla e_a = \nabla e_a$, 
we find
\begin{equation}
d_\nabla J = \left( - i d X^I \wedge \frac{1}{2} d F_{IJ} 
+ c.c. \right) 
\otimes \frac{\partial}{\partial y_J}  \;.
\end{equation}
Note the rewriting
\[
dX^I \wedge dF_{IJ} = dX^I \wedge F_{IJ\Upsilon} d\Upsilon = 
- d(F_{IJ} d X^I ) = d(F_{I\Upsilon} d\Upsilon) \;,
\]
where we used symmetry of $F_{IJ}$ and the chain rule. 
Therefore
\begin{equation}
\label{DeformJ}
d_\nabla J = \left( -i d(F_{I\Upsilon} d\Upsilon) + c.c.\right)
\otimes \frac{\partial}{\partial y_I} =
\left( - i F_{IJ\Upsilon} dX^J \wedge d\Upsilon + c.c. \right)
\otimes \frac{\partial}{\partial y_I} \;,
\end{equation}
which is non-vanishing.
As a consistency check, observe that
$d_\nabla^2=0$, which must hold because $\nabla$ is flat.
Note that the non-vanishing of $d_\nabla J $ is expressed in terms of 
an exact form constructed out of the function $F_\Upsilon$.

In summary, $(\hat{M} = M \times \mathbb{C}, J, g)$ is a K\"ahler 
manifold with K\"ahler form $\omega$, 
equipped with a flat, torsion-free connection $\nabla$, with non-vanishing  $\nabla \omega$
and $d_\nabla J$ given by (\ref{DeformOmega}) and (\ref{DeformJ}).

For completeness we remark that the pullback of the complex 
symplectic form $\Omega$ of $V$ is non-vanishing,\footnote{Obviously,
$\hat{M}$ cannot be a (locally immersed) Lagrangian submanifold
of $V$ on dimensional grounds.}
\begin{equation}
\phi^* \Omega = F_{I\Upsilon} dX^I \wedge d\Upsilon =
- d(\Upsilon d F_\Upsilon) \;,
\end{equation}
where the right hand side is exact and controlled by $F_\Upsilon$.

\subsubsection{Holomorphic anomaly equation from the Hessian structure \label{hessianstr1}}

Next, we turn to the study of the 
integrability condition  for the existence of a Hesse
potential $H$ on $\hat{M}$, and we reinterpret it as a holomorphic anomaly
equation for a hierarchy of symplectic functions constructed from $F_{\Upsilon}$.

In \eqref{nabdv} we showed that $\hat{M}$ can be equipped with a Hessian
structure $(\nabla, g^H)$.
Then,  in $\nabla$-affine coordinates
$Q^a= (x^I, y_I, \Upsilon, \bar{\Upsilon})$, the 
totally symmetric covariant rank three tensor $S = \nabla g^H$
has components $S_{abc} = \partial_a g_{bc}$ which
satisfy the integrability condition
$ \partial_a g_{bc} =  \partial_b g_{ca} =  \partial_c g_{ab}$.
One particular integrability relation is
\begin{equation}
S_{x^I \Upsilon \Upsilon} = S_{\Upsilon x^I \Upsilon} \;,
\end{equation}
i.e.
\begin{equation}
\label{HA1}
\left. \partial_{x^I} g^H_{\Upsilon \Upsilon} \right|_y=  \left.
\partial_\Upsilon g^H_{x^I \Upsilon}  \right|_{x,y} \;,
\end{equation}
with metric components given by (c.f. \eqref{metricceffH} and  \eqref{metrigHuu})
\begin{eqnarray}
g^H_{x^I \Upsilon} &=& 2 \bar{F}_{IJ} N^{JK} F_{K\Upsilon} \;, \nonumber\\
g^H_{\Upsilon \Upsilon} &=& - iD_\Upsilon F_\Upsilon \;, 
\end{eqnarray}
where the derivative $D_\Upsilon$,
\begin{equation}
\label{D_Upsilon_holo}
D_\Upsilon = \left. \frac{\partial }{\partial \Upsilon}\right|_{X} 
+ i N^{IJ} F_{J\Upsilon} \frac{\partial }{\partial X^I} \;,
\end{equation}
 is the symplectic covariant
derivative that was introduced in \eqref{eq:cov-der-multiple}, and which takes the form
\eqref{D_Upsilon_holo} when acting on a holomorphic $F(X, \Upsilon)$.

We now evaluate equation (\ref{HA1}) 
in coordinates
$(X^I, \bar{X}^I, \Upsilon, \bar{\Upsilon})$, using the 
Jacobian \eqref{jacxuxyups}, to obtain
\begin{eqnarray}
\label{SconvS}
S_{x^I \Upsilon \Upsilon} 
&=& \left.  \frac{\partial g^H_{\Upsilon \Upsilon}}{\partial {x^I}}\right|_{y} = 
\left. \frac{\partial g^H_{\Upsilon \Upsilon}}{\partial {x^I}}\right|_{u}
+
 \frac{\partial g^H_{\Upsilon \Upsilon} }{
\partial u^K} \, \frac{\partial u^K}{\partial x^I} \;,\;\;\;\mbox{where}\;\;\;
\left. \frac{\partial}{\partial {x^I}}\right|_{u}
= \frac{\partial}{\partial X^I} + 
\frac{\partial}{\partial \bar{X}^I} \;, \nonumber\\
S_{\Upsilon x^I \Upsilon}  &=& \left. \frac{\partial g^H_{x^I \Upsilon}}{\partial
\Upsilon}\right|_{x,y} =  \left. \frac{\partial g^H_{x^I \Upsilon}}{\partial
\Upsilon}\right|_{x,u} +  \frac{\partial
g^H_{x^I \Upsilon}}{\partial u^K} \, \frac{\partial u^K}{\partial \Upsilon} \;.
\end{eqnarray}
We compute
\begin{eqnarray}
\left. \frac{\partial g^H_{\Upsilon \Upsilon} }{\partial {x^I}}\right|_{u}
& = & -i \frac{\partial}{\partial \bar{X}^I}
D_\Upsilon F_\Upsilon -i
{\bar F}_{\bar I}{}^{KL} F_{K \Upsilon} F_{L \Upsilon} \;, \nonumber\\
\left. \frac{\partial g^H_{\Upsilon \Upsilon} }{
\partial u^K}\right|_{x} &=&  \left(\frac{\partial}{\partial X^K} - \frac{\partial}{\partial {\bar X}^K} \right)
\left( F_{\Upsilon \Upsilon} + i N^{KL} F_{K \Upsilon} F_{L \Upsilon} \right) \nonumber\\
&=& F_{K \Upsilon \Upsilon} - F_K{}^{PQ} F_{P \Upsilon} F_{Q \Upsilon} - 2 N^{PQ} F_{KP \Upsilon} F_{Q \Upsilon}
- {\bar F}_K{}^{PQ} F_{P \Upsilon}  F_{Q \Upsilon} \;, \nonumber\\
\left. \frac{\partial g^H_{x^I \Upsilon}}{\partial
\Upsilon}\right|_{x,u} &=& - i F_{I \Upsilon \Upsilon} + F_{\Upsilon I}{}^J F_{J \Upsilon} + \left(F_{IJ} + {\bar F}_{IJ}
\right)
 \left(i F_{\Upsilon}{}^{JK} F_{K \Upsilon} + F_{\Upsilon \Upsilon}{}^J \right) \;,
\nonumber\\
\left. \frac{\partial
g^H_{x^I \Upsilon}}{\partial u^K}\right|_{x}  &=& F_{IK \Upsilon} + i F_{IK}{}^L F_{L \Upsilon} + i 
\left(F_{IL} + {\bar F}_{IL} \right) \left( i F_K{}^{LP} F_{P \Upsilon} + F_{\Upsilon K}{}^L 
\right)  \nonumber\\
&& - i {\bar F}_{IK}{}^L F_{L \Upsilon} - \left(F_{IL} + {\bar F}_{IL} \right) {\bar F}_{K}{}^{LP} F_{P \Upsilon} 
\;,
\end{eqnarray}
where indices are raised using $N^{IJ}$. 
Then, the integrability condition 
(\ref{HA1}) 
results in
\begin{equation}
\label{HA2}
\frac{\partial}{\partial \bar{X}^I}
D_\Upsilon F_\Upsilon = 
\bar{F}_{{I}{J}{K}} N^{JP} N^{KQ} F_{P\Upsilon}
F_{Q\Upsilon} \;.
\end{equation}

We now explore the consequences of \eqref{HA2}. To this end, we first define \cite{deWit:1996gjy}
 a hierarchy of 
symplectic functions through covariant derivatives of the holomorphic
symplectic function $F_\Upsilon(X,\Upsilon)$,
\begin{equation}
\Phi^{(n)}(X,\bar{X},\Upsilon, \bar{\Upsilon}) 
= \frac{1}{n!} D^{n-1}_\Upsilon F_\Upsilon \;\;\;,\;\;\; n \in \mathbb{N} \;,
\end{equation}
and $\Phi^{(0)}=0$.
Note that $\Phi^{(1)}$ is the only holomorphic function in this hierarchy. 
Then, \eqref{HA2} can be expressed as
\begin{equation}
\frac{\partial \Phi^{(2)}}{\partial \bar{X}^I} = 
\frac{i}{2} \frac{\partial N^{JK}}{\partial \bar{X}^I} F_{J\Upsilon} 
F_{K\Upsilon} = \frac{1}{2} \bar{F}_{I}^{JK} \partial_J \Phi^{(1)} \partial_K
\Phi^{(1)} \;,
\label{holophi2}
\end{equation}
where $ \bar{F}_{I}^{JK} =  \bar{F}_{IPQ} N^{PJ} N^{QK}$.
Thus, the integrability condition \eqref{HA1} results in \eqref{holophi2}
which captures the non-holomorphicity of $\Phi^{(2)}$.

Using \eqref{holophi2} as a starting point, one derives, by complete induction, 
the following holomorphic anomaly equation,
\begin{equation}
\label{A0}
\frac{\partial \Phi^{(n)}}{\partial \bar{X}^I} = \frac{1}{2} 
\bar{F}_{I}^{JK} \sum_{r=1}^{n-1} \partial_J \Phi^{(r)} \partial_K
\Phi^{(n-r)} \;,\;\;n\geq 2 \;,
\end{equation}
which captures the departure from holomorphicity of the $\Phi^{(n)}$, with $n \geq 2$.
In doing so, one uses \cite{Cardoso:2012nh}
\begin{equation}
\partial_{\bar{I}} F_\Upsilon = 0 \;,\;\;\;
D_\Upsilon \bar{F}_{IJK} = 0 \;\;\;,\;\;\; 
[D_\Upsilon, N^{IJ}\partial_J]= 0 \;.
\end{equation}
For example,
to derive the anomaly equation for $\Phi^{(3)}$,
we need to evaluate
\begin{eqnarray}
\partial_{\bar{I}} D^2_\Upsilon F_\Upsilon &=&
D_\Upsilon \partial_{\bar{I}} D_\Upsilon F_\Upsilon + 
i (\partial_{\bar{I}} N^{JK}) F_{J\Upsilon} \partial_K D_\Upsilon F_\Upsilon \nonumber\\
&=& 3 \bar{F}_{I}^{JK} \partial_J F_\Upsilon \partial_K D_\Upsilon
F_\Upsilon \;.
\end{eqnarray}
Using that $D^{n-1}_\Upsilon F_\Upsilon = n! \Phi^{(n)}$ this becomes 
\begin{equation}
\partial_{\bar{I}} \Phi^{(3)} = \bar{F}_{I}^{JK} \partial_J \Phi^{(1)} 
\partial_K \Phi^{(2)} = \frac{1}{2}  \bar{F}_{I}^{JK}  
\sum_{r=1}^2 \partial_J \Phi^{(r)} 
\partial_K \Phi^{(3-r)}  \;.
\end{equation}

Next, we define
\begin{equation}
F^{(n)}({X},\bar{ X}) = \Phi^{(n)}(X,\bar{X}, \Upsilon=\bar \Upsilon = 0) \;.
\label{Fnhol}
\end{equation}
The $F^{(n)}({X},\bar{ X})$ satisfy
the
holomorphic anomaly equation
\begin{equation}
\label{A1}
\frac{\partial F^{(n)}}{\partial \bar{X}^I} = \frac{1}{2} 
\bar{F}_{I}^{(0)JK} \sum_{r=1}^{n-1} \partial_J F^{(r)} \partial_K
F^{(n-r)} \;,\;\;n\geq 2 \;.
\end{equation}
Here, $\bar{F}_{I}^{(0)JK}$ is computed from the undeformed function $F^{(0)} (X) = F(X, \Upsilon)_{
\vert \Upsilon =0}$, i.e. $\bar{F}_{I}^{(0)JK}=  \bar{F}_{I}^{JK}{_{\vert \Upsilon =0}}$.

The hierarchy of equations (\ref{A0}) 
can be re-organized into a master anomaly equation, by introducing
\begin{equation}
G(X,\bar{X}, \Upsilon, \bar{\Upsilon}, \mu) = \sum_{n=0}^\infty \mu^{n+1} \Phi^{(n+1)}(X,\bar{X}, 
\Upsilon, \bar{\Upsilon}) \;,
\end{equation}
where $\mu$ denotes an expansion parameter.
Then, 
the function $G$ satisfies the master anomaly equation
\begin{equation}
\frac{\partial} {\partial \bar{X}^I} G = \frac{1}{2} \bar{F}_I^{JK}
\partial_J G \partial_K G \;.
\end{equation}

Finally, 
one may ask whether other components of $S = \nabla g^H$ will give rise to additional non-trivial
differential equations. To investigate this, we now consider the component $S_{x^I \Upsilon \bar \Upsilon}
= \left. \partial_{x^I} g^H_{\Upsilon \bar \Upsilon}\right|_{y}$,
which is constructed out of the 
metric component
$g^H_{\Upsilon \bar \Upsilon} = N^{IJ} F_{I \Upsilon} {\bar F}_{J \Upsilon}$.
Evaluating the relation 
$S_{x^I \Upsilon  \bar \Upsilon} = S_{\bar \Upsilon x^I \Upsilon} =\left.  \partial_{\bar \Upsilon} \,  g^H_{x^I \Upsilon}\right|_{x,y}$
in supergravity variables we find that it is identically satisfied. Thus, the only non-trivial differential equation resulting from 
$g^H_{\Upsilon \Upsilon}$ and $g^H_{\Upsilon \bar \Upsilon}$ is encoded in the relation $S_{x^I \Upsilon \Upsilon} = S_{\Upsilon x^I \Upsilon}$.


\subsection{Non-holomorphic deformation \label{nholskg}}

Next, we extend the discussion to a non-holomorphic generalized prepotential
$F=F(X,\bar{X}, \Upsilon, \bar{\Upsilon})$ by considering a non-holomorphic map $\phi: \hat{M} \rightarrow V$.

Since $F$ and $F_\Upsilon$ are no longer holomorphic, they will have 
non-vanishing derivatives with respect to $\bar{X}^I$ and $\bar{\Upsilon}$. To distinguish
between these various derivatives, we will, 
in the following, use a notation that involves `unbarred' indices
$I,J,\ldots $ and `barred' indices $\bar{I}, \bar{J}, \ldots$.

\subsubsection{Non-holomorphic deformation of the prepotential}

We generalize the map (\ref{phi_holo})
to 
\begin{equation}
\label{phi_non_holo}
\phi\;: \hat{M} = M \times \mathbb{C} \rightarrow V \;,\;\;\;
(X^I, \Upsilon) \mapsto (X^I, F_I(X,\bar{X},\Upsilon,\bar{\Upsilon})) \;,
\end{equation}
where $F_I = \partial F/\partial X^I$, can be obtained from a 
generalized prepotential $F$. We assume that $F$ 
has the form \cite{LopesCardoso:2006ugz}
\begin{equation}
\label{gF=F+Omega}
F(X,\bar{X}, \Upsilon, \bar{\Upsilon}) = 
F^{(0)}(X) + 2 i \Omega(X,\bar{X}, \Upsilon, \bar{\Upsilon}) \;,
\end{equation}
where $F^{(0)}$ is the undeformed prepotential, and 
where $\Omega$ is a real-valued function that describes the deformation.\footnote{This function is not to be confused with the complex symplectic form on the vector
space $V$ introduced in subsection \ref{sec:kaehimm}.}

 The holomorphic deformation is recovered
when $\Omega$ is harmonic. This makes use of the observation 
that the complex symplectic vector $(X^I, F_I)$ does not uniquely 
determine the prepotential $F$ \cite{Cardoso:2014kwa}. 
If we make a transformation 
\begin{eqnarray}
F^{(0)}(X) &\mapsto&  F^{(0)}(X) + g (X,\Upsilon) \;,\;\;\; \nonumber\\
\Omega(X,\bar{X}, \Upsilon, \bar{\Upsilon}) &\mapsto & 
\Omega(X,\bar{X}, \Upsilon, \bar{\Upsilon}) - \frac{1}{2i}
(g(X,\Upsilon) - \bar{g}(\bar{X}, \bar{\Upsilon}))  \;,
\end{eqnarray}
where $g(X,\Upsilon)$ is holomorphic, then $F$
changes by an antiholomorphic function, $F\mapsto F + \bar{g}$,
and the symplectic vector $(X^I,F_I)$ and the map $\phi$ are 
invariant. If $\Omega$ is harmonic, 
\begin{equation}
\Omega(X,\bar{X},\Upsilon,\bar{\Upsilon}) = f(X,\Upsilon) 
+ \bar{f} (\bar{X}, \bar{\Upsilon}) 
\;,
\end{equation}
we can make a transformation with $g =  2i f$ and obtain
\begin{equation}
F \mapsto F^{(0)}(X) + 2 i f(X,\Upsilon) =: F(X,\Upsilon)\;,
\end{equation}
which is a holomorphically deformed prepotential, as considered in subsection \ref{sec:holo-def-prep}.
If, however, $\Omega$ is not harmonic, then we have
a genuine generalization which requires us to consider non-holomorphic
generalized prepotentials.

\subsubsection{Non-holomorphic deformation and geometry}

We proceed by analysing the geometry induced by pulling back 
the standard Hermitian form $\gamma_V$ of $V$ given by \eqref{hermV}
to $\hat{M}$ using (\ref{phi_non_holo}),
\begin{eqnarray}
\gamma &=& - i (F^{(0)}_{IJ} - \bar{F}^{(0)}_{\bar{I}\bar{J}}) dX^I \otimes
d \bar{X}^J 
+ 2 (\Omega_{IJ} + \Omega_{\bar{I} \bar{J}}) dX^I \otimes d \bar{X}^J
+ 2 \Omega_{\bar{I}J} dX^I \otimes dX^J \nonumber\\
&& + 2 \Omega_{I \bar{ J}} d\bar{X}^I 
\otimes d \bar{X}^J 
+ 2 \Omega_{\bar{I} \bar{\Upsilon}} dX^I \otimes d \bar{\Upsilon}
+ 2 \Omega_{I \Upsilon} d \Upsilon \otimes d \bar{X}^I
+ 2 \Omega_{\bar{I}\Upsilon} dX^I \otimes d\Upsilon 
\nonumber\\
&& + 2 \Omega_{I \bar{\Upsilon}} d \bar{\Upsilon} \otimes d \bar{X}^I\;.
\end{eqnarray}
By decomposing $\gamma = g + i \omega$, we obtain the following 
metric on $\hat{M}$,
\begin{eqnarray}
\label{metricgpull}
g &=& - i (F^{(0)}_{IJ} - \bar{F}^{(0)}_{\bar{I}\bar{J}}) dX^I d\bar{X}^J
+ 2 (\Omega_{IJ} + \Omega_{\bar{I} \bar{J}}) dX^I d \bar{X}^J \nonumber\\
&& + 2 \Omega_{\bar{I}J} dX^I dX^J  
+ 2 \Omega_{I\bar{J}} d\bar{X}^I  d \bar{X}^J 
 + 2 \Omega_{\bar{I} \bar{\Upsilon}} dX^I  d \bar{\Upsilon}
+ 2 \Omega_{I \Upsilon} d \Upsilon  d \bar{X}^I \nonumber\\
&& + 2 \Omega_{\bar{I}\Upsilon} dX^I  d\Upsilon 
+ 2 \Omega_{I \bar{\Upsilon}} d \bar{\Upsilon}  d \bar{X}^I \;.
\end{eqnarray}
This expression shows that $g$ is not Hermitian, and
hence not K\"ahler with respect to the natural complex structure $J$.
The non-Hermiticity is encoded in the mixed derivatives $\Omega_{I\bar{J}}$,
which makes it manifest that it is related to the non-harmonicity
of $\Omega$. This metric occurs in the sigma model discussed
in \cite{Cardoso:2012mc}. 

The imaginary part of $\gamma$ defines a two-form on $\hat{M}$, 
\begin{eqnarray}
\omega &=& \frac{1}{2i} (- i (F^{(0)}_{IJ} - \bar{F}^{(0)}_{\bar{I} \bar{J}}))
dX^I \wedge d\bar{X}^J - i (\Omega_{\bar{I} \bar{J}}  + \Omega_{IJ})
dX^I \wedge d\bar{X}^J  \nonumber \\
&& - i \Omega_{\bar{I} J} dX^I \wedge d X^J 
+ i \Omega_{I\bar{J}} d\bar{X}^I \wedge d \bar{X}^J 
- i \Omega_{\bar{I}\bar{\Upsilon}} dX^I \wedge d \bar{\Upsilon}
- i \Omega_{I \Upsilon} d \Upsilon \wedge d\bar{X}^I 
\nonumber \\
& & 
- i \Omega_{\bar{I} \Upsilon} dX^I \wedge d \Upsilon
+ i \Omega_{I\bar{\Upsilon}} d\bar{X}^I \wedge d \bar{\Upsilon} \;.
\label{omega_non_holo}
\end{eqnarray}
This two-form is no longer of type $(1,1)$ with respect to the 
standard complex structure, which is consistent with the
non-Hermiticity of $g$. However, $\omega$ is still closed
\begin{equation}
d\omega=0 \;,
\end{equation}
and hence $(\hat{M},\omega)$ is a symplectic manifold. 

 The difference between the symplectic forms $\omega$ 
of $\hat{M}$ and $2dx^I \wedge dy_I$ of $M$ is exact,
\begin{equation}
\omega = 2 dx^I \wedge d y_I  + \frac{1}{2} d (\Upsilon d F_\Upsilon)
+ \frac{1}{2} d ( \bar{\Upsilon} d {\bar F}_{\bar{\Upsilon}}) 
+  \partial \overline{\partial} F \;,
\end{equation}
where $\partial = dX^I \otimes \partial_{X^I} + d \Upsilon \otimes 
\partial_\Upsilon$. Compared to (\ref{Diff_holo}) there is 
an additional term which measures the non-holomorphicity of the generalized 
prepotential.

\subsubsection{
The Hesse potential \label{realHe}}

We introduce real coordinates $(q^a) = (x^I, y_I)$ by
\begin{equation}
X^I = x^I + i u^I(x,y,\Upsilon, \bar{\Upsilon}) \;,\;\;\;
F_I (X, \bar X, \Upsilon, \bar \Upsilon) = y_I + i v_I(x,y, \Upsilon, \bar{\Upsilon}) \;.
\label{XFqnonh}
\end{equation}
We introduce the combinations
\cite{Cardoso:2012mc}
\begin{equation}
N_{\pm IJ} = N_{IJ} \pm 2 \mbox{Im} F_{I\bar{J}} = - i (F_{IJ} - \bar{F}_{\bar{I}
\bar{J}} \pm  F_{I\bar{J}} \mp \bar{F}_{\bar{I} J} )
\end{equation}
and 
\begin{equation}
R_{\pm IJ} = R_{IJ}  \pm 2 \mbox{Re} F_{I\bar{J}} = 
F_{IJ} + \bar{F}_{\bar{I} \bar{J}} \pm F_{I\bar{J}} \pm \bar{F}_{\bar{I} J} \;.
\end{equation}
Note that $N_-^T = N_-$, while $R_{\pm}^T = R_{\mp}$.

In the presence of 
a non-holomorphic deformation, the Hesse potential
is defined as 
the Legendre transform of  
\begin{equation}
\label{lforh}
L = 2 \mbox{Im} F - 2 \Omega = 2 \mbox{Im} F^{(0)} + 2 \Omega \;,
\end{equation}
c.f.  \eqref{eq:H-sympl} (the normalization used here differs
from the one in \eqref{eq:H-sympl} by a factor $2$).
As explained in section \ref{sec:ubiquity},
the function $L$ can be interpreted as a 
Lagrange function, and the Hesse potential as the corresponding 
Hamilton function.
Thus, the Hesse potential associated to $F(X, \bar X, \Upsilon, \bar \Upsilon)$ is
\begin{equation}
H(x,y,\Upsilon,\bar{\Upsilon}) = 
-i (F - \bar{F}) - 2 \Omega 
- 2 u^I y_I 
\;.
\end{equation}

We now compute the associated Hessian metric $g^H$
by taking derivatives of $H$ with respect to the coordinates
$(Q^A)=(q^a, \Upsilon, \bar{\Upsilon})$, where $(q^a)=(x^I, y_I)$.
To convert from coordinates $(x^I, u^I, \Upsilon, \bar \Upsilon)$ to 
coordinates $(Q^A)$ and back, we use the Jacobians \eqref{jacnonh} and \eqref{acnonhinv}.
We 
obtain for the components of the Hessian metric $g^H$,
\begin{eqnarray}
\frac{\partial H}{\partial q^a \partial q^b} &=&
\left( 
\begin{array}{cc}
N_+  + R_- N_-^{-1} R_+  \; &\;  - 2 R_- N_-^{-1}  \\
- 2 N_-^{-1} R_+ & 4 N_-^{-1} \\
\end{array} \right) \;, \\
 \frac{\partial^2 H}{\partial x^I \partial \Upsilon} &=&
- i (F_{I\Upsilon} - \bar{F}_{\bar{I} \Upsilon})
+ R_{-IK} N_-^{KJ} ( F_{J\Upsilon} + \bar{F}_{\bar{J} \Upsilon})
\;,\;\;\; \nonumber\\
\frac{\partial^2 H}{\partial y_I \partial \Upsilon} &=& 
- 2 N_-^{IK} (F_{K\Upsilon} + \bar{F}_{\bar{K} \Upsilon}) \;,
\nonumber
\end{eqnarray}
together with their complex conjugates, and
\begin{eqnarray}
\frac{\partial^2 H}{\partial \Upsilon \partial \bar{\Upsilon}} &=&
- i F_{\Upsilon \bar{\Upsilon}} + N_-^{IJ} ( \bar{F}_{\bar{I} \bar{\Upsilon}}
- \bar{F}_{I \bar{\Upsilon}})(F_{\Upsilon J} - F_{\Upsilon \bar{J}})
= - i D_{\Upsilon} F_{\bar{\Upsilon}}  \;,\;\;\; \nonumber\\
\frac{\partial^2 H}{\partial \Upsilon \partial \Upsilon} &=& 
- i D_{\Upsilon} F_\Upsilon \;,\;\;\;
\frac{\partial^2 H}{\partial \bar{\Upsilon} \partial \bar{\Upsilon}} = 
i \overline{D_{\Upsilon} F_\Upsilon} \;,\;\;\;
\end{eqnarray}
where 
\begin{equation}
D_\Upsilon = \partial_\Upsilon + i {N}_-^{IJ} ( F_{\Upsilon J} - 
F_{\Upsilon \bar{J}} ) \left( \frac{\partial}{\partial X^I} -
\frac{\partial}{\partial \bar{X}^I} \right)
\label{D_Upsilon_non_holo}
\end{equation}
is the symplectically covariant derivative introduced in \eqref{eq:cov-der-multiple}.

As before (c.f. subsection \ref{sec:real-hess-pot}), 
the Hessian metric $g^H$ differs from the metric $g$ in \eqref{metricgpull}
(induced by pulling back $g_V$ using $\phi$) by differentials involving
derivatives of $H$ with respect to $\Upsilon, \bar{\Upsilon}$,
\begin{equation}
g^H = g + \left. \partial^2H \right|_{x,y} \;,
\end{equation}
where 
\begin{equation}
\left. \partial^2H \right|_{x,y} = 
\frac{\partial^2 H}{\partial \Upsilon \partial \Upsilon} d\Upsilon d\Upsilon +
2 \frac{\partial^2 H}{\partial \Upsilon \partial \bar{\Upsilon}} d\Upsilon 
d\bar{\Upsilon} +
\frac{\partial^2 H}{\partial \bar{\Upsilon} \partial \bar{\Upsilon}} 
d\bar{\Upsilon} d\bar{\Upsilon}  \;.
\end{equation}


\subsubsection{Hierarchy of non-holomorphic symplectic functions}

The function $ F_\Upsilon  = \partial_{\Upsilon} F$ is a non-holomorphic symplectic function, c.f. \eqref{tilFAFA}.
Using 
the symplectically covariant derivative $D_\Upsilon$ given  in \eqref{D_Upsilon_non_holo}, 
we construct a hierarchy of symplectic functions by
\begin{equation}
\Phi^{(n+1)} (X,\bar{X},\Upsilon, 
\bar{\Upsilon}) 
 = \frac{1}{(n+1)!} D^n_\Upsilon F_\Upsilon (X,\bar{X},\Upsilon, 
\bar{\Upsilon}) \;\;\;,\;\;\; n \in \mathbb{N}_0 \;.
\label{Phihier}
\end{equation}
Then, we define symplectic functions $F^{(n)}(X, \bar{X}) $ by
\begin{equation}
F^{(n)}(X, \bar{X}) = 
\left. \Phi^{(n)}(X,\bar{X},\Upsilon,\bar{\Upsilon}) \right|_{\Upsilon=\bar \Upsilon =0} 
\;,\;\;\; n\geq 1 \;.
\label{freeenergsympl}
\end{equation}
The functions $F^{(n)}(X, \bar{X}) $ with $n \geq 2$ will satisfy a holomorphic anomaly
equation, whose precise form depends on the details of the 
non-holomorphic deformation. 

\subsubsection{The holomorphic anomaly equation of perturbative topological string theory}

For a specific deformation, the resulting holomorphic anomaly equation
is the one of perturbative topological string theory \cite{Bershadsky:1993cx}. Namely, 
let us first rescale $F^{(n)} \mapsto 2i F^{(n)}$, for convenience.  Now we take $\Upsilon$ to be real, 
and $F^{(1)}$ to be 
\begin{equation}
    F^{(1)} =\, f^{(1)} + \bar
   f^{(1)} + \alpha \ln \det N_{IJ}^{(0)}\;.
  \end{equation}
 Here, $\alpha \in \mathbb{R}$ is the deformation parameter, and $N_{IJ}^{(0)}$ equals $N_{IJ}^{(0)} = -i (F_{IJ}^{(0)} - {\bar F}_{IJ}^{(0)} )$.
 When $\alpha =0$, $F^{(1)}$ is the real part of a holomorphic function $f^{(1)} (X)$. 
 For the $\alpha$-deformation, the holomorphic anomaly equation satisfied by the  $F^{(n)}$ with $n\geq 2$ is given by (see \cite{Cardoso:2014kwa})
  \begin{equation}
\frac{\partial}{\partial \bar{X}^K} F^{(n)} 
= i {\bar F}^{(0)IJ}_K \left( \sum_{r=1}^{n-1} \partial_I F^{(r)} 
\partial_J F^{(n-r)} - 2 \alpha D_I \partial_J F^{(n-1)} \right) \;\;\;,\;\;\; n \geq 2 \;,
\label{fullhaeg2}
\end{equation}
where ${\bar F}^{(0)IJ}_K = {\bar F}^{(0)}_{KQP} N^{(0) QI}ÊN^{(0) PJ}$. 
The covariant derivative $D_I$, when acting on a vector $V_J$, takes the form
\begin{equation}
D_I V_J = \partial_I V_J - \Gamma_{IJ}{}^K V_K \;,
\end{equation}
where $\Gamma_{IJ}{}^K$ is the Levi-Civita connection associated with 
the K\"ahler metric of the undeformed theory (i.e. the K\"ahler metric computed
from $F^{(0)} (X)$).
When $\alpha =0$, this anomaly equation reduces to the one given in \eqref{A1}, upon undoing the rescaling $F^{(n)} \mapsto 2i F^{(n)}$ performed above.
When $ \alpha=- 1/2$, \eqref{fullhaeg2}
is the holomorphic anomaly equation of perturbative topological string theory 
\cite{Bershadsky:1993cx,Grimm:2007tm}. Let us display the expression for $F^{(2)}$ obtained by solving the anomaly equation 
\cite{Bershadsky:1993cx,Aganagic:2006wq,Grimm:2007tm},
\begin{eqnarray}
  && F^{(2)}(X, \bar X) = f^{(2)}  - N^{IJ}_{(0)}
   \big( f_I^{(1)}  - \mathrm{i} \alpha \, F^{(0)}_{IKL} N_{(0)}^{KL} \big)
   \big( f_J^{(1)} - \mathrm{i} \alpha \, F^{(0)}_{JPQ} N_{(0)}^{PQ} \big)
   \nonumber\\
   && \qquad +  2 \alpha \, N_{(0)}^{IJ} f^{(1)}_{IJ} - \alpha^2 \big[\mathrm{i}
    N_{(0)}^{IJ} N_{(0)}^{KL} F^{(0)}_{IJKL}
     - \tfrac23  N_{(0)}^{IJ} F^{(0)}_{IKL} N_{(0)}^{KP} N_{(0)}^{LQ} F^{(0)}_{JPQ} \big]
   ,\nonumber\\
   \end{eqnarray}
with holomorphic input data $f^{(1)} (X)$ and $f^{(2)} (X)$.
 
 The expressions for the higher $F^{(n)}(X, \bar X)$ become very lengthy quickly, see the expression
for  $F^{(3)}(X\, \bar X)$ given in Appendix D of  \cite{Cardoso:2014kwa}.
 The 
non-holomorphicity of $F^{(n)}(X, \bar X)$ is
entirely contained in the quantities $N_{IJ}^{(0)}, N_{(0)}^{IJ}$.  Observe that $F^{(1)}$
is real, while the higher $F^{(n)}$ ($n \geq 2$) are not.

\subsubsection{Holomorphic anomaly equation from the Hessian structure }

The holomorphic anomaly equation \eqref{fullhaeg2} 
is encoded in the underlying Hesse structure, namely
in the relation 
\begin{equation}
S_{x^I \Upsilon \Upsilon} = S_{\Upsilon x^I \Upsilon} \;,
\end{equation}
which the 
totally symmetric
rank three tensor $S= \nabla g^H$ has to satisfy, where $g^H$ denotes the Hessian metric computed in subsection \ref{realHe}.
We refer to \cite{Cardoso:2015qhq} for the somewhat technical verification of this assertion, where this was shown for the case of the anomaly equation for $F^{(2)}$.
Thus, the holomorphic anomaly equation \eqref{fullhaeg2}  is intimately related to the existence of a Hessian structure on $\hat M$.


\section{Dimensional reduction over space and time. Euclidean special geometry\label{sec:dim_red}}

In this section we will review how the special geometries of five- and four-dimensional
vector multiplets are related to each other, and to the special geometry of hypermultiplets,
by dimensional reduction. We take this opportunity to also discuss how special geometry
gets modified for theories defined on a Euclidean space`time,' by including time-like 
dimensional reductions. We will focus on presenting and discussing key facts and
results while referring to the literature for details. 

\subsection{Space-like and time-like dimensional reductions \label{sect:dimred_general}}

Space-like and time-like dimensional reductions of Lagrangians differ by 
specific relative signs between terms. We illustrate this with a simple example,
a theory involving a free massless scalar $\sigma$ and an
abelian vector field $A_\mu$ in $n+1$ dimensions,
\begin{equation}
S = \int d^{n+1} x \left( - \frac{1}{2} \partial_\mu \sigma \partial^\mu \sigma
- \frac{1}{4} F_{\mu \nu} F^{\mu \nu} \right) \;.
\end{equation}
Upon dimensional reduction, the vector field $A_\mu$ decomposes into
a vector field $A_m$ and a scalar $b=A_*$, where $*$ is the index of the
direction we reduce over. The reduced Lagrangian, where we only keep 
the massless modes, is
\begin{equation}  
\label{Lag_reduced}
S = \int d^{n} x \left( - \frac{1}{2} \partial_m \sigma \partial^m \sigma
+ \frac{1}{2} \varepsilon \partial_m b \partial^m b 
- \frac{1}{4} F_{mn} F^{mn } \right) \;,
\end{equation}
where $\varepsilon=-1$ for a space-like reduction and $\varepsilon=1$ for
a time-like reduction.\footnote{Note that part of the literature on dimensional
reduction defines $\varepsilon$ with the opposite sign.} Thus in a Euclidean theory obtained by time-like
dimensional reduction, the sign of the kinetic term of the scalar $b$ 
is inverted and the Euclidean action is indefinite. This distinguishes 
such Euclidean theories from Euclideanized theories obtained by 
Wick rotation, see section \ref{Sect:Euc_disc} for discussion. 

For space-like reductions we can combine the real scalars $\sigma$ and
$b$ into a complex scalar $X=\sigma+ib$. For time-like reductions there
are two ways to proceed. Either we can use {\em adapted real coordinates}
which are lightcone coordinates with respect to the scalar target space,
$X_\pm = \sigma \pm b$, or we can introduce {\em para-complex coordinates}
by employing para-complex numbers $z=x+ey$, $x,y\in \mathbb{R}$, where
the para-complex unit $e$ satisfies
\begin{equation}
e^2=1 \;,\;\;\;\bar{e} = -e\;.
\end{equation}
The anti-linear involution 
$\bar{\cdot}$ is called {\em para-complex conjugation}. The para-complex
numbers $C:= \mathbb{R} \oplus e \mathbb{R}$ form a real algebra, but not
a number field, and not even a division algebra. Zero divisors correspond 
to `lightcone directions', 
since $(1+e)(1-e) = 0$. Nevertheless one can use para-complex numbers 
to define various types of structures on 
differentiable manifolds, which are analogous to those based on complex numbers, such 
as complex, Hermitian and K\"ahler
structures. Para-complex geometries are useful to formulate 
special geometry in Euclidean signature 
\cite{Cortes:2003zd,Cortes:2005uq,Cortes:2009cs,Cortes:2015wca}, and more recently have taken on a role
in generalized and doubled geometry as well \cite{Vaisman:2012ke,Vaisman:2012px,Freidel:2017yuv,Freidel:2018tkj,Marotta:2018myj}. We provide some background information 
in \ref{App:para_complex}, and refer to \cite{Cruc:1996} for a historical review. 

One advantage of working with para-complex scalar fields is that it makes the
similarities between space-like and time-like reductions manifest. In particular one
can introduce an {\em $\varepsilon$-complex notation} by
\begin{equation}
i_\varepsilon = \left\{ \begin{array}{cc}
i & \mbox{for}\;\;\varepsilon =-1  \;,\\
e & \mbox{for}\;\; \varepsilon = 1  \;,\\
\end{array} \right \} \;\;\Rightarrow 
i_\varepsilon^2 = \varepsilon \;,\;\;\;
\overline{i_\varepsilon} = - i_\varepsilon \;.
\end{equation}
In $\varepsilon$-complex notation, the reduced Lagrangian \eqref{Lag_reduced}
becomes
\begin{equation}
S = \int d^n x \left( - \frac{1}{2} \partial_m X \partial^m \bar{X} - \frac{1}{4} 
F_{mn} F^{mn} \right) \;, \;\;\;\mbox{where}\;\;\;X=\sigma + i_\varepsilon b \;.
\end{equation}

\subsection{Euclidean and Euclideanized theories \label{Sect:Euc_disc}} 

Before proceeding we need to clarify the distinction between Euclidean 
and Euclideanized theories. In this review a `Euclidean supersymmetric theory' 
or `Euclidean supergravity theory' is a theory with a Lagrangian which is
invariant under the Euclidean supersymmetry algebra. This is true in 
particular for theories which are obtained by a time-like dimensional reduction, but
one can also construct Euclidean theories ab initio, starting from the Euclidean
supersymmetry algebra, see for example \cite{Gall:2018ogw}, or 
by analytical continuation of Killing spinor equations, see for example 
\cite{Gutowski:2012yb}. In contrast by a `Euclideanized theory' 
we refer to a theory which has been obtained from a theory in 
Lorentz signature by applying a Wick rotation. From the previous section 
it is clear that for four-dimensional theories which can be
obtained by dimensional reduction from five dimensions, the Euclidean
and Euclideanized theory will in general have bosonic Lagrangians which 
differ by relative signs for some of the scalars.
For theories containing fermions the additional complication arises that
reality condition are signature dependent, which can lead to a doubling
of the fermionic degrees of freedom upon Euclideanization. In four
dimensions Majorana spinor exist in Lorentzian, but not in Euclidean
signature, which in particular implies that there is no Euclidean 
`${\cal N}=1$' supersymmetry algebra with four real supercharges. One can
still define a meaningful Euclideanization of four-dimensional ${\cal N}=1$
theories within the Osterwalder-Schrader formalism \cite{Nicolai:1978vc}. In this approach 
one uses a modified Hermiticity condition in the Euclidean theory, 
and supersymmetry is encoded in Euclidean Ward identities which 
become the standard supersymmetric Ward identities upon 
continuation to Lorentz signature. 
An alternative proposal for the Euclideanization of supersymmetric theories
with extended supersymmetry, where there is no issue with the doubling of
fermionic degrees of freedom,  
is to modify the Wick rotation such that the resulting theory has an
action which is invariant under Euclidean supersymmetry
\cite{vanNieuwenhuizen:1996ip,vanNieuwenhuizen:1996tv,Theis:2001ef}.
For a certain class of theories, which include the bosonic parts of four-dimensional
vector multiplet theories, 
Euclidean and Euclideanized actions can be mapped
to each other using that the Hodge dualization of 
axion-like scalars does not commute with a Wick rotation \cite{Cortes:2009cs}.

Since Euclidean actions obtained by a time-like reduction can be 
indefinite, while a well-behaved Euclidean functional integral 
requires an 
Euclidean action which is  bounded from
below, one 
might think that only 
Euclideanized theories can provide the proper starting 
point for defining supersymmetric theories. However, the situation 
is more complicated for various reasons. Firstly, in Euclidean signature
the Hodge-dualization of $p$-form fields changes the sign of 
their `kinetic term' and thus relates definite and indefinite 
actions.\footnote{See section \ref{Red_to_three}.}
Secondly real integrals can be dominated 
by complex saddle points and the functional integral of a supersymmetric
theory can be dominated by real BPS solutions to an indefinite Euclidean
action. In particular, this is the case for the D-instanton solutions of type-IIB
string theory \cite{Gibbons:1995vg}. At least in simple examples one can 
show explicitly that 
Euclidean and Euclideanized actions can be used alternatively to 
perform a saddle point evaluation of the same functional integral,
using different `integration contours' in complexified field space 
\cite{Mohaupt:2010du}. This suggest to construct theories
on space-times of different signatures
as different real forms of 
master theories on with a complexified field space on 
a complexified space-time.  In this context it is natural 
to also consider space-time signature other than Euclidean
and Lorentzian, see below.

Euclidean actions also serve a practical role as part of generating techniques
for stationary solutions of theories in Lorentzian signature  \cite{Galtsov:1998yu,Stelle:1998xg}. 
Upon time-like dimensional 
reduction one obtains an auxiliary Euclidean theory, whose field equations are 
often easier to solve. Solutions of the reduced Euclidean theory can then 
be lifted to stationary solutions of the Lorentzian theory. This can be viewed
as generating `solitons' (stationary finite energy solutions of a Lorentzian theory)
from `instantons' (finite action solutions of Euclidean theory). With proper
attention to boundary terms one can show that the instanton action of 
certain Euclidean solutions agrees exactly with the ADM mass of 
the black hole solutions obtained by lifting \cite{Cortes:2009cs}. This `reduction/oxidation' method
is not limited to BPS solutions and can be used to generate non-extremal
solutions. 

\subsection*{Some remarks on general space-time signatures}

Once time-like 
T-dualities are admitted, the web of string dualities 
relates theories in different space-time signatures \cite{Hull:1998ym,Hull:1998vg,Hull:1998fh}.
The maximally supersymmetric supergravity theories in ten and
eleven dimensions can all be related to real forms of a single complex
ortho-symplectic Lie superalgebra \cite{Bergshoeff:2000qu,Bergshoeff:2007cg}.
Five- and four-dimensional vector multiplets for all possible 
space-time signatures have been obtained in \cite{Sabra:2017xvx,Gall:2018ogw,Cortes:2019mfa}.

\subsection{Reduction from five to four dimensions: the r-map}

\subsubsection{Reduction without gravity: the rigid r-map}

We now turn to the dimensional reduction of the five-dimensional bosonic
vector multiplet Lagrangian \eqref{5d_rigid_VM_Lagrangian}, following \cite{Cortes:2003zd},
and treating space-like and time-like reduction in parallel. 
Upon reduction, the five-dimensional vector fields $A^I_\mu$ decompose into
four-dimensional vector fields $A^I_m$ and scalars $b^I$, which we combine with
five-dimensional scalars $\sigma^I$ to $\varepsilon$-complex scalars
$X^I=\sigma^I + i_\varepsilon b^I$. The four-dimensional field strengths are
decomposed into selfdual and antiselfdual parts according to 
\begin{equation}
\label{Fpm_universal}
F^{\pm}_{mn}= \frac{1}{2} ( F_{mn} \pm \tilde{F}_{mn}) =
\frac{1}{2} (F_{mn} \pm \frac{1}{2i_\varepsilon} \varepsilon_{mnpq} F^{pq}) \;.
\end{equation}
The couplings of the four-dimensional theory are encoded in an
$\varepsilon$-holomorphic prepotential $F(X)$, which up to a 
constant factor is obtained by extending the Hesse potential 
$h(\sigma)$ of the five-dimensional theory from real to $\varepsilon$-complex
values:
\begin{equation}
\label{prepo_universal}
F(X) = - \frac{1}{2i_\varepsilon} h(\sigma + i_\varepsilon b) \;.
\end{equation}
We extend our previous definitions according to\footnote{In 
\cite{Cortes:2003zd} $N_{IJ}$ and $K$ were defined with the opposite sign. 
This has been compensated for by changing the overall sign of $F$. Apart from this,
some fields have to be rescaled by constant factors.} 
\begin{eqnarray}
&& R_{IJ} = F_{IJ} + \bar{F}_{IJ} \;,\;\;\;
N_{IJ} = - i_\varepsilon (F_{IJ} - \bar{F}_{IJ})  = \frac{\partial^2 K}{\partial X^I 
\partial \bar{X}^J}
\;,\;\;\;  \nonumber \\
&& K = i_\varepsilon (X^I \bar{F}_I - F_I \bar{X}^I )\;.
\end{eqnarray}
The resulting four-dimensional bosonic Lagrangian takes the form
\begin{eqnarray}
L &=& - N_{IJ} \partial_m X \partial^m \bar{X} + \left( \frac{i_{\varepsilon}}{4} F_{IJ} F^{I-}_{mn} F^{J-mn} 
+ h.c.\right)  + \cdots \label{Lagr_4dVM_epsilon} \\
&=&  -N_{IJ} \partial_m X^I  \partial^m \bar{X}^J 
- \frac{1}{8} N_{IJ} F^{I}_{mn} F^{Jmn} - \frac{1}{16} R_{IJ} \varepsilon^{mnpq} F^I_{mn} F^J_{pq}
+ \cdots \;, \nonumber
\end{eqnarray}
where we have omitted the auxiliary fields $Y^I_{ij}$.

We now turn to the relation between the scalar manifolds $M$ of the five-dimensional theory
and $N$ of the four-dimensional theory. 
The ASR metric $g_M = h_{IJ} d\sigma^I d\sigma^J$ is mapped
to the {\em affine special $\varepsilon$-K\"ahler metric}
\begin{equation}
\label{metric4dVM_universal}
g_N = N_{IJ}(\sigma) d\sigma^I d\sigma^J - \varepsilon N_{IJ}(\sigma) db^I db^J =
N_{IJ} d X^I d\bar{X}^J \;,
\end{equation}
with $\varepsilon$-holomorphic prepotential \eqref{prepo_universal}. Since the 
fields $b^I$ take values in $\mathbb{R}^n$, where $n$ is the number of 
five-dimensional vector multiplets, we can identify $N$ with the tangent bundle
of $M$, that is $N\cong TM$. The metric \eqref{metric4dVM_universal} 
only depends on the scalars $\sigma^I$ and therefore has an isometry
group which contains the constant shifts $b^I \mapsto b^I + \beta^I$, where
$(\beta^I) \in \mathbb{R}^n$. These isometries are relicts of the five-dimensional
abelian gauge symmetry. Moreover, the metric \eqref{metric4dVM_universal} is 
block-diagonal with respect to $\sigma^I$ and $b^I$. This decomposition has
an invariant meaning, because the special connection $\nabla$ of the ASR
manifold $M$ can be used to decompose 
\begin{equation}
T_X N = T_XN^{\mathrm{vert}} \oplus T_XN ^{\mathrm{hor}}_\nabla 
\cong T_\sigma M \oplus T_\sigma M 
\;,\;\;\;X\in N=TM \;,\;\;\;\sigma = \pi(X)\in M  \;,
\end{equation}
where $\pi: N=TM \rightarrow M$ is the canonical projection. The vertical 
space can be identified with $T_{\pi(X)} M=T_\sigma M$ using the projection 
\begin{equation}
T_XN^{\mathrm{vert}} := \mbox{ker}( d\pi_X) \cong T_\sigma M \;.
\end{equation}
While in general there is no canonical complement of $T_XN^{\mathrm{vert}}
\subset T_X N$, the connection $\nabla$ defines a horizontal subbundle
$TN^{\mathrm{hor}}_\nabla$, which is spanned by vectors tangent to the horizontal lifts
of curves on $N$. This can be used to identify the horizontal
subspace $T_XN^{\mathrm{hor}}_\nabla$ 
with the tangent space $T_\sigma M$ using the projection:
\begin{equation}
\left. d\pi_X  \right|_{T_X N^{\mathrm{hor}}_\nabla} \;:
T_X N^{\mathrm{hor}}_\nabla   \xrightarrow{\cong} T_\sigma M\;.
\end{equation}

A similar construction based on  the Levi-Civita
connection $D$ is used to define the so-called Sasaki metric on 
the tangent bundle $N=TM$ of a Riemannian manifold $M$, which has
a block-diagonal structure like in \eqref{metric4dVM_universal}. The
`Sasaki-like' metric $g_N$ on the tangent bundle $N=TM$ of an ASR manifold with
metric $g_M$ is defined by the special connection $\nabla$ instead of $D$, and
it comes in two versions, labelled by $\varepsilon$, 
which differ by a relative sign of the metric 
along the horizontal and vertical distribution. It has been shown 
in \cite{Cortes:2003zd} that if $(M,g_M,\nabla)$ is an ASR manifold, then
$N=TM$ carries the structure of an affine special $\varepsilon$-K\"ahler 
manifold $(N,J_N, g_N, \nabla_N)$, where the metric $g_N$, the $\varepsilon$-complex structure
$J_N$ and the special connection $\nabla_N$ can be 
constructed out of the ASR data. The map
\begin{equation}
r_\varepsilon :  \{ \mbox{ASR manifolds} \} \rightarrow 
\{ \mbox{AS}\varepsilon\mbox{K manifolds}  \}\;: M \mapsto N=TM
\end{equation}
is called the {\em rigid r-map}. 

While we have omitted the supersymmetry transformations and fermionic
terms of the Lagrangian, these can be found in \cite{Cortes:2003zd}. 
We remark that by dimensional reduction one can only obtain a subset of
the four-dimensional vector multiplet theories, namely those where the 
prepotential is a cubic polynomial in the $\varepsilon$-complex special coordinates $X^I$.
Such prepotentials are called {\em very special}. However, the only terms not obtained
by dimensional reduction are four-fermion terms which are proportional 
to the fourth derivatives $F_{IJKL}$, $\bar{F}_{IJKL}$ of the prepotential.
To obtain the general four-dimensional Lagrangian one takes $F$ to 
be a general $\varepsilon$-holomorphic function. Then the Lagrangian
is only invariant up to terms generated by variation of terms involving
the third derivatives of the prepotential. The four-fermion terms are
determined by imposing that their variation restores the supersymmetry
invariance of the Lagrangian \cite{Cortes:2003zd}.

We remark that Euclidean supersymmetric theories, and in fact supersymmetric
theories on space-times of arbitrary signature can also be constructed
ab initio, rather than by dimensional reduction. In particular,  five-dimensional rigid
off shell vector multiplets and their Lagrangians have been obtained 
for all signatures $(t,s)$, $t+s=5$ in \cite{Gall:2018ogw}.

\subsubsection{Reduction with gravity: the supergravity r-map}

We now turn to the more interesting case of performing the reduction 
in supergravity. When starting in five dimensions with $n_{(5)}$ 
vector multiples coupled to Poincar\'e supergravity, we end up in 
four dimensions with $n_{(4)} = n_{(5)}+1$ vector multiplets coupled
to Poincar\'e supergravity, because the five-dimensional supergravity 
multiplet decomposes into the four-dimensional supergravity multiplet and
an additional Kaluza-Klein vector multiplet. 
The five-dimensional metric decomposes as 
\begin{equation}
g_{\mu \nu} dx^\mu dx^\nu = - \varepsilon e^{2\sigma} \left( dx^* + {\cal A}_m dx^m \right)^2
+ g_{mn} dx^m dx^n \;,
\end{equation}
where $g_{mn}$ is the four-dimensional metric with signature $(\varepsilon,+,+,+)$, 
${\cal A}_m$ is the Kaluza-Klein vector and $\sigma$ is the Kaluza-Klein scalar. 

We start from \eqref{Lagr5dsugra1}, \eqref{g2a}, \eqref{Lagr5dsugra2} with $\kappa=1$
and relabel $I = 0, \ldots, n_{(5)}$ into 
$a =1, \ldots, n_{(5)}+1 = n_{(4)}$, so that we can use $I,J=0, \ldots n_{(4)}$ 
to label four-dimensional vector multiplets. It is convenient to work with the
constrained scalars $h^a$, subject to ${\cal V}(h) = C_{abc}h^ah^bh^c=1$, instead of the physical
scalars $\phi^x$. Upon reduction, one can then define new scalars
\begin{equation}
y^a := 6^{1/3} e^{\sigma}  h^a  \;.
\end{equation} 
These are $n_{(4)}$ unconstrained real scalars which encode 
the Kaluza-Klein scalar through
\begin{equation}
{\cal V}(y) = C_{abc} y^a y^b y^c = 6 e^{3\sigma} \;,
\end{equation}
while the physical five-dimensional scalars $\phi^x$ can be parametrized by 
the independent ratios $h^x/h^{n_{(4)}}$, $x=1, \ldots, n_{(5)}$. 
The real scalars $y^a$ are combined with the scalar components
$x^a \propto A^a_*$ of the five-dimensional gauge fields into 
$\varepsilon$-complex scalars $z^a := x^a + i_\varepsilon y^a$. 
With this convention the five-dimensional gauge symmetry induces
an invariance of the four-dimensional theory under real shifts of 
the scalars $z^a$, that is under $z^a \mapsto z^a + r^a$, with $(r^a) 
\in \mathbb{R}^{n_{(4)}}$. Thus the scalar manifold looks locally like
a higher-dimensional version of the upper half plane.\footnote{There are other 
conventions in the supergravity literature where 
the axion-like scalars are taken to be the imaginary rather than real parts,
in particular the fields $S,T,U$ of the much studied STU-model are defined that way.}
For the vector fields it is necessary to take field dependent ($x^a$-dependent)
linear combinations in order to make the four-dimensional gauge symmetry 
manifest. Moreover, to arrive at standard four-dimensional conventions, 
fields need to be rescaled by constant factors, see \cite{Cortes:2009cs} for details.
The resulting bosonic Lagrangian takes the form\footnote{Compared to 
\cite{Cortes:2009cs}, there is an explicit factor $\varepsilon$ in the last term to
account for the different definition of the $\epsilon$-tensor in Lorentzian signature.}
\begin{eqnarray}
 L &=& \frac{1}{2} R - \bar{g}_{ab} \partial_m z^a \partial^m \bar{z}^b
+ \frac{1}{4} \mbox{Im} {\cal N}_{IJ} F^I_{mn} F^{Jmn}
+ \frac{\varepsilon}{4} \mbox{Re} {\cal N}_{IJ} F^I_{mn} \frac{1}{2} \varepsilon^{mnpq} F^J_{pq} 
\nonumber \\
 &=& \frac{1}{2} R - \bar{g}_{ab} \partial_m z^a \partial^m \bar{z}^b
+ \frac{1}{4} \mbox{Im} {\cal N}_{IJ} F^I_{mn} F^{Jmn}
+ \frac{\varepsilon i_\varepsilon}{4} \mbox{Re} {\cal N}_{IJ} F^I_{mn} \tilde{F}^{J|pq} \nonumber \\
&=& \frac{1}{2} R - \bar{g}_{ab} \partial_m z^a \partial^m \bar{z}^b
+ \left( \frac{1}{4 i_\varepsilon} {\cal N}_{IJ}  F^{+I}_{mn} F^{+Jmn} + h.c.\right) \;,
\label{4dVM_onshell_epsilon}
\end{eqnarray}
which generalizes \eqref{efflag_poincare} to the $\varepsilon$-complex case.
As in the rigid case only a subclass of four-dimensional theories can be obtained
by reduction. Given a five-dimensional Hesse potential of the form $h=C_{abc} h^a h^b h^c$
the $\varepsilon$-holomorphic prepotentials resulting from reduction have the very 
special form
\begin{equation}
F = - \frac{1}{6} C_{abc} \frac{X^a X^b X^c}{X^0} \;,
\end{equation}
where $X^I$, $I=0, \ldots, n_V^{(4)}$ are related to the physical scalars 
$z^a$ by $z^a = X^a/X^0$. It was shown in \cite{Cortes:2009cs} that the
superconformal quotient admits an $\varepsilon$-complex generalization, which
for $\varepsilon=1$ connects conical affine special para-K\"ahler manifolds
$N$ to projective special para-K\"ahler manifolds $\bar{N}$. In the 
para-complex version of the quotient, $\mathbb{C}^*=\mathbb{R}^{>0} \times U(1)$ 
is replaced by $C^* = \mathbb{R}^{>0} \times SO(1,1)$. The group $SO(1,1)$ 
replacing $U(1)$ is the abelian factor of the R-symmetry group $SO(1,1) \times SU(2)$ 
of the four-dimensional Euclidean supersymmetry algebra \cite{Cortes:2003zd,Cortes:2009cs}.
With suitable conventions, all local formulae of special K\"ahler geometry have
$\varepsilon$-complex extensions. In particular 
\begin{equation}
{\cal N}_{IJ} = \bar{F}_{IJ} - i \varepsilon \frac{ N_{IK} X^K N_{JL} X^L}{ N_{MN} X^M X^N}
\end{equation}
generalizes \eqref{calN-N} while
\begin{equation}
\bar{K} = - \log \left( -i_\varepsilon (X^I \bar{F}_I - F_I \bar{X}^I)\right) = - \log (-K)
\end{equation}
generalizes \eqref{KL_potential}. The expressions \eqref{N0}
for the projectable tensor, and $K=N_{IJ} X^I \bar{X}^J=-1$ for the D-gauge are valid for
both values of $\varepsilon$.

Dimensional reduction relates the scalar manifolds of the two theories by 
assigning to every PSR manifold $\bar{M}$ of real dimension $n=n_{(5)}$ 
a PS$\varepsilon$K manifold $\bar{N}$ of real dimension $2n+2 = 2n_{(4)}$. 
The additional two scalars come from the reduction of the five-dimensional
supergravity multiplet, one from the metric, one from the graviphoton. The resulting 
map
\begin{equation}
\bar{r}_\varepsilon \;: \{ \mbox{PSR  manifolds} \} \rightarrow \{ \mbox{PS}\varepsilon\mbox{K 
manifolds} \} \;,\;\;\; \bar{M}_{n} \mapsto  \bar{N}_{2n+2} 
\end{equation}
is called the {\em supergravity r-map}.

For dimensional reasons
$\bar{N} \not\cong T\bar{M}$, which raises the question how to understand the
geometry of this map. So far we have considered Poincar\'e supergravity in 
an on-shell formulation. For many purposes, including having full manifest
symplectic covariance, working off-shell, and including higher derivatives,
one needs to have the superconformal off-shell version of the dimensional 
reduction, and for the $\bar{r}$-map. Here we focus on the scalar geometry.
The full off-shell reductions of five-dimensional superconformal
vector and hypermultiplets coupled to the Weyl multiplet can be found in 
\cite{Banerjee:2011ts,deWit:2017cle}.
In the superconformal
setting we have an $(n+1)$-dimensional real cone $M_{n+1}$ over $\bar{M}_n$ 
and a $(2n+4)$-dimensional $\varepsilon$-complex cone $N_{2n+4}$ 
over $\bar{N}_{2n+2}$. Since the superconformal theories are gauged
versions of rigid superconformal theories, it is natural to apply the rigid
r-map to $M_{n+1}$. This yields an AS$\varepsilon$K manifold 
$\hat{N}_{2n+2}$, which is not conical. Note that the dimensional reduction of 
a rigid superconformal symmetry breaks conformal symmetry, as follows immediately
from our results on the r-map. The cubic Hesse potential of $\bar{M}_{n+1}$ 
maps to a cubic prepotential for $\hat{N}_{2n+2}$, but rigid superconformal
symmetry requires a prepotential which is homogeneous of degree two.
To lift the supergravity r-map $\bar{M}_n \mapsto \bar{N}_{2n+2}$ 
to a map $M_{n+1} \mapsto N_{2n+4}$ between the associated conical
manifolds, one needs to combine the rigid r-map $M_{n+1} \mapsto
\hat{N}_{2n+2}$ with another map called the `conification map'
$\mathrm{con}:\hat{N}_{2n+2} \mapsto N_{4n+2}$, which
canonically, that is without arbitrary choices and only using 
given data, assigns a cone $N_{4n+2}$ to the non-conical manifold $\hat{N}_{2n+2}$. 

Such a conification map 
has been constructed, for the case of space-like reduction ($\varepsilon=1$) 
in \cite{Cortes:2017utn},\cite{Dietrich:2017}.\footnote{While it should be straightforward to extend this to the
para-complex setting, we restrict ourselves to reviewing published work.}
This conification map induces a map $\hat{N}_{2n+2} \mapsto \bar{N}_{2n+2}$
between ASK manifolds and PSK-manifolds of the
same dimension, called the  ASK/PSK
correspondence. The situation is summarized in the following diagram
\begin{equation}
\label{Diagram:r-map}
\xymatrix{
{M} \ar@{|->}[r]^{r} \ar@{|->}[d]_{\mathrm{SC}}
& \hat{N}  \ar@{|->}[rrr]^{\mathrm{con}} \ar@{|->}[drrr]^{\mathrm{ASK/PSK}}
& & & N  \ar@{|->}[d]^{\mathrm{SC}}\\
\bar{M} \ar@{|->}[rrrr]_{\bar{r}} & & &  &\bar{N} \\
}
\end{equation}
where `SC' indicates a superconformal quotient. Since the rigid $r$-map relates
a cubic Hesse potential $h(\sigma^a)$ to a cubic prepotential $F_{\hat{N}}(X^a)$, one
expects that the conification map yields a prepotential of the form  $F_N(X^I) = F_{\hat{N}}(X^a)/X^0$ 
for the CAS$\varepsilon$K manifold $N$. While this turns out to be correct, we 
stress that it is not clear a priori how to formulate the relation between $\hat{N}$ and
$N$ in a way that is independent of a choice of coordinates. Note that the special
coordinates $X^a$ on $\hat{N}$ are unique up to transformations in 
$Sp(2n+2,\mathbb{R}) \ltimes \mathbb{C}^{2n+2}$, while the conical special coordinates
$X^I$ on $N$ are unique up to transformations in 
$Sp(2n+4, \mathbb{R})$. Understanding the geometric meaning of the conification of 
$\hat{N}$ into $N$ requires in particular to relate these two group action to one another.

\subsubsection*{The conification map}

The concepts of Lagrangian pairs and of special K\"ahler pairs, which 
were introduced in section \ref{sect:SK_vector_bundle},
 are needed for defining the conification of ASK manifolds. It turns out that  
the conification map can be formulated such that
it applies to any ASK-manifold, not only to those which can be
obtained using the rigid r-map:
\begin{equation}
\mathrm{con} \;: \{ \mathrm{ASK}\;\;\mathrm{manifolds} \} \rightarrow
\{ \mathrm{CASK}\;\;\mathrm{manifolds} \} \;,\;\;
\hat{N}_{2n} \mapsto N_{2n+2} \;.
\end{equation}
Compared to the previous paragraphs we have shifted $n\mapsto n-1$  in order to stress that this construction is
valid for any ASK manifold.\footnote{The case  $n=0$ can be interpreted as mapping
the zero-dimensional ASK manifold $\{pt\}$  
consisting of a single point
to the CASK manifold $\mathbb{C}$ with its standard flat metric, corresponding to a 
quadratic prepotential.}

 Consider the complex 
symplectic vector space $\mathbb{C}^{2n+2}$ with Darboux coordinates 
$(X^I,W_I)$, where $I=0,\ldots, n$. The vector field $\partial_{W_0}$ is Hamiltonian with moment
map $X^0$ and the symplectic reduction\footnote{See \ref{app:symp_man} for a review
of Hamiltonian vector fields, moment maps and symplectic reductions.} with respect to $\partial_{W_0}$ 
can be identified with the symplectic vector space $\mathbb{C}^{2n}$ with 
Darboux coordinates $(X^a,W_a)$, $a=1,\ldots, n$:
\begin{equation}
\{ X^0= 1 \} / \langle \partial_{W_0}\rangle \cong \mathbb{C}^{2n} \;.
\end{equation}
In section \ref{sect:SK_vector_bundle} we introduced the group $G_\mathbb{C}=
Sp(\mathbb{C}^{2n}) \ltimes \mbox{Heis}_{2n+1} (\mathbb{C})$ which acts
on Lagrangian pairs by the affine representation $\bar{\rho}: G_\mathbb{C} 
\rightarrow \mbox{Aff}_{Sp(\mathbb{C}^{2n})}(\mathbb{C}^{2n})$. As shown 
in \cite{Cortes:2017utn} and reviewed in \ref{app:groups} this affine
representation can be extended to a linear symplectic representation of $G_\mathbb{C}$ 
on $\mathbb{C}^{2n+2}$. Based on this observation, the conification of ASK manifolds
can be formulated locally using Lagrangian pairs and special K\"ahler pairs, 
and then globalized using a principal bundle based on the subgroup $G_{SK}\subset G_{\mathbb{C}}$. 
Recall from section \ref{sect:SK_vector_bundle} that any ASK manifold can be 
described locally by a special K\"ahler pair $(\phi, F)$, that is an embedding
$\phi: \hat{N}\supset U \rightarrow \mathbb{C}^{2n}$
defined by a prepotential $F$, where $\phi=dF$. The special K\"ahler pair $(\phi,F)$ 
determines a Lagrangian pair $(L,f)$, consisting
of a Lagrangian submanifold $L\subset \mathbb{C}^{2n}$ together with a Lagrange potential $f$. 
To describe CASK manifolds in this approach one needs to add the condition that the
embedding  $\phi$ is conical, as defined in section \ref{caskg}. The corresponding
Lagrangian submanifolds are called {\em regular Lagrangian cones}. 
Proposition 3.4 
of \cite{Cortes:2017utn} establishes a one-to-one correspondence 
between Lagrangian pairs in $\mathbb{C}^{2n}$ and regular 
Lagrangian cones in $\mathbb{C}^{2n+2}$, provided by two maps
called conification con and reduction $\mbox{red}=\mbox{con}^{-1}$. The action of the group $G_{SK}\subset G_\mathbb{C}$ is equivariant with 
respect to these maps, which allows to define the conification of special
K\"ahler pairs. Up to the action of $G_{SK}$ the conification works by 
`homogenization' of the prepotential,
\begin{equation}
\label{F_reduction}
F_{\hat{N}}(X^1, \ldots, X^n) \mapsto F_N(X^0,X^1, \ldots X^n) = (X^0)^2 
F_{\hat{N}}(X^1/X^0, \ldots, X^n/X^0) \;.
\end{equation}
Interestingly, only the action of the subgroup $G=Sp(\mathbb{R}^{2n}) \ltimes
\mbox{Heis}_{2n+1}(\mathbb{R})$ preserves the induced K\"ahler metric on 
the Lagrangian cone. This means that the supergravity r-map admits
non-trivial deformations, which at the level of the prepotential correspond
to adding terms of the form
\begin{equation}
\label{delta_F}
\delta F = i (a_{0a} X^0 X^a + c (X^0)^2) \;,\;\;\;a_{0a}, c \in \mathbb{R} \;.
\end{equation}
We will discuss the physical interpretation of these deformations below. Having
defined the conification of special K\"ahler pairs, the extension to the
conification of general ASK manifolds uses the flat $G_{SK}$-principal bundle
of special K\"ahler pairs introduced in section \ref{sect:SK_vector_bundle}.
Roughly speaking, starting with a local conification of $\hat{N}$ using
a special K\"ahler pair $(\phi,F)$ one obtains the global conification $N$ of $\hat{N}$ by 
maximal analytical extension of $(\phi,F)$. We refer to \cite{Cortes:2017utn}
for details.

\subsubsection{The deformed supergravity r-map}

Using the conification $\hat{N}_{2n+2} \mapsto N_{2n+4}$ we 
obtain the ASK/PSK correspondence $\hat{N}_{2n+2} \mapsto \bar{N}_{2n+2}$,
while composing the rigid r-map with the conification map 
we can 
lift the supergravity r-map to the superconformal level,
$M_{n+1} \mapsto N_{2n+4}$. More precisely, while the 
homogenized prepotential \eqref{F_reduction} matches with the result 
of the reduction  of five-dimensional vector multiplets, the conification map
allows to include the non-trivial deformations \eqref{delta_F}. Such terms
are allowed for four-dimensional vector multiplets, but disappear when 
a decompactification limit to five-dimensions is performed
\cite{Antoniadis:1995vz,Louis:2003gj}.
Terms of the form \eqref{delta_F} with $a_{0a}=0$ but $c\not=0$ 
do actually occur in string theory. In type-II compactifications on 
Calabi-Yau three-folds they arise as worldsheet instantons with a coefficient
proportional to the Euler number $\chi$ of the three-fold \cite{Grisaru:1986kw,Hosono:1993qy},
while in heterotic
compactifications on $K3\times T^2$ they are part of the one-loop 
corrections and proportional to an expansion coefficient of a (model dependent)
modular form \cite{Ceresole:1995jg,deWit:1995zg,Antoniadis:1995ct,Harvey:1995fq}. We remark that deformations where $a_{0a}\not=0$ and $c=0$ do not have a 
known realization in string  theory. Note that $\delta F$ in \eqref{delta_F} has purely imaginary
coefficients,
and is therefore distinct from terms of the form $\hat{\delta} F = \frac{1}{24} c_{2I} X^0 X^I$,
which arise in IIA-compactifications, where $c_{2I}$ are the components
of the second Chern class. Terms of the form $\hat{\delta}F$ have real
coefficients, and can be absorbed by a symplectic transformation. Thus they do not
provide a non-trivial deformation, while  \eqref{delta_F} does.

\subsection{Reduction from four to three dimensions: the c-map}

\subsubsection{Reductions to three dimensions \label{Red_to_three}}

Compared to the generic situation considered in section \ref{sect:dimred_general},
reductions to three dimensions have an enhanced number of scalar fields,
because abelian vector fields can be dualized into scalars. Consider the generalized 
Maxwell Lagrangian 
\begin{equation}
L(A) = - \frac{1}{2}  F \wedge * F
\end{equation}
for a $p$-form field strength $F=dA$ in $n=t+s$ dimensions, where 
$t$ is the number of time-like dimensions, more precisely, the number
of negative eigenvalues of the metric. By promoting the Bianchi identity
$dF=0$ to a field equation using a Lagrange multiplier $(n-p-1)$-form 
$B$, and subsequently eliminating $F$ by its algebraic equation of motion,
one arrives, after dropping any boundary terms resulting from integration by parts,
 at the dual Lagrangian
\begin{equation}
\tilde{L}(B) = (-)^t  \frac{1}{2} G \wedge * G\;,
\end{equation}
where $G=dB$ is the Hodge dual of $F$. Note that the sign 
of the generalized Maxwell term flips whenever the number of 
time-like dimensions (negative eigenvalues of the metric)
is even, in particular in Euclidean signature, while it remains the same 
for an odd number of time-like dimensions, in particular in Lorentzian 
signature.\footnote{Irrespective of whether we choose a mostly plus or 
mostly minus convention for the metric, the sign
of the kinetic energy is preserved in Lorentzian and reversed in Euclidean
signature. }

Consider now starting with a four-dimensional action with 
one $\varepsilon_1$-complex scalar and one abelian gauge field.
\begin{equation}
S = \int d^4 x \left( -\frac{1}{2} \partial_\mu X \partial^\mu \bar{X} - \frac{1}{4}
F_{\mu \nu} F^{\mu \nu} \right) \;.
\end{equation}
Upon reduction to three dimensions we end up with four real scalars:
the real and imaginary parts\footnote{Here and in the following `real part' and `imaginary part'
is short for `$\varepsilon$-real part' and `$\varepsilon$-imaginary part,' respectively.} of $X=\sigma + i_{\varepsilon_1} b$, the component 
$p = A_*$ of the four-dimensional vector field along the direction we reduce over,
and the scalar $s$ we gain by dualizing the three-dimensional abelian vector field
$A_m$. For $\varepsilon_1=-1$ we take the four-dimensional theory to have
signature $(-+++)$ and consider both a space-like reduction, $\varepsilon_2=-1$,
and a time-like reduction, $\varepsilon_2=1$. For $\varepsilon_1=1$, we take the
four dimensional theory to have signature $(++++)$,  and only a space-like
reduction, $\varepsilon_2=-1$ is possible.

The corresponding three-dimensional
actions are
\begin{equation}
S = \int d^3 x \left( - \frac{1}{2} \partial_m \sigma \partial^m \sigma 
+ \frac{\varepsilon_1}{2} \partial_m b \partial^m b 
+ \frac{\varepsilon_2}{2} \partial_m p \partial^m p 
- \frac{\varepsilon_1 \varepsilon_2}{2} \partial_m s \partial^m s \right) \;.
\end{equation}
For $\varepsilon_1=\varepsilon_2=-1$ we can combine the four
real scalars into one scalar valued in the quaternions $\mathbb{H}_{-1} := \mathbb{H}$,
\begin{equation}
\label{quaternion}
q = \sigma + i b + j p + k q \;,\;\;\;\bar{q} = \sigma - i b - j p - k q \;,
\end{equation}
where $i,j,k$  anticommute pairwise, and  where $i^2=j^2=k^2=-1$. In the
other cases we can combine them into a scalar valued
in the algebra $\mathbb{H}_1$ of para-quaternions, where two of the
complex units are replaced by para-complex units. For example, for
$\varepsilon_1=-1$, $\varepsilon_2=1$ we can use \eqref{quaternion}
with $j^2=k^2=1$.\footnote{See \ref{app:epsilon_quat} for a brief review
of quaternions, para-quaternions, and the related `$\varepsilon$-quaternionic'
geometric structures.} 
To treat both cases in parallel we use
an $\varepsilon$-quaternionic notation where $\mathbb{H}_{\varepsilon}$
denotes the quaternions for $\varepsilon=-1$ and the para-quaternions
for $\varepsilon=1$.\footnote{We write $\varepsilon_1, \varepsilon_2$ for 
signs related to the four-dimensional Lagrangian and its reduction to
three dimensions, respectively, while using $\varepsilon$ when talking
about $\varepsilon$-complex structures in general.} The resulting action takes the form
\begin{equation}
S = \int d^3 x \left( - \frac{1}{2} \partial_m q \partial^m \bar{q} \right) \;.
\end{equation}
For theories with several interacting scalars this type of rewriting is not
practical, but it illustrates that the target space geometries that one 
obtains by dimensionally reducing four-dimensional vector multiplets
are $\varepsilon$-quaternionic geometries. More specifically, 
when dimensionally reducing vector multiplets and dualizing all the
three-dimensional vector fields, the resulting supersymmetry representations
are hypermultiplets, and the target space geometry is $\varepsilon$-hyper-K\"ahler ($\varepsilon$-HK) 
in the rigid and $\varepsilon$-quaternionic K\"ahler 
($\varepsilon$-QK), a.k.a.
$\varepsilon$-quaternion-K\"ahler  in the
supergravity case. 

\subsubsection{The rigid c-map}

We now turn to the reduction of a four-dimensional bosonic
on-shell vector multiplet Lagrangian of the form 
\eqref{Lagr_4dVM_epsilon}. This section is based on 
\cite{Cortes:2005uq}, to which we refer for details.\footnote{The 
conventions used in \cite{Cortes:2005uq} are slightly different
from those used in this review, which leads to various constant 
rescalings of fields.}  The parameter
$\varepsilon_1=\pm 1$ 
labels the four-dimensional scalar target geometry, which is
affine special K\"ahler for Lorentzian and affine special para-K\"ahler  for Euclidean space-time 
signature, with 
a general $\varepsilon_1$-holomorphic prepotential. 
The second parameter $\varepsilon_2=\pm 1$ distinguishes
between space-like reduction and time-like reduction, where the
latter is only possible if we start in Lorentzian signature. 
After dualization of the three-dimensional vector fields, the Lagrangian
takes the form
\begin{eqnarray}
L &=& - N_{IJ} \partial_m X^I \partial^m \bar{X}^J 
+ \varepsilon_2 (N_{IJ} - \varepsilon_1 R_{IK} N^{KL} R_{LJ} ) \partial_m p^I \partial^m p^J \\
&& + 4 \varepsilon_1 \varepsilon_2 R_{IK} N^{KJ} \partial_m p^I \partial^m s_J 
- 4 \varepsilon_2 \varepsilon_2 N^{IJ} \partial_m s_I \partial^m s_J  \;.\nonumber
\end{eqnarray}
Here $p^I \propto A^I_*$ are the scalar components of the four-dimensional vector fields
and $s_I$ the scalars obtained from dualizing the three-dimensional vector fields. 

The target space geometry of the three-dimensional theory is 
hyper-K\"ahler for $\varepsilon_1=\varepsilon_2=-1$ 
\cite{Cecotti:1988qn}
and para-hyper-K\"ahler for $\varepsilon_1 \varepsilon_2 = -1$ \cite{Cortes:2005uq}.
It is possible to combine the real fields $(p^I,s_I)$ into $\varepsilon$-complex
coordinates $W_I$ (where $\varepsilon=-\varepsilon_1 \varepsilon_2$)
and to make the $\varepsilon$-hyper-K\"ahler geometry 
of the target space manifest by finding explicit expressions for 
the three $\varepsilon$-complex structures and for an $\varepsilon$-K\"ahler
potential in terms of the special geometry data of the four-dimensional theory
\cite{Cortes:2005uq}. 
Alternatively, one can work in real coordinates. 
The $\varepsilon_1$-complex version of the expression \eqref{Hessian_RN} for the Hessian 
metric on the four-dimensional scalar target space is
\begin{equation}
(H_{ab} ) = \left( \begin{array}{cc}
N_{IJ} - \varepsilon_1 R_{IK} N^{KJ} R_{JL} & 2 \varepsilon_1 R_{IK} N^{KJ} \\
2 \varepsilon_1 N^{IK} R_{KJ} & - 4 \varepsilon_1 N^{IJ} \\
\end{array} \right)  \;.
\end{equation}
Replacing the $\varepsilon_1$-complex scalars $X^I$ by special real
coordinates $q^a$, and combining the remaining real scalars into the symplectic
vector $\hat{q}^a = (p^I, s_I)$, we obtain
\begin{equation}
L = - H_{ab}(q) \partial_m q^a \partial^m q^b + \varepsilon_2 H_{ab}(q) \partial_m \hat{q}^a 
\partial^m \hat{q}^b \;.
\end{equation}
From this expression it is manifest that we can interpret the target space $N$
of the three-dimensional theory as the tangent bundle $N=TM$ of the 
AS$\varepsilon_1$K target manifold $M$ of the four-dimensional theory, 
equipped with the Sasaki-like metric
\begin{equation}
ds_{N=TM}^2 = H_{ab} (dq^a dq^b - \varepsilon_2 d\hat{q}^a d\hat{q}^b)\;.
\end{equation}
Similar to the case of the rigid r-map, the special connection $\nabla$ 
of the AS$\varepsilon_1$K manifold $M$ can be used to perform
a canonical splitting of $TN$ into a horizontal and a vertical distribution. 
Moreover, the special geometry data of $M$ can be used to show that 
$N=TM$ globally carries the structure of an $\varepsilon$-HK manifold. 
The map induced by dimensional 
reduction of four-dimensional vector multiplets is called the rigid 
$c$-map:
\begin{equation}
c_{\varepsilon_1, \varepsilon_2} \;: 
\{ \mathrm{AS}\varepsilon_1\mathrm{K}\;\;\mathrm{manifolds} \} 
\rightarrow
\{ \varepsilon-\mathrm{HK} \;\;\mathrm{manifolds} \} \;,\;\;\;
M_{2n} \mapsto N_{4n} \cong TM \;.
\end{equation}
Depending on $\varepsilon_1, \varepsilon_2$ there are three subcases:
\begin{enumerate}
\item
The spatial c-map, or simply, the (rigid) c-map: $\varepsilon_1=\varepsilon_2=-1$,
and $\varepsilon= -\varepsilon_1 \varepsilon_2 =1$. This corresponds to
the standard, space-like reduction of vector multiplets in Lorentzian signature,
and was first described in \cite{Cecotti:1988qn}.
All involved scalar target space geometries are positive definite.\footnote{That is,
if we impose positive kinetic energy for all fields. Mathematically we can 
also consider scalar target spaces with indefinite metrics.}
\item
The temporal c-map, $\varepsilon=-1, \varepsilon_2 = 1$ and 
$\varepsilon = - \varepsilon_1 \varepsilon_2=1$. This corresponds to the
time-like reduction of a Lorentzian vector multiplet theory and relates a
positive definite scalar geometry to one with neutral signature. 
\item
The Euclidean c-map, $\varepsilon_1=1, \varepsilon_2 =-1$ and 
$\varepsilon=-\varepsilon_1 \varepsilon_2 =1$. This corresponds to the
space-like reduction of a Euclidean vector multiplet theory and relates
two target space geometry with neutral signature.
\end{enumerate}
We remark that instead of setting $N=TM$, we can alternatively 
take $N=T^*M$, since the metric allows to identify tangent 
spaces with cotangent spaces. Then
\begin{equation}
ds_{N=T^*M}^2 = H_{ab} dq^a dq^b - \varepsilon_2 H^{ab} d\hat{q}_a d\hat{q}_b \;,
\end{equation}
where $H^{ab}$ is the inverse of $H_{ab}$ and $d\hat{q}_a = H_{ab} dq^b$.\footnote{The integrability condition for the local 
existence of the functions $\hat{q}_a$, which are fibre coordinates on $T^*M$, 
follows from $H_{ab}$ 
being the components of a Hessian metric on $M$.}
Thus the cotangent bundle of an AS$\varepsilon_1$K manifold is 
an $\varepsilon$-HK manifold \cite{Cecotti:1988qn,Cortes:2005uq}. 
This is a stronger result than for generic K\"ahler manifolds, where it is
known that the cotangent bundle admits the structure of an HK manifold locally,  in a 
neighbourhood of its zero section  \cite{1997alg.geom.10026K,Gates:1999ea}.

\subsubsection{The supergravity c-map and its deformation}

We finally turn to the reduction of four-dimensional vector multiplets coupled 
to supergravity to three dimensions. Our starting point is the bosonic
on-shell Lagrangian \eqref{4dVM_onshell_epsilon} in Lorentzian or
Euclidean space-time signature, with a general $\varepsilon_1$-holomorphic
prepotential. The four-dimensional metric is decomposed according to
\begin{equation}
ds_4^2 = g_{\mu \nu} dx^\mu dx^\nu = - \varepsilon_2 e^\phi (dx^* + V_m dx^m)^2 
+ e^{-\phi} g_{mn} dx^m dx^n  \;,
\end{equation}
where $V_m$ is the Kaluza-Klein vector and $\phi$ the Kaluza-Klein scalar. After
reduction to three dimensions, all abelian vector fields are dualized into scalars. 
The bosonic field content of the resulting three-dimensional theory is:
\begin{itemize}
\item
The three-dimensional metric $g_{mn}$.
\item
The $n= n_{(4)}$ $\varepsilon_1$-complex four-dimensional scalars $z^A$, where
$n_{(4)}$ is the number of four-dimensional vector multiplets.
\item
The $n+1$ real scalars $\zeta^I \propto A^I_*$ obtained by reducing the $n+1$
four-dimensional vector fields $A^I_\mu$.
\item
The $n+1$ real scalars $\tilde{\zeta}_I$ obtained by dualizing the $n+1$
three-dimensional vector fields $A^I_m$.
\item
The Kaluza-Klein scalar $\phi$ and the scalar $\tilde{\phi}$ obtained by 
dualizing the Kaluza-Klein vector $V_m$. 
\end{itemize}
The three-dimensional metric does not carry local degrees of freedom while the 
$4n+4$ real scalars $\mbox{Re}(z^A), \mbox{Im}(z^A) , \zeta^I, \tilde{\zeta}_I, \phi, \tilde{\phi}$
are the bosonic components of $4n+4$ hypermultiplets, coupled to three-dimensional 
Poincar\'e supergravity. The three-dimensional Lagrangian is 
\cite{Ferrara:1989ik,Cortes:2015wca}
\begin{eqnarray}
L_3^{(\varepsilon_1, \varepsilon_2)}  
&=& \frac{1}{2} R_3 - \bar{g}_{A\bar{B}}
\partial_m z^A \partial^m \bar{z}^{\bar{B}} - \frac{1}{4} \partial_m \phi \partial^m \phi 
 \label{lag_3d_local}   \\
& & + \varepsilon_1 e^{-2\phi}
\left[ \partial_m \tilde{\phi} + \frac{1}{2}
\left( \zeta^I \partial_m \tilde{\zeta}_I - \tilde{\zeta}_I \partial_m \zeta^I \right) \right]^2
\nonumber  \\ &&-
 \frac{\varepsilon_2}{2} e^{-\phi} \left[ 
{\cal I}_{IJ} \partial_m \zeta^I \partial^m \zeta^J - \varepsilon_1 {\cal I}^{IJ}
\left( \partial_m \tilde{\zeta}_I - {\cal R}_{IK} \partial_m \zeta^K\right)^2
\right]\;. \nonumber
\end{eqnarray}

The target space geometry of hypermultiplets coupled to supergravity is quater\-nio\-nic-K\"ahler
\cite{Bagger:1983tt}.
For Euclidean hypermultiplets obtained by dimensional
reduction the target space geometry is para-quaternionic K\"ahler \cite{Cortes:2003zd,Cortes:2015wca}.
The map between scalar geometries induced by dimensional reduction of 
four-dimensional vector multiplets coupled to supergravity is called the
supergravity c-map:
\begin{equation}
\bar{c}_{(\varepsilon_1,\varepsilon_2)} \;: \{ \mathrm{PS}\varepsilon_1\mathrm{K}\;\;
\mathrm{manifolds} \} \rightarrow \{ \varepsilon-\mathrm{QK}\;\;\mathrm{manifolds} \}
\;,\;\;\bar{M}_{2n} \mapsto \bar{N}_{4n+4}\;.
\end{equation}
The properties of the three types of supergravity c-maps are summarized in Table \ref{Table:c-maps}.
\begin{table}
\begin{tabular}{l|l|l|l}
c-map  & Space-time & scalar geometry & scalar manifold signature\\
 & signature & & \\ \hline
spatial& $(1,3) \mapsto (1,2)$ & PSK $\mapsto$ QK & $(2p,2q) \mapsto (4p+4, 4q) $ \\
temporal  & $(1,3) \mapsto (0,3)$ & PSK $\mapsto$ PQK & $(2p,2q) \mapsto (2d,2d)$ \\
Euclidean & $(0,4) \mapsto (0,3)$ & PSPK $\mapsto$  PQK & $(r,r) \mapsto (2d,2d)$\\
\end{tabular}
\caption{This table summarizes the relations between the space-time signatures, 
target space geometries and target space-signatures for the 3 types of supergravity
c-maps. We include the case where the PSK manifold has indefinite signature,
which is mathematically well defined, but corresponds to a vector multiplet theory
where some of the fields have negative kinetic energy. In this case the QK manifold
obtained by the spatial supergravity c-map is also indefinite. Para-K\"ahler and 
para-QK manifolds always have neutral signature. 
Manifolds of dimension $2n$ map to manifolds of dimension $4n+4$, therefore
$d=p+q+2$ in row 2 and $d=r+1$ in row 3. \label{Table:c-maps}}
\end{table}

Showing that the scalar target manifold $\bar{N}_{4n+4}$  of the Lagrangian \eqref{lag_3d_local}
is $\varepsilon$-quaternionic K\"ahler is somewhat involved, in particular if one 
wants to have a global description of $\bar{N}_{4n+4}$. There are various
ways to describe the geometry of $\bar{N}$, which we discuss in turn.

\subsubsection*{Supergravity c-map spaces as group bundles}

The first description of the geometry of $\bar{N}$ 
 is based on an observation of \cite{Ferrara:1989ik} 
for the case  $\varepsilon_1=\varepsilon_2=-1$:
when restricting to constant values of $z^i$, 
the metric on the corresponding subspace is K\"ahler, only depends on the 
number of vector multiplets, and is in fact the metric of a Riemannian symmetric
space. It was shown in \cite{Cortes:2011aj} that if the underlying PSK
manifold $\bar{M}$ is a PSK domain, then the image under the supergravity c-map 
is a QK domain of the form $\bar{N} = \bar{M}\times G$, where $G$ is a solvable
Lie group, and where the QK metric $g_{\bar{N}}$ is a `bundle metric' 
$g_{\bar{N}} = g_{\bar{M}} + g_{G}(p)$, where $g_G(p)$ is a family of 
left-invariant metric on $G$ parametrized by $p\in \bar{M}$. It was also 
shown in \cite{Cortes:2011aj} that this construction can be `globalized,' that 
is one can apply the supergravity c-map domain-wise and then glue together
the resulting QK domains consistently and uniquely to obtain a QK manifold.
Moreover, it was shown that the supergravity c-map preserves 
geodesic (and hence metric) completeness, that is, if $\bar{M}$ is complete
so is its image $\bar{N}$ under the supergravity c-map. 
Except for the completeness result (which heavily relies on the involved metrics
being definite), the description of $\bar{N}$ by gluing domains
should also apply to the case where $\varepsilon_1
\varepsilon_2=-1$. In \cite{Cortes:2015wca} it was shown that 
the image of a PS$\varepsilon_1$K
domain $\bar{M}$ under $c_{(\varepsilon_1, \varepsilon_2)}$ takes the
form $\bar{N}=\bar{M} \times G$, with a bundle metric $g_{\bar{N}} = g_{\bar{M}} + g_G(p)$,
where $G$ is a solvable Lie group. The solvable Lie groups $G$ and left-invariant 
metrics on $G$ were found to be the following:
\begin{enumerate}
\item
$\varepsilon_1=\varepsilon_2=-1$. This is the standard (spatial) supergravity c-map which
was already considered in \cite{Ferrara:1989ik}.
The solvable Lie group is the Iwasawa subgroup of $U(n+2,1)$ and can be identified globally 
with the complex hyperbolic space 
\begin{equation}
\mathbb{C}H^{n+2} = \mathrm{U}(n+2,1)/ ( U(n+2) \times U(1)) 
\end{equation}
equipped with a positive definite K\"ahler metric of constant holomorphic sectional
curvature $-1$.\footnote{The $\varepsilon$-holomorphic sectional curvature of an 
$\varepsilon$-K\"ahler manifold $(M,J,g)$ is $\langle(R(X,JX) JX, X\rangle/\langle X \wedge
JX, X \wedge X\rangle$, where $X$ is a vector field, where  $\langle\cdot ,\cdot \rangle$ is
the scalar product between tensors induced by the metric, and where $R$ is the curvature tensor.
It can be interpreted as
the sectional curvature of the $\varepsilon$-complex line $X\wedge JX$ \cite{Ballmann,Cortes:2015wca}. } 
The metric on the resulting QK manifold $\bar{N} = \bar{M} \times G$
is positive definite.\footnote{We assume, here and in the following case, that the target space metric of 
the four-dimensional vector multiplet theory is positive definite. }
\item
$\varepsilon_1=-1$, $\varepsilon_2=1$. This is the temporal supergravity c-map. 
The group $G$ is again the Iwasawa subgroup of $U(n+2,1)$, but with a 
different, indefinite left-invariant metric. It can be identified locally with the
indefinite complex hyperbolic space
\begin{equation}
\mathbb{C}H^{1,n+1} \cong U(1,n+2) / ( U(1,n+1) \times U(1)) 
\end{equation}
equipped with a pseudo-K\"ahler metric of complex signature $(1,n+1)$ and 
constant holomorphic sectional curvature $-1$. Note that for non-compact
symmetric spaces of indefinite signature the Iwasawa subgroup does not
act transitively, so that we cannot identify $G$ globally with the above symmetric
space. However, it can be shown that $G$ acts with an open orbit, thus allowing the
identification of $G$ with an open subset of the symmetric space. The signature of the resulting
space $\bar{M}\times G$ is neutral $(2n+2, 2n+2)$, as required for a 
para-quaternionic K\"ahler manifold.
\item
$\varepsilon_1=1$, $\varepsilon_2=-1$. This is the Euclidean supergravity
c-map. The solvable Lie group $G$ is the Iwasawa subgroup of $SL(n+3,\mathbb{R})$ 
and can be identified locally with para-complex hyperbolic space
\begin{equation}
CH^{n+2} \cong SL(n+3,\mathbb{R}) / S (GL(1) \times GL(n+2))
\end{equation}
equipped with a para-K\"ahler metric of real signature $(n+2,n+2)$ and
of constant para-holomorphic sectional curvature $-1$. The signature
of $\bar{M}\times G$ is $(2n+2,2n+2)$, as required for a para-quaternionic
K\"ahler manifold.
\end{enumerate}
The simplest examples for $\bar{N}$ are the `universal hypermultiplets' obtained
by reducing pure four-dimensional ${\cal N}=2$ supergravity. In this case 
$\bar{M} = \{ pt \}$ and $\bar{N}$ is locally isometric to one of the 
following manifolds:
\begin{enumerate}
\item
For the spatial supergravity c-map, $\varepsilon_1=\varepsilon_2=-1$, the
target space is globally isometric to
\begin{equation}
\mathbb{C}H^2 \cong SU(2,1)/(U(2)\times U(1)) \;.
\end{equation} 
This is the `universal hypermultiplet' which is obtained by the reduction of
pure ${\cal N}=2$ supergravity. In general c-map spaces the universal 
hypermultiplet spans a distinguished subspace. Note however, that 
once string corrections to the hypermultiplet metric are taken into account
the universal hypermultiplet ceases to be an identifiable, `universal' part of
the scalar manifold \cite{Aspinwall:2000fd}.
\item
For the temporal supergravity c-map, $\varepsilon_1=-1, \varepsilon_2 =1$, 
the target space is locally isometric to
\begin{equation}
\mathbb{C}H^{1,1} \cong U(2,1)/(U(1,1)\times U(1)) \;.
\end{equation}
This is target space for a time-like reduction of pure 
${\cal N}=2$ supergravity. 
\item
For the Euclidean supergravity c-map, the target space is
\begin{equation}
CH^2 \cong SL(3,\mathbb{R}) / S(GL(1) \times GL(2)) \;.
\end{equation}
This target space does not only arise in the dimensional reduction of
pure ${\cal N}=2$ Euclidean supergravity \cite{Theis:2001ef}, but
also when dualizing the so-called double tensor multiplet in 
Euclidean signature \cite{Theis:2002er}. This reflects that Euclidean
actions which differ by sign flips can be related by using that 
dimensional reduction/lifting, Wick rotation and Hodge dualization do 
not commute which each other, see also \cite{Cortes:2009cs}, as 
discussed in section \ref{Sect:Euc_disc}.
\end{enumerate}

\subsubsection*{Conification of $\varepsilon$-HK manifolds}

We now turn to another way of describing the scalar manifold $\bar{N}$.
As in the case of the supergravity r-map, one can lift the supergravity c-map to
the superconformal level. Within the superconformal formalism, it is not 
possible to formulate hypermultiplets off-shell with a finite number of auxiliary fields. 
However, as long as the hypermultiplet manifold has sufficiently many isometries,
hypermultiplets can be dualized into tensor multiplets, which admit a 
superconformal off-shell representation \cite{deWit:2006gn}.
Alternatively, the projective superspace formalism
can be used to describe hypermultiplets off-shell,
see also section \ref{sec:twistor_approach}.
 Off-shell formulations 
of the supergravity c-map were obtained in \cite{Rocek:2005ij} using projective
superspace, and in
\cite{deWit:2007qz,Banerjee:2015uee} using the superconformal formalism.

We will 
review the global geometric construction of the supergravity c-map given in 
\cite{Cortes:2015wca}, which is inspired by the superconformal approach and which 
provides all the data necessary for describing 
the theory at the superconformal level. This description also allows a 
complete and relatively short proof that spaces in the image of the
supergravity c-map are global 
$\varepsilon$-quaternionic K\"ahler manifolds. Moreover, this proof also applies 
to a one-parameter family of non-trivial deformations of the metric obtained from 
the supergravity c-map. 

When working with hypermultiplets 
the situation regarding the scalar target spaces is the 
$\varepsilon$-quaternionic analogon of the real and complex settings for five- and four-dimensional
vector multiplets. To each $\varepsilon$-QK manifold $\bar{N}_{4n+4}$ describing 
$n+1$ hypermultiplets coupled to Poincar\'e supergravity, one can associate
an $\varepsilon$-HK cone $N_{4n+8}$, that is an $\varepsilon$-HK manifold
with a homothetic action of the group $\mathbb{H}^*_\varepsilon$ of 
invertible $\varepsilon$-quaternions, such that $\bar{N} \cong N/\mathbb{H}^*_\varepsilon$.
Conversely $N$ is an $\mathbb{H}^*_\varepsilon$-bundle over $\bar{N}$. 
We remark that while it would be more in line with our terminology for vector multiplets
to use the term `conical $\varepsilon$-HK manifold', we follow the literature in 
using `$\varepsilon$-HK cone' instead. 

One can obtain the superconformal lift 
$\tilde{c}: M_{2n+2} \mapsto N_{4n+8}$ of the supergravity c-map 
$c: \bar{M}_{2n} \mapsto \bar{N}_{4n+4}$ by composing the rigid
c-map $c: M_{2n+2} \mapsto \hat{M}_{4n+4} \cong TM$ with 
a conification map $\mathrm{con}: \hat{M}_{4n+4} \mapsto M_{4n+8}$. 
The situation is summarized by the following diagram:
\begin{equation}
\label{Diagram:c-map}
\xymatrix{
{M} \ar@{|->}[r]^{c} \ar@{|->}[d]_{\mathrm{SC}} \ar@/^1.5pc/[rrrr]^{\tilde{c}} 
& \hat{N}  \ar@{|->}[rrr]^{\mathrm{con}} \ar@{|->}[drrr]^{\varepsilon\mathrm{HK/QK}}
& & & N  \ar@{|->}[d]^{\mathrm{SC}}\\
\bar{M} \ar@{|->}[rrrr]_{\bar{c}} & & &  &\bar{N} \\
}
\end{equation}
This diagram induces a correspondence between 
$\varepsilon$-HK and $\varepsilon'$-QK manifolds of the same
dimension.\footnote{Note that $\varepsilon\not= \varepsilon'$ can occur, 
see Table \ref{Table:c-maps}.}

\subsubsection*{The $\varepsilon$-HK/QK correspondence}

The correspondence can be formulated independently 
of the supergravity c-map, and then also applies to $\varepsilon$-HK manifolds 
$\hat{N}$ which 
are not in the image of the rigid c-map, but specify the conditions stated below. 
The resulting $\varepsilon$-HK/QK correspondence
generalizes the HK/QK correspondence of \cite{Haydys:2008}, which was 
applied to the space-like supergravity c-map in \cite{Alexandrov:2011ac} in the context of
the twistor approach, see also section \ref{sec:twistor_approach} below. We follow 
\cite{Alekseevsky:2012fu,Alekseevsky:2013nua,Dyckmanns:2016qoe},
who have extended the HK/QK correspondence to arbitrary signature and to the para-complex
setting. 

For an $\varepsilon$-HK manifold $\hat{N}$ 
to admit a conification $N$
the following conditions must hold: 
\begin{enumerate}
\item
$\hat{N}$ admits a time-like or space-like Killing vector field $Z$, which
is $\varepsilon$-holomorphic with respect to an $\varepsilon$-complex structure $J_1$,
which is part of $\varepsilon$-HK structure. The Killing vector field $Z$ is 
Hamiltonian with respect to the corresponding $\varepsilon$-K\"ahler form 
$\omega_1$, that is, there exists a function  $f$ such that $df = -\omega_1(Z\cdot, \cdot)$. 
\item
The functions $f$ and  $f_1 = f - \frac{1}{2} g(Z,Z)$ are nowhere zero.
\item
The Killing vector field $Z$ rotates the other two $\varepsilon$-complex structures, $J_2$ and $J_3$,
of the $\varepsilon$-HK structure, 
that is, $L_Z J_2 = 2 \varepsilon J_3$.
\end{enumerate}
Having constructed an $\varepsilon$-HK cone $N_{4n+8}$, one obtains
a corresponding $\varepsilon$-QK manifold $N_{4n+4}$ by a superconformal
quotient.\footnote{Since the construction involves the moment map of $Z$ explicitly, 
the correspondence allows a one-parameter deformation to be discussed in 
\ref{sect:deformed_sugra_c-map}.} 
Conversely, given any $\varepsilon$-QK manifold $\bar{N}_{4n+4}$, there always
exist the associated $\varepsilon$-HK cone (which for $\varepsilon=-1$ is also 
known as the Swann bundle) $N_{4n+8}$.
One can then obtain an $\varepsilon$-HK manifold $N_{4n+4}$ by taking an 
$\varepsilon$-HK quotient, provided that $N_{4n+8}$ admits a tri-holomorphic
Killing vector field $X_N$ which commutes with the Euler field $\xi$ of the cone 
$N_{4n+8}$, acts freely and satisfies a technical condition regarding the level
sets of its moment map. The existence of such a vector field follows from the 
existence of a space-like or time-like Killing vector field $X$ on $\bar{N}$,
again subject to a technical condition.\footnote{We refer to \cite{Dyckmanns:2016qoe} for
the details.}

It turns out that the $\varepsilon$-HK/QK correspondence 
can be formulated without using the $\varepsilon$-HK cone 
$N_{4n+8}$ explicitly. Roughly 
speaking, three of the four extra dimensions of the cone do not play
an essential role, so that one can take a shortcut and 
relate $\hat{N}_{4n+4}$ and $\bar{N}_{4n+4}$
via a manifold $P_{4n+5}$ of real dimension $4n+5$.
 The manifold $P_{4n+5}$
is a rank one principal bundle over $N_{4n+4}$ with principal action 
generated by a vector field $X_P$,
and simultaneously a rank one principal
bundle over $\bar{N}_{4n+4}$ with principal action generated by a 
vector field $Z_P$. The vector fields $X_P$ and $Z_P$ 
are lifts of the Killing vector fields $X$ on $\bar{N}_{4n+4}$ and 
$Z$ on $\hat{N}_{4n+4}$ that we mentioned before. The manifold $P_{4n+5}$ 
is a submanifold of the $\varepsilon$-HK cone $N_{4n+8}$, and taking
quotients of $P_{4n+5}$ with respect to the principal actions of
$X_P$ and $Z_P$ is consistent with taking an $\varepsilon$-HK quotient
and an $\varepsilon$-QK quotient of $N_{4n+8}$, respectively. 
The situation is summarized in the following diagram:
\[
\xymatrix{
 & (N,X_N, Z_N) \ar@/_1pc/[ddl]_{/\mathbb{H}^*_\varepsilon}  
  \ar@/^1pc/[ddr]^{/\mathbb{H}^*_\varepsilon}& \\
 & (P,X_P,Z_P)  \ar[dl]_{/\langle X_P\rangle} \ar[dr]^{/\langle Z_P\rangle}
 \ar@{^{(}->}[u]&  \\
(\hat{N}, Z) \ar@<0.5ex>[rr]^{\varepsilon\mathrm{HK}/\mathrm{QK}}& & 
\ar@<0.5ex>[ll]^{\varepsilon\mathrm{QK}/\mathrm{HK}}(\bar{N},X) \\  
}
\]
Explicit expressions for the all relevant geometric data on $\hat{N}, \bar{N}, N$ and $P$ 
can be found in \cite{Alekseevsky:2012fu,Alekseevsky:2013nua,Dyckmanns:2016qoe}.

\subsubsection*{A symplectic parametrization of the supergravity c-map}

The space $P_{4n+5}$ appears naturally in the dimensional reduction 
of four-dimensional vector multiplets, if we use special real coordinate
for the CASK manifold $M$ and insist on maintaining manifest symplectic
 invariance after dimensional reduction. This leads to 
a reformulation of \eqref{lag_3d_local} in terms of a gauged sigma model
with target space $P_{4n+5}$, which is equivalent to a sigma model
with target space $\bar{N}_{4n+4}$ \cite{Mohaupt:2011aa}.

This reformulation requires a couple of steps. 
First we 
replace the four-dimensional scalars $z^A$ by the projective scalars $X^I$ which
take values in $M_{2n+2}$:
\begin{equation}
\bar{g}_{A\bar{B}} \partial_m z^A \partial^m \bar{z}^{\bar{B}} =
\tilde{g}^{(0)}_{IJ} \partial_m X^I \partial^m \bar{X}^J \;.
\end{equation}
 Here $\tilde{g}^{(0)}_M=\pi^* \bar{g}_{\bar{M}}$ is the lift of the PSK metric to the
 CASK manifold. Since $\tilde{g}^{(0)}_M$ has a two-dimensional kernel, this rewriting does not increase 
 the number of propagating degrees of freedom. The right hand side can be viewed as
 a gauged sigma model, where the connection gauging the action of $\mathbb{C}^*_{\varepsilon_1}
 \cong \mathbb{R}^{>0} \times U(1)_{\varepsilon_1}$ on $M$ 
 has been integrated out.\footnote{This
 proceeds by imposing the K-gauge  $b_\mu=0$ on \eqref{covX} and then eliminating
the $U(1)$ gauge field  $A_\mu$ by its equation of motion, see section \ref{4dvecmult}.}
 Here 
 \begin{equation}
 U(1)_{\varepsilon_1} = \left\{ \begin{array}{ll}
 U(1)\;, & \mbox{for}\;\; \varepsilon_1=-1 \;, \\
 GL(1,\mathbb{R}) \;, &\mbox{for}\;\;\varepsilon_1 =1\;, \\
 \end{array} \right.
 \end{equation}
 is the $\varepsilon_1$-unitary group which is part of the R-symmetry group of the
 supersymmetry algebra. Rewriting the vector field couplings in terms of $X^I$ is trivial 
 since the matrix ${\cal N}_{IJ} = 
 {\cal R}_{IJ} + i_{\varepsilon_1} {\cal I}_{IJ}$ is homogeneous of degree zero.

 The second step is to make the field redefinition $Y^I := e^{\phi/2} X^I$, 
 which absorbs the Kaluza-Klein scalar $\phi$ into the superconformal scalars $X^I$.
If we impose the D-gauge on $X^I$, then $\phi$ can be expressed as a function 
of the new scalars $Y^I$:
\begin{equation}
-i_{\varepsilon_1} (X^I \bar{F}_I - F_I \bar{X}^I ) =  1 \Rightarrow 
-i_{\varepsilon_1} (Y^I \bar{F}_I - F_I \bar{Y}^I ) =  e^\phi \;.
\end{equation}
From now on we do not regard $\phi$ as an independent field, 
but as a function of the fields $Y^I$. Since the fields $Y^I$ are
subject to $U(1)_{\varepsilon_1}$-gauge transformations, 
the $(n+1)$ $\varepsilon_1$-complex scalars $Y^I$ represent
$2n+1$ propagating degrees of freedom. Geometrically, 
$2n$ scalars correspond to excitations transverse to the
$\mathbb{C}^*_{\varepsilon_1}$-action on $M$ and thus to
the independent four-dimensional scalars $z^i$, while the additional scalar corresponds to 
the radial direction of the real cone $M = \mathbb{R}^{>0} \times S$,
where $S$ is the $\varepsilon_1$-Sasakian submanifold of $M$ defined by the D-gauge. 

The third step is to use special real coordinates on $M$. Since 
$Y^I$ can be interpreted as special $\varepsilon$-holomorphic coordinates
on $M$, we can define associated special real coordinates $q^a=(x^I,y_I)$,
\begin{equation}
Y^I = x^I + i_{\varepsilon_1} u^I (x,y) \;,\;\;\;
F_I(Y) = y_I + i_{\varepsilon_1} v_I (x,y)\;,
\end{equation}
which compared to the usual special real coordinates have been rescaled by a 
factor $e^{\phi/2}$ involving the Kaluza-Klein scalar $\phi$. 

The fourth step is to express the vector field coupling matrix ${\cal N}_{IJ}$ 
in terms of the tensor field $\hat{H}_{ab}$ using \eqref{H_hat_cal_N}. 
Finally,  instead of using the tensors $H_{ab}$ and $\hat{H}_{ab}$ it is convenient to express all couplings 
in terms of the Hessian metric
\begin{equation}
\tilde{H}_{ab} = \partial^2_{a,b} \left[ - \frac{1}{2} \log (-2H) \right] =
- \frac{1}{2H} H_{ab} + \frac{1}{2H^2} H_a H_b 
\end{equation}
where
\begin{equation}
H(q^a) = -\frac{1}{2} e^\phi = \frac{i_\varepsilon}{2} (Y^I \bar{F}_I - F_I \bar{Y}^I)
\end{equation}
is the Hesse potential for the CAS$\varepsilon_1$K-metric on $M$.
Defining $\hat{q}^a := \frac{1}{2} (\zeta^I, \tilde{\zeta}_I)$ and 
\begin{equation}
(\Omega_{ab}) = \left( \begin{array}{cc}
0 & \mathbbm{1}_{n+1} \\
- \mathbbm{1}_{n+1} & 0 \\
\end{array} \right) 
\end{equation}
we can rewrite \eqref{lag_3d_local} in the form \cite{Mohaupt:2011aa,Cortes:2015wca}
\begin{eqnarray}
\label{lag:3d_gauged} 
 L_3^{(\varepsilon_1, \varepsilon_2)} 
&=& \frac{1}{2} R_3 - \tilde{H}_{ab} \left( \partial_m q^a \partial^m q^b - \varepsilon_2
\partial_m \hat{q}^a \partial^m \hat{q}^b \right)\nonumber  \\
&& + \frac{\varepsilon_1}{H^2} \left( q^a \Omega_{ab} \partial_m q^b \right) 
\left( q^a \Omega_{ab} \partial^m q^b \right)  \nonumber \\
&& - \frac{2\varepsilon_1 \varepsilon_2}{H^2} \left( q^a \Omega_{ab} \partial_m \hat{q}^b \right)
\left( q^a \Omega_{ab} \partial^m \hat{q}^b \right) \nonumber \\
&& + \frac{\varepsilon_1}{4 H^2}
 \left( \partial_m \tilde{\phi} + 2 \hat{q}^a \Omega_{ab} \partial_m \hat{q}^b \right)
  \left( \partial^m \tilde{\phi} + 2 \hat{q}^a \Omega_{ab} \partial^m \hat{q}^b \right) \;.
\end{eqnarray}
This is a non-linear sigma model for $4n+5$ real scalars $q^a, \hat{q}^a, \tilde{\phi}$ 
coupled to gravity. Its target space  $P_{4n+5}$ is the total space of the
rank one principal bundle $\pi: P_{4n+5} \rightarrow \bar{N}_{4n+4}$ which occurs when 
constructing the supergravity c-map using the $\varepsilon$-HK/QK correspondence. 
Since the scalar fields $q^a$ are subject to $U(1)_{\varepsilon_1}$ gauge transformations,
there are only $4n+4$ propagating degrees of freedom. 
The symmetric tensor 
\begin{eqnarray}
g_P &=& \tilde{H}_{ab} (dq^a dq^b - \varepsilon_2 d\hat{q}^a d\hat{q}^b) 
- \frac{\varepsilon_1}{H^2} (q^a \Omega_{ab} dq^b)^2 
+ \frac{2\varepsilon_1 \varepsilon_2}{H^2} (q^a \Omega_{ab} d\hat{q}^b)^2 
 \nonumber \\
&&- \frac{\varepsilon_1}{4H^2} (d\tilde{\phi}^2
 + 2 \hat{q}^a \Omega_{ab} d\hat{q}^b)^2 \label{g_P}
\end{eqnarray}
defined by the Lagrangian \eqref{lag:3d_gauged} 
has a one-dimensional kernel and 
is projectable with respect to the $U(1)_{\varepsilon_1}$-action. Thus \eqref{lag:3d_gauged}
is  a gauged non-linear sigma model (with the $U(1)_{\varepsilon_1}$-connection integrated out),
and defines, by projection onto orbits,  a non-linear sigma model with target space $\bar{N} = P/U(1)_{\varepsilon_1}$ and $\varepsilon$-QK metric $g_{\bar N}$, where $g_P = \pi^* g_{\bar{N}}$. 

As explained in section \ref{sec:PSR}, 
there is no natural choice of an  $U(1)_{\varepsilon_1}$-gauge
which realizes the PSK manifold 
$\bar{M}$ canonically as an embedded submanifold of the CASK manifold $M$, because
the distribution orthogonal to the $U(1)_{\varepsilon_1}$-action is not integrable. 
Similarly, there is no preferred way to identify $\bar{N}$ with a submanifold of $P$. 
Instead of making a conventional
choice, it is possible and advantageous to work with the $P$-valued gauged sigma
model. The coordinates we have constructed on $P$ are either symplectic vectors,
$q^a, \hat{q}^a$, or symplectic scalars, $\tilde{\phi}$. Fixing a $U(1)_{\varepsilon_1}$ 
gauge requires to impose a condition on $q^a$ and symplectic covariance is lost. 
However for many purposes, including to prove that $(\bar{N}_{4n+4}, g_{\bar{N}})$ 
is $\varepsilon$-QK, one can work on $P_{4n+5}$ and maintain symplectic
covariance.

Describing the supergravity c-map using a gauged sigma model with target 
$P_{4n+5}$ amounts to replacing the diagram \eqref{Diagram:c-map}
by 
\begin{equation}
\label{Diagram:c-map2}
\xymatrix{
{M} \ar@{|->}[r]^{c} \ar@{|->}[d]_{\mathrm{SC}}
& \hat{N} \cong TM  \ar@{|->}[rrr] \ar@{|->}[drrr]^{\varepsilon\mathrm{HK/QK}}
& & & P \cong TM \times \mathbb{R} \ar@{|->}[d]^{/U(1)_{\varepsilon_1}}\\
\bar{M} \ar@{|->}[rrrr]_{\bar{c}} & & &  &\bar{N} \\
}
\end{equation}
Defining the one-forms
\begin{equation}
\rho = H^{-1} q^a \Omega_{ab} dq^b \;,\;\;\;
\sigma = H^{-1} q^a \Omega_{ab} d\hat{q}^b \;,\;\;\;
\tau = H^{-1} \hat{q}^a \Omega_{ab} d\hat{q}^b 
\end{equation}
the projectable tensor \eqref{g_P} takes the form
\begin{equation}
\label{g_P2}
g_P = \tilde{g}_{TM} - \varepsilon_1 \rho^2 + 2 \varepsilon_1 \varepsilon_2 \sigma^2
- \varepsilon_1 (d\tilde{\phi} + \tau)^2 
\end{equation}
where 
\begin{equation}
\tilde{g}_{TM} = \tilde{H}_{ab} (dq^a dq^b - \varepsilon_2 d\hat{q}^a d\hat{q}^b) 
\end{equation}
is the image of $\tilde{g}_M = \tilde{H}_{ab} dq^a dq^b$ under the rigid c-map. 
The manifold $P_{4n+5}$ is defined as $TM \times \mathbb{R}$, where 
$\mathbb{R}$ is parametrized by $\tilde{\phi}$. The tensor $g_P$ is obtained
by twisting the product metric $\tilde{g}_{TM} - \varepsilon_1 d\tilde{\phi}^2$ 
using the one forms $\rho, \sigma, \tau$. The vector field ${\partial}/{\partial \tilde{\phi}}$ 
leaves $g_P$ invariant and generates a principal action on $P$ which allows to recover
$TM$ as a quotient $TM\cong P/\mathbb{R}$. We remark that one can replace 
$\mathbb{R}$ by $S^1$, which is indeed the choice usually made in the 
$\varepsilon$-HK/QK correspondence. The choice $\mathbb{R}$ is suitable for the
supergravity c-map, where $\tilde{\phi}$ is the dualized Kaluza-Klein vector. 
The principal $\mathbb{C}^*_{\varepsilon_1}$-action on $M$ can be lifted to $TM$
and to $P$, which then allows to take a quotient of $P$ by the principal action 
of $U(1)_{\varepsilon_1}\subset \mathbb{C}^*_{\varepsilon_1}$. The tensor $g_P$
is invariant under and transversal with respect to this group action and
defines a non-degenerate metric $g_{\bar{N}}$ on $\bar{N}=P/U(1)_{\varepsilon_1}$.

Alternatively, we interpret the diagram \ref{Diagram:c-map2} 
such that the rigid c-map is applied to the CAS$\varepsilon_1$K metric 
$g_M= H_{ab} dq^a dq^b$ to obtain the $\varepsilon$-HK metric
$g_{TM} = H_{ab} (dq^a dq^b - d\hat{q}^a d\hat{q}^b)$. 
The tensor $g_P$ is then obtained by a conformal rescaling and twisting
by the one-forms $d\tilde{H}=\tilde{H}_a dq^a$ and $\tilde{H}_a d\hat{q}^a$ 
in addition to the modifications which relate $\tilde{g}_{TM}$ to $g_P$. 
Note that for both $g_{M}$ and $\tilde{g}_M$ their relation to $g_{P}$ is
determined by the $\varepsilon$-HK/QK correspondence (equivalently, by
conification), and therefore is canonical.

Proving that a metric is an $\varepsilon$-QK metric is usually  
difficult, because an $\varepsilon$-QK manifold need not admit any
globally defined and integrable $\varepsilon$-complex structures. 
One advantage of constructing $\bar{N}$ as a quotient of $P\cong TM\times \mathbb{R}$
is that $TM$ is $\varepsilon$-HK. This can be used to construct data on $P$ which
by projection define an $\varepsilon$-QK structure on $\bar{N}$, thus providing 
a concise proof that $\bar{N}$ is $\varepsilon$-QK. 
We refer to \cite{Cortes:2015wca} for details. As shown there, calculations on $P$
can be translated into calculations on $\bar{N}$ using local sections. The original
proof of \cite{Ferrara:1989ik} that spaces in the image of spatial supergravity 
c-map are QK uses an adapted co-frame on $\bar{N}$. 
The approach of  \cite{Cortes:2015wca}
also allows
to show that $\varepsilon$-QK manifolds obtained from the supergravity c-map
admit integrable $\varepsilon$-complex structures. In particular, the $\varepsilon_1$-complex
structure of $M$ induces an integrable $\varepsilon_1$-complex structure on $\bar{N}$ \
which is part of the $\varepsilon$-QK structure.\footnote{For the spatial 
supergravity c-map this was first shown in \cite{Cortes:2011ut}.} 
There also always exists a second
integrable $\varepsilon_1$-complex structure, which is not part of the 
$\varepsilon$-QK structure, and which differs from the first integrable structure 
by a sign flip on a 
two-dimensional distribution. A third integrable structure only exists if 
the Hessian metric $g_M$ on $M$ has a quadratic Hesse potential, 
$\nabla g_M =0$. 

The parametrization \eqref{lag:3d_gauged} of the c-map has turned out
to be useful for obtaining explicit non-extremal black hole
and black brane in solutions, as well as cosmological
solutions, for four-dimensional ${\cal N}=2$ vector multiplets
coupled to Poincar\'e supergravity, without and with gauging 
\cite{Errington:2014bta,Dempster:2015xqa,Dempster:2016rra,Gutowski:2019iyo}.\footnote{Non-extremal solutions for five-dimensional vector multiplets can be obtained
in a similar way using the r-map \cite{Mohaupt:2010fk}.}

\subsubsection*{From Griffiths to Weil flags}

It was observed in \cite{MR2494168} and \cite{Aalst:2008} that the spatial supergravity 
c-map involves the so-called Weil intermediate Jacobian, which parametrizes
Hodge structures on Calabi-Yau three-folds. Similarly, the rigid c-map 
involves the so-called Griffiths intermediate Jacobian  \cite{Cortes:1998}.
While this can be interpreted in the context of Calabi-Yau compactifications,
where the scalar manifold are related to the moduli spaces of complex 
and K\"ahler structures,  the supergravity c-map is well defined for any 
theory of ${\cal N}=2$ vector multiplets coupled to supergravity. 
Therefore,  one should be able to understand the appearance of the Griffiths and Weil 
Jacobians without reference to Calabi-Yau manifolds. 
In \cite{Cortes:2011aj} a geometrical interpretation was given based on the realization 
of CASK manifolds as Lagrangian cones in $V=\mathbb{C}^{2n+2} = T^* \mathbb{C}^{n+1}$
We have already noted that besides the CASK metric $g_M = H_{ab} dq^a dq^b$ 
the CASK manifold $M$ admits another metric $\hat{g}_{M} = \hat{H}_{ab} dq^a dq^b$, 
which, up to an overall factor, differs by a sign flip along the distribution spanned
by the vector fields $\xi$ and $J\xi$ which generate the $\mathbb{C}^*$-action. 
This operation can be viewed as a reflection on $V$ which induces an
$Sp(\mathbb{R}^{2n+2})$ equivariant diffeomorphism between certain flag manifolds
defined over $V$. These flag manifolds are of the same type as the Griffiths and
and Weil intermediate Jacobians, and have therefore been dubbed Griffiths and
Weil flags, respectively. 

From our description of the supergravity c-map it is clear why it involves a map
from Griffiths to Weil flags. In a rigid vector multiplet theory the 
matrix encoding the vector field couplings is $H_{ab}$, while in a local vector multiplet
theory it is $\hat{H}_{ab}$. The rigid c-map generates a term of the form
$H_{ab} d\hat{q}^a d\hat{q}^b$ in the metric on $TM$, which in the 
three-dimensional Lagrangian corresponds to the dimensional reduction of 
the vector field of a rigid vector multiplet theory. The twisting relating 
$g_{TM}$ to $g_{P}$ involves (among other things) the replacement
$H_{ab} d\hat{q}^a d\hat{q}^b \mapsto \hat{H}_{ab} d\hat{q}^a d\hat{q}^b$, where 
the latter term corresponds in the three-dimensional Lagrangian to the
dimensional reduction of the vector fields of local vector multiplets. Thus
the HK/QK part of the supergravity c-map acts as a reflection on $V$ which 
replaces Griffiths flags by Weil flags.

\subsubsection*{The deformed supergravity c-map \label{sect:deformed_sugra_c-map}}

Similar to the ASK/PSK correspondence, the $\varepsilon$-QK-metrics obtained from the 
$\varepsilon$-HK/QK correspondence depend explicitly on the choice of a moment 
map for the $\varepsilon$-holomorphic vector field $Z$ on $\hat{N}$. This results 
in a one-parameter family of metrics $g^{(c)}_{\bar{N}}$, with $c=0$
corresponding to the supergravity c-map \cite{Alexandrov:2011ac}. 
It has been shown directly, that is without invoking supersymmetry, that while the
deformation is non-trivial, the metrics $g^{(c)}_{\bar{N}}$ with $c\not=0$ 
are still $\varepsilon$-QK 
\cite{Alekseevsky:2012fu,Alekseevsky:2013nua,Dyckmanns:2016qoe}. In the QK case the deformation corresponds, for a 
specific value of $c$, to the one-loop correction to the hypermultiplet metric in 
type-II Calabi-Yau compactifications \cite{Antoniadis:2003sw,RoblesLlana:2006ez}. 
Explicit expressions for the generalization of \eqref{g_P}, \eqref{g_P2} can be 
found in \cite{Alekseevsky:2013nua,Dyckmanns:2016qoe}.

\subsubsection{Results on completeness, classification  and symmetries of PSR, PSK and QK manifolds}

In this section we collect results on the geodesic 
completeness, classification and isometries of PSR, PSK  and QK manifolds. 
Recall that a pseudo-Riemannian 
manifold is called {\em homogeneous} if its group of isometries acts transitively, and
{\em globally symmetric} if every point is the fixed point of an involutive isometry. 
A pseudo-Riemannian manifold is called {\em geodesically complete}, if
any geodesic can be extended to infinite affine parameter. If the metric
is positive definite, geodesic completeness is equivalent to metric completeness. 
Pseudo-Riemannian symmetric spaces are in particular homogeneous, and homogeneous spaces
are geodesically complete. Locally, 
symmetric spaces are characterized by their Riemann tensor being parallel.
A manifold of {\em co-homogeneity} $k$ is a manifold where the minimal co-dimension 
of orbits of the isometry group is $k$.

It was proved in \cite{Cortes:2011aj} that for metrics of positive signature 
the supergravity r-map and c-map preserve geodesic completeness. This
is useful for obtaining new results in Riemannian geometry, because 
it allows to 
generate complete PSK manifolds from complete PSR manifolds, and 
complete QK manifolds from complete PSK manifolds.

The r-map and c-map do not only preserve completeness, but also preserve 
isometries and in fact create new ones. The obvious induced isometries
are those descending from higher-dimensional gauge symmetries whenever
a vector field is reduced dimensionally. But there are additional `hidden'
symmetries as well.  Without aim for completeness, some 
relevant references are \cite{Ferrara:1989ik,deWit:1990na,deWit:1992wf,deWit:1995tf}.
There is also a relation between symmetric PSR manifolds and 
Jordan algebras, as was already observed in \cite{Gunaydin:1983bi}. This has
been studied extensively in the literature, but lies outside the topic of this review. 
We refer the interested reader  to \cite{Gunaydin:2007qq,Bellucci:2009qv,Gunaydin:2009pk} and references 
therein.

All homogeneous (and thus in particular all symmetric) PSR manifolds 
have been classified in \cite{deWit:1991nm}.
A simple criterion for the completeness of PSR manifolds was
proved in \cite{2014arXiv1407.3251C}. Complete PSR manifolds of dimension one 
and two have been classified in \cite{Cortes:2011aj}
and \cite{2013arXiv1302.4570C}, respectively. Already in dimension two there 
are continuous families of non-isomorphic PSR spaces. Complete PSR manifolds based 
on reducible cubic polynomials have been classified in \cite{Cortes:2017yrp}
and belong to four infinite series, two of which consist of homogenous spaces 
while the other two consist of spaces of co-homogeneity one. 

Homogeneous (pseudo-)PSK manifolds of the form $G/K$, where $G$ is a semi-simple 
Lie group and $K$ a compact subgroup are automatically symmetric spaces
\cite{Alekseevski:2000}. Examples of PSK manifolds with co-homogeneity one
have been constructed by applying the r-map to non-homogeneous PSR manifolds
\cite{Cortes:2017yrp}. A general criterion of the geodesic completeness of PSK
manifolds has been proved in \cite{Cortes:2016wjn}.

The spatial c-map is a powerful tool for the construction and classification
of quaternionic K\"ahler manifolds, which are the most complicated
non-exceptional types of Riemannian manifolds with special holonomy.
The hypermultiplet manifolds occurring in supergravity always have 
negative scalar curvature \cite{Bagger:1983tt}. 
Alekseevskian spaces, that is homogeneous QK spaces of
negative scalar curvature which admit a completely solvable\footnote{A solvable Lie group
action is called completely solvable if the generators in the adjoint representation have
real eigenvalues.} and simply
transitive group of isometries have been classified in \cite{MR0402649}.
The classification of homogeneous QK manifolds generated by the
supergravity c-map \cite{deWit:1991nm} contains a class of spaces 
not contained in the original list of \cite{MR0402649}. It was then 
shown in \cite{MR1395026} that these were the only cases missing,
thus completing the classification. With the exception of the
quaternionic hyperbolic spaces $\mathbb{H}H^{n+1}$, all Alekseevsky 
spaces can be obtained from the supergravity c-map. 

Mathematically the supergravity c-map is extremely useful because 
it allows the explicit construction of non-homogeneous quaternionic-K\"ahler
spaces. Moreover, since it preserves completeness, 
one can use complete, non-homogeneous PSK manifolds to obtain
complete, non-homogeneous QK manifolds.
Two infinite families of complete
QK manifold of co-homogeneity 1 have been constructed 
in \cite{Cortes:2017yrp}.

While the supergravity  c-map preserves completeness, this is no longer true 
for the deformed supergravity c-map, that is if one includes a non-trivial 
constant $c\not=0$ in the choice of the moment map for the vector field 
$Z$ on the partner manifold under the HK/QK correspondence. However, 
one can show that every PSK manifold which exhibits so-called regular
boundary behaviour is complete, and that its image under the deformed
supergravity c-map is a complete QK manifold for $c\geq 0$ \cite{Cortes:2016wjn}. The same
is true for complete PSK manifolds with a cubic prepotential, irrespective
of their boundary behaviour \cite{Cortes:2016wjn}. This allows to construct a huge class of complete
non-homogeneous  QK manifolds. The results of \cite{Cortes:2016wjn} 
have a curious implication for physics, where in type-II Calabi-Yau 
compactifications the parameter $c$ corresponds to the one-loop correction 
to the hypermultiplet metric and is proportional to the Euler number 
of the Calabi-Yau three-fold. Given that mirror symmetry is a symmetry
of string theory and maps the Euler number to its negative, is it surprising 
that whether the one-loop correction preserves completeness depends
on the sign of the Euler number. Understanding this observation 
will likely involve to also consider the effect of further (instanton) corrections
to the hypermultiplet metric. 

The supergravity c-map also allows to construct homogeneous and non-homogeneous
pseudo-QK and para-QK spaces. Due to the lack of a completeness
result comparable to \cite{Cortes:2011aj} much less is known. The classification
of symmetric pseudo-QK and para-QK spaces can be obtained from the 
classification of pseudo-Riemannian symmetric spaces
by analysing their isotropy representations \cite{Cortes:2004aa}. 
The Hesse potentials of symmetric PSK take the form $H = \sqrt{ Q }$, where
$Q$ is a homogeneous polynomial of degree four. These polynomials have been
determined in \cite{Baues:2003}. They have an immediate geometric interpretation
for the associated QK manifold, because they determine the so-called quartic
Weyl tensor, which is the traceless part of the curvature tensor of a 
QK manifold. Explicit descriptions for the homogeneous PSK manifolds in the
image of the supergravity c-map, including their prepotentials, K\"ahler potentials
and realizations as bounded open domains  can be found in 
\cite{Cecotti:1988ad}.

\subsubsection{The c-map in string theory}

In this review we have focussed on the c-map as a construction in supergravity. 
Originally the c-map was formulated in the context of string theory, more 
precisely type-II compactifications on Calabi-Yau three-folds \cite{Cecotti:1988qn}. By T-duality
type-IIA string theory on $X\times S^1_R$, where $X$ is Calabi-Yau three-fold
and $S^1_R$ is a circle of radius $R$ (in string units), is equivalent to 
type-IIB string theory on $X\times S^1_{R^{-1}}$. By taking the limits
$R\rightarrow 0$ and $R\rightarrow \infty$ one obtains a relation between
type-IIA and type-IIB string theory compactified on the same Calabi-Yau manifold, thus
somewhat complementary to mirror symmetry. This form of T-duality 
is often referred to as the c-map, though using the terminology of this review
it actually combines the supergravity c-map
and its inverse, as follows. Given a type-IIA compactification on $X$, 
we have an ${\cal N}=2$ supergravity theory with $n_V^{(A)}$ vector multiplets
and $n_H^{(A)}$ hypermultiplets and a scalar manifold $\bar{M}_{(A)} \times 
\bar{N}_{(A)}$ which is the product of a PSK and a QK manifold. Upon
reduction to three dimensions, this becomes supergravity coupled 
to $n_V^{(A)} + 1 +n_H^{(A)}$ hypermultiplets, with scalar manifold
$\bar{N}'_{(A)} \times \bar{N}_{(A)}$, where $\bar{N}'_{(A)}$ is the QK
manifold obtained by applying the c-map to $\bar{M}_{(A)}$. 

Applying T-duality and lifting back to four dimensions results in an
effective 
IIB-theory with $n_V^{(B)} = n_H^{(A)}-1$ vector multiplets and 
$n_H^{(B)} = n_V^{(A)} +1$ hypermultiplets. The scalar manifold 
is $\bar{M}_{(B)} \times \bar{N}_{(B)}$, where 
$\bar{M}_{(B)}$ is the image of 
$\bar{N}_{(B)} = \bar{N}'_{(A)}$ under the inverse 
of the supergravity c-map.  Note the shifts $\pm 1$ in the number of multiplets which accounts for
the degrees of freedom residing in the Poincar\'e supergravity multiplet. 

This construction allows the obtain the tree-level hypermultiplet metrics
for the type-IIA/B theory from the vector multiplet metrics of type-IIB/A. However
type-II hypermultiplet metrics are subject to perturbative and 
non-perturbative corrections. The perturbative corrections arise at the
one-loop level and have been discussed above. Non-perturbative
corrections have been studied extensively by combining string dualties
with the twistor approach, see below. For Calabi-Yau three-folds $X$ which are 
K3-fibrations, the type-II compactification on $X$ is believed to be 
dual to a heterotic compactification on $K3\times T^2$ with a suitable
choice of an $E_8\times E_8$ or $\mbox{Spin}(32)/\mathbb{Z}_2$ 
 vector bundle $V\rightarrow K3$. While heterotic 
hypermultiplet metrics are believed to be exact at string tree level, 
they are hard to compute because they are related to the moduli
spaces of vector bundles (instantons) on a K3 surfaces. 

\subsubsection{The twistor approach and instanton corrections to hypermultiplet metrics\label{sec:twistor_approach}}

Every quaternonic K\"ahler space $\bar{N}_{4n}$ admits an associated 
twistor space ${\cal Z}_{4n+2}$, which, roughly speaking, is the 
$S^2\cong P_\mathbb{C}^1$ bundle obtained by attaching to each point
of $\bar{N}_{4n+2}$ the sphere $\{ a J_1 + b J_2 + cJ_3 | a^2 + b^2 + c^2 =1\}$
of complex structures generated by the endomorphisms $J_1,J_2, J_3$ which
locally spann the quaternionic structure. The twistor space can be embedded
into the HK cone (or Swann bundle) $N_{4n+4}$ and thus `sits half-ways'
between $\bar{N}_{4n}$ and $N_{4n+4}$. 

Twistor spaces have been used extensively to study the supergravity c-map
and to obtain perturbative and non-perturbative corrections to 
hypermultiplet metrics. One advantage of this approach is that it allows
to describe quaternionic K\"ahler spaces in terms of holomorphic data
on the twistor space. 
The twistor approach is closely related to 
the projective superspace formulation of supersymmetry. This is 
complementary to the approach underlying this review, which 
focusses on Hessian structures and the superconformal formalism.
We refer the interested reader to the
literature, in particular to \cite{Gunaydin:2007qq,Alexandrov:2011va}
and references therein.


\section{Static BPS black holes and entropy functions 
in five dimensions  \label{secdic} } 

The equations of motion of ${\cal N}=2$ supergravity coupled to abelian vector multiplets 
in four and five space-time dimensions admit static, single-centre, extremal black hole solutions.
These are black hole solutions whose near-horizon geometry is $AdS_2 \times S^p$, with
$p=2$  ($p=3$) in four (five) space-time dimensions. These solutions are supported by the Maxwell charges
as well as by the scalar fields of the theory. Asymptotically, these scalar fields take arbitrary values. When approaching the
event horizon of the black hole (which at the two-derivative level is a Killing horizon \cite{Hawking:1973uf}), the scalar fields 
flow to specific values that are entirely determined by the
charges of the black hole \cite{Ferrara:1995ih,Ferrara:1996dd,Ferrara:1996um}. This is the so-called attractor mechanism
for extremal black holes, 
which can be explained by rewriting the equations of motion as gradient flow equations:
regardless of their asymptotic values, the scalar fields are driven to specific values
at the horizon. When these extremal black hole solutions are also supersymmetric, they are called 
BPS black holes.

In this section we study static BPS black holes in five dimensions. They are electrically charged.
The gradient flow equations for these BPS black holes were
originally obtained by studying the supersymmetry preserved by these solutions \cite{Gibbons:1993xt,Chamseddine:1998yv}.
Here, we derive them by 
 performing a suitable rewriting
of the underlying action \cite{LopesCardoso:2007qid,Larsen:2006xm} .

The near-horizon geometry of these five-dimensional black hole solutions is described by an $AdS_2 \times S^3$ space-time.
In this geometry, 
the attractor values for the scalar fields at the horizon can be obtained  from a variational principle
based on the so-called entropy function for extremal black holes \cite{Sen:2005wa}. Evaluating this entropy function at the extremum then yields
the entropy of the black hole. For BPS black holes, the horizon values of the scalar fields 
can also be derived from a variational principle based on a different entropy function, called BPS entropy function.
 The BPS entropy function is constructed from the Hesse potential ${\cal V}$ of the CASR manifold discussed in section
\ref{sect:special-real-local-coordinates}.

 The above considerations based on the entropy function can be extended
 to the case where one considers BPS black hole solutions in ${\cal N}=2$ supergravity theories 
 in the presence of higher-derivative terms
 proportional to the square of the Weyl tensor \cite{Castro:2007hc,Castro:2007ci,Castro:2008ne,deWit:2009de}.
We discuss the effect of Weyl square interactions on the entropy of static BPS black holes.

\subsection{Single-centre BPS black hole solutions through gradient flow equations}

\subsubsection{Action and line element ansatz}

The action for ${\cal N}=2$ Poincar\'e supergravity at the two-derivative level is given in \eqref{lagpoinc5d}.
Here we set 
$\kappa^2 =1$, i.e. $G^{-1} = 8 \pi$, and we will denote the scalar fields $h^I$ ($I = 0, \dots, n$)
by $X^A$ ($A = 0, \dots, n$), so that now 
\begin{equation}
C_{ABC} X^A X^B X^C = 1 \;.
\label{X5dconst}
\end{equation}
Correspondingly, we will denote the quantities $h_I$ and $\stackrel{\circ}a_{IJ}$ introduced in \eqref{aijCh}
by $X_A$ and $G_{AB}$, respectively,\footnote{We note
that the normalizations used in this section differ slightly from those used in \cite{LopesCardoso:2007qid}.}
\begin{eqnarray}
X_A &=& C_{ABC} X^B X^C = G_{AB} \, X^B \;, \nonumber\\
G_{AB} &=& - 2  \, C_{ABC} X^C + 3 X_A X_B \;.
\label{XAGAB}
\end{eqnarray}
We note the following useful relations, which will be used in the following,
\begin{eqnarray}
X_A \, X^A &=& 1 \;, \nonumber\\
X_A \, \partial_{\mu} X^A &=& X^A \partial_{\mu} X_A = 0 \;, \nonumber\\
G^{AB} \, \partial_{\mu} X_B &=& -  \partial_{\mu} X^A \;, \nonumber\\
G_{AB} \partial_{\mu} X^A \partial^ {\mu} X^ B &=&  G^{AB} \partial_{\mu} X_A \partial^ {\mu} X_B \:.
\label{CXXCXX}
\end{eqnarray}

Next, we display the part of the ${\cal N}=2$ Poincar\'e supergravity action that is relevant for the purpose of obtaining 
gradient flow equations for static extremal black hole solutions,
\begin{eqnarray}
	S =  \int d^5 x \sqrt{-g} \left( \frac12 R - \frac34 G_{AB} \, \partial_{\mu} X^A \, \partial^{\mu} X^B- \frac38 \, G_{AB} F^A_{\mu \nu} \, F^{B \mu \nu} \right) \;.
		\label{action5} 
\end{eqnarray}
We are interested in static solutions to the equations of motion, and hence we take
the five-dimensional line element, the one-form gauge fields $A^A$ and
the scalar fields $X^A$ to have the following form in adapted coordinates,
\begin{eqnarray}
	\label{metric} ds_5^2 &=& g_{\mu \nu} \, dx^{\mu} \, dx^{\nu} = - f^2 (r) \, dt^2 + f^{-1}(r) \, ds^2_{\rm GH} \;, \nonumber\\
	A^A &=& \chi^A (r) \, dt \;, \nonumber\\
	X^A &=& X^A(r) \;,
	\label{AA} 
\end{eqnarray}
where $ds^2_{\rm GH}$ describes four-dimensional Euclidean flat space, which we write in the form
\begin{eqnarray}
	ds^2_{\rm GH} = r^{-1}  \left(dr^2 + r^2 (d \theta^2 + \sin^2 \theta \, d \varphi^2) \right) +  r  \,( d\psi +  \cos \theta \, d\varphi )^2 \;.
	\label{tnhyper} 
\end{eqnarray}
Here $\theta \in [0, \pi],\, \varphi \in [0, 2 \pi), \, \psi \in [0, 4 \pi)$.
Indeed, by
changing the radial coordinate to
\begin{equation}
\rho^2 = 4 \,  r \;,
\end{equation}
one obtains
\begin{equation}
ds^2_{\rm GH} = d \rho^2 + \frac{\rho^2}{4} \left( \sigma_1^2 + \sigma_2^2 + \sigma_3^2 \right) = dx^m dx^m \;\;\;,\;\;\;
m=1, \dots, 4 \;,
\label{hkflat}
\end{equation}
where 
\begin{eqnarray}
\sigma_1 &=& - \sin \psi \, d \theta + \cos \psi \, \sin \theta \, d \varphi \;,
\nonumber\\
\sigma_2 &=& \cos \psi \, d \theta + \sin \psi \, \sin \theta \, d \varphi \;, 
\nonumber\\
\sigma_3 &=&  d\psi +  \cos \theta \, d\varphi
\;.
\end{eqnarray}

We take the electric field $\partial_{\rho} \chi^A$ to be sourced by electric charges which we denote by $q_A$,
up to a normalization constant, 
so that $\partial_{\rho} \chi^A \sim f^2 G^{AB} q_B / \rho^3$, and hence
\begin{equation}
\partial_r \chi^A = - \frac23 \frac{f^2}{r^2} \, G^{AB} q_B \;.
\label{electricEq}
\end{equation}

\subsubsection{Gradient flow equations 
\label{sec:flow5dw5}}

Here we derive first-order  flow equations for solutions of the form \eqref{AA}. These solutions
describe single-centre static extremal black holes in a five-dimensional asymptotically  flat space-time.
We follow
 \cite{LopesCardoso:2007qid}.

Inserting the ansatz \eqref{AA} into the action \eqref{action5} yields,
\begin{eqnarray}
\label{bulkeva} 
	S &=& \frac14 \int dt \, dr \, d \theta \, d \varphi \, d \psi\, \sin \theta \\
	&& \Big[ -3 r^2 f^{-2} (f')^2 - 3 r^2 G_{AB} (X^A)' (X^B)'
	+ 3 r^2 f^{-2} G_{AB} {\chi'^A} \, {\chi'^B} \nonumber\\
&&  \quad   + 2 \partial_r \left(r^2 f^{-1} f' \right) \Big] \;, \nonumber
\end{eqnarray}
where $\; '= \partial_r \;$. 
Introducing the radial coordinate
\begin{equation}
	\tau = \frac{1}{r} \;,
\end{equation}
and using \eqref{CXXCXX} as well as \eqref{electricEq}, 
this can can be rewritten into
\begin{eqnarray}
	\label{stationarySBPS} 
	S &=& \frac14 \int dt \, dr \, d \theta \, d \varphi \, d \psi\, \sin \theta
	\nonumber\\
	&& \biggl[
	- 3 \,  \tau^2 \, G^{AB} \left( \partial_{\tau}  X_A + f \,  \partial_{\tau}  f^{-1}  \, X_A - \frac23 \, s \,  f \, q_A \right) \nonumber\\
	&& \qquad \qquad \qquad 
	 \left( \partial_{\tau}  X_B + f \, \partial_{\tau} f^{-1}  \, X_B - \frac23 \, s \,  f \, q_B \right) 
	\nonumber\\
		&& \left. 
	+ 3 \tau^2 f^{-2} G_{AB} \left(\partial_{\tau} {\chi^A} -  \frac23 f^2 G^{AC} q_C \right) \left( \partial_{\tau} {\chi^B} - \frac23  f^2 G^{BD} q_D \right) \right. 
	\nonumber\\
	&& + 2 \partial_r \left(r^2 f^{-1} f' -2 q_A \, \chi^A - 2 \, s \, f \, q_A X^A \right) \biggr] \;,
\end{eqnarray}
where $s = \pm 1$.

The last line in \eqref{stationarySBPS} denotes a total derivative.
Thus, up to a total derivative term, $S$ is expressed in terms of squares of first-order terms which, when requiring stationarity of $S$ with respect to variations of the fields, results in 
\begin{eqnarray}
	\label{BPSflow5d}
	\partial_{\tau}  X_A + f \,  \partial_{\tau}  f^{-1}  \, X_A & = &  \frac23 \, s \,  f \, q_A \;,
	\nonumber\\
			\partial_{\tau} {\chi^A} &= & \frac23  f^2 G^{AB} q_B \;.
\end{eqnarray}
Contracting the first equation with $X^A$ yields the flow equation for the warp factor $f$,
\begin{equation}
  \partial_{\tau}  f^{-1}   = \frac23 \, s \, q_A X^A \;.
  \end{equation}
  The gradient flow equations \eqref{BPSflow5d} then take the equivalent form
\begin{eqnarray}
	\label{BPSflow5d2}
	\partial_{\tau}  \left( f^{-1} \, X_A \right) & = & \frac23 \, s \,  \, q_A \;,
	\nonumber\\
	  \partial_{\tau}  f^{-1}   & = & \frac23 \, s  \, q_A X^A \;, \nonumber\\
			\partial_{\tau} {\chi^A} &= & \frac 23 f^2 G^{AB} q_B \;.
\end{eqnarray}
It can be checked that the five-dimensional Einstein-, Maxwell- and scalar field equations of motion derived from \eqref{action5} are satisfied 
by the solutions to the flow equations \eqref{BPSflow5d2}.

The first flow equation in \eqref{BPSflow5d2} is solved by
\begin{equation}
f^{-1} \, X_A = \frac23 \, s \, H_A \;\;\;,\;\;\; H_A = h_A + q_A \, \tau \;,
\label{XAH}
\end{equation}
where $h_A$ denote integration constants.
Contracting this with $X_A$ results in
\begin{equation}
f^{-1} = \frac23 \, s \, H_A  \, X^A \;.
\end{equation}
One then verifies that this solves the flow equation for $f^{-1}$ by virtue of the relation $X_A \partial_{\tau} X^A =0$, c.f. \eqref{CXXCXX}.
Thus, the flow equations \eqref{BPSflow5d2} are solved by
\begin{eqnarray}
f^{-1} \, X_A &=& \frac23 \, s \, H_A \;, \nonumber\\
f^{-1} &=& \frac23 \, s \, H_A  \, X^A \;, \nonumber\\
\chi^A &=& - s \,  f \, X^A  \;.
\label{sol5dXA}
\end{eqnarray}

In the following, we take $s =1$ and we assume that the $C_{ABC}$ in \eqref{X5dconst} are all positive, so that 
$X^A > 0$.
Demanding $f^{-1} > 0$ along the flow, we infer that $H_A > 0$ along the flow, and hence also $h_A, q_A > 0$.
The solution describes a static, electrically charged extremal black hole solution in five dimension,
which is BPS \cite{Chamseddine:1998yv}. The latter can be deduced as follows. The Lagrangian \eqref{stationarySBPS} contains
the term $q_A G^{AB} q_B$, also called black hole potential $V_{\rm BH}$.
It can be expressed in terms of the five-dimensional central charge,
\begin{equation}
Z_5 = q_A \, X^A \;,
\end{equation}
as
\begin{equation}
V_{\rm BH} = q_A G^{AB} q_B =  Z_5^2  + G^{AB} \left( D_A Z_5 \right) \left( D_B Z_5 \right) \;,
\label{5dbhpot}
\end{equation}
where 
\begin{equation}
D_A Z_5 = q_A - X_A Z_5 \;.
\end{equation}
Likewise, the gradient flow equations for $f^{-1}$ and $X^A$ can be expressed in terms of $Z_5$ and $D_A Z_5$ \cite{Ortin:2011vm,Larsen:2006xm},
\begin{eqnarray}
\partial_{\tau} X^A &=& -  \frac23 \, f\, G^{AB} \, D_B Z_5 \;, \nonumber\\
\partial_{\tau} f^{-1} &=& \frac23  \, Z_5 \;.
\label{flowZ5}
\end{eqnarray}
The scalar fields $X^A$ stop flowing when  $D_A Z_5 =0 \, \forall A=1, \dots, n$. The latter corresponds to a critical point
of the black hole potential. If at this critical point $Z_{5, {\rm crit} }  \neq 0$, then the scalar fields $X^A$ attain the constant values $X_A = q_A/Z_{5, \rm crit}$,
and the warp factor $f^{-1}$ becomes $f^{-1} =  \frac23  \, Z_{5, \rm crit} \, \tau$. The associated line element describes the geometry of $AdS_2 \times S^3$,
which is the near-horizon geometry of a static extremal black hole in five dimensions.
Thus, when approaching the horizon of the black hole, the 
scalar fields $ X^A$ flow to a critical point of the black hole potential \eqref{5dbhpot}  satisfying $D_A Z_5 =0$ with $Z_5 \neq 0$. Such a 
a critical point is a BPS critical point \cite{Ferrara:1996dd}.

The black hole potential may have other critical points that are not BPS.
Suppose that the black hole potential
admits a second  decomposition, in terms of a real quantity $W_5 = Q_A X^A$, 
\begin{equation}
V_{\rm BH} = q_A G^{AB} q_B =  W_5^2  + G^{AB} D_A W_5 D_B W_5 \;,
\label{bhW}
\end{equation}
with $W_5 \neq Z_5$, and that it possesses a critical point $D_A W_5 =0$ with $W_5 \neq 0$. Then this 
critical point  is non-BPS, and it is associated to a non-BPS static extremal black hole solution
that can be obtained by solving first-order flow equations of the form
\eqref{flowZ5}, but now with $Z_5$ replaced by $W_5$. This is so, because the rewriting of the action 
\eqref{bulkeva} using \eqref{bhW} proceeds in exactly the same manner as the one discussed above.
Thus, in certain cases, non-BPS static extremal black holes solutions may be obtained by solving first-order flow equations
 \cite{LopesCardoso:2007qid}.

\subsection{Entropy functions for static BPS black holes }


\subsubsection{Entropy function  at the two-derivative level}

We consider the solution \eqref{sol5dXA} with $s=1$.
In the coordinates \eqref{tnhyper}, the near-horizon geometry of the BPS black hole
is obtained by sending $r \rightarrow 0$. Inspection of \eqref{sol5dXA} shows that in this limit
the $X^A$ become constant, while
$f^{-1} (r) \propto 1/r$.
Setting
\begin{equation}
f^{-1} (r) = \frac{v_2}{4 \, r} \;,
\end{equation}
with $v_2$ a positive constant,
and inserting
this into \eqref{AA} shows that the 
near-horizon geometry of a static BPS black hole is
$AdS_2 \times S^3$,
\begin{eqnarray}
ds^2_5 = v_1 (- r^2 dt^2  + \frac{dr^2}{r^2}) 
+ \frac{v_2}{4}  \left(d \theta^2 + \sin^2 \theta \, d\varphi^2 \right)
+ \frac{v_2 }{4} \left(d \psi + \cos \theta \, d \varphi \right)^2  \;,
\label{5dline2}
\end{eqnarray}
with $v_1 = v_2/4$.
In this near-horizon geometry, 
the gauge potentials $\chi^A$ behave as $\chi^A(r) \propto r$, and hence we
set
\begin{equation}
\chi^A(r) = e^A \, r \;,
\label{electe}
\end{equation}
with constant $e^A $. The near-horizon solution is thus specified by 
\eqref{5dline2} and \eqref{electe}, and supported by constant $X^A$.
The values of the $X^A$ at the horizon are, according to the attractor mechanism for extremal black holes,
 specified by the charges carried by the black hole.  These values can be determined by 
 extremizing the so-called entropy function, to which we now turn.

We consider the reduced Lagrangian ${\cal F}_5$ which is obtained
by evaluating the Lagrangian \eqref{action5}  in the near-horizon 
BPS black hole background \eqref{5dline2}, 
\eqref{electe}
and integrating over the horizon {\cite{Cardoso:2007rg},
\begin{eqnarray}
{\cal F}_5 &=& \frac{1}{8 \pi} \,  \int d \psi \, d\theta \, d \phi \, \sqrt{-g} \,
{L}_5 \;, \nonumber\\
&=& \pi \, 
\frac{v_1 \, (v_2^{3} )^{1/2}}{4}  
\left[
 - \frac{1}{v_1} + \frac{3}{v_2} 
+ \frac34 \frac{G_{AB} \, e^A \, e^B }{ v_1^2}  \right] 
\;.
\label{redact45}
\end{eqnarray}
The entropy function is then given by the Legendre transform 
\cite{Sen:2005wa}
\begin{equation}
{\cal E}_5 = 2 \pi \left( 2 \pi \, q_A \, e^A - {\cal F}_5 \right) \;.
\end{equation}
The entropy function is a function of the constant parameters $e^A, v_1, v_2, X^A$.
Extremizing the entropy function with respect to these parameters and evaluating
the entropy function at the extremum, yields the entropy of the static BPS black hole
expressed in terms of the charges $q_A$.

Varying the entropy function ${\cal E}_5$ with respect
to the electric fields $e^A$ and setting 
$\partial_e {\cal E}_5 =0$ yields
\begin{eqnarray}
 \frac{3 \pi}{8}  \, \frac{(v_2^3 )^{1/2}}{v_1} \, G_{AB} \, e^B
= 2 \pi \, q_A 
\;.
\label{varef}
\end{eqnarray}
Varying ${\cal E}_5$  with respect to $v_1, v_2$ and setting the variations to zero 
yields
\begin{equation}
v_1 = \frac{v_2}{4} =  G_{AB} \, e^A \, e^B \;.
\label{v1v2q}
\end{equation}
Inserting \eqref{v1v2q} into 
${\cal E}_5$ yields
\begin{eqnarray}
{\cal E}_5 = \frac{\pi^2}{2 } \left(v_2^3  \right)^{1/2}\;,
\label{extrent}
\end{eqnarray}
which equals the macroscopic entropy ${\cal S}_{\rm macro} = 
A_5/4$ of the static black hole, where $A_5$ denotes the horizon
area. Using \eqref{varef}, we infer
\begin{equation}
\frac94 \, v_2 \, G_{AB} \, e^A \, e^B = \frac{q_A \, G^{AB}Ê\, q_B}{ 4 \pi^2} \;,
\end{equation}
and hence
\begin{equation}
v_2^2 = \frac{4}{9 \pi^2} \, q_A \, G^{AB}Ê\, q_B \;.
\label{v2qq}
\end{equation}
The horizon values of the $X^A$ are determined in terms of the charges $q_A$
by varying ${\cal E}_5$ with respect to the $X^A$ and setting 
$\delta_X {\cal E}_5 =0$. In doing so, one has to take into account
the constraint \eqref{X5dconst}, which implies
\begin{equation}
C_{ABC}ÊX^A X^B \delta X^C = 0
\label{varCXXX}
\end{equation}
for arbitrary variations $\delta X^C$. Using 
the relation for $G_{AB}$ given in \eqref{XAGAB},
one obtains for 
$\delta_X {\cal E}_5 =0$,
\begin{equation}
e^A e^B \delta X^C \left( - C_{ABC} + 3 \, C_{ACE} X^E X_B
+ 3 \, C_{BCE} X^E X_A \right) = 0 \;.
\label{eqeeX}
\end{equation}
Setting $e^A = \gamma \, X^A$, as required for a BPS solution, solves \eqref{eqeeX} by virtue of 
 \eqref{varCXXX}. 
The scale factor $\gamma$ is determined by inserting this expression into
\eqref{v2qq}, which results in
\begin{equation}
\gamma = \frac12 \ \sqrt{v_2} \;.
\end{equation}
Then, using \eqref{varef}, we infer
\begin{equation}
v_2 \, X_A = \frac83 \, q_A \;.
\label{v2Xq}
\end{equation}
This is the so-called attractor equation, whose solution determines the values
of the scalar fields $X^A$
at the horizon in terms of the charges carried by the BPS black hole.
Contracting \eqref{v2Xq} with $X^A$ yields $v_2 = \frac83 q_A X^A$, and using \eqref{X5dconst} one infers $v_2 \sim q$, and hence
${\cal S}_{\rm macro} \sim q^{3/2}$  \cite{Ferrara:1996dd}.

\subsubsection{BPS entropy function at the two-derivative level, the Hesse potential and its dual}

The attractor equation \eqref{v2Xq}
can also be derived from a variational principle based on  a different entropy function,
which we call the BPS entropy function. The BPS entropy function is constructed from 
 the Hesse potential ${\cal V}$ of the CASR manifold discussed in section \ref{sect:special-real-local-coordinates},
 \begin{equation}
H ({\cal Y} ) = \frac{1}{2} \, {\cal V} ( {\cal Y} ) = 
\frac{1}{2} \, C_{ABC} \, {\cal Y}^A \, {\cal Y}^B \, {\cal Y}^C \;,
\label{HcalV}
\end{equation}
where we have introduced
\begin{equation}
{\cal Y}^A = v_2^{1/2} \, X^A 
\;.
\label{claYv2}
\end{equation}
The BPS entropy function reads
\begin{equation}
\Sigma ({\cal Y}, q) = 4 \, q_A \, {\cal Y}^A -  H ({\cal Y} ) \;.
\end{equation}
Extremizing with respect to ${\cal Y}^A $ yields
\begin{equation}
 C_{ABC} \,  {\cal Y}^B \, {\cal Y}^C = \frac83 \, q_A \;,
 \end{equation}
 which expresses the ${\cal Y}^A$ in terms of the charges $q_A$.
The value of $\Sigma$ at this extremum
is
\begin{equation}
\Sigma (q) = C_{ABC} \, {\cal Y}^A \, {\cal Y}^B \, {\cal Y}^C = v_2^{3/2} \;,
\end{equation}
and hence
\begin{equation}
{\cal S}_{\rm macro} = \frac{\pi^{2} }{2} \, \Sigma (q) \;.
\end{equation}
Thus, upon extremization, the electric charges $q_A$ become proportional to the dual special real coordinates, while the BPS entropy
is proportional to the dual Hesse potential, evaluated on the background, c.f. \eqref{Hqq}.

\subsubsection{$R^2$-corrected BPS entropy function, the Hesse potential and its dual}

Now we allow for the presence of a specific class of $R^2$ terms in the ${\cal N}=2$ supergravity Lagrangian,
namely those arising from the coupling of vector multiplets to the square of the Weyl multiplet.
The effect of these higher derivative terms on the near-horizon region of static BPS black hole solutions
and on the associated BPS entropy has been thoroughly discussed in \cite{Castro:2007hc,Castro:2007ci,Castro:2008ne,deWit:2009de}.
We follow \cite{deWit:2009de}. 

The coupling of vector multiplets to the square of the Weyl multiplet can be conveniently described 
using the superconformal approach to supergravity. This is reviewed in \ref{sec:5dscg}.
One salient feature is that the 
Lagrangian describing the couplings of vector multiplets to the square
of the Weyl multiplet contains a term proportional to the square of the Weyl tensor,
with coupling function $c_A \, X^A$, where $c_A$ are constant coefficients 
\cite{Hanaki:2006pj}.

We focus on solutions to the associated equations of motions that have full supersymmetry.  These field configurations satisfy \cite{deWit:2009de},
\begin{eqnarray}
\label{horsol}
\partial_{\mu} X^A = 0  \;\;,\;\;F_{ab}^A = 4 \, X^A  \, T_{ab}  \;\;,\;\;  Y^{ij} = 0  \;\;,\;\; D = 0 \;\;,\;\;
T_{ab} T^{ab}Ê= {\rm constant} \;.
\end{eqnarray}
The associated line element describes a circle fibred over an $AdS_2 \times S^2$
base,
\begin{eqnarray}
ds^2 &=& \frac{1}{16 v^2} \left( - r^2 dt^2 + \frac{dr^2}{r^2} + d \theta^2 + \sin^2 \theta
d \varphi^2 \right) + e^{2g} \left( d \psi + B \right)^2 \;, \nonumber\\
B &=& - \frac{1}{4 v^2} e^{-g} \left( T_{23} \, r \, dt - T_{01} \, \cos \theta \, d \varphi 
\right) \;, \nonumber\\
v &=& \sqrt{ (T_{01})^2 + (T_{23})^2} \;.
\label{line5dgen}
\end{eqnarray}
Here, $T_{01}$ and $T_{23}$ denote the 
non-vanishing components of $T_{ab}$, and they are associated
with $(t, r, \theta, \varphi)$. $v$ and $e^g$ are constants.
In the following, we focus on static configurations, and hence set $T_{23} =0$.
Introducing  the notation ($T_{01} \neq 0$) 
\begin{equation} 
p^0 = \frac{e^{-g}}{4 v^2}Ê\, T_{01} 
\end{equation}
and using $v^2 = (T_{01})^2$, 
the line element \eqref{line5dgen} may be brought into the form
\begin{eqnarray}
ds^2 = \frac{1}{16 v^2} \left( - r^2 dt^2 + \frac{dr^2}{r^2} + d \theta^2 + 
d \varphi^2 
+ 
\frac{1}{(p^0)^2} d \psi^2 + \frac{2}{p^0} \cos \theta d \varphi  d \psi\right) \;.
\label{linebps}
\end{eqnarray}
Then, demanding $p^0=1$, in which case $e^{-g} = 4 T_{01} = 4 v > 0$, and fixing the
periodicity of $\psi$ to $\psi \in [0, 4 \pi)$, the line element becomes the
line element for $AdS_2 \times S^3$ given in \eqref{5dline2}, with $v_2 = 1/(4 v^2)$.
This is the near-horizon
geometry of a static BPS black hole supported by electric charges $q_A$
and constant scalar fields $X^A$. The latter are expressed in terms of
the charges through 
the attractor equation 
\begin{equation}
q_A =  \frac{3 e^g}{8 T_{01}} \left( C_{ABC} X^B X^C - c_A (T_{01})^2 
\right) = \frac{3 }{32 v^2} \left( C_{ABC} X^B X^C - c_A \, v^2
\right)
 \;,
\label{attr5dst}
\end{equation}
where we normalized the charges as in \eqref{v2Xq}.

In this background, the equation of motion for the auxiliary $D$-field takes the form
\begin{equation}
\chi = - 2 C_{ABC}ÊX^A X^B X^C + 4  c_A X^A (T_{01})^2Ê\;.
\label{efnorm}
\end{equation}
Then, imposing the normalization of the 
Einstein-Hilbert term (c.f. \eqref{lagR2}),
\begin{equation}
C_{ABC}ÊX^A X^B X^C - \frac32 \chi= 4 \;,
\end{equation}
yields
the constraint
\begin{equation}
C_{ABC}ÊX^A X^B X^C = 1 + \frac32 c_A X^A (T_{01})^2Ê\;.
\end{equation}
Introducing ${\cal Y}^A$ as in \eqref{claYv2},
\begin{equation}
{\cal Y}^A = \frac{1}{2 v} X^A \;,
\end{equation}
we obtain
\begin{equation}
C_{ABC}Ê{\cal Y}^A {\cal Y}^B {\cal Y}^C = \frac{1}{8 v^3} + \frac{3}{16} c_A {\cal Y}^A Ê\;.
\label{constrcaly}
\end{equation}
The attractor equation \eqref{attr5dst} becomes 
\begin{equation}
\hat{q}_A \equiv q_A + \frac{3}{32} c_A  =  \frac{3 }{8}  C_{ABC} {\cal Y}^B {\cal Y}^C 
 \;.
\label{attr5dst2}
\end{equation}

The entropy of these static BPS black holes is computed using Wald's entropy formula \eqref{entronoether}.
Using the $R^2$-corrected Lagrangian \eqref{lagR2} in the background 
\eqref{horsol}, \eqref{linebps}, we obtain
\begin{equation}
 \frac{\partial L}{\partial R_{\mu \nu \rho \sigma} }\, \varepsilon_{\mu \nu } \varepsilon_{\rho \sigma} = - C_{ABC}  X^A X^B X^C  \;.
 \end{equation}
 Note that the contributions proportional to $c_A$ have cancelled out.
 Then, using the line element \eqref{linebps}
 \begin{equation}
 \int_{S^3} \sqrt{h} \, d \Omega = \frac{\pi^2}{4 v^3} \;,
 \end{equation}
 we obtain for the macroscopic entropy of a static BPS black hole, 
  \begin{equation}
{\cal S}_{\rm macro} = \frac{\pi^2 }{2}  C_{ABC} {\cal Y}^A {\cal Y}^B {\cal Y}^C \;,
\label{entro5dbpsst}
\end{equation}
with the ${\cal Y}^A$ expressed in terms of the charges through \eqref{attr5dst2}.
The constant $v$ in the line element \eqref{linebps} is determined through \eqref{constrcaly} in terms
of the charges. This fully determines the near-horizon geometry of the static BPS black hole.

The attractor equation \eqref{attr5dst2}  can be obtained by extremizing
the following BPS entropy function,
\begin{equation}
\Sigma ({\cal Y}, q) = 4 \, {\hat q}_A \, {\cal Y}^A -  H ({\cal Y} ) \;,
\end{equation}
with $H ({\cal Y} )$ given as in \eqref{HcalV}. The BPS entropy function is thus given in terms
of the dual Hesse potential, c.f. \eqref{Hqq}.
The value of $\Sigma$ at the extremum
yields the entropy \eqref{entro5dbpsst},
\begin{equation}
{\cal S}_{\rm macro} = \frac{\pi^{2} }{2} \, \Sigma (q) \;.
\end{equation}

\section{Static BPS black holes and entropy functions in four dimensions \label{bpsentro4dbh}}

In four dimensions, 
the equations of motion of ${\cal N}=2$ supergravity coupled to abelian vector multiplets 
(without or with higher-derivative terms proportional to the square of the Weyl tensor)
admit single-centre, dyonic, extremal black hole solutions. These are spherically symmetric 
solutions. When they are supersymmetric, they are called BPS solutions.

In so-called isotropic coordinates $(t, r,  \theta,  \phi)$, the line element of a spherically symmetric space-time
takes the form
\begin{equation}
ds^2 = - e^{2 g(r)} \, dt^2 + e^{2 f(r)} \left( dr^2 + r^2 \, ( d \theta^2 + \sin^2 \theta \, d \phi^2 )\right) \;.
\end{equation}
At the two-derivative level, BPS black hole solutions satisfy $f = - g$  \cite{Tod:1983pm,Tod:1995jf}.
In the following, we will restrict the discussion to the class of solutions with $f = - g$, and we will write
their line element as
\begin{equation}
ds^2 = - e^{2 U(r)} \, dt^2 + e^{-2 U(r)} \left( dr^2 + r^2 \, ( d \theta^2 + \sin^2 \theta \, d \phi^2 )\right) \;.
\label{eq:line-blackh}
\end{equation}
 
Extremal black hole solutions  
carry electric and magnetic charges $(q_I, p^I)$ associated with the abelian gauge fields $A_{\mu}^I$ of the theory,
\begin{eqnarray}
\int_{S^2_{\infty}} d \theta d \phi \, F_{\theta\phi}{}^I = p^I \;\;\;,\;\;\;
\int_{S^2_{\infty}} d \theta d \phi \, G_{\theta\phi \,I} = q_I \;,
\label{chargesqp}
\end{eqnarray}
where we integrate over an asymptotic two-sphere $S^2_{\infty}$. Here,
$G_{\theta\phi \,I}$ denotes 
the dual field strength defined in \eqref{eq:dual-F}.

These black holes are furthermore supported by complex
scalar fields $X^I$ that reside in the vector multiplets.  These scalar fields will, generically, have a non-trivial profile, i.e.
$X^I = X^I(r)$.
Asymptotically, the scalar fields take arbitrary values.
When approaching the event horizon, the scalar fields flow to fixed values  that are entirely determined by the
charges of the black hole. 
This is the attractor mechanism for extremal black holes  \cite{Ferrara:1995ih,Ferrara:1996dd,Ferrara:1996um,Goldstein:2005hq}:
the values of the scalar fields at the event horizon are attracted to specific values given in terms of
the charges of the black hole, irrespective of their asymptotic values at spatial infinity.  
For BPS black holes at the two-derivative level, 
the flow to the event horizon
is described by gradient flow equations for the scalar fields and for the metric factor $e^U$.
These first-order flow equations can be obtained from a reduced action in one dimension, 
see subsection \ref{sec:first-order-rew}. 

In the near-horizon region $r \approx 0$, the metric \eqref{eq:line-blackh} takes the form
\begin{equation}
ds^2 = v_1 \,\left( - r^2 \, dt^2 + \frac{dr^2}{r^2}\right)  + v_2 \, ( d \theta^2 + \sin^2 \theta \, d \phi^2 ) \;,
\label{eq:line-hor}
\end{equation}
with $v_1 = v_2$,
and describes the line element of an $AdS_2 \times S^2$ space-time.  Here, $v_1$ denotes a constant
whose value is entirely specified by the charges carried by the black hole, through the attractor mechanism. The attractor values for the scalar fields at the horizon can be obtained  from a variational principle
based on the so-called entropy function. Evaluating this entropy function at the extremum then yields
the entropy of the black hole. This entropy function can
  be derived from the reduced action evaluated in the near-horizon
  geometry \eqref{eq:line-hor}, as we will discuss in subsection \ref{sec:entrofunc}.

 The above considerations based on the entropy function can be extended
 to the case where one considers extremal black hole solutions in ${\cal N}=2$ supergravity theories 
 in the presence of higher-derivative terms
 proportional to the square of the Weyl tensor. Then, for BPS black holes, one
 still finds $v_1 = v_2$ \cite{LopesCardoso:1998tkj}, while for non-BPS black holes, one generically has 
  $v_1 \neq v_2$ \cite{Sahoo:2006rp,Cardoso:2006xz}. For further reading on these topics, we refer to 
  \cite{Mohaupt:2000mj,Mohaupt:2007mb,deWit:2007dn,Sen:2007qy,Mohaupt:2008gt}.

There are many other interesting aspects about black hole attractors which we do not describe in this review.
These include: relations 
between topics in number theory and BPS black holes 
\cite{Moore:1998zu,Moore:1998pn,Dabholkar:2012nd}; multicenter bound states of BPS black holes \cite{Denef:2000nb};
the OSV conjecture \cite{Ooguri:2004zv,LopesCardoso:2006ugz};
the $4D/5D$ connection between black objects
\cite{Gaiotto:2005xt,Behrndt:2005he,Ceresole:2007rq}; attractors and cosmic censorship \cite{Bellorin:2006xr};
rotating attractors \cite{Astefanesei:2006dd}; the quantum entropy function \cite{Sen:2008vm,Dabholkar:2010uh};
attractor flows and CFT \cite{Nampuri:2010um}; a Riemann-Hilbert approach to rotating attractors
\cite{Camara:2017hez};
 hot attractors \cite{Goldstein:2018mwt}.

\subsection{Single-centre BPS black hole  solutions through gradient flow equations \label{sec:first-order-rew}}

In the following, we will derive gradient flow equations for BPS black hole solutions
in ${\cal N}=2$ supergravity theories at the two-derivative level in four
dimensions \cite{Ferrara:1997tw,DallAgata:2010ejj}. These are first-order flow equations that will be obtained from a reduced action
based on the Lagrangian \eqref{efflag_poincare}. The latter is evaluated in the
background \eqref{eq:line-blackh} and subsequently rewritten in terms of squares of first-order terms. 
We use the relation \eqref{eq:relgN} to perform the 
rewriting in terms of (rescaled) scalar fields $X^I$, rather than in terms of scalar fields $z^a$, 
as follows \cite{Barisch:2011ui}.

We introduce
rescaled scalar fields $Y^I$ defined by
\begin{equation}
Y^I = {e}^{-U} \, {\tilde X}^I = {e}^{-U} \, {\bar \varphi} \, X^I \;,
\label{YUX}
\end{equation}
where $U$ denotes the metric factor in \eqref{eq:line-blackh}.
Here, $\bar \varphi$ denotes a phase, with a
$U(1)$-weight that is opposite to the one of $X^I$.  Thus, ${\tilde X}^I = {\bar \varphi} \, X^I$
denotes a $U(1)$ invariant variable.

Using  \eqref{eq:relgN}, 
we evaluate the Lagrangian \eqref{efflag_poincare} in the background \eqref{eq:line-blackh},
taking $X^I $ and $Y^I$ to be functions of $r$, only.
We evaluate the covariant derivative \eqref{eq:kah-conn},
\begin{equation}
{\bar \varphi} \, {\cal D}_{r}ÊX^I= \partial_{r}  {\tilde X}^I  + i  \, \hat{A}_{r} \, {\tilde X}^I  \;,
\label{covX_fix2}
\end{equation}
where
\begin{equation}
\hat{A}_{r} = A_r + i \varphi \, \partial_r {\bar  \varphi} \;,
\end{equation}
where $\varphi$ is the complex conjugate of $\bar{\varphi}$,
and where  $A_{r}$ is given by \eqref{U1_fix}. Then,
\begin{equation}
N_{IJ} \, {\cal D}_r X^I \, {\cal D}_r {\bar X}^J = N_{IJ} \, \tilde{X}'^I \bar{\tilde{X}}'^J + \hat{A}_{r}^2
\;,
\label{eq:kin-scalars}
\end{equation} 
where $\tilde{X}'^I = \partial_r \tilde{X}^I$.
Observe that in view of $N_{IJ} X^I \, {\bar X}^J = -1$, we have
\begin{equation}
{e}^{-2 U} = - N_{IJ} \, Y^I \, {\bar Y}^J \;,
\label{eq:A-Y}
\end{equation}
as well as
\begin{equation}
{e}^{-2 U} \, U' =  \tfrac12 \, N_{IJ} \, \left( Y'^I \, {\bar Y}^J + Y^I \, {\bar Y}'^J \right) \;,
\label{eq:relAp-Y}
\end{equation}
where we used the homogeneity property $F_{IJK} X^K =0$. Similarly, using the
homogeneity property of $F_I$, the connection $\hat{A}_{r} $ can be expressed in terms of 
the $Y^I$ as
\begin{equation}
\hat{A}_{r} = - \tfrac12 \, e^{2U} \left[ (F_I - {\bar F}_I) \, \partial_r (Y^I - {\bar Y}^I ) 
- (Y^I - {\bar Y}^I ) \, \partial_r (F_I - {\bar F}_I ) \right] \;.
\label{exphatA}
\end{equation}

Extremal black holes carry electric and magnetic charges $(q_I, p^I)$.
Electric fields $E^I (r)$ and magnetic charges $p_I$ are introduced as (c.f. \eqref{chargesqp})
\begin{equation}
  F_{rt}{}^I = E^I\,,\qquad F_{\theta\phi}{}^I = \frac{p^I}{4\pi}\,
  \sin\theta \,.
  \label{fieldstrengths}
\end{equation}
The $\theta$-dependence of
$F_{\theta\phi}{}^I$ is fixed by rotational invariance.

Rather than using a description based on $(p^I, E^I)$, we seek a description in terms
of magnetic/electric charges $(p^I, q_I)$. To introduce electric charges $q_I$, we consider the dual field
strengths $G_{\mu\nu I} $ defined in \eqref{eq:dual-F}.
Adopting the conventions where
$x^\mu=(t,r,\theta,\phi)$ and $\varepsilon_{tr\theta\phi}= 1$,
 it follows that, in the background \eqref{eq:line-blackh},
\begin{eqnarray}
  \label{eq:G}
  G_{\theta\phi\,I}\! = - e^{- 2U(r)}Ê\, r^2 \,  \,\sin\theta\,
  \frac{\partial L} {\partial F_{rt}{}^I}= - e^{- 2U(r)}Ê\, r^2 \, 
  \,\sin\theta\,   \frac{\partial L} {\partial E^I}  \,.
 \end{eqnarray}
Writing  (c.f. \eqref{chargesqp})
\begin{equation}
G_{\theta\phi \,I} = \frac{q_I}{4\pi}\,
  \sin\theta \,,
  \end{equation}
  where the $\theta$-dependence
 is again fixed by rotational invariance,
  we infer
\begin{eqnarray}
  q_I  =- 4\pi \, e^{- 2U(r)}Ê\, r^2 \, \,
  \frac{\partial L} {\partial E^I}  \;.
  \label{q_R_rel}
\end{eqnarray}

We pass from a description based on $(p^I, E^I)$ to a description based on $(p^I, q_I)$ by means of 
the Legendre transform
\begin{equation}
L_{1d} = \left(  \int \, d \theta \, d \phi \, \sqrt{-g} \, L \right)  + q_I \, E^I \;,
\label{Eqlegendre}
\end{equation}
with the Lagrangian $L$, given in \eqref{efflag_poincare}, evaluated in the background \eqref{eq:line-blackh}.

To keep the discussion as simple as possible, let us first consider the case of
electrically
charged black holes. Subsequently we extend the discussion to the case of dyonic black holes.

Using \eqref{q_R_rel}, we obtain
\begin{equation} 
\label{Eq}
E^I =  {e}^{2 U } \, \frac{ \left[\left( {\rm Im} {\cal N}\right)^{-1}\right]^{IJ} \, q_J}{4 \pi \, r^2} \;.
\end{equation}
The resulting one-dimensional action $4 \pi \, S_{1d} \equiv \int dr \, L_{1d}$ reads,
\begin{eqnarray} \label{s1d}
- S_{1d} &=& \int dr \, r^2 \left[ U'^2 + 
N_{IJ} \, {\tilde X}'^I \, \bar{\tilde X}'^J + \hat{A}_{r}^2
- \ft12 \, {e}^{2 U } \, \frac{q_I \left[ \left({\rm Im} {\cal N}\right)^{-1}\right]^{IJ}
\, q_J }{ (4 \pi \, r^2)^2}  \right]
\nonumber\\ && 
- \int dr \, \frac{d}{dr} \left( r^2 \,   U' \right).
\label{eq:action-1d}
\end{eqnarray}
In the following, we will, for notational simplicity, 
absorb a factor $4 \pi$ into $q_I$, i.e.
$q_I/(4 \pi) \rightarrow q_I$.

Next, we rewrite \eqref{s1d} in terms of the rescaled variables $Y^I$. Using \eqref{eq:relAp-Y},
we obtain the intermediate result
\begin{eqnarray}
- S_{1d} &=& \int dr \,  r^2 \, \Big\{
2 \left[ U' +  e^{2 U}  {\rm Re} \Big( \frac{Y^I \, q_I }{r^2}  \Big) \right]^2 
\nonumber\\ && \qquad \qquad 
+ {e}^{2 U} \, N_{IJ} \, \left(Y'^I + \, N^{IK} \, \frac{q_K}{r^2} \right)
 \, \left({\bar Y}'^J + N^{JL} \, \frac{q_L}{r^2}\right) + \hat{A}_{r}^2
  \nonumber\\
&&  \qquad  \qquad 
- \ft12 \, \frac{{e}^{2 U }}{r^4} \, q_I \left[ \left({\rm Im} {\cal N}\right)^{-1}\right]^{IJ}
\, q_J  -   \frac{e^{2 U}}{r^4} \, q_I \, N^{IJ} \, q_J  \nonumber\\
&& \qquad \qquad -  2 e^{4 U} \left[
 {\rm Re}  \left( \frac{Y^I \, q_I }{r^2} \right) \right]^2
  \Big\}
   \nonumber\\
&&
-  \int dr  \, \frac{d}{dr} \left[  r^2 \, U' + 
2
{e}^{2U} \, {\rm Re} \left(  Y^I \, q_I   \right) 
\right]
   \;.
\label{eq:action-1d-interm2}
\end{eqnarray}
Then, using the second identity in \eqref{eq:N-id-X}
we obtain
\begin{eqnarray}
- S_{1d} = S_{\rm square} + S_{\rm TD} \;,
\label{SsumSS}
\end{eqnarray}
where
\begin{eqnarray}
\label{eq:action-square}
S_{\rm square} &=& 
\int dr \, r^2  \Big\{   2  \Big[U' + {e}^{2U }
{\rm Re} \left(  \frac{Y^I \, q_I}{r^2}    \right) 
\Big]^2 
 \nonumber\\ && \qquad \qquad 
+ 
{e}^{2 U} \, N_{IJ} \,  \left(Y'^I + N^{IK} \, \frac{q_K }{r^2} \right)
\,  \left({\bar Y}'^J + N^{JL} \, \frac{q_L}{r^2} \right) 
 \nonumber\\
&&  \qquad \qquad 
+ \hat{A}_{r}^2 + 2   {e}^{4U } \,  \left[ {\rm Im} \left(  \frac{Y^I \, q_I }{ r^2} \right) \right]^2
 \Big\} \;, 
\end{eqnarray}
and
\begin{eqnarray}
\label{eq:action-TD}
S_{\rm TD} = - \int dr  \, \frac{d}{dr} \left[ 
   r^2 \, U' +
2 \, 
{e}^{2U} \,{\rm Re} \left(   Y^I \, q_I   \right) 
\right] \;. 
\end{eqnarray}

The above results can be easily extended to the case of dyonic black holes, as follows.
First, we view the term $q ({\rm Im} {\cal N})^{-1} q$
in the action \eqref{eq:action-1d} as part of the black hole potential \eqref{Vbhpq},
\begin{eqnarray}
V_{\rm BH} = - \tfrac12 \, q_I \, \left[({\rm Im} {\cal N})^{-1}\right]^{IJ} \, q_J 
= g^{i {\bar \jmath}} \, {\cal D}_i Z \, \bar{\cal D}_{\bar \jmath} {\bar Z} + |Z|^2 \;,
\label{blackpotq}
\end{eqnarray}
where
$Z(X) = - q_I \, X^I$. Turning on magnetic charges $p^I$ amounts to extending $Z(X)$ as in \eqref{ZXhatq},
\begin{eqnarray}
Z(X) = p^I \, F_I(X) - q_I \, X^I = \left(p^I \, F_{IJ} - q_J \right) X^J = - \hat{q}_I \, X^I \;,
\label{zpq}
\end{eqnarray}
where\footnote{Here we subject $p^I$ to the same rescaling as the $q_I$, i.e. 
$p^I/(4 \pi) \rightarrow p^I$. }
\begin{eqnarray}
\hat{q}_I = q_I - F_{IJ} \, p^J \;.
\end{eqnarray}
Then, 
the extension to the dyonic case proceeds 
by replacing $q_I$ with ${\hat q}_I$ in \eqref{SsumSS}, which results in 
\begin{eqnarray}
\label{eq:action-square-dyon}
S_{\rm square} &=& 
\int dr \, r^2 \Big\{   2  \Big[U' + {e}^{2U }
{\rm Re} \left(  \frac{Y^I \, {\hat q}_I }{r^2}   \right) 
\Big]^2  
\nonumber\\ && \qquad \qquad
 + 
{e}^{2 U} \, N_{IJ} \,  \left(Y'^I + N^{IK} \,\frac{ \bar{\hat{q}}_K }{ r^2} \right)
\,  \left({\bar Y}'^J + N^{JL} \, \frac{{\hat q}_L }{ r^2} \right) 
 \nonumber\\
&&   \qquad \qquad  + 
 \hat{A}_{r}^2 + 2   {e}^{4U } \,  \left[ {\rm Im} \left(  \frac{Y^I \, \hat{q}_I }{r^2} \right) \right]^2
 \Big\} \;, 
\end{eqnarray}
and
\begin{eqnarray}
\label{eq:action-TD-dyon}
S_{\rm TD} =  - \int dr  \, \frac{d}{dr} \left[ 
   r^2 \, U' +
2 \, 
{ e}^{2U} \,{\rm Re} \left(   Y^I \, \hat{q}_I   \right) 
\right] \;. 
\end{eqnarray}

Now we vary $S_{\rm square}$  with respect to $U$ and to $Y^I$, respectively.
The vanishing of these variations can be achieved by 
setting the variation of the individual squares in $S_{\rm square}$ to zero,
\begin{eqnarray}
U' &=&-  {e}^{2U }
{\rm Re} \left(  \frac{Y^I \, \hat{q}_I }{r^2} \right)  \;,\nonumber\\
Y'^I &=& -  N^{IK} \,\frac{\bar{\hat{q}}_K }{r^2}  \;, \nonumber\\
{\rm Im} \left(  Y^I \, \hat{q}_I \right) &=& 0 \;, \nonumber\\
 \hat{A}_{r} &=& 0 \;.
\label{eq:flow-dyonic}
\end{eqnarray}
This yields first-order flow equations for $Y^I$ and for $U$. Note that these gradient
equations 
are consistent with one another: the latter is a consequence of the former by virtue of
 \eqref{eq:relAp-Y}.  
 
It is convenient to introduce a rescaled version of $Z(X)$, namely
\begin{eqnarray}
Z(Y) = p^I \, F_I (Y) - q_I \, Y^I \;,
\label{eq:Z-Y}
\end{eqnarray}
in terms of which the first-order flow equations become
 \begin{eqnarray}
r^2 \, U' &=& {e}^{2U } \, 
{\rm Re} \, Z(Y)  \;,\nonumber\\
r^2 \, Y'^I &=&   N^{IK} \, \frac{\partial}{\partial {\bar Y}^K}Ê\bar{Z} (\bar Y)  \;, \nonumber\\
{\rm Im} \, Z(Y)  &=& 0 \;, \nonumber\\
 \hat{A}_{r} &=& 0 \;.
\label{eq:flow-dyonic2}
\end{eqnarray}
The gradient flow equations for the $Y^I$ can be rewritten as
 \begin{eqnarray}
\begin{pmatrix} (Y^I - {\bar Y}^I)' \\ (F_I - {\bar F}_I)' 
\end{pmatrix} = 
 2 i  \, {\rm Im}
\begin{pmatrix} 
N^{IK} \, {\hat q}_K/ r^2  \\  \,{\bar F}_{IK} \, N^{KJ} \, {\hat q}_J /r^2 
\end{pmatrix} = - i  \, 
\begin{pmatrix} 
p^I / r^2  \\  q_I / r^2
\end{pmatrix} \;,
 \label{eq:attrac-Q-P}
\end{eqnarray}
where here $F_I = \partial F(Y)/\partial Y^I$.
Each of the vectors appearing in this expression transforms as a symplectic vector under
 $Sp(2n+2, \mathbb{R})$ transformations.
These gradient flow equations can be readily integrated, 
  \begin{eqnarray}
\begin{pmatrix} Y^I - {\bar Y}^I \\ F_I - {\bar F}_I 
\end{pmatrix}  =  i  \, 
\begin{pmatrix} 
h^I +  p^I/r  \\  h_I + q_I/r
\end{pmatrix} = i \, 
\begin{pmatrix} 
H^I (r)   \\  H_I (r)
\end{pmatrix} \;,
 \label{eq:attrac-int}
\end{eqnarray}
where $(h^I, h_I)$ denote integration constants.  These integration constants are constrained
by the third equation in \eqref{eq:flow-dyonic2}, which
 yields the condition
\begin{equation}
p^I \, h_I - q_I \, h^I = 0 \;.
\label{constraphqh}
\end{equation}
 The metric factor $e^{-2U}$ is then determined by \eqref{eq:A-Y},
\begin{equation}
e^{-2U} = H^I \, F_I(Y) - H_I \, Y^I = H^I \, {\bar F}_I( \bar Y) - H_I \, {\bar Y}^I \;,
\label{eUYH}
\end{equation}
where we used \eqref{eq:attrac-int}.
And finally, using \eqref{exphatA}, it follows that the fourth equation in \eqref{eq:flow-dyonic2} is automatically satisfied by 
\eqref{eq:attrac-int} with \eqref{constraphqh}.

The integrated flow equations \eqref{eq:attrac-int} and the constraint \eqref{constraphqh} 
give rise to
BPS black hole solutions \cite{Behrndt:1997ny}.
The equations \eqref{eq:attrac-int} are called attractor equations: the scalar fields $Y^I$ flow to specific values 
at the horizon of the black hole, irrespective of their asymptotic values at $r = \infty$. 
These horizon values are 
entirely determined by the charges carried by the BPS black hole.
In the
near-horizon region 
$r \approx 0$, the metric \eqref{eq:line-blackh} and the scalar
fields $Y^I$ take the form (c.f. \eqref{eq:line-hor})
\begin{equation}
e^{-2U} = \frac{v_2}{r^2} \;\;\;,\;\;\; Y^I = \frac{Y^I_{\rm hor}}{r} \;,
\end{equation}
with 
\begin{equation}
v_1 = v_2 = Z( Y_{\rm hor})  = {\bar Z}( {\bar Y}_{\rm hor}) = p^I F_I (Y_{\rm hor}) - q_I Y^I_{\rm hor} \;,
\end{equation}
where the horizon values $Y^I_{\rm hor}$ are determined by solving the equations
${\cal P}^I = {\cal Q}_ I =0$, with 
\begin{eqnarray}
  \label{eq:P-Q-2d}
  \mathcal{P}^I &\equiv& p^I + {i}(Y^I_{\rm hor}-{\bar Y}^I_{\rm hor})  \,,
  \nonumber \\ 
  \mathcal{Q}_I &\equiv& q_I + {i} \left(F_I(Y_{hor})- {\bar F}_I ({\bar Y}_{\rm hor}) \right) \,.
\end{eqnarray}
Using \eqref{YUX}, one infers the relation
\begin{equation}
Y^I_{\rm hor} = {\bar Z} ({\bar X}_{\rm hor}) \, X^I_{\rm hor} \;,
\label{YZX}
\end{equation}
so that 
\begin{equation}
v_1 = v_2 =Z( Y_{\rm hor}) =  |Z(X_{\rm hor})|^2 \;.
\end{equation}

The gradient flow equations that we obtained were derived from the reduced action
\eqref{eq:action-1d} (with $q_I$ replaced by ${\hat q}_I$) . The equations of motion in four dimensions impose one more condition on the solutions
to the field equations derived from the reduced action, namely the so-called 
Hamiltonian constraint.
For a Lagrangian density $\sqrt{-g}\, (\,\frac12 \, R\,+\,{\cal L}_M)\,$, it is given by the variation of the action with
respect to $g^{00}$,
\begin{equation} \label{hamilton}
\tfrac12  \, R_{00}\,+\,\frac{\delta{\cal L}_M}{\delta g^{00}}\,-\tfrac12 \,g_{00}\,\Big(\tfrac12  \, R\,+\,{\cal L}_M
\Big)\,=\,0\ .
\end{equation}
Then, using the Lagrangian \eqref{efflag_poincare} as well as the metric ansatz \eqref{eq:line-blackh},
and replacing the gauge fields by their charges, as in \eqref{Eq},  results in 
\begin{eqnarray}
 r^2  \left\{ U'^2\, \,+\, N_{IJ}\,{\tilde X}^I\;\!\! ' \,\bar{\tilde X}^J\; \!\! '
+ \frac{e^{2U}}{r^4} \, V_{\rm BH}
\right\} 
-2 \,\Big[\, r^2 \,U'\,\Big]' = 0\;,
\label{eq:hamilton-s1d}
\end{eqnarray}
where $V_{\rm BH}$ denotes the black hole potential \eqref{bhpot1}. We rewrite this as 
\begin{eqnarray}
 r^2  \left\{ U'^2\, \,+\, N_{IJ}\,{\tilde X}^I\;\!\! ' \,\bar{\tilde X}^J\; \!\! '
-  \frac{e^{2U}}{r^4} \, V_{\rm BH}
\right\} 
= 2 \,\Big[\, r^2 \,U'\,\Big]' - 2 \frac{e^{2U}}{r^2} \, V_{\rm BH} \;.
\end{eqnarray}
Using the first-order flow equation \eqref{eq:flow-dyonic2}, one readily verifies that the right hand side of the equation vanishes. This yields the Hamiltonian constraint 
in the form \cite{Ferrara:1997tw}
\begin{eqnarray}
 U'^2\, \,+\, N_{IJ}\,{\tilde X}^I\;\!\! ' \,\bar{\tilde X}^J\; \!\! '  
= \frac{e^{2U}}{r^4} \, V_{\rm BH} \;,
\end{eqnarray}
which is satisfied by virtue of \eqref{eq:flow-dyonic2}.
Thus, the Hamiltonian constraint does not lead to any further restriction.

The black hole potential $V_{\rm BH}$ may have several critical points.
Critical points $*$ that satisfy $( {\cal D}_a Z )|_*= 0  \; \forall  \,   a = 1, \dots, n$ with $Z|_* \neq 0$
correspond to BPS black hole solutions, whose macroscopic entropy is given by
${\cal S}_{\rm macro} (p,q) = \pi v_2 = \pi V_{\rm BH}|_* = \pi |Z(X_{\rm hor})|^2 $, c.f. \eqref{entro=bhpot}. These BPS solutions
are obtained by solving the flow equations \eqref{eq:flow-dyonic2}.  Critical points satisfying ${\cal D}_a Z \neq 0$
do not correspond to BPS solutions. However,
if the black hole potential  $V_{\rm BH}$  admits a second decomposition in terms of a quantity 
$W(X)$ (possibly only when restricting to a subset of charges),
\begin{equation}
V_{\rm BH} = g^{a \bar{b}} {\cal D}_a W  \bar{\cal D}_{\bar b} {\bar W} + |W|^2  \;,
\end{equation}
with $ W  \neq  Z$, such that a critical point that is non-BPS satisfies 
$( {\cal D}_a W)|_*= 0  \, \forall  \,   a = 1, \dots, n$ with $W|_* \neq 0$, then
this non-BPS critical point describes a non-BPS black hole solution that can be obtained by solving
first-order flow equations of the form \eqref{eq:flow-dyonic2}, but now with $Z$ replaced by $W$
\cite{Ceresole:2007wx,Ceresole:2009iy,DallAgata:2011zkh}.
The macroscopic entropy of this non-BPS black hole is given by
${\cal S}_{\rm macro} (p,q) = \pi v_2 = \pi |W(X_{\rm hor})|^2 $.
Thus, in certain cases, 
non-BPS solutions may be obtained by solving first-order flow equations 
\cite{Ceresole:2007wx,Ceresole:2009iy,DallAgata:2011zkh}.


\subsection{Entropy functions for static BPS black holes \label{sec:entrofunc}}

The scalar fields supporting an extremal black hole flow to specific values at the horizon. These
values are entirely specified by the charges carried by the black hole, and they can be obtained
by means of a variational principle based on a so-called entropy function \cite{Sen:2005wa,Sen:2005iz}.

  BPS black holes constitute a subset of extremal black holes, and hence their entropy
  can be obtained from the entropy function mentioned above.  However, their entropy
  can also be inferred from a so-called  BPS entropy function \cite{Behrndt:1996jn,LopesCardoso:2006ugz} associated with 
  supersymmetry enhancement at the
  horizon. Both notions of entropy functions give identical results at the semi-classical level. 
  In the following, we review both notions of entropy functions and their relation, 
  with/without
    higher-curvature terms proportional to the square of the Weyl tensor  \cite{Cardoso:2006xz}.

  \subsubsection{ Reduced action and entropy function}  
  
We consider a local, gauge and
general coordinate invariant
Lagrangian $L$ that 
describes a general system of abelian vector gauge
fields, scalar and matter fields coupled to gravity, with or without higher-derivative terms.
We focus on field configurations
in the near-horizon  geometry \eqref{eq:line-hor}. These field configurations have
the symmetries of $AdS_2 \times S^2$.
We introduce the associated reduced action and derive the entropy function from it.

We denote the scalar and matter fields collectively by $u_{\alpha}$.
The field strengths $F_{\mu\nu}{}^I$ of the abelian gauge fields $A_{\mu}^I$ are given by
\eqref{fieldstrengths}: they are given in terms of the electric field $E^I$ and the magnetic charge $p^I$.
In the geometry \eqref{eq:line-hor}, $v_1, v_2, E, u_{\alpha}$ 
take constant values, since
they are invariant under the $AdS_2 \times S^2$ isometries.

Proceeding as 
in \eqref{Eqlegendre} and \eqref {eq:G},   we pass from 
a description based on $(p^I, E^I)$ to a description based
of magnetic/electric charges $(p^I, q_I)$, 
\begin{eqnarray}
  \label{eq:def-q-f}
  q_I  =- 4\pi\, v_1v_2 \,
  \frac{\partial {L}} {\partial E^I}  \;.
\end{eqnarray}

Defining the reduced Lagrangian by the integral of the
full Lagrangian $L$ over $S^2$, 
\begin{equation}
  \label{eq:reduced-action}
  \mathcal{F}(E,p,v,u) = \int  d \theta\, d \phi\;
  \sqrt{ - g } \, {L} \,,
\end{equation}
we infer
\begin{equation}
  \label{eq:q-F-f-F}
  q_I = - \frac{\partial\mathcal{F}}{\partial E^I}\;. 
\end{equation}

This reduced Lagrangian does not transform as a function under electric-magnetic duality transformations  \eqref{FGdual}.
A quantity that does transform as a function under electric-magnetic duality transformations is the so-called
entropy function \cite{Sen:2005wa},
\begin{equation}
  \label{eq:entropy-function}
  \mathcal{E}(q,p,v,u) =- \mathcal{F}(E,p,v,u) -E^I q_I \,,
\end{equation}
which takes the form of a Legendre transform in view of (\ref{eq:q-F-f-F}). Thus, $ \mathcal{E}$ is the
analogue of the Hamiltonian density associated with the reduced
Lagrangian density \eqref{eq:reduced-action}, as far as the
vector fields are concerned. Under 
electric-magnetic duality, it transforms according to $\tilde {\mathcal{E}}(\tilde
q,\tilde p, v,u) = {\mathcal{E}}(q,p, v,u)$.

The constant values of the fields $v_{1,2}$ and $u_{\alpha}$ are
determined by demanding $\mathcal{E}$ to be stationary under
variations of $v$ and $u$,
\begin{eqnarray}
  \label{eq:field-eqs}
  \frac{\partial \mathcal{E}}{\partial v} =  \frac{\partial
  \mathcal{E}}{\partial u} = 0\,.
 \end{eqnarray}
The  equations 
\eqref{eq:field-eqs} are  the attractor equations that determine the values of 
$v$ and $u$ at the horizon of the black hole.
The Wald entropy is directly proportional to the value of
$\mathcal{E}$ at the stationary point \cite{Sen:2005wa},
\begin{equation}
  \label{eq:wald}
  \mathcal{S}_{\mathrm{macro}}(p,q) \propto \mathcal{E}
  \Big\vert_{\mathrm{attractor}}\,.  
\end{equation}

Note
that the entropy function need not depend on all  the fields at
the horizon. The values of some of the fields will then be left
unconstrained, but they will not appear in the expression for the
Wald entropy.

\subsubsection{Entropy function and black hole potential at the two-derivative level}

 Consider the Maxwell terms in the two-derivative Lagrangian \eqref{efflag_poincare}, which
 is part of the Lagrangian describing Poincar\'e supergravity.
 The associated reduced
Lagrangian \eqref{eq:reduced-action} reads
\begin{equation}
  \label{eq:quadr-reduced-Lagr}
  \mathcal{F} = \tfrac14\left\{ \frac{{i} v_1\, p^I (\bar
  {\mathcal{N}} - \mathcal{N})_{IJ}\,  p^J}{4\pi\,v_2} - 
  \frac{4 {i} \pi\,v_2\, E^I (\bar
  {\mathcal{N}} - \mathcal{N})_{IJ}\, E^J}{v_1}\right\} - \tfrac12 E^I
  (\mathcal{N} + \bar{\mathcal{N}})_{IJ}\, p^J \;,
\end{equation}
and the entropy function
\eqref{eq:entropy-function} is given by (setting $v_1 = v_2$)
\begin{equation}
  \label{eq:entropy-quadratic}
  \mathcal{E}= - 
   \frac{1}{8\pi} \; (q_I - \mathcal{N}_{IK}\,p^K)\,
  [(\mathrm{Im}\,\mathcal{N})^{-1}]^ {IJ}\,  
  (q_J - \bar{\mathcal{N}}_{JL}\,p^L) \;, 
\end{equation}
which equals the black hole potential given in  \eqref{bhpef}
\cite{Ferrara:1997tw}, up to an overall constant.
$ \mathcal{E}$ transforms as  a function under electric-magnetic duality, as can be verified
by noting  the transformation property  \eqref{transf_cal_N} of 
${\cal N}$.

\subsubsection{The BPS entropy function}
\label{sec:bps-entropy-function}

The isometries of the near-horizon geometry \eqref{eq:line-hor}
played a crucial role 
in defining the entropy function \eqref{eq:entropy-function}.
On the other hand, when dealing with BPS black holes,
it is 
supersymmetry
enhancement at the horizon that plays a crucial role in constraining 
fields in the 
 near-horizon geometry. This gives rise to a different form of the entropy function
 for BPS black holes \cite{Behrndt:1996jn,LopesCardoso:2006ugz}, as follows.

 We consider ${\cal N}=2$ supergravity theories coupled to vector multiplets,  and allow
 for the presence of 
  higher-order derivative interactions
involving the square of the Weyl tensor.
As reviewed in section \ref{sec:couchiralback}, the associated Wilsonian effective action is encoded in a holomorphic function 
$F( X, \hat A)$  that is homogeneous of degree two under complex rescalings. Introducing rescaled variables $(Y^I, \Upsilon)$, we have
\begin{equation}
  \label{eq:homogeneous-F}
  F(\lambda Y,\lambda^2 \Upsilon) = \lambda^2\, F( Y,\Upsilon) \;\;\;,\;\;\; \lambda \in \mathbb{C}^* \;.
\end{equation}
Here the $Y^I$ are related to the $X^I$ by a uniform rescaling, and
$\Upsilon$ is a complex scalar field related to 
the square ${\hat A} = 4 (T_{ab}^- )^2 $ by the uniform rescaling, c.f. \eqref{eq:rescaledX}.

At the horizon, the fields $Y^I$ and $\Upsilon$ flow to constant values $Y^I_{\rm hor}$ and $\Upsilon = -64$, with the  $Y^I_{\rm hor}$ determined by the
 BPS attractor equations \cite{LopesCardoso:1998tkj},
 \begin{equation}
  \label{eq:BPS-attractor}
  \mathcal{P}^I=0\,,\qquad \mathcal{Q}_I= 0\,, 
  \end{equation}
where 
\begin{eqnarray}
  \label{eq:P-Q-def}
  \mathcal{P}^I &\equiv& p^I + {i}(Y^I-\bar Y^I)  \,,
  \nonumber \\ 
  \mathcal{Q}_I &\equiv& q_I + {i}(F_I (Y, \Upsilon)-\bar F_I (\bar Y, \bar \Upsilon))  \,.
\end{eqnarray}
These equations are those given in \eqref{eq:P-Q-2d}, but now in the presence of a chiral background field $\Upsilon$.

The BPS attractor equations \eqref{eq:BPS-attractor} can be obtained from a variational
principle based on an entropy function
\cite{Behrndt:1996jn,LopesCardoso:2006ugz}
\begin{equation}
  \label{eq:Sigma-simple}
  \Sigma(Y,\bar Y,p,q) =  \mathcal{F}(Y,\bar Y,\Upsilon,\bar\Upsilon)
  - q_I   (Y^I+\bar Y^I) + p^I (F_I+\bar F_I)  \;,
\end{equation}
where $p^I$ and $q_I$ couple to the corresponding magneto- and
electrostatic potentials at the horizon (c.f.
\cite{LopesCardoso:2000qm}) in a way that is consistent with
electric-magnetic duality. 
The quantity $\mathcal{F}(Y,\bar
Y,\Upsilon,\bar\Upsilon)$, which will be denoted as BPS free energy,
is defined by
\begin{equation}
  \label{eq:free-energy-phase}
  \mathcal{F}(Y,\bar Y,\Upsilon,\bar\Upsilon)= - i \left( {\bar Y}^I
  F_I - Y^I {\bar F}_I
  \right) - 2 i \left( \Upsilon F_\Upsilon - \bar \Upsilon
  \bar F_{\Upsilon}\right)\,,
\end{equation}
where $F_\Upsilon= \partial F/\partial\Upsilon$. Also this expression
is compatible with electric-magnetic duality, 
i.e. it transforms as a function under electric-magnetic duality, 
c.f. \eqref{tilFAFA}
 \cite{deWit:1996gjy}. Varying
the BPS entropy function $\Sigma$ with respect to the $Y^I$, while keeping
the charges and $\Upsilon$ fixed, yields the result,
\begin{equation}
  \label{eq:Sigma-variation-1}
 \delta \Sigma = \mathcal{P}^I
 \, \delta ( F_{I} + \bar F_I)  -\mathcal{Q}_I\,
 \delta (Y^I+ \bar Y^I)  \;,
\end{equation}
where we made use of  the
homogeneity of the function $F(Y,\Upsilon)$.
Assuming that the matrix 
 $ N_{IJ} = \mathrm{i}(\bar F_{IJ}-F_{IJ}) $
is non-degenerate, it  follows that stationary points of
$\Sigma$ satisfy the BPS attractor equations \eqref{eq:BPS-attractor}. 

The macroscopic entropy  ${\cal S}_{\rm macro}$ is
equal to the entropy function evaluated at the attractor point,
and hence it is 
 Legendre transform of the
free energy $\mathcal{F}$.  
It is given by
 \cite{LopesCardoso:1998tkj},
\begin{equation}
  \label{eq:W-entropy}
  {\cal S}_{\rm macro}(p,q) = \pi \, \Sigma\Big\vert_{\rm attractor}
    = \pi \Big[ p^I F_I - q_I Y^I   - 256\,
  {\rm Im}\, F_\Upsilon \Big ]_{\rm attractor}
  \,.
\end{equation}
Here the term $\pi ( p^I F_I - q_I Y^I )|_{\rm attractor}$ equals
a quarter of the horizon area (in units where $G_N =1, \kappa^2 = 8 \pi$),
i.e. $v_1 = v_2 =  ( p^I F_I - q_I Y^I )|_{\rm attractor}$. The contribution proportional to  $F_\Upsilon$ denotes
the deviation from the
Bekenstein-Hawking area law, and is 
subleading in the limit of large charges. In addition, the
area also depends on
$\Upsilon$, and hence it also contains subleading terms.
In the absence of $\Upsilon$-dependent terms, the
homogeneity of the function $F(Y)$ implies that the area scales
quadratically with the charges.

In subsection \ref{sec:bps-black-holes}, we will show that for BPS black holes, the BPS entropy \eqref{eq:W-entropy}
coincides with the one calculated from entropy function \eqref{eq:entropy-function}.

\subsubsection{The BPS entropy function, the generalized Hesse potential and its dual}

The BPS free energy $\cal F$ and the BPS entropy function $\Sigma$ can be expressed in terms of the generalized Hesse potential $H$
and its dual, as follows \cite{LopesCardoso:2006ugz}.

The generalized Hesse potential $H$ is expressed in terms of real variables $(x^I, y_I)$ (c.f. \eqref{XFxyups}),
\begin{equation}
Y^I = x^I + i u^I(x,y,\Upsilon, \bar{\Upsilon}) \;,\;\;\;
F_I = y_I + i v_I(x,y, \Upsilon, \bar{\Upsilon}) \;,
\end{equation}
and defined by a Legendre transform with respect to $u^I$, 
\begin{equation}
H(x,y,\Upsilon, \bar{\Upsilon}) = 2 \, {\rm Im}Ê\, F(x + i u(x,y,\Upsilon,\bar{\Upsilon}), \Upsilon)
- 2 y_I u^I(x,y,\Upsilon,\bar{\Upsilon}) \;.
\label{hessgen}
\end{equation}
Using the homogeneity relation
\eqref{scalingrelA}
 which, in the present context reads
\begin{equation}
2 \, F(Y, \Upsilon) = Y^I \, F_I (Y, \Upsilon)+ 2 \, \Upsilon  \, F_{\Upsilon }  (Y, \Upsilon)\;, 
\end{equation}
one obtains
\begin{equation}
H(x, y, \Upsilon, \bar \Upsilon)  = \tfrac12 \, {\cal F}(Y, \bar Y, \Upsilon, \bar \Upsilon) \;.
\end{equation}
The BPS entropy function $\Sigma$ can then be expressed as 
\begin{equation}
\Sigma (x, y, \Upsilon, \bar \Upsilon) = 2 \, H(x, y, \Upsilon, \bar \Upsilon) - 2 \, q_I \, x^I  + 2 \, p^I \, y_I \;.
\end{equation}
The macroscopic BPS entropy \eqref{eq:W-entropy} is given by
\begin{equation}
  {\cal S}_{\rm macro}(p,q) = 2 \pi \left( H - x^I \, \frac{\partial H}{\partial x^I} -y_I \, \frac{\partial H}{\partial y_I}
  \right)_{\rm attractor} \;.
  \end{equation}
Thus, upon extremization,  the charges $(p^I, q_I)$ become proportional to the dual affine coordinates, while the BPS entropy 
is proportional to the dual Hesse potential, evaluated on the background,  c.f. \eqref{Hqq}.

\subsubsection{Entropy functions for ${\cal N}=2$ supergravity theories}
\label{sec:N=2sugra-applic}

In this section, we follow \cite{Cardoso:2006xz}. We use the normalization
$G_N = 1, \kappa^2 = 8 \pi$, as in \cite{Cardoso:2006xz}.

We consider the Wilsonian effective action describing ${\cal N}=2$ vector multiplets
coupled to ${\cal N}=2$ supergravity,  in the presence of interactions proportional to the 
square of the Weyl multiplet,  reviewed in section \ref{sec:couchiralback}.
This requires the presence of a second 
compensating supermultiplet, which we take to be a hypermultiplet.
Additional
hypermultiplets may also be added, but play a passive role in the
following. 
The relevant Lagrangian $L$ is given by \eqref{action_back}, \eqref{background-def}
\cite{LopesCardoso:2000qm}. The components of the Weyl, vector and hypermultiplets
are displayed in Tables \ref{weyl}   and \ref{vechyp}.

We impose 
 spherical symmetry and derive the
reduced Lagrangian \eqref{eq:reduced-action}. 
In
a spherically symmetric configuration the
field $T_{ab}{}^{ij}$
 can be expressed in
terms of a single complex scalar $w$ \cite{Sahoo:2006rp},
\begin{equation}
  \label{eq:T-w}
  T_{\underline{r}\underline{t}}^-=- i \,
  T_{\underline{\theta}\underline{\phi}}^- =
  \tfrac12 \, w\,, 
\end{equation}
where underlined indices denote tangent-space indices. Consequently we
have $\hat A= -4 w^2$. The field strengths $F_{\mu\nu}{}^I$ of the abelian gauge fields $A_{\mu}^I$ are given
 in terms of electric fields $E^I$ and magnetic charges $p^I$, as in \eqref{fieldstrengths}.

We restrict to a class of solutions by assuming the following consistent set of constraints,
\begin{eqnarray}
  \label{eq:restricted-solution}
  R(\mathcal{V})_{\mu\nu}{}^i{}_j =  R({A})_{\mu\nu} =
  \mathcal{D}_\mu X^I =  \mathcal{D}_\mu A_i{}^\alpha =  0 \,,  
\end{eqnarray}
where the first two tensors denote the $SU(2) \times U(1)$ R-symmetry field strengths.
These constraints are in accord with those that follow
from requiring supersymmetry enhancement at the horizon
\cite{LopesCardoso:2000qm}. Then, 
since $\hat B_{ij}$ is proportional to
$R(\mathcal{V})_{\mu\nu}{}^i{}_j$, this field vanishes as
well. Furthermore the auxiliary fields $Y_{ij}{}^I$ can be dropped
as a result of their equations of motion.

Then, 
in the $AdS_2$ background \eqref{eq:line-hor},  the resulting Lagrangian only depends on the 
field variables
$v_1$, $v_2$, $w$, $D$, $E^I$, $X^I, \chi$, which are all constant, and on the magnetic charges. We refer to \cite{Cardoso:2006xz} for
the somewhat lengthy expression for the Lagrangian. We trade these field variables for scale invariant variables
\begin{eqnarray}
  \label{eq:rescaledX}
  && 
  Y^I = \ft14 v_2 \, {\bar w} \, X^I \,,\quad
  \Upsilon = \ft{1}{16} v_2^2 \, {\bar w}^2 \, {\hat A} = - \ft14
  v_2^2 \, \vert w \vert^4 \,, \quad 
  U = \frac{v_1}{v_2} \,,\nonumber\\ 
  &&
  {\tilde D} = v_2 \, D + \tfrac23(U^{-1} -1) \,,\quad 
  \tilde\chi=  v_2 \, \chi \;.  
\end{eqnarray}
Observe that $\Upsilon$ is real and negative, and that
$\sqrt{-\Upsilon}$ and $U$ are real and positive. Note also that the
hypermultiplets contribute only through the hyperk\"ahler potential
$\chi$. 

We compute the entropy function \eqref{eq:entropy-function}, adopting the normalization
of the Lagrangian used in  \cite{Cardoso:2006xz}. 
Next, 
we require that $\mathcal{E}$ be
stationary with respect to variations of $\tilde D$ and $\tilde\chi$. This yields 
$\tilde D = 0$, and expresses  $\tilde\chi$ in terms of the other fields.
Upon substitution of these two equations into the entropy function, 
the expression for $\mathcal{E}$ simplifies considerably  \cite{Cardoso:2006xz},
\begin{eqnarray}
  \label{eq:entropy-total-2}
   \mathcal{E}(Y,\bar Y,\Upsilon,U)  &=& 
   \ft12 U\,\Sigma(Y, \bar Y,p,q) +\ft12 U\,N^{IJ} (\mathcal{Q}_I-F_{IK}
   \mathcal{P}^K)\, (\mathcal{Q}_J-\bar F_{JL} \mathcal{P}^L) 
   \nonumber\\
   &&
   -\frac{4\mathrm{i}}{\sqrt{-\Upsilon}} (\bar Y^IF_I -Y^I\bar F_I)
   (U -1)   \\
  &&
  - \mathrm{i}(F_\Upsilon-\bar F_\Upsilon) \Big [- 2 U\Upsilon + 32
   (U+U^{-1} -2) - 8 (1+U) \sqrt{-\Upsilon} \Big] \;. \nonumber
\end{eqnarray}
This result that is consistent with electric-magnetic duality \cite{Sahoo:2006rp,Cardoso:2006xz}.

 The entropy function
\eqref{eq:entropy-total-2} depends on the variables $U$, $\Upsilon$
and $Y^ I$, whose values are determined by demanding stationarity of 
 $\mathcal{E}$. These values are the attractor values.
 The macroscopic entropy is
proportional to the entropy function taken at the attractor values,
\begin{equation}
  \label{eq:entropy<function}
  \mathcal{S}_{\mathrm{macro}}(p,q) = 2\pi
  \mathcal{E}\Big\vert_{\mathrm{attractor}} \,. 
\end{equation}
In the following, we will discuss the extremization of $\cal E$ with
respect to these variables, first in the absence of $R^2$-terms, and
then for BPS black holes in the presence of $R^2$-terms.

\subsubsection{Variational equations without $R^2$-interactions}
\label{sec:variation-without-R2}
In the absence of $R^2$-interactions, the function $F$ does not depend
on $\Upsilon$, so that the entropy function \eqref{eq:entropy-total-2}
reduces to
\begin{eqnarray}
  \label{eq:entropy-total-3}
   \mathcal{E}(Y,\bar Y,\Upsilon,U)  &=& 
   \ft12 U\,\Sigma(Y, \bar Y,p,q) +\ft12 U\,N^{IJ} (\mathcal{Q}_I-F_{IK}
   \mathcal{P}^K)\, (\mathcal{Q}_J-\bar F_{JL} \mathcal{P}^L) 
   \nonumber\\
   &&
   -\frac{4\mathrm{i}}{\sqrt{-\Upsilon}} (\bar Y^IF_I -Y^I\bar F_I)
   (U -1)   \,.
\end{eqnarray}
Varying \eqref{eq:entropy-total-3} with respect to $\Upsilon$ yields
\begin{equation}
  \label{eq:eq-Upsilon}
  U = 1 \,.
\end{equation}
The latter implies that the Ricci scalar of the four-dimensional
space-time vanishes. Here we assumed that $\left(\bar Y^I F_I - Y^I
  \bar F_I \right )$ is non-vanishing, which is required so that
Newton's constant remains finite, c.f. \eqref{efflagtotal_DY}.
Varying with respect to $U$ yields,
\begin{eqnarray}
  \label{eq:eq-U}
  \Sigma +  \left( {\cal Q}_I - F_{IK} \, {\cal P}^K\right) N^{IJ} 
  \left( {\cal Q}_J - {\bar F}_{JL} \, {\cal P}^L \right)
  - \frac{8\mathrm{i}}{\sqrt{-\Upsilon}} 
  \left(\bar Y^I F_I -Y^I \bar F_I \right ) =0 \;,
\end{eqnarray}
which determines the value of $\Upsilon$ in terms of the $Y^I$. This
is consistent with the fact that when 
the function $F$ depends exclusively
on the $Y^I$, the field equation for $T_{ab}^-$ is algebraic, c.f. \eqref{T_2d}.

The resulting effective entropy function reads
\begin{equation}
  \label{eq:effective-entropy-function}
  \mathcal{E}(Y,\bar Y,\Upsilon,1) = \ft12 \Sigma(Y, \bar Y,p,q)
   +\ft12 N^{IJ} (\mathcal{Q}_I-F_{IK} 
   \mathcal{P}^K)\, (\mathcal{Q}_J-\bar F_{JL} \mathcal{P}^L)\,,
\end{equation}
which is independent of $\Upsilon$.
Note that \eqref{eq:effective-entropy-function} is
homogeneous under uniform rescalings of the charges $q_I$ and $p^I$
and the variables $Y^I$. This implies that the entropy will be
proportional the the square of the charges. Under infinitesimal
changes of $Y^I$ and $\bar Y^I$ the entropy function
\eqref{eq:effective-entropy-function} changes according to
\begin{eqnarray}
  \label{eq:attractorytree}
  \delta\mathcal{E} &=& 
  \mathcal{P}^I \,\delta(F_I+\bar F_I) -\mathcal{Q}_I \,\delta(Y^I+\bar
  Y^I) \\
  && 
  + \tfrac1{2} i  \left({\cal Q}_K -\bar F_{KM} \,{\cal P}^M
  \right) 
  N^{KI} \,\delta F_{IJ} \, N^{JL} \left( {\cal Q}_L - {\bar F}_{LN}
  \,{\cal P}^N \right)\nonumber \\
  && 
  - \tfrac1{2} i  \left({\cal Q}_K -F_{KM} \,{\cal P}^M
  \right) 
  N^{KI} \,\delta \bar F_{IJ} \, N^{JL} \left( {\cal Q}_L -  F_{LN}
  \, {\cal P}^N \right) = 0 \;, \nonumber
\end{eqnarray}
where $\delta F_I= F_{IJ}\,\delta Y^J$ and
$\delta{F}_{IJ}=F_{IJK}\,\delta{Y}^K$. This equation determines the
horizon value of the $Y^I$ in terms of the black hole charges $(p^I,
q_I)$. Because the function $F(Y)$ is homogeneous of second degree, we
have $F_{IJK}Y^K=0$. Using this relation one deduces from
\eqref{eq:attractorytree} that $\left({\cal Q}_J - F_{JK} \, {\cal
    P}^K \right) Y^J = 0$, which is equivalent to
\begin{equation}
  \label{eq:Z-Kappa}
  i ({\bar Y}^I F_I - Y^I{\bar F}_I) = p^I F_I - q_I Y^I \;. 
\end{equation}
Therefore, at the attractor point, we have 
\begin{equation}
\label{eq:expsig}
  \Sigma = i ( {\bar Y}^I F_I - Y^I {\bar F}_I ) \;.
\end{equation}
Inserting this result into (\ref{eq:eq-U}) yields
\begin{equation}
  \label{eq:valueups}
  \sqrt{- \Upsilon}  = \frac{8\,\Sigma}{\Sigma + N^{IJ} \left(
        {\cal Q}_I - F_{IK} \, {\cal P}^K\right) 
      \left( {\cal Q}_J - {\bar F}_{JL} \, {\cal P}^L \right)} \;,
\end{equation}
which gives the value of $\Upsilon$ in terms of the attractor values
of the $Y^I$. Using (\ref{eq:valueups}) we obtain
\begin{eqnarray}
  \label{eq:entrotree}
  {\cal S}_{\rm macro} (p,q) = 2 \pi \, 
  {\cal E} \Big\vert_{\rm attractor} = 
  \frac{8 \pi\, \Sigma}{\sqrt{-\Upsilon}}\Big\vert_{\rm attractor}  \;.
\end{eqnarray}
Observe that, for a BPS black hole, ${\cal Q}_I = {\cal P}^J =0$ and
$\Upsilon = - 64$, so that ${\cal S}_{\rm macro} = \pi \,
\Sigma\vert_{\rm attractor}$ in accord with \eqref{eq:W-entropy}. 

The entropy function \eqref{eq:effective-entropy-function}
can be written as
\begin{equation}
  \label{eq:Epsilon-U=1} 
  \mathcal{E} = - q_I (Y^I+ {\bar Y}^I) + p^I (F_I +{\bar F}_I )  +
  \ft12 N^{IJ} (q_I-F_{IK}p^K) (q_J-\bar F_{JL}p^L)+ N_{IJ} Y^I\bar
  Y^J  \,,
\end{equation}
where we used the homogeneity of the function $F(Y)$. Expressing the
$Y^I$ as in \eqref{YZX}, one obtains (using $N_{IJ} X^I {\bar X}^J = -1$)
\begin{equation}
  \label{eq:Epsilon-U=1-X} 
  \mathcal{E} =   \ft12 \left( N^{IJ}+ 2\, X^I
  \bar X^J\right)  (q_I-F_{IK}p^K) (q_J-\bar F_{JL}p^L)\,, 
\end{equation}
where $F_{IJ}$ is now the second derivative of $F(X)$ with respect to
$X^I$ and $X^J$. Comparision with the black hole potential \eqref{bhpot1} gives
$ \mathcal{E} = \frac12 \, V_{\rm BH}$, and hence, 
\begin{equation}
{\cal S}_{\rm macro} (p,q) = 
2 \pi \, 
  {\cal E} \Big\vert_{\rm attractor} = \pi \, V_{\rm BH}\Big\vert_{\rm attractor} = \pi \, V_{\rm BH} (p,q)
  \;.
  \label{entro=bhpot}
  \end{equation}

\subsubsection{BPS black holes with $R^2$-interactions}
\label{sec:bps-black-holes}
In the presence of $R^2$ interactions, the horizon values of $U$ and
$\Upsilon$ for extremal BPS black holes are $U =1$ and $\Upsilon =
-64$ \cite{LopesCardoso:2000qm}.  Inserting these values into
\eqref{eq:entropy-total-2} results in
\begin{eqnarray} 
  \label{eq:entropy-total-2-BPS}
  {\cal E} (Y, {\bar Y}, -64 ,1) 
  =  \ft12  \Sigma (Y, {\bar Y}, p, q) 
  + \ft12 N^{IJ} \left( {\cal Q}_I - F_{IK} \, {\cal P}^K\right) 
  \left( {\cal Q}_J - {\bar F}_{JL} \, {\cal P}^L \right) \;.
\end{eqnarray}
Observe that the variational principle based on 
\eqref{eq:entropy-total-2-BPS} is only consistent
with the one based on \eqref{eq:entropy-total-2}
provided that 
\eqref{eq:entropy-total-2-BPS} is supplemented by
the extremization equations for $U$,
\begin{eqnarray}
  \label{eq:valueU} 
  && \Sigma + \left( {\cal Q}_I - F_{IK} \, {\cal P}^K\right) 
  N^{IJ} \left({\cal Q}_J -{\bar F}_{JL} \, {\cal P}^L\right)
   -\frac{8\mathrm{i}}{\sqrt{-\Upsilon}} (\bar Y^IF_I -Y^I\bar F_I)
     \nonumber\\
  &&
  - \mathrm{i}(F_\Upsilon-\bar F_\Upsilon) \Big [- 4 \Upsilon + 64
   (1-U^{-2}) - 16  \sqrt{-\Upsilon} \Big] =0 \;,
\end{eqnarray}
 and for $\Upsilon$,
 \begin{eqnarray}
  \label{eq:homo-delta}
  &&
  U\Sigma - \mathrm{i} (\bar Y^I F_I -  Y^I \bar F_I) \left[U + 4 
  (-\Upsilon)^{-1/2} (U-1) \right]\nonumber\\
  &&
  +2\mathrm{i}U  \left[ \Upsilon F_{I\Upsilon} N^{IJ} (\mathcal{Q}_J -
  \bar F_{JK}\mathcal{P}^K) - \mathrm{h.c.} \right] \nonumber \\
  &&
  +2\mathrm{i} (F_\Upsilon - \bar F_\Upsilon) \left[ 2 U\Upsilon + 4
  \sqrt{-\Upsilon} (1+U)\right]  = 0\;. 
\end{eqnarray}
For BPS solutions it can be readily checked
that the latter are indeed satisfied.
Using ${\cal Q}_I = {\cal P}^J =0$, we obtain ${\cal S}_{\rm macro} = \pi \,
\Sigma\vert_{\rm attractor}$, in accord with \eqref{eq:W-entropy}.

\subsection{Large and small BPS black holes: examples   }

As an application \cite{LopesCardoso:1999fsj,LopesCardoso:2004law} 
of the above, let us consider BPS black holes in an ${\cal N}=2$
supergravity theory coupled to Weyl-square terms, whose Wilsonian action is
encoded in the holomorphic function
$F(Y, \Upsilon) = F^{(0)} (Y) +  F^{(1)} (Y) \, \Upsilon$ given by
\bea
F(Y,\Upsilon) = - \frac{Y^1 Y^a \eta_{ab} 
Y^b}{Y^0} + c_1 \; \frac{Y^1}{Y^0} \; \Upsilon \;\;.
\label{hetprep}
\eea
Here 
\bea
 Y^a \eta_{ab} Y^b = Y^2 Y^3 - \sum_{a=4}^{n} (Y^a)^2 \;,\qquad 
a=2, \ldots, n \;,
\eea
with real constants $\eta_{ab}$ and 
$c_1$.  We define $S = -i Y^1/Y^0$.

We introduce 
the charge vectors $N^I$ and $M_I$,
\begin{eqnarray}
N^I &=& (p^0, q_1, p^2, p^3, \dots, p^n) \;, \nonumber\\
M_I &=& ( q_0, - p^1, q_2, q_3, \dots, q_n) \;.
\end{eqnarray}
There are 
three bilinear charge combinations
that are invariant under $SO(n-1,2;  \mathbb{Z})$-transformations
\cite{LopesCardoso:1996yk}, also referred to as target space duality transformations, 
\bea
\langle M,M \rangle 
&=& 2 \Big(M_0 M_1 + \ft{1}{4} M_a \eta^{ab} M_b\Big) = 
2\Big( - q_0 p^1 + \ft{1}{4}
 q_a \eta^{ab} q_b \Big) \;\;,
\nonumber\\
\langle N,N \rangle &=& 2\Big( N^0 N^1 + N^a \eta_{ab} N^b \Big) = 
2 \Big(  p^0 q_1 +  p^a \eta_{ab} p^b \Big) \;\;,
\nonumber\\
M \cdot N &=& M_I N^I = q_0 p^0 - q_1 p^1 + q_2 p^2 + \cdots + q_{n}
p^{n} \;\;.
\label{bilincomb}
\eea
For instance, the charge bilinears are clearly invariant under the $SO(n-1,2;  \mathbb{Z})$-transformation 
\begin{equation}
      \begin{array}{rcl}
      p^0 &\to& q_1 \;,\\
      p^1 &\to& - q_0  \;,\\
      p^a &\to&  p^a  \;,
    \end{array}
    \quad
    \begin{array}{rcl}
      q_0 &\to& -p^1 \;, \\
      q_1 &\to& p^0 \;, \\
      q_a &\to& q_a \;.
    \end{array}
\end{equation}

\subsubsection{Large BPS black holes}
\noindent

\begin{defi} {\bf Large BPS black holes.}
A large single-centre BPS black hole in four dimensions is a dyonic spherically symmetric BPS black hole carrying electric/magnetic charges $(q_I, p^I)$, such that 
the charge bilinears $\langle M, M \rangle, \langle N, N \rangle$ are positive
and 
$ \langle M,M \rangle  \langle N,N \rangle  - ( M \cdot N )^2 \gg 1$.
\end{defi}

This ensures that at the two-derivative level, the black hole
has a non-vanishing horizon area $A$ \cite{Behrndt:1996jn},   $A= 2\pi (S + \bar S) \,  \langle N, N  \rangle $, 
c.f. \eqref{entros}
and \eqref{entroc1} below. \\

Using (c.f. \eqref{eq:P-Q-def})
\begin{equation}
Y^I-\bar Y{}^I= i\, p^I \;\;,\qquad 
F_I(Y,\Upsilon) -\bar F_I(\bar Y, \bar \Upsilon) = i\, q_I \;,
\label{attracgen}
\end{equation}
one obtains for $I= a$,
\begin{equation}
Y^a = {1\over S+\bar S} \Big[ -\ft12 \eta^{ab} q_b +i\bar S \,p^a
\Big]\, , 
\label{valueYa}
\end{equation}
where $\eta^{ab}\,\eta_{bc} = \delta^a_{\,c}$. Similarly, one finds
\bea
p^I F_I  - q_I Y^I
&=& i (\bar Y^I F_I - Y^I \bar F_I)  \\
&=& (S+\bar S) \Big(\,\bar Y^a \eta_{ab} Y^b + {\bar Y^0\over Y^0}
\Big[- Y^a \eta_{ab} Y^b   + c_1\, \Upsilon \Big]
+ \mbox{ h.c.}\Big) \,, \nonumber
\eea
as well as
\bea
q_1\,p^0 &=& - (Y^0- \bar Y^0 ) (F_1 - \bar F_1) \nonumber\\
&=& \Big({\bar Y^0 \over Y^0}-1\Big)   \Big[- Y^a
\eta_{ab} Y^b   + c_1\, \Upsilon  \Big]
+ \mbox{ h.c.}\,.
\eea
Combining these two equations and using \eqref{valueYa} 
yields 
\bea
p^I F_I  - q_I Y^I
= (S + {\bar S}) \Big( \ft{1}{2} \, \langle N,N \rangle
+ (c_1 \,\Upsilon + \mbox{ h.c.}) \, \Big)
\;,
\label{hzz}
\eea
where the bilinear charge combination $\langle N,N \rangle$
is defined in \eqref{bilincomb}.

Using \eqref{eq:W-entropy}, we obtain for Wald's entropy (with $\Upsilon = - 64$)
\bea
{\cal S}_{\rm macro} = \ft{1}{2}\, \pi \, (S + {\bar S})
\, \Big(\langle N,N \rangle - 512 \, c_1 \,   \Big)
\;,
\label{entros}
\eea
where $S$ is evaluated at the horizon.  We now determine its value.

Using \eqref{attracgen}, one finds that the combinations 
$S\bar S \, q_1\,p^0 + q_0\,p^1$ and $i(\bar S -S) \, q_1\,p^0+
q_1\,p^1 - q_0 \,p^0$ do not explicitly depend on $Y^0$.
This results in the following equations for $S$,
\bea
S\bar S \,\langle N,N\rangle &=&  \langle M,M\rangle  -2(S + {\bar
S}) ( c_1\,\Upsilon \,S + \mbox{ h.c.})\,, \nonumber\\
(S- {\bar S}) \,\langle N,N\rangle &=& 2i\, M\cdot N + 2 \,(S + {\bar S})
\,(c_1\,\Upsilon - \mbox{ h.c.} )\,,
\eea
from which one infers the value of $S$ at the horizon in terms of the charges,
\bea
S  =
\sqrt{ {\langle M,M \rangle \langle N,N \rangle - ( M \cdot N )^2}\over
{\langle N,N \rangle\,(\langle N,N \rangle  - 512 \,c_1) } } \; + \; i \;
\frac{ M \cdot N }{ \langle N,N \rangle} \;.
\label{Sc1}
\eea
The resulting entropy is expressed in terms of the charges as
\bea
{\cal S}_{\rm macro} = \pi \; \sqrt{ \langle M,M \rangle 
 \langle N,N \rangle - ( M \cdot N )^2 }
\;  \sqrt{1 - \frac{512 \,c_1 \,}{\langle N,N \rangle}}
\;.
\label{entroc1}
\eea
When $c_1 = 0$, this equals one quarter of the area of the horizon.

\subsubsection{Small BPS black holes}

\noindent

\begin{defi} {\bf Small BPS black holes.}
A small BPS black hole in four dimensions is a BPS black hole carrying electric/magnetic charges $(q_I, p^I)$ such that 
the charge
combination $ \langle M,M \rangle  \langle N,N \rangle  - ( M \cdot N )^2$ vanishes, and such that its macroscopic (Wald) entropy ${\cal S}_{\rm macro}$
is, for large charges, given by ${\cal S}_{\rm macro} \propto \sqrt{Q^2}$. Here $Q^2$ denotes a linear combination of
charge bilinears.
\end{defi}

At the two-derivative level,
a small BPS black hole is a null-singular solution to the equations of motion
of ${\cal N}=2$ supergravity theory.
For a small BPS black hole to have a non-vanishing area of the event horizon, 
higher-curvature corrections need to be taken into account \cite{Sen:1995in,Dabholkar:2004yr,Sen:2005kj,Dabholkar:2004dq}.\footnote{For a recent discussion and a different viewpoint, see \cite{Cano:2018hut,Faedo:2019xii}.}
When $c_1 \neq 0$, a horizon forms, leading to the cloaking of a null singularity that is present when $c_1 =0$
\cite{Dabholkar:2004dq}.
This requires $c_1 < 0$, as we will see below.

In the following, we will consider small black holes with charges $N^I=0$ in the model \eqref{hetprep}.
To compute 
the horizon value of $S$ as well as the entropy  \eqref{eq:W-entropy} of such a small
black hole, we proceed s follows. We start by considering
a large BPS black hole which is axion free, i.e. one for which ${\rm Im}Ê\, S = 0$. 
We thus set $M \cdot N =0$ in \eqref{Sc1} and in \eqref{entroc1}, which yields
\bea
S  + {\bar S}=
2 \sqrt{ {\langle M,M \rangle }\over
{\langle N,N \rangle  - 512 \,c_1 } } \;,\;
{\cal S}_{\rm macro} = \pi \; \sqrt{ \langle M,M \rangle 
 \langle N,N \rangle - 512 \,c_1 \, \langle M,M \rangle } .
\eea
Next, we set $\langle N,N \rangle =0$ in these expressions, which results in
\begin{equation}
S+\bar S = \sqrt{ - \langle M,M \rangle/(128\,c_1)} > 0 
 \;\;,\;\;
{\cal S}_{\rm macro} = 
2\,\pi\, \sqrt{-128\,c_1\,  \langle M,M \rangle}    \;.
\end{equation}
Using  \eqref{hzz},
\bea
\pi \left( p^I F_I  - q_I Y^I \right)
= - \pi \, 128 c_1 \, (S + {\bar S}) \;,
\eea
which equals 
a quarter of the horizon area and 
needs to be positive, we infer $c_1 < 0$, and hence $\langle M,M \rangle >0$. Thus,  $ \mathcal{S}_{\text{macro}}$
equals one half of the horizon area  \cite{Dabholkar:2004dq}.
Note that $S$ and the entropy only have finite values due to $c_1 \neq 0$.

\section{Born-Infeld-dilaton-axion system and $F$-function \label{bidaf}}

In subsection \ref{sec:ubiquity} we discussed how to recast point-particle Lagrangians in terms of functions $F$ of the form 
\eqref{FOm}. Here, we will consider the example of a homogenous function $F(x, \bar x, \eta)$ of degree $2$, with $\eta$ having scaling weight $m = -2$
(c.f. subsection  \ref{sec-F-hom}) and 
show that this describes the Born-Infeld-dilaton-axion system in an $AdS_2 \times S^2$
background. We follow \cite{Cardoso:2012nh}.

\subsection{ Homogeneous function $F$ }

We consider a function $F$  that depends on three complex scalar fields $X^I$ (with $I = 0, 1 ,2$), 
as well as on an external real
parameter $\eta$, 
\begin{align}
F(X, {\bar X}, \eta) = - \ft12 \,  \frac{X^1 (X^2)^2}{X^0} + 2 i\,  \Omega(X,{\bar X}, \eta) \;.
\label{eq:F-Y-eta}
\end{align}
We demand $F$ to be homogeneous of degree 2 under 
rescalings $ X^I \mapsto \lambda \, X^I \;,\;
\eta \mapsto \lambda^{m} \, \eta $, with $\lambda \in \mathbb{R}\backslash\{0\}$, as in \eqref{xeta}.
We leave the scaling weight $m$ arbitrary, for the time being.

 Duality
transformations are represented by ${\rm Sp}(6, \mathbb{R})$ matrices (which are $6 \times 6$ matrices of the form
\eqref{eq:matrix-sympl})
acting on $(X^I, F_I)$, 
where $F_I = \partial F(X, \bar X, \eta) / \partial X^I$. The external parameter $\eta$ is inert under these transformations.

Let us assume that 
the model based on \eqref{eq:F-Y-eta} is invariant under 
S-duality as well as under a particular T-duality transformation. These symmetry transformations belong to 
an 
${\rm SL}(2, \mathbb{R}) \times {\rm SL}(2, \mathbb{R})$
subgroup of ${\rm Sp}(6, \mathbb{R})$.  
The first ${\rm SL}(2, \mathbb{R})$ subgroup acts as follows on 
$(X^I, F_I)$, 
\begin{equation}
        \begin{array}{rcl}
      X^0 &\mapsto& d \, X^0 + c\, X^1 \;,\\
      X^1 &\mapsto& a \, X^1 + b \, X^0 \;,\\
      X^2 &\mapsto& d\, X^2 - c \,F_2  \;,
    \end{array}
    \quad
    \begin{array}{rcl}
      F_0 &\mapsto& a\,  F_0 -b\,F_1 \;, \\
      F_1 &\mapsto& d \,F_1 -c\, F_0 \;,\\
      F_2 &\mapsto& a \, F_2 -b\,  X^2 \;,
    \end{array}
    \label{eq:electro-magn-dual}
\end{equation}
where $a, b, c, d$ are real parameters that satisfy $a d - b c = 1$. 
This symmetry is referred to as S-duality. Let us describe its action on two complex scalar fields
$S$ and $T$ that are given by the scale invariant combinations
$S = - i X^1/X^0$ and $T = - i X^2/X^0$. 
The S-duality transformation \eqref{eq:electro-magn-dual} acts as 
\begin{equation}
S \mapsto \frac{ a S - i b }{ i c S + d} \;\;\;,\;\;\;
T \mapsto T + \frac{ 2 i \, c }{\Delta_{\rm S} \, (Y^0)^2} \, \frac{\partial \Omega}{\partial T} 
\;\;\;,\;\;\; X^0 \mapsto \Delta_{\rm S} \, X^0\;,
\label{eq:S-dual-T-Y}
\end{equation}
where we view $\Omega$ as a function of $S, T, X^0$ and their complex conjugates, and where
\begin{equation}
\Delta_{\rm S} = d + i c \, S \;.
\end{equation}
The second ${\rm SL}(2, \mathbb{R})$ subgroup is referred to as 
T-duality group. Here we focus on a particular T-duality transformation given by
\begin{equation}
     \begin{array}{rcl}
      X^0 &\mapsto& F_1 \;,\\
      X^1 &\mapsto& - F_0  \;,\\
      X^2 &\mapsto&  X^2  \;,
    \end{array}
    \quad
    \begin{array}{rcl}
      F_0 &\mapsto& -X^1 \;, \\
      F_1 &\mapsto& X^0 \;, \\
      F_2 &\mapsto& F_2 \;,
    \end{array}
    \label{eq:T-dual-inv}
\end{equation}
which results in
\begin{equation}
  S\mapsto S+ \frac2{\Delta_{\mathrm{T}}(X^0)^2}
  \,\left[- X^0 \frac{\partial\Omega}{\partial X^0} + 
  T\frac{\partial\Omega}{\partial T} \right]  \;\;\;,\;\;\;
 T \mapsto \frac{T}{\Delta_{\mathrm{T}}}  \;\;\;,\;\;\;
  X^0 \mapsto  \Delta_{\mathrm{T}} \, X^0 \;,
  \label{eq:ST-T-full}
\end{equation} 
where
\begin{equation}
  \label{eq:Delta-T}
  \Delta_{\mathrm{T}} = \tfrac12 T^2 +\frac{2}{(X^0)^2}
  \frac{\partial\Omega}{\partial S}\;. 
\end{equation}

When a symplectic transformation describes a symmetry of the system, a convenient method for verifying
this consists in 
performing the substitution $X^I \mapsto \tilde{X}^I$ in the derivatives $F_I$, and checking that this 
substitution 
correctly induces the symplectic transformation of $F_I$.
This will impose restrictions on
the form of $F$, and hence also on $\Omega$.
Imposing that  S-duality \eqref{eq:electro-magn-dual} constitutes a symmetry of 
the model \eqref{eq:F-Y-eta}
results in the following 
conditions on
the transformation behaviour of the derivatives of $\Omega$ \cite{Cardoso:2008fr}, 
\begin{align}
 \left( \frac{\partial \Omega}{\partial T} \right)^\prime_\mathrm{S}
=&\, \frac{\partial \Omega}{\partial T} \;, \nonumber\\
\left( \frac{\partial \Omega}{\partial S} \right)^\prime_\mathrm{S}  
=&\,
\Delta_{\rm S}{}^2 \,
\left( \frac{\partial \Omega}{\partial S} \right) + 
 \frac{\partial\left( \Delta_{\rm S}{}^2 \right)}{\partial S} \left[ - \ft12 X^0 \, \frac{\partial \Omega}{\partial X^0}   
 - i c  \, {2 \Delta_{\rm S} \, (X^0)^2} \left(\frac{\partial \Omega}{\partial T}\right)^2
 \right] \;, \nonumber\\
 \left(X^0 \, \frac{\partial \Omega}{\partial X^0} \right)^\prime_\mathrm{S}
=&\, X^0 \, \frac{\partial \Omega}{\partial X^0} + \frac{2i c}{\Delta_{\rm S} \, (X^0)^2} \left(\frac{\partial \Omega}{\partial T}\right)^2\;,
\label{eq:Oprime-O-rel}
\end{align}
while requiring the particular T-duality transformation \eqref{eq:T-dual-inv} to constitute a symmetry imposes the transformation behaviour \cite{Cardoso:2008fr}
\begin{align}
  \label{eq:T-invariance}
  \left(\frac{\partial\Omega}{\partial S}\right)^\prime_\mathrm{T} =&\, 
  \frac{\partial\Omega}{\partial S} \;, \nonumber\\
  \left(\frac{\partial\Omega}{\partial T}\right)^\prime_\mathrm{T} =&\,
  \left(\Delta_{\mathrm{T}}  - T^2
  \right) \,\frac{\partial\Omega}{\partial T}
  + T\;X^0 \frac{\partial\Omega}{\partial X^0} \;,
  \nonumber\\ 
  \left(X^0 \frac{\partial\Omega}{\partial X^0}\right)^\prime_\mathrm{T} =&\,
  X^0 \frac{\partial\Omega}{\partial X^0} + 
  \frac{4}{\Delta_{\mathrm{T}}\,(X^0)^2}
  \,\frac{\partial\Omega}{\partial S} 
  \left[- X^0 \frac{\partial\Omega}{\partial X^0} +
  T \frac{\partial\Omega}{\partial T }\right]\;. 
\end{align}

Solutions to both
\eqref{eq:Oprime-O-rel} and \eqref{eq:T-invariance} may be constructed iteratively by assuming that $\Omega$
possesses a power series expansion in $\eta$,
\begin{equation}
\Omega (X, \bar X, \eta) = \sum_{n=1}^{\infty} {\eta}^{n} \, \Omega^{(n)} (X, \bar X) \;.
\label{ex:om-pseries}
\end{equation}
Note that since $\Omega$ and $\eta$ are real, so are the expansion functions $\Omega^{(n)}$.
The latter have to scale as $\lambda^{-m n +2}$. 
Once a solution $\Omega^{(1)}$ to \eqref{eq:Oprime-O-rel} and \eqref{eq:T-invariance} has been found, the full expression  \eqref{ex:om-pseries}
 can be constructed by solving 
\eqref{eq:Oprime-O-rel}  and \eqref{eq:T-invariance}
iteratively starting from 
$\Omega^{(1)}$.

So far, we have not made any assumptions about the scaling weight $m$ in \eqref{xeta}.  Depending on the choice
of $m$, the expansion \eqref{ex:om-pseries} will have different properties.  For concreteness, let us 
take $m=-2$, which implies that the expansion functions $\Omega^{(n)}$ in \eqref{ex:om-pseries} will have
to scale as $\lambda^{2 n +2}$. The lowest function $\Omega^{(1)}$ will therefore scale as $\lambda^4$.
We make an ansatz for $\Omega^{(1)}$ that is consistent with this scaling behaviour, 
\beq
\Omega^{(1)} (X, \bar X) = |X^0|^{4} \, g(S, T, \bar S, \bar T) \;.
\eeq
The equations \eqref{eq:Oprime-O-rel} and \eqref{eq:T-invariance}  require $\Omega^{(1)}$ to be invariant under
the S-duality and T-duality transformations given above, and determine it to be given by
\begin{equation}
\Omega^{(1)} = \ft18 \, |X^0|^4 \, (S+\bar S)^2 \, |T|^4 \;,
\label{eq:om-1-g2}
\end{equation}
where we have chosen a particular normalization, for later convenience.
  
We may now proceed iteratively to determine the higher $\Omega^{(n)}$, solving \eqref{eq:Oprime-O-rel} and \eqref{eq:T-invariance}
order by order in $\eta$, using the transformation laws \eqref{eq:S-dual-T-Y} and \eqref{eq:ST-T-full}).
Rather than proceeding in this way, we present an exact solution to  \eqref{eq:Oprime-O-rel} and \eqref{eq:T-invariance}
that, to lowest order in $\eta$, reduces to \eqref{eq:om-1-g2},
\bea
 \Omega(X,\bar X, \eta) &=& \, \tfrac18 \,\eta^{-2} \left[ \; \sqrt{1 - \ft12
       \eta^2 \, (S + \bar S) \,  (T X^0 - {\bar T} {\bar X}^0)^2}  \right. \nonumber\\
 && \qquad   \qquad  \left.   - \sqrt{1 - \ft12 \eta^2 \, (S + \bar S) \,
       (T X^0 + {\bar T} {\bar X}^0)^2} \; \right]^2 \,. 
 \label{fullOm}      
\eea
It can be verified that \eqref{fullOm} satisfies \eqref{eq:Oprime-O-rel} and \eqref{eq:T-invariance}.
Note that \eqref{fullOm} scales correctly as $\Omega(\lambda X, \lambda {\bar X},  \lambda^{-2} \, \eta) = 
\lambda^2 \, \Omega(X, \bar X, \eta ) $.

In the next subsection, we
turn to the interpretation of the function $F$ based on \eqref{fullOm}.

\subsection{Interpretation: the Born-Infeld-dilaton-axion system in an $AdS_2 \times S^2$ background}

The function $F$ based on \eqref{fullOm} describes a
Born-Infeld-dilaton-axion system in an $AdS_2 \times S^2$ background, as we proceed to explain.

We consider the Born-Infeld Lagrangian in the presence of a dilaton-axion field $S = \Phi + i \, B $
\cite{Gibbons:1995ap},
\begin{equation}
  \label{eq:BI-dil-gen}
 {L}= -g^{-2} \left[\sqrt{\vert \det[g_{\mu\nu} +
        g\,\Phi^{1/2} \,  F_{\mu\nu}] \vert } - \sqrt{\vert \det g_{\mu \nu} \vert } \right] 
        + \sqrt{\vert \det g_{\mu \nu} \vert } \, \tfrac14 \, B \, F_{\mu \nu} {\tilde F}^{\mu \nu} \;,
\end{equation}
where here ${\tilde F}_{ab} = \tfrac12 \varepsilon_{abcd} \, F^{cd}$ with $\varepsilon_{0123} =1$.
In this Lagrangian, the gauge coupling $g$ appears multiplied by the dilaton field $\Phi$, while
 the term $ B F_{\mu \nu } {\tilde F}^{\mu \nu}$ introduces a scalar field degree of freedom
called the axion.  The Born-Infeld-dilaton-axion system described by \eqref{eq:BI-dil-gen} has
duality symmetries that will be described below.

Let us consider the system \eqref{eq:BI-dil-gen} in an $AdS_2 \times S^2$ background
\begin{align}
ds^2 =&\, v_1 \left( - r^2 \, dt^2 + \frac{dr^2}{r^2} \right) + v_2 \left(d \theta^2 + \sin^2 \theta \, d\phi^2 \right) \;, \nonumber\\
F_{rt} =&\, v_1 \, e \;\;\;,\;\;\; F_{\theta\phi} = v_2 \, p \, \sin \theta \;,
\label{eq:static-spher-sym}
\end{align}
i.e., let us restrict
to field
configurations that have the 
$SO(2,1) \times SO(3)$ symmetry of $AdS_2 \times S^2$, in which case $v_1, v_2, e, p, \Phi, B$ are constants.
Integrating over
the angular variables and setting $v_1 v_2 \, 4 \pi =1$, for convenience, yields 
\begin{equation}
{L} (e, p, \Phi, B) = - g^{-2} \left[\sqrt{1-g^2 \, \Phi \,
      e^2}\;\sqrt{1+g^2 \, \Phi \, p^2} -1\right] + B \, e \, p \,,
      \label{eq:red-lag-S-bS}
\end{equation}
where we assume $g^2 \, \Phi \, e^2 < 1$.
 To obtain the associated Hamiltonian ${H}$,
\begin{equation}
{H}(p,q, \Phi, B) = q \, e - \mathcal{L} (e,p, \Phi, B) \;,
\label{eq:leg-H-L-S}
\end{equation}
we first compute $q=\partial {L}/\partial e$,
\begin{equation}
  \label{eq:def-q-S}
  q=
    e \, \Phi \;\sqrt{\frac{1+g^2 \, \Phi \,
      p^2} {1-g^2 \, \Phi \, e^2}} +  B \, p \;.
 \end{equation}
Inverting this relation yields
\begin{equation}
e = \frac{q- B \, p }{ \sqrt{\Phi^2  + g^2 \, \Phi \, \left[ \Phi^2  \, p^2 + (q - B \, p )^2 \right] }} \;,
\label{elece}
\end{equation}
and substituting in \eqref{eq:leg-H-L-S} gives 
\begin{equation}
  \label{eq:red-H-Phi-B}
  {H} (p,q, \Phi, B) =
  g^{-2}\left[\sqrt{1+g^2[\Phi \, p^2+ \Phi^{-1} \, \left(q - B \, p \right)^2]}
    -1\right ] \,. 
\end{equation}
Then, expressing $\Phi$ and $B$ in terms of $S$ and $\bar S$ results in 
\begin{equation}
  \label{eq:red-H-S}
 {H} (p,q, S, \bar S) =
  g^{-2}\left[\sqrt{1+ 2 \, g^2 \, \Sigma(p, q, S, \bar S)} 
      -1\right ] \,,
\end{equation}
where 
\begin{equation}
\Sigma (p, q, S, \bar S) = \frac{q^2 + i p \,q (S - \bar S) + p^2 \, |S|^2}{S + \bar S} \;.
\label{eq:Sigma}
\end{equation}
The Hamiltonian \eqref{eq:red-H-S} depends on canonical coordinates $(p,q)$, on an external parameter $g^2$
as well as on the dilaton-axion field, which describes a background field. We observe that ${H}$ scales as 
${H} \mapsto \lambda^2 {H}$
under  $(p, q) \mapsto \lambda (p, q) \,,\,
g^2 \mapsto \lambda^{-2} \, g^2 \,,\, S \mapsto S$, with $\lambda \in \mathbb{R}\backslash\{0\}$.
The electric field \eqref{elece} scales as  $e  \mapsto \lambda e$.

Let us now return to the reduced Lagrangian \eqref{eq:red-lag-S-bS} and recast it in the form
${ L} = 4 \left[ {\rm Im} F - \Omega \right]$, c.f. \eqref{eq:H-sympl},
where we introduce the complex variable
$x = \ft12 (p + i e)$, which scales as  $x \mapsto \lambda x$. The function $F$ will now depend on the two complex scalar fields $x$ and $S$, 
\begin{align}
F(x, {\bar x}, S, {\bar S}, g^2) = F^{(0)}(x, S)
+ 2 i \Omega(x, {\bar x}, S, {\bar S},  g^2) \;,
\label{eq:F-S-x}
\end{align}
and is determined as follows. The holomorphic function $F^{(0)}$ encodes all the contributions 
that are independent of $g^2$, while $\Omega$, which is real, accounts for all the terms in the reduced Lagrangian that
depend on $g^2$.  This yields,
\bea
  \label{eq:def-Omega-dil}
  F^{(0)}(x, S) &=&  - \ft12 i \, S \, x^2 \;, \\
     \Omega(x,\bar x, S, {\bar S}, g^2) &=& \tfrac18 \,g^{-2} \left(\sqrt{1+\ft12
       g^2 \, (S + \bar S) \, (x+\bar x)^2} \right. \nonumber\\
&& \left. \qquad   \qquad       - \sqrt{1+\ft12 g^2 \, (S + \bar S) \,
       (x-\bar x)^2}\right)^2 \,. \nonumber
\eea
Under the scaling 
 $x \mapsto \lambda x$, $g^2 \mapsto \lambda^{-2} \, g^2 \,,\, S \mapsto S$, 
$F$ scales as $F \mapsto \lambda^2 F$.

Now we note that 
the function $F$ given in \eqref{eq:def-Omega-dil} precisely matches the one given in \eqref{eq:F-Y-eta} and
\eqref{fullOm} upon identifying
\beq
S = -i \frac{X^1}{X^0} \;\;\;,\;\;\; x = X^2 = i \, T \, X^0 \;\;\;,\;\;\; g = \eta \;.
\eeq
The Hamiltonian \eqref{eq:red-H-S} is invariant under the S- and T-duality transformations
discussed in the previous subsection. We proceed to verify this.
The external parameter $g^2$ is inert under these transformations.
Using \eqref{eq:theorem-prop} we infer that the canonical pair $(p,q)$ is given by 
$(2 \, {\rm Re} \, x, 2 \,  {\rm Re} \, F_x)$.
The T-duality transformation \eqref{eq:T-dual-inv} leaves $(x, F_x)$ invariant.  
Since $\Omega$
given in 
\eqref{eq:def-Omega-dil}, or equivalently in \eqref{fullOm}, satisfies
$X^0 \partial \Omega / \partial X^0 = T \partial \Omega / \partial T$, $S$ is inert under  \eqref{eq:T-dual-inv}.  Consequently,
the Hamiltonian \eqref{eq:red-H-S} is invariant under the T-duality transformation \eqref{eq:T-dual-inv}.
The 
S-duality transformation \eqref{eq:electro-magn-dual},
\begin{eqnarray}
S \mapsto \frac{a S - i b}{i c S + d} \;,
\label{eq:transf-S}
\end{eqnarray}
 induces the following transformation of the canonical pair $(p,q)$,
\begin{equation}
\begin{pmatrix}
p \\ q 
\end{pmatrix}
\mapsto
\begin{pmatrix}
\tilde{p} \\ \tilde{q} 
\end{pmatrix}
=
\begin{pmatrix}
d & - c\\ - b & a 
\end{pmatrix}
\begin{pmatrix}
p \\ q 
\end{pmatrix} \;,
\label{eq:em-duality-pq}
\end{equation} 
where $a, b, c, d \in \mathbb{R}$ and $ad-bc=1$. 
Hence, $\Sigma$ given in \eqref{eq:Sigma}
is invariant under 
S-duality, and so is $H$.

\section{$F$-function for an STU-model \label{FSTUmod}}

As an application of electric-magnetic duality in a chiral background, discussed in section \ref{sec:emchirback},  let us 
consider the STU-model of Sen and Vafa (referred to as $N=2$ Example D in \cite{Sen:1995ff})
 in the presence of higher curvature interactions proportional to the 
square of the Weyl tensor. This model possesses duality symmetries which were used recently in \cite{Cardoso:2019rjs}
to determine the function $F$. The holomorphic function $F$ takes the form
\begin{eqnarray}
F(X, \hat A) = - \frac{X^1 X^2 X^3}{X^0} + 2 i \Omega (X, \hat A) \;,
\end{eqnarray}
with $\hat A$ given in \eqref{background-def}. Note that $\hat A$ has scaling weight $2$. The model possesses S-, T- and U-duality symmetries $\Gamma_0(2)_S \times
\Gamma_0(2)_T \times \Gamma_0(2)_U$  as well as triality symmetry.
$\Gamma_0(2)$ is the
subgroup of the group ${\rm SL}(2, \mathbb{Z})$ defined by
restricting its integer-valued matrix elements $a,b,c,d$ (with
$ad-bc=1$) to $a, d\in 2\,\mathbb{Z}+1$, $c\in 2\,\mathbb{Z}$ and
$b\in \mathbb{Z}$.  Triality symmetry refers to the invariance of the model 
under exchanges
of the scalar fields $S = -i X^1/ X^0, \,T = -i X^2/ X^0$ and $U = -i X^3/ X^0$.
The duality and triality symmetries of the model are very restrictive and allow for the determination of the function $F$.
For instance, under S-duality, the derivatives of $\Omega$ are required to transform in the following way,
\begin{align}
  \label{eq:S-invariance-STU}
  \bigg(\frac{\partial\Omega}{\partial T}\bigg)^\prime_\mathrm{S} =&\;
  \frac{\partial\Omega}{\partial T} \;,\qquad\quad
  \left(\frac{\partial\Omega}{\partial U}\right)^\prime_\mathrm{S} =~
  \frac{\partial\Omega}{\partial U} \;,
 \nonumber\\[1mm]
  \bigg(\frac{\partial\Omega}{\partial S}\bigg)^\prime_\mathrm{S} -
  \Delta_{\mathrm{S}}{\!}^2\,\frac{\partial\Omega}{\partial S} =&\;
  \frac{\partial\Delta_{\mathrm{S}}}{\partial{S}} 
  \bigg[- \Delta_{\mathrm{S}} \, X^0\, \frac{\partial\Omega}{\partial X^0}
  -\frac{2}{(X^0)^2}\,
  \frac{\partial\Delta_{\mathrm{S}}}{\partial{S}} 
  \,\frac{\partial\Omega}{\partial T} 
  \frac{\partial\Omega}{\partial U} \bigg] \;,
  \nonumber\\[1mm] 
  \bigg(X^0\frac{\partial\Omega}{\partial X^0}\bigg)^\prime_\mathrm{S} =&\;
  X^0 \frac{\partial\Omega}{\partial X^0} 
  +\frac{4}{\Delta_{\mathrm{S}}\,(X^0)^2}
  \,\frac{\partial\Delta_{\mathrm{S}}}{\partial{S}}\, 
  \,\frac{\partial\Omega}{\partial T}
  \frac{\partial\Omega}{\partial U}\;.
\end{align}
Using triality, one obtains similar equations under T- and U-duality.

The equations \eqref{eq:S-invariance-STU} are non-linear in $\Omega$, and where solved  \cite{Cardoso:2019rjs}
by iteration using the fact that $\Omega (X, \hat A) $ must be a
homogeneous function of second degree, c.f. \eqref{homFA}. This was achieved by expanding $\Omega (X, \hat A) $ 
in a series expansion in powers of ${\hat A}\,(X^0)^{-2}$ (which has scaling weight zero),
with coefficient functions that depend on $S,T,U$ and on an overall factor $\hat A$,
\begin{equation}
  \label{eq:Ups-holo-expansion}
  \Omega(X,\hat A) = {\hat A}\, \bigg[
  \gamma\,\ln\frac{(X^0)^2}{\hat A} + \omega^{(1)}(S,T,U) +
  \sum_{n=1}^{\infty} \, 
  \Big(\frac{\hat A}{(X^0)^2} \Big)^{n} \, \omega^{(n+1)}(S,T,U)\bigg]
  \,.
\end{equation}
Note the presence of the logarithmic term, whose inclusion allowed to implement the duality symmetries
of the model, leading to the determination of the gravitational coupling
functions $\omega^{(n)}(S,T,U)$ by iteration. Additional important information about the structure of $F$ was gleaned from the Hesse
potential for the model and the associated holomorphic anomaly equation. We refer to 
 \cite{Cardoso:2019rjs} for a detailed discussion thereof.

\subsection*{Acknowledgements}
We would like to thank Vicente Cort\'es, Bernard de Wit
and Swapna Mahapatra for the long-time collaboration which 
created the work we have been reporting on in this review. 
We thank Antoine Van Proeyen and Edoardo Lauria for making available to us the draft
of a set of Lecture Notes on `Supergravity and its matter couplings:
An introduction to ${\cal N}=2$ in $D=4,5,6$.' 
The work of GLC was supported by FCT/Portugal through UID/MAT/04459/2019.
We thank the referee for helpful comments on the first version of this article.

\begin{appendix}
  
\section{Mathematics background}

\subsection{Manifolds, group actions, submanifolds, immersions and embeddings \label{app:manifolds}}

In this article, manifolds $M$ are understood to be smooth, Hausdorff
and second countable. The Hausdorff separation property requires that
any two points on $M$ can be separated by non-intersecting open 
neighbourhoods. The second countability property requires that the topology 
(set of open subsets) is generated by a countable collection of open subsets.  

The (left) action 
\begin{equation}
G \times M \rightarrow M  \;,\;\;\; (g,x) \mapsto g\cdot x 
\end{equation}
of a group $G$ on a manifold $M$ is called
\begin{itemize}
\item 
{\em transitive}, if any two $x,y\in M$ are related by the action of $G$,
\item
{\em effective (faithful)}, if every $g \in G$ acts non-trivially on $M$,
\item
{\em free}, if all group elements different from the identity act on $M$ without
fixed points,
\item
{\em principal (regular, simply transitively)}, if $G$ acts both freely and transitively. 
\item
{\em proper}, if $G$ is a topological group and $G\times M \rightarrow M \times M,
(g,x) \mapsto (g\cdot x, x)$ is a proper map in the topological sense, that is, 
pre-images of compact sets are compact.
\end{itemize}
Since the orbits of $G$ on $M$ need not all have the same dimension, the 
{\em space of orbits} $M/G$ is in general not a manifold. Moreover, even if $M/G$
is a smooth manifold and $M$ is Hausdorff, it can happen that $M/G$ is not
Hausdorff. A sufficient condition for 
$M/G$ to be Hausdorff is that the action of $G$ is proper, which is satisfied
in particular for compact groups $G$ . If the action of $G$ is both free and proper,
then $M \rightarrow M/G$ is a {\em $G$-principal bundle}, see \ref{app:bundles}.
Since the group actions we are interested in involve non-compact groups, 
we will impose that quotients are Hausdorff as an explicit condition. Actions
of Lie groups on manifolds can be described using generating vector fields,
see \ref{App:integral_curves}.

The rank of a smooth map $F:M\rightarrow N$ between
manifolds $M,N$ is the rank of the induced linear map 
$F_*:T_pM \rightarrow T_{F(p)} N$ between tangent space. 
A smooth map $F$ is called
an {\em immersion (submersion)} if $F_*$ is injective (surjective) at
every point, that is if $\mbox{rank}(F) = \dim M$ ($\mbox{rank}(F) = 
\dim N$). A smooth {\em embedding} is an immersion that is also a
topological embedding, that is, a homeomorphism $F: M \rightarrow F(M)
\subset N$, where $F(M)$ carries the topology induced by $N$ through
restriction. Embedded
submanifolds are precisely the images of smooth embeddings.
An immersed submanifold $S\subset N$ is a subset which is 
a manifold such that $\iota: S \rightarrow N$ is an injective immersion. 
Immersed submanifold are precisely the images of injective
immersions. 

Note that the image of an immersion need not be a submanifold, 
since immersions are not required to be invertible. Thus they can 
have self-intersection points, for example.  Moreover, just requiring an immersion to be
invertible does not make it an embedding, because the topology of the
image need not agree with the submanifold topology induced by $N$.
However, locally an immersion is an embedding, and if one is
interested in local problems one can choose the domain of 
an immersion small enough, so that it becomes an embedding. 
This is used frequently in the main part of this review. 

As an example consider a smooth immersion which maps the real line
onto a `figure eight' shaped figure in $\mathbb{R}^2$, such that 
points $x \in \frac{1}{2} \mathbbm{Z}$ on the line
are mapped to the self-intersection point of the image. Now
restrict to an open interval $a< x < b$, equipped with the subspace 
topology induced by $\mathbb{R}$. For $a<0, b>1$ the self-intersection point appears at least
twice as an image, and the immersion is not invertible. For $a=0,
b=1$, the immersion is invertible, but not a topological embedding:
if we take a Cauchy sequence accumulating at, say, $a=0$, this does 
not converge to a point in the interval, but the image of this
sequence will converge to the self-intersection point in the topology
induced by $\mathbb{R}^2$. For $0<a < b <1$ the topology
induced by $\mathbb{R}^2$ is the standard topology of an 
open one-dimensional interval, and the immersion becomes an 
embedding.

For further reading we refer to \cite{Lee:Smooth_Manifolds}, on which
this section is partly based. 

\subsection{Fibre bundles and sections \label{app:bundles}}

The material in \ref{app:bundles} -- \ref{App:Pull_back} is mostly standard. Our
presentation is based on various sources, including \cite{Frankel,GHL,ChoquetBruhat}.

A smooth {\em fibre bundle}
\begin{equation}
F \xrightarrow{} E \xrightarrow{\pi} M
\end{equation}
is a smooth manifold $E$ which locally looks like the product $M\times F$
of two smooth manifolds, the base $M$ and the fibre $F$. More precisely,
there is a smooth surjective map $\pi\;: E \rightarrow M$ such that for all $x\in M$
there exists a neighbourhood $U$ such that $\pi^{-1}(U)$ is diffeomorphic to 
$U \times F$. Given an open cover $\{ U_{(i)} | i \in I\} $ of $M$ a fibre bundle
can be described in terms of an atlas with charts $(U_{(i)}, \varphi_{(i)}) $ that are glued together 
consistently by transitions functions 
\begin{equation}
\phi_{(ij)} = \varphi_{(i)} \varphi_{(j)}^{-1}  \;:  U_{(ij)} \times F \rightarrow U_{(ij)} \times F 
\end{equation}
on overlaps $U_{(ij)} = U_{(i)} \cap U_{(j)}$ .
The inverse image $F_x = \pi^{-1}(x) \cong F$ of $x$ is called the
fibre over $x\in M$. Most of the fibre bundles relevant for us are {\em vector bundles}, 
where $F$ is a vector space. Particular cases are the tangent bundle $TM$,
the cotangent bundle $T^*M$, and tensor bundles
\begin{equation}
TM \otimes \cdots \otimes TM \otimes T^*M \otimes \cdots \otimes T^*M \;.
\end{equation}
A smooth {\em section} of a fibre bundle is a smooth map $s: M \rightarrow E$ such that 
$\pi \circ s = \Id_M$. In addition to global sections, that is sections defined 
over all of $M$, one can consider local sections 
over domains $U\subset M$. Local sections need not to extend to global sections. 
By considering all open subsets $U\subset M$ together with all sections of $E$
over subsets $U$, one obtains the {\em sheaf of sections} of $E$.
In our applications it will be clear from context whether sections of vector bundles
(vector fields, tensor fields) are required to exist locally or globally.

An {\em affine bundle} modelled on a vector bundle  $V \rightarrow M$ is a fibre bundle $A \rightarrow M$
such that:
\begin{itemize}
\item
The fibres $A_p$ of $A$ over $p\in M$ 
are affine spaces over the vector spaces $V_p$, which are the fibres of the vector bundle 
$V$.
\item
The transition functions of a bundle atlas of $A$ are affine isomorphisms whose linear
parts are the transition functions of $V\rightarrow M$. 
\end{itemize}

Another important class of bundles are {\em principal bundles}. For a Lie group $G$ a $G$-principal
bundle $P$ over a manifold $M$ is a manifold $P$ equipped with a principal action of $G$.
Since the $G$-action on $P$ is free and transitive, each orbit of $G$ on $P$ can be identified
with $G$ upon choosing one point on the orbit, which is identified with the unit element. Thus orbits are loosely speaking copies 
of $G$ where we forget where the unit element is located (similar to passing from a vector space
to the associated affine space, or from vector bundles to affine bundles). The base manifold $M$
of the fibre bundle $P\rightarrow M$ is the space of orbits, $M=P/G$. A principal bundle is trivial, 
that is $P = M\times G$ is a product, if and only if $P$ admits a global section (which identifies,
in each fibre, which point corresponds to the unit element of the group). 
By picking a representation $\rho: G \rightarrow V$ of $G$ on a vector space $V$ one can
associate to the principal bundle $G$ a vector bundle with fibre $V$ and $G$-action defined
by $\rho$. One then says that the vector bundle is associated to the principal bundle. 
A $U(1)$ principal bundle is also called a circle bundle. By choosing the representation 
of $U(1)$ by the action of $SO(2)$ on the complex plane, one obtains an associated
complex line bundle, that is a vector bundle with fibre $\mathbb{C}$. We refer to
\ref{App:ComplexVBdl} for more material on complex vector bundles.

\subsection{Vector fields and differential forms \label{app:vector-fields}}

\subsubsection{Vector fields and frames}

Let $M$ be a smooth manifold. 
Vector fields are denoted $X,Y, \ldots \in \mathfrak{X}(M) =
\Gamma(TM)$.\footnote{Where convenient or required by 
consistency with the physics literature, we will also use
symbols, like $\xi, \eta, \ldots$, or $t,s,\ldots$ for vector fields.
} 
The local expansion of a vector field with respect to coordinates
$x^m$ is
\begin{equation}
X = X^m \frac{\partial}{\partial x^m} = X^m\partial_m \;.
\end{equation}
Vector fields operate on functions as first order differential
operators (directional derivatives):
\begin{equation}
X(f) = X^m \partial_m f \;.
\end{equation}
The {\em Lie bracket} $[X,Y]$ of two vector fields
\begin{equation}
[X,Y](f) = XY(f) - YX(f) = 
\left(X^m(\partial_m Y^n) - Y^m (\partial_m X^n)\right) \partial_n f
\end{equation}
is again a first order differential operator. The Lie bracket gives
the space of vector fields the structure of a Lie algebra.

Instead of a {\em coordinate frame} $\partial_m$, we can more generally expand a vector
field with respect to a local {\em frame} $e_m$, that is a set of vector fields
which form a basis of $T_xM$ for all $x\in U \subset M$, where $U$ is
an open neighbourhood,
\begin{equation}
X = X^m e_m  \;.
\end{equation}
The local sections $e_m$ are generators for
the Lie algebra of vector fields, $[e_m, e_n] = c^p_{mn} e_p$. A frame
$\{ e_m \}$ is locally a coordinate frame if and only if
$c^p_{mn}=0$ \cite{Frankel}. 

The expression of a Lie bracket with respect to frame is:
\begin{equation}
[X,Y] = \left(  X^m e_m (Y^p) - Y^m e_m(X^p) + X^m Y^n c^p_{mn}
\right) e_p \;.
\end{equation}

\subsubsection{Differential forms, dual frames, exterior derivative}

Given a frame $\{ e_m\}$, 
the dual {\em co-frame} $\{e^m\}$, which forms a basis for the one-forms
$\omega \in \Omega^1(M) = \Gamma(T^*M)$ is defined by 
$e^m(e_n)=\delta^m_n$. In the following `choosing a frame' (or
co-frame) always means that we choose a dual pair $\{ e^m, e_n\}$. 
Given a coordinate system, the coordinate differentials
$dx^m$ form the 
frame dual  to the coordinate vector fields $\partial_m$.  A co-frame
is locally a coordinate co-frame if $de^m = 0$. 
The expansion of a one-form in a coordinate co-frame is
\begin{equation}
\omega= \omega_mdx^m \;.
\end{equation}
The {\em wedge product} of one-forms is defined by
\begin{equation}
\label{asym}
\alpha \wedge \beta = \alpha \otimes \beta - \beta \otimes \alpha  \;.
\end{equation}
Our convention for the components of a $p$-form $\omega \in
\Omega^p(M) =\Gamma(\Lambda^p T^*M)$ is
\begin{equation}
\omega = \frac{1}{p!} \omega_{m_1 \cdots m_p} 
dx^{m_1} \wedge \cdots \wedge dx^{m_p} \;.
\end{equation}
Therefore the evaluation of a $p$-form on vector fields gives:
\begin{equation}
\omega(X_1, \ldots, X_p) = 
\omega_{m_1 \cdots m_p}  X_1^{m_1} \cdots X_p^{m_p} \;. 
\end{equation}

\subsubsection{Exterior derivative and dual Lie algebra structure of co-frames}

The coordinate expression for the {\em exterior derivative} $d\omega 
\in \Omega^{p+1}(M)$ of a $p$-form $\omega$ is:
\begin{equation}
d\omega = \frac{1}{p!} \partial_m \omega_{m_1 \cdots m_p}
dx^m \wedge dx^{m_1} \wedge \cdots dx^{m_p}
\Leftrightarrow (d\omega)_{m m_1 \cdots m_p} =
(p+1) \partial_{[m} \omega_{m_1 \cdots m_p]} \;.
\end{equation}
Note that we distinguish by brackets between the component
$(d\omega)_{m m_1 \cdots m_p}$ of the form $d\omega$ (a notation
used by physicists) and the exterior derivative 
$(d\omega_{m_1\cdots m_p})$ of the component $\omega_{m_1 \cdots m_p}$
regarded as a function (a notation used by mathematicians),
\begin{equation}
d \omega_{m_1 \cdots m_p} = \partial_m \omega_{m_1 \cdots m_p} dx^m \;.
\end{equation}
Our convention for the {\em antisymmetrization symbol}
$[\cdots ]$ is such that it includes a weight factor $1/p!$:
\begin{equation}
T_{[m_1 \cdots m_p]} = \frac{1}{p!} \sum_{\sigma \in S_p}
(-1)^{\mathrm{sign}(\sigma)}  T_{\sigma(m_1) \cdots \sigma(m_p)} \;,
\end{equation}
where $S_p$ is the permutation group of $p$ objects. 

The generators $e^m$ of a co-frame
satisfy the {\em dual Lie algebra}, $de^m = - \frac{1}{2} c^m_{np} e^n
\wedge e^p$.

The exterior derivative is a natural map $\Omega^p(M) \rightarrow \Omega^{p+1}(M)$ 
in the sense that it commutes with pullbacks of smooth maps $f: M \rightarrow N$, 
that is
\begin{equation}
f^* d\omega = d(f^* \omega) \;.
\end{equation}

\subsubsection{Interior product and contraction}

The {\em interior product} $\iota_X$ between a vector field $X\in \mathfrak{X}(M)$
and a $p$-form $\omega\in\Omega^p(M)$ 
is defined by substituting $X$ into
the first argument of the form, that is by contraction over the first
index:
\begin{equation}
(\iota_X \omega)(X_1, \ldots, X_{p-1})  = \omega(X,X_1, \ldots, X_{p-1} )
\Leftrightarrow
(\iota_X \omega)_{m_1 \cdots m_{p-1}} 
 = X^m \omega_{m m_1 \cdots m_{p-1}} \;.
\end{equation}
We will often write $\omega(X, \cdot) :=  \iota_X \omega(\cdot)$.

\subsubsection{Lie derivatives}

The {\em Lie derivative} $L_XT$ of a tensor field
$T \in {\cal T}^p_{\;q}(M) := 
\Gamma(\bigotimes^p TM \otimes \bigotimes^q T^*M)$ with respect to a vector
field $X$  is a directional derivative
which is defined using the flow of the vector field $X$. The Lie
derivative is additive and satisfies the Leibnitz rule,
\begin{equation}
L_X(T+S) = L_X T + L_X S \;,\;\;\;
L_X (T\otimes S) = L_X T \otimes S + T \otimes L_X S \;,
\end{equation}
where $T,S$ are tensor fields. 
To compute the components $(L_X T)^{m_1 \cdots m_p}_{\;\;\;\;\;\;\;\;\;\;\;\;n_1 \cdots n_q}$
of the Lie derivative $L_XT$ of a tensor field $T$ it is therefore
sufficient to know the action of $L_X$ on functions $f$, coordinate
vector fields $\partial_p$ and coordinate differentials $dx^p$:
\begin{equation}
L_X f = X^m \partial_m f \;,\;\;\;
L_X \partial_p = - (\partial_p X^n) \partial_n  \;,\;\;\;
L_X dx^p = (\partial_n X^p) dx^n \;.
\end{equation}
For vector fields $Y$ and one-forms $\omega$ one obtains:
\begin{eqnarray*}
L_X Y &=& [X,Y] = (X^m \partial_m Y^n - Y^m \partial_m X^n) \partial_n \;,
\nonumber  \\
L_X \omega &=& i_X d\omega + d(i_X \omega) =
(X^m \partial_m \omega_n + \omega_m \partial_n X^m) dx^n \;.
\end{eqnarray*}
The second formula remains valid for $p$-forms, and is known as
Cartan's magic formula
\begin{equation}
\label{Cartan_magic}
L_X \omega = i_X d\omega + d(i_X \omega) \;,\;\;
X \in \mathfrak{X}(M)\;,\;\;\omega\in \Omega^p(M) \;.
\end{equation}
For computations it is useful to note that
\begin{equation}
L_X f = X(f) = df(X) \;.
\end{equation}

\subsection{Pseudo-Riemannian manifolds \label{app:pseudo-Riemannian}}

A {\em pseudo-Riemannian} manifold is a manifold equipped with a symmetric, 
non-degenerate rank two co-tensor field, called the metric. Pseudo-Riemannian
manifolds are also referred to as {\em semi-Riemannian} manifolds. 

Our convention for the {\em symmetrized tensor product} of one-forms is
\begin{equation}
\label{sym}
\alpha \beta = \frac{1}{2} (\alpha \otimes \beta + \beta \otimes
\alpha) \;.
\end{equation}
Therefore the local expression for the metric is
\begin{equation}
g= g_{mn} dx^m dx^n   =\frac{1}{2}
 g_{mn} (dx^m \otimes dx^n + dx^n \otimes dx^m) \;.
\end{equation}
The metric provides a natural isomorphism between vector fields and
one forms. We use the `musical' notation:
\begin{equation}
X = X^m \partial_m \Rightarrow X^\flat = X_m dx^m \;,\;\;\;X_m =
g_{mn} X^n \;,
\end{equation}
\begin{equation}
\omega = \omega_m dx^m \Rightarrow \omega^\sharp =
\omega^m \partial_m\;,\;\;\; \omega^m = g^{mn} \omega_n \;,
\end{equation}
where $g^{mn}$ are the components of the matrix inverse of $g_{mn}$.

We do not 
require that the metric is positive definite, and consider general
signatures $(t,s)$, where $t$ is the number of time-like and $s$ the number
of space-like dimensions. Since we adopt a `mostly plus convention', 
$t$ is the number of negative eigenvalues, and $s$ the number of positive
eigenvalues of the matrix $g_{mn}$. For completeness we define
that a Riemannian manifold is a 
pseudo-Riemannian manifold with definite signature.

\subsection{Connections \label{app:connections}}

\subsubsection{Connections on the tangent bundle}

A {\em connection} $\nabla$ on $TM$ (also called a connection on $M$, or an
affine or linear connection on $TM$) 
is a bilinear map\footnote{Alternatively, one can view $\nabla$ as a map
$\mathfrak{X}(M) \rightarrow \Omega^1(M) \otimes \mathfrak{X}(M)$
which assigns to a vector field $X$ the vector-valued one-form $\nabla X$.} 
\begin{equation}
\nabla\;:
\mathfrak{X}(M) \times \mathfrak{X}(M) \rightarrow \mathfrak{X}(M) \;:
(X,Y)\mapsto \nabla_X Y \;,
\end{equation}
which satisfies 
\begin{equation}
\nabla_{fX} Y = f \nabla_{X} Y \;,\;\;\;
\nabla_X(fY) = X(f) Y + f \nabla_X Y \;, 
\end{equation}
for all $f\in C^\infty(M)$. The {\em covariant derivative} 
\begin{equation}
\nabla_X \;: \mathfrak{X}(M) \rightarrow \mathfrak{X}(M) \;:  \;\; Y \mapsto 
\nabla_X Y
\end{equation}
is extended to general tensor fields,
\begin{equation}
\nabla_X \;: {\cal T}^p_{\;q}(M) 
\rightarrow {\cal T}^p_{\;q}(M) 
\end{equation}
by imposing linearity and the Leibnitz rule 
in ${\cal T}^p_{\;q}(M)$ and $C^\infty(M)$-linearity in $X$.

We remark that in the literature the expressions `covariant
derivative' and `connection' are used variably for $\nabla$ and
$\nabla_X$. If one needs to distinguish $\nabla$ from $\nabla_X$, then
the first is called the absolute covariant derivative and the second
the directional covariant derivative.

The {\em connection coefficients} $\gamma^p_{mn}$ and {\em connection one-form}
$\omega^p_n = \gamma^p_{mn} e^m$ with respect to a frame are defined by
\begin{equation}
\nabla_{e_m} e_n = \gamma^p_{mn} e_p 
\end{equation}
or in terms of the dual frame
\begin{equation}
\nabla_{e_p} e^m = -\gamma^m_{pn} e^n \;.
\end{equation}
If the frame $e_m$ is a coordinate frame, the connection coefficients are
denoted $\Gamma^p_{mn}$:
\begin{equation}
\nabla_{\partial_m} \partial_n = \Gamma^p_{mn} \partial_p 
\;.
\end{equation}
The {\em torsion} and {\em curvature} of a connection 
are the following multilinear maps
\begin{eqnarray}
T^\nabla(X,Y)& =& \nabla_X Y - \nabla_Y X - [X,Y]\;, \\
R^\nabla_{X,Y}Z &=& \nabla_X \nabla_Y Z - \nabla_Y \nabla_X Z -
                    \nabla_{[X,Y]} Z \;,
\end{eqnarray}
where $X,Y,Z\in \mathfrak{X}(M)$. 
The torsion and curvature tensor are defined by
\begin{eqnarray}
T(\alpha, X,Y) &=& \alpha (T^\nabla(X,Y) )\;, \\
R(\alpha, Z, X,Y) &=& \alpha(R^\nabla_{X,Y} Z)  \;
\end{eqnarray}
where $\alpha \in \Omega^1(M)$. 
The components with respect to a frame are:
\begin{eqnarray}
T^m_{\;\;np} &=& T(e^m,e_n,e_p) = \gamma^m_{np} - \gamma^m_{pn} 
- c^m_{np} \;, \\
R^{m}_{\;npq} &=& R(e^m, e_n, e_p, e_q) = 
e_p (\gamma^m_{qn}) - e_q (\gamma^m_{pn} ) \\
&& + \gamma^m_{pa} \gamma^a_{qn} - \gamma^m_{qa} \gamma^a_{pn}
- c^a_{pq} \gamma^m_{an} \;.
\end{eqnarray}
For a coordinate frame these expression reduce to
\begin{eqnarray}
T^m_{\;\;pq} &=& \Gamma^m_{pq} - \Gamma^m_{qp}  \;,  \label{TRcomp}\\
R^m_{\;\;npq} &=& \partial_p \Gamma^m_{qn} - \partial_q \Gamma^m_{pn}
+ \Gamma^m_{pa}\Gamma^a_{qn} - \Gamma^m_{qa} \Gamma^a_{pn}  \nonumber\;.
\end{eqnarray}

\subsubsection{The Levi-Civita Connection}

The {\em Levi-Connection} $D$ on a Riemannian manifold $(M,g)$ is the unique
connection on the tangent bundle $TM$ which is both metric (compatible) and
torsion free:
\begin{equation}
D_X g = 0 \;,\;\;\;T^D(X,Y) = 0 \;,\;\;\;\forall X,Y \in \mathfrak{X}(M) \;.
\end{equation}
Our conventions for the Levi-Civita connection and the Christoffel symbols 
are summarized in \ref{app:not+con}.

\subsubsection{Flat, torsion-free connections and affine manifolds \label{app:flat-connections}}

If a connection is flat, $R^\nabla=0$, it is possible to choose a
frame consisting of parallel vector fields \cite{Frankel}, i.e. 
\begin{equation}
\nabla_{e_m} e_n = 0 \Rightarrow \gamma^p_{mn} = 0  \;.
\end{equation}
If the connection is in addition torsion-free, then this
parallel frame is a coordinate frame, since 
\begin{equation}
\left.
\begin{array}{l}
T(e^m, e_n, e_p) = 0 \\
\gamma^m_{np} = 0 \\
\end{array} \right\} \Rightarrow c^m_{np} = 0 \;.
\end{equation}

Alternatively, we note that the expression for the torsion
tensor with respect to a frame is
\begin{equation}
\label{TinFrame}
T = e_m \otimes de^m + e_n \otimes  \omega_m^n \otimes e^m \;.
\end{equation}
If $\nabla$ is flat, we can choose a basis of parallel sections, so
that  $\omega_m^n=0$, and then 
\begin{equation}
\left. \begin{array}{l}
T = 0 \\
\omega_m^n=0 \\
\end{array} \right\} 
\Rightarrow de^m = 0 \Rightarrow e^m = dq^m \;,
\end{equation}
where $q^m$ are local functions that provide coordinates underlying
the parallel frame. Such coordinates are called $\nabla$-affine
coordinates and are unique up to affine transformations. The condition
on a coordinate system to be affine is $\nabla dq^m =0$, that is, that
the coordinates define  a parallel co-frame. 

If a manifold admits a flat, torsion-free connection, it can be covered
with $\nabla$-affine coordinate charts which are related by affine 
transition functions. Such an atlas is called an {\em affine structure}. 
A manifold $M$ equipped with a flat, torsion-free connection $\nabla$ is called an
{\em affine manifold}.

\subsubsection{Connections on vector bundles \label{App:ConnVB}}

Let $E\rightarrow M$ be a vector bundle over a manifold $M$. 
A {\em connection} on $E$ is a map
\begin{equation}
\nabla \;: \mathfrak{X}(M) \times \Gamma(E)  \rightarrow \Gamma(E)
\;,\;\;\;
(X,s) \mapsto \nabla_X s \;,
\end{equation}
which is linear and satisfies the product rule with respect to
sections $s\in \Gamma(E)$, while being $C^\infty(M)$-linear 
with respect to vector fields $X\in \mathfrak{X}(M)$.

Let $E\rightarrow M$ be a vector bundle with connection $\nabla$, and let 
$D$ be a linear connection on $M$. If $s\in \Gamma(E)$ is a section of 
$E$, then $\nabla s$ is a section of $T^*M \otimes E$. One can then use the
connection induced by $D$ and $\nabla$ to define the {\em second covariant
derivative} $\nabla^2 s$, 
which is a section of $T^*M \otimes T^*M \otimes E$:
\begin{equation}
\nabla^2 s(X,Y) = \nabla_X(\nabla_Y s) - \nabla_{D_XY} s \;.
\end{equation}
Alternative notations are $\nabla^2_{X,Y}s$ or $(\nabla^2 s)_{X,Y}$. 

If $E=TM$, denoting the connection induced by $D$ on tensor bundles again by 
$D$, we obtain the following formula for the second covariant derivative of a vector field:
\begin{equation}
\label{covD}
D^2_{X,Y} Z = D_X (D_Y Z) - D_{D_XY} Z \;.
\end{equation}
In local coordinates, the relevant expression are, using the notation 
$D_m = D_{\partial_m}$:
\begin{eqnarray}
(D^2_{X,Y} Z)^p &=& X^m Y^n D_m D_n Z^p  \;, \\
(D_X (D_Y Z))^p &=& X^m D_m (Y^n D_n Z^p) \;,  \\
(D_{D_X Y} Z)^p &=& X^m (D_m Y^n) D_n Z^p \;.
\end{eqnarray}
We can define the {\em Hessian} $Dd f$ of a function $f$ with respect to the linear connection 
$D$:
\begin{eqnarray}
Ddf(X,Y) &=& X Y (f) - (D_X Y) f  = X^m D_m (Y^n \partial_n f) - 
X^m (D_m Y^n) \partial_n f  \nonumber \\ &=& X^m Y^n D_m \partial_n f \;.
\end{eqnarray}
If the connection $D$ is torsion-free, 
\begin{equation}
D_X Y  - D_Y X = [X,Y] \;,
\end{equation}
then the Hessian is symmetric, and the 
curvature of $D$ can be written
\begin{equation}
R^D_{X,Y}Z = [D_X , D_Y]Z  - D_{[X,Y]} Z = D^2_{X,Y} Z - D^2_{Y,X} Z\;.
\end{equation}

For the bundle $\Omega^p(M,E)=\Gamma(\Lambda^pT^*M \otimes E)$ 
of vector-valued $p$-forms, one defines the {\em exterior covariant
derivative}
\begin{equation}
d_\nabla\;: \Omega^p(M,E) \rightarrow \Omega^{p+1}(M,E)
\end{equation}
by its action on sections of $E$. For a basis $\{ s_a \}$ of sections
one sets
\begin{equation}
d_\nabla s_a := \nabla s_a = \omega^b_a \otimes s_b \;,
\end{equation}
where $\omega^b_a$ is the connection one-form of $\nabla$. The
exterior covariant derivative of a general section $s = f^a s_a \in
\Omega^0(M,E)=\Gamma(E)$ is determined by the product rule
\begin{equation}
d_\nabla s = df^a \otimes s_a + f^a \omega^b_a \otimes s_b \;.
\end{equation}
The extension of $d_\nabla$ to forms of degree $p>0$ is uniquely 
determined by linearity and the product rule:
\begin{equation}
d_\nabla (\alpha \otimes s) = d\alpha \otimes s +
(-1)^{\mathrm{deg}(\alpha)}
\alpha \wedge d_\nabla s \;,\;\; \alpha \in \Omega^p(M) \;.
\end{equation}
The exterior covariant derivative of a vector valued $p$-form
$\rho \in \Omega^p(M,E)$ 
can be expressed in terms of the
covariant derivative by
\begin{eqnarray}
(d_\nabla \rho)(X_0, \ldots, X_p) &=& \sum_{l=0}^p (-1)^l 
\nabla_{X_l} ( \rho (\ldots, \hat{X}_l, \ldots))   \\
&& + \sum_{i<j} (-1)^{i+j} \rho( [X_i, X_j], \ldots, \hat{X}_i,
   \ldots, \hat{X}_j, \ldots ) \;,  \nonumber
\end{eqnarray}
where $X_0, \ldots, X_p$ are vector
fields, and where $\hat{X}$ indicates that the vector field $\hat{X}$
is omitted as an argument.  The second exterior derivative of
a section $s\in \Gamma(E)$ is related to the curvature of the connection
$\nabla$ by
\begin{equation}
d_\nabla^2 s(X,Y) = R^\nabla_{X,Y} \;,  \;\; \forall X,Y \in \mathfrak{X}(M)\;, s\in \Gamma(E) \;.
\end{equation}
Thus $d_\nabla$ satisfies $d_\nabla^2=0$ if and only if the connection is flat. 
If this is the case, a version of the Poincar\'e lemma holds which allows to write
a $d_\nabla$-closed vector-valued $p$-form locally as the $d_\nabla$ derivative
of a vector valued $(p-1)$-form. In general the Bianchi identity 
$d_\nabla R_\nabla =0$ for the curvature implies that $d_\nabla^3=0$. 
We refer to \cite{LN_Diff_Geom_Fernandez}, \cite{Besse} for more details
on the exterior covariant derivative.

In the case when $E=TM$, $\nabla$ is a connection on $TM$ and we can define
its torsion. It is useful to note that the
torsion tensor can be expressed as
\begin{equation}
\label{TasDerivative}
T^\nabla = d_\nabla \mbox{Id} \;,
\end{equation}
where $\mbox{Id} = e^m \otimes e_m \in
\Gamma(\mbox{End}(TM)) \simeq \Gamma( T^*M \otimes TM) 
\simeq \Omega^1(M,TM)$ is the
identity endomorphism on $TM$, regarded as a vector-valued one-form.
Equation (\ref{TasDerivative}) can be verified using that
\begin{equation}
d_\nabla \left( e^a \otimes e_a\right) = de^a \otimes e_a + e^a \wedge \omega^b_a
\otimes e_b 
\end{equation}
and evaluating both
sides of the equation on vector fields, that is by showing that $T^\nabla(X,Y)=
(d_\nabla \mbox{Id})(X,Y)$.  Instead of general vector fields $X,Y$,
one can choose $X=e_a$, $Y=e_b$ with arbitrary $a,b$, thus comparing the 
components with respect to a frame.

\subsection{Pull-back bundles \label{App:Pull_back}}

If $f:M\rightarrow N$ is a smooth map between smooth manifolds $M,N$, then one can
pull back any vector bundle $\pi_E: E \rightarrow N$ to a vector bundle $f^*E \rightarrow M$
over $M$, called the {\em pull-back bundle of $M$ by $f$}, which is constructed as follows:
\begin{itemize}
\item
The total space of $f^*E$ is
\begin{equation}
f^* E := \{ (m,e) \in M \times E |  f(m) = \pi_E(e)  \} 
\end{equation}
\item
The bundle projection is the restriction of the canonical projection
$\pi_1 : M \times E \rightarrow M$
to $f^*E$:
\begin{equation}
\pi_{f^*E}(m,e) = m \;.
\end{equation}
\end{itemize}
By construction the fibres of $f^*E$ are mapped to fibres of $E$, more precisely
$(f^*E)_m \cong E_{f(m)}$ for all $m \in M$. By restricting the canonical 
projection $\pi_2: M\times E \rightarrow E$
to $f^*E$ we obtain the so-called covering morphism
\begin{equation}
F: f^*E \rightarrow E \;:  (m,e) \mapsto  F(m,e) = e  \;,
\end{equation}
which completes the commutative diagram
\begin{equation}
\xymatrix{
f^*E \ar[rr]^F  \ar[d]^{\pi_{f^*E}} && E \ar[d]^{\pi_E} \\
M \ar[rr]^f && N \\
}
\end{equation}
The pull-back
$f^*s \in \Gamma(f^*E)$ of a section $s\in \Gamma(E)$ is defined by
\begin{equation}
(f^*s)(m) = s(f(m)) \;.
\end{equation}
We can also pull back a connection $D$ on $E$ to a connection 
$f^*D$ on $f^*E$ . This pull-back connection is defined by
\begin{equation}
(f^*D)_X f^*s := D_{df X} s \;,
\end{equation}
for all vector fields $X$ on $M$.

\subsection{The Frobenius theorem, hypersurfaces, and hypersurface orthogonal vector fields
   \label{app:Frobenius}}

This section is partly based on  \cite{Frankel} and on \cite{Wald:1984rg}, Appendix B.

A $p$-dimensional {\em distribution} $V=\cup_{x\in M} V_x$ on the tangent bundle $TM$ of a smooth
manifold is a map
\begin{equation}
M \ni x \mapsto V_x \subset T_xM \;,
\end{equation}
where $V_x$ is a $p$-dimensional subspace of $T_xM$. A distribution 
is called smooth if it depends smoothly on $p$. This means that for each
$x\in M$ there exists a neighbourhood $U$ and $p$ linearly independent
smooth vector fields defined on $U$ which span $V_x$ for $x\in U$. 
One may then ask whether there exist on $M$ smooth $p$-dimensional
submanifolds which are tangent to $V$. Such submanifolds are called the {\em integral
manifolds} of the distribution, and provide a {\em foliation} of $M$, that is a 
disjoint decomposition into submanifolds, called the leafs of the foliation.

According to the Frobenius theorem 
a distribution is integrable if and only if it is involutive,
that is if the Lie bracket of any two tangent vector fields is again a
tangent vector field, for all points $x\in M$. Distributions which 
possess integral manifolds are called (Frobenius-)integrable.

The Frobenius theorem
can be given a dual formulation in terms of differential forms. Given
a distribution $V\subset TM$ one can consider the dual
distribution $V^* \subset T^*M$ on the cotangent bundle defined by
\begin{equation}
\omega \in V^* \Leftrightarrow \omega(X) = 0 \;,\;\;\forall X \in V\;.
\end{equation}
For differential forms, the integrability condition is
\begin{equation}
d\omega = \sum_i \alpha_{(i)} \wedge \beta_{(i)} \;,
\end{equation}
where $\alpha_{(i)} \in V^*$, and where $\beta_{(i)}\in \Omega^1(M)$. 

A vector field $\xi$ is called {\em hypersurface orthogonal} if it is
orthogonal to a foliation of $M$ by hypersurfaces. This is equivalent
to the statement that the distribution $V = \langle \xi \rangle^\perp$
is Frobenius integrable. The dual distribution $V^*$ on the cotangent bundle
is spanned by the one-form $\xi^\flat$, that is $V^* = (\langle \xi \rangle^\perp)^*  = 
\langle \xi^\flat \rangle$, because $\xi^\flat (\cdot) = g(\xi,\cdot)$. Specializing the dual 
version of the Frobenius theorem to the case of a hypersurface distribution we obtain
\begin{equation}
d \xi^\flat = \xi^\flat \wedge \beta  \;,
\end{equation}
for some one-form $\beta$,
where we used that the distribution $V^*$ 
is one-dimensional. This equation is 
equivalent to
\begin{equation}
\xi^\flat \wedge d \xi^\flat = 0 \Leftrightarrow \xi_{[m} \partial_n
\xi_{p]} =0 \;,
\end{equation}
which is the standard criterion used in the literature for verifying 
the hypersurface orthogonality of a
vector field. Note that due to the antisymmetrization the expression
$\xi_{[m} \partial_n \xi_{p]}$ is covariant, since we can replace
$\partial_n$ by any torsion free covariant derivative.
Also note that the integrability condition is satisfied in particular if 
the vector field is closed, that is if $d\xi^\flat =0$.

Foliations by hypersurfaces can be described locally as level sets of
a function $F: M \rightarrow \mathbb{R}$:
\begin{equation}
M \simeq \cup_{\gc \in \mathbb{R}} \{ x\in M | F(x) = \gc \} \;.
\end{equation}
The standard normal vector field to such a foliation is $n =
\mbox{grad}(F) = (dF)^\sharp$, with components $n^m = g^{mn} \partial_n F$.
Tangent vectors $t$ to the foliation are characterized
by any of the following relations:
\begin{equation}
g_{mn} n^m t^n = g(n,t) = 0  = dF(t) = t^m \partial_m F\;.
\end{equation}
The most general vector field $\xi$ normal to the foliation can
differ from the standard normal $n$ by a function $f:M\rightarrow \mathbb{R}$, that is
$\xi = f (dF)^\sharp$. Such a vector field clearly satisfies the
integrability condition we derived earlier, since $\xi^\flat = f dF$. 
The standard normal vector field $n$ is distinguished by being
`closed', more precisely by $d n^\flat = 0$. This is a stronger 
condition than Frobenius integrability.

\subsection{Integral curves, one-parameter groups and quotient manifolds \label{App:integral_curves}}

A one-dimensional distribution on the tangent bundle is always integrable, because
the integrability condition becomes trivial. Such a distribution defines a smooth vector field $X$, 
and its integrability corresponds to the existence of a family of so-called integral curves, whose
tangent vectors are given by $X$. The integral curve 
$C_{x_0} : t \mapsto x(t)$ 
through a given point $p\in M$ with coordinate $x_0$
is found by solving the initial value problem
\begin{equation}
\frac{dx}{dt} = X(t) \;,\;\;t \in I \subset \mathbb{R}\;,\;\;\;
x(0) = x_0 \;.
\end{equation}
The flow of the vector field $X$ is defined by
\begin{equation}
\sigma: I \times M \rightarrow  M \;,\;\;\;(t,x) \mapsto  \sigma(t,x)= x(t)
\end{equation}
where $x(t)=\sigma_x(t)$ is the integral curve of $X$ with initial condition $x(0)=x_0$. 

Further defining
\begin{equation}
\sigma_t \;:\;\;\; M \rightarrow M \;, \;\;\; x(0) = \sigma_0(x) = \sigma(x,0) \mapsto \sigma(x,t) = 
\sigma_t(x) = x(t)
\end{equation}
we see that $\sigma_t$ moves the points of $M$ along the integral curves of $X$.  
Since  $\sigma_{s+t} = \sigma_t \circ \sigma_s$ and
$\sigma_0 = \Id$, these transformations form a group, called the one-parameter 
transformation group generated by $X$. If this action is a globally defined group action of 
$G=U(1)$ or $G=\mathbb{R}$ on $M$, then the integral curves are called the orbits of $G$,
and denoted $\langle X \rangle$. As already discussed in \ref{app:manifolds},
the space of orbits, denoted $M/\langle X \rangle = M/G$,
need not be a manifold, in particular it need not satisfy the Hausdorff 
separation axiom. However, in many cases, including those
relevant for this review, the quotient is a (Hausdorff) manifold, and various structures, such as the metric,
complex or symplectic structure project to the quotient manifolds. Quotient manifolds can
also be defined with respect to the action of higher-dimensional groups. 
Examples relevant for this review are the action of the group $\mathbb{C}^*$ on CASK manifolds and the action of the group $\mathbb{H}^*$ on 
hyper-K\"ahler manifolds.

\subsection{Metric cones and metric products\label{app:metric_cones}}

In this section we elaborate on some standard definitions, and in particular adapt
them to the pseudo-Riemannian setting. 

If $({\cal H},h)$ is a pseudo-Riemannian manifold, then the {\em metric cone} (or {\em Riemannian cone})
$(M,g)$ over $({\cal H},h)$ is the manifold $M= \mathbb{R}^{>0} \times {\cal H}$ equipped with the
metric
\begin{equation}
g  = \pm d\rho^2 + \rho ^2 h \;.
\end{equation}
We note that $\xi=\partial_\rho$ is a closed homothetic Killing vector field:
\begin{equation}
L_\xi g = 2 g \;,\;\;\;d\xi^\flat = 0 \;.
\end{equation}
Since $\xi$ is closed, it is gradient vector field:
\begin{equation}
\label{HS1}
\xi^\flat = dH \Leftrightarrow \xi = \mbox{grad} H
\end{equation}
or, in local coordinates $x^m$ on $M$:
\begin{equation}
\xi_m =\partial_m H \Leftrightarrow \xi^m = g^{mn} \partial_n H  \;.
\end{equation}
$M$ is foliated by the level surfaces $H = \gc$ which are orthogonal to $\xi$, and
${\cal H}$ can be identified with the hypersurface $H=1$. 

The two equations (\ref{HS1}) are the symmetric and anti-symmetric part of
\begin{equation}
D \xi = \Id_{TM} \Leftrightarrow D_m \xi_n =  g_{mn} \;,
\end{equation}
where $D$ is the Levi-Civita connection of $g$. This equation provides a 
local characterization of a metric cone:
\begin{rmk}
Let $(M,g)$ be a pseudo-Riemannian manifold of dimension $n+1$, equipped with a vector field $\xi$, which is 
nowhere isotropic, that is $g(\xi,\xi)\not=0$ everywhere, and which satisfies
\begin{equation}
D\xi = \Id_{TM} \;.
\end{equation}
Then there exist local coordinates $(r, x^i)$, $i=1, \ldots, n$  such that  metric $g$  takes the form 
\begin{equation}
g = \pm dr^2 + r^2 h_{ij} dx^i dx^j \;,
\end{equation}
where $h_{ij}$ only depend on the coordinates $x^i$.
\end{rmk}
This is a special case of the standard form of an {\em $n$-conical Riemannian metric},
which we derive in section \ref{sect:ConHess}.

If $({\cal H}_1, h_1)$ and $({\cal H}_2, h_2)$ are two pseudo-Riemannian manifolds, 
their {\em metric product} or {\em Riemannian product} $(M,g)$ is defined by $M = {\cal H}_1 \times {\cal H}_2$
equipped with the product metric
\begin{equation}
g= h_{{\cal H}_1} + h_{{\cal H}_2} \;.
\end{equation}
In local coordinates $(x^m, y^i)$ on ${\cal H}_1\times {\cal H}_2$, this takes the form
\begin{equation}
g = (h_1)_{mn}(x) dx^m dx^n + (h_2)_{ij} (y)dy^i dy^j \;.
\end{equation}
In applications we encounter product manifolds of the special form 
\begin{equation}
M = \mathbb{R} \times {\cal H} \cong \mathbb{R}^{>0} \times {\cal H} \;,
\end{equation}
for which the metric takes the form
\begin{equation}
\label{product_local}
g = \pm d\rho^2 + h_{ij} dx^i dx^j = \pm \frac{dr^2}{r^2} + h_{ij} dx^i dx^j  \;,
\end{equation}
where the coordinates $r,\rho$ are related by $r = e^\rho$. The vector
field $\xi = \partial_\rho = r \partial_r$ is a Killing vector field, $L_\xi g= 0$, 
which is closed $d\xi^\flat =0$,
and therefore hypersurface orthogonal, 
and which in addition has constant norm $g(\xi,\xi)=\pm 1$. The manifold
$M$ is foliated by hypersurfaces where $\rho=\mbox{const.}$, and all
these hypersurfaces are isometric to each other and to $({\cal H},h)$. 

The Killing equation can be combined with the closed-ness condition to 
\begin{equation} 
D\xi = 0  \;.
\end{equation}
Note that this equation does not by itself imply that a metric $g$ locally 
takes the form (\ref{product_local}) of a product. This requires in addition 
that the norm of $\xi$ is constant, so that surfaces of constant $\rho$ are 
isometric to each other. The proof that this is sufficient to bring the metric to the form
(\ref{product_local}) is given in section \ref{sect:ConHess}.

\subsection{Affine hyperspheres and centroaffine hypersurfaces \label{App:affine_hyperspheres}}

Here we review some facts about {\em affine hyperspheres} 
and {\em centroaffine hypersurfaces}, following
\cite{1999math.....11079B,Cortes:2001bta,2014arXiv1407.3251C}.
Consider $\bR^{m+1}$ equipped with the
standard connection $\partial$ (given by the partial derivative with respect
to standard linear coordinates), and
the standard volume form $\mbox{vol}$, which is parallel with respect to $\partial$.
Let $M$ be a connected manifold which is immersed as a hypersurface
\begin{equation}
\varphi: M \rightarrow \bR^{m+1}  \;.
\end{equation}
Assume that there exists a vector field $\xi$ which is transversal along $M$.
Then $\mbox{vol}_M = \mbox{vol}(\xi, \cdots)$ is a volume form on $M$,
and by decomposing
\begin{eqnarray}
\partial_X  Y &=& \nabla_X Y + g(X,Y) \xi  \;, \label{induced_metric}\\
\partial_X \xi &=& S X + \theta(X) \xi \;,
\end{eqnarray}
where $X,Y$ are tangent to $M$, one obtains on $M$:  
(i) a torsion-free connection $\nabla$, (ii) a symmetric co-tensor $g$,
(iii) an endomorphism field $S$ and (iv) a one-form $\theta$. If $g$ is
non-degenerate, it defines a pseudo-Riemannian metric on $M$. 
It can be shown that once the orientation of $M$ has been fixed
there is a unique choice for $\xi$, called the {\em affine normal}
such that the induced volume form $\mbox{vol}_M$ of $M$ 
coincides with the volume form defined
by the metric $g$. If $\xi$ is chosen to be the affine normal, then  $\theta=0$ and 
$S$ can be expressed in terms of the so-called Blaschke data $(g,\nabla)$. 

There are two special cases:
\begin{enumerate}
\item
A hypersurface is called a {\em parabolic} (or {\em improper}) {\em affine
hypersphere} if the affine normals are parallel, $\partial \xi =0$, and thus
only intersect `at $\infty$.' One can show that
\begin{equation}
\partial \xi = 0 \Leftrightarrow S=0 \Leftrightarrow \nabla \;\;\mbox{flat} \;.
\end{equation}
Thus parabolic affine hyperspheres carry a flat torsion-free connection.

\item
A hypersurface is called a {\em proper affine hypersphere} if the lines generated by the affine normals
intersect at a point $p\in \bR^{m+1}$. For a proper affine hypersphere 
$S=\lambda \mbox{Id}$, $\lambda \in \mathbb{R}^*$.
\end{enumerate} 
The ASK manifolds of four-dimensional
vector multiplets are parabolic affine hyperspheres with additional structure,
called {\em special parabolic hyperspheres}, see section \ref{Sect:Aff_HS}.

The PSR manifolds of five-dimensional vector multiplets coupled to supergravity,
which are discussed in  section \ref{sect:PAHS}, are, in general, not (proper) 
affine hyperspheres, but {\em centroaffine hypersurfaces}. According to 
section 1.1 of \cite{2014arXiv1407.3251C} a hypersurface immersion 
$\varphi: M \rightarrow \mathbb{R}^{m+1}$ is called a {\em centroaffine hypersurface
immersion} if the position vector field $\xi$ is transversal to the image of $M$. 
The equation
\begin{equation}
\partial_X  Y =  \nabla_X Y + g(X,Y) \xi \;,
\end{equation}
for $X,Y \in \mathfrak{X}(M)$ induces on $M$ a connection $\nabla$, 
a symmetric tensor field $g$,  
and a $\nabla$-parallel volume form $\mbox{vol}_M = \det( \xi, \ldots )$.
The data $(\nabla, g, \mbox{vol}_M)$ are called the induced centroaffine
data on $M$. The hypersurface $M$ is called {\em non-degenerate} if $g$ is
non-degenerate, {\em definite} if $g$ is definite, {\em elliptic} if $g$ is negative definite,
and {\em hyperbolic} if $g$ is positive definite. Every homogeneous function 
defines a centroaffine hypersurface embedding, and every centroaffine hypersurface
immersion is locally generated by a homogeneous function. Centroaffine structures
can be characterized intrinsically: a {\em centroaffine manifold} $(M,\nabla, g, \mbox{vol}_M)$
is a manifold equipped with a torsion-free connection $\nabla$, a pseudo-Riemannian metric
$g$ and a volume form $\mbox{vol}_M$, subject to three compatibility conditions:
(i) the volume form is $\nabla$-parallel, (ii) the cubic form $C:= \nabla g$ is completely
symmetric, and (iii) the curvature tensor $R$ of $\nabla$ is given by
\begin{equation}
R(X,Y) Z = - (g(Y,Z) X - g(X,Z))Y) 
\end{equation}
for $X,Y,Z \in \mathfrak{X}(M)$. By Theorem 1.6 of  \cite{2014arXiv1407.3251C} 
a centroaffine immersion $\varphi: M \rightarrow \mathbb{R}^{m+1}$ induces on 
$M$ the structure of a centroaffine manifold. Conversely, every connected and
simply connected centroaffine manifold can be realized as a centroaffine 
immersion, which is unique up to $SL(m+1,\mathbb{R})$ transformations. 
Note that in contradistinction to affine hyperspheres, the position vector field
$\xi$ of a centroaffine hypersurfaces is in general not the affine normal of 
$M$.\footnote{We thank the referee for pointing this out to us.}

PSR manifolds, which are the scalar manifolds of five-dimensional vector multiplets
coupled to supergravity, were discussed in section \ref{sect:PAHS}. We now 
review how they fit into the theory of centroaffine hypersurfaces, following section
2.1 of \cite{2014arXiv1407.3251C}. A PSR manifold is a smooth hypersurface
$\bar{M} \cong {\cal H} \subset \mathbb{R}^{m+1}$, which is realized as the level set ${\cal V}=1$ of a homogeneous cubic polynomial ${\cal V}$, such that $\partial^2 {\cal V}$ is negative definite
on $T{\cal H}$. This induces a centroaffine structure $(\nabla, g, \mbox{vol}_{\bar{M}})$ on  $\bar{M}$.

According to definition 2.2 of \cite{2014arXiv1407.3251C} an {\em intrinsic projective 
special real manifold} is a centroaffine manifold 
$(\bar{M}, \nabla, g, \mbox{vol}_{\bar{M}})$ with a positive definite metric $g$ such that the
covariant derivative of the cubic form $C=\nabla g$ is given by
\begin{equation}
(\nabla_X C)(Y,Z,W) = g(X,Y) g(Z,W) + g(X,Z) g(W,Y) + g(X,W) g(Y,Z) 
\end{equation}
for all $X,Y,Z,W\in \mathfrak{X}(\bar{M})$. 

Theorem 2.3 of  \cite{2014arXiv1407.3251C} relates the extrinsic and intrinsic definitions 
of PSR manifolds. The induced centroaffine structure on a PSR manifold gives it the 
structure of an intrinsic PSR manifold, and any connected and simply connected 
intrinsic PSR manifold can be realized by an immersion $\varphi: \bar{M} \rightarrow 
\mathbb{R}^{m+1}$ which is unique up to $SL(m+1,\mathbb{R})$ transformations.

\subsection{Complex manifolds \label{App:cplxmfds}}

An {\em almost complex
manifold} $(M,J)$ is a real manifold $M$ together with an almost
complex structure $J$. An {\em almost complex structure} $J$ is a section
of $\mbox{End}(TM) \simeq TM \otimes T^*M$, which 
satisfies $J^2 = - \mathbbm{1}_{TM}$. 
Note that an almost complex
manifold is always of even dimension. 
A {\em complex manifold} $N$ of 
complex dimension $n$ is a manifold which is locally biholomorphic to
$\mathbb{C}^n$. 
A complex manifold automatically carries an almost
complex structure (with additional properties, see below) which is
called its complex structure. In terms of local holomorphic coordinates
$z^i = x^i + i y^i$, the complex structure acts on $TM$ as
\begin{equation}
J X_i = Y_i \;,\;\;\; J Y_i = - X_i
\end{equation}
where $X_i, Y_i$ is the coordinate frame
\begin{equation}
X_i = \frac{\partial}{\partial x^i} \;,\;\;\;
Y_i = \frac{\partial}{\partial y^i}  \;.
\end{equation}
$X_i,Y_i$ is called a holomorphic frame on $(M,J)$. 

As a
consequence of the Newlander-Nirenberg theorem, an almost
complex manifold $(M,J)$ is a complex manifold if and only if the
Nijenhuis tensor (or torsion tensor)  associated to $J$, defined by
\begin{equation}
N_J(X,Y) := 2 \left( [JX,JY] - [X,Y] - J[X,JY] - J[JX,Y] \right) \;,
\end{equation}
vanishes.
An almost complex structure with vanishing
torsion tensor is called an  integrable almost complex
structure, or simply a complex structure. 

For further reading we refer to \cite{Ballmann}, on which 
\ref{App:cplxmfds} -- \ref{App:Herm_Mfd} are mostly based.

\subsection{Complex vector bundles \label{App:ComplexVBdl}}

A {\em complex vector bundle} $E$ over a manifold $M$ is a vector
bundle whose fibres are complex vector spaces. A one-dimensional
complex vector bundle is called a {\em complex line bundle}.  A {\em Hermitian 
metric} $\gamma$ on $E$ is a family of Hermitian scalar products
$\gamma_x$ on the fibres $E_x$, which
varies smoothly with $x\in M$. Our convention for Hermitian forms is that
they are complex linear in the first and complex anti-linear in the second argument.
A {\em Hermitian vector bundle} $(E,M,\gamma)$ is
a  complex vector bundle $(E,M)$ equipped with a Hermitian metric.
A connection $D$ on a Hermitian vector bundle is called
{\em metric compatible}, or {\em metric}, or {\em Hermitian} if
\begin{equation}
d \left( \gamma(s,t) \right) = \gamma(Ds, t) + \gamma(s,Dt)
\end{equation}
for all sections $s,t$.

A {\em holomorphic vector bundle} $E$ is a complex vector bundle 
over a complex manifold $M$ such that the projection $\pi: E
\rightarrow M$ is holomorphic. Every complex manifold comes
equipped with a standard holomorphic vector bundle, the
tangent bundle $TM$ equipped with the complex structure $J$. 
Another canonical complex vector bundle over $M$ is the
{\em complexified tangent bundle} 
$T_{\mathbb{C}}M= TM \otimes_{\mathbb{R}} \mathbb{C}$, 
equipped with the complex linear extension of $J$. The
complexified tangent bundle can then be split into the 
eigen-distributions of $J$, called the holomorphic and 
anti-holomorphic tangent bundle,
\begin{equation}
T_{\mathbb{C}} M = T^{(1,0)}M + T^{(0,1)}M\;.
\end{equation}
The maps
\begin{equation}
TM \rightarrow T^{(1,0)}M \;: X \mapsto \frac{1}{2} (X-iJX) \;,\;\;\;
TM \rightarrow T^{(0,1)}M \;: X \mapsto \frac{1}{2} (X+iJX) \;,\;\;\;
\end{equation}
are complex linear and complex anti-linear isomorphisms, respectively,
of complex
vector bundles. Since $TM$ is a holomorphic vector bundle over $M$, 
so is $T^{(1,0)}M$, but the smooth complex vector bundle
$T^{(0,1)}M$ is not a holomorphic vector bundle in a natural way. 

A {\em complex vector field} $Z$ is a section 
of $T_\mathbb{C} M$ 
and can  be decomposed into its $(1,0)$ and $(0,1)$ parts
\begin{equation}
Z^{(1,0)} = \frac{1}{2}\left( Z - i J Z \right) \;,\;\;\;
Z^{(0,1)} = \frac{1}{2}\left( Z + i J Z \right) \;.
\end{equation}

Given local holomorphic coordinates
$z^i = x^i + i y^i$ we can define local complex frames
\begin{equation}
Z_i = \frac{\partial}{\partial z^i} = \frac{1}{2} (X_i - i Y_i) \;,\;\;\;
\overline{Z}_i = \frac{\partial}{\partial \bar{z}^i} = \frac{1}{2} (X_i + i Y_i) \;,\;\;\;
\end{equation}
on $T^{(1,0)}M$ and $T^{(0,1)}M$, where $X_i = \frac{\partial}{\partial x^i}, Y_i= JX_i = \frac{\partial}{\partial y^i}$ is a coordinate frame on $TM$.

Like the complexified tangent bundle, 
all associated complex tensor bundles 
admit decompositions into `holomorphic' and `anti-holomorphic' components.
For example complex $n$-forms 
can be decomposed into $(p,q)$-forms, $p+q=n$, 
$\Omega^n(M) = \bigoplus_{p+q=n} \Omega^{p,q}(M)$.
The de-Rham differential can be decomposed as 
\begin{equation}
d = \partial + \bar{\partial} \;,\;\;\;
\partial=d^{(1,0)} \;,\;\;
\bar{\partial}= d^{(0,1)} \;.
\end{equation}
If the complex structure is integrable, then
\begin{equation}
\partial: \Omega^{p,q}(M) \rightarrow \Omega^{p+1,q}(M) \;,\;\;\;
\bar{\partial}: \Omega^{p,q}(M) \rightarrow \Omega^{p,q+1}(M) \;,\
\end{equation}
and since $\partial^2=0=\bar{\partial}^2$, the  de-Rham cohomology
admits a refinement called Dolbeault cohomology:
\begin{equation}
H^n(M) = \sum_{n=p+q} H^{p,q}_{\bar{\partial}}(M) \;.
\end{equation}

A connection $D$ on a holomorphic vector bundle is called a {\em holomorphic
connection} if it is 
compatible with the holomorphic structure, that is 
if $\pi^{0,1} D s = \overline{\partial} s=0$ for all holomorphic
sections $s$, where $\pi^{0,1}$ is the projection
onto the anti-holomorphic co-tangent bundle, and where
$\overline{\partial} = \pi^{0,1}d $ is the standard anti-holomorphic
partial derivative, i.e. the anti-holomorphic projection of the
exterior derivative $d$. Equivalently, the
$(0,1)$-part of the connection one-form vanishes, 
$\omega^{0,1} = \pi^{0,1} \omega = 0$. Equivalently, for holomorphic sections $s$ the 
covariant derivative along a complex  vector field of type $(0,1)$ vanishes, $D_{\bar{X}} s = 0$ for all
$X \in \Gamma(T^{(1,0)} M)$ and $s \in \Gamma_{\rm holom} (E)$.

On a holomorphic Hermitian vector bundle there is a unique connection,
called the {\em Chern 
connection}, which is simultaneously Hermitian and holomorphic.
As an example consider the trivial holomorphic Hermitian 
vector bundle $\mathbb{C}^n \times \mathbb{C}^m \rightarrow
\mathbb{C}^n$, where the Hermitian metric $\gamma$ is defined by
choosing a Hermitian
inner product on $\mathbb{C}^m$. This vector bundle carries a 
canonical flat connection $d$ which is defined by the standard partial derivative, 
that is by declaring that any frame defined by 
a basis $(e_i)$ of $\mathbb{C}^m$ is parallel, $d_X e_i=0$ for all
complex vector fields $X$ on $\mathbb{C}^n$.  
The covariant derivative $d_Xv$ of a section 
$v(P) = v^i(P) e_i$, $P\in M$ along a complex vector field 
$X=X^a \partial_a + X^{\bar{a}} \partial_{\bar{a}} \in
\Gamma(T_{\mathbb{C}}\mathbb{C}^n)$ is
\begin{equation}
d_X v= X(v^i) e_i =  (X^a \partial_a v^i+ X^{\bar a} \partial_{\bar a}
v^i)e_i = \partial_X v  + \overline{\partial}_X v \;.
\end{equation}
The connection $d$ is manifestly holomorphic, and it is 
also Hermitian since
\begin{equation}
d_X \gamma(v,w) = X \gamma(v,w) = \gamma(d_X v, w) +
\gamma(v,d_{\bar{X}} w) \;.
\end{equation}

\subsection{Hermitian manifolds \label{App:Herm_Mfd}}

An {\em (almost) 
Hermitian manifold} $(M,J,g)$ is an (almost) complex manifold $(M,J)$
equipped with a $J$-invariant pseudo-Riemannian metric $g$, 
\begin{equation}
(J^*g)(X,Y) = g(JX,JY) = g(X,Y) \;,\;\;\;\forall X,Y \in
\mathfrak{X}(M) \;.
\end{equation}
Note that we allow the Riemannian metric to be indefinite. Such
manifolds are often called (almost) pseudo-Hermitian. Also note that
the positive and negative eigenvalues of an (almost) Hermitian metric
always come in pairs. One therefore says that a 
pseudo Hermitian metric has {\em complex signature}
$(m,n)$ if the underlying Riemannian metric has real signature $(2m,2n)$.

The metric $g$ can be extended complex-linearly to the complexified tangent bundle
$T_\mathbb{C} M$. The resulting complex bilinear form has the following properties,
where $Z,W$ are complex vector fields:
\begin{eqnarray}
&& g(Z,W) = g(W,Z) \;,\\
&& g(\bar{Z}, \bar{W}) = \overline{ g(Z,W)} \;,\;\;\; \\
&& g(Z,  W) = 0 \;,\mbox{if}\;\;\;Z,W \in \Gamma(T^{(1,0)}M) \;,\\
&& g(\bar{Z},Z) > 0 \;,\mbox{unless}\;\;Z=0  \;.
\end{eqnarray}
Assume that  $J$ is integrable, and let $z^i = x^i + i y^i$ be local complex coordinates, 
with associated holomorphic frame  $Z_i =\frac{1}{2}(X_i -  i Y_i)$ .  
Then the components of the metric are
\begin{eqnarray}
g_{jk} &:=& 2 g(Z_j, Z_k) = 0 \;, \\
g_{\bar{j}\bar{k}} &:=& 2 g(\overline{Z}_j, \overline{Z}_k) = 0 \;, \\
g_{j\bar{k}} &:=& 2 g(Z_j, \overline{Z}_k) = 2 g(\overline{Z}_k, Z_j) =
g_{\bar{k}j}\;, \\
\overline{g_{j\bar{k}}} &=& g_{\bar{j}k} = g_{k\bar{j}}   \;.
\end{eqnarray}
Here we used properties of the complex-linear extension of the metric
to $T_\mathbb{C}M$, and choose the normalization for later 
convenience. Note that the coefficients can be arranged as a Hermitian matrix.
The metric can be written
\begin{equation}
g = g_{j\bar{k}} dz^j d\overline{z}^k  = \frac{1}{2}  g_{j\bar{k}} 
\left( dz^j \otimes d \overline{z}^k + d \overline{z}^k \otimes
d z^j \right)  \;.
\end{equation}
Note that $g(Z_i, \bar{Z}_j) = \frac{1}{2}
g_{i\bar{j}}$, which explains our normalization of $g_{i\bar{j}}$.

Given a  metric and a compatible (almost) complex structure, 
one defines the {\em fundamental two-form} $\omega$ by 
\begin{equation}
\omega(X,Y) := g (X, JY) \;.
\end{equation}
The coefficients of the fundamental two-form with respect to the 
holomorphic frame $Z_i$ are
\begin{eqnarray}
\omega_{i\bar{j}} &=& 2 \omega(Z_i, \bar{Z}_j) = 
2 g(Z_i, J \bar{Z}_j) = 2 g (Z_i, -i  \bar{Z}_j) = - i g_{i\bar{j}} \nonumber \\
\omega_{\bar{j}i} &=& 2 \omega(\bar{Z}_j, Z_i) = 
 2 g(\bar{Z}_j , J Z_i) = i g_{\bar{j} i} = 
i g_{i \bar{j}} =  - \omega_{i\bar{j}} \;. \label{components_omega}
\end{eqnarray}
Therefore the fundamental two-form has the expansion
\begin{equation}
\omega = -\frac{i}{2}  g_{i\bar{j}} \left( dz^i  \otimes d\bar{z}^j - d\bar{z}^j \otimes dz^i \right)
= - \frac{i}{2} g_{i\bar{j}} dz^i \wedge d\overline{z}^j 
= \frac{1}{2} \omega_{i\bar{j}} dz^i \wedge \overline{z}^j \;.
\end{equation}

The fundamental two-form is non-degenerate. Given $\omega$ and $J$ we can
therefore solve for the metric using that
\begin{equation}
g(X,Y) = \omega(JX,Y) 
\end{equation}
Moreover the complex structure 
\begin{equation}
J \in \Gamma(\mbox{End}(TM)) \;: TM \rightarrow TM
\end{equation}
is determined by $g$ and $\omega$ as
\begin{equation}
J = g^{-1} \omega  \in \Gamma(T^*M \otimes TM) \cong \Gamma( \mbox{End}(TM))  \;,
\end{equation}
where the map $g^{-1} \omega$ is defined by
\begin{equation}
Y = (g^{-1}\omega)(X) \Leftrightarrow g(\cdot, Y ) = \omega(\cdot, X) \;.
\end{equation}
Thus any two of the three compatible 
data $(g,\omega,J)$ suffice to determine the third.
To provide the corresponding local formulae, we introduce the components of the
inverse metric by
\begin{equation}
g^{i\bar{k}} g_{\bar{k}j} = \delta^i_j \;,\;\;\;
g^{\bar{i}k} g_{k\bar{j}} = \delta^{\bar{i}}_{\bar{j}} \;.
\end{equation}
The components \eqref{components_omega}
of the fundamental form are determined by antisymmetry.
These relations are consistent with complex conjugation,
$\overline{\omega_{i\bar{j}}} =  \omega_{\bar{i}j}$. 
Evaluating $J^m_{\;\;n} = g^{mp}\omega_{pn}$ in complex coordinates,
we obtain the components of the complex structure:
\begin{equation}
J^i_{\;\;j} = g^{i\bar{k}} \omega_{\bar{k}j}= i \delta^{i}_{j} \;,\;\;\;
J^{\bar{i}}_{\;\;\bar{j}} = g^{\bar{i}k}\omega_{k\bar{j}} =
 - i \delta^{\bar{i}}_{\bar{j}} \;.
\end{equation}

The metric $g$ and the fundamental form $\omega$ can be combined into
a Hermitian form $\gamma$, which defines
a Hermitian metric on the complex vector bundle $TM$. 
Its components with respect to the holomorphic frame $Z_i = \frac{\partial}{\partial z^i}$ are
\begin{equation}
\gamma = g_{i \bar{j}} dz^i \otimes d \overline{z}^j = g + i \omega \;.
\end{equation}
We remark that our conventions differ from \cite{Ballmann}, on which 
\ref{App:cplxmfds},
\ref{App:ComplexVBdl},
\ref{App:Herm_Mfd}, and \ref{app:Kaehler} are partly based. 
 In particular, 
we avoid a factor $\frac{1}{2}$ between the coefficients of the 
metric $g$ on $M$ and the
Hermitian metric $\gamma$ on the complex 
vector bundle $TM$, we include a factor $\frac{1}{2}$ in 
the definition of the symmetrized tensor product, we define $\omega$ in terms of $g,J$ 
with a relative minus sign, and we take Hermitian forms complex anti-linear
in the second rather than in the first argument.

\subsection{Symplectic manifolds \label{app:symp_man}}

A  {\em symplectic manifold}
$(M,\omega)$ is a real manifold equipped with a closed non-degenerate
two-form $\omega$, called the symplectic form. 
Symplectic
manifolds are even dimensional. 
The tangent spaces $(T_p M, \omega_p)$
are symplectic vector spaces isomorphic to $\mathbb{R}^{2n}$ with its standard
symplectic form $\omega$. Let $W$ be a linear subspace, $\iota: W \rightarrow V$  
be the canonical embedding, and
\begin{equation}
W^\perp = \{ v\in V| \omega(v,w)=0 \;,\;\forall w \in W \}
\end{equation}
be the `symplectically perpendicular' subspace.
Then
\begin{itemize}
\item
$W\subset V$ is called {\em isotropic} if $W\subset W^\perp$. This
implies $\dim W \leq \frac{1}{2} \dim V$ and $\iota^* \omega$ is
totally degenerate,  $\iota^* \omega=0$.
\item
$W\subset V$ is called {\em co-isotropic} if $W^\perp \subset W$. This
implies $\dim W \geq \frac{1}{2} \dim V$ and $W/W^\perp$ 
inherits a symplectic structure from $V$. 
\item
$W\subset V$ is called {\em Lagrangian} if it is isotropic and co-isotropic, that is 
 if $W^\perp = W$. This
implies $\dim W=\frac{1}{2} \dim V$ and $W$ is an isotropic subspace of maximal 
dimension. 
\item
$W\subset V$ is called {\em symplectic} if $W\cap W^\perp = \{ 0 \}$. 
\end{itemize}
Consider the following example of a co-isotropic subspace. 
Let $\{ \xi, \eta, X_1, \ldots, X_n,$ $Y_1, \ldots, Y_n\}$ be a basis 
of $V\cong \mathbb{R}^{2n+2}$, such that
\begin{equation}
\omega(\xi,\eta) = 1 \;,\;\;\;\omega(X_i,Y_j) = \omega_{ij} \;,
\end{equation}
with all other components determined by antisymmetry, or else being zero.
Define $W$ as the linear
subspace $W=\langle \eta, X_1, \ldots, X_n, Y_1, \ldots , Y_n \rangle$.
Then $W^\perp = \mbox{ker} (\iota^* \omega )= \langle \eta \rangle \subset W$,
so that $W$ is co-isotropic. The quotient $\bar{W}:= W/W^\perp$ is defined by 
the equivalence relation 
\begin{equation}
\label{EquivRel}
w \sim w' \Leftrightarrow w - w' = \alpha \eta \;.
\end{equation}
The projection map onto the quotient is
\begin{equation}
\pi: W \rightarrow \bar{W}\;,\;\;\;w \mapsto \bar{w} = \pi(w)\;,
\end{equation}
where $\bar{w}= \pi(w)$ denotes the equivalence class of $w$ with respect to 
(\ref{EquivRel}). On $\bar{W}$ we can define a two-form 
$\bar{\omega}$ by
\begin{equation}
\bar{\omega}(\bar{X}, \bar{Y}) = (\pi^* \bar{\omega})(X,Y) = (\iota^* \omega)(X,Y) \;,
\end{equation}
which is non-degenerate because we have factored out the kernel of $\iota^* \omega$. 
Choosing the basis $\{ \bar{X}_1, \ldots, \bar{X}_n, \bar{Y}_1, \ldots, \bar{Y}_n \}$
for $\bar{W}$, the components of $\bar{\omega}$ are
\begin{equation}
\bar{\omega}_{ij} = \bar{\omega}(\bar{X}_i,  \bar{Y}_j) = \omega_{ij} \;.
\end{equation}

A submanifold $\iota: S \rightarrow M$ is called a(n) {\em isotropic, co-isotropic,
Lagrangian and symplectic submanifold}, respectively, if all its tangent spaces  
are  isotropic, co-isotropic,
Lagrangian and symplectic, respectively. The pullback $\iota^* \omega$ of the
symplectic form is thus totally degenerate on isotropic and symplectic submanifolds,
and isotropic submanifolds have maximal dimension $\frac{1}{2} \dim M$.

An immersion $\iota: S \rightarrow M$ is called a {\em Lagrangian immersion} if
its image is a Lagrangian submanifold. 
A vector field $X$ on $(M,\omega)$ is called {\em a Hamiltonian vector field} if
\begin{equation}
\omega(X,\cdot) = - d H(\cdot) 
\end{equation}
for a function $H$, called the Hamiltonian or moment(um) map(ping) of $X$.

{\bf Example of a symplectic quotient.}
We now give a simple example of a {\em symplectic quotient} 
(or {\em symplectic reduction}),
which is useful for understanding the complex version of the superconformal
quotient relating affine conical to projective special K\"ahler manifolds. 
Let $(M,\omega)$ be a symplectic manifold, and let
$X$ be a Hamiltonian vector field which generates a $U(1)$-action on $M$. The
level surfaces ${{\cal H}_\gc = \{ H = \gc \} = H^{-1}(\gc})$ of the moment 
map $H$\footnote{Here $H^{-1}(\gc)$ denotes the inverse image of $\gc$ under
$H$, that is, the level set. This notation is common in the literature about
symplectic quotients.} are invariant
under the action of $X$, since $L_X H = dH(X) = - \omega(X,X)=0$. We assume
that the resulting $U(1)$-action on ${\cal H}_c$ is such that the orbit space
$\bar{M} = {\cal H}_\gc/ \langle X \rangle = {\cal H}_{\gc} / U(1)$ 
is a smooth manifold. We note that any vector field $T$ which is tangent to ${\cal H}_\gc$
must be symplectically perpendicular to $X$, that is
\begin{equation}
\omega(X,T) = -dH(T) = 0  \;.
\end{equation}
In particular $X$ itself is tangent to ${\cal H}_\gc$. We choose a vector field $\xi$
transversal to ${\cal H}_\gc$ by imposing the condition
$\omega(\xi, X) = dH(\xi) = 1$.
Thus in a coordinate system where we
use $H$ as one of the coordinates, $\xi= \partial_H$. 
The restriction $\omega_\gc: = \iota_\gc^* \omega$
of $\omega$ to the immersed hypersurface $\iota_{\gc}: {\cal H}_\gc \rightarrow M$
is degenerate. From the above it is clear that its kernel is spanned by $X$, 
therefore ${\cal H}_{\gc}$ is a co-isotropic
submanifold. The two-form $\omega_\gc$ is invariant under $X$,
\begin{equation}
L_X \omega_\gc = d( \omega_\gc(X, \cdot)) + (d \omega_\gc)(X, \cdot,\cdot) = 0 \;,
\end{equation}
because $\omega_\gc(X,\cdot)=0$ and $d\omega_\gc=d\iota^*_{\gc} \omega =
\iota^*_\gc d \omega = 0$. Since $\omega_\gc$ is
also transversal to the action of $X$ (that is, its components in the $X$-direction vanish),
$\omega_\gc$ can be projected to the quotient $\bar{M}= {\cal H}_{\gc}/U(1)$ to define
a two-form $\bar{\omega}$ by $\pi^* \bar{\omega} = \omega_\gc$,
where $\pi : {\cal H}_\gc \rightarrow \bar{M}$ is the projection onto the quotient. Since
we take a quotient with respect to the kernel of $\omega_\gc$, the two-form 
$\bar{\omega}$ is non-degenerate. It is also smooth because all maps entering into
its construction are by assumption smooth. To verify that $\bar{\omega}$ is closed we note
that $d \omega=0$ 
implies
\begin{equation}
0 = d \omega_\gc = d (\pi^* \bar{\omega}) = \pi^* d \bar{\omega}  \;.
\end{equation}
Since the projection map is surjective, every tangent vector $\bar{X}$ of $\bar{M}$ 
can be lifted to a tangent vector $X$ of ${\cal H}_\gc$. Therefore
\begin{equation}
 d \bar{\omega}( \bar{X}, \bar{Y}) = (\pi^* d\bar{\omega})(X,Y) = 0 \;,
 \end{equation}
 for all $\bar{X}, \bar{Y} \in \mathfrak{X}(\bar{M})$, and thus $d\bar{\omega}=0$. 
This shows that $({\cal H}_\gc/U(1), \bar{\omega})$ is a 
symplectic manifold. The construction by  which it is obtained from $(M,\omega)$ is called a symplectic quotient, denoted $M//U(1)$. 
Symplectic quotients can more generally be defined for symplectic actions of Lie groups $G$
on symplectic manifolds, and are denoted $M//G$ \cite{Marsden:1974dsb}.

\subsection{K\"ahler manifolds \label{app:Kaehler}}

An {\em (almost)
K\"ahler manifold} $(M,g,\omega)$  is an (almost) Hermitian manifold $(M,J,g)$ where the
fundamental two-form is closed. 
We will restrict ourselves to K\"ahler manifolds, that is to the case where $J$
is integrable and $(M,J)$ is a complex manifold. As for Hermitian manifolds we
include cases where the metric is indefinite. We remark that for
Hermitian manifolds the 
condition $d\omega=0$ is equivalent to $J$ being parallel with respect
to the Levi-Civita connection, $D  J = 0$.  The fundamental form of a
K\"ahler manifold is called its {\em K\"ahler form}. 
Note that $(M,\omega)$
is a symplectic manifold. Thus K\"ahler manifolds are pseudo-Riemannian
manifolds which simultaneously admit a compatible complex
structure and a compatible symplectic structure. 
Another equivalent characterization of a K\"ahler manifold is that
the Chern connection of the Hermitian metric $\gamma = g+i\omega$ 
on $TM$ is equal to the Levi-Civita connection $D$ of $g$. Evaluating the 
condition $d\omega=0$ in local holomorphic coordinates we obtain
the integrability condition
\begin{equation}
\partial_l g_{j\bar{k}} = \partial_j g_{l\bar{k}} \;,
\end{equation}
or, equivalently,
\begin{equation}
\partial_{\bar{l}} g_{j\bar{k}} = \partial_{\bar{k}} g_{j\bar{l}} \;.
\end{equation}
Another equivalent characterization is the local existence of a 
K\"ahler potential $K$, that is of a smooth real function such that
\begin{equation}
\omega = - \frac{i}{2} \partial \overline{\partial} K \;.
\end{equation}
This follows by combining Poincar\'e's lemma with the 
decomposition of forms into types ($\partial \bar{\partial}$-lemma):
locally $\omega = d\alpha$, where $\alpha = \beta + \bar{\beta}$, with $\beta\in \Omega^{1,0}(M)$. 
Since $\omega \in \Omega^{1,1}(M)$ and $d=\partial + \bar{\partial}$, 
where $\partial^2 = 0 = \bar{\partial}^2$, and where $\partial, \bar{\partial}$
act consistently with type (since we assume that the complex structure is integrable):
\begin{equation}
d\alpha = \partial( \beta + \bar{\beta}) + \bar{\partial} (\beta + \bar{\beta}) 
\in \Omega^{1,1}(M)  \Rightarrow 
\partial \beta = 0 \;,\;\omega=\bar{\partial} \beta  + \partial \bar{\beta} \;,\;
\bar{\partial}\bar{\beta} = 0  \;.
\end{equation}
Hence $\beta = \partial \varphi$ by the $\partial$-version of the Poincar\'e lemma,
where $\varphi$ is a smooth complex function. Therefore
\begin{equation}
\omega = \partial \bar{\partial} (\bar{\varphi} - \varphi) = 
-\frac{i}{2} \partial \bar{\partial} K  \;,
\end{equation}
where $K= -2i (\varphi - \bar{\varphi})$. 
This provides a real potential for the metric,
\begin{equation}
g_{i\bar{k}}  = \partial_i \partial_{\bar{k}} K  \;,
\end{equation}
called the K\"ahler potential. Note that the K\"ahler potential 
is only determined up to adding the real part of a harmonic 
function, since $K$ and $K+f+\bar{f}$ with $\bar{\partial}f=0$ 
define the same metric. For further reading on K\"ahler manifolds
we refer to \cite{Ballmann} on which this section is partly based.

Since a K\"ahler manifold is in particular a symplectic manifold, one 
can apply symplectic reduction. 
If the
symplectic group action is in addition holomorphic and isometric,
it preserves the extra structures which distinguish a K\"ahler manifold
from a symplectic manifold, and the quotient carries an induced 
K\"ahler structure. The Hamiltonian vector fields
generating such a  group action must be holomorphic Killing vector fields. 
Symplectic quotients of K\"ahler manifolds by symplectic, holomorphic and
isometric group actions are called {\em K\"ahler quotients} \cite{Hitchin:1986ea}. 
One uses the same
notation $M//G$ as for symplectic quotients.

\subsection{Contact manifolds \label{App:ContactGeometry}}

A one-form $\theta$ on a manifold $M$ of odd dimension $2n+1$ is called
a {\em contact form} if the $(2n+1)$-form $\theta \wedge (d\theta)^n$ is 
a volume, that is, if it is nowhere vanishing,
\begin{equation}
(\theta \wedge d\theta \wedge \cdots \wedge d\theta)_p \not=0 \;,\;\;\;\forall p \in M \;.
\end{equation}
 A {\em contact manifold} $(M,\theta)$ 
is an odd-dimensional manifold equipped with a contact form. A {\em contact structure}
on an odd-dimensional manifold $M$ is defined by the choice of a hyperplane 
distribution $V=\cup_{\in M} V_x$ on its tangent bundle $TM$, which is 
maximally non-integrable, that is non-integrable at every point.

To relate the concepts of contact form and contact structure, we note that the
kernel $\mbox{ker} (\theta)$ of the one-form $\theta$ defines a hyperplane distribution 
on $TM$. By the dual version of the  Frobenius theorem, the integrability condition
for this distribution is $\theta \wedge d\theta =0$, which implies
$\theta \wedge (d\theta)^n=0$. Thus by definition a contact
distribution is not integrable, and in fact maximally non-integrable, since
the integrability condition does not hold at any point of the manifold. 
Consequently, a contact form determines a contact structure. 
Since any two one-forms $\theta$, $\theta'$, which differ by multiplication with
a nowhere vanishing function $f$, $\theta' = f \theta$ have the same kernel, 
a contact structure corresponds to an equivalence class of contact forms. Since
$\theta \wedge (d\theta)^n$ is nowhere vanishing, the 
kernel of $d\theta$ defines a one-dimensional
distribution on $TM$ which is complementary to the contact distribution, 
that is $TM = \mbox{ker}(\theta) \oplus \mbox{ker}(d\theta)$. 

To each contact form there is an associated vector field, called the {\em Reeb vector field}
$R$, which is the unique vector field on $M$ such that 
\begin{equation}
\theta(R) = 1 \;,\;\;\;\ d\theta(R,\cdot) = 0 \;.
\end{equation}
Thus $R$ spans the kernel of $d\theta$ and extends any given frame on 
$V=\mbox{ker} (\theta)$ to a frame on $TM$.

Contact manifolds can be regarded as the odd-dimensional analogues of 
symplectic manifolds. Contact and symplectic manifolds can be related 
by constructions which change the number of dimensions by one. 

{\bf The symplectification of a contact manifold.} Let $(M,\theta)$ be a contact
manifold of dimension $2n+1$. Consider the cone $\mathbb{R}^{>0} \times M$ over $M$
with coordinate $r$ on $\mathbb{R}^{>0}$. Then $(\mathbb{R}^{>0} \times M, \omega)$,
with $\omega = r^2 d\theta + 2 r dr \wedge \theta$ is a symplectic manifold, because
$d\omega = 2 r dr \wedge d\theta - 2 r dr \wedge d\theta =0$, and because $\omega$
is non-degenerate, as can be verified using a frame consisting of $\partial_r$ and a frame
for $M$. Note that we have seen above that the kernels of $\theta$ and $d\theta$ define
complementary distributions on $TM$. Using the variable $\rho$, defined by $r^2 =e^\rho$, 
we can write the cone in `product form': $(\mathbb{R}^{>0} \times M, \omega) \cong
(\mathbb{R} \times M, \omega')$, where $\omega' = d (e^\rho \theta)$. In this parametrization
we see that the symplectic form is exact. The symplectic manifold $(\mathbb{R}\times M, d(e^\rho \theta))
\cong (\mathbb{R}^{>0}, r^2 d\theta + 2 r dr \wedge \theta)$ is called the {\em symplectification} 
(or symplectization) of the contact manifold $(M,\theta)$.

{\bf Legendrian submanifolds.}
If $\theta$ is a contact form on a manifold $M$ of dimension $2n+1$, 
then $d\theta_{|V}$ is a symplectic form on the
contact distribution $V=\mbox{ker}\theta$. Therefore a subdistribution $L \subset V$ 
can only be integrable if it isotropic with respect to $d\theta_{|V}$. This implies
that $2 \dim L \leq \dim M  -1 = 2n$, that is $\dim L\leq n$. Integral manifolds of 
dimension $n$ which saturate this bound are called {\em Legendrian submanifolds}, 
and are the counterparts of 
Lagrangian submanifolds in symplectic geometry. In particular, the Legendrian submanifolds
of a contact manifold lift to Lagrangian submanifolds of its symplectification. 
An immersion $\iota: {\cal H} \rightarrow M$ into a contact manifold $(M,\theta)$ is 
called a {\em Legendrian immersion} if the image of ${\cal H}$ is a Legendrian
submanifold.

For further reading on contact geometry we refer to \cite{Rudolph:2013}.

\subsection{Sasakian Manifolds \label{app:sasakian}}

The following  section is based on various sources, including 
\cite{Baues:2003,Alekseevsky:2012fu,Cortes:2017utn,Dietrich:2017}.

K\"ahler manifolds can be thought of as symplectic manifolds with an additional
pseudo-Riemannian metric subject to compatibility conditions, which determine a
complex structure. Sasakian manifolds are the `contact analogue' of K\"ahler manifolds,
that is contact manifolds equipped with a metric which satisfies certain compatibility 
conditions. One way to characterize Sasakian manifolds is by requiring that their
symplectification is K\"ahler:
 A {\em Sasakian manifold} $(S, \theta, g)$ is a contact manifold $(S,\theta)$ equipped with
 a (pseudo-)Riemannian metric $g$, such that the Riemannian cone 
 $(M,g_M)= (S\times  \mathbb{R}^{>0}, r^2 g_S + dr^2)$ is a K\"ahler manifold with K\"ahler
 form $\omega=r^2 d\theta + 2r dr \wedge \theta$. 
   Comparing to the previous section we see that the Riemannian cone is indeed the
 symplectification of the contact manifold $(M,\theta)$. We remark that the complex
 structure $J$ relates the homothetic Killing vector field  $\xi = r \partial_r$ to the
 Reeb vector field $R=-J\xi$. 
 
 If in addition the Reeb vector field generates a $U(1)$-action on $S$ such that
 $\bar{M}=S/U(1)$ is a smooth manifold, then $\bar{M}$ is the K\"ahler quotient
 of $M$, $\bar{M} = M//U(1) = S/U(1)$.  Moreover, if a K\"ahler manifold $M$ admits a
 homothetic Killing vector field $\xi$, which satisfies $D \xi = \Id$, then $M$
 is a Riemannian cone over the Sasakian $S= \{ g(\xi,\xi)=1 \}$. If the 
 quotient $M/\mathbb{C}^*$ by the holomorphic and homothetic action 
 generated by $\xi, J\xi$ defines a smooth manifold, this manifold is precisely
 the symplectic quotient with respect to the action of $J\xi$. Finally, given
 a K\"ahler manifold $\bar{M}$ we can construct a `complex cone' or 
 `conical K\"ahler manifold' $M$ as the total space of a $\mathbb{C}^*$
 bundle over $\bar{M}$, such that ${\bar M} = M//\mathbb{C}$.\footnote{Another natural
 name for $M$ would be `K\"ahler cone' in analogy to Riemannian cone, but 
 K\"ahler cone is also use for the cone of K\"ahler structures on a Calabi-Yau manifold.}

\subsection{Complex symplectic manifolds and complex contact manifolds \label{app:complex_symplectic}}

The concepts of symplectic and contact geometry, which we have formulated 
for real manifolds, can be formulated analogously for complex manifolds. We illustrate
this by examples. 

The vector space $V=T^*\mathbb{C}^n 
\cong \mathbb{C}^{2n}$ equipped with the complex symplectic form 
$\Omega = dz^i \wedge dw_i$ is the standard example for a complex 
symplectic vector space of complex dimension $2n$. Its projectivization
$P(V')$, where $V' =  \{ (z,w)  \in \mathbb{C}^{2n} | (z,w)\not= (0,0) \}$,
is the space $V' / \sim $, 
where $\sim$  denotes the equivalence relation
\begin{equation}
(z,w) \sim (z',w') \Leftrightarrow (z',w') = \lambda (z,w) \;,\;\;\, \exists \lambda \in \mathbb{C}^* \;.
\end{equation}
$P(V')$ is a complex contact space with complex symplectification $V$. In special 
geometry projective special K\"ahler manifolds $\bar{M}$ can be realized as holomorphic
Legendrian immersions into $P(V')$, which lift to holomorphic Lagrangian immersion
of the corresponding conical affine special K\"ahler manifold $M$ into $V$.

\subsection{Some groups and their actions \label{app:groups}}

This section is based on  \cite{Cortes:2017utn,Dietrich:2017}.
The Heisenberg group  $\mbox{Heis}_{2n+1}(\mathbb{R})$ is the nilpotent Lie 
group obtained as a central
extension of the translation group $\mathbb{R}^{2n}$, with group law
\begin{equation}
(s,v) \circ (s',v') = \left( s+s' + \frac{1}{2} \Omega(v,v') , v+v' \right)  \;,
\end{equation}
where $s,s'\in \mathbb{R}$ are central, $v,v'\in \mathbb{R}^n$ 
are translations, and where $\Omega$ is the standard
symplectic form on $\mathbb{R}^{2n}$. The standard generators $p_i,q_i,z$, $i=1, \ldots, n$
for the Lie algebra
$\mathfrak{heis}_{2n+1}(\mathbb{R})$ satisfy 
\begin{equation}
[p_i,q_j] = \delta_{ij} z\;,
\end{equation}
with all other commutators vanishing. The group $G= \mbox{Sp}(\mathbb{R}^{2n}) \ltimes
\mbox{Heis}_{2n+1}(\mathbb{R})$  is the semi-direct extension of the real Heisenberg group
by its group $\mbox{Sp}(\mathbb{R}^{2n})$ of automorphisms, with group law
\begin{equation}
g \cdot g' = \left( MM', s+s' + \frac{1}{2} \Omega(v, Mv'), v + Mv' \right) \;,
\end{equation}
where $M,M'\in \mbox{Sp}(\mathbb{R}^{2n})$ and $(s,v) \in \mbox{Heis}_{2n+1}(\mathbb{R})$. 
We use the same notation $g=(M,s,v)$ for elements of the complexification $G_\mathbb{C} = \mbox{Sp}(\mathbb{C}^{2n}) \ltimes \mbox{Heis}_{2n+1}(\mathbb{C})$. The quotient map
\begin{equation}
G_\mathbb{C} \rightarrow \mbox{Aff}_{{\rm Sp}(\mathbb{C}^{2n}) }= G_\mathbb{C}/Z(G_\mathbb{C})\;:
(M,s,v) \mapsto (M,v)
\end{equation}
induces an affine representation $\bar{\rho}$ of $G_\mathbb{C}$, whose restriction to 
the real subgroup $G$ provides an affine representation of $\mbox{Aff}_{{\rm Sp}(\mathbb{R}^{2n})}(\mathbb{R}^{2n})$.

On the complex vector space $\mathbb{C}^{2n}$ we choose Darboux 
coordinates $(X^I,W_I)$, such that the complex symplectic form is $\Omega = dX^I\wedge dW_I$.
We can embed $\mathbb{C}^{2n}$ into $\mathbb{C}^{2n+2}=\mathbb{C}^2\oplus \mathbb{C}^{2n}$ 
with standard coordinates $(X^0, W_0, X^i, W_i)$. 
A linear representation $\rho: G_{\mathbb{C}} \rightarrow Sp(\mathbb{C}^{2n+2})$  is
defined by
\begin{equation}
g=(M,s,v) \mapsto \rho(x) = \left( \begin{array}{ccc}
1 & 0 & 0 \\
-2s & 1 & \hat{v}^T \\
v & 0 & M \\
\end{array}\right) \;,\;\;\;
\hat{v} := M^T \Omega_0 v = \Omega_0 M^{-1} v \;,
\end{equation}
where $\Omega_0$ is the standard representation matrix for the symplectic form 
on $\mathbb{C}^{2n}$. According to Proposition 3.2.2 of \cite{Dietrich:2017} this is a 
faithful representation which induces the affine representation $\bar{\rho}: G_\mathbb{C} \rightarrow
\mbox{Aff}_{{\rm Sp}(\mathbb{C}^{2n})} (\mathbb{C}^{2n})$ because 
it preserves the affine hyperplane $\{ X^0=1\} \subset \mathbb{C}^{2n+2}$
and the distribution $\partial_{W_0}$. The orbit space $\{ X^0=1\}/\langle \partial_{W_0}\rangle$ 
is the symplectic reduction of $\mathbb{C}^{2n+2}$ with respect to the holomorphic
Hamiltonian group action generated by $\partial_{W_0}$, and $\rho$ induces $\bar{\rho}$
under this quotient.
Similarly, the real symplectic
affine space $\mathbb{R}^{2n}$ is the symplectic reduction of the real symplectic
vector space $\mathbb{R}^{2n+2}$, with $G_\mathbb{C}$ replaced by its real 
subgroup $G$. 

Finally we define the group $G_{SK}=\mbox{Sp}(\mathbb{R}^{2n}) \ltimes \mbox{Heis}_{2n+1}(\mathbb{C})\subset G_\mathbb{C}$. Note that $G\subset G_{SK}$ and that $G_{SK}$ is 
a central extension of $\bar{\rho}(G_{SK}) = \mbox{Aff}_{{\rm Sp}(\mathbb{R}^{2n})}(\mathbb{C}^{2n})=Sp(\mathbb{R}^{2n}) \ltimes \mathbb{C}^{2n}$. The latter group 
acts simply transitively on K\"ahlerian Lagrangian immersions of simply connected
ASK manifolds, in other words, it is the duality group of ASK geometry. 
 
\subsection{Para-complex geometry \label{App:para_complex}}

Here we collect some definitions and statements about para-complex 
geometry. More details can be found in 
\cite{Cortes:2003zd,Cortes:2005uq,Cortes:2009cs,Cortes:2015wca}.

A {\em para-complex structure} $J$ on a finite-dimensional
real vector space $V$ is a non-trivial involution $J\in \mbox{End}(V)$, $J\not= \Id$, 
$J^2 = \Id$, such that the eigenspaces $V^\pm := \mbox{ker}(\Id\mp J)$ of $J$ are of
the same dimension. A {\em para-complex vector space} $(V,J)$  is a real vector space $V$ 
endowed with a para-complex structure $J$. A homomorphism of para-complex vector
spaces is a linear map $\Phi: (V,J) \rightarrow (V',J')$ such that $\Phi \circ J = J' \circ \Phi$.
Para-complex vector spaces have even dimension and admit bases $e^\pm_i$ such that
$Je^\pm _i = \pm e^\pm_i$. It is easy to see that for $\dim_\mathbb{R} V =2n$
a para-complex structure is invariant under the group
$\mbox{Aut}(V,J) := \{ L \in GL(V) | LJ L^{-1}\}$, where
\begin{equation}
\mbox{Aut}(V,J) \cong GL(n,\mathbb{R}) \times
GL(n,\mathbb{R}) \subset GL(V) \cong GL(2n,\mathbb{R})\;.
\end{equation}

An {\em almost para-complex structure}
on a smooth manifold $M$ is an endomorphism field $J \in \mbox{End}(TM) :
p \mapsto J_p$ such that $J_p$ is a para-complex structure on $T_pM$ for
all $p\in M$. An {\em almost para-complex manifold} $(M,J)$ is a smooth manifold
$M$ endowed with an almost para-complex structure.

If one relaxes the condition that the eigenspaces of $J_p$ have equal dimension,
one obtains the concept of an {\em almost product structure.} Thus almost 
para-complex structures are almost product structure where the dimensions
of the eigendistributions `balance.' This creates many analogies with almost complex
manifolds.

An almost para-complex structure $J$ is called
{\em integrable} if the eigendistributions $T^\pm M := \mbox{ker} (\Id \mp J)$ are 
both integrable. An integrable almost para-complex structure is called a 
{\em para-complex structure}. A {\em para-complex manifold} $(M,J)$ is a manifold $M$
endowed with a para-complex structure $J$.
The Frobenius theorem implies that an almost  para-complex structure is
integrable if and only if its Nijenhuis tensor
$N_J(X,Y) = [X,Y]+[JX,JY]-J[X,JY]-J[JX,Y]$ vanishes for all vector fields
$X,Y$ on $M$.

A smooth map $\Phi: (M,J_M) \rightarrow (N,J_N)$ between para-complex manifolds
is called a {\em para-holomorphic map} if $d\Phi J_M = J_N d\Phi$.

It can be shown that the integrability of the
almost para-complex structure $J$ is equivalent to the existence of 
local para-complex coordinate systems.
This uses the algebra
$C$ of {\em para-complex numbers}, which are also known as split complex
numbers of hyperbolic complex numbers. As a real algebra $C$
is generated by $1$ and the symbol $e$, subject to the relation $e^2=1$. 
The map
\begin{equation}
\bar{\cdot} \;: C\rightarrow C \;, x+ey \mapsto x -e y \;,\;\;\;x,y\in \mathbb{R}
\end{equation}
is called para-complex conjugation and is a $C$-antilinear involution, which 
allows to regard $x,y$ as the real and imaginary part of $z=x+ey$. 
The algebra $C$ has zero-divisors, its group of invertible elements
is isomorphic to $O(1,1)$ and has four connected components 
separated by the light cone $z\bar{z}=x^2-y^2=\pm 1$. The algebra $C$ and 
the free $C$-module $C^n$ are para-complex vector spaces of real dimensions
$2$ and  $2n$, respectively,  with a
para-complex structure given by multiplication with $e$. 
One can show that a smooth manifold $M$ endowed with an atlas of $C^n$-valued
coordinate maps related by para-holomorphic coordinate transformations
admits an integrable para-complex structure. Conversely, any real manifold with 
an integrable para-complex structure  admits a para-complex atlas.

\begin{rmk}
For almost para-complex manifolds it is interesting to consider the case
where only one of the eigendistributions $T^\pm M$ is integrable. This
has applications in particular in doubled/generalized geometry. 
Here we focus on the case were both eigendistributions 
are integrable, which is relevant for Euclidean special geometry. 
\end{rmk}

A para-holomorphic map 
$\Phi: (M,J) \rightarrow C$ is called a para-holomorphic function.

A {\em para-holomorphic vector bundle} 
of rank $r$ is a smooth real vector bundle $W\rightarrow M$ of rank $2r$ 
whose total space $W$ and base $M$ are para-complex manifolds and 
whose projection $\pi$ is a para-holomorphic map.
On a para-holomorphic vector bundle we have a canonical splitting $W=W^+ \oplus W^-$
induced by the para-complex structure. 
The tangent bundle $TM \rightarrow M$ over any para-complex manifold $M$
is a para-holomorphic vector bundle. The splitting $TM = T^+M \oplus T^-M$ can 
be used to define a real version of Dolbeault cohomology on any para-complex manifold.

The {\em para-complexified tangent bundle} $T_C M = TM \otimes C$ can be equipped with 
the $C$-linear extension of the para-complex structure $J$.
It decomposes canonically into eigenbundles of $J$ with eigenvalues $\pm e$,
\begin{equation}
T_CM = T^{1,0}M \oplus T^{0,1}M \;.
\end{equation}
There is a canonical isomorphism
\begin{equation}
TM \xrightarrow{\cong} T^{1,0} M \;,\;\;\; X \mapsto \frac{1}{2}( X + eJX)
\end{equation}
of real vector bundles which is compatible with the para-complex structures on the 
fibre. $C$-valued differential forms admit a decomposition into types in analogy
with complex-valued differential forms on complex manifolds, which allows to define
a para-complex version of Dolbeault cohomology.

A {\em para-Hermitian vector space} $(V,J,g)$ is 
a para-complex vector space $(V,J)$, equipped with a pseudo-Euclidean
scalar product $g$ for which $J$ is an anti-isometry,
\begin{equation}
J^* g= g(J\cdot, J \cdot) = - g\;.
\end{equation}
Then $g$ is called a {\em para-Hermitian scalar product}, and $(J,g)$ a {\em para-Hermitian
structure} on $V$.
A para-Hermitian scalar product always has neutral signature. 
The {\em standard para-Hermitian structure on}  $\mathbb{R}^{2n} = \mathbb{R}^n \oplus \mathbb{R}^n$
is given by 
\begin{equation}
I e^\pm_{i} = \pm e^\pm_i \;,\;\;\;
g(e^\pm_i, e^\pm_j) = 0 \;,\;\;\;
g(e^\pm_i, e^\mp_j) = \delta_{ij} \;,\;\;\;
\end{equation}
where $e^+_i = e_i \oplus 0$, and $e^-_i = 0 \oplus e_i$. 

The {\em standard para-Hermitian structure on} $C^n = \mathbb{R}^n \oplus e \mathbb{R}^n$
with basis $e_i, f_i:= e e_i$ is given by
\begin{equation}
J e_i = f_i \;,\;\;\;
J f_i = e_i \;,\;\;\;
g(e_i,e_j) = - g(f_i, f_j) = \delta_{ij}\;.
\end{equation} 

Any two para-Hermitian vector spaces of the same dimension are isomorphic. 
Using any of the two standard realizations  $\mathbb{R}^{2n}$ or $C^n$,
it is straightforward to show that the {\em para-unitary group} 
\begin{equation}
U^\pi(V) := \mbox{Aut}(V,J,g)
=\{ L \in GL(V) | LJL^{-1} = J \;,L^* g = g\} 
\end{equation}
of a para-Hermitian vector space of real dimension $2n$ is
\begin{equation}
U^\pi (V)  \cong GL(n,\mathbb{R}) \subset
\mbox{Aut}(V,J) \cong GL(n,\mathbb{R}) \times GL(n,\mathbb{R}) \subset
GL(V) \cong GL(2n,\mathbb{R})\;.
\end{equation}
Note that $J$ itself is not an element of the para-unitary group, though
it is an element of the para-unitary Lie algebra.

An ({\em almost) para-Hermitian 
 manifold} $(M,J,g)$ is an (almost) para-complex manifold
 $(M,J)$ endowed with a pseudo-Riemannian metric $g$ such that
 $J^* g = g(J\cdot, J\cdot) = - g$. The two-form $\omega=g(\cdot, J \cdot)
 = - g(J\cdot, \cdot)$ is called the {\em fundamental two-form} of the 
 (almost) para-Hermitian manifold $(M,J,g)$. 
 Compared to \cite{Cortes:2003zd} we have changed the sign of $\omega$
 to be consistent with our conventions. Note that it is essential that $J$
 is an anti-isometry, and not an isometry, for $\omega$ to be antisymmetric.

 A {\em para-K\"ahler manifold} is an almost para-Hermitian
 manifold $(M,J,g)$ such that $J$ is parallel with respect to the Levi-Civita connection,
 $DJ=0$. 
 Note that $DJ=0$ implies both $d\omega=0$ and the integrability condition 
 $N_J=0$. 
 Alternatively, a para-K\"ahler manifold is a para-Hermitian manifold with 
 closed fundamental form. The symplectic form $\omega$ is called
 the {\em para-K\"ahler form}. 
 It can be shown that for a para-K\"ahler metric
 there exists around any point a real valued function $K$, called 
 a {\em para-K\"ahler potential}, such that the coefficients of $\omega$
 and $g$ are given by the mixed second derivatives with respect
 to para-holomorphic coordinates.

 An {\em affine special para-K\"ahler manifold} $(M,J,g,\nabla)$ is a para-K\"ahler
 manifold $(M,J,g)$ endowed with a flat, torsion-free connection such that
 $\nabla$ is symplectic, i.e. $\nabla \omega=0$, and such that $d_\nabla J=0$. 
  One can show that any simply connected affine special para-K\"ahler manifold
 can be realized by a para-K\"ahlerian Lagrangian immersion $\phi: M \rightarrow V$
 into the standard para-complex vector space $V=T^*C^n \cong C^{2n}$
 endowed with the $C$-valued symplectic form $\Omega=dX^I \wedge dW_I$,
 standard para-complex structure $I_V$, para-complex conjugation $\tau$ and
 para-Hermitian form $\gamma = g + e \omega = e \Omega(\cdot, \tau \cdot)$. 
 For a generic choice of para-complex symplectic coordinates $X^I,W_I$,
 the image of $\phi$ is the graph of a map $C^n \rightarrow C^n$, and therefore
 $\phi$ has a para-holomorphic prepotential $F$, i.e. $\phi = dF$.
 
 A {\em conical affine special
 para-K\"ahler manifold} $(M,J,g,\nabla,\xi)$ is an affine special 
 para-K\"ahler manifold $(M,J,g,\nabla)$ endowed with a vector field
 $\xi$ such that
 \begin{equation}
 \nabla \xi = D \xi = \Id \;,
 \end{equation}
 where $D$ is the Levi-Civita connection.

One can show that near any point $p\in M$ there exist 
coordinates $(q^a)=(x^I,y_I)$ such that
\begin{equation}
\xi = q^a \partial_a = x^I \partial_{x^I} + y_I \partial_{y_I} \;.
\end{equation}
Such coordinates are unique up to linear symplectic transformations, and are
called conical special real coordinates. On conical special para-K\"ahler manifolds
it is understood that `special coordinates' means `conical special coordinates.'
A para-holomorphic immersion $\phi \rightarrow V=C^{2n}$ 
is called a {\em conical para-holomorphic immersion} if the position vector field
$\xi^V = p \in V \cong T_p V$ is tangent along $\phi$, that is,
if $\xi^V \in d\phi_p T_p M$. Every simply connected conical special 
para-K\"ahler manifold can be realized by a conical para-K\"ahlerian Lagrangian
immersion, which is unique up to linear symplectic transformations. The corresponding
para-holomorphic prepotential can be chosen to be homogeneous of degree two. 

The vector fields $\xi$ and $J\xi$ generate an infinitesimal $C^*$-action on 
the conical affine special para-K\"ahler manifold $M$. To be able to take a 
quotient which defines a para-K\"ahler manifold, one needs to make additional
assumptions. 
A conical affine special
para-K\"ahler manifold $(M,J,g,\nabla,\xi)$ 
is called  a {\em regular conical affine special K\"ahler manifold} if the norm $g(\xi,\xi)$ of $\xi$ does not vanish on $M$ and 
if the quotient map $\pi \;: M \rightarrow \bar{M} = M / C^*$ is a para-holomorphic
submersion onto a Hausdorff manifold. 
Under these assumptions, the symmetric tensor field 
\begin{equation}
\tilde{g}^{(0)} = - \partial \bar{\partial}
\left(- e (X^I \bar{F}_I - F_I \bar{X}^I)\right) \;,
\end{equation}
which projects onto the orbit space $\bar{M}$, induces a para-K\"ahler metric
$\bar{g}$ on $\bar{M}$, such that $\tilde{g}^{(0)}=\pi^* \bar{g}$. 
A {\em projective special para-K\"ahler manifold} $(\bar{M}, \bar{J},\bar{g})$ is 
a para-K\"ahler manifold that can be realized locally as the quotient 
of a regular conical affine special para-K\"ahler manifold $M$ by 
its $C^*$-action.

 With a proper choice of conventions, local formulae for affine special
 K\"ahler and affine special para-K\"ahler manifolds are related by the
 replacement $i\rightarrow e$. Therefore one can use an $\varepsilon$-complex
 terminology which employs the notation $i_\varepsilon =e,i$ for $\varepsilon =\pm 1$. All statements
 in this section remain true when omitting `para' or replacing it by `$\varepsilon$-'
 and applying the appropriate substitutions for $e$.

\subsection{$\varepsilon$-quaternionic geometries \label{app:epsilon_quat}}

This section is based on
\cite{Cortes:2003zd,Cortes:2005uq,Cortes:2009cs,Cortes:2015wca}.

Hypermultiples contains four real scalars and their scalar geometries 
are related to the algebra $\mathbb{H}_{-1} = \mathbb{H}$ of the 
quaternions, or, for Euclidean  space-time signature, to the algebra $\mathbb{H}_1$
of para-quaternions. We treat both cases in parallel by writing 
$\mathbb{H}_\varepsilon$, where $\varepsilon=\pm 1$. 

The algebra $\mathbb{H}_\varepsilon$ 
of $\varepsilon$-quaternions is the four-dimensional real algebra
generated by three $\varepsilon$-complex units $i_1, i_2,i_3$, which pairwise
anticommute and satisfy the $\varepsilon$-quaternionic algebra
\begin{equation}
\label{epsilon_quat_algebra}
i_1^2 = i_2^2 = -\varepsilon i_3^2 = \varepsilon \;,\;\;\;i_1 i_2= i_3.
\end{equation}
An {\em $\varepsilon$-quaternionic structure} on a real vector space of dimension $4n$
is a Lie subalgebra $Q\subset \mathrm{End}(V)$ spanned by three pairwise
anticommuting endomorphisms $J_1, J_2, J_3$ which satisfy the $\varepsilon$-quaternionic
algebra \eqref{epsilon_quat_algebra}. The Lie group generated by the Lie algebra generated by $J_\alpha$, $\alpha=1,2,3$
is 
\begin{equation}
Sp_\varepsilon(1) = \left\{ \begin{array}{ll}
SU(2) \cong Sp(1)\;, & \mathrm{if}\;\;\varepsilon=-1 \;,\\
SU(1,1) \cong Sp(2,\mathbb{R}) \cong SL(2, \mathbb{R})\;, & \mathrm{if}\;\;\varepsilon=1 \;.\\
\end{array} \right.
\end{equation}
Our notation for symplectic groups is such that 
 $Sp(2n,\mathbb{R}) = Sp(\mathbb{R}^{2n})$,
$Sp(2n, \mathbb{C}) = Sp(\mathbb{C}^{2n})$ 
and
\begin{equation}
Sp(n) = Sp(2n,\mathbb{C}) \cap U(2n) \;,\;\;\;
Sp(k,l) = Sp(2n,\mathbb{C}) \cap U(2k,2l) \;.
\end{equation}
In particular, $Sp(1) = Sp(2,\mathbb{C}) \cap U(2) = SU(2)$ is the group 
often denoted $USp(2)$ in the physics literature. Also note that
$Sp(1,1) = Sp(2n,\mathbb{C}) \cap U(1,1) = SU(1,1)\cong Sp(2,\mathbb{R})$.

While there are various types of $\varepsilon$-quaternionic geometries,
we will only use two types which can be viewed as generalizations of 
$\varepsilon$-K\"ahler geometry. The first is realized by rigid hypermultiplets.

An {\em $\varepsilon$-hyper-K\"ahler} manifold 
($\varepsilon$-HK manifold)
is a pseudo-Riemannian manifold $(N,g)$
of dimension $4n=4k+4l$ 
whose holonomy group $\mathrm{Hol}(N)$ is contained in $Sp_\varepsilon(k,l)$, where 
\begin{equation}
Sp_\varepsilon(k,l) = \left\{ \begin{array}{ll}
Sp(k,l) \subset SO(4k,4l) \;,& \mathrm{if}\;\;\varepsilon =-1  \;, \\
Sp(2n,\mathbb{R}) \subset SO(2n,2n) \;, & \mathrm{if}\;\;\varepsilon=1  \;.
\end{array} \right.
\end{equation}
An $\varepsilon$-K\"ahler manifold has three pairwise anticommuting integrable
$\varepsilon$-complex structures $I_\alpha$, such that $\omega_\alpha = g(I_\alpha \cdot, \cdot)$
are antisymmetric and closed, and therefore form an $Sp_\varepsilon(1)$-triplet of
$\varepsilon$-K\"ahler forms. The $\varepsilon$-K\"ahler metric $g$ admits
$\varepsilon$-K\"ahler potentials with respect to any of the three $\varepsilon$-complex
structures, though in general there is no `$\varepsilon$-hyper-K\"ahler potential,' that is 
a potential which is $\varepsilon$-K\"ahler with respect to all three $\varepsilon$-complex
structures simultaneously. It is useful to note that the closed-ness of the three forms
$\omega_\alpha$ implies the integrability of the $\varepsilon$-complex structures
\cite{Hitchin:1986ea}. The construction of symplectic and K\"ahler quotients has 
been extended to the so-called hyper-K\"ahler quotient \cite{Hitchin:1986ea}, which can 
be adapted to para-K\"ahler manifolds. 

Hypermultiplets coupled to supergravity display another type of $\varepsilon$-quaternionic
geometry. 
An {\em $\varepsilon$-quaternionic K\"ahler manifold}
($\varepsilon$-QK manifold)
 of real dimension $4n = 4k+4l>4$ 
is a pseudo-Riemannian manifold $(N,g)$ whose holonomy group 
$\mathrm{Hol}(N)$ is contained in $Sp_\varepsilon(1) \cdot Sp_\varepsilon(k,l)$. 
An {\em $\varepsilon$-quaternionic K\"ahler manifold of real dimension 4} is an Einstein manifold
equipped with an $\varepsilon$-quaternionic structure under which 
the curvature tensor is invariant.
In this definition it is assumed implicitly that the $\varepsilon$-HK case is excluded,
that is that the holonomy group is not contained in $Sp_\varepsilon(k,l)$. Due to the
presence of the additional factor $Sp_\varepsilon(1)$, an $\varepsilon$-QK 
manifold need not admit any global $\varepsilon$-complex structure, and in particular
need not be $\varepsilon$-K\"ahler. Instead it possesses an $\varepsilon$-quaternionic
structure, that is the tangent bundle $TN$ carries a fibre-wise $\varepsilon$-quaternionic
structure, which is parallel with respect to a torsion-free connection (here: the Levi-Civita 
connection). In addition the
locally defined $\varepsilon$-complex structures $J_\alpha$ are skew with respect
to the metric $g$, and the distribution spanned by them is parallel with respect 
to the Levi-Civita connection. Note that only the distribution $\langle J_\alpha |
\alpha =1,2,3\rangle$ is invariant under parallel transport, while the individual structures
undergo $Sp_\varepsilon(1)$-transformations which mix them. Using the locally defined fundamental
forms $\omega_\alpha=g(J_\alpha \cdot, \cdot)$ one can define the four-form 
$\Lambda = \sum_{\alpha=1}^3 \omega_\alpha \wedge \omega_\alpha$, which is globally
defined and closed. It is useful to know that for manifolds of dimension $4n \geq 12$ 
the closed-ness of the  four-form $\Lambda$ already implies 
that the manifold $\varepsilon$-QK (for $\varepsilon=-1$ this is known from 
\cite{Swann:1990,Swann:1991}). The `generic' definition given 
for dimension $4n>4$ is not satisfactory for dimension 4, since $\mbox{Hol}(N) \subset 
Sp_\varepsilon(1) \cdot Sp_\varepsilon(1)$ only implies that $N$ is 
orientable: $\mathrm{Hol}(N) \subset SU(2) \cdot SU(2) \cong SO(4)$, or $\mathrm{Hol}(N) \subset
SL(2,\mathbb{R}) \cdot SL(2,\mathbb{R}) \cong SO(2,2)$. The property of the
curvature tensor used in the above definition for dimension $4n=4$ is non-trivial
and natural, since it follows for dimension $4n>4$ from the `generic' definition.

Every $\varepsilon$-QK manifold of dimension $4n$ can be obtained as the
quotient of a conical $\varepsilon$-HK manifold of dimension $4n+4$ by the
action of the invertible $\varepsilon$-quaternions $\mathbb{H}^*_\varepsilon$.
Here `conical $\varepsilon$-HK manifold' is defined analogously to 
CASR and CASK manifolds, and every such manifold defines a $\varepsilon$-QK
manifold. In the physics literature conical $\varepsilon$-HK manifolds are
usually called $\varepsilon$-HK cones, while in the mathematical literature
they are known, for $\varepsilon=-1$, as the Swann bundle associated to 
a QK manifold \cite{Swann:1990,Swann:1991}. One interesting property of $\varepsilon$-HK cones is
that they admit an $\varepsilon$-HK-potential, that is a potential which 
is an $\varepsilon$-K\"ahler potential for all three $\varepsilon$-complex
structures simultaneously. For the case $\varepsilon=-1$ we have 
encountered the HK potential $\chi$ in the context of the superconformal
construction of the four-dimensional 
Poincar\'e supergravity Lagrangian. We mention for completeness that
there also is a quotient construction which relates QK manifolds to 
QK manifolds, called quaternionic reduction or quaternionic quotient \cite{Galicki:1986ja,Galicki:1987}.

\section{Physics background}

\subsection{Non-linear sigma models and maps between manifolds \label{App:sigma_models}}

Supergravity theories with scalars involve non-linear sigma models coupled
to gravity. Sigma models are theories of massless scalars on a pseudo-Riemannian
space-time $(N,h)$, which are valued in another 
pseudo-Riemannian manifold $(M,g)$, called the target space. More precisely, 
scalar fields are components of a map
\begin{equation}
f: (N,h) \rightarrow (M,g)
\end{equation}
between two pseudo-Riemannian manifolds. When expressed in local 
coordinates $x=(x^1, \ldots, x^n)$ on $N$ and $\varphi=(\varphi^1, \ldots, 
\varphi^m)$ on $M$, scalar fields become real-valued local functions 
on space-time, which are the pull-backs to space-time of the composition of the
map $f$ with coordinate maps. In this section we explain the relation 
between the global geometrical description in terms of the map $f$ and the
local description used in the physics literature in some detail. We remark that
we do not aim for the highest degree of generality. In particular, one can 
define scalar fields more generally as sections of a pseudo-Riemannian 
submersion $\pi: P \rightarrow N$. We refer to \cite{Lazaroiu:2017qyr}
for a detailed discussion of this generalization and its potential implications. 
We also remark that both $N$ and $M$ can have various signatures, 
so that it makes sense to discuss sigma models in the general pseudo-Riemannian
set-up. Space-time $N$ has Lorentzian signature, but in the Euclidean formulation
of quantum field theories it is replaced by an `Euclidean' manifold, that is a
Riemannian manifold (pseudo-Riemannian manifold of definite signature). 
 Also, string dualities and the idea that space-time signature might be 
 dynamical in quantum gravity
 motivate the study of exotic space-times with
multiple time-like directions. Similarly, in standard cases the target manifold
has positive signature, to ensure that all scalar fields have positive signature.
However, dimensional reduction over time, which is used frequently to find
stationary solutions, sometimes leads to indefinite signature target spaces.
Moreover, supersymmetric theories on Euclidean space-times
sometimes also require target spaces of indefinite signature.

The standard action of a non-linear sigma model coupled to gravity is the sum of 
the Einstein-Hilbert action and of the energy
functional (or Dirichlet functional) for a map
between two pseudo-Riemannian manifolds $(N,h)$ and $(M,g)$,
\begin{equation}
S[h,f] = 
\int d \, \mbox{vol}_h \left( \frac{1}{2} R[h] - \langle df, df \rangle \right) \;.
\end{equation}
Here $R[h]$ and $d\,\mbox{vol}_h$ are the Ricci scalar and the volume form 
of $(N,h)$. Since we have coupled the
sigma model to gravity, the metric $h$ is a dynamical field, while the metric
$g$ is fixed and part of the definition of the model. The vector valued one-form 
$df\in \Omega^1(N,f^*TM)$  is the differential of the map $f:N\rightarrow M$, and 
$\langle \cdot , \cdot \rangle$ is the scalar product 
induced by the metrics $h$ and $g$ on the vector bundle
$T^*N \otimes f^{*} TM$ over $N$ whose fibre over $p\in N$ is 
$T^*_p N \otimes T_{f(p)}M$.

We introduce the following coordinate maps:
\begin{equation}
\begin{array}{ll}
\psi : N \supset V \rightarrow {\cal V}  \subset \bR^n  \;,\;\;\;\ & 
p \mapsto \psi(p) = (x^1(p), \ldots, x^n(p)) \;, \\
 \varphi: M \supset U  \rightarrow {\cal U}  \subset \bR^m \;,\;\;\; &
q \mapsto \varphi(q) = (\varphi^1(q), \ldots, \varphi^m(q)) \;.
\end{array}
\end{equation}
By restricting $f$ to $V$ and composing with the coordinate
maps we obtain a local representation of $f$ as a vector-valued function $\phi$,
\begin{eqnarray}
\phi = \varphi \circ f \circ \psi^{-1} \;: \mathbb{R}^n\supset {\cal V} &\rightarrow &{\cal U} 
\subset \mathbb{R}^m\;, \\
x &\mapsto& \phi(x) := (\phi^a(x^\mu)) := ( \varphi^a( f (\psi^{-1}(x^\mu)))) \;.\nonumber
\end{eqnarray}
The physical scalar fields as defined in the physics literature are 
the components $\phi^a(x)$ of the map $f:N\rightarrow M$ with respect to the
local coordinates $\{x^\mu\}, \{\varphi^a\}$.

Each of the above maps has a differential, which assigns to each point of its domain
a linear map between the tangent spaces of domain and target:
\begin{eqnarray}
df \;: p &\mapsto&  df_p \;: T_p N \rightarrow T_{f(p)} M \;,\\
d\varphi\;: q &\mapsto&  
d\varphi_q\;: T_q M \rightarrow T_{\varphi(q)} {\cal U} \;,\\
d \psi\;: p &\mapsto & d\psi_p \;: T_p N \rightarrow T_{\psi(p)} {\cal V} \;.
\end{eqnarray}
The linear maps  $d\varphi_q$ and $d\psi_p$ are invertible at all points. The differential 
of the local coordinate expression $d\phi: x \mapsto d\phi_x$ of $df$ at the point $x$ is
\[
d\phi_x   = (d\varphi \circ df \circ (d\psi)^{-1})_x   \;:
 T_x{\cal V} \cong \mathbb{R}^n \rightarrow T_{\phi(x)}{\cal U} \cong \mathbb{R}^m \;.
\]
Since $d\phi_x \in \mbox{Hom}(T_x {\cal V} , T_{\phi(x)}{\cal U}) \cong
T^*_x{\cal V} \otimes T_{\phi(x)} {\cal U}$, we interpret
$d\phi \in \Gamma({\cal V}, T^*{\cal V} \otimes \phi^* T {\cal U}) =
\Omega^1({\cal V}, \phi^{*} T{\cal U})$ as a vector-valued one-form on
${\cal V}$,
\begin{equation}
d\phi =
\frac{\partial \phi^a}{\partial x^\mu} dx^\mu \otimes \partial_a  
\in \Omega^1({\cal V}, \phi^{*} T{\cal U})
\;.
\end{equation}
The local coordinate expression for the metric $g$ restricted to $U \cong {\cal U}$ is
\begin{equation}
g_{\cal U} = g_{ab}(\varphi)  d\varphi^a d\varphi^b \;.
\end{equation}
Using that the pull-back is given by $\phi^a(x) = \varphi^a(f(\psi^{-1}(x)))$,
the corresponding expression for the pull-back metric $f^* g$ is
\[
\phi^* g = g_{ab} (\phi(x)) d\phi^a(x) d \phi^b(x) = g_{ab}(\phi(x)) 
\partial_\mu \phi^a \partial_\nu \phi^b dx^\mu dx^\nu\;.
\]

The local expression for $\langle df, df \rangle$ is 
\begin{equation}
\langle df , df \rangle = \langle d\phi, d \phi \rangle = 
h^{\mu \nu}(x) \left( g_{ab}(\phi(x))  \partial_\mu \phi^a \partial_\nu \phi^b \right)
= \mbox{tr}_h (f^* g) \;,
\end{equation}
where $\mbox{tr}_h$ is the trace defined by contraction with the metric $h$,
and where $f^*g$ is the pullback by $f$ to $N$ of the metric $g$. 
The Lagrangian ${\cal L}$ is defined by 
\begin{equation}
S = \int d\,\mbox{vo}l_h {\cal L}\;.
\end{equation}
In local coordinates it takes the form
\begin{equation}
{\cal L}_{|{\cal V}} = \frac{1}{2} R[h] - g_{ab}(\phi(x)) \partial_\mu \phi^a 
\partial^\mu \phi^b \;,
\end{equation}
and the corresponding equations of motion are
\begin{eqnarray}
&& R[h]_{\mu \nu} - \frac{1}{2} R[h] h_{\mu \nu} = T_{\mu \nu} \;,\\
&& \Delta_h \phi^a  + \Gamma^a_{\;\;bc} \partial_\mu \phi^b \partial^\mu \phi^c = 0 \;,
\end{eqnarray}
where $R[h]_{\mu \nu}$ is the Ricci tensor of $(N,h)$. We denote by 
$\Delta_h$ the pseudo-Riemannian Laplace operator
\begin{equation}
\Delta_h = \mbox{tr}_h ( D^{(h)}D^{(h)} ) = h^{\mu \nu} D^{(h)}_\mu D^{(h)}_\nu \;,
\end{equation}
where $D^{(h)}$ is the Levi-Civita connection on $(N,h)$. 
$\Gamma^a_{\;\;bc}$ are
the Christoffel symbols with respect to the Levi-Civita connection on $(M,g)$. Finally
\begin{equation}
T_{\mu \nu} := \frac{-2}{\sqrt{|\det h|}} \frac{\delta {\cal L}_{\rm Matter}}{\delta h^{\mu \nu}} 
= 2 g_{ab} \partial_\mu \phi^a \partial_\nu \phi^b 
- h_{\mu \nu} g_{ab} \partial_\rho \phi^a \partial^\rho \phi^b 
\end{equation}
is the energy momentum tensor, which is proportional to the variation of the 
matter Lagrangian
\begin{equation}
{\cal L}_{\rm Matter} =  - \sqrt{|\det(h)|} g_{ab} \partial_\mu \phi^a \partial^\mu \phi^b
\end{equation}
with respect to the metric $h$.

The coordinate-free version of the equations of motion is:
\begin{eqnarray}
&& \mbox{Ric}[h] - \frac{1}{2} R[h]  h = T \;, \;\;\mbox{where} \;
T:= 2 f^* g - \langle df, df \rangle  h \;,\\
&& \mbox{tr}_h D df = 0 \;, \;\;\; \label{harmonic-map}
\end{eqnarray} 
where $D$ is the covariant derivative on $T^* N \otimes f^* TM$ induced
by the Levi-Civita connections on $(N,h)$ and $(M,g)$.
Equation (\ref{harmonic-map}) is the equation satisfied by a harmonic
map $f: (N,h) \rightarrow (M,g)$ 
between two pseudo-Riemannian manifolds.

To obtain the local coordinate form of $\mbox{tr}_h Ddf$, we start with
$Ddf$ and evaluate it in local coordinates:
\begin{equation}
D_\mu \partial_\nu \phi^a = D^{(h)}_\mu \partial_\nu \phi^a +
\partial_\mu \phi^b \Gamma^a_{\;\;bc} \partial_\nu \phi^c\;,
\end{equation}
where $D^{(h)}$ is the Levi-Civita connection on $(N,h)$, and where
$\partial_\mu \phi^b \Gamma^a_{\;\;bc}$ is the pullback by $f$ to $N$ of the
connection coefficients $\Gamma^a_{\;\;bc}$ of the connection on $M$. 
Taking the trace using the metric $h$ we obtain
\begin{equation}
\mbox{tr}_h(Ddf) = h^{\mu \nu} \left( D^{(h)}_\mu \partial_\nu \phi^a 
+ \Gamma^a_{\;\;bc} \partial_\mu \phi^b \partial_\nu \phi^c \right) 
= \Delta_h \phi^a + \Gamma^a_{\;\;bc} \partial_\mu \phi^b \partial^\mu \phi^c\;.
\end{equation}

We remark that this expression does not require the existence of a metric 
on $M$: the metric $g$ is not used explicitly, and instead of the 
Levi-Civita connection we could use any other connection on $M$.

\subsection{Notation and Conventions \label{app:not+con}}

Our notation and conventions for space-times with Minkowski signature 
in four and in five dimensions are as follows.

We denote space-time indices by $\mu, \nu, \dots$, and local Lorentz indices by $a, b, \dots = 0, 1,2, 
\dots$. Indices $i, j, k, \dots =1,2$ are reserved for $SU(2)_R$ indices.

Our (anti)symmetrization conventions
are 
\begin{equation}
[ab] = \tfrac12 (ab - ba ) \;\;\;,\;\;\; (ab) = \tfrac12 (ab + ba) 
\end{equation}
and (c.f. \eqref{sym} and \eqref{asym})
\begin{eqnarray}
da \, db &=& \tfrac12 \left( da \otimes db + db \otimes da \right) \;, \nonumber\\
da \wedge db &=& da \otimes db - db \otimes da \:. 
\end{eqnarray}

We take the Lorentz metric $\eta_{ab}$ to have signature $(- + + \dots +)$. We denote
the vielbein by $e_{\mu}{} ^a$, and its inverse by $e_a{}^{\mu}$,
\begin{equation}
e_{\mu}{}^a \, e_a{}^{\nu} = \delta_{\mu}{}^{\nu} \;\;\;,\;\; e_a{}^{\nu} \, e_{\nu}{}^b = \delta_a{}^b \;.
\end{equation}
The space-time metric $g_{\mu \nu}$ and the Lorentz metric $\eta_{ab}$ are related by
\begin{equation}
g_{\mu \nu} = e_{\mu}{}^a \, \eta_{ab} \, e_{\nu}{}^b \;.
\end{equation}
The Christoffel symbols of the Levi-Civita connection read 
\begin{equation}
\Gamma^{\mu}{}_{\nu \rho} = \tfrac12 \, g^{\mu \lambda} \left( 2 \, \partial_{(\nu } g_{\rho) \lambda} - 
\partial_{\lambda} g_{\nu \rho} 
\right) \;.
\end{equation}
We note
\begin{equation}
\Gamma^{\rho}{}_{\rho \mu} = \tfrac12 \, g^{\lambda \rho} \partial_{\mu} \, g_{\lambda \rho} =  \tfrac12 \, \partial_{\mu} \ln | g | \;,
\end{equation}
where $g =  \det g_{\mu \nu} $.

The Riemann tensor is (c.f. \eqref{TRcomp})  
\begin{equation}
R_{\mu \nu}{}^{\rho}{}_\sigma = 2 \partial_{[ \mu} \Gamma^{\rho}{}_{\nu] \sigma} + 
\Gamma^{\rho}{}_{\mu \lambda} \, \Gamma^{\lambda}{}_{\nu \sigma} - 
\Gamma^{\rho}{}_{\nu \lambda} \, \Gamma^{\lambda}{}_{\mu \sigma} \;.
\label{riemann-gamma}
\end{equation}
We raise and lower space-time indices by contracting with the space-time metric, i.e.
\begin{equation}
R_{\mu \nu \rho \sigma} = g_{\rho \lambda} \, R_{\mu \nu}{}^{\lambda}{}_\sigma \;.
\label{Riemann_lower}
\end{equation}
The Riemann tensor satisfies the pair exchange property $R_{\mu \nu \rho \sigma} = 
R_{\rho \sigma \mu \nu}$.

We define covariant derivatives (c.f. \eqref{covD})
\begin{eqnarray}
D_{\mu} V_{\nu} &=& \partial_{\mu} V_{\nu} - \Gamma^{\lambda}{}_{\mu \nu} \, V_{\lambda} \;, 
\nonumber\\
D_{\mu} V^{\nu} &=& \partial_{\mu} V^{\nu} + \Gamma^{\nu}{}_{\mu \lambda} \, V^{\lambda} \;.
\label{LCconn}
\end{eqnarray}
We have
\begin{equation}
[ D_{\mu} , D_{\nu} ] V_{\rho} = - R_{\mu \nu}{}^{\lambda}{}_{\rho} \, V_{\lambda} \;.
\end{equation}
We define the Ricci tensor by
\begin{equation}
R_{\mu \nu} = R^{\lambda}{}_{\mu \lambda \nu} =  R_{\lambda \mu}{}^{\lambda}{}_{ \nu} =
R_{\mu}{}^{\lambda}{}_{\nu \lambda }
\;.
\label{Ricci}
\end{equation}
It satisfies the property
\begin{equation}
R_{\mu \nu} = R_{\nu \mu} \;.
\end{equation}
The Ricci scalar is
\begin{equation}
R = g^{\mu \nu} \, R_{\mu \nu} \;.
\end{equation}
With these conventions, the kinetic terms for physical fields in a gravitational action take the form (we set $\kappa^2 = 8 \pi G_ N = 1$)
\begin{equation}
\label{protlag}
L = \tfrac12 R -  \tfrac12 \partial_{\mu} \phi \partial^{\mu} \phi  - \tfrac14 F^{\mu \nu} F_{\mu \nu} \;.
\end{equation}

We define covariant derivatives of vectors $V^a $ by
\begin{equation}
{\cal D}_{\mu} V^{a} = \partial_{\mu} V^{a} + \omega_{\mu}{}^{ab} \, V_b \:,
\label{covvect}
\end{equation}
where $\omega_{\mu}{}^{ab}$ denotes the
spin connection, 
\begin{equation}
\omega_{\mu}{}^{ab} = 2 e^{\nu [ a} \, \partial_{[\mu} e_{\nu]}{}^{b]} - 
 e^{\nu [ a} \,  e^{b] \sigma} \, e_{\mu c} \partial_{\nu} e_{\rm \sigma}{}^c \;,
\label{spine}  
\end{equation}
and 
satisfies the compatibility requirement
\begin{equation}
0 = {\cal D}_{\mu} e_{\nu}{}^a = \partial_{\mu} e_{\nu}{}^a + \omega_{\mu}{}^{ab} \, e_{\nu b} - 
\Gamma^{\rho}{}_{\mu \nu} \, e_{\rho}{}^a \;.
\end{equation}
Defining
\begin{equation}
\Omega_{ab}^{\ \ c}  =  2 e_{[a}^{\ \mu} e_{b]}^{\ \nu} \partial_\mu
e_\nu^{\ c} ,  
\end{equation}
we obtain
\begin{equation}
  \omega_{a\,bc} =  \ft12 \left(\Omega_{abc} + \Omega_{cab} 
- \Omega_{bca} \right)\;.
\end{equation}
The associated Riemann tensor reads
\begin{equation}
R_{\mu \nu}{}^{ab} = 2 \, \partial_{[\mu} \omega_{\nu]}{}^{ab} + 2 \, \omega_{[\mu}{}^{ac} \, \omega_{\nu] c}{}^b \;,
\label{Rome}
\end{equation}
and it is related to the one in \eqref{riemann-gamma} by
\begin{equation}
R_{\mu \nu}{}^{\rho}{}_\sigma = R_{\mu \nu}{}^{ab} \, e_a{}^{\rho} \, e_{\sigma b} \;.
\end{equation}

We define the completely antisymmetric Levi-Civita tensor as follows. In four space-time dimensions, we 
take
\begin{eqnarray}
 \varepsilon^{\mu \nu \lambda \sigma} &=& e \, \varepsilon^{abcd} \, e_a{}^{\mu} 
 \, e_b{}^{\nu} \, e_c{}^{\lambda} \, e_d{}^{\sigma}  \;\;\;,\;\;\; \varepsilon_{0123} = 1\;, \nonumber\\
  \varepsilon_{\mu \nu \lambda \sigma} &=& e^{-1} \, \varepsilon_{abcd} 
  \, e_{\mu}{}^a \, e_{\nu}{}^b
\, e_{\lambda}{}^c \, e_{\sigma}{}^d  \;,
 \label{defvareps}
\end{eqnarray}
where $e^{-1} = | g|^{-1/2} $. Similarly,  in five space-time dimensions we take
\begin{eqnarray}
 \varepsilon^{\mu \nu \lambda \sigma \rho} &=& e \, \varepsilon^{abcde} \, e_a{}^{\mu} 
 \, e_b{}^{\nu} \, e_c{}^{\lambda} \, e_d{}^{\sigma}  \, e_e{}^{\rho} 
  \;\;\;,\;\;\; \varepsilon_{01235} = 1\;, \nonumber\\
  \varepsilon_{\mu \nu \lambda \sigma \rho } &=& e^{-1} \, \varepsilon_{abcde} 
  \, e_{\mu}{}^a \, e_{\nu}{}^b
\, e_{\lambda}{}^c \, e_{\sigma}{}^d \, e_{\rho}{}^e  \;,
 \label{defvareps5d}
\end{eqnarray}
where $e^{-1} = | g|^{-1/2} $.

In four dimensions, 
we define
the dual of an antisymmetric tensor field $F_{ab}$ by
\begin{equation}
{\tilde F}_{ab} = - \tfrac{i}{2}   \, \varepsilon_{abcd} \, F^{cd} \;.
\label{dualF}
\end{equation}
We denote the selfdual part of $F_{ab}$ by $F^+_{ab}$, and the anti-selfdual part by $F^-_{ab}$,
\begin{equation}
F^{\pm}_{ab} = \tfrac12 \left( F_{ab} \pm {\tilde F}_{ab} \right) \;.
\end{equation}  

In four dimensions, in the context of 
${\cal N}=2$ special geometry, we will encounter the $SU(2)_R$
valued selfdual tensor field $T_{ab ij}$
and the  $SU(2)_R$ valued anti-selfdual tensor field $T_{ab}^{ij}$. Accordingly, we introduce the
notation
\begin{eqnarray}
T_{ab}^+ &=& \tfrac12 \, \varepsilon^{ij} \, T_{ab ij} \;, \nonumber\\
T_{ab}^- &=&  \tfrac12 \, \varepsilon_{ij} \, T_{ab}^{ ij} \;,
\label{def-T-}
\end{eqnarray}
where the Levi-Civita symbol $\varepsilon_{ij} = - \varepsilon_{ji}$ satisfies
\begin{equation}
\varepsilon_{ij} \varepsilon^{jk} = - \delta_i{}^k \;,
\end{equation}
with 
\begin{equation}
\varepsilon_{12} = \varepsilon^{12} = 1 
\end{equation}
and
$\varepsilon_{ij} \varepsilon^{ji} = -2 $.
Under Hermitian conjugation (${\rm h.c.}$), selfdual becomes 
anti-selfdual
and vice-versa. Any $SU(2)_R$ index $i$ 
changes position under ${\rm h.c.}$, for instance
\beq (T_{ab\, ij})^* = T_{ab}^{ij} \,.
\eeq

\subsection{Jacobians \label{jacob_conv}}

The Jacobians for the coordinate transformations \eqref{conv_coordinv} take the form
\begin{eqnarray}
\frac{D(x,u,\Upsilon, \bar{\Upsilon})}{D(x,y,\Upsilon,\bar{\Upsilon})} =
\left( \begin{array}{cccc}
\mathbbm{1} & 0 & 0 & 0 \\
\left. \frac{\partial u}{\partial x} \right|_y &
\left. \frac{\partial u}{\partial y} \right|_x &
\left. \frac{\partial u}{\partial \Upsilon} \right|_{x,y} &
\left. \frac{\partial u}{\partial \bar{\Upsilon}} \right|_{x,y} \\
0 & 0 & \mathbbm{1} & 0 \\
0 & 0 & 0 & \mathbbm{1} \\
\end{array} \right)
\end{eqnarray}
and
\begin{eqnarray}
\frac{D(x,y,\Upsilon, \bar{\Upsilon})}{D(x,u,\Upsilon,\bar{\Upsilon})} =
\left( \begin{array}{cccc}
\mathbbm{1} & 0 & 0 & 0 \\
\left. \frac{\partial y}{\partial x} \right|_u &
\left. \frac{\partial y}{\partial u} \right|_x &
\left. \frac{\partial y}{\partial \Upsilon} \right|_{x,u} &
\left. \frac{\partial y}{\partial \bar{\Upsilon}} \right|_{x,u} \\
0 & 0 & \mathbbm{1} & 0 \\
0 & 0 & 0 & \mathbbm{1} \\
\end{array} \right) \;.
\end{eqnarray}
By the chain rule it is straightforward to evaluate
\begin{eqnarray}
\label{jachol}
\frac{D(x,y,\Upsilon,\bar{\Upsilon})}{D(x,u,\Upsilon,\bar{\Upsilon})} =
\left( \begin{array}{cccc}
\mathbbm{1}  & 0 & 0 & 0 \\
\frac{1}{2} R \; & \;  - \frac{1}{2} N \; & \; \frac{1}{2} F_{I\Upsilon} \; &
\; \frac{1}{2} \bar{F}_{I\Upsilon} \\
0 & 0 & \mathbbm{1} & 0 \\
0 & 0 & 0 & \mathbbm{1} \\
\end{array} \right) \;,
\end{eqnarray}
where $2 F_{IJ} = R_{IJ} + i N_{IJ}$. 
This matrix can easily be inverted, 
\begin{equation}
\frac{D(x,u,\Upsilon,\bar{\Upsilon})}{D(x,y,\Upsilon,\bar{\Upsilon})} =
\left( \begin{array}{cccc}
\mathbbm{1} & 0 & 0 & 0 \\
N^{-1} R \; &\;  -2 N^{-1} \; & \; N^{-1} F_{I\Upsilon} \; & \; N^{-1} \bar{F}_{I\Upsilon} \\
0 & 0 & \mathbbm{1} & 0 \\
0 & 0 & 0 & \mathbbm{1} \\
\end{array} \right) \;.
\label{jacxuxyups}
\end{equation}
In order to transform the K\"ahler metric \eqref{kahgups}
to special real
coordinates (c.f. \eqref{gwithups}), the following relations are useful,
\begin{equation}
\frac{\partial H}{\partial x^I} = 2 v_I \;,\;\;\;
\frac{\partial H}{\partial y_I} = - 2 u^I \;.
\end{equation}
Moreover, using the chain rule, one computes
\begin{eqnarray}
\left. \frac{\partial v_I}{\partial x^J}\right|_y  &=&
\frac{1}{2} \left( N + RN^{-1}R \right)_{IJ} \;, \nonumber\\
\left. \frac{\partial v_I}{\partial y_J}\right|_x  &=& \left. - \frac{\partial u^J}{\partial x^I}\right|_y = 
2 \left( N^{-1} \right)^{IJ} \;, \nonumber\\
\left. \frac{\partial v_I}{\partial u^J}\right|_x &=& \frac{1}{2} R_{IJ} \;. 
\end{eqnarray}

The Jacobians for the coordinate transformations \eqref{XFqnonh} are given by
\begin{equation}
\label{jacnonh}
\frac{D(x,y,\Upsilon, \bar{\Upsilon})}{D(x,u,\Upsilon,\bar{\Upsilon})} =
\left( \begin{array}{cccc}
\mathbbm{1} & 0 & 0 & 0 \\
\frac{1}{2} R_+ \; & \; - \frac{1}{2} N_- \; & \; \frac{1}{2} (F_{I\Upsilon} + 
\bar{F}_{\bar{I}\Upsilon}) \; & \;  \frac{1}{2} ( \bar{F}_{\bar{I} \bar{\Upsilon}}
+ F_{I\bar{\Upsilon}}) \\
0 & 0 & \mathbbm{1} & 0 \\
0 & 0 & 0 & \mathbbm{1} \\
\end{array} \right)
\end{equation}
and
\begin{equation}
\label{acnonhinv}
\frac{D(x,u,\Upsilon,\bar{\Upsilon})}{D(x,y,\Upsilon, \bar{\Upsilon})} =
\left( \begin{array}{cccc}
\mathbbm{1} & 0 & 0 & 0 \\
N_-^{-1}  R_+ \; & \;  - 2 N_-^{-1}  \; & \;  N_-^{-1} (F_{I\Upsilon} + 
\bar{F}_{\bar{I}\Upsilon}) \;  & \; N_-^{-1} ( \bar{F}_{\bar{I} \bar{\Upsilon}}
+ F_{I\bar{\Upsilon}}) \\
0 & 0 & \mathbbm{1} & 0 \\
0 & 0 & 0 & \mathbbm{1} \\
\end{array} \right) \;.
\end{equation}
This reduces to the results for the Jacobians \eqref{jachol} and \eqref{jacxuxyups}
when switching off the non-holomorphic
deformation.



\subsection{Superconformal formalism in four dimensions  \label{sec:4dscg}} 

The idea behind the superconformal approach to supergravity consists in using
the superconformal symmetry as a powerful tool for constructing matter-coupled
theories with local Poincar\'e supersymmetry, and in doing so to gain insights into
the structure of Poincar\'e supergravity \cite{deWit:1979dzm,Bergshoeff:1980is,deWit:1980lyi,deWit:1984wbb,deWit:1984rvr}.
We refer to \cite{Freedman:2012zz}
for a recent detailed discussion.

To illustrate this construction, we begin by reviewing the formulation of Einstein gravity in four dimensions
based on the bosonic conformal algebra.

\subsubsection{Gravity as a conformal gauge theory}

Consider the following action in four dimensions,
\begin{equation}
S = \int d^4 x \, \sqrt{-g} \left( \tfrac12 \, (\partial_{\mu} \phi) (\partial^{\mu} \phi) + \tfrac{1}{12} \, R \, \phi^2
\right) \;,
\label{confgrav}
\end{equation}
where $\phi(x)$ denotes a real scalar field. Note that the sign of the kinetic energy term of
the scalar field is opposite from the one of a physical scalar field, c.f. \eqref{protlag}. This Lagrangian is invariant
under local scale transformations, also called local dilatations or local Weyl transformations, given by
\begin{equation}
\phi \mapsto e^{\lambda_D} \, \phi \;\;\;,\;\; g_{\mu \nu} \mapsto e^{-2 \lambda_D } \,
g_{\mu \nu} \;.
\end{equation}
Here, $\lambda_D (x)$ denotes the local parameter of Weyl transformations. 

The field $\phi$ 
is called a compensating field (or, compensator), because it compensates for the non-invariance of the Einstein-Hilbert term under local scale transformations 
caused  by the transformation properties of the metric,  thus resulting in a Weyl-invariant action. 
We can eliminate
the compensating field $\phi$ by performing the gauge-fixing
\begin{equation}
\phi \mapsto  e^{\lambda_D} \, \phi \equiv \frac{\sqrt{6}}{\kappa^2} \;.
\end{equation}
Inserting this into \eqref{confgrav} results in the Einstein-Hilbert action, 
\begin{equation}
S = \frac{1}{2 \kappa^2} \, \int d^4 x \, \sqrt{-g} \, R  \;.
\label{EH_action}
\end{equation}
Here $\kappa^{2} = 8 \pi G_N$, where $G_N$  denotes the Newton's constant. 
Thus, the Einstein-Hilbert action can be obtained by starting from an action that possesses invariance under
local scale transformations due to the presence of a compensating field, and then eliminating the compensating
field by going to a particular Weyl gauge. That is, Einstein gravity emerges from a theory that is initially
invariant under transformations associated with the generators of the bosonic conformal algebra. We review
this algebra next.

\subsubsection{The bosonic conformal algebra }

The bosonic conformal algebra in four dimensions is isomorphic to $so(4,2)$, and contains generators $P_a, M_{ab},K_a, D$
associated with translations, Lorentz transformations, special conformal transformations and dilations, respectively.
These generators satisfy the algebra (we only give the commutators that are non-vanishing)
\begin{eqnarray}
\big[M_{ab}, M_{cd} \big] &=& 4 \, \eta_{[a[c} M_{d]b]} = 
\eta_{ac} M_{db} - \eta_{bc} M_{da} - \eta_{ad} M_{cb} + \eta_{b d} M_{ca} \;,
 \nonumber\\
\big[P_a, M_{bc}\big] &=& 2 \, \eta_{a [b} P_{c]} \;, \nonumber\\
\big[K_a, M_{bc}\big] &=& 2 \, \eta_{a [b} K_{c]} \;, \nonumber\\
\big[ P_a, K_b \big] &=& 2 \left( \eta_{ab} \, D + M_{ab} \right) \;, \nonumber\\
\big[ D, P_a \big] &=& P_a \;,\nonumber\\
\big[ D, K_a \big] &=& -K _a \;.
\end{eqnarray}
To each of these generators, we assign a local parameter, as well as a gauge field. This is summarized
in the Table \ref{conf_gen_fie} below.

\begin{table}[h]
\begin{center}
\begin{tabular}{|c||cccc||cc|}
\hline
generator        &
   $P_a$         &
   $M_{ab}$      &
   $K_a$     &
   $D$   \\
   \hline
\hline 
parameter       & $\xi^a$ & $\lambda^{ab}$  & $\lambda_K^a$ & $\lambda_D$ \\[.5mm]
\hline
gauge field        & $e_{\mu}{}^a$ & $\omega_{\mu}^{ab}$ & $f_{\mu}{}^a$ & $b_{\mu}$
            \\[.5mm]
            \hline
Weyl weight $w$     & $-1$ & $0$ & $1$ & $0$
 \\[.5mm]
\hline
\end{tabular}
\vskip 5mm
\caption{The bosonic conformal algebra: generators, local parameters, gauge fields, Weyl weights. \label{conf_gen_fie}}
\end{center}
\end{table}

The translations $P_a$, which are gauged by $e_{\mu}^a$, play a special rule, and will be considered
separately. Under infinitesimal conformal transformations generated by  $M_{ab},K_a, D$, the gauge
fields transform as follows,
\begin{eqnarray}
\delta e_{\mu}{}^a &=& - \lambda^{ab} \, e_{\mu b} - \lambda_D \, e_{\mu}{}^a \;, \nonumber\\
\delta \omega_{\mu}^{ab} &=& \partial_{\mu} \lambda^{ab} + 2 \omega_{\mu c}{}^{[a} \, \lambda^{b]c} - 4 \lambda_K^{[a} \,
e_{\mu}{}^{b]} \;, \nonumber\\
\delta f_{\mu}{}^a &=&   \partial_{\mu} \lambda_K^a - b_{\mu} \, \lambda_K^a + \omega_{\mu}{}^{ab} \, \lambda_{K b}
- \lambda^{ab} \, f_{\mu b} + \lambda_D \, f_{\mu}{}^a \;, \nonumber\\
\delta b_{\mu} &=& \partial_{\mu} \lambda_D + 2 \lambda_K^a \, e_{\mu a} \;.
\label{var-gauge}
\end{eqnarray}
The commutators of two transformations \eqref{var-gauge} yield a realization of the conformal algebra.
The transformation behaviour under dilatations is specified by the Weyl weight $w$ of each of the gauge fields.
The vielbein has weight $w = -1$, the field $f_{\mu}{}^a$ has weight $w=1$, while the other gauge fields have
$w=0$. Note that all the gauge fields, with the exception of the vielbein, transform under special conformal
transformations.

Next, we introduce a field strength for each of the generators of the conformal algebra.
These field strengths, of the form $R_{\mu \nu}{}^A$, transform covariantly under
conformal transformations.  They are given by
\begin{eqnarray}
R_{\mu \nu}{}^a (P) &=& 2 \left(\partial_{[\mu} + b_{[\mu} \right) e_{\nu]}{}^a + 2 \omega_{[\mu}{}^{ab} 
\, e_{\nu] b} = 2 \, {\cal D}_{[\mu} e_{\nu]}{}^a \;, \nonumber\\
R_{\mu \nu}{}^{ab} (M) &=& 2 \partial_{[\mu} \omega_{\nu]}{}^{ab} + 2  \omega_{[\mu}{}^a{}_c \,
 \omega_{\nu]}{}^{cb}+ 8 f_{[\mu}{}^{[a} \, e_{\nu]}{}^{b]} \;, \nonumber\\
R_{\mu \nu}{}^a (K) &=& 2 \left( \partial_{[\mu} - b_{[\mu} \right) f_{\nu]}{}^a + 2 \omega_{[\mu}{}^{ab} \,
f_{\nu ] b} \;, \nonumber\\
R_{\mu \nu} (D) &=& 2 \partial_{[\mu} b_{\nu]} - 4  f_{[\mu}{}^{a} \, e_{\nu] a } \;.
\end{eqnarray}
It can checked that these field strengths transform covariantly under the transformations 
\eqref{var-gauge}.

Since we are interested in the construction of Einstein gravity as a gauge fixed version 
of a gravitational theory that is invariant under conformal transformations, not all of 
the gauge fields associated to the conformal algebra can describe independent gauge fields.
To be able to identify the translations generated by $P_a$ with space-time diffeomorphisms, one needs to impose
a constraint on the associated field strength $R_{\mu \nu}{}^a (P)$, so as to ensure that the
translation gauge field  $e_{\mu}^a$ becomes a vielbein field (frame) over space-time.
In addition, the  gauge field $f_{\mu}{}^a$ for special conformal transformations needs to be eliminated
as an independent gauge field. This is achieved by 
imposing the constraints 
\begin{eqnarray}
R_{\mu \nu}{}^a (P) &=&0 \;, \nonumber\\
R_{\mu \nu}{}^{ab} (M) \, e_{b}{}^{\nu} &=& 0 \;.
\label{curvature_constraints}
\end{eqnarray}
In this way, 
two of the gauge fields,
namely the spin connection $\omega_{\mu}^{ab}$ and 
the gauge field $f_{\mu}{}^a$ for special conformal transformations, become composite fields,
\begin{eqnarray}
 \omega_{\mu}^{ab} &=& \omega_{\mu}^{ab} (e) + 2 e_{\mu}{}^{[a} \, e^{b] \nu} \, b_{\nu} \;, \nonumber\\
 f_{\mu}{}^a &=& - \tfrac14 \, R_{\mu}{}^a + \tfrac{1}{24} e_{\mu}{}^a \, R \;,
\label{om-b-f}
\end{eqnarray}
with $\omega_{\mu}^{ab} (e)$ given by \eqref{spine}. The constraint $R_{\mu \nu}{}^a (P) = 2 \, {\cal D}_{[\mu} e_{\nu]}{}^a =0$
is the condition for metric compatibility, but now in the presence of the dilational connection $b_{\mu}$.
Note that the Riemann tensor computed from the
spin connection \eqref{om-b-f} does not  have the pair exchange property mentioned
below \eqref{Riemann_lower}.
To obtain the relation for $f_{\mu}{}^a$, we expressed the second constraint in \eqref{curvature_constraints}
as
\begin{equation}
\left( R_{\mu \nu}{}^{ab} + 8 f_{[\mu}{}^{[a} \, e_{\nu]}{}^{b]} \right) e_{b}{}^{\nu} =0 \;,
\end{equation}
where $R_{\mu \nu}{}^{ab}$ is the Riemann tensor constructed out of the spin connection 
$ \omega_{\mu}^{ab}$ that also contains the gauge field $b_{\mu}$, c.f. \eqref{om-b-f}.
Then, using the definitions for the Ricci tensor\footnote{We use the last equation given
in \eqref{Ricci}.} and the Ricci scalar,
\begin{equation}
R_{\mu \nu} = R_{\mu \rho \nu }{}^{\rho}{} \;\;\;,\;\;\; R = g^{\mu \nu} \, R_{\mu \nu} \;,
\label{defRR}
\end{equation}
we obtain
\begin{equation}
R_{\mu }{}^{a} + 2 \left(2 f_{\mu}{}^a +  f_{\nu}{}^{\nu} \, e_{\mu}{}^a \right)= 0 \;,
\label{Rf1}
\end{equation}
where $ f_{\nu}{}^{\nu} =  f_{\nu}{}^{a}\, e_a{}^{\nu}$.
Contracting this relation with $e_a{}^{\mu}$ gives
\begin{equation}
 f_{\mu}{}^{\mu} = - \tfrac{1}{12} \, R \;,
 \end{equation}
Inserting this into \eqref{Rf1} gives the relation in \eqref{om-b-f}.

As a check of \eqref{om-b-f}, one verifies that when inserting
the transformation law for $e_{\mu}{}^a$ and for $b_{\mu}$ into 
\eqref{om-b-f}, one correctly reproduces the transformation laws for $\omega_{\mu}^{ab} $ and
$f_{\mu}{}^{a} $ given in \eqref{var-gauge}.

Upon imposing the constraints \eqref{curvature_constraints},
the independent gauge fields in \eqref{om-b-f} are the vielbein $e_{\mu}{}^a$ and the gauge field for dilations
$b_{\mu}$. Inspection of the transformation law for the field $b_{\mu}$ given in \eqref{var-gauge}
shows that the value of $b_{\mu}$ can be arbitrarily changed by performing a special conformal
transformation.  Therefore, we fix $b_{\mu}$ to the value
\begin{equation}
b_{\mu} =0 \;\;\;,\;\;\; {\rm K-gauge} \;,
\label{b0K}
\end{equation}
by means of a special conformal transformation. Since this represents a gauge-fixing of
special conformal transformations with gauge parameter $\lambda_{K \mu} \equiv \lambda_K^a \, 
e_a{}^{\mu}$, this is called the K-gauge.
In this gauge, special conformal transformations are no longer independent transformations.
Inspection of \eqref{var-gauge} shows that in order to stay in the K-gauge \eqref{b0K},
the allowed residual special conformal transformations are
\begin{equation}
\lambda_{K  \mu} = - \tfrac12 \, \partial_{\mu} \lambda_D \;.
\end{equation}

\subsubsection{Weyl multiplet}

The extension of the above to supergravity is called the superconformal
approach to supergravity \cite{deWit:1979dzm,Bergshoeff:1980is,deWit:1980lyi,deWit:1984wbb,deWit:1984rvr}.
The standard superconformal approach to ${\cal N}=2$ supergravity in four dimensions is based
on the Weyl multiplet. In its standard formulation, 
the Weyl multiplet is a supermultiplet with 24+24 bosonic and fermionic off-shell
degrees of freedom.\footnote{Recently, a new Weyl multiplet was constructed in \cite{Butter:2017pbp}, called the
dilaton Weyl multiplet, with $24+24$ off-shell degrees of freedom.}
Let us briefly describe this multiplet.

The ${\cal N}=2$ superconformal algebra contains the following bosonic generators:
it contains
the bosonic conformal algebra discussed in the previous
subsection as well as two bosonic generators $T$ and $U_i{}^j$ that generate
$U(1)_R$ and $SU(2)_R$ R-symmetry transformations, respectively. As before, we assign a local parameter and a
gauge field to each of these bosonic generators. The 
gauge fields associated with $U(1)_R$ and $SU(2)_R$ R-symmetry transformations will be 
denoted by $(A_{\mu}, {\cal V}_{\mu}{}^i{}_{j})$.
This is summarized in Table \ref{conf_gen_fie_N2} below.\\

\begin{table}[h]
\begin{center}
\begin{tabular}{|c||cccccc|}
\hline
bosonic generator        &
   $P_a$         &
   $M_{ab}$      &
   $K_a$     &
   $D$ &
   $T$
   &
   $U^j{}_i$ 
     \\
   \hline
\hline 
parameter       & $\xi^a$ & $\lambda^{ab}$  & $\lambda_K^a$ & $\lambda_D$ & $\lambda_T$ & $\lambda{}^i{}_j $ \\[.5mm]
\hline
gauge field        & $e_{\mu}{}^a$ & $\omega_{\mu}^{ab}$ & $f_{\mu}{}^a$ & $b_{\mu}$ &  $A_{\mu}$ & 
${\cal V}_{\mu}{}^i{}_{ j}$
            \\[.5mm]
\hline
\end{tabular}
\vskip 5mm
\caption{The ${\cal N}=2$ bosonic subalgebra: generators, local parameters, gauge fields.
\label{conf_gen_fie_N2}}
\end{center}
\end{table}

The bosonic components of the Weyl multiplet are given by the gauge fields displayed in table \ref{conf_gen_fie_N2},
together with a complex anti-selfdual tensor field $T_{ab}^-$ and a real scalar field $D$:
\begin{equation}
(e_{\mu}{}^a,
\omega_{\mu}^{ab}, f_{\mu}{}^a, b_{\mu}, A_{\mu}, {\cal V}_{\mu}{}^i{}_{j}, T_{ab}^{-}, D) \;.
\label{weylcomp}
\end{equation}
These describe $24$ independent bosonic degrees of freedom, as depicted in Table \ref{tabledof}.

\begin{table}[h]
\begin{center}
\begin{tabular}{|c||c||c|}
\hline
field & subtraction by gauge transformations & number of degrees of freedom left \\[.5mm]
\hline
\hline
$e_{\mu}{}^a$        & $P_a, M_{ab}, D$ & $16 -(4 + 6 + 1) = 5 $ \\[.5mm]
\hline
$\omega_{\mu}{}^{ab}$ & &  composite field         \\[.5mm]
\hline
$f_{\mu}{}^a$ &  & composite field \\ [.5mm]
\hline
$b_{\mu}$       & $K_a$& 0
 \\[.5mm]
 \hline
$ A_{\mu} $  & $U(1)_R$ & $4 - 1 = 3$ \\ [.5mm]
\hline
${\cal V}_{\mu}{}^i{}_{j} $  & $SU(2)_R$ &  $12 - 3 = 9$  \\[.5mm]
\hline
$ T_{ab}^{-} $ &  & 6 \\ [.5mm]
\hline
$ D $  &  & 1 \\ [.5mm]
\hline
\end{tabular}\\ [.13in]
\caption{Counting of bosonic off-shell degrees of freedom: $5 + 9 + 3 + 6 + 1 =24$.
}
\label{tabledof}
\end{center}
\end{table}

The component fields of the Weyl multiplet carry a Weyl weight $w$ and a chiral ($U(1)_R$) weight $c$.
This is summarized for the bosonic components in table \ref{tablewcWeyl}. \\

\begin{table}[h]
\begin{center}
\begin{tabular}{|c||cccccc|cc|}
\hline
field          &
   $e_\mu{}^a$   &
    $b_\mu$       &
   $A_\mu$       &
   ${\cal V}_{\mu}{}^i{}_{ j}$ &
   $T_{ab}^-$    &
   $D$           &
   $\omega_\mu{}^{ab}$ &
   $f_\mu{}^a$  
\\[.5mm]
\hline
\hline     
$w$         & $-1$    
& $0$      & $0$      & $0$
  & $1$      
   & $2$      & $0$       & $1$    
 \\[.5mm]
 \hline
$c$         & $0$    
 & $0$      & $0$      & $0$
  & $-1$    
  & $0$      & $0$       & $0$     
\\[.5mm]
\hline
\end{tabular}\\[.13in]
\caption{Weyl and chiral weights
($w$ and $c$, respectively)
                of the Weyl multiplet bosonic component fields. 
             \label{tablewcWeyl}}
\label{weyl}
\end{center}
\end{table}

As indicated in table \ref{tablewcWeyl}, 
the gauge fields  $\omega_\mu{}^{ab}$ 
and $f_\mu{}^a$ are composite fields. Their expressions are obtained by imposing 
constraints, as in \eqref{curvature_constraints}. While we still impose 
$R_{\mu \nu}{}^a (P) =0$, which results in the expression for the spin connection given in \eqref{om-b-f},
we impose the following constraint on the curvature $ R_{\mu \nu}{}^{ab} (M) $, taking into account that there are
additional fields in the Weyl multiplet,\footnote{Note that there
are additional fermionic terms in this expression which we have suppressed.}
\begin{equation}
R_{a c}{}^{bc} (M) + i {\tilde R}_a{}^b (T) - \tfrac14 T_{ac}^- \, T^{+ bc} + \tfrac32 \, \delta_a{}^b D  = 0  \;,
\label{curvature_constraint-f}
\end{equation}
where ${\tilde R}^{ab} (T)$ denotes the dual of the $U(1)_R$ field strength $R_{ab} (T)$, 
c.f. \eqref{dualF},
where
$R_{ab} (T) = e_a{}^{\mu} e_b{}^{\nu}
\, R_{\mu \nu} (T)$ with
\begin{equation}
R_{\mu \nu}(T) = 2 \partial_{[\mu} A_{\nu]} \;.
\end{equation}
Note that all the terms  in the linear combination \eqref{curvature_constraint-f}
have Weyl weight $2$.

The constraint \eqref{curvature_constraint-f} results in 
\begin{equation}
R_{a}{}^{b} + 2 \left(2 f_a{}^b + f_c{}^c \, \delta_a{}^b \right)
+ i {\tilde R}_a{}^b (T) - \tfrac14 T_{ac}^- \, T^{+ bc} + \tfrac32 \, \delta_a{}^b D  = 0 
 \;,
 \label{f2}
\end{equation}
where $R_a{}^b = R_{ac}{}^{bc}$, with 
$R_{\mu \nu}{}^{ab}$ the Riemann tensor constructed out of the spin connection 
$ \omega_{\mu}^{ab}$ that also contains the gauge field $b_{\mu}$, c.f. \eqref{om-b-f}.
Then, contracting \eqref{f2} gives 
\begin{equation}
R + 12 f_a{}^a + 6 D =0 \;,
 \label{f3}
\end{equation}
where $R = R_a{}^a$. Therefore, we infer
\begin{equation}
 f_{a}{}^{a} = - \tfrac{1}{12} \, R - \tfrac12 \, D \;.
 \end{equation}
Inserting this into \eqref{f2} gives the relation 
\begin{eqnarray}
f_{\mu}{}^a = \tfrac12 \left(-
\tfrac12 R_{ \mu}{}^a - \tfrac14 \left(D - \tfrac13 \, R \right) e_{\mu}{}^a - \tfrac12 i 
{\tilde R}_{ \mu}{}^a (T) + \tfrac{1}{8} T_{ \mu b}^{-} T^{ab +}  \right) \;.
\label{f-FP4} 
\end{eqnarray}


\subsubsection{Covariant derivatives}

In the superconformal approach one introduces covariant derivatives  ${\cal D}_{\mu}$ and $D_{\mu}$.
The first one, 
 ${\cal D}_{\mu}$,  denotes a covariant derivative with respect to Lorentz, dilatations,
$U(1)_R$ and $SU(2)_R$ transformations.
The second one, 
$D_{\mu}$, denotes a covariant derivative
with respect to these transformations as well as with 
respect to special conformal transformations,\footnote{Here, $D_{\mu}$ should not be confused with the
Levi-Civita connection \eqref{LCconn}.}
and it is used to construct actions
that are invariant under superconformal transformations.
Let us illustrate this.

Consider a scalar field $\phi$ with Weyl weight $w$ and chiral weight $c$. It transforms
as 
\begin{eqnarray}
\delta_D \phi &=& w \, \lambda_D \, \phi \;,\nonumber\\
\delta_T \phi &=& i c \, \lambda_T \, \phi
\end{eqnarray}
under infinitesimal dilatational and $U(1)_R$ transformations. 
The associated covariant derivative of $\phi$ is
\begin{equation}
{\cal D}_{\mu}Ê\phi = \left( \partial_{\mu} - w \, b_{\mu} - i c \, A_{\mu} \right) \phi \;.
\label{covphi}
\end{equation}
Note that $D_{\mu} \phi = {\cal D}_{\mu} \phi$.
Since the dilational
connection $b_{\mu}$ transforms as in \eqref{var-gauge} under special conformal transformations, 
${\cal D}^{a} \phi$ undergoes a $K$-transformation, 
\begin{equation}
\delta_K {\cal D}^{a} \phi = - 2 w \, \lambda_K^a \phi \;,
\end{equation}
that needs to be compensated for when constructing an invariant action. To this end, consider
evaluating 
\begin{equation}
D_{\mu} D^{a} \phi = D_{\mu} {\cal D}^{a} \phi = 
 {\cal D}_{\mu} {\cal D}^{a} \phi + 2 w \, f_{\mu}{}^a \, \phi \;,
\end{equation}
where
\begin{equation}
{\cal D}_{\mu} {\cal D}^{a} \phi 
= \partial_{\mu} {\cal D}^a \phi - (w +1) b_{\mu} {\cal D}^a \phi
- i c \, A_{\mu} {\cal D}^a \phi + \omega_{\mu}{}^{ab}  {\cal D}_b \phi \;.
\end{equation}
Here we used that the covariant derivative ${\cal D}_{\mu}$ of a vector $V^a$ of Weyl weight $w$ and chiral weight $c$ is
\begin{equation}
{\cal D}_{\mu} V^a = \partial_{\mu} V^a - w \, b_{\mu} V^a - i c \, A_{\mu} V^a + \omega_{\mu}{}^{ab} V_b \;,
\end{equation}
c.f.  \eqref{covvect}. Then, under $K$-transformations, $D_{\mu} D^{\mu} \phi$ transforms as 
\begin{equation}
\delta_K \left( D_{\mu} D^{\mu} \phi \right) = 4( 1 - w) \lambda_K^a \, {\cal D}_a \phi \;.
\end{equation}
Choosing $w=1$ renders $D_{\mu} D^{\mu} \phi$ invariant under $K$-transformations. Then, the quantity $e \phi D_{\mu} D^{\mu} \phi$, which
has Weyl weight zero, is invariant under both $K$-transformations and under local dilations.  It can thus be used
as a Lagrangian that is invariant under the transformations associated with the bosonic conformal algebra discussed earlier.
It contains the term $\phi^2 \, f_{\mu}{}^{\mu} \propto \phi^2 R$, as in \eqref{confgrav}.

Similarly, consider evaluating $D_{\mu} D_{c} T_{ab}^+$, where $T_{ab}^+$ has Weyl and
chiral weights $w = c = 1$, so that
\begin{equation}
{\cal D}_{\nu}   T^{ab+} = \left(\partial_{\nu} - b_{\nu} - i A_{\nu} \right) T^{ab+}
+ \omega_{\nu}{}^{ad} T_d{}^{b+} + \omega_{\nu}{}^{bd} T^a{}_d{}^{+} \;.
\end{equation}
Taking into account that both $b_{\nu}$ and $\omega_{\nu}{}^{ab}$ transform under
$K$-transformations, c.f. \eqref{var-gauge},
we infer
\begin{equation}
\delta_K {\cal D}_{\nu}   T^{ab+} = - 2 \lambda_{K \nu} \,  T^{ab+} - 4 \lambda_K^{[a} \, e^{d]}{}_{\nu} \,
T_d{}^{b +} - 4 \lambda_K^{[b} \, e^{d]}{}_{\nu} \,
T^a{}_d{}^{+}  \;.
\end{equation}
This needs to be compensated for in $D_{\mu} D_{\nu} T_{ab}^+$,
\begin{eqnarray}
D_{\mu} D_{c} T_{ab}^+ &=& {\cal D}_{\mu} {\cal D}_{c} T_{ab}^+
+ 2 f_{\mu c} T_{ab}^+  + 4 f_{\mu}{}^{[a} \, \delta^{d]}_c \, T_d{}^{b +} + 
4 f_{\mu}{}^{[b} \, \delta^{d]}_c \, T^a{}_d{}^{+} \nonumber\\
&=&  {\cal D}_{\mu} {\cal D}_{c} T_{ab}^+ + 2 f_{\mu c} T_{ab}^+  -4
f_{\mu [a} \,T_{b] c}^+ + 4 f_{\mu}{}^{d} \, \eta_{c [ a} T_{b] d}^+  \;.
\end{eqnarray}
Hence
\begin{equation}
D_{\mu} D^c T_{cb}^+ = {\cal D}_{\mu} {\cal D}^c T_{cb}^+ - 2  f_{\mu}{}^c T_{cb}^+ \;.
\end{equation}
It follows that
\begin{equation}
T^{ab -} D_{a} D^c T_{cb}^+ = T^{ab -} {\cal D}_{a} {\cal D}^c T_{cb}^+ - 2 f_a{}^c \, T^{ab -} T_{cb}^+
\;.
\label{TboxT}
\end{equation}
This relation will be used in the main text.

\subsubsection{Vector multiplets}

The field content of a four-dimensional abelian vector multiplet is
given by  a complex scalar field $X$, an abelian gauge field\footnote{Not to be confused with the $U(1)$ gauge field in the Weyl multiplet 
\eqref{weylcomp}.} $A_{\mu}$, an $SU(2)_R$ triplet of scalar fields $ Y_{ij}$,
and  an $SU(2)$ doublet of chiral
fermions $\Omega_i$, i.e.  $(X,  \Omega_i,  A_{\mu}, Y^{ij})$, where 
$Y_{ij}$ is a symmetric
matrix satisfying the reality condition
\begin{equation}
Y_{ij} = \varepsilon_{ik} \, \varepsilon_{jl} \, Y^{kl} \;\;\;,\;\;\; Y^{ij} = (Y_{ij} )^* \;.
\end{equation}
Here, 
$i=1,2$ is an $SU(2)_R$ index.
Thus, off-shell, an abelian vector multiplet has
eight bosonic and eight fermionic real degrees of freedom.

The component fields of a vector multiplet carry a  Weyl weight $w$ and a chiral weight $c$.
This is summarized for the bosonic components in Table \ref{vechyp4d}.

\begin{table}[h]
\begin{center}
\begin{tabular}{|c||ccc||c|}
\hline
&
   \multicolumn{3}{c||}{vector multiplet}  &
   \multicolumn{1}{l|}{\parbox[]{1.8cm}{\begin{center}hyper- \linebreak 
multiplet \end{center}}} \\[-2mm]
\hline
\hline
field          &
   $X^I$         &
   $A_\mu^{\,I}$     &
   $Y_{ij}^{\,I}$    &
   $A_i^\alpha$   
     \\
     \hline
\hline 
$w$         & $1$  & $0$ & $2$ &
            $1$  \\[.5mm]
\hline
$c$         & $-1$ & $0$ & $0$ &
            $0$ \\[.5mm]
\hline
\end{tabular}\\[.13in]
\caption{Weyl and chiral weights ($w$ and $c$,
  respectively) of the vector and hypermultiplet bosonic component fields.}
\label{vechyp4d}
\end{center}
\end{table}

\subsubsection{Hypermultiplets \label{sec:hypmul}}

The bosonic degrees of freedom of $r$  hypermultiplets are described by $4r$ real scalar fields $\phi^A$ ($A = 1, \dots, 4 r$) that 
can be 
conveniently described in terms of 
local sections $A_i{}^{\alpha} (\phi)$ of an ${\rm Sp} (r) \times {\rm Sp} (1)$ bundle ($\alpha = 1, \dots, 2r; i = 1,2$) \cite{deWit:1999fp}.
In the main text we set $r = n_H + 1$.
The hypermultiplets provide one of the compensating multiplets for obtaining Poincar\'e supergravity. 
In this review, we will not be concerned with physical hypermultiplets, and hence set $n_H =0$.

The hyper-K\"ahler potential  $\chi$ and the covariant
derivative ${\cal D}_{\mu} A_i{}^{\alpha}(\phi)$ are defined by
\begin{eqnarray}
\varepsilon_{ij}\,\chi &=& {\bar  \Omega}_{\alpha\beta} A_i{}^{\alpha} A_j{}^{\beta}
\,,\nonumber\\  
{\cal D}_{\mu} A_i{}^{\alpha} &=& \partial_{\mu} A_i{}^{\alpha} - 
b_{\mu} A_i{}^{\alpha} +\ft12V_{\mu i}{}^jA_j{}^{\alpha}  +  
\partial_{\mu}\phi^A\,{\Gamma_A}^{\alpha}{}_{\beta} \, A_i{}^{\beta}\,,
\label{Dhyp4d}
\end{eqnarray}
in accordance with the Weyl weight given in Table \ref{vechyp4d}.
The connection $\Gamma_A{}^{\alpha}{}_{\beta}$ takes values in ${\rm sp} (n_H +1)$,
and ${\bar \Omega}_{\alpha \beta} $ is a covariantly constant antisymmetric tensor
\cite{deWit:1999fp}.

\subsection{Superconformal formalism in five dimensions
\label{sec:5dscg}}

\subsubsection{Weyl multiplet}

The superconformal approach to ${\cal N}=2$ supergravity in five space-time dimensions
\cite{Bergshoeff:2001hc,Fujita:2001kv,Bergshoeff:2004kh,Hanaki:2006pj,Butter:2017pbp}
 is based
on the Weyl multiplet. In its standard formulation, 
the Weyl multiplet in five dimensions is a supermultiplet with 32+32 bosonic and fermionic off-shell
degrees of freedom. When reduced to four space-time dimensions  \cite{Banerjee:2011ts}, it decomposes into
the Weyl multiplet in four dimensions with 24+24  bosonic and fermionic off-shell
degrees of freedom, and a  vector multiplet with 8+8 bosonic and fermionic off-shell
degrees of freedom.

The algebra underlying the superconformal approach is the
${\cal N}=2$ superconformal algebra. In five dimensions, 
this superalgebra contains the bosonic generators
$P_a, M_{ab}, K_a, D, U_i{}^j$
associated with translations, Lorentz transformations, special
conformal transformations, dilations and 
$SU(2)_R$ R-symmetry transformations, respectively. One assigns a local parameter and a
gauge field to each of these bosonic generators. The 
gauge fields associated with $SU(2)_R$ R-symmetry transformations will be 
denoted by $ {\cal V}_{\mu}{}^i{}_{j} $, which is an anti-hermitian, traceless matrix in the indices $i,j$.
This is summarized
in Table \ref{conf_gen_fie_N2_5d} below.\\

\begin{table}[h]
\begin{center}
\begin{tabular}{|c||ccccc|}
\hline
bosonic generator        &
   $P_a$         &
   $M_{ab}$      &
   $K_a$     &
   $D$ 
   &
   $U^j{}_i$ 
     \\
   \hline
\hline 
parameter       & $\xi^a$ & $\lambda^{ab}$  & $\lambda_K^a$ & $\lambda_D$ & $\lambda{}^i{}_j $ \\[.5mm]
\hline
gauge field        & $e_{\mu}{}^a$ & $\omega_{\mu}^{ab}$ & $f_{\mu}{}^a$ & $b_{\mu}$ &  
${\cal V}_{\mu}{}^i{}_{ j}$
            \\[.5mm]
\hline
\end{tabular}
\vskip 5mm
\caption{The ${\cal N}=2$ bosonic subalgebra: generators, local parameters, gauge fields.
\label{conf_gen_fie_N2_5d}}
\end{center}
\end{table}

The bosonic components of the Weyl multiplet are given by the gauge fields displayed in Table \ref{tabledof5d}
together with a real anti-symmetric tensor field $T_{ab}$ and a real scalar field $D$:
\begin{equation}
(e_{\mu}{}^a, 
\omega_{\mu}^{ab}, f_{\mu}{}^a, b_{\mu}, {\cal V}_{\mu}{}^i{}_{j}, T_{ab}, D) \;.
\end{equation}
These describe $32$ independent bosonic degrees of freedom.

\begin{table}[h]
\begin{center}
\begin{tabular}{|c||c||c|}
\hline
field & subtraction by gauge transformations & number of degrees of freedom left \\[.5mm]
\hline
\hline
$e_{\mu}{}^a$        & $P_a, M_{ab}, D$ & $25 -(5 + 10 + 1) = 9 $ \\[.5mm]
\hline
$\omega_{\mu}{}^{ab}$ & &  composite field         \\[.5mm]
\hline
$f_{\mu}{}^a$ &  & composite field \\ [.5mm]
\hline
$b_{\mu}$       & $K_a$& 0
 \\[.5mm]
\hline
${\cal V}_{\mu}{}^i{}_{j} $  & $SU(2)_R$ &  $15 - 3 = 12$  \\[.5mm]
\hline
$ T_{ab}$ &  & 10 \\ [.5mm]
\hline
$ D $  &  & 1 \\ [.5mm]
\hline
\end{tabular}\\ [.13in]
\caption{Counting of bosonic off-shell degrees of freedom: $9 + 12 + 10 + 1 =32$.
\label{tabledof5d}}
\end{center}
\end{table}

The component fields of the Weyl multiplet carry a Weyl weight $w$.
This is summarized for the bosonic components in Table \ref{tablewcWeyl_5d}. \\

\begin{table}[h]
\begin{center}
\begin{tabular}{|c||ccccc|cc|}
\hline
field          &
   $e_\mu{}^a$   &
     $b_\mu$       &
    ${\cal V}_{\mu}{}^i{}_{ j}$ &
   $T_{ab}$    &
    $D$           &
   $\omega_\mu{}^{ab}$ &
   $f_\mu{}^a$  
 \\[.5mm]
\hline
\hline     
$w$         & $-1$    
& $0$          & $0$
  & $1$      
     & $2$      & $0$       & $1$    
   \\[.5mm]
   \hline
  \end{tabular}\\[.13in]
\caption{Weyl weights
                    of the Weyl multiplet bosonic component fields \cite{deWit:2009de}.
             \label{tablewcWeyl_5d}}
\end{center}
\end{table}

As indicated in Table \ref{tabledof5d}, 
the gauge fields  $\omega_\mu{}^{ab}$ 
and $f_\mu{}^a$ are composite fields. Their expressions are obtained by imposing 
constraints on the associated field strengths $R_{\mu \nu}{}^a (P) $ and 
$R_{\mu \nu}{}^{ab} (M)$ \cite{deWit:2009de},\footnote{Note that our definition of $ {R}_{\mu \nu}{}^{ab} (M) $ differs from 
the one in \cite{deWit:2009de} by an overall minus sign.}
\begin{eqnarray}
\label{confconst5d}
R_{\mu \nu}{}^a (P) &=& 2 {\cal D}_{[\mu} e_{\nu]}{}^a 
= 
0 \;, \\
e_a{}^{\mu} \, {R}_{\mu \nu}{}^{ab} (M) &=& 
e_a{}^{\mu} \left(2 \partial_{[\mu} \omega_{\nu]}{}^{ab} + 2 \omega_{[\mu}{}^{ac} \omega_{\nu ]c}{}^b +
8 e_{[\mu}{}^{[a} f_{\nu]}{}^{b]}
\right)
=
0 \;. \nonumber
\end{eqnarray}
Here,  the covariant derivative ${\cal D}_{\mu}$ of a vector $V^a$ of Weyl weight $w$  is
\begin{equation}
{\cal D}_{\mu} V^a = \partial_{\mu} V^a - w \, b_{\mu} V^a + \omega_{\mu}{}^{ab} V_b \;.
\end{equation}
We infer from \eqref{confconst5d},
\begin{eqnarray}
f_a{}^a &=& - \frac{1}{16} \, R \;, \nonumber\\
f_{\mu}{}^a &=& \frac16 \left(
- R_{ \mu}{}^a + \frac18 \, e_{\mu}{}^a \, R 
  \right) \;.
\label{f-FP} 
\end{eqnarray}


\begin{table}[h]
\begin{center}
\begin{tabular}{|c||ccc||c|}
\hline
&
   \multicolumn{3}{c||}{vector multiplet}  &
   \multicolumn{1}{l|}{\parbox[]{1.8cm}{\begin{center}hyper- \linebreak 
multiplet \end{center}}} \\[-2mm]
\hline
\hline
field          &
   $\sigma^I$         &
     $A_\mu^{\,I}$     &
   $Y_{ij}^{\,I}$    &
   $A_i{}^\alpha$   
   \\
   \hline
\hline 
$w$         & $1$ &
$0$ & $2$ &
            $\ft32$
             \\[.5mm]
\hline
\end{tabular}\\[.13in]
\caption{Weyl weights $w$ 
  of the vector and hypermultiplet bosonic component fields.
\label{vechyp} }
\end{center}
\end{table}

\subsubsection{Vector multiplets}

The field content of a five-dimensional abelian vector multiplet is given by
a real scalar field $\sigma$, an abelian gauge field $A_{\mu}$, an $SU(2)_R$ triplet of scalar fields $ Y_{ij}$,
and an $SU(2)_R$ doublet of symplectic Majorana fermions
$\lambda^i$, 
i.e. $(\sigma, \lambda^i, A_{\mu}, Y^{ij})$, where
$Y^{ij}$ is a symmetric matrix satisfying the reality condition
\begin{equation}
Y_{ij} = \varepsilon_{ik} \, \varepsilon_{jl} \, Y^{kl} \;\;\;,\;\;\; Y^{ij} = (Y_{ij} )^* \;.
\end{equation}
Here, 
$i=1,2$ is an $SU(2)_R$ index.
Thus, off-shell, an abelian vector multiplet has
eight bosonic and eight fermionic real degrees of freedom.

The component fields of a vector multiplet carry a Weyl weight $w$.
This is summarized for the bosonic components in Table \ref{vechyp}.

\subsubsection{Hypermultiplets}

As we mentioned in \ref{sec:hypmul}, 
the bosonic degrees of freedom of $r$  hypermultiplets are described by $4r$ real scalar fields $\phi^A$ ($A = 1, \dots, 4 r$) that 
can be 
conveniently described in terms of 
local sections $A_i{}^{\alpha} (\phi)$ of an ${\rm Sp} (r) \times {\rm Sp} (1)$ bundle ($\alpha = 1, \dots, 2r; i = 1,2$) \cite{deWit:1999fp}.
In the main text we set $r = n_H + 1$.

In five space-time dimensions, 
the hyper-K\"ahler potential  $\chi$ and the covariant
derivative ${\cal D}_{\mu} A_i{}^{\alpha}(\phi)$ are defined by
\begin{eqnarray}
\varepsilon_{ij}\,\chi &=& \Omega_{\alpha\beta} A_i{}^{\alpha} A_j{}^{\beta}
\,,\nonumber\\  
{\cal D}_{\mu} A_i{}^{\alpha} &=& \partial_{\mu} A_i{}^{\alpha} - \tfrac32
b_{\mu} A_i{}^{\alpha} +\ft12V_{\mu i}{}^jA_j{}^{\alpha}  +  
\partial_{\mu}\phi^A\,{\Gamma_A}^{\alpha}{}_{\beta} \, A_i{}^{\beta}\,,
\label{Dhyp5d}
\end{eqnarray}
in accordance with the Weyl weight given in Table \ref{vechyp}.
The connection $\Gamma_A{}^{\alpha}{}_{\beta}$ takes values in ${\rm sp} (n_H +1)$,
and $\Omega_{\alpha \beta} $ is a covariantly constant antisymmetric tensor \cite{deWit:2009de}.

\subsection{Special holomorphic coordinates}
\label{specialgeo}

As discussed in subsection \ref{sec:PSR},
the PSK manifold $( {\bar M}, g_{\bar M})$ can be
obtained by a superconformal
quotient of a regular CASK manifold $(M, g_M)$.
As mentioned in 
subsection \ref{sec:formlinebund}, one may choose special holomorphic coordinates
$z^a = X^a/X^0 $ ($a = 1, \dots, n$) on the PSK manifold $( {\bar M}, g_{\bar M})$.
Here, we provide a few more details on the relation of these coordinates to the special holomorphic 
coordinates $X^I$ ($I = 0, \dots, n$) on the CASK manifold $(M, g_M)$.
We give various conversion formulae that facilitate the construction of the space-time two-derivative
Lagrangian for the $z^a$ when viewed
as components of a map ${\cal Z}: N \rightarrow {\bar M}$
from space-time $N$ into the PSK manifold ${\bar M}$.

The superconformal quotient proceeds by
first restricting the $X^I$ to the hypersurface
\begin{equation}
i \left( {\bar X}^I \, F_I -  {\bar F}_I \, X^I \right) = 1 \;.
\label{eq:einstein-norm}
\end{equation}
Setting
\begin{equation}
(X^I, F_I) = \frac{(X^I(z), F_I(z)) }{ || (X^I(z), F_I(z))|| } \;,
\end{equation}
the constraint \eqref{eq:einstein-norm} imposes that $(X^I, F_I)$ has unit norm.
Here
\begin{equation}
 || (X^I(z), F_I(z))|| = \sqrt{| i \left( {\bar X}^I (\bar z) \, F_I(z) -  {\bar F}_I(\bar z) \, X^I (z)\right) |} \;.
 \end{equation}
As discussed in subsection  \ref{sec:formlinebund}, the vector $(X^I(z), F_I(z))$ denotes the components of 
the holomorphic section $s^* \phi: {\bar M} \rightarrow {\cal U}^{\bar M}$
of the line bundle ${\cal U}^{\bar M} \rightarrow {\bar M}$, which depends holomorphically on $z^a$. 
The norm of the vector $(X^I(z), F_I(z))$  yields the 
K\"ahler potential $K(z, {\bar z})$ of $g_{\bar M}$,
\begin{equation}
{ e}^{- K(z , \bar z) } =  i \left( {\bar X}^I (\bar z) \, F_I(z) -  {\bar F}_I(\bar z) \, X^I (z)\right) \;,
\label{kaehpot}
\end{equation}
so that
\begin{equation}
X^I = {e}^{\tfrac12 K(z, {\bar z})} \, X^I (z) \;.
\end{equation}
The $\mathbb{C}^*$-action 
\begin{equation}
X^I(z) \mapsto {e}^{- f(z)} \, X^I (z) 
\label{eq:projectransf}
\end{equation}
 induces the K\"ahler transformation
\begin{equation}
K \mapsto K + f + \bar f 
\label{eq:kahlertransf}
\end{equation}
on the K\"ahler potential, while on the symplectic vector $(X^I, F_I (X))$
it induces the $U(1)$-transformation
\begin{equation}
(X^I, F_I (X)) \mapsto {\rm e}^{- \ft12 (f - \bar f)} \, (X^I, F_I (X)) \;.
\label{eq:U1transf}
\end{equation}

The K\"ahler potential \eqref{kaehpot} can be written as 
\begin{equation}
{ e}^{- K(z , \bar z) } = |X^0 (z)|^2 \, \left( - N_{IJ} \, Z^I \, {\bar Z}^J \right) \;,
\end{equation} 
with $Z^I (z) = (Z^0, Z^a) = (1, z^a)$, and 
\begin{equation}
N_{IJ} = -i \left( F_{IJ} - {\bar F}_{IJ} \right) \;.
\label{eq:N-big}
\end{equation}
Using the homogeneity of $F(X)$,
\begin{equation}
F(X) = \left(X^0\right)^2 \, {\cal F}(z) \;,
\end{equation}
we get
\begin{equation}
F_{0} = X^0 \left(2 {\cal F}(z) - z^a \, {\cal F}_a \right) \;,
\end{equation}
where ${\cal F}_a= \partial {\cal F}/\partial z^a$.  Using
\begin{eqnarray}
F_{00} &=& 2 {\cal F} - 2 z^a \, {\cal F}_a + z^a \, z^b\, {\cal F}_{ab} \;, \nonumber\\
F_{0b} &=& {\cal F}_b - z^a \, {\cal F}_{ab} \;, \nonumber\\
F_{ab} &=& {\cal F}_{ab} \;,
\label{eq:F-calF}
\end{eqnarray}
where ${\cal F}_{ab}= \partial^2 {\cal F}/\partial z^a \partial z^b$,
we obtain
\begin{equation}
 - N_{IJ} \, Z^I \, {\bar Z}^J = i \left[ 2\left( {\cal F} - {\bar {\cal F}} \right) - \left( z^a - {\bar z}^a \right)
 \left( {\cal F}_a + {\bar {\cal F}}_a \right) \right] \;, 
 \end{equation}
and hence
\begin{equation}
{e}^{- K(z , \bar z) } = i \, |X^0 (z)|^2 \,
\left[ 2\left( {\cal F} - {\bar {\cal F}} \right) - \left( z^a - {\bar z}^a \right)
 \left( {\cal F}_a + {\bar {\cal F}}_a \right) \right] \;.
\label{eq:kahler-pot} 
 \end{equation}
 The metric $g_{\bar M}$ on the PSK manifold is, locally, given by
 \begin{equation}
g_{a  \bar{b}} = \frac{\partial^2 K(z, \bar z)}{\partial z^a\, \partial {\bar z}^b} \;.
\label{eq:kaehler-metric}
\end{equation}

Next, we relate the PSK metric \eqref{eq:kaehler-metric} 
to the CASK metric \eqref{eq:N-big}.
Differentiating ${e}^{-K}$ yields 
\begin{eqnarray}
\label{eq:ddK}
\partial_a \partial_{\bar b} {e}^{- K} &=& 
[- g_{a \bar b} + 
\partial_a K \, \partial_{\bar b} K ]\, {e}^{-K} \\
&=& i \, |X^0 (z)|^2 \,
\left( {\cal F}_{ab} - {\bar {\cal F}}_{ab} \right) 
- \Big[\partial_a \ln X^0(z) \, \partial_{\bar b} \ln {\bar X}^0 (\bar z)
\nonumber\\
&& \qquad \qquad \qquad \qquad - \partial_a \ln X^0(z) \, \partial_{\bar b} K
- \partial_a K \, \partial_{\bar b} \ln {\bar X}^0 (\bar z) \Big] { e}^{-K} \;. \nonumber
\end{eqnarray}
Using \eqref{eq:F-calF} we have
\begin{equation}
N_{ab} = - i \left({\cal F}_{ab} - {\bar {\cal F}}_{ab} \right) \;,
\end{equation}
and hence we infer from \eqref{eq:ddK} that
\begin{equation}
g_{a \bar{b}} = N_{ab} \, |X^0|^2 + \frac{1}{|X^0 (z)|^2} {\cal D}_a X^0(z) {\cal D}_{\bar b} \, {\bar X}^0 (\bar z) \;,
\label{eq:relgNX0}
\end{equation}
where
\begin{equation}
{\cal D}_a X^0 (z)= \partial_a X^0(z) - i  A_a^h  \, X^0(z) = \partial_a X^0(z) + \partial_a K \, X^0(z)
\end{equation}
denotes the connection given in \eqref{connsecMbarM}, i.e. the 
covariant derivative under the transformation \eqref{eq:projectransf}.

Next, using the connection given in  \eqref{connunit}, 
\begin{equation}
{\cal D}_a X^I =  
 \partial_a X^I + \tfrac12 \partial_a K \, X^I \;,
\end{equation}
we introduce 
the space-time covariant derivative
\begin{equation}
{\cal D}_{\mu} X^I = \partial_{\mu} X^I + i {A}_{\mu} X^I = 
\partial_{\mu} X^I + \tfrac12 \left(\partial_a K \, \partial_{\mu} z^a - \partial_{\bar a} K
\partial_{\mu} {\bar z}^a \right) X^I \;,
\label{eq:kah-conn}
\end{equation}
which is a covariant derivative for $U(1)$ transformations \eqref{eq:U1transf}. 
Observe that
\begin{equation}
{\cal D}_{\mu} X^0 = {\rm e}^{K/2} \, {\cal D}_a X^{0}(z) \, \partial_{\mu} z^a \;.
\label{eq:DDX0}
\end{equation}
Now we evaluate the $U(1)$ invariant combination 
$N_{IJ} \, {\cal D}_{\mu} X^I \, {\cal D}^{\mu} {\bar X}^J$ subject to the constraint \eqref{eq:einstein-norm},
\begin{eqnarray}
&& N_{IJ} \, {\cal D}_{\mu} X^I \, {\cal D}^{\mu} {\bar X}^J |_{-N_{IJ} X^I {\bar X}^J = 1}
= |X^0|^2 \, N_{ab} \, \partial_{\mu} z^a \,
\partial^{\mu} {\bar z}^b
- \frac{1}{|X^0|^2} \, {\cal D}_{\mu} X^0 \,
{\cal D}^{\mu} \bar{X}^0 \nonumber\\
&& \qquad + \frac{X^0}{{\bar X}^0} \, N_{a J} {\bar X}^J \, \partial_{\mu} z^a\, {\cal D}^{\mu} {\bar X}^0
+ \frac{{\bar X}^0}{X^0} \, N_{I a} X^I \, \partial_{\mu} {\bar z}^a \, {\cal D}^{\mu} X^0 \;.
\end{eqnarray}
Using
\begin{equation}
X^0 \, {\bar X}^J \, N_{a J} = \frac{1}{X^0(z)} \, {\cal D}_a X^0(z) \;,
\end{equation}
as well as \eqref{eq:relgNX0} and \eqref{eq:DDX0}
we establish
\begin{equation}
N_{IJ} \, {\cal D}_{\mu} X^I \, {\cal D}^{\mu} {\bar X}^J |_{-N_{IJ} X^I {\bar X}^J = 1} = g_{a \bar b} \, \partial_{\mu} z^a \,
\partial^{\mu} {\bar z}^b \;.
\label{eq:relgN}
\end{equation}

We close with the following useful relations.
First, we note the relation \cite{deWit:1996ag}
\begin{equation}
N^{IJ} = g^{a \bar b} \, {\cal D}_a X^I \, \bar{\cal D}_{\bar b} {\bar X}^J  - X^I \, {\bar X}^J \
\;.
\end{equation}
Then, we recall the definition of ${\cal N}_{IJ}$ in \eqref{calN-N},
and we note the relations
\begin{eqnarray}
{\cal N}_{IJ} \, X^J &=& F_I \;, \nonumber\\
 - \tfrac12 \, \left[\left({\rm Im} {\cal N}\right)^{-1}\right]{}^{IJ}
& =&  
N^{IJ} + 
{ X}^I \, {\bar {X}}{}^J + {X}^J \, {\bar {X}}{}^I \;.
\label{eq:N-id-X}
\end{eqnarray}

\subsection{The black hole potential}

We consider the Maxwell terms in the two-derivative 
 Lagrangian \eqref{efflag_poincare}, and define
 $\mu_{IJ} =  {\rm Im}  \, {\cal N}_{IJ}$ and $\nu_{IJ} =  {\rm Re}  \, {\cal N}_{IJ}$.

The black hole potential in four dimensions is defined by \cite{Ferrara:1996dd},
\begin{equation}
V_{\rm BH} = g^{a \bar{b}} {\cal D}_a Z \bar{\cal D}_{\bar b} {\bar Z} + |Z|^2 = 
\left( N^{IJ} + 2 {X}^I \bar{{X}}^J \right) {\hat q}_I  \, \bar{\hat q}_J \;,
\label{bhpot1}
\end{equation}
where
\begin{equation}
Z(X) = p^I \, F_I (X) - q_I \, X^I = 
- \hat{q}_I \, X^I \;\;\;,\;\;\; \hat{q}_I = q_I - F_{IJ} \, p^J \;.
\label{ZXhatq}
\end{equation}
Here, $(p^I, q_I)$ denote magnetic/electric charges as in \eqref{chargesqp}.
The black hole potential transforms as  a function under symplectic transformations \eqref{FGdual}.

Using \eqref{eq:N-id-X},
the black hole potential can also be written as
\begin{equation}
 V_{\rm BH} =  - \frac{1}{2} \; (q_I - \mathcal{N}_{IK}\,p^K)\,
  [(\mathrm{Im}\,\mathcal{N})^{-1}]^ {IJ}\,  
  (q_J - \bar{\mathcal{N}}_{JL}\,p^L) \;.
  \label{bhpef}
\end{equation}
This equals  \cite{Ferrara:1996dd,Ferrara:1997tw}
 \begin{eqnarray}
V_{\rm BH} = -  \tfrac12 \left( p \quad q \right) 
\begin{pmatrix}
\mu + \nu \mu^{-1} \nu & -  \nu \mu^{-1} \\
- \mu^{-1} \nu  & \mu^{-1}\\
\end{pmatrix}
\begin{pmatrix}
p \\
 q \\
\end{pmatrix} \;,
\label{Vbhpq}
\end{eqnarray}
where we have suppressed the indices $I, J$ for notational simplicity. The black hole potential can be expressed 
\cite{Mohaupt:2011aa} in 
terms of the tensor field ${\hat H}_{ab}$ defined in \eqref{hatg},
 \begin{eqnarray}
V_{\rm BH} = -  \tfrac12 Q^a \, {\hat H}_{ab} \,  Q^b \;,
\end{eqnarray}
where $Q^a = (p^I, q_I)^T$.

Extrema of the black hole potential $V_{\rm BH}$  may either correspond to BPS black holes or to
non-BPS black holes. If an extremum satisfies
${\cal D}_a Z =0 \; \forall \, a=1, \dots, n$ with $Z \neq 0$, then it corresponds to a  BPS black hole \cite{Ferrara:1996dd}.
Conversely, if ${\cal D}_a Z \neq 0$ at the extremum, then the black hole
is non-supersymmetric.  

\subsection{Wald's entropy}

In a general classical theory of gravity with higher-curvature terms, based on a diffeomorphism invariant Lagrangian, the entropy of a stationary black hole 
is computed using 
Wald's definition of black hole entropy \cite{Wald:1993nt,Jacobson:1993vj,Iyer:1994ys,LopesCardoso:1999cv}. If the higher-curvature
terms involve the Riemann tensor, but not derivatives of the Riemann tensor, Wald's entropy is given by
\begin{equation}
{\cal S}_{\rm macro} = - \tfrac14 \int_{\Sigma_{\rm hor}} \frac{\partial L}{\partial R_{\mu \nu \rho \sigma} }\, \varepsilon_{\mu \nu} \varepsilon_{\rho \sigma} \;,
\label{entronoether}
\end{equation}
where $\varepsilon_{\mu \nu} $ denotes the bi-normal tensor associated with a cross-section of the Killing horizon ${\Sigma_{\rm hor}}$, normalized such that
$\varepsilon_{\mu \nu}  \varepsilon^{\mu \nu}  = -2$. In tangent space indices, the non-vanishing components are $\varepsilon_{01} = \pm 1$.
We have normalized \eqref{entronoether} in such a way that when $L = \tfrac12 \, R$, we obtain the area law ${\cal S}_{\rm macro} = A/4$.

\end{appendix}

\newpage 
\noindent

\end{document}